%% file: phdthesis.tex
\documentclass[11pt,oneside,onehalfspaced]{monash-thesis}

\usepackage{graphicx}
\usepackage{amsmath}
\usepackage{amssymb}
\usepackage{mathtools}
\usepackage{natbib}
\usepackage{subfigure}
\usepackage{tabu}
\usepackage{array}
\usepackage{setspace}
\usepackage{fancybox,fancyhdr}
\usepackage{epigraph}
\usepackage{rotating}
\usepackage{multirow}
\usepackage{pdfpages}
\usepackage{tabulary}

\usepackage{aas_macros}
\usepackage{mdwlist}
\usepackage{caption}
\usepackage{multirow}
\usepackage{adjustbox}

\allowdisplaybreaks

\defcitealias{pm05}{PM05}
\defcitealias{rj95}{RJ95}
\defcitealias{pf10}{PF10}

\fancypagestyle{declaration}{
\fancyhf{}
\fancyhead[R]{asdf}
\fancyhead[C]{}
\fancyhead[L]{}
\fancyfoot[R,L]{}
\fancyfoot[C]{\thispage}
}

\begin{document}

\title{\bf Simulating Astrophysical Magnetic Fields with Smoothed Particle Magnetohydrodynamics}
\submitdate{\today}
\degree{Doctor of Philosophy}
\school{School of Mathematical Sciences}
\faculty{Faculty of Science}
\author{Terrence Stanislaus Tricco}
\prevdegrees{B.Sc.(Hons), M.Sc.}

\titlepage

\begin{preliminary}

\maketitle

\tableofcontents

\include{copyright}

\include{abstract}

\include{publications}

\include{acknowledgments}

\listoffigures
\addcontentsline{toc}{chapter}{List of Figures}

\listoftables
\addcontentsline{toc}{chapter}{List of Tables}

\end{preliminary}

% Chapters

\include{introduction}

\include{spmhd}

\include{declaration-chapter3}
\include{cleaning}

\include{declaration-chapter4}
\include{switch}

\include{declaration-chapter5}

\include{turbcomp}

\include{conclusion}

\appendix

\include{cleaning-appendix}

\include{turbcomp-appendix1}

\include{turbcomp-appendix2}

% References

{
\singlespacing

\bibliographystyle{klunamed}
\bibliography{fullbib}
}

\end{document}

%% file: copyright.tex
\preface{Copyright Notice}

Under the Copyright Act 1968, this thesis must be used only under the normal conditions of scholarly fair dealing. In particular no results or conclusions should be extracted from it, nor should it be copied or closely paraphrased in whole or in part without the written consent of the author. Proper written acknowledgement should be made for any assistance obtained from this thesis.

I certify that I have made all reasonable efforts to secure copyright permissions for third-party content included in this thesis and have not knowingly added copyright content to my work without the owner's permission.

%% file: abstract.tex
\preface{Summary}

Numerical methods to improve the treatment of magnetic fields in smoothed field magnetohydrodynamics (SPMHD) are developed and tested. A mixed hyperbolic/parabolic scheme is developed which ``cleans'' divergence error from the magnetic field. The method introduces a scalar field which is coupled to the magnetic field. A conservative form for the hyperbolic equations is obtained by first defining the energy content of the new field, then using it in the discretised Lagrangian to obtain equations which manifestly conserve energy. This is shown to require conjugate first derivative operators in the SPMHD cleaning equations. Average divergence error is shown to be an order of magnitude lower for all test cases considered, and allows for the stable simulation of the gravitational collapse of magnetised molecular cloud cores. The effectiveness of the cleaning may be improved by explicitly increasing the hyperbolic wave speed or by cycling the cleaning equations between timesteps. In the latter, it is possible to achieve $\nabla \cdot {\bf B}=0$ in SPMHD. The method is adapted to work with a velocity field, demonstrating that it can reduce density variations in weakly compressible SPH simulations by a factor of 2.

A switch to reduce dissipation of the magnetic field from artificial resistivity is developed. Discontinuities in the magnetic field are located by monitoring jumps in the gradient of the magnetic field at the resolution scale relative to the magnitude of the magnetic field. This yields a simple yet robust method to reduce dissipation away from shocked regions. Compared to the existing switch in the literature, this leads to sharper shock profiles in shocktube tests, lower overall dissipation of magnetic energy, and importantly, is able to capture magnetic shocks in the highly super-Alfv\'enic regime. 

These numerical methods are compared against grid-based MHD methods by comparison of the small-scale dynamo amplification of a magnetic field in driven, isothermal, supersonic turbulence. We use the SPMHD code, {\sc Phantom}, and the grid-based code, {\sc Flash}. We find that the growth rate of {\sc Flash} is largely insensitive to the numerical resolution, whereas {\sc Phantom} shows a resolution dependence that arises from the scaling of the numerical dissipation terms. The saturation level of the magnetic energy in both codes is about $2$--$4\%$ of the mean kinetic energy, increasing with higher magnetic Reynolds numbers. {\sc Phantom} requires lower resolution to saturate at the same energy level as {\sc Flash}. The time-averaged saturated magnetic spectra have a similar shape between the two methods, though {\sc Phantom} contains twice as much energy on large scales. Both codes have PDFs of magnetic field strength that are log-normal, which become lopsided as the magnetic field saturates. We find encouraging agreement between grid- and particle methods for ideal MHD, concluding that SPMHD is able to reliably simulate the small-scale dynamo amplification of magnetic fields. We note that quantitative agreement on growth rates can only be achieved by including explicit, physical terms for viscosity and resistivity, because those are the terms that primarily control the growth rate and saturation level of the turbulent dynamo.

%% file: publications.tex
\preface{Declaration of Published Material}

In accordance with Monash University Doctorate Regulation 17.2 Doctor of Philosophy and Research MasterÕs regulations the following declarations are made:

I hereby declare that this thesis contains no material which has been accepted for the award of any other degree or diploma at any university or equivalent institution and that, to the best of my knowledge and belief, this thesis contains no material previously published or written by another person, except where due reference is made in the text of the thesis. 

This thesis includes 3 original papers published in or submitted to peer reviewed journals and 2 unpublished conference proceedings. The ideas, development and writing up of all the papers in the thesis were the principal responsibility of myself, the candidate, working within the School of Mathematical Sciences under the supervision of Dr. Daniel Price.

The inclusion of co-authors reflects the fact that the work came from active collaboration between researchers and acknowledges input into team-based research.

I have renumbered sections of submitted or published papers in order to generate a consistent presentation within the thesis.

The following chapters and/or sections have appeared in conference proceedings, have been published in peer reviewed journals, or have been submitted for publication:
\begin{itemize}
\item Sections~\ref{sec:hyperbolic}--\ref{sec:cleaningtests}, \ref{sec:cleaningsummary}, and Appendix~\ref{sec:appendix-cleaning-dissipation} appear in \\ Tricco, T.~S. and Price, D.~J.: 2012, `Constrained hyperbolic divergence cleaning for smoothed particle magnetohydrodynamics'. {\it J. Comput. Phys.} {\bf 231}, 7214--7236.
\item Section~\ref{sec:velclean} appears in \\ Tricco, T.~S. and Price, D.~J.: 2012, `Hyperbolic divergence cleaning for SPH'. {\it Proceedings of the 7th International SPHERIC Workshop.}
\item Sections~\ref{sec:formulation}, \ref{sec:switchtests}, and \ref{sec:switchsummary} appear in \\ Tricco, T.~S. and Price, D.~J.: 2013, `A switch to reduce resistivity in smoothed particle magnetohydrodynamics'. {\it MNRAS} {\bf 436}, 2810--2817.
\item Section~\ref{sec:switch-generalisation} appears in \\ Tricco, T.~S. and Price, D.~J.: 2013, `A switch for artificial resistivity and other dissipation terms', {\it Proceedings of the 8th International SPHERIC Workshop.}
\item Chapter~\ref{sec:chapter-mhdturb} and Appendices~\ref{sec:gridinterp} and \ref{sec:prandtl} to appear in \\ Tricco, T.~S., Price, D.~J., and Federrath, C.: 2014, `A comparison between grid and particle methods on small-scale dynamo amplification of magnetic fields in supersonic turbulence', {\it Submitted to ApJ}.
\end{itemize}

% left bottom right top
\hspace{3.5cm} \includegraphics[trim=0cm 3cm 0.6cm 23cm, clip, width=\textwidth, angle=-1.2, scale=0.8]{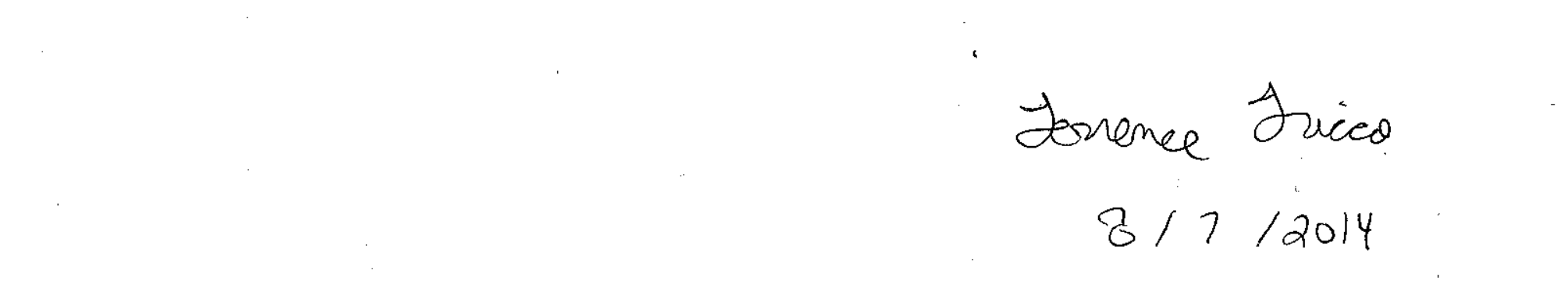}

%% file: acknowledgments.tex
\preface{Acknowledgments}

I remember the moment I realised I should go to Australia. I enjoyed working with SPMHD and thought there was a lot of room for innovation in that area. At first I hadn't considered doing a Ph.D. outside of Canada, but then the decision became obvious --- I had been reading primarily the work of Daniel Price, and he seemed like a good guy: I should do my Ph.D. with him.

Moving to Melbourne in general has really been an positive influence on my life. I've thoroughly enjoyed my time at Monash, in Melbourne, and in Australia. I have had more opportunities for my Ph.D. than I could have hoped for, and I have made a lot of great friends and had great experiences. I'm sad to see this chapter of my life coming to a close.

I first and foremost sincerely wish to thank my supervisor, Daniel Price. He has been an incredible teacher, and there are many anecdotes I would like to share to demonstrate that, but then this would end up becoming very long. Suffice to say, he has really taught me a lot --- directly and indirectly --- on what takes to be a good researcher. The support, encouragement, and understanding that he has given me exceeded what I could have expected. The opportunities that he has given me have made a significant impact on the quality and enjoyment of my Ph.D. life. Many times I wished to nominate him for supervisor of the year, but was unable to because the regulations for these awards required he had to have supervised a number of students already. He definitely deserves it.

On a practical note, I would like to acknowledge the financial support of the following: The Australian government for supporting my candidature through an Australian Postgraduate Award and an Endeavour International Postgraduate Research Scholarship. The Astronomical Society of Australia for their travel grant supporting my attendance at Protostars and Planets VI in Heidelberg, Germany. Monash University, the Faculty of Science, the School of Mathematical Sciences, and the Monash Centre for Astrophysics for their respective contributions for travel funding throughout my candidature. Daniel Price for supplying a top-up to my stipend scholarship and funding conference travel through his ARC Discovery Grant DP1094585.

Computational resources have been utilised at the Multi-modal Australian ScienceS Imaging and Visualisation Environment (MASSIVE) through the National Computational Merit Allocation Scheme supported by the Australian Government (project NCIdz3), gStar through Astronomy Australia Ltd's Astronomy Supercomputer Time Allocation Committee, and the Leibniz-Rechenzentrum (grant pr32lo).

Thank you to Joe Monaghan, Matthew Bate, Christoph Federrath, and Guillaume Laibe for their help throughout various stages of my Ph.D. I would also like to acknowledge my fellow Ph.D. students at Monash who have overlapped their time with mine, both ones finished and ones recently started. I would like to make particular mention of Tim Dolley, Nicolas Bonne, and David Palamara, the original trio who started at the same time as me. As well, Joelene Buntain and Hauke Worpel for being great officemates. 

Thank you to my friends and loved ones back home: Kate Murphy; Michael Healey, Stephen Hinchey; John Hawkin, Hugh Newman, Marek Bromberek, and the rest of the MUN physics crowd; Geoff Barnes, Deanna Norman, and the rest of the `newfs'; Sharon Griffin, Donna Acorn, and all of the Griffins; Jack \& Doreen Tricco, Doug Tricco, Sheila Wadland, and all of the Triccos. I've missed you guys a lot, and you've all helped in your own ways whether you've known it or not. Thank you to my brother, Jon, and a sincere thank you to my mother, Brenda. Reaching this point is in no small part due to her continual support. To my father, Paul, wish you were still around so you could see how far I've got.

%% file: introduction.tex
%%
%% Introduction chapter
%%

\chapter{Introduction}
\label{sec:chapter-introduction}

\setlength{\epigraphrule}{0pt}
\setlength{\epigraphwidth}{0.54\columnwidth}
\epigraph{... magnetic fields may be included without difficulty ...}{\citet{gm77}}

%Well, here we are.

Magnetic fields are ubiquitous throughout the Universe. It is believed that even if the Universe began unmagnetised, battery effects would lead to an initial magnetisation of baryonic matter (e.g., the Biermann battery, \citealt{biermann50}, see also the review by \citealt{widrowetal12}). Since magnetic monopoles do not exist in nature, there are no `sinks' of magnetic field and it is difficult to destroy them. Therefore, once an initial magnetisation is present, dynamo processes can lead to ever stronger magnetic fields.

Nearly all current theoretical problems in astrophysics involve magnetic fields to some degree. Neutron stars have some of the strongest magnetic fields in the Universe. The magnetic fields of galaxies are thought to be dynamically relevant for their evolution, and are responsible for determining the propagation of cosmic rays. The magnetic field of the Sun is responsible for sunspots and solar flares. 

Magnetic fields also play an important role in all stages of the star formation process. Stars form in cold (${\sim}10$K) clouds of molecular gas (primarily ${\rm H}_{2}$), which contain between $10^3$--$10^7$ $M_\odot$ of material. Supersonic turbulence in these clouds plays a key role in regulating star formation \citep[see review by][]{mo07}. As the supersonic shock waves collide in the cloud, they create dense filaments which act as the nucleation sites along which stars can form. These provide the dense cores that begin the star formation process \citep{larson81}. The extra pressure from magnetic fields help guard against gravitational collapse, and numerical studies has shown that this can reduce star formation rates \citep[e.g.,][]{nl08, pb08, pb09, pn11, fk12}.

On the scale of individual protostars, magnetic fields are responsible for driving jets and outflows --- a signature of star formation. As a molecular cloud core collapses under its gravitational weight, conservation of angular momentum leads to an increase in angular velocity, winding up magnetic field lines. There are two ways in which the magnetic field may drive an outflow. One occurs when the tension in the field lines becomes too strong, driving material outwards as the magnetic field pops out of the plane of the disc \citep[the `magnetic tower',][]{lynden-bell96, lynden-bell03}. The other is when material is centrifugally accelerated along poloidal magnetic field lines, essentially being `sling shotted' away from the protostar \citep{bp82}. Outflows are important sources of removing angular momentum from the star-disc system and in reducing the efficiency of gas conversion into stars.

Magnetic fields also play an important role in the accretion discs around young stars. Magnetised, differentially rotating flows are well known to be susceptible to the magneto-rotational instability \citep{bh91}. Consider two pieces of material on nearby orbits that are joined by a magnetic field line. As they drift apart, the tension in the magnetic field line resists the motion. This pulls the two pieces towards each other, causing the material in the inner orbit to slow down, and the material in the outer orbit to speed up. However, this only causes the inner material to drop to a lower orbit, and the outer material to drift outwards, exacerbating the problem. By this process, angular momentum is transported outwards through the disc. This instability leads to turbulence, and is thought to play a key role in driving accretion onto young stars.

Observations of magnetic fields may be obtained directly through Zeeman splitting measurements of spectral lines, or indirectly by the linear polarisation of thermal emission from dust grains. However, Zeeman measurements only yield information about the magnetic field along the line of sight, and the polarisation of dust grains only about its orientation in the plane of the sky. Therefore, the full information about the magnetic field is difficult to obtain. Furthermore, performing these observations may require a considerable amount of time, for example, \citet{tc08}'s survey of Zeeman measurements of magnetic fields in molecular cloud cores involved {$\sim$}500 hours of observing time.

An important approach to test astrophysical theories is through the use of numerical simulation, and it is crucial that these numerical experiments reflect reality as closely as possible in order to yield meaningful results. This is accomplished through careful design and calibration of numerical methods.

This thesis is focused on improving the treatment of magnetic fields in smoothed particle magnetohydrodynamics (SPMHD), a Lagrangian particle based numerical method built on smoothed particle hydrodynamics (SPH). The general picture of SPH is to solve the equations of hydrodynamics by discretising a fluid into a collection of particles that mimic fluid behaviour. SPH has many advantages for astrophysics. One, the resolution is tied to the mass. Regions of higher mass have more particles, thus more resolution, which is advantageous as the densest areas are typically the most interesting (e.g., stars forming in a molecular cloud). Two, it is trivial to incorporate gravitational N-body methods since SPH is a particle based scheme. Three, advection is done perfectly, that is, without any dissipation, since it is a Lagrangian method. Four, it can easily handle complex geometries. Five, the Courant timestep does not depend upon the fluid velocity, thus allowing larger timesteps. And six, perhaps its strongest attribute, it has exact simultaneous conservation of mass, momentum, angular momentum, energy, and entropy to the precision of the time-stepping algorithm. This makes SPH significantly robust and stable since it reflects the conservation properties of nature.

SPH is widely used in astrophysics for the preceding reasons. For example, the cosmological code {\sc Gadget 2} \citep{gadget2} has over 1900 citations at the time of writing. The impetus to include other physics in SPH, such as magnetic fields, is clear. The foundation of SPMHD has been laid substantially through the Ph.D. research of Daniel Price (see \citealt{pm04a, pm04b, pm05} and also the recent review by \citealt{price12}), building on the earlier work of \citet{pm85} and \citet{morris96}. This thesis follows as its spiritual successor, shoring up the remaining deficiencies to build a method that is able to accurately simulate a wide range of astrophysical problems. 

The thesis is structured as follows: In Chapter~\ref{sec:chapter-spmhd}, the current state of SPMHD is reviewed. First, the continuum equations of ideal MHD are derived, which is more than mere exercise as this will elucidate some of the numerical issues to be discussed. The numerical method is built up step-by-step. We provide an overview of how to estimate the density for the set of SPH particles, how to adaptively set the resolution based on density, how to perform basic interpolation of quantities, and how to obtain the discretised form of the induction, energy, and momentum equations. The numerical instability present in the equations of motion will be discussed, along with strategies for its treatment. Methods for the capturing of shocks and discontinuities are outlined.

In Chapter~\ref{sec:chapter-cleaning}, the constrained hyperbolic divergence cleaning method to uphold the divergence-free constraint of the magnetic field in SPMHD is developed and tested. Past approaches to treat the divergence-free constraint in SPMHD are summarised first. In Chapter~\ref{sec:chapter-switch}, a method to reduce numerical dissipation of the magnetic field is presented and tested. Chapter~\ref{sec:chapter-mhdturb} presents a comparison of the SPMHD methods developed against grid-based methods on the simulation of small-scale dynamo amplification of a magnetic field. The thesis is summarised in Chapter~\ref{sec:chapter-conclusion}.

%% file: spmhd.tex
%%
%% Background chapter
%%

%% ``...magnetic fields may be included without difficulty...'' Gingold, Monaghan 1977.

\chapter{Smoothed particle magnetohydrodynamics}
\label{sec:chapter-spmhd}

Smoothed particle magnetohydrodynamics (SPMHD) is a numerical method for solving the equations of magnetohydrodynamics (MHD) based on the smoothed particle hydrodynamics (SPH) method \citep{gm77, lucy77}. The basic premise is to discretise the fluid by mass into a set of particles. To recover continuum behaviour from the collection of point particles, a weighting kernel is used to {\it smooth} their quantities over a local volume. Fluid properties can be reconstructed at any point in space by summing the weighted contributions of nearby particles. 

The first attempts to include magnetic fields in SPH were performed by \citet{gm77} who considered magnetic polytropes, though in a form which did not conserve momentum or angular momentum. The basic SPMHD method has its roots in the work by \citet{pm85}, who formulated equations of motion that conserve momentum, and applied the method to three-dimensional simulations of gravitationally collapsing gas clouds \citep{phillips86a, phillips86b}. The modern SPMHD method was developed by \citet{pm04a, pm04b, pm05}, who constructed fully conservative equations incorporating varying resolution, formulated magnetic shock capturing terms, and investigated approaches to treat the divergence-free constraint on the magnetic field. Since then, SPMHD has been applied to studies of protostar formation \citep{pb07, burzleetal11b, burzleetal11a, ptb12, btp14}, star cluster formation \citep{pb08, pb09}, neutron star mergers \citep{pr06}, and magnetic fields in galaxies and galaxy clusters \citep{pb08, donnertetal09, kotarbaetal09, kotarbaetal10, kotarbaetal11, bonafedeetal11, becketal12, becketal13}.

Technical difficulties in SPMHD arise from the divergence-free constraint on the magnetic field -- an issue faced by any numerical MHD method. Magnetic monopoles are introduced if this constraint is not upheld, which is not only physically inaccurate, but leads to spurious monopole accelerations. This causes numerical instability when it exceeds the isotropic pressure. The details of the issues surrounding $\nabla \cdot {\bf B}=0$ in SPMHD will be discussed as the method is presented throughout this chapter and in Chapter~\ref{sec:chapter-cleaning}.

We begin by deriving the continuum equations of ideal MHD along with the MHD wave solutions. This process is instructive and leads to further understanding of some of the finer points of the numerical scheme. The SPMHD discretised version of the MHD equations will then be constructed. It begins with a method to estimate density in SPH, and a review of basic SPH interpolation theory. With this, the discretised induction equation used to evolve the magnetic field can be obtained. Using the density estimate and discretised induction equation, the conservative equations of motion are built through a Lagrangian approach. The instability present in these equations will be discussed, along with approaches for removing it, making particular note of how it is related to the divergence-free constraint of the magnetic field. Dissipation terms for capturing shocks are presented, followed by methods to reduce dissipation. 

\section{Continuum magnetohydrodynamics}

Ideal MHD is the merger of fluid dynamics with electromagnetic theory. Several useful textbooks for magnetohydrodynamic theory are \citet{choudhuri98, griffiths99, batchelor00, bellan06}. The relevant equations are given by Euler fluid flow, describing the motion of an inviscid fluid, 
\begin{align}
 \frac{\partial \rho}{\partial t} &= - \nabla \cdot (\rho {\bf v}) , \\
 \frac{\partial {\bf v}}{\partial t} &= - ({\bf v} \cdot \nabla) {\bf v} - \frac{\nabla P}{\rho} , \label{eq:continuummomeq}
\end{align}
Maxwell's equations of electromagnetism,
\begin{align}
\nabla \times {\bf E} &= - \frac{\partial {\bf B}}{\partial t} , \label{eq:gausslaw} \\
\nabla \cdot {\bf E} &= \frac{\tau}{\epsilon_0} , \\
\nabla \times {\bf B} &= \mu_0 \left( {\bf J} + \epsilon_0 \frac{\partial {\bf E}}{\partial t} \right) , \label{eq:ampereslaw} \\
\nabla \cdot {\bf B} &= 0 ,
\end{align}
and the Lorentz force law,
\begin{equation}
\rho \frac{\partial {\bf v}}{\partial t} = \tau {\bf E} + {\bf J} \times {\bf B} . \label{eq:lorentzforce}
\end{equation}
Here, ${\bf v}$ is the fluid velocity, $\rho$ is the density, $P$ is the thermal pressure, ${\bf E}$ is the electric field, ${\bf B}$ is the magnetic field, $\tau$ is the charge density, ${\bf J}$ is the current density, $\mu_0$ is the permeability of free space, and $\epsilon_0$ is the permittivity. 

One key assumption is made: The fluid is highly ionised. This means that while the fluid can carry a magnetic field, on macroscopic scales (relevant for astrophysical systems), the positive and negative charges will average out and the fluid will be electrically neutral. Furthermore, since there is a significant number of free electrons, the fluid can be treated as an ideal conductor. Both of these conditions imply that the stationary electric field inside the fluid can be treated as negligible.

\subsection{Momentum equation}
\label{sec:mhdmomeq}

Forces from the magnetic field are due to the Lorentz force law (Equation~\ref{eq:lorentzforce}). Assuming that ${\bf E} = 0$ and does not vary with time, we can use Equation~\ref{eq:ampereslaw} to define the current density in terms of the magnetic field, such that the Lorentz force becomes
\begin{equation}
 \frac{\partial {\bf v}}{\partial t} = \frac{1}{\mu_0 \rho} (\nabla \times {\bf B}) \times {\bf B} .
\end{equation}
This may be rewritten as
\begin{equation}
 \frac{\partial {\bf v}}{\partial t} = - \frac{1}{2 \mu_0 \rho} \nabla B^2 + \frac{1}{\mu_0 \rho} ({\bf B} \cdot \nabla) {\bf B} , \label{eq:lorentzforcerewrite}
\end{equation}
from which it becomes clear that the magnetic field exerts two forces on the fluid. One is an isotropic magnetic pressure, which pushes fluid down gradients of magnetic field strength. The second is an attractive force directed along magnetic field lines, which functions like a tension in the magnetic field lines.

The total force on the fluid is the combination of pressure and magnetic forces. The momentum equation is therefore the addition of Equation~\ref{eq:continuummomeq} and \ref{eq:lorentzforcerewrite}, yielding
\begin{equation}
\frac{{\rm d}{\bf v}}{{\rm d}t} = - \frac{1}{\rho} \nabla \left( P + \frac{B^2}{2 \mu_0}\right) + \frac{1}{\mu_0 \rho} ({\bf B} \cdot \nabla) {\bf B} .\label{eq:momeqcontinuum}
\end{equation}
Here we have introduced the material derivative, ${\rm d}/{\rm d}t = \partial / \partial t + ({\bf v} \cdot \nabla) {\bf v}$, which follows the frame of reference of a parcel of fluid along its streamline. As SPH is a particle based method, it is natural to write equations using the material derivative.

The momentum equation can be written in terms of a stress tensor. Assuming that the magnetic field is divergence-free, the stress tensor can be defined as
\begin{equation}
 S^{ij} = - \delta^{ij} \left(P + \frac{B^2}{2 \mu_0}\right) + \frac{B^i B^j}{\mu_0} ,
\end{equation}
which leads to
\begin{equation}
 \frac{{\rm d} v^i}{{\rm d} t} = \frac{1}{\rho} \frac{\partial S^{ij}}{\partial x^j} .
\end{equation}
Expanding this, the momentum equation becomes
\begin{equation}
 \frac{{\rm d} {\bf v}}{{\rm d} t} = - \frac{1}{\rho} \nabla \left( P + \frac{B^2}{2 \mu_0}\right) + \frac{1}{\rho} \nabla \cdot \left( \frac{ {\bf BB}}{\mu_0} \right) . \label{eq:momeqstressexpanded}
\end{equation}

This is similar to Equation~\ref{eq:momeqcontinuum} except in the magnetic tension term. It contains an extra tensional force, ${\bf B} ( \nabla \cdot {\bf B} ) / \mu_0 \rho$, which appears due to the assumption that $\nabla \cdot {\bf B}=0$. The conservative form of the SPMHD momentum equation is obtained by using the stress tensor, though since it may be unsafe to assume the magnetic field is divergence-free when solving the equations numerically, this extra force term requires careful consideration. This is discussed in Sections~\ref{sec:spmhd-mom} and \ref{sec:spmhd-tensileinstability}.

\subsection{Induction equation}

The current density may be defined using Ohm's law, 
\begin{equation}
 {\bf J}' = \sigma {\bf E}' ,
\end{equation}
which expresses the current density, ${\bf J}'$, in terms of the electric field in the co-moving frame of an observer and the electrical conductivity, $\sigma$, of the material, which is treated as constant. In a fixed frame of reference, the electric field is given by
\begin{equation}
 {\bf E}' = {\bf E} + {\bf v} \times {\bf B} ,
\end{equation}
giving Ohm's law as
\begin{equation}
 {\bf J} = \sigma \left( {\bf E} + {\bf v} \times {\bf B} \right) , \label{eq:ohmslaw}
\end{equation}
which is the combination of current induced by an electric field and by moving through a magnetic field. This permits the electric field in to be expressed as
\begin{equation}
 {\bf E} = - {\bf v} \times {\bf B} + \frac{{\bf J}}{\sigma} . \label{eq:electricfield}
\end{equation}
Taking the curl of Equation~\ref{eq:electricfield}, the electric field may be replaced using Equation~\ref{eq:gausslaw} to obtain an evolution equation for the magnetic field as
\begin{equation}
 \frac{\partial {\bf B}}{\partial t} = \nabla \times ({\bf v} \times {\bf B}) - \frac{1}{\sigma} \nabla \times {\bf J} .
\end{equation}
In the limit of ideal MHD (infinite conductivity, $\sigma \to \infty$), the term involving the current density drops out. For non-ideal MHD, ${\bf J}$ may be replaced using Equation~\ref{eq:ampereslaw} (neglecting the displacement current, $\partial {\bf E} / \partial t$), obtaining
\begin{equation}
 \frac{\partial {\bf B}}{\partial t} = \nabla \times \left( {\bf v} \times {\bf B} \right) - \eta \nabla \times \left( \nabla \times {\bf B} \right) , \label{eq:induction1}
\end{equation}
where $\eta = 1 / \sigma \mu_0$ is the magnetic resistivity. 

In ideal MHD, the conductivity of the fluid is taken to be infinite, therefore $\eta=0$. Expanding the first term of Equation~\ref{eq:induction1}, we can write
\begin{equation}
 \frac{\partial {\bf B}}{\partial t} = - ({\bf v} \cdot \nabla) {\bf B} - (\nabla \cdot {\bf v}) {\bf B} + ({\bf B} \cdot \nabla) {\bf v} + (\nabla \cdot {\bf B}) {\bf v} ,
\end{equation}
or by using $\nabla \cdot {\bf B}=0$ and the material derivative,
\begin{equation}
 \frac{{\rm d} {\bf B}}{{\rm d} t} = ({\bf B} \cdot \nabla) {\bf v} - (\nabla \cdot {\bf v}) {\bf B} .
\end{equation}
The first term affects the magnetic field through shearing motion, while the second will increase the magnetic field when undergoing compression.

\subsection{Summary of ideal MHD equations}
\label{sec:ideal-mhd-summary}

The concise set of ideal MHD equations to be solved are
\begin{align}
\frac{{\rm d}\rho}{{\rm d}t} &= - \rho \nabla \cdot {\bf v} \label{eq:mhdcty} , \\
\frac{{\rm d}{\bf v}}{{\rm d}t} &= - \frac{1}{\rho} \nabla \left( P + \frac{B^2}{2 \mu_0} \right) + \frac{1}{\rho} \nabla \cdot \left( \frac{ {\bf BB}}{\mu_0} \right) \label{eq:mhdmomentumeqn} , \\
\frac{{\rm d}{\bf B}}{{\rm d}t} &= - \left( {\bf B} \cdot \nabla \right) {\bf v} + {\bf B} \left( \nabla \cdot {\bf v} \right) \label{eq:induction} , \\
\nabla \cdot {\bf B} &= 0 . \label{eq:divbconstraint}
\end{align}

\subsection{Wave solutions}
\label{sec:mhdwaves}

The ideal MHD wave equations permit three wave modes, not just sound waves as found in simple fluids (i.e., Euler fluids). Understanding the ideal MHD wave solutions will be useful when introducing shock capturing schemes to our numerical method.

The ideal MHD wave modes can be obtained as follows. Assume a uniform density fluid at rest with a constant magnetic field. The equation of state is taken to be isothermal, $P = c_{\rm s}^2 \rho$, where $c_{\rm s}$ is the speed of sound. Small perturbations are introduced to the density, velocity, and magnetic fields such that
\begin{align}
\rho &= \rho_0 + \delta \rho , \label{eq:mhdwaves-perturb-rho} \\
{\bf v} &= \delta {\bf v} , \label{eq:mhdwaves-perturb-v} \\
{\bf B} &= {\bf B}_0 + \delta {\bf B} , \label{eq:mhdwaves-perturb-B}
\end{align}
where $\rho_0$ and ${\bf B}_0$ are the background density and magnetic fields, with $\delta \rho$, $\delta {\bf v}$, and $\delta {\bf B}$ perturbations to each field. The perturbations are taken to be sufficiently small so as to not disturb the equilibrium values of the fluid, thus the background fields remain constant in time. Inserting Equations~\ref{eq:mhdwaves-perturb-rho}--\ref{eq:mhdwaves-perturb-B} into the ideal MHD equations, the set of linearised equations are then
\begin{align}
 \frac{\partial \delta \rho}{\partial t} &= - \rho_0 \nabla \cdot \delta {\bf v} , \label{eq:mhdwaves-linearised-rho} \\
 \frac{\partial \delta {\bf v}}{\partial t} &= - \frac{c_{\rm s}^2 \nabla \delta \rho}{\rho_0} + \frac{1}{\mu_0 \rho_0} (\nabla \times \delta {\bf B} ) \times {\bf B}_0 , \label{eq:mhdwaves-linearised-v} \\
 \frac{\partial \delta {\bf B}}{\partial t} &= \nabla \times (\delta {\bf v} \times {\bf B}_0) . \label{eq:mhdwaves-linearised-B}
\end{align}
Second order effects are assumed negligible so those terms involving multiple perturbations are discarded.

We assume that the perturbations have wave-like solutions of the form $\exp({\it i} {\bf k} \cdot {\bf r} - {\it i} \omega t)$, where $k$ is the wave vector. Equations~\ref{eq:mhdwaves-linearised-rho} and \ref{eq:mhdwaves-linearised-v} become
\begin{align}
 \omega \delta \rho &= \rho_0 {\bf k} \cdot \delta {\bf v} , \label{eq:mhdwaves-wavelike-rho} \\
 \omega \delta {\bf v} &= c_{\rm s}^2 \left( \frac{{\bf k} \cdot \delta {\bf v}}{\omega}\right) {\bf k} - \frac{\left({\bf B}_0 \cdot {\bf k}\right) \delta {\bf B}}{\mu_0 \rho_0} + \frac{\left({\bf B}_0 \cdot \delta {\bf B}\right) {\bf k}}{\mu_0 \rho_0}, \label{eq:mhdwaves-wavelike-v} %\frac{1}{\mu_0 \rho} ({\bf k} \times \delta {\bf B}) \times {\bf B} , \\
% \omega \delta {\bf B} &= ({\bf k} \cdot \delta {\bf v}) {\bf B}_0 - ({\bf k} \cdot {\bf B}_0) \delta {\bf v} .%- {\bf k} \times (\delta {\bf v} \times {\bf B}) . 
\end{align}
where in deriving Equation~\ref{eq:mhdwaves-wavelike-v}, we have made use of Equation~\ref{eq:mhdwaves-wavelike-rho} to substitute $\delta \rho$. Taking the time derivative of Equation~\ref{eq:mhdwaves-wavelike-v}, and using Equation~\ref{eq:mhdwaves-linearised-B}, we obtain
\begin{equation}
 \left[ \omega^2 - ({\bf v}_{\rm A} \cdot {\bf k})^2 \right] \delta {\bf v} = \left[ (c_{\rm s}^2 + v_{\rm A}^2) ({\bf k} \cdot \delta {\bf v}) - ({\bf v}_{\rm A} \cdot \delta {\bf v}) ({\bf k} \cdot {\bf v}_{\rm A}) \right] {\bf k} - \left[ ({\bf v}_{\rm A} \cdot {\bf k}) ({\bf k} \cdot \delta {\bf v}) \right] {\bf v}_{\rm A} , \label{eq:mhdwaves-timederiv}
% \left[ \omega^2 - \frac{({\bf B} \cdot {\bf k})^2}{\mu_0 \rho} \right] \delta {\bf v} = \left[ \left(c_{\rm s}^2 + \frac{B^2}{\mu_0 \rho}\right) ({\bf k} \cdot \delta {\bf v}) - \frac{({\bf B} \cdot \delta {\bf v}) ({\bf k} \cdot {\bf B})}{\mu_0 \rho} \right] {\bf k} - \frac{1}{\mu_0 \rho} \left[ ({\bf B} \cdot {\bf k}) ({\bf k} \cdot \delta {\bf v}) \right] {\bf B} ,
\end{equation}
where we have defined ${\bf v}_{\rm A} \equiv {\bf B}_0 / \sqrt{\mu_0 \rho_0}$, known as the Alfv\'en speed.

Taking the magnetic field to be in the $z$-direction and $k$ vector in the $y$-$z$ plane, that is ${\bf B}=B \hat{\bf z}$ and ${\bf k} = k_y \hat{\bf y} + k_z \hat{\bf z}$, Equation~\ref{eq:mhdwaves-timederiv} yields the following set of equations,
\begin{equation}
\left(
\begin{tabu}{ccc}
\omega^2 - v_{\rm A}^2 k_z^2 & 0 & 0 \\
0 & \omega^2 - c_{\rm s}^2 k_y^2 - v_{\rm A}^2 k^2 & - c_{\rm s}^2 k_y k_z \\
0 & -c_{\rm s}^2 k_y k_z & \omega^2 - c_{\rm s}^2 k_z^2 
\end{tabu}
\right)
\delta {\bf v} =
\left(
\begin{tabu}{c}
0 \\
0 \\
0
\end{tabu}
\right) .
\end{equation}

The first component is
\begin{equation}
 ( \omega^2 - v_{\rm A}^2 k_z^2 ) \delta v_x = 0 .
\end{equation}
Since this is directed along $v_x$, orthogonal to both the direction of wave propagation and magnetic field, this is a transverse oscillation known as an Alfv\'en wave. These have phase velocity $\omega / k_z = v_{\rm A}$, directed along the orientation of the magnetic field. Alfv\'en waves can be understood as occurring from the tension present in magnetic field lines, operating in a similar fashion to vibrating strings in a string instrument.

Waves in the $y$-$z$ plane are found from the determinant,
\begin{equation}
 \omega^4 - (c_{\rm s}^2 + v_{\rm A}^2) k^2 \omega^2 + c_{\rm s}^2 v_{\rm A}^2 k_z^2 k^2 .
\end{equation}
Using the quadratic formula, we can solve for $\omega^2 / k^2$, finding phase velocity
\begin{equation}
 \frac{\omega^2}{k^2} = \frac{1}{2} \left( c_{\rm s}^2 + v_{\rm A}^2 \right) \pm \frac{1}{2} \left( \left( c_{\rm s}^2 + v_{\rm A}^2 \right) - 4 c_{\rm s}^2 v_{\rm A}^2 \frac{k_z^2}{k^2} \right)^{1/2} .
\end{equation}
The wave velocity is in the same plane as the wave vector, therefore these are longitudinal waves. If no magnetic field is present, that is $v_{\rm A}=0$, these reduce to ordinary sound waves. These two wave types occur from the combination of magnetic and thermal pressure. One wave mode has its speed boosted by the addition of magnetic pressure (called {\it fast} MHD wave). The second wave mode is preferentially guided by the magnetic field, such that propagation across magnetic field lines is halted ({\it slow} MHD waves).

\section{Discretised magnetohydrodynamics}

The beauty of SPH is in its simplicity and intuitiveness. The basic method can be derived from first principles such that the discretised equations which are solved are the physical equations governing the system of discrete particles. Therefore, SPH inherently has the conservation properties of real physics, giving the method numerical stability and robustness. 

In this section, an overview of SPMHD is presented. Focus will be on how the MHD equations (\ref{eq:mhdcty}--\ref{eq:divbconstraint}) are solved numerically, highlighting how the discretised equations are obtained and the numerical challenges specific to SPMHD. For a complete background on interpolation theory, properties of smoothing kernels, stability analysis, and other deeper technical issues, the reader is referred to the PhD thesis of \citet{morris96} and the reviews by \citet{monaghan05} and \citet{price12}.

\subsection{Estimating the density}

The first step is to obtain an estimate of the density. This is accomplished by taking a weighted summation of the mass of neighbouring particles within a characteristic radius $h$, known as the smoothing length. The density in SPH is estimated as 
\begin{equation}
\rho({\bf r}_a) = \sum_b m_b W\left(\vert {\bf r}_a - {\bf r}_b \vert, h_a\right) , \label{eq:densitysum}
\end{equation}
where $W$ is the weighting function known as the smoothing kernel. We assume that each particle is allowed its own smoothing length and that it is spatially varying.

The density sum of Equation~\ref{eq:densitysum} can be used to calculate the density of a particle whenever required. It is unnecessary under general circumstances to evolve the density of a particle using its time derivative. However, for constructing the SPMHD equations, the discretised version of the continuity equation is useful. It can be obtained by taking the time derivative of Equation~\ref{eq:densitysum}, yielding
\begin{equation}
 \frac{{\rm d} \rho_a}{{\rm d}t} = \sum_b m_b \left( \frac{\partial W_{ab}(h_a)}{\partial {\bf r}_a} \cdot \frac{{\rm d}{\bf r}_a}{{\rm d} t} + \frac{\partial W_{ab}(h_a)}{\partial {\bf r}_b} \cdot \frac{{\rm d}{\bf r}_b}{{\rm d}t} + \frac{\partial W_{ab}(h_a)}{\partial h_a}\frac{\partial h_a}{\partial \rho_a}\frac{{\rm d}\rho_a}{{\rm d}t} \right) ,
\end{equation}
where we have made use of the chain rule and introduced the shorthand notation $W_{ab}(h_a) \equiv W\left(\vert {\bf r}_a - {\bf r}_b \vert, h_a\right)$. Using ${\rm d}{\bf r}/{\rm d}t = {\bf v}$ and the anti-symmetry of the kernel gradient, that is $\nabla_a W_{ab}(h_a) \equiv - \nabla_b W_{ab}(h_a)$, this can be simplified to yield
\begin{equation}
 \frac{{\rm d} \rho_a}{{\rm d}t} = \frac{1}{\Omega_a} \sum_b m_b ({\bf v}_a - {\bf v}_b) \cdot \nabla_a W_{ab}(h_a) , \label{eq:discretisedcty}
\end{equation}
where $\nabla_a$ is the gradient taken respect to the coordinates of particle $a$ and
\begin{equation}
 \Omega_a = 1 - \sum_b m_b \frac{\partial W_{ab}(h_a)}{\partial h_a} \frac{\partial h_a}{\partial \rho_a} .
\end{equation}
The $\Omega$ term is important to correctly account for spatially varying smoothing lengths \citep[see][]{sh02, monaghan02}. In general, they should be used when computing any derivative estimate. In particular, inclusion of these terms into the SPMHD momentum, induction, and energy equations has been shown to improve the representation of wave propagation and shocks \citep{pm04b}. Obtaining $\partial h_a/\partial \rho_a$ may be done through Equation~\ref{eq:h} below, as given by Equation~\ref{eq:dhdrho}.

\subsection{Setting the smoothing length}
\label{sec:spmhd-h}

The smoothing length is individually set per particle by mutually solving
\begin{equation}
h_a = \eta_{\rm h} \left(\frac{m_a}{\rho_a}\right)^{1/\nu}  \\ \label{eq:h}
\end{equation}
with the density summation (Equation~\ref{eq:densitysum}). The derivative is given by
\begin{equation}
 \frac{\partial h_a}{\partial \rho_a} = - \frac{h_a}{\nu \rho_a}, \label{eq:dhdrho}
\end{equation}
and Equation~\ref{eq:h} leads to an expression for the density as
\begin{equation}
 \rho_a = m_a \left( \frac{h_a}{\eta_{\rm h}}\right)^{-\nu} . \label{eq:rho-exact}
\end{equation}
Here, $\nu=1$, $2$, $3$ is the dimension of the system and $\eta_{\rm h}$ is a dimensionless quantity specifying the ratio of smoothing length to particle spacing. For the spline family of kernels \citep{schoenberg46, monaghan85, ml85}, this is typically chosen to be $\eta_{\rm h}=1.2$. Since the density itself is a function of smoothing length, this requires iteration until both quantities converge. This is an expensive process, as the density summation needs to be re-evaluated for each iteration, which may further necessitate performing a neighbour search.

A root finding technique can be used to find the smoothing length and density for each particle. The function to find the root of is
\begin{equation}
 f(h) \equiv m_a \left( \frac{h_a}{\eta_{\rm h}} \right)^{-\nu} - \sum_b m_b W_{ab}(h_a) .
\end{equation}
That is, the root, $f(h)=0$, occurs when the expected density (from Equation~\ref{eq:rho-exact}) agrees with the density as calculated through summation (Equation~\ref{eq:densitysum}). 

The root can be found with the Newton-Raphson technique \citep[e.g.][]{pm04b, pm07}. The smoothing length may be iterated according to
\begin{equation}
 h^\text{new} = h^\text{old} - \frac{f(h)}{f'(h)} , \label{eq:newtonraphson-h2}
\end{equation}
where $f'(h)$ is the first derivative of $f(h)$,
\begin{equation}
 f'(h) = -\frac{\nu}{h_a} m_a \left(\frac{h_a}{\eta_{\rm h}} \right)^{-\nu} - \sum_b m_b \frac{\partial W_{ab}(h_a)}{\partial h_a} .
\end{equation}
For convenience of implementation, this may be rewritten using $\Omega_a$ as
\begin{equation}
%  f'(h) = - \frac{\nu}{h_a} \left( m_a \left(\frac{h_a}{\eta_{\rm h}} \right)^{-\nu} + \rho_a \left( \Omega_a - 1 \right) \right) .
 f'(h) = -\frac{\nu \rho_a \Omega_a}{h_a} ,
\end{equation}
where $\rho_a$ is the expression in Equation~\ref{eq:rho-exact}. By using the tangent of $f(h)$ to iterate towards the root, the method has second order convergence and as such is an efficient means to solve for $h$ and $\rho$. However, it may fail if the tangent of $f(h)$ is nearly parallel, leading to the next iteration to significantly overshoot the root and diverge. This risk can be curbed by including a check to restrict modification of $h$ between iterations by no more than, say, $20\%$. For a typical SPMHD calculation, the risk of the method diverging is minimal (and usually indicative of a more serious problem elsewhere).

An approach guaranteed to converge, though slower with only linear convergence, is to use a bisection method. The method is simple. If the converged value of $h$ is known to lie within a specified interval, then the interval can be halved until it is found. Which half of the interval to reject can be determined through $f(h)$, where if $f(h)>0$, then $h$ should be decreased from its current value, otherwise $h$ should be increased on the next iteration.  

Convergence can be determined by monitoring the relative difference in $h$ (or $\rho$) between iterations. This is determined according to
\begin{equation}
\frac{\vert h^\text{new} - h^\text{old} \vert}{h^\text{0}} < \epsilon , \label{eq:hconvergencecriterion}
\end{equation}
where $\epsilon \sim 10^{-4}$. Note that $h^\text{0}$, the value of the smoothing length before the first iteration, is used so that the denominator remains fixed and convergence occurs only when $h^\text{new}$ agrees with $h^\text{old}$. 

A simple approach to reduce the number of overall iterations is to time integrate the smoothing length, predicting a value close to the root before beginning the root finding technique. Taking the time derivative of Equation~\ref{eq:h}, and using the continuity equation (Equation~\ref{eq:mhdcty}), we obtain
\begin{equation}
 \frac{{\rm d} h_a}{{\rm d}t} = \frac{h_a}{\nu} \nabla \cdot {\bf v}_a . \label{eq:dhdt}
\end{equation}
Given that $\nabla \cdot {\bf v}$ is usually calculated in an SPH code, this adds almost no additional computational cost, yet can significantly increase the overall efficiency of the code.

At the beginning of each timestep, the step-by-step approach to setting the smoothing length per particle is:
\begin{enumerate*}
\item Compute the density using summation.
\item Iterate to $h^{\rm new}$ using Equation~\ref{eq:newtonraphson-h2} (or other root-solving technique).
\item Do $h^{\rm new}$ and $h^{\rm old}$ agree within the specified tolerance (Equation~\ref{eq:hconvergencecriterion})?
 \begin{enumerate*}
 \item {\bf Yes}, accept and proceed to step 4.
 \item {\bf No}, begin again from step 1
 \end{enumerate*}
\item Predict $h$ for the next timestep using Equation~\ref{eq:dhdt}.
\end{enumerate*}

\subsection{Interpolation basics}
\label{sec:interpolation}

In order to define the discretised induction equation, it is necessary to understand the basics of SPH interpolation theory. This is not a comprehensive review, but will introduce the basics necessary to formulate the SPMHD equations. Throughout this section, we assume that the smoothing length is uniform and constant so that the presentation may be clearer. Inserting $\Omega$ terms to account for variable smoothing lengths may be appropriately inserted where gradients of the smoothing kernel are taken. 

In the continuum limit, the value of a quantity $A$ can be obtained by using a delta function to pluck that value at a specified location, that is
\begin{equation}
 A({\bf r}) = \int A({\bf r}') \delta({\bf r} - {\bf r}') d{\bf r}' .
\end{equation}
In SPH, the smoothing kernel plays the role of the delta function. It has property such that
\begin{equation}
 \lim_{h \to 0} W(\vert {\bf r} - {\bf r}' \vert, h) = \delta({\bf r} - {\bf r}') .
\end{equation}
The kernel is assumed to be spherically symmetric, and normalised such that $\int W dV = 1$. In a discretised system, the integral is replaced by a summation over elements. The quantity $A$ can be obtained through
\begin{equation}
 A_a = \sum_b \frac{m_b}{\rho_b} A_b W_{ab} , \label{eq:basicinterpolation2}
\end{equation}
where $m_b/\rho_b$ acts like the volume element of the integral. This reduces to the density summation if $A\equiv\rho$, and is the traditional way to introduce SPH. In this thesis however, we use the mass weighted summation
\begin{equation}
 A_a = \frac{1}{\rho_a} \sum_b m_b A_b W_{ab}, \label{eq:basicinterpolation} 
\end{equation}
where $\rho_a$ acts like the normalisation on the summation. If $A$ is a constant, this returns $A$ exactly. Though for clarity of presentation, we continue using Equation~\ref{eq:basicinterpolation2}.

The derivative of Equation~\ref{eq:basicinterpolation2} is 
\begin{equation}
 \nabla_a A_a = \sum_b \frac{m_b}{\rho_b} A_b \nabla_a W_{ab} . \label{eq:interp-grad}
\end{equation}
However, this yields a poor estimate of the gradient. For example, constant functions will yield a non-zero result. Higher accuracy gradient estimates may be obtained by taking a Taylor series expansion to obtain error terms, then subtracting those errors from the gradient \citep{morris96, price04}. The Taylor series expansion of $A_b$ in Equation~\ref{eq:interp-grad} about ${\bf r}_a$ is
\begin{align}
 \nabla_a A_a &= A_a \sum_b \frac{m_b}{\rho_b} \nabla_a W_{ab} + \frac{\partial A_a}{\partial {\bf r}^\alpha} \sum_b \frac{m_b}{\rho_b} ({\bf r}_b - {\bf r}_a)^\alpha \nabla_a W_{ab} \nonumber \\
& \hspace{20mm} + \frac{1}{2} \frac{\partial^2 A_a}{\partial {\bf r}^\beta \partial {\bf r}^\gamma} \sum_b \frac{m_b}{\rho_b} ({\bf r}_b - {\bf r}_a)^\beta ({\bf r}_b - {\bf r}_a)^\gamma \nabla_a W_{ab} + \mathcal{O}(h^3) .
\end{align}
Thus, the gradient estimate can be made first order by subtracting the first error term in the Taylor series, yielding
\begin{equation}
 \nabla_a A_a = - \sum_b \frac{m_b}{\rho_b} (A_a - A_b) \nabla_a W_{ab} + \mathcal{O}(h) , \label{eq:interp-grad-firstorder}
\end{equation}
which is exact for constant functions. This is the most common form for calculating gradients in SPH. Obtaining divergence and curl estimates for vector quantities may be obtained through a similar procedure as the preceding, with the divergence and curl of the magnetic field presented in Section~\ref{sec:divb-curlb}.

A second order estimate can be obtained by subtracting the second error term of the Taylor series, yielding 
\begin{equation}
 \frac{\partial A_a}{\partial {\bf r}^\alpha} = - \chi^{\alpha \beta} \sum_b \frac{m_b}{\rho_b} (A_a - A_b) \nabla^\beta W_{ab} + \mathcal{O}(h^2) , \label{eq:interp-grad-secondorder}
\end{equation}
where 
\begin{equation}
 \chi^{\alpha \beta} = \left( - \sum_b \frac{m_b}{\rho_b} ({\bf r}_b - {\bf r}_a)^\alpha \nabla^\beta W_{ab} \right)^{-1}  \label{eq:interp-grad-secondorder-matrix}
\end{equation}
is a matrix that acts as a correction to the kernel gradient. This approach has been used by \citet{bl99}. This second order derivative estimate adds more computational expensive since it requires a matrix inversion and storage of the $3 \times 3$ matrix elements.

While second derivatives may be estimated by taking the derivative of Equation~\ref{eq:interp-grad-firstorder}, this is quite sensitive to particle disorder and leads to a poor estimate. A less noisy estimate is to take two first derivatives, applying Equation~\ref{eq:interp-grad-firstorder} (or Equation~\ref{eq:interp-grad-secondorder}) twice in succession. This is more expensive as it requires two loops over particle neighbours. An alternative approach is that of \citet{brookshaw85}, whereby a second derivative estimate is obtained from
\begin{equation}
 \nabla^2_a A_a = 2 \sum_b \frac{m_b}{\rho_b} (A_a - A_b) \frac{{\bf r}_{ab} \cdot \nabla_a W_{ab}}{r_{ab}^2} . \label{eq:brookshaw}
\end{equation}
This may also be written as
\begin{equation}
\nabla^2_a A_a = 2 \sum_b \frac{m_b}{\rho_b} (A_a  - A_b) \frac{F_{ab}}{\vert r_{ab} \vert} ,
\end{equation}
making use of the definition $\nabla_a W_{ab} \equiv \hat{\bf r}_{ab} F_{ab}$, where $F_{ab}$ is the scalar portion of the kernel derivative. We note that there is inconsistent usage of this definition in the literature. Notably the SPH review by \citet{price12} uses the aforementioned definition, whereas the SPH review by \citet{monaghan05} instead uses $\nabla_a W_{ab} \equiv {\bf r}_{ab} F_{ab}$, differing by a factor $1/\vert r_{ab} \vert$. A reader should be careful of this difference in the literature. In this thesis, we adopt usage consistent with \citet{price12} ($\nabla_a W_{ab} \equiv \hat{\bf r}_{ab} F_{ab}$).
% \begin{equation}
%  \nabla^2_a A_a = 2 \sum_b \frac{m_b}{\rho_b} (A_b - A_a) \frac{F_{ab}}{\vert {\bf r}_{ab} \vert} \nabla_a W_{ab} ,
% \end{equation}
% where we have used the definition $\nabla_a W_{ab} \equiv {\bf r}_{ab} F_{ab}$. 

This second derivative estimate may be obtained through Taylor series expansion of $A({\bf r}')$ about ${\bf r}$ in the integral approximation,
\begin{equation}
 \int \left( A({\bf r}) - A({\bf r}') \right) \frac{({\bf r} - {\bf r}') \cdot \nabla_a W({\bf r})}{({\bf r} - {\bf r}')^2} {\rm d}{\bf r}' = \frac{1}{2} \nabla^2 A({\bf r}) + \mathcal{O}(({\bf r} - {\bf r}')^2) .
\end{equation}
This functionally approximates a second derivative by dividing the first derivative of the smoothing kernel by the particle spacing, ${\bf r}_{ab}$.

\subsection{$\nabla \cdot {\bf B}$ and $\nabla \times {\bf B}$}
\label{sec:divb-curlb}

The divergence and curl of the magnetic field may be obtained in a similar manner to the first derivative estimates of scalar quantities. The divergence and curl equivalents of Equation~\ref{eq:interp-grad} are given by 
\begin{equation}
 \nabla \cdot {\bf B}_a = \sum_b \frac{m_b}{\rho_b} {\bf B}_b \cdot \nabla_a W_{ab}
\label{eq:divb-zeroth}
\end{equation}
and
\begin{equation}
 \nabla \times {\bf B}_a = - \sum_b \frac{m_b}{\rho_b} {\bf B}_b \times \nabla_a W_{ab} .
\label{eq:curlb-zeroth}
\end{equation}
It is possible to obtain first order accurate estimates of the first derivatives through other means than the Taylor series expansion presented in the preceding section. For the divergence operator, consider the identity
\begin{equation}
 \nabla \cdot {\bf A} = \frac{1}{\rho} \left[ \nabla \cdot (\rho {\bf A}) - {\bf A} \cdot (\nabla \rho) \right] .
\end{equation}
Inserting the simple first derivative operators from Equations~\ref{eq:interp-grad}, \ref{eq:divb-zeroth} and \ref{eq:curlb-zeroth} will yield first order accurate estimates. In this thesis, we use the mass weighted summations, with the first order accurate divergence and curl of the magnetic field given by (including variable smoothing length terms) 
\begin{equation}
 \nabla \cdot {\bf B}_a = - \frac{1}{\Omega_a \rho_a} \sum_b m_b ({\bf B}_a - {\bf B}_b) \cdot \nabla_a W_{ab}(h_a)
\label{eq:divb-diff}
\end{equation}
and
\begin{equation}
 \nabla \times {\bf B}_a = \frac{1}{\Omega_a \rho_a} \sum_b m_b ({\bf B}_a - {\bf B}_b) \times \nabla_a W_{ab}(h_a) .
\label{eq:curlb-diff}
\end{equation}

Other identities lead to other first derivative estimates. The identity
\begin{equation}
 \nabla \cdot {\bf A} = \rho \left[ \nabla \cdot \left( \frac{{\bf A}}{\rho} \right) + \frac{{\bf A}}{\rho^2} \cdot \nabla \rho \right] .
\end{equation}
may be used to obtain an entirely different form for the first derivative. The divergence and curl with this operator (including variable smoothing length terms) are
\begin{equation}
 \nabla \cdot {\bf B}_a = \rho_a \sum_b m_b \left[ \frac{{\bf B}_a}{\Omega_a \rho_a^2} \cdot \nabla_a W_{ab}(h_a) + \frac{{\bf B}_b}{\Omega_b \rho_b^2} \cdot \nabla_a W_{ab}(h_b) \right] , \label{eq:divb-symm}
\end{equation}
and
\begin{equation}
 \nabla \times {\bf B}_a = - \rho_a \sum_b m_b \left[ \frac{{\bf B}_a}{\Omega_a \rho_a^2} \times \nabla_a W_{ab}(h_a) + \frac{{\bf B}_b}{\Omega_b \rho_b^2} \times \nabla_a W_{ab}(h_b) \right] . \label{eq:curlb-symm}
\end{equation}
The error for these estimates is large ($\mathcal{O}(1)$), and yield non-zero results for constant functions. 

We refer to the first derivative operators as the `difference' measure (Equation~\ref{eq:divb-diff} and \ref{eq:curlb-diff}) and the `symmetric' measure (Equation~\ref{eq:divb-symm} and \ref{eq:curlb-symm}). It is noteworthy that the symmetric measure of $\nabla \cdot {\bf B}$ is what will appear in the equations of motion, and the implications of this derivative estimate will be discussed in Section~\ref{sec:spmhd-tensileinstability}.

\subsection{Energy equation}

The equations of motion need to be coupled to an equation of state to determine the thermal pressure. If the pressure is a function of internal energy, $u$, a suitable equation must be used to evolve $u$ in time. Consider the thermodynamic relation,
\begin{equation}
 {\rm d}u = T{\rm d}s - P{\rm d}V , \label{eq:thermodynamicrelation}
\end{equation}
where $T$ is the temperature, $s$ is the entropy, $V$ is the volume, and quantities are expressed per unit mass. Since SPMHD is inherently dissipationless, ${\rm d}s$ may be taken to be zero. (Alternatively, if the pressure is a function of entropy, the entropy per particle may be passively advected and increased only from added sources of dissipation.) Converting ${\rm d}V$ to be per unit mass, the time derivative of $u$ is 
\begin{equation}
 \frac{{\rm d}u}{{\rm d}t} = \frac{P}{\rho^2} \frac{{\rm d}\rho}{{\rm d}t} . \label{eq:thermodynamicrelation2}
\end{equation}
Using the SPH continuity equation (Equation~\ref{eq:discretisedcty}), the discretised internal energy equation is thus
\begin{equation}
 \frac{{\rm d}u_a}{{\rm d}t} = \frac{P_a}{\Omega_a \rho_a^2} \sum_b m_b {\bf v}_{ab} \cdot \nabla_a W_{ab}(h_a) . \label{eq:discretisedenergyeq}
\end{equation}

\subsection{Induction equation}

Using basic interpolation theory, the induction equation (Equation~\ref{eq:induction}) may be discretised as
\begin{equation}
\frac{{\rm d}{\bf B}_a}{{\rm d}t} = - \frac{1}{\Omega_a \rho_a} \sum_b m_b \left[ {\bf v}_{ab} ({\bf B}_a \cdot \nabla_a W_{ab}(h_a)) - {\bf B}_a ({\bf v}_{ab} \cdot \nabla_a W_{ab}(h_a)) \right] . \label{eq:spmhd-ind}
\end{equation}
Alternatively, the quantity ${\bf B}/\rho$ could be evolved. Rewriting the induction equation as
\begin{equation}
 \frac{{\rm d}}{{\rm d}t} \left( \frac{{\bf B}}{\rho} \right) = \left( \frac{{\bf B}}{\rho} \cdot \nabla \right) {\bf v} ,
\end{equation}
the SPMHD form is
\begin{equation}
 \frac{{\rm d}}{{\rm d}t}\left(\frac{{\bf B}_a}{\rho_a}\right) = -\frac{1}{\Omega_a \rho_a^2} \sum_b m_b {\bf v}_{ab} \left({\bf B}_a \cdot \nabla_a W_{ab}(h_a)\right) .
\end{equation}
Both approaches are utilised throughout this thesis, depending on the code used. Neither approach confers any significant advantage over the other \citep{price12}.

\subsection{Conservative equations of motion}
\label{sec:spmhd-mom}

The equations of motion for SPMHD will be obtained by using the Lagrangian for the discretised system \citep{price04, pm04b}. This will yield the equations of motion that physically govern the system, providing a method that has exact conservation of momentum, energy, and entropy. Consider the SPMHD Lagrangian, 
\begin{equation}
 L_{\rm SPH} = \sum_a m_a \left( \tfrac{1}{2} v_a^2 - u_a - \frac{B_a^2}{2 \mu_0 \rho_a} \right) .
\end{equation}
The action integral, $S = \int L {\rm d}t$, is stationary. Therefore, small perturbations must not change the solution, that is
\begin{equation}
\delta S = \int \delta L {\rm d}t = 0 . \label{eq:actionint}
\end{equation}
If small deviations are introduced into the Lagrangian about $\delta {\bf r}_a$, then
\begin{equation}
% \delta L_a = m_a v_a^i \delta v_a^i - \sum_b m_b \left. \frac{\partial u}{\partial \rho} \right\vert_{s} \delta \rho_b + \sum_b m_b \frac{B_b^2}{2 \mu_0 \rho_b^2} \delta \rho_b - \sum_b m_b \frac{B_b^i}{\mu_0 \rho_b} \cdot \delta B_b^i . \label{eq:variational-lagrangian}
 \delta L_a = m_a {\bf v}_a \cdot \delta {\bf v}_a - \sum_b m_b \left. \frac{\partial u}{\partial \rho} \right\vert_{s} \delta \rho_b + \sum_b m_b \frac{B_b^2}{2 \mu_0 \rho_b^2} \delta \rho_b - \sum_b m_b \frac{{\bf B}_b}{\mu_0 \rho_b} \cdot \delta {\bf B}_b . \label{eq:variational-lagrangian}
\end{equation}
Using the density summation (Equation~\ref{eq:densitysum}) and induction equation (Equation~\ref{eq:spmhd-ind}) as constraints, the variations $\delta \rho$ and $\delta {\bf B}$ can be written in terms of $\delta {\bf r}$ according to
\begin{align}
% \delta \rho_b &= \frac{1}{\Omega_b} \sum_c m_c \left( \delta r_b^i - \delta r_c^i \right) \nabla_b^i W_{bc}(h_b) , \\
% \delta B_b^i &= \frac{1}{\Omega_b \rho_b} \sum_c m_c \left[ B_b^i \left( \delta r_b^j - \delta r_c^j \right) \nabla_b^j W_{bc}(h_b) - \left( \delta r_b^i - \delta r_c^i \right) B_b^j \nabla_b^j W_{bc}(h_b) \right] .
 \delta \rho_b &= \frac{1}{\Omega_b} \sum_c m_c \left( \delta {\bf r}_b - \delta {\bf r}_c \right) \cdot \nabla_b W_{bc}(h_b) , \\
 \delta {\bf B}_b &= \frac{1}{\Omega_b \rho_b} \sum_c m_c \left[ {\bf B}_b \left( \delta {\bf r}_b - \delta {\bf r}_c \right) \cdot \nabla_b W_{bc}(h_b) - \left( \delta {\bf r}_b - \delta {\bf r}_c \right) {\bf B}_b \cdot \nabla_b W_{bc}(h_b) \right] .
\end{align}
Inserting these into Equation~\ref{eq:variational-lagrangian} and using the thermodynamic relation (Equation~\ref{eq:thermodynamicrelation2}), we obtain
\begin{align}
% \delta L_a = &m_a v_a^i \delta v_a^i - \sum_b m_b \frac{P_b}{\Omega_b \rho_b^2} \sum_c m_c \left( \delta r_b^i - \delta r_c^i \right) \cdot \nabla_b^i W_{bc}(h_b) \nonumber \\
%& + \sum_b m_b \frac{B_b^2}{2 \mu_0 \Omega_b \rho_b^2} \sum_c m_c \left( \delta r_b^i - \delta r_c^i \right) \nabla_b^i W_{bc}(h_b) \nonumber \\
%& - \sum_b m_b \frac{{\bf B}_b}{\mu_0 \Omega_b \rho_b^2} \cdot \sum_c m_c \left[ {\bf B}_b \left( \delta {\bf r}_b - \delta {\bf r}_c \right) \cdot \nabla_b W_{bc}(h_b) - \left( \delta {\bf r}_b - \delta {\bf r}_c \right) {\bf B}_b \cdot \nabla_b W_{bc}(h_b) \right] .
 \delta L_a = &m_a {\bf v}_a \cdot \delta {\bf v}_a - \sum_b m_b \frac{P_b}{\Omega_b \rho_b^2} \sum_c m_c \left( \delta {\bf r}_b - \delta {\bf r}_c \right) \cdot \nabla_b W_{bc}(h_b) \nonumber \\
& + \sum_b m_b \frac{B_b^2}{2 \mu_0 \Omega_b \rho_b^2} \sum_c m_c \left( \delta {\bf r}_b - \delta {\bf r}_c \right) \cdot \nabla_b W_{bc}(h_b) \nonumber \\
& - \sum_b m_b \frac{{\bf B}_b \cdot {\bf B}_b}{\mu_0 \Omega_b \rho_b^2} \cdot \sum_c m_c \left( \delta {\bf r}_b - \delta {\bf r}_c \right) \cdot \nabla_b W_{bc}(h_b) \nonumber \\
& - \sum_b m_b \frac{{\bf B}_b}{\mu_0 \Omega_b \rho_b^2} \cdot \sum_c m_c \left( \delta {\bf r}_b - \delta {\bf r}_c \right) {\bf B}_b \cdot \nabla_b W_{bc}(h_b) .
\end{align}

The perturbations in $\delta {\bf r}_b$ and $\delta {\bf r}_c$ can be removed by multiplying the latter terms by $\delta {\bf r}_a / \delta {\bf r}_a$, introducing delta functions into the equation and yielding
\begin{align}
 \delta L_a = &m_a {\bf v}_a \cdot \delta {\bf v}_a - \delta {\bf r}_a \cdot \Bigg[ \sum_b m_b \frac{P_b}{\Omega_b \rho_b^2} \sum_c m_c \left( \delta_{ab} - \delta_{ac} \right) \nabla_b W_{bc}(h_b) \nonumber \\
& - \sum_b m_b \frac{B_b^2}{2 \mu_0 \Omega_b \rho_b^2} \sum_c m_c \left( \delta_{ab} - \delta_{ac} \right) \nabla_b W_{bc}(h_b) \nonumber \\
& + \sum_b m_b \frac{{\bf B}_b}{\mu_0 \Omega_b \rho_b^2} \sum_c m_c \left( \delta_{ab} - \delta_{ac} \right) {\bf B}_b \cdot \nabla_b W_{bc}(h_b) \Bigg] . \label{eq:variationlagrange1}
\end{align}
Simplifying out the delta functions and using the anti-symmetry of the kernel gradient ($\nabla_a W_{ab} = - \nabla_b W_{ab}$), we obtain 
\begin{align}
 \delta L_a = &m_a {\bf v}_a \cdot \delta {\bf v}_a - m_a \delta {\bf r}_a \cdot \Bigg\{ \sum_b m_b \left[ \frac{P_a}{\Omega_a \rho_a^2} \nabla_a W_{ab}(h_a) + \frac{P_b}{\Omega_b \rho_b^2} \nabla_a W_{ab}(h_b) \right] \nonumber \\
& - \frac{m_a}{2\mu_0} \sum_b m_b \left[ \frac{B_a^2}{\Omega_a \rho_a^2} \nabla_a W_{ab}(h_a) + \frac{B_b^2}{\Omega_b \rho_b^2} \nabla_a W_{ab}(h_b) \right]  \nonumber \\
& + \frac{m_a}{\mu_0} \sum_b m_b \left[ \frac{{\bf B}_a}{\Omega_a \rho_a^2} {\bf B}_a \cdot \nabla_a W_{ab}(h_a) + \frac{{\bf B}_b}{\Omega_b \rho_b^2} {\bf B}_b \cdot \nabla_a W_{ab}(h_b) \right] \Bigg\} . \label{eq:variationlagrange2}
\end{align}

Inserting Equation~\ref{eq:variationlagrange2} into \ref{eq:actionint} and integrating the velocity term by parts, the equations of motion are found to be
\begin{align}
 \frac{{\rm d}{\bf v}_a}{{\rm d}t} = &- \sum_b m_b \left[ \frac{P_a}{\Omega_a \rho_a^2} \nabla_a W_{ab}(h_a) + \frac{P_b}{\Omega_b \rho_b^2} \nabla_a W_{ab}(h_b) \right] \nonumber \\
& - \frac{1}{2\mu_0} \sum_b m_b \left[ \frac{B_a^2}{\Omega_a \rho_a^2} \nabla_a W_{ab}(h_a) + \frac{B_b^2}{\Omega_b \rho_b^2} \nabla_a W_{ab}(h_b) \right] \nonumber \\
& + \frac{1}{\mu_0} \sum_b m_b \left[ \frac{{\bf B}_a}{\Omega_a \rho_a^2} {\bf B}_a \cdot \nabla_a W_{ab}(h_a) + \frac{{\bf B}_b}{\Omega_b \rho_b^2} {\bf B}_b \cdot \nabla_a W_{ab}(h_b) \right] . \label{eq:spmhd-momeq}
\end{align}
This is equivalent to writing the momentum equation in terms of the stress tensor. The implication of this is that the tension force contains a component due to monopole moments. The issue of monopole forces is complicated in SPMHD, in that even for a field which is constant and uniform (i.e., $\nabla \cdot {\bf B}=0$), the discretisation used in the momentum equation may produce monopole forces. We discuss the implications of this below.

\subsection{Removing the tensile instability}
\label{sec:spmhd-tensileinstability}

\citet{pm85} noted that the conservative form of the SPMHD equations contain an instability when the magnetic tension exceeds the isotropic pressure, causing the particles to unphysically clump. This arises due to monopole forces. Describing the momentum equation in terms of the stress tensor assumes that the magnetic field contains no monopole moments -- a condition which may not be upheld numerically. As noted in Section~\ref{sec:mhdmomeq}, the tension force is equivalent to $({\bf B} \cdot \nabla){\bf B} + {\bf B}(\nabla \cdot {\bf B})$ when writing the momentum equation in terms of the stress tensor. The force contributions proportional to $\nabla \cdot {\bf B}$ are present in order to be momentum conserving in the presence of monopoles.

Several approaches have been taken to counteract this instability. \citet{pm85} used the simple and effective technique of adding a constant pressure to the system to ensure the total force between particles was always repulsive. This preserves conservation of momentum, though no longer conserves energy and incurs a computational cost to determine the amount of stress to add. 

Non-momentum conserving approaches have also been investigated. \citet{mwd95} solved the Lorentz force directly using ${\bf J}\times{\bf B}$, rather than using the stress tensor, though \citet{morris96} showed that this approach is poor at capturing shocks due to its poor conservation properties. \citet{morris96} formulated an approach which uses the conservative form for the magnetic pressure term, but with a more accurate estimate for the magnetic tension. \citet{bot01} used the conservative form for both the magnetic pressure and tension, but explicitly subtracted the non-physical force arising from monopole contribution. This is the approach we use. 

The premise is to subtract the monopole force contribution from the conservative equations of motion using the same discretisation for $\nabla \cdot {\bf B}$ as in the momentum equation, that is,
\begin{equation}
% \frac{{\rm d}{\bf v}_a}{{\rm d}t}
 -{\bf B} ( \nabla \cdot {\bf B}) = - {\bf B}_a \sum_b m_b \left(\frac{{\bf B}_a}{\Omega_a \rho_a^2} \cdot \nabla_a W_{ab}(h_a) + \frac{{\bf B}_b}{\Omega_b \rho_b^2} \cdot \nabla_a W_{ab}(h_b) \right) . \label{eq:borve-divb-subtract}
\end{equation}
This yields a numerically stable solution. Stability analysis by \citet{bot04} showed that the instability manifests only when $\tfrac{1}{2} B^2 > P$, therefore they introduce an adjustable parameter $\hat{\beta}$, showing that multiplying the force correction (Equation~\ref{eq:borve-divb-subtract}) by $\hat{\beta} = \tfrac{1}{2}$ is still sufficient to correct the instability in the magnetic pressure-dominated regime. Indeed, recently \cite{bkw12} have recommended using $\hat{\beta} = \tfrac{1}{2}$ for general SPMHD calculations. However, we find in this thesis (Section~\ref{sec:halfdivb}) that using $\hat{\beta} < 1$ can produce numerical artefacts (c.f.~Figure~\ref{fig:halfdivb}). While using $\hat{\beta} = \tfrac{1}{2}$ is technically sufficient to correct for the instability, it leaves the particles in a near-pressureless state. We therefore strongly recommend using $\hat{\beta} = 1$ and adopt this throughout unless otherwise specified. Note that with $\hat{\beta} = 1$ the induction and momentum equations are formally equivalent to Powell's eight wave approach \citep{powell94}. 

The discretisation used to calculate $\nabla \cdot {\bf B}$ in the momentum equation slightly complicates the issue of momentum loss as it is a rather poor estimate of the divergence (having errors $\mathcal{O}(1)$). For example, even if the magnetic field is constant and uniform, thus $\nabla \cdot {\bf B}=0$ should be true, monopole forces may arise due to particle disorder alone. Typical errors introduced from particle disorder are minimal, however if the magnetic field is highly unphysical, the momentum loss can become quite significant. It is important to use a method to maintain the solenoidal constraint on the magnetic field to minimise spurious momentum injection. For example, when the divergence cleaning method developed in Chapter~\ref{sec:chapter-cleaning} is used in simulations of protostar formation, the momentum drift is only 1\% of that when no divergence control is used (Figure~\ref{fig:star-mom}).

\subsection{Summary of discretised MHD equations}

The SPMHD equations which are solved (including the tensile instability correction) are
\begin{align}
\rho_a = &\sum_b m_b W_{ab}(h_a) , \label{eq:spmhd-summary-density} \\
h_a = &\eta_{\rm h} \left(\frac{m_a}{\rho_a}\right)^{1/\nu} , \label{eq:spmhd-summary-h} \\
\frac{{\rm d}{\bf v}_a}{{\rm d}t} = &- \sum_b m_b \left[ \frac{P_a}{\Omega_a \rho_a^2} \nabla_a W_{ab}(h_a) + \frac{P_b}{\Omega_b \rho_b^2} \nabla_a W_{ab}(h_b) \right] \nonumber \\
& - \frac{1}{2\mu_0} \sum_b m_b \left[ \frac{B_a^2}{\Omega_a \rho_a^2} \nabla_a W_{ab}(h_a) + \frac{B_b^2}{\Omega_b \rho_b^2} \nabla_a W_{ab}(h_b) \right] \nonumber \\
& + \frac{1}{\mu_0} \sum_b m_b \frac{({\bf B}_b - {\bf B}_a)}{\Omega_b \rho_b^2} {\bf B}_a \cdot \nabla_a W_{ab}(h_b) \label{eq:spmhd-summary-mom}, \\
\frac{{\rm d}{\bf B}_a}{{\rm d}t} = &-\frac{1}{\Omega_a \rho_a} \sum_b m_b \{ {\bf v}_{ab} \left[ {\bf B}_a \cdot \nabla_a W_{ab}(h_a)\right] - {\bf B}_a \left[ {\bf v}_{ab} \cdot \nabla_a W_{ab}(h_a) \right] \} , \label{eq:spmhd-summary-ind}  \\
\frac{{\rm d}u_a}{{\rm d}t} = &\frac{P_a}{\Omega_a \rho_a^2} \sum_b m_b {\bf v}_{ab} \cdot \nabla_a W_{ab}(h_a) . \label{eq:spmhd-summary-energyeq}
\end{align}

\subsection{Capturing shocks and discontinuities}
\label{sec:spmhd-shock-capturing}

%Derive the artificial dissipation terms. These are used to smooth discontinuities so that they remain single valued.

Discontinuities in the fluid require special treatment in numerical hydrodynamics. The SPMHD equations assume that the evolved quantities are differentiable, which no longer is true when the quantity becomes multi-valued at shocks and discontinuities. Artificial dissipation terms are used in SPMHD to smooth discontinuities over the resolution scale so that the fluid remains single valued.

\subsubsection{Artificial viscosity}
\label{sec:artvisc}

Artificial viscosity is used to not just to smooth shocks, but to damp the oscillations in particles which have been shocked. Since SPH particles behave similar to a molecular system, oscillations are introduced in the particle motion on the length scale of the inter-particle separation. It is therefore important to damp these oscillations.

\citet{monaghan97} derived a form of artificial viscosity by analogy with Riemann solvers. Treating a pair of particles as the left and right states of the Riemann problem, an artificial viscosity can be obtained of the form 
\begin{equation}
  \frac{{\rm d}{\bf v}_a}{{\rm d}t} = - \sum_b \frac{m_b}{\overline{\rho}_{ab}} v_{\rm sig} {\bf v}_{ab} \cdot \hat{\bf r}_{ab} \nabla_a W_{ab} , \label{eq:artvisc97}
\end{equation}
where overscored quantities are averages. It utilises a signal velocity representing the speed of information propagation between the two states, $v_{\rm sig} = \tfrac{1}{2}  \alpha ( v_{{\rm mhd},a} + v_{{\rm mhd},b}  - \beta {\bf v}_{ab} \cdot \hat{\bf r}_{ab})$, where $v_{\rm mhd}$ is the fast MHD wave speed, and $\alpha=1$ and $\beta=2$ are dimensionless constants. This will generate heat according to
\begin{equation}
 \frac{{\rm d}u_a}{{\rm d}t} = - \frac{1}{2} \sum_b \frac{m_b}{\overline{\rho}_{ab}} v_{\rm sig} ({\bf v}_{ab} \cdot \hat{\bf r}_{ab})^2 \hat{{\bf r}}_{ab} \cdot \nabla_a W_{ab} .
\end{equation}
\citet{hfg13} found that using the harmonic mean instead of the arithmetic mean for $\overline{\rho}_{ab}$ may confer an advantage when there is a large density contrast, as this will give more weight to the lower density region. 

The $\beta$ term accounts for the relative motion of particle pairs, and is important for preventing penetration of particles through each other and maintaining the coherency of shockfronts \citep{monaghan89}. %For hydrodynamics simulations, the sound speed, $c_{\rm s}$, is used, though for MHD, it is more appropriate to replace this by the fast MHD wave speed, $v_{\rm mhd}$ (see Section~\ref{sec:mhdwaves}). 

\subsubsection{Artificial resistivity}
\label{sec:artresis}

Artificial resistivity was developed by \citet{pm04a, pm05}. It adds dissipation to the magnetic field according to
\begin{equation}
 \frac{{\rm d}{\bf B}_a}{{\rm d}t} = \rho_a \sum_b m_b \frac{v_\text{sig}^{\rm B}}{\overline{\rho}_{ab}^2} \left( {\bf B}_a - {\bf B}_b \right) \hat{{\bf r}}_{ab} \cdot \nabla_a W_{ab} , \label{eq:artificialresistivity}
\end{equation}
where the signal velocity for artificial resistivity may have its own dimensionless parameter $\alpha_{\rm B}$, which is analogous to $\alpha$ in artificial viscosity. In Section~\ref{sec:artresis-signalvelocity}, we find that using $v_{\rm sig}^{\rm B} = \tfrac{1}{2} \alpha_{\rm B} ( v_{{\rm mhd}, a} + v_{{\rm mhd}, b})$ is sufficient for capturing magnetic discontinuities, with no need for the $\beta$ term. By inspection with Equation~\ref{eq:brookshaw}, this is equivalent to adding a physical dissipation term $\nabla^2 {\bf B}$ with dissipation parameter $\eta\sim \tfrac{1}{2} v_{\rm sig}^{\rm B} h$. Dissipated energy may be added to the internal energy through
\begin{equation}
 \frac{{\rm d}u_a}{{\rm d}t} = - \sum_b m_b \frac{v_\text{sig}^{\rm B}}{\overline{\rho}_{ab}^2} \left( {\bf B}_a - {\bf B}_b \right)^2 \hat{{\bf r}}_{ab} \cdot \nabla_a W_{ab} .
\end{equation}
Artificial resistivity is applied to both approaching and receding particles, since discontinuities in the magnetic field can occur during both compression and rarefaction, and to all components of the magnetic field (rather than just along the line of sight like artificial viscosity) since magnetic discontinuities can occur oblique to the motion \citep{pm04a, pm05}.

\subsubsection{Thermal conductivity}

In deriving the internal energy equation (Equation~\ref{eq:discretisedenergyeq}), it is assumed that the density is differentiable, and this assumption is carried onto the internal energy (or entropy). Unless treated, this leads to a multivalued pressure at contact discontinuities, causing an artificial surface tension to appear. This can prevent fluid mixing, for example, stifling the formation of Kelvin-Helmholtz instabilities and the breakup of cold clumps of gas falling into a warm environment \citep{agertzetal07}. In some sense, the problem arises because the conservation of SPH is too good. It has no inherent numerical dissipation. This has lead to the belief that SPH cannot handle contact discontinuities \citep[e.g.,][]{sijackietal12, haywardetal13}, however this issue is no different than running SPH without an artificial viscosity and saying it cannot capture shocks. The issue of contact discontinuities can be treated through a simple fix.

\citet{monaghan97} \citep[see also][]{cm97} introduced an artificial conductivity term, where internal energy is diffused according to
\begin{equation}
 \frac{{\rm d}u_a}{{\rm d}t} = - \sum_b \frac{m_b}{\overline{\rho}_{ab}} \alpha_u v_{\rm sig}^u (u_a - u_b) \hat{{\bf r}}_{ab} \cdot \nabla_a W_{ab}(h_a) ,
\end{equation}
where $\alpha_u$ is a dimensionless parameter. Section~\ref{sec:thermcondswitches} discusses choices for the thermal conductivity signal velocity, $v_{\rm sig}^u$. This form is similar to earlier work on heat diffusion \citep{brookshaw85}. Using this will mix energy (or entropy) between particles, mitigating the surface tension effect at contact discontinuities.

Developments have been made towards implementations of SPH that inherently handle pressure discontinuities without the need for artificial dissipation \citep{rt02, sm13, hopkins13}. The idea is to calculate the pressure through an integral representation thereby making no assumptions about its differentiability. For example, the pressure can be obtained through summation of internal energy according to
\begin{equation}
 P_a = \sum_b (\gamma-1) m_a u_a W_{ab}(h_a) ,
\end{equation}
from which suitable equations of motion may be derived which utilise not the density summation, but rather the pressure summation. 

\citet{price08} compared the \citet{rt02} method to standard SPH with an artificial conductivity term, finding that while the \citet{rt02} approach has a more continuous pressure distribution when simulating Kelvin-Helmholtz instabilities, it leads to more particle noise. However, \citet{hopkins13} constructed equations of motion though a Lagrangian derivation which incorporate `pressure $\Omega$' terms to account for variable smoothing lengths, and found that the method works well at simulating Kelvin-Helmholtz and Rayleigh-Taylor instabilities. Notably, \citet{hopkins13} find that it is still better to estimate the density from the standard mass summation, rather than solving for it from the pressure summation. Using the latter may lead to multivalued densities in mixed regions near contact discontinuities.

\citet*{rha10} used the integral representation of \citet{rt02} to set pressures in their OSPH method (Optimised SPH). However, in \citet{rh12} they argue that while this will avoid multivalued pressures by construction and has excellent performance for multiphase flows, it performs poorly for strong shocks (such as a Sedov blast wave). For this reason, their updated SPHS method (the second S is for `switch') uses an artificial conductivity to treat contact discontinuities.

Overall, the pressure summation formulations hold promise as a way to formulate the SPH equations of motion that inherently handles contact discontinuities, though requires more investigation.

\subsection{Reducing artificial dissipation}

The artificial dissipation terms are intended for the smoothing of fluid quantities at the location of shocks and discontinuities. It is unnecessary (and typically undesirable) to add dissipation to regions of the fluid away from discontinuities. Therefore, this lends to the idea of a switch, where if the location of discontinuities can be determined, the dissipation can be activated only in those regions. Most methods regulate the applied dissipation by varying the $\alpha$ and $\alpha_{\rm B}$ parameters.

\subsubsection{Artificial viscosity switches}
\label{sec:artvisc-switches}

The most widely used artificial viscosity switch is that from \citet{mm97}. The idea is to set $\alpha$ individually per particle, which is time integrated according to
\begin{equation}
 \frac{{\rm d}\alpha_a}{{\rm d}t} = \max(- \nabla \cdot {\bf v}_a, 0) - \frac{\alpha - \alpha_0}{\tau} . \label{eq:mm97switch}
\end{equation}
This increases $\alpha$ in regions undergoing compression, reducing it post-shock to its minimum value $\alpha_0$ in a timescale $\tau=h/\sigma_{v} v_{\rm mhd}$. It is important to enforce $\alpha \in [\alpha_0, 1]$, and it is common to use $\alpha_0=0.1$. It is typical to choose $\sigma_{v}=0.1$, which corresponds to decay over five smoothing lengths. This slow decay is important in order to apply dissipation behind the shock front and damp out post-shock oscillations.  The $\alpha$ term in the artificial viscosity is replaced by the average between particle pairs to maintain conservation. In this thesis, we exclusively use the \citet{mm97} switch to reduce artificial viscosity.

\citet{balsara95} introduced an artificial viscosity limiter which reduces dissipation in the presence of shearing flows, and as such is well suited for accretion discs. Defining
\begin{equation}
 f_a = \frac{\vert \nabla \cdot {\bf v}_a \vert}{\vert \nabla \cdot {\bf v}_a \vert + \vert \nabla \times {\bf v}_a \vert + 0.0001 c_{\rm s}/h_a} ,
\end{equation}
the artificial viscosity is reduced by multiplying it by the average $\overline{f}_{ab}$ between each particle pair. The limiter is designed to approach unity in regions of strong compression, yet tend towards zero when strong shearing motions are present. 

\citet{cd10} designed a switch to improve on the \citet{mm97} switch. They found that using ${\rm d}(\nabla \cdot {\bf v})/{\rm d}t$ as the shock detector is better able to distinguish between converging flows and weak shocks. In their method, $\alpha$ is not increased through time integration, but directly by setting
\begin{equation}
\alpha_a = \frac{h_a^2 A_a}{v_{\rm sig}^2 + h_a^2 A_a} , \label{eq:cullendehnenswitch}
\end{equation}
where $A = \xi \max(- {\rm d}(\nabla \cdot {\bf v})/{\rm d}t, 0)$. To reduce dissipation when shearing flows are dominant over convergent flows, they use a limiter, $\xi$, which is similar in form to that of the \citet{balsara95} limiter. Furthermore, $\nabla \cdot {\bf v}$ and its time derivative are estimated with higher order operators in order to avoid false compression detection in strong shear flows. The value of $\alpha$ is set via Equation~\ref{eq:cullendehnenswitch} whenever it exceeds the current value, otherwise it is decayed by integrating
\begin{equation}
\frac{{\rm d}\alpha_a}{{\rm d}t} = - \frac{\alpha - \alpha_0}{\tau} .
\end{equation}
This slow decay is necessary to retain significant values of $\alpha$ behind the shock front to damp post-shock oscillations. An advantage to the \citet{cd10} switch is that $\alpha$ is increased immediately when a shock is detected. Since the \citet{mm97} switch increases $\alpha$ through time integration, \citet{cd10} found that this leads to $\alpha$ peaking behind the shock front. Additionally, \citet{cd10} suggest that their method allows for $\alpha_0=0$, letting artificial viscosity be completely removed in regions away from shocks. 

\citet{rh12} used a switch in their SPHS method that utilises $\nabla (\nabla \cdot {\bf v})$ to locate discontinuities and shocks. It operates similarly to the \citet{cd10} switch. Defining 
\begin{equation}
 A_{\rm SPHS} = 
\begin{dcases}
 \frac{h_a^2 \vert \nabla (\nabla \cdot {\bf v}_a) \vert}{h_a^2 \vert \nabla (\nabla \cdot {\bf v}_a) \vert + h_a \vert \nabla \cdot {\bf v}_a \vert + 0.05 c_{\rm s}} & \nabla \cdot {\bf v}_a < 0 , \\
 0 & {\rm otherwise,}
\end{dcases}
\end{equation}
$\alpha$ is immediately increased to
\begin{equation}
 \alpha_a = A_{\rm SPHS}
\end{equation}
when $A_{\rm SPHS}$ exceeds the current value of $\alpha$, otherwise it is slowly decayed according to
\begin{equation}
 \frac{{\rm d}\alpha_a}{{\rm d}t} =
\begin{dcases}
 - \frac{\alpha_a - A_{\rm SPHS}}{\tau} & \alpha_0 < A_{\rm SPHS} < \alpha_a , \\
 - \frac{\alpha_a - \alpha_0}{\tau} & A_{\rm SPHS} < \alpha_0 .
\end{dcases}
\end{equation}
They enforce $\alpha \in [0, 1]$, letting artificial viscosity be completely switched off similar to \citet{cd10}. Since second derivatives are sensitive to particle disorder, this method requires a good estimate in order to avoid unnecessarily triggering dissipation. \citet{rh12} use a polynomial fit to obtain both the first and second derivatives. This requires the inversion of a $10\times10$ matrix in 3D, with each element requiring a summation over neighbouring particles, and therefore adds significant computational expense. \citet{rh12} use the \citet{balsara95} limiter formed from these higher order derivative estimates. Notably, they use this switch for all other dissipation terms, with the polynomial fit adjusted to the particular fluid quantity.

\subsubsection{Artificial resistivity switches}
\label{sec:artresis-switches}

\citet{pm05} added a switch for artificial resistivity based on analogy to the \citet{mm97} switch for artificial viscosity. In this case,
\begin{equation}
 \frac{{\rm d}\alpha_{{\rm B}, a}}{{\rm d}t} = \max\left( \frac{\vert \nabla \times {\bf B}_a \vert}{\sqrt{\mu_0 \rho_a}} , \frac{\vert \nabla \cdot {\bf B}_a \vert}{\sqrt{\mu_0 \rho_a}} \right) - \frac{\alpha_{\rm B} - \alpha_{{\rm B}, 0}}{\tau} . \label{eq:pm05switch}
\end{equation}
\citet{bkw12} found that $\alpha_{{\rm B},0} = 0$ may be used, as this still lead to satisfactory results in their shock tube and other two-dimensional MHD tests. Using the cosmological Santa Barbara cluster simulation \citep{frenketal99} that included magnetic fields, they found that this choice is optimal to reduce spurious dissipation. \citet{tp13} formulated a new artificial resistivity switch that supersedes the \citet{pm05} switch, the details of which are presented in Chapter~\ref{sec:chapter-switch}.

\subsubsection{Thermal conductivity switches}
\label{sec:thermcondswitches}

Real systems transfer heat between two states of unequal temperature. However, the primary purpose of the thermal conductivity term in SPH is to treat numerical errors, diffusing heat across discontinuities to avoid discontinuous pressures. Various switches have been developed for thermal conductivity. They either vary $\alpha_u$ (typically using the sound speed for the signal velocity) or keep a constant $\alpha_u=1$ and vary the signal velocity.

\citet{price08} introduced a thermal conductivity switch that defines the signal velocity to be
\begin{equation}
 v_{\rm sig}^u = \sqrt{\frac{\vert P_a - P_b \vert}{\overline{\rho}_{ab}}} ,
\end{equation}
such that conductivity is applied only where there is a difference in pressure. In this manner, for jumps in internal energy which are counterbalanced by jumps in density (i.e., the pressure is constant across the interface), internal energy is diffused only until the pressure is equalised. 

\citet{valckeetal10} commented that this signal velocity assumes that regions of lower internal energy will have lower pressure, thereby as energy is transferred into the low energy region, the pressure difference will close. However, if this is not the case, for example with an ideal gas equation of state (i.e., $P = [\gamma - 1] \rho u$) where the low internal energy region may have higher pressure due to a high density, then transferring energy into the low energy will only cause the pressure difference to increase. They suggest modifying the signal velocity according to
\begin{equation}
v_{\rm sig}^u = {\rm sign}[ (P_a - P_b) (u_a - u_b)] \sqrt{\frac{\vert P_a -P_b \vert}{\overline{\rho}_{ab}}} ,
\end{equation}
such that its sign is determined by the pressure and internal energy differences. This may, however, cause heat to transfer from low to high energy states.

\citet{pm05} introduced an artificial conductivity switch based on the second derivative of $u$, whereby $\alpha_{\rm u}$ is time integrated according to
\begin{equation}
 \frac{{\rm d}\alpha_{{\rm u},a}}{{\rm d}t} = 0.1 h_a^2 \vert \nabla^2 u_a \vert - \frac{(\alpha_{{\rm u},a} - \alpha_{{\rm u},0})}{\tau} .
\end{equation}
However, as discussed for artificial viscosity switches based on second derivatives, this requires a good estimate in order to limit the sensitivity to particle disorder. 

The limitation of these approaches is that they do not recognise systems for which pressure gradients are balanced by external forces (such as gravity). This can lead to continual diffusion despite being in hydrostatic equilibrium. In consideration of this, \citet{valdarnini12} set the artificial conductivity's signal velocity to be 
\begin{equation}
v_{\rm sig}^u = \vert {\bf v}_{ab} \cdot \hat{\bf r}_{ab} \vert ,
\end{equation}
which is essentially a measure of the divergence of the velocity. They additionally use a (slightly modified) form of the \citet{pm05} $\alpha_{\rm u}$ switch.

\citet{rh12} set $\alpha_u$ according to the same higher order SPHS switch for artificial viscosity, such that diffusion is only applied in regions of converging flow. This may be more well suited for gravitational systems. In this case, they set the artificial conductivity signal velocity to be the artificial viscosity signal velocity multiplied by a pressure limiter, equivalent to using
\begin{equation}
v_{\rm sig}^u = \frac{\vert P_a - P_b \vert}{P_a + P_b} \left( c_{{\rm s},a} + c_{{\rm s}, b} -  3 {\bf v}_{ab} \cdot \hat{\bf r}_{ab} \right) .
\end{equation}
They enforce the signal velocity to be positive definite by setting it to 0 whenever $c_{{\rm s},a} + c_{{\rm s}, b} - 3 {\bf v}_{ab} \cdot \hat{\bf r}_{ab} < 0$.

\citet{wvc08} used artificial thermal conductivity to model turbulent mixing at the sub-resolution scale, following the assumption of \citet{smagorinsky63} that these effects are primarily diffusive. Their method utilises the absolute value of the velocity difference, and is equivalent to using
\begin{equation}
v_{\rm sig}^u = \vert {\bf v}_a - {\bf v}_b \vert \frac{\overline{h}_{ab}}{\vert r_{ab} \vert}.
\end{equation}
\citet{sws10} modified the method to instead use the trace-free shear tensor, setting 
\begin{equation}
v_{\rm sig}^u = \frac{1}{2} \frac{\left( \vert S^{ij}_a \vert h_a^2 + \vert S^{ij}_b \vert h_b^2 \right)}{\vert r_{ab} \vert} ,
\end{equation}
where
\begin{equation}
S^{ij}_a = \frac{1}{2} \left( \widetilde{S}^{ij}_a + \widetilde{S}^{ji}_a \right) - \delta_{ij} \frac{1}{3}~{\rm Trace}({\bf \widetilde{S}}_a) 
\end{equation}
and
\begin{equation}
\widetilde{S}^{ij}_a = - \frac{1}{\Omega_a \rho_a} \sum_b m_b (v^j_a - v^j_b) \nabla_a^i W_{ab} .
\end{equation}
Using this measure of velocity promotes mixing in shearing flows, with no mixing for compressive or purely rotating flows. They use $\alpha^u=0.1$.

\subsection{Leapfrog time integration}
\label{sec:leapfrog}

The SPMHD equations are time integrated in this thesis using leapfrog integration. This integrator is time reversible and symplectic. Despite being only second order accurate, it has several desirable properties. One is that is it cost effective, requiring only one force evaluation per time step. Contrast that to the two force evaluations required by second order Runge-Kutta methods. Another is that the method is explicit when accelerations have no velocity dependence (such as accelerations arising from pressure or gravity). Perhaps most importantly, it has excellent stability properties as a consequence of its time reversibility. It inherently conserves the energy of the system. Each timestep is a canonical transformation of the discrete Hamiltonian, so that even though the discrete Hamiltonian is only an approximation to the true energy of the system, the area of its phase space is preserved (what is called, `symplectic'). This means that while the errors are second order and it may not produce the exact solution, it does reproduce the qualitative behaviour of the system. This may be of substantially more benefit. For a thorough introduction on time integration schemes and their properties, see \citet{hlw06}. 

In practice, the desirable properties of the leapfrog scheme are not exactly upheld when performing SPMHD simulations. Using variable size timesteps break the time reversibility of the scheme, though there have been attempts to design time-stepping methods which are reversible \citep[e.g.][]{hmm95, pt99}. Letting particles evolve on individual timestep sizes also break its symplectic nature. Furthermore, artificial viscosity introduces velocity-dependent accelerations, and the magnetic field introduces a third variable that depends both upon the velocity and itself. The scheme cannot be fully explicit in such a scenario.

The leapfrog scheme is often written in a staggered way,
\begin{align}
{\bf x}^1 &= {\bf x}^0 + {\bf v}^{1/2} \Delta t, \nonumber \\
{\bf v}^{3/2} &= {\bf v}^{1/2} + {\bf a}({\bf x}^1) \Delta t \nonumber , 
\end{align}
where ${\bf x}$, ${\bf v}$, and ${\bf a}$ are the positions, velocities, and accelerations with superscripts referring to the time step. The timestep size is $\Delta t$. Each position and velocity update utilises the velocity and acceleration, respectively, at the midpoint of the timestep and these are always explicitly available due to the staggered nature in which the variables are updated. 

For SPMHD, the magnetic field is integrated alongside the velocity. The scheme is by necessity modified to be implicit because accelerations arise from both the magnetic field and, due to the artificial viscosity, the velocity. Thus, a predictor-corrector type scheme is used to update the velocity and magnetic field. Starting from the initial state ${\bf x}^0$, ${\bf v}^0$, and ${\bf B}^0$, they are first updated to
\begin{align}
{\bf v}^{1/2} &= {\bf v}^0 + {\bf a}({\bf x}^0, {\bf v}^0, {\bf B}^0) \frac{\Delta t}{2}, \nonumber \\
{\bf B}^{1/2} &= {\bf B}^0 + \dot{\bf B}({\bf x}^0, {\bf v}^0, {\bf B}^0) \frac{\Delta t}{2}, \nonumber \\
{\bf x}^{1} &= {\bf x}^0 + {\bf v}^{1/2} \Delta t, \nonumber 
\end{align}
where $\dot{\bf B} \equiv \partial {\bf B} / \partial t$. The velocity and magnetic field at the end of the timestep are predicted according to 
\begin{align}
{\bf v}^{*} &= {\bf v}^0 + {\bf a}({\bf x}^0, {\bf v}^0, {\bf B}^0) \Delta t, \nonumber \\
{\bf B}^{*} &= {\bf B}^0 + \dot{\bf B}({\bf x}^0, {\bf v}^0, {\bf B}^0) \Delta t. \nonumber
\end{align}
Using the predicted values, the derivatives ${\bf a}$ and $\dot{\bf B}$ are calculated, with the corrector step given by
\begin{align}
{\bf v}^1 &= {\bf v}^{1/2} + {\bf a}({\bf x}^1, {\bf v}^*, {\bf B}^*) \frac{\Delta t}{2}, \nonumber \\
{\bf B}^1 &= {\bf B}^{1/2} + \dot{\bf B}({\bf x}^1, {\bf v}^*, {\bf B}^*) \frac{\Delta t}{2}. \nonumber
\end{align}
The predictor-corrector is iterated until ${\bf v}^*$ and ${\bf v}^1$ converge. 

This scheme corresponds to the Kick-Drift-Kick (KDK) update: velocities are updated half a step, positions a full step, then velocities half a step \citep{quinnetal97, gadget2}. Though equivalent to a Drift-Kick-Drift (DKD) scheme for constant timesteps, it has been demonstrated that there are advantages to using the KDK scheme when using variable timestep sizes based on acceleration. The KDK scheme will base the timestep on ${\bf a}^0$, whereas the DKD update will use the acceleration from half a timestep behind. This leads to the DKD scheme growing errors at four times the rate of the KDK \citep{gadget2}. Additionally, when using individual particle timesteps in a hierarchical block scheme, the KDK scheme will synchronise accelerations for all active particles.

The timestep criterion used in this work is the Courant-Friedrichs-Lewy (CFL) condition \citep{cfl28}, given by
\begin{equation}
\Delta t = C_{\rm cour} \min_a \left(\frac{h_a}{v_{{\rm sig},a}}\right) ,
\end{equation}
where $v_{\rm sig}$ is the maximum signal velocity used in the artificial viscosity given in Section~\ref{sec:artvisc}. We use $C_{\rm cour}\sim0.3$. Physically, this condition ensures that the time resolution is sufficient to capture sound and MHD wave propagation. We also impose the following criterion based on the acceleration,
\begin{equation}
\Delta t = C_{\rm f} \min_a \left(\frac{h_a}{\vert {\bf a}_a \vert}\right)^{1/2} ,
\end{equation}
with $C_{\rm f}\sim0.25$.

%% file: declaration-chapter3.tex
\newpage
{
\chapter*{}

\vspace{-40mm}

\section*{Declaration for Chapter 3}

\subsection*{Declaration by Candidate}

\noindent In the case of Chapter 3, the nature and extent of my contribution to the work was the following:

\vspace{4mm}

\noindent \begin{tabular}{| >{\raggedright}p{11.55cm} | >{\raggedright}p{3.45cm} |}
\hline
{\bf Nature of Contribution} & {\bf Extent of Contribution (\%)} \tabularnewline
\hline
First author of 2012, ``Constrained hyperbolic divergence cleaning for smoothed particle magnetohydrodynamics'', {\it J. Comput. Phys.} {\bf 231}, 7214--7236.
 & 90 \tabularnewline
\hline
\end{tabular}

\vspace{6mm}
\noindent The following co-authors contributed to the work:

\vspace{4mm}

{
\noindent \begin{tabular}{| >{\raggedright}p{3cm} | >{\raggedright}p{8.1cm} | >{\raggedright}p{3.45cm} |}
\hline 
{\bf Name} & {\bf Nature of Contribution} & {\bf Extent of Contribution (\%) for student co-authors} \tabularnewline
\hline
Daniel Price & PhD supervisor & \tabularnewline
\hline
\end{tabular}
}

\vspace{6mm}

\noindent The undersigned hereby certify that the above declaration reflects the nature and extent of the candidate's and co-author's contributions to this work.

% left bottom right top
\noindent \includegraphics[trim=1.2cm 14cm 0.6cm 12cm, clip, width=\textwidth, angle=-1, scale=1.03]{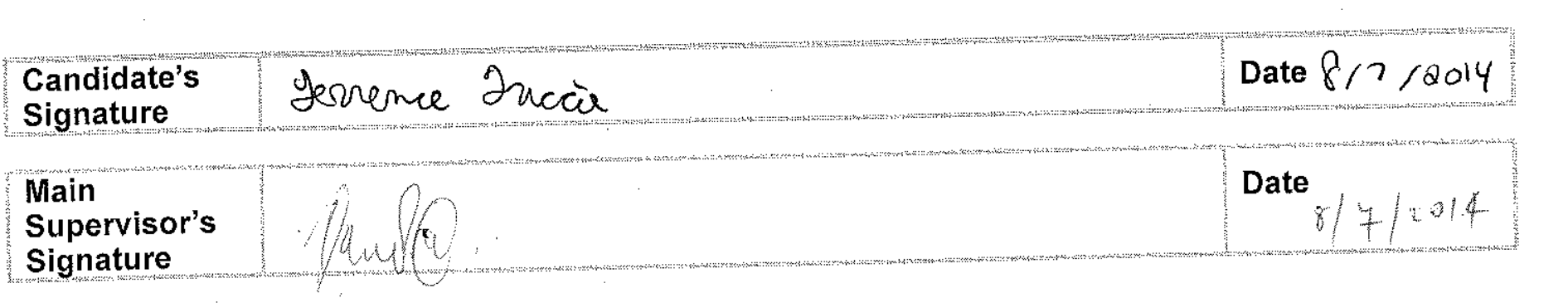}

}

%% file: cleaning.tex
%%
%% Divergence cleaning
%%

\chapter{Constrained hyperbolic divergence cleaning}
\label{sec:chapter-cleaning}

A key problem in numerical magnetohydrodynamics (MHD) is maintenance of the divergence constraint, $\nabla \cdot {\bf B} = 0$ from Maxwell's equations.  If this is not maintained, a spurious force parallel to the magnetic field appears which can lead to numerical instability.  A variety of methods have been developed to combat numerical divergence error, including \citet{bb80} projection method, \citet{eh88} constrained transport, and \citet{powell94} and  \citet{powelletal99}'s eight wave approach or variants thereof.  In general, these methods either aim to ``clean'' any divergence of the magnetic field that has been generated, or to alter the MHD formulation so that the divergence constraint is satisfied by construction.  \citet{toth00} provides an excellent comparison of these schemes.

It is important to consider in which discretisation the magnetic field is considered divergence-free. Even methods such as constrained transport which guarantee divergence-free magnetic fields only do so in a particular discretisation, though if the order of the method is sufficient, a low divergence error in one discretisation will correspond to a low divergence error in other discretions. As such, it is not just the goal of methods to not just keep $\nabla\cdot{\bf B}$ exactly zero in one discretisation, but to prevent the growth of numerical artefacts in different discretisations --- such as those used in the force terms.

\citet{dedneretal02}'s hyperbolic divergence cleaning scheme has found popular use in both Eulerian \citep[i.e.,][]{mt10, wa09} and Lagrangian codes \citep{gn11, pbs11}.  To facilitate cleaning of divergence errors, an additional field $\psi$ is coupled to the magnetic field. The \citet{dedneretal02} scheme was originally adapted to SPMHD by \citet{pm05} (hereafter PM05), but was not adopted for two main reasons: i) the reduction in divergence error was relatively small (a factor of $\sim 2$--$3$) and ii) on certain test cases it was found that it could lead to an increase in the divergence error. As such, its use was not recommended \citep[c.f.][]{price12}.
 
Our aim here is to provide a formulation of hyperbolic divergence cleaning for SPMHD that is guaranteed to be stable and ensures a negative definite contribution to the magnetic energy. This means that the divergence cleaning is guaranteed to decrease the errors associated with non-zero divergence of the magnetic field, in turn leading to a method that is suitable for general use in SPMHD simulations.
 
We begin with a summary of past approaches to handle $\nabla \cdot {\bf B}=0$ in SPMHD. In Section~\ref{sec:hyperbolic}, hyperbolic cleaning as part of the ideal MHD equations is introduced, followed by defining an energy term associated with the $\psi$ field (Section~\ref{sec:continuum-energy-conservation}). Using this energy term, we derive a new form for the $\psi$-evolution equation which conserves total energy (Section~\ref{sec:idealmhdenergy}).  In Section~\ref{sec:discretised-hyperbolic}, the discretisation of hyperbolic cleaning into SPMHD is discussed and we show how the constraint of energy conservation can be used to construct a formulation that is numerically stable.  In particular, this leads to a requirement for the discretisation of $\nabla\cdot{\bf B}$ and $\nabla\psi$ used in the induction and $\psi$-evolution equations to form a conjugate pair (Section~\ref{sec:spmhd-clean-diff} and Section~\ref{sec:spmhd-clean-symm}).  Importantly, we prove that the dissipative (parabolic) term in the evolution of $\psi$ gives a negative definite contribution to magnetic energy (Section~\ref{sec:negdef}).  Our new, constrained formulation of hyperbolic cleaning in SPMHD is then applied to a suite of test problems designed to evaluate all aspects of the algorithm and to derive parameter ranges suitable for general use (Section~\ref{sec:cleaningtests}). The final test (Section~\ref{sec:jet}) is drawn from our current work on applying the method to star formation problems and shows that our technique performs well in practice, dramatically improving the accuracy and robustness of realistic SPMHD simulations in three dimensions. Approaches to enhance the cleaning method are investigated in Section~\ref{sec:cleaning-enhanced}. The cleaning scheme is adapted for use on velocity fields in conjunction with simulations of weakly compressible SPH for the modelling of incompressible fluids (Section~\ref{sec:velclean}). The results are discussed and summarised in Section~\ref{sec:cleaningsummary}.

\section{Previous approaches to treat $\nabla \cdot {\bf B}=0$ in SPMHD}

The divergence-free constraint of the magnetic field has been the main technical difficulty in SPMHD. Evolving the magnetic field directly via the induction equation \citep[as in][]{pm85} places no restriction on the divergence of the magnetic field. Even for a magnetic field that is initially divergence-free, numerical errors will introduce divergence in the field. Therefore, approaches are required to explicitly handle the divergence-free constraint on the magnetic field. 

One class of techniques is to evolve the magnetic field in a way that enforces the divergence-free constraint by construction. Use of the Euler potentials, ${\bf B} = \nabla \alpha \times \nabla \beta$, were proposed as early as \citet{pm85}. Due to the Lagrangian nature of SPH, the scalar variables are advected exactly, which means the magnetic field can be reconstructed simply from the particle positions relative to the initial conditions. This approach has been used in simulations of protostar formation \citep{pb07}, star cluster formation \citep{pb08, pb09}, neutron star mergers \citep{pr06}, and the magnetic fields of galaxies \citep{dp08, kotarbaetal09}. However, the Euler potentials cannot represent certain magnetic field topologies, and winding motions cannot be modelled past one rotation as the field is essentially ``reset'' with each turn \citep{brandenburg10}. It is also difficult to incorporate Ohmic dissipation. 

\citet{price10} investigated use of the vector potential formulation of the magnetic field (${\bf B}=\nabla\times{\bf A}$) as a way to overcome these limitations while still retaining the guarantee of zero physical divergence in the field. However, this results in an even larger instability in the equations of motion, and significant difficulties were found with the time evolution of the vector potential. \citet{price10} concluded that this was not a viable approach. 

The constrained transport method \citep{eh88} enforces $\nabla \cdot {\bf B}=0$ by reconstructing the magnetic field from the electric flux across surfaces. The flux on one side of a surface is exactly balanced by the flux on the other side, therefore if the initial magnetic field is divergence-free, it will remain so. \citet{mvh14} recently proposed a constrained transport implementation for unstructured meshes. However, it is not clear how to adapt constrained transport to SPMHD since there are no clearly defined surfaces.

A second class of techniques is to evolve the magnetic field as normal with the induction equation, then ``clean'' errors out of the field. \citet{morris96} added parabolic diffusion terms to smooth the magnetic field at the resolution scale. The artificial resistivity formulation of \citet{pm04a, pm05} has been used for the same purpose \citep[e.g.][]{burzleetal11b}. However, artificial resistivity is intended for shock capturing, and dissipates both physical and unphysical components of the field. \citet{bot01} used periodic smoothing of the magnetic field to remove fluctuations below the resolution limit, though this adds computational expense and is time resolution dependent. 

\citet{bb80} used a projection method to obtain a divergence-free magnetic field. Considering an ``unclean'' magnetic field, ${\bf B}^*$, it can be written in terms of its physical and unphysical components according to ${\bf B}^* = \nabla \times {\bf A} + \nabla \phi$, where ${\bf A}$ is the vector potential and is the physical portion of the field (since the divergence of the curl is zero). From this, we can state that $\nabla \cdot {\bf B}^* = \nabla^2 \phi$, and then by solving for $\phi$, the divergence-free magnetic field can be obtained from ${\bf B} = {\bf B}^* - \nabla \phi$. \citetalias{pm05} tested this approach for SPMHD, finding that it worked well for simple test problems. The disadvantage to this approach is the computational cost in solving the elliptic set of equations. Even using a tree for efficiency will still have $\mathcal{O}(N \lg N)$ algorithmic complexity. However, it would be worth revisiting this approach and testing it further.

\citet{dedneretal02} coupled parabolic diffusion terms with a hyperbolic set of equations. This improves the effectiveness of the parabolic diffusion, and remains computationally inexpensive. \citetalias{pm05} first investigated its use for SPMHD, however found that it could cause the divergence of the magnetic field to increase in some situations. We here present a conservative implementation of mixed hyperbolic / parabolic cleaning that is constrained to always decrease the divergence of the magnetic field.

\section{Hyperbolic divergence cleaning}
\label{sec:hyperbolic}

\subsection{Hyperbolic divergence cleaning for the MHD equations}

Hyperbolic divergence cleaning involves the introduction of a new scalar field, $\psi$, that is coupled to the magnetic field by a term appearing in the induction equation,
\begin{equation}
 \left( \frac{{\rm d}\bf{B}}{{\rm d}t} \right)_\psi = - \nabla \psi,
\label{eq:induction_psi}
\end{equation}
and the field $\psi$ evolves according to
\begin{equation}
 \frac{{\rm d} \psi}{{\rm d} t} = -c_{\rm h}^2 \nabla \cdot {\bf{B}} - \frac{\psi}{\tau}.
\label{eq:psi_evolution}
\end{equation}
In the comoving frame of the fluid, Equation~\ref{eq:induction_psi} and \ref{eq:psi_evolution} combine to produce a damped wave equation
\begin{equation}
 \frac{\partial^{2} (\nabla \cdot {\bf B})}{\partial t^2} - c_{\rm h}^{2} \nabla^2 (\nabla \cdot {\bf B}) + \frac{1}{\tau} \frac{\partial (\nabla \cdot {\bf B})}{\partial t} = 0. \label{eq:cleaning-waveeqn}
\end{equation}
 The equation above shows that this approach spreads divergence of the magnetic field like a wave away from a source, diluting the initial divergence over a larger area, enabling the parabolic (diffusion) term, $-\psi/\tau$, to act more effectively in reducing it to zero. The wave speed, $c_{\rm h}$, is chosen to be the fastest speed permissible by the time step, typically equal to the speed of the fast MHD wave. A key consideration is setting the damping strength correctly to achieve critical damping of the wave, which maximises the benefit of wave propagation without damping being too weak.  \citet{dedneretal02} suggested using $1/\tau = c_{\rm h} c_{\rm r}$ where $c_{\rm r} = 0.18$, though this is problematic as $c_{\rm r}$ is not a dimensionless quantity.  Instead, PM05 define 
\begin{equation}
\frac{1}{\tau} \equiv \frac{\sigma_\psi c_{\rm h}}{h},
\label{eq:sigmadef}
\end{equation}
where $h$ is the smoothing length and $\sigma_\psi$ is a dimensionless quantity specifying the damping strength. \citetalias{pm05} found that optimal cleaning was obtained for $\sigma_\psi \in [0.4,0.8]$ in their tests.  A similar form was also adopted by \citet{mt10} in their Eulerian code, who suggested values $\sigma_\psi \in [0,1]$.

\subsection{Energy associated with the $\psi$ field}
\label{sec:continuum-energy-conservation}

For later purposes it will be useful to define an energy term associated with the $\psi$ field, $e_{\psi}$ (here defined as the energy per unit mass). Specifically, the energy should be defined such that, in the absence of damping terms, any change in magnetic energy should be balanced by a corresponding change in $e_{\psi}$. This is not merely a book-keeping exercise, as it will enable us to construct a formulation of hyperbolic divergence cleaning in SPMHD that is guaranteed to be stable.

If we consider the closed system of equations formed by Equations~\ref{eq:induction_psi} and \ref{eq:psi_evolution}, the total energy of the system can be specified according to
\begin{equation}
E = \int \left[ \frac{B^2}{2 \mu_0 \rho} + e_\psi \right] \rho {\rm d}V.
\label{eq:tote}
\end{equation}
Conservation of energy in this subsystem implies
\begin{equation}
\frac{{\rm d}E}{{\rm d}t} = \int \left[ \frac{\bf B}{\mu_0 \rho} \cdot \left( \frac{{\rm d} \bf B}{{\rm d}{t}} \right)_\psi + \frac{{\rm d} e_\psi}{{\rm d}t} \right]  \rho {\rm d}V = 0,
\end{equation}
where we have used the fact that ${\rm d}(\rho {\rm d}V)/{\rm d}t = 0$.
We assume that the time derivative of $e_{\psi}$ can be related to the time derivative of $\psi$, giving
\begin{equation}
\int \left[ \frac{\bf B}{\mu_0\rho} \cdot \left( \frac{{\rm d} \bf B}{{\rm d}{t}} \right)_\psi + \chi \frac{{\rm d} \psi}{{\rm d} t}  \right] \rho{\rm d}V = 0,
\end{equation}
where $\chi$ is an unspecified variable to be determined. Using Equations~\ref{eq:induction_psi} and \ref{eq:psi_evolution} in the absence of damping gives
\begin{equation}
\label{eq:psi_energy_equal}
 \int \left[ -\frac{\bf{B}}{\mu_0\rho} \cdot \nabla \psi - \chi c_{\rm h}^2 \nabla \cdot {\bf{B}} \right] \rho {\rm d}V = 0.
\end{equation}
Integrating the first term by parts, we obtain
\begin{equation}
 \int \left[ \frac{\psi}{\mu_0\rho}  - \chi c_{\rm h}^2 \right] (\nabla \cdot {\bf{B}}) \rho {\rm d}V -  \frac{1}{\mu_0} \int_s \psi {\bf{B}} \cdot {\rm d}\hat{\bf{s}} = 0.
\label{eq:psi_energy_derivation}
\end{equation}
We take the surface integral in Equation~\ref{eq:psi_energy_derivation} to be zero.  If the bounding surface is taken to be at infinity, then this assumption is reasonable as it should be expected that the amplitude of a divergence wave would be diluted to zero at such a limit.  For closed systems, it is not clear how the surface term should be treated. However, similar surface terms appear in the standard SPH formulation and are treated by the addition of diffusion terms to capture discontinuities \citep{price08}. For this reason we investigated adding an artificial $\psi$-diffusion term to account for $\psi$-discontinuities, but found no particular advantage to using this in practice (see Appendix~\ref{sec:appendix-cleaning-dissipation}).

From Equation~\ref{eq:psi_energy_derivation} we conclude that energy conservation requires $\chi \equiv \psi / \mu_0 \rho c_{\rm h}^2$, and therefore that the specific energy of the $\psi$ field should be defined according to
\begin{equation}
e_\psi \equiv \frac{\psi^2}{2 \mu_0 \rho c_{\rm h}^2}.
\end{equation}

\subsubsection{Energy conservation as part of the ideal MHD equations}
\label{sec:idealmhdenergy}

Considering total energy conservation with hyperbolic divergence cleaning included as part of the set of ideal MHD equations, additional terms relating to ${\rm d}\rho/{\rm d}t$ appear in the preceding analysis (along with kinetic and other energy terms). Any terms not involving $\psi$ do not need to be considered as they conserve energy together \citep[see][]{pm04b}, so energy conservation reduces to the condition
\begin{equation}
%\label{eq:psi_energy_discrete_density_variation}
%\sum_a m_a \left[ \frac{{\bf B}_a}{\mu_0 \rho_a} \cdot \left( \frac{{\rm d}{\bf B}_a}{{\rm d}t} \right)_\psi + \frac{\psi_a}{\mu_0 \rho_a c_{\rm h}^2} \frac{{\rm d}\psi_a}{{\rm d}t} - \frac{\psi_a^2}{2 \mu_0 \rho_a^2 c_{\rm h}^2} \frac{{\rm d}\rho_a}{{\rm d}t} \right ] = 0 .
\label{eq:psi_energy_density_variation}
\int \left[ \frac{\bf B}{\mu_0 \rho} \cdot \left( \frac{{\rm d} \bf B}{{\rm d}{t}} \right)_\psi + \frac{\psi}{\mu_0 \rho c_{\rm h}^2} \frac{{\rm d}\psi}{{\rm d}t} - \frac{\psi^2}{2 \mu_0 \rho_a^2 c_{\rm h}^2} \frac{{\rm d}\rho}{{\rm d}t} \right]  \rho {\rm d}V = 0.
\end{equation}
The first two terms balance each other, however, the third term remains. There are several possible approaches to ensuring total energy conservation with respect to this term. One approach which we explored was to derive the MHD$+$cleaning equations from a Lagrangian that includes the $e_{\psi}$ term. The result is that an additional isotropic pressure term, $-\tfrac12 \psi^{2}/(\mu_{0} c_{\rm h}^{2})$, appears in the momentum equation. Since it is undesirable to change the physical forces in the system, we instead adopt a simpler approach, which is to slightly modify the evolution equation for $\psi$.

From the continuity equation (Equation~\ref{eq:mhdcty}), we can deduce that the third term in Equation~\ref{eq:psi_energy_density_variation} will be balanced by replacing Equation~\ref{eq:psi_evolution} with
\begin{equation}
 \frac{{\rm d} \psi}{{\rm d} t} = -c_{\rm h}^2 \nabla \cdot {\bf{B}} - \frac{\psi}{\tau} - \tfrac{1}{2} \psi \nabla \cdot {\bf v}.
\label{eq:psi_evolution_halfdivv}
\end{equation}

\section{Hyperbolic divergence cleaning in SPMHD}
\label{sec:discretised-hyperbolic}

\subsection{Hyperbolic divergence cleaning in SPMHD}

Hyperbolic divergence cleaning in SPMHD can be constructed for either the difference (Equation~\ref{eq:divb-diff}) or symmetric (Equation~\ref{eq:divb-symm}) measure of $\nabla \cdot {\bf B}$ by using the appropriate operator in Equation~\ref{eq:psi_evolution_halfdivv}. While both measure the divergence of the magnetic field, they do not provide the same measurement.  For example, if a random distribution of particles is given a uniform magnetic field, the difference form will measure precisely zero --- since the magnetic field is equal for all particles --- but the symmetric form will not because it will reflect the disordered particle arrangement. Thus, it may be expected that the difference operator in general gives a more accurate measure of $\nabla\cdot{\bf B}$ and should be the operator used for cleaning.
On the other hand, it is the symmetric form which is used in the momentum equation (Equation~\ref{eq:spmhd-momeq}) and correspondingly in the tensile instability correction (Equation~\ref{eq:borve-divb-subtract}), and cleaning in this operator may be more effective at improving the conservation of momentum. Thus in the context of divergence cleaning, it is not clear {\it a priori} which of the two should be preferred. This is one of the questions we will seek to answer in our tests.

It is also not clear how the operator for $\nabla \psi$ should be chosen. In \citetalias{pm05} a difference operator was used for both $\nabla\cdot{\bf B}$ and $\nabla \psi$. However, the choice of operator for $\nabla\psi$ turns out to be an important issue in ensuring a stable method.

\subsection{Energy conservation of discretised hyperbolic divergence cleaning}
\label{sec:sph_energy_conserv}

The key constraint we wish to impose on our divergence cleaning scheme is that the total magnetic energy should never increase due to cleaning. That is, any magnetic energy transferred into the $\psi$-field should either be conserved or dissipated. Specifically, in the absence of damping terms, the propagation of divergence waves should conserve energy, not only in the continuum limit but also in the discrete system. We can thus use the $e_{\psi}$ derived in Section~\ref{sec:continuum-energy-conservation} to derive stable formulations of hyperbolic divergence cleaning for SPMHD --- for either difference or symmetric $\nabla\cdot{\bf B}$ operators.
  
As in Section~\ref{sec:continuum-energy-conservation}, we first consider only the subsystem formed by Equations~\ref{eq:induction_psi} and \ref{eq:psi_evolution}. This means that for the moment we do not consider additional terms related to ${\rm d}\rho/{\rm d}t$ (these are discussed in Section~\ref{sec:spmhdenergy}). The total energy of the subsystem (Equation~\ref{eq:tote}) can be discretised by writing the integral as a sum and replacing the mass element $\rho {\rm d}V$ with the particle mass $m$, giving
\begin{equation}
 E = \sum_a m_a \left[ \frac{B_a^2}{\mu_0 \rho_{a}} + \frac{\psi_a^2}{\mu_0 \rho_{a} c_{\rm h}^2} \right] .
\end{equation}
Assuming that the total energy of the subsystem is conserved, we have
\begin{equation}
\frac{{\rm d}E}{{\rm d}t} = \sum_a m_a \left[ \frac{{\bf B}_a}{\mu_0 \rho_{a}} \cdot \left( \frac{{\rm d}{\bf B}_a}{{\rm d}t} \right)_\psi + \frac{\psi_a}{\mu_0 \rho_{a} c_{\rm h}^2} \frac{{\rm d}\psi_a}{{\rm d}t} \right ] = 0.
\label{eq:dedtspmhd}
\end{equation}

\subsubsection{Hyperbolic cleaning with difference operator for $\nabla\cdot{\bf B}$}
\label{sec:spmhd-clean-diff}
If we choose to clean using the difference operator for $\nabla \cdot {\bf B}$, then the SPMHD version of Equation~\ref{eq:psi_evolution} in the absence of the damping term is given by
\begin{equation}
\frac{d\psi_{a}}{{\rm d}t} = c_{\rm h}^{2} \frac{1}{\Omega_a \rho_a} \sum_b m_b \left( {\bf B}_a - {\bf B}_b \right) \cdot \nabla_a W_{ab}(h_a).
\label{eq:psievo-divb-diff}
\end{equation}
Using this in Equation~\ref{eq:dedtspmhd}, we have
\begin{equation}
\sum_a \frac{m_a}{\mu_0 \rho_a} {\bf{B}}_a \cdot \left( \frac{{\rm d}{\bf B}_a}{{\rm d}t} \right)_\psi  = - \sum_a \frac{m_a}{\mu_0 \rho_a} \psi_a \frac{1}{\Omega_a \rho_a} \sum_b m_b \left( {\bf B}_a - {\bf B}_b \right) \cdot \nabla_a W_{ab}(h_a) .
\end{equation}
Expanding the right hand side into two separate terms gives
\begin{align}
 \sum_a \frac{m_a}{\mu_0 \rho_a} {\bf B}_a \cdot \left( \frac{{\rm d}{\bf B}_a}{{\rm d}t} \right)_\psi  = &- \sum_a \sum_b \frac{m_a m_b}{\mu_0 \Omega_a \rho_a^2} \psi_a {\bf B}_a \cdot \nabla_a W_{ab}(h_a) \nonumber \\
&+ \sum_a \sum_b \frac{m_a m_b}{\mu_0 \Omega_a \rho_a^2} \psi_a {\bf B}_b \cdot \nabla_a W_{ab}(h_a) ,
\end{align}
where by swapping the arbitrary summation indices $a$ and $b$ in the second term on the right hand side and using the anti-symmetry of the kernel gradient ($\nabla_{a} W_{ab} = -\nabla_{b} W_{ba}$), we can simplify to find
\begin{equation}
 \sum_a \frac{m_a}{\mu_0 \rho_a} {\bf B}_a \cdot \left( \frac{{\rm d}{\bf B}_a}{{\rm d}t} \right)_\psi = - \sum_a \frac{m_a}{\mu_0 \rho_a} {\bf B}_a \cdot \left\{ \rho_a \sum_b m_b \left[ \frac{\psi_a}{\Omega_a \rho_a^2} \nabla_a W_{ab}(h_a) + \frac{\psi_b}{\Omega_b \rho_b^2} \nabla_a W_{ab}(h_b) \right] \right\} .
\end{equation}
This gives the SPMHD version of Equation~\ref{eq:induction_psi} in the form
\begin{equation}
\left( \frac{{\rm d}{\bf B}_a}{{\rm d}t} \right)_\psi = -\rho_a \sum_b m_b \left[ \frac{\psi_a}{\Omega_a \rho_a^2} \nabla_a W_{ab}(h_a) + \frac{\psi_b}{\Omega_b \rho_b^2} \nabla_a W_{ab}(h_b) \right].
\label{eq:gradpsisym}
\end{equation}

Thus, by choosing the difference operator for $\nabla \cdot {\bf B}$, the symmetric operator for $\nabla \psi$ is imposed. That is, the total energy of the hyperbolic divergence cleaning scheme is only conserved if the operators are chosen to form a conjugate pair \citep[c.f.][]{cr99, price10, price12, wpa14}. This is an important improvement over the \citetalias{pm05} implementation which used a difference operator for both. We demonstrate in Section~\ref{sec:cleaningtests} that indeed the use of conjugate operators significantly improves the robustness and stability of our cleaning algorithm in practice.

\subsubsection{Hyperbolic cleaning with symmetric operator for $\nabla\cdot{\bf B}$}
\label{sec:spmhd-clean-symm}
 An energy-conserving formulation can also be constructed for divergence cleaning with the symmetric operator. That is, with Equation~\ref{eq:psi_evolution} discretised according to
\begin{equation}
\frac{d\psi_{a}}{{\rm d}t} = -c_{\rm h}^{2} \rho_a \sum_b m_b \left[ \frac{{\bf B}_a}{\Omega_a \rho_a^2} \cdot \nabla_a W_{ab}(h_a) + \frac{{\bf B}_b}{\Omega_b \rho_b^2} \cdot \nabla_a W_{ab}(h_b) \right],\
\label{eq:psievo-divb-symm}
\end{equation}
the discrete version of Equation~\ref{eq:induction_psi} which must be used to conserve energy is constrained to be
\begin{equation}
\left( \frac{{\rm d}{\bf B}_a}{{\rm d}t} \right)_\psi = \frac{1}{\Omega_a \rho_a} \sum_b m_b \left( \psi_a - \psi_b \right) \nabla_a W_{ab}(h_a),
\label{eq:gradpsidiff}
\end{equation}
which again forms a conjugate pair.

\subsubsection{Hyperbolic cleaning as part of the SPMHD equations}
\label{sec:spmhdenergy}

In Section~\ref{sec:idealmhdenergy}, the evolution equation for $\psi$ was modified to include a $-\tfrac{1}{2} \psi (\nabla\cdot{\bf v})$ term (Equation~\ref{eq:psi_evolution_halfdivv}).  This was done to conserve energy in the presence of ${\rm d}\rho/{\rm d}t$ terms.  The discretised form of $\nabla\cdot{\bf v}$ in Equation~\ref{eq:psi_evolution_halfdivv} should therefore be the same as that used in the SPH continuity equation (Equation~\ref{eq:discretisedcty}), which leads to
\begin{equation}
 - \tfrac{1}{2} \psi_a (\nabla \cdot {\bf v}_a) = \frac{\psi_a}{2\Omega_a\rho_a} \sum_b m_b ({\bf v}_a - {\bf v}_b) \cdot \nabla_a W_{ab} (h_{a}).
\end{equation}

\subsection{Energy loss due to damping}
\label{sec:negdef}
For completeness, it is important to prove that the damping term in Equation~\ref{eq:psi_evolution_halfdivv} will result in a negative definite energy change. Inserting the damping term into the total change of $\psi$ energy, we see that
\begin{equation}
\left(\frac{{\rm d}E}{{\rm d}t}\right)_{\rm damp} = \sum_a m_a \frac{\psi_a}{\mu_0 \rho_a c_{\rm h}^2} \left( \frac{{\rm d}\psi_a}{{\rm d}t} \right)_{\text{damp}} = - \sum_a m_a \frac{\psi_a^2}{\mu_0 \rho_a c_{\rm h}^2 \tau} ,
\end{equation}
which is indeed negative definite. These energy changes could be balanced with equivalent increases in thermal energy to keep the total energy constant. The issue with doing this is that the heat generated is not necessarily deposited in the same location as it was removed from the magnetic field, due to the transport of divergence errors inherent in the hyperbolic cleaning scheme. Thus, we do not add such heat gains as part of our method, although the term above can be used to keep track of the energy loss due to divergence cleaning.

\section{Tests}
\label{sec:cleaningtests}
 We have designed our numerical tests to examine the following key aspects of our constrained hyperbolic divergence cleaning algorithm:
 \begin{enumerate}
 \item[i)] The importance of the energy-conserving, ``constrained'' formulation compared to a non-conservative, or ``unconstrained'', approach,
 \item[ii)] Whether or not cleaning using the symmetric $\nabla\cdot{\bf B}$ operator (Section~\ref{sec:spmhd-clean-symm}) provides any advantage over use of the difference operator (Section~\ref{sec:spmhd-clean-diff}), e.g. by improving momentum conservation,
 \item[iii)] Optimal parameter choices for $\sigma_{\psi}$, 
\item[iv)] The practical effect of including the $-\frac12\psi \nabla\cdot {\bf v}$ term (Equation~\ref{eq:psi_evolution_halfdivv}).  \end{enumerate}
 In particular, we have investigated these aspects both in isolation using simple idealised setups, as well as their combined effects in more realistic 2 and 3D simulations. Our goal is to verify the robustness of the algorithm for practical application in astrophysics, though it offers a general solution to maintaining the divergence constraint in SPMHD. 
 
 As well as examining the divergence of the magnetic field using the operators given by Equation~\ref{eq:divb-diff} and \ref{eq:divb-symm}, we measure the divergence error in the standard manner for SPMHD with the dimensionless quantity,
 \begin{equation}
\frac{h \vert \nabla\cdot{\bf B} \vert}{\vert {\bf B} \vert}.
 \end{equation}
To prevent artificially high values where $\vert {\bf B} \vert \to 0$, a small parameter $\epsilon$ is added to $\vert {\bf B} \vert$ in the denominator, where $\epsilon \sim 1\%$ of the maximum B-field value.  We find this is only necessary for the Orszag-Tang vortex problem.

 All of the tests have been performed using a Leapfrog integrator with magnetic field integrated alongside the velocity (Section~\ref{sec:leapfrog}) and timesteps set according to the standard condition $\Delta t < \min_a (C_{\rm cour} h_{a}/v_{{\rm mhd},a})$, where $C_{\rm cour} = 0.2$ and $v_{{\rm mhd},a}$ is the MHD fast wave speed on each particle. We therefore use $c_{\rm h} = \max_a (v_{{\rm mhd},a})$ in the hyperbolic cleaning, except for the final test (Section~\ref{sec:jet}) where $c_{\rm h}$ is individual to each particle.  The damping parameter is chosen to be $\sigma_\psi=0.4$ ($\sigma_\psi=0.8$ for the star formation test), except for cases when it is varied to find optimal values. Unless otherwise indicated, we use the standard SPH cubic spline kernel for all tests with $\eta_{\rm h} = 1.2$ in Equation~\ref{eq:h} corresponding to $\sim 18$ neighbours in 2D and $\sim 58$ neighbours in 3D. The magnetic field is specified in units such that $\mu_{0} = 1$ in the code \citep[c.f.][]{pm04a}.  Artificial resistivity is only used where noted, in which case it is applied as described in Section~\ref{sec:artresis}.

 \subsection{Divergence advection}
\label{sec:divBadvection}

  The simplest test we consider consists of divergence in the magnetic field artificially induced in the initial conditions and advected by a uniform flow. The test is performed in a two dimensional periodic domain with three dimensional magnetic and velocity fields (2.5D). The first version of this test is identical to the `divergence advection problem' from \citet{dedneretal02}, as generalised by \citetalias{pm05}. We use this to illustrate the basic features of the hyperbolic/parabolic cleaning approach and to examine the optimal choice of $\sigma_\psi$ when the divergence error has a scale comparable to the numerical resolution.

\begin{figure}
 \centering
\includegraphics[width=\textwidth]{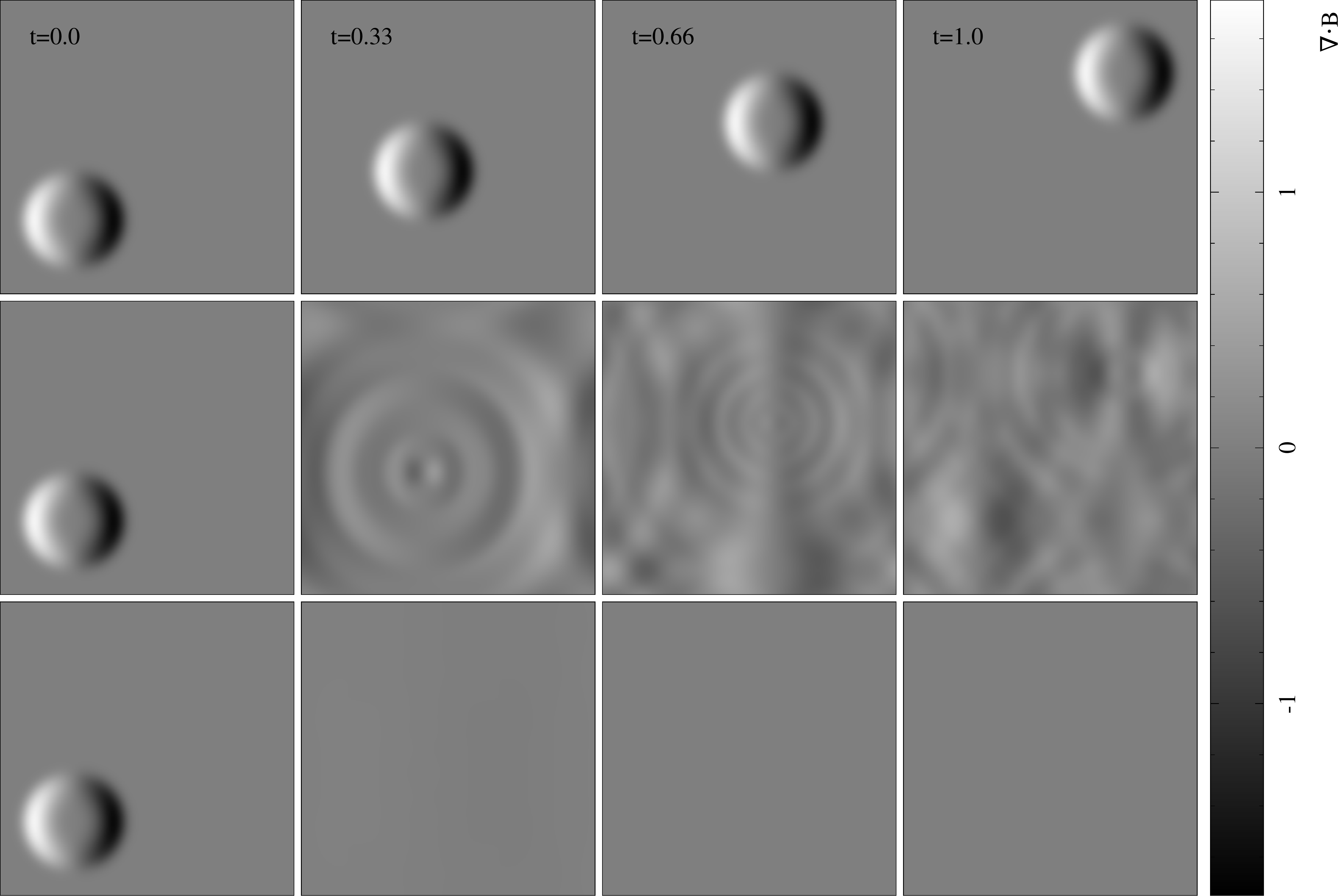}
\caption{A fluid with uniform velocity has a blob of divergence introduced to the initial conditions. In the top row, no cleaning is applied and the divergence blob is advected exactly with the flow.  Undamped cleaning (purely hyperbolic) is applied to the centre row and the divergence in the magnetic field is spread through the system as a system of interacting waves.  In the bottom row, damped cleaning (mixed hyperbolic/parabolic) is utilised and the divergence in the magnetic field is rapidly removed.}
\label{fig:advection}
 \end{figure}

\subsubsection{Setup}
%% The tolerance set on h iterations is 10^-8. This is important! At 10^-4, the divergence will start creeping upwards, whereas at 10^-8, it will continue decreasing. 
The domain is a square area of fluid in the region $x,y \in [-0.5, 1.5]$.  The system has uniform density $\rho = 1$, with pressure $P = 6$ and $\gamma = 5/3$.  The velocity field is ${\bf v} = [1,1]$ and $B_z = 1 / \sqrt{4 \pi}$.  A perturbation is created in the x-component of the magnetic field of the form
\begin{equation}
\label{eq:adv-divergence-perturbation}
B_x = \frac{1}{\sqrt{4 \pi}} \left[ \left(r / r_0\right)^8 - 2 \left( r/r_0 \right)^4 + 1 \right]; \hspace{1cm} r < r_{0},
\end{equation}
where $r \equiv \sqrt{x^{2} + y^{2}}$ and $r_0$ specifies the radial extent. We set up the problem using $50 \times 50$ particles on a square lattice, giving $h = 1.2 \Delta x = 0.048$.

\subsubsection{Results}
Figure~\ref{fig:advection} shows renderings of $\nabla\cdot{\bf B}$ at various times from three calculations: no cleaning, undamped cleaning (purely hyperbolic), and damped cleaning (mixed hyperbolic/parabolic) with $r_{0} = 1/\sqrt{8}$, following \citet{dedneretal02}. These three calculations illustrate the basic ideas behind the divergence cleaning scheme: In the absence of any cleaning (top), the magnetic field and its divergence perturbation is advected without change on the particles. With the addition of hyperbolic cleaning (middle), the divergence errors are spread in a wave-like manner throughout the domain. Finally, the addition of the parabolic damping term (bottom row) acts to rapidly diffuse the divergence error to zero.

\begin{figure}
\centering
 \includegraphics[width=0.45\textwidth]{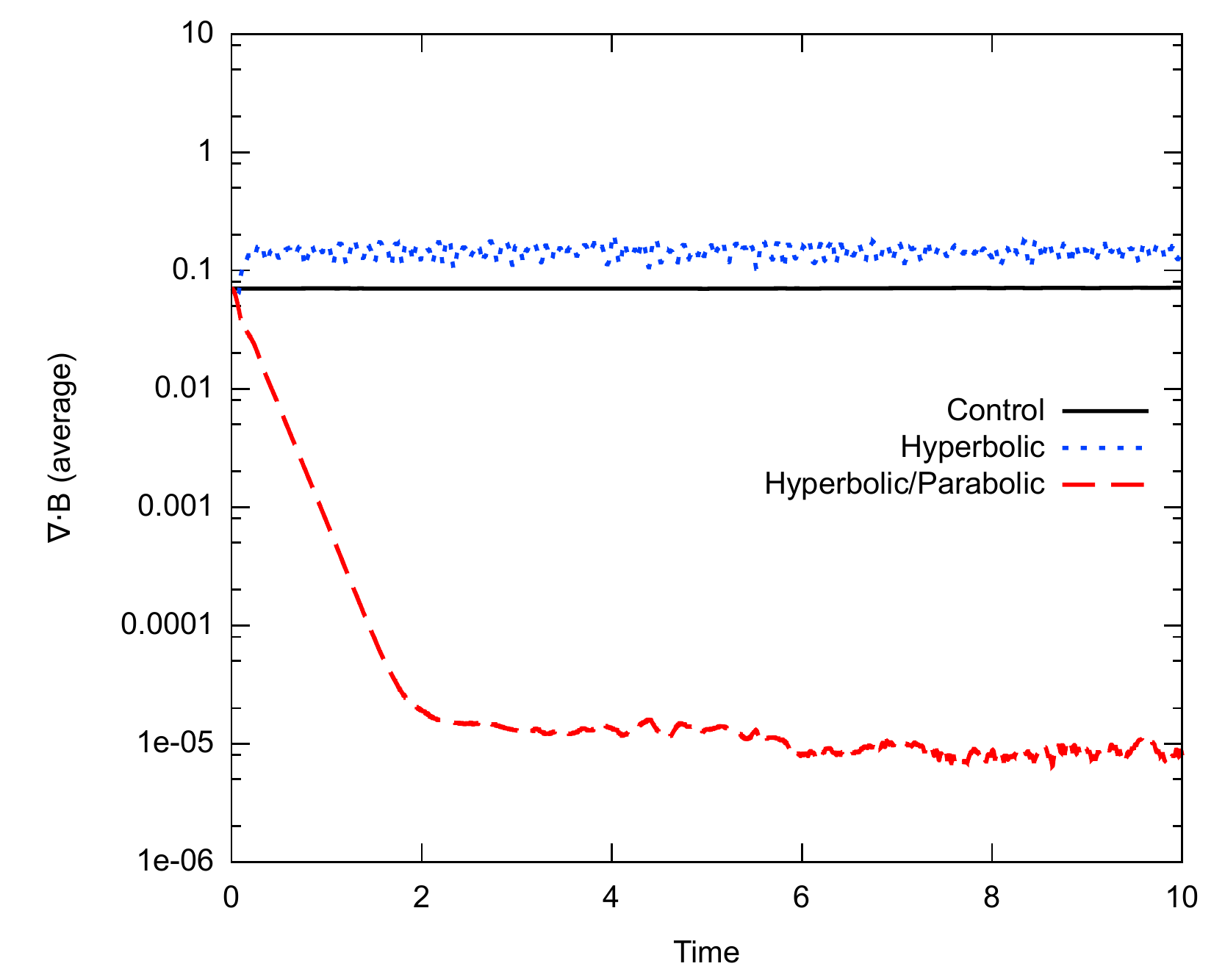}
 \includegraphics[width=0.45\textwidth]{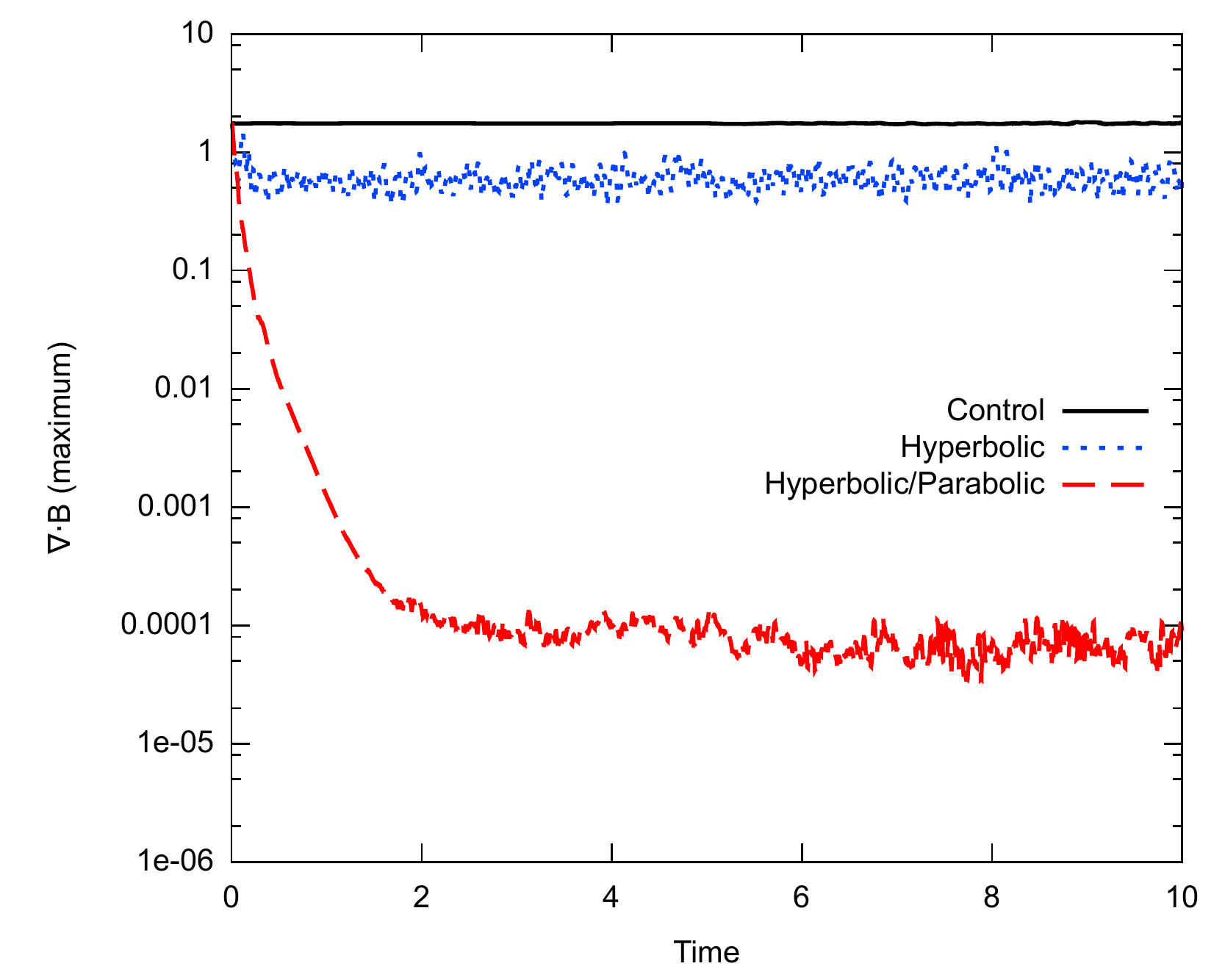}
\caption{Average and maximum $\nabla \cdot {\bf B}$ in code units, measured using the difference operator (Equation~\ref{eq:divb-diff}), as a function of time for the divergence advection test with $r_{0} = 1/\sqrt{8}$.  Without cleaning, the divergence for this simple problem remains constant.  Using undamped cleaning (purely hyperbolic), the maximum divergence is reduced with an increase in average throughout the system.  With damped cleaning (mixed hyperbolic/parabolic), both average and maximum are rapidly reduced.}
\label{fig:advection-divb}
\end{figure}

This is demonstrated more quantitatively in Figure~\ref{fig:advection-divb}, which shows the average and maximum values of $\vert\nabla\cdot{\bf B}\vert$ as a function of time for the three calculations. While purely hyperbolic cleaning can be seen to quickly reduce the maximum divergence error, the average error increases. The parabolic damping means that both the average and maximum values are reduced by an order of magnitude in roughly the time it takes for the hyperbolic waves to cross the simulation domain ($t\sim 0.3$), and by roughly 5 orders of magnitude after several crossing times ($t \gtrsim 2$). After this time the divergence error continues to decrease, but at a much slower rate (this is more obvious in Figure~\ref{fig:tuning} for the case $r_{0} = h$).  We attribute the turnover in the decay rate to the rapid removal of the short wavelength errors by the cleaning scheme, leaving only slowly decaying long-wavelength modes.  This result suggests that such long-wavelength modes will decay on the order of the wave crossing time. We have confirmed this interpretation by verifying that the transition to a slow decay is independent of timestepping, resolution and is similar using the quintic \citep[see][]{price12} instead of the cubic spline kernel. See also Section~\ref{sec:itvovc-static-test} for further examination of this.
  
\subsubsection{Optimal choice of damping parameter in 2D}
\begin{figure}
\centering
 \includegraphics[width=0.45\textwidth]{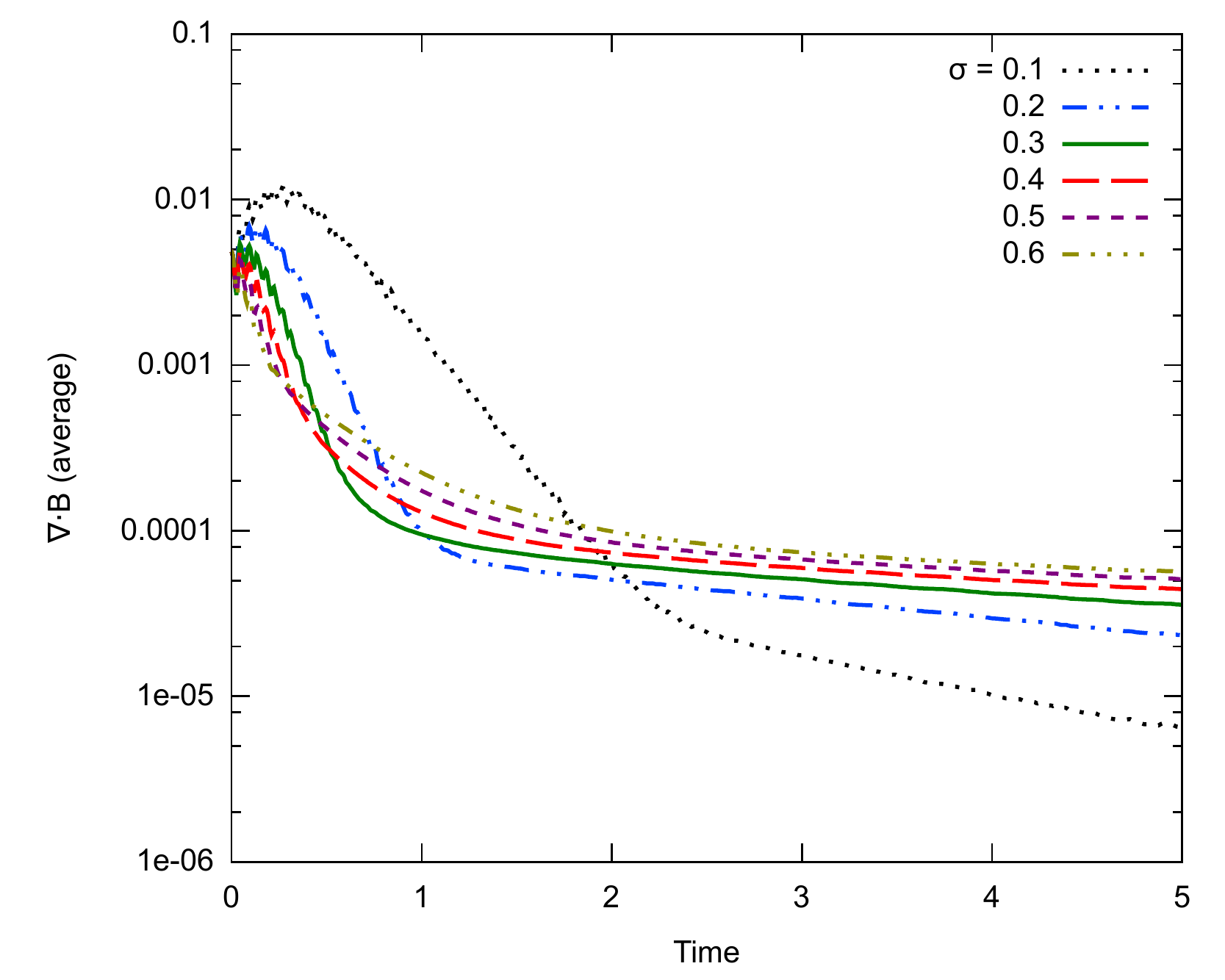}
 \includegraphics[width=0.45\textwidth]{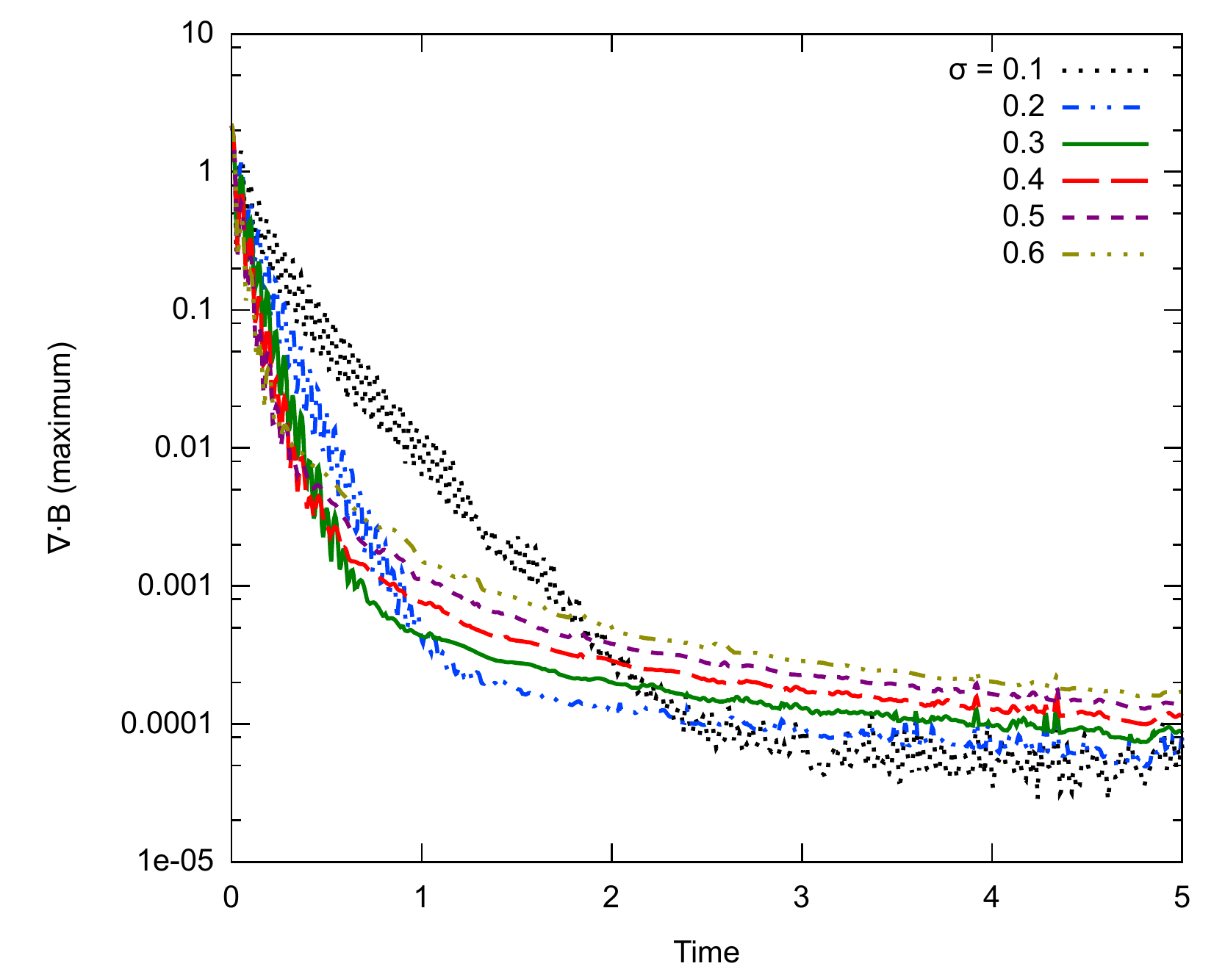}
\caption{The effect of varying the damping parameter $\sigma_\psi$ on the average and maximum $\nabla \cdot {\bf B}$ for the divergence advection test with $r_{0} = h$.  The best results for 2D are obtained for values between 0.2--0.3.}
\label{fig:tuning}
\end{figure}

 As noted by \citetalias{pm05}, the optimal choice of damping parameter, $\sigma_\psi$, for this problem with $r_{0} = 1/\sqrt{8}$ is somewhat misleading, since in reality one expects divergence errors arising in simulations to have length scales of order the smoothing length. Thus, Figure~\ref{fig:tuning} shows the average and maximum $\nabla\cdot{\bf B}$ in a series of calculations employing $r_{0} = h$ and values of $\sigma_\psi$ between 0.1 and 0.6. The results are similar to those shown in Figure~\ref{fig:advection-divb}, with best results obtained in this 2D case using $\sigma_\psi \sim$ 0.2--0.3.

\subsection{Static cleaning test: density jump}
\label{sec:test-density-jump}

\begin{figure}
 \centering
\includegraphics[width=\textwidth]{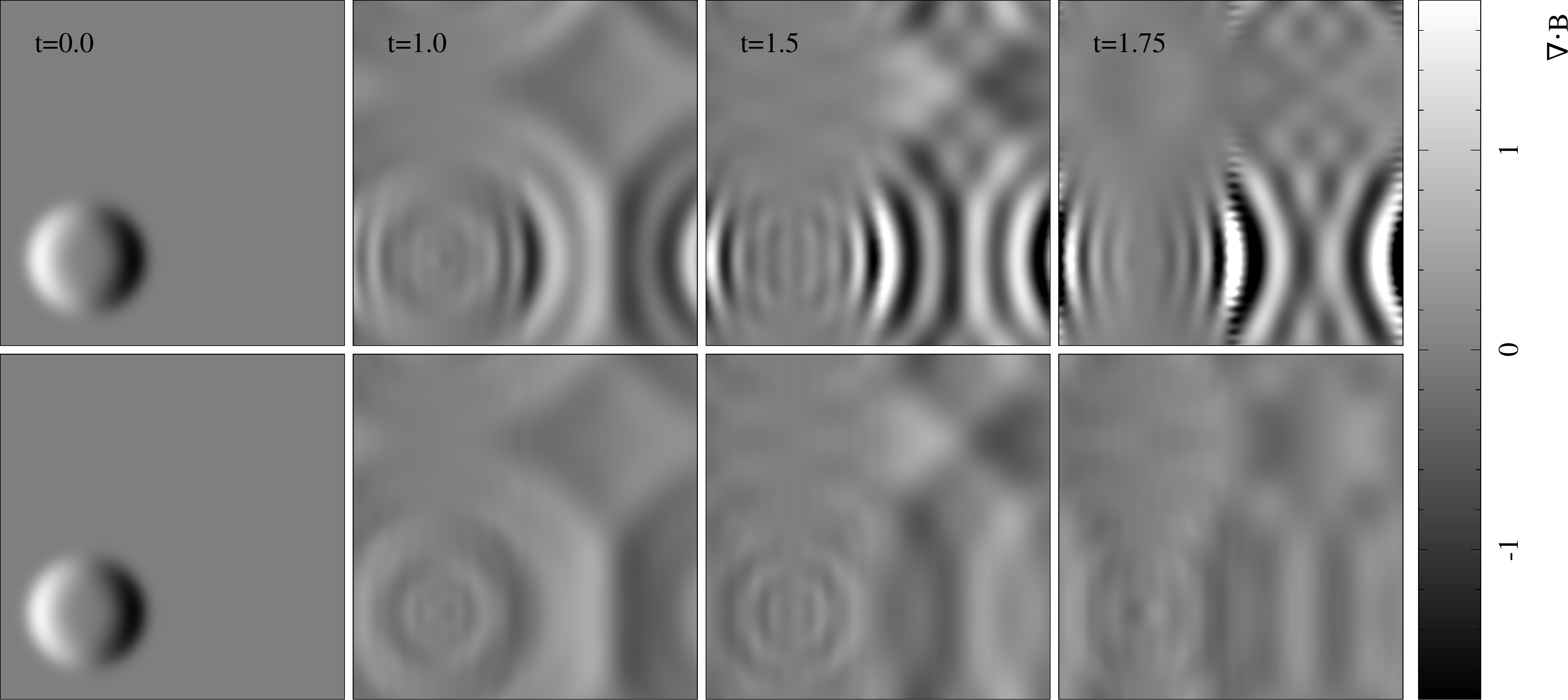}
\caption{Results of the static cleaning test across a 2:1 density jump. Undamped non-conservative cleaning (top) increases the divergence of the magnetic field at the density jump, in turn leading to numerical instability (Figure~\ref{fig:leftright-div-plots}). Using our constrained divergence cleaning method (bottom), the waves cross the density boundary without issue and the scheme remains stable.}
\label{fig:leftright}
 \end{figure}
 
 Our second test is a variant on the divergence advection problem, with identical setup ($r_{0} = 1/\sqrt{8}$) except that the right half of the domain has its density increased by a factor of two. The idea is to examine the reflection and refraction of the divergence waves as they transition between media of differing densities, as may frequently occur in applications of SPMHD. To simplify the test, we solve only the subset of equations given by Equations~\ref{eq:induction_psi}--\ref{eq:psi_evolution} --- that is, the system can only evolve due to divergence cleaning.

\subsubsection{Setup}
 The setup is performed in 2D with $25 \times 50$ particles on a square lattice in the left half of the domain ($x < 0.5$, $\rho = 1$), and $35 \times 70$ particles placed in the right half of the domain ($x > 0.5$, $\rho = 2$), with all particles of equal mass, giving a 2:1 density jump at $x=0.5$. The actual density on the particles is found in the usual manner by iterating the smoothing length and density self-consistently as described in Section~\ref{sec:spmhd-h}. The velocity field is set to zero, all other system parameters are set as previously for the divergence advection test (Section~\ref{sec:divBadvection}), and periodic boundary conditions are employed.

\subsubsection{Results}
\begin{figure}
\centering
 \includegraphics[width=0.45\textwidth]{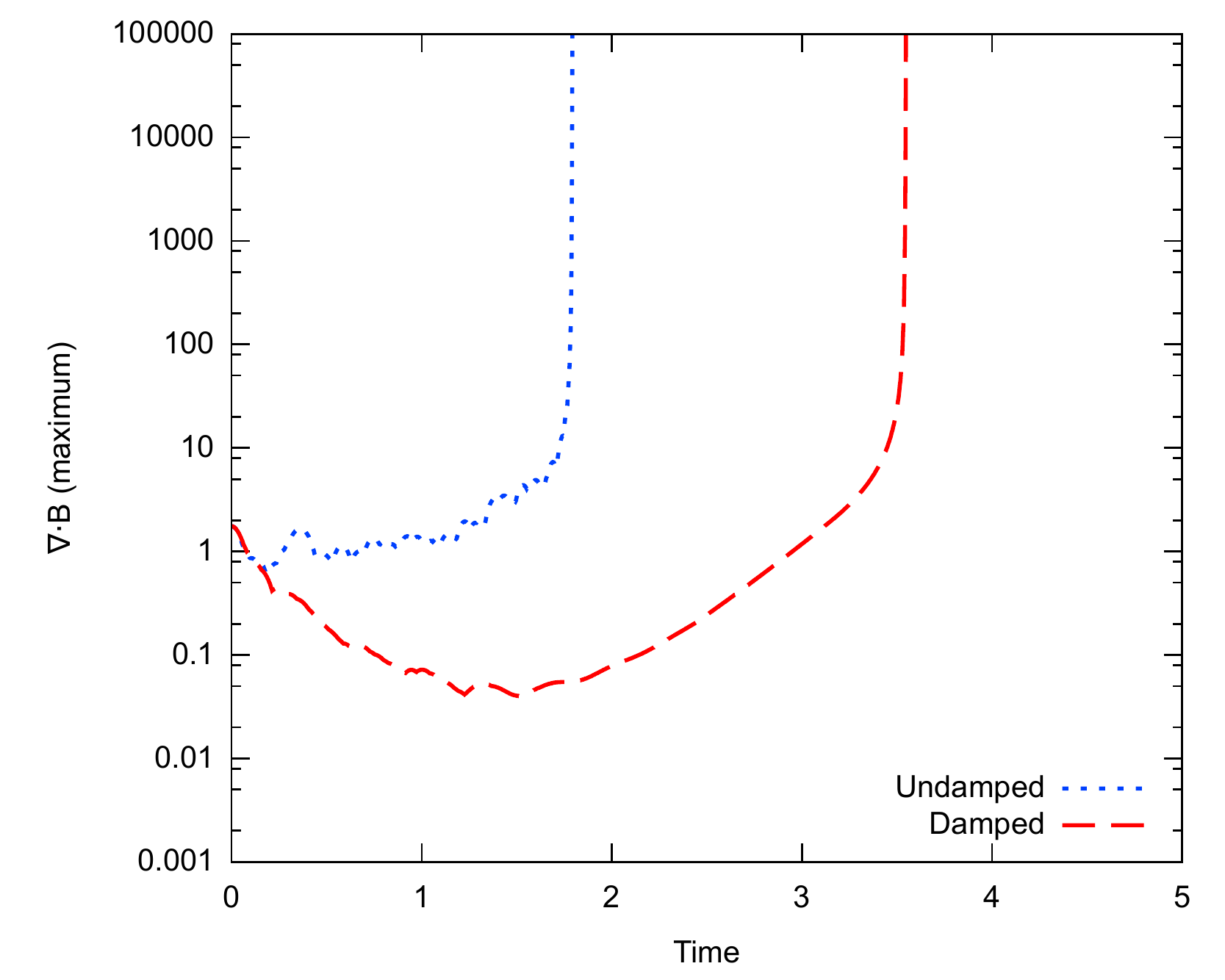}
 \includegraphics[width=0.45\textwidth]{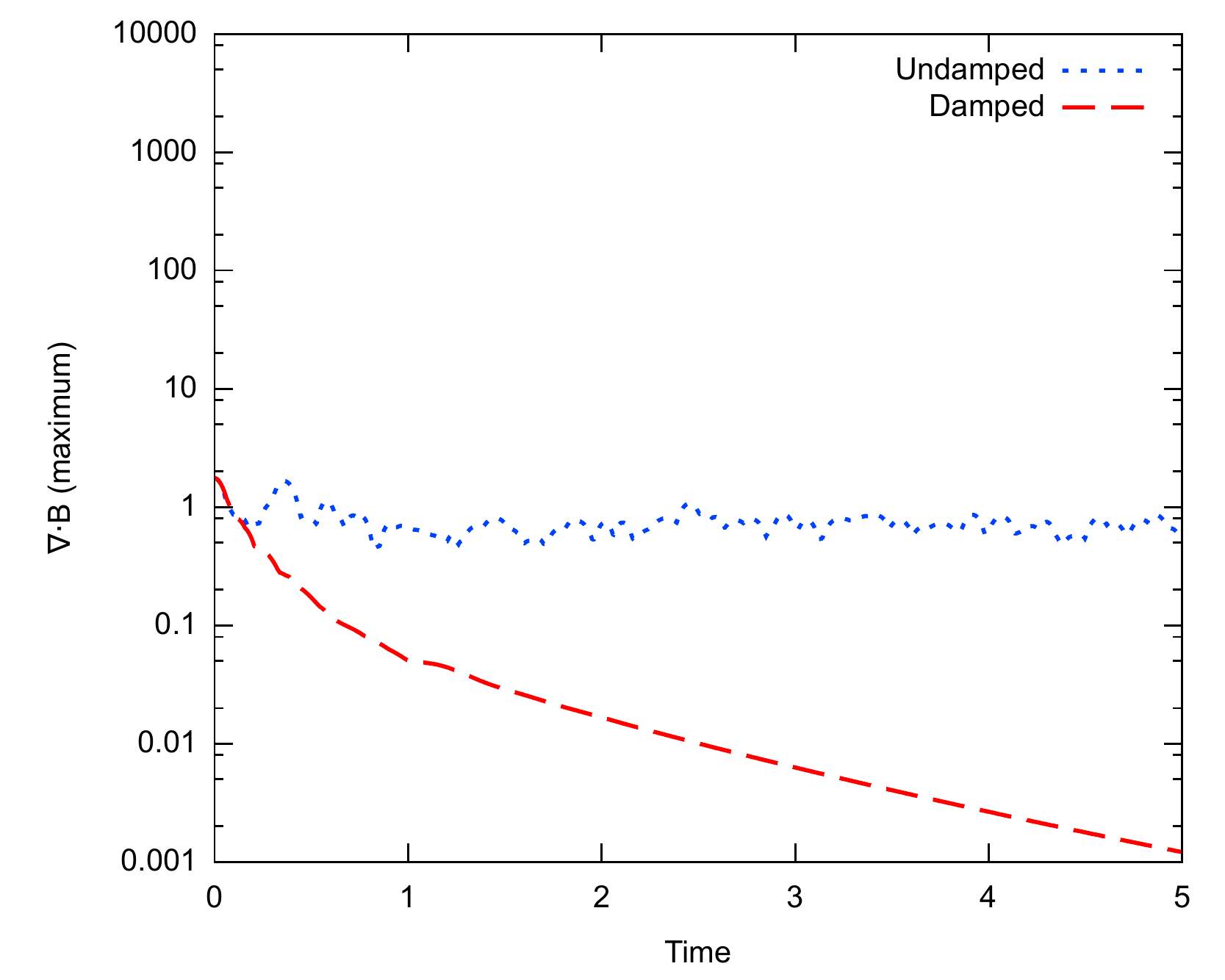}
\caption{Maximum values of $\nabla \cdot {\bf B}$ (difference) for the density jump test for the non-conservative formulation (left) and the new constrained divergence cleaning (right).  The interaction between the divergence waves and the density jump for the non-conservative formulation is unstable, for both damped and undamped cleaning.  Using constrained divergence cleaning is stable across the density jump, with damped cleaning reducing $\nabla \cdot {\bf B}$ as in previous tests.}
\label{fig:leftright-div-plots}
\end{figure}

 Figure~\ref{fig:leftright} shows the propagation of purely hyperbolic ($\sigma_\psi = 0$) divergence waves in this test using i) the non-energy conserving formulation with difference operators for both $\nabla \cdot {\bf B}$ (Equation~\ref{eq:divb-diff}) and $\nabla \psi$ (Equation~\ref{eq:gradpsidiff}), and ii) our new constrained hyperbolic divergence cleaning scheme with a difference operator for $\nabla \cdot {\bf B}$ and the conjugate, symmetric operator for $\nabla \psi$ (Equation~\ref{eq:gradpsisym}). The corresponding time evolution of the maximum $\vert\nabla\cdot{\bf B}\vert$ is shown in Figure~\ref{fig:leftright-div-plots}. Using the unconstrained formulation, the interaction of the divergence wave with the density jump causes amplification of the divergence errors (top row of Figure~\ref{fig:leftright}), in turn leading to exponential growth in the total energy and numerical instability (left panel of Figure~\ref{fig:leftright-div-plots}). By contrast, our new conservative formulation remains stable and continues to reduce the divergence error throughout the domain (bottom row of Figure~\ref{fig:leftright} and right panel of Figure~\ref{fig:leftright-div-plots}).

\subsection{Static cleaning test: free boundaries}
\label{sec:test-free-boundaries}

\begin{figure}
 \centering\includegraphics[width=\textwidth]{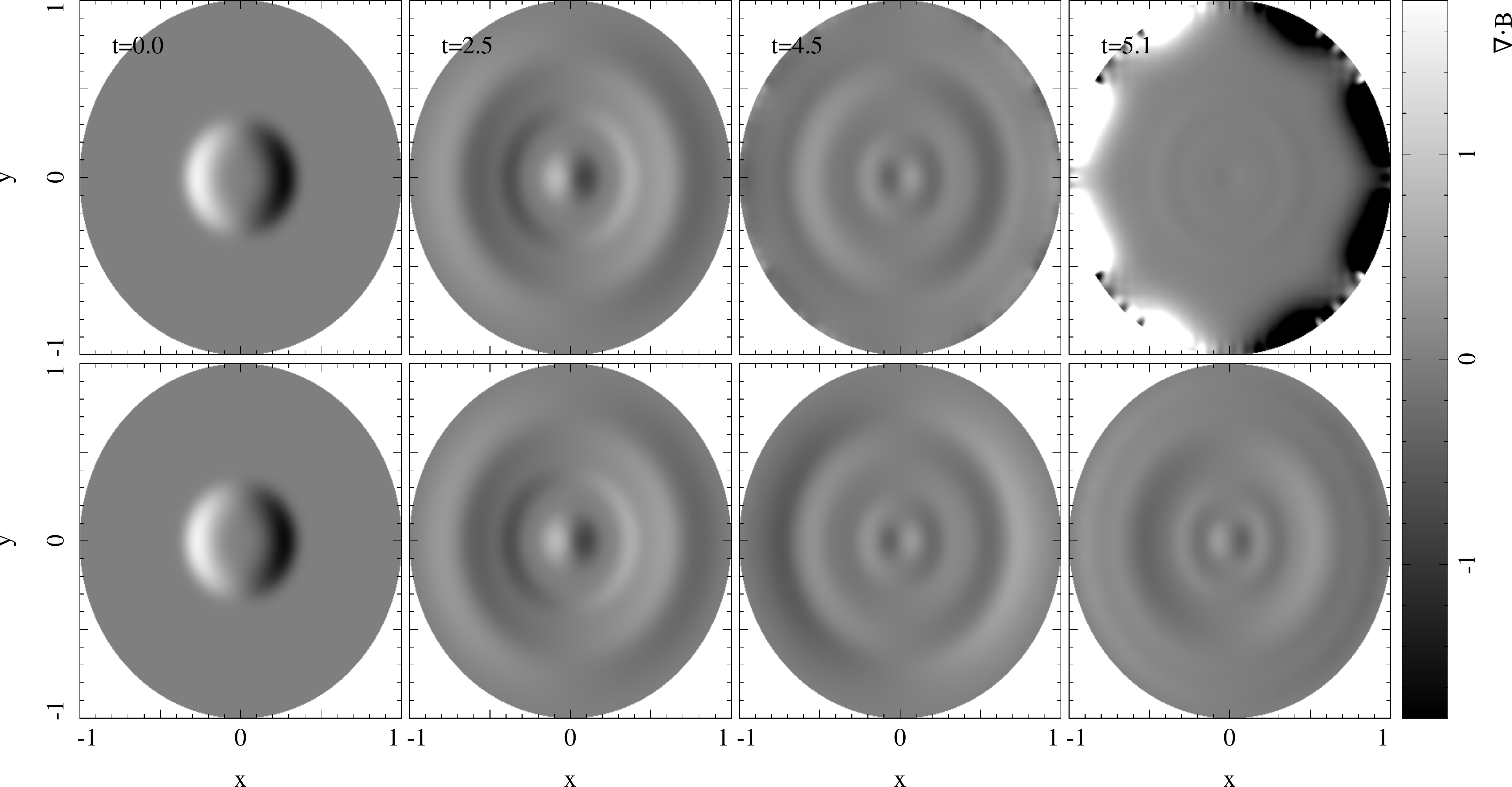}
\caption{$\nabla \cdot {\bf B}$ of the static cleaning test using free boundaries.  In the case of non-conservative cleaning (top row), the interaction of the divergence waves with the boundary cause unchecked divergence growth.  Using constrained cleaning (bottom row), the boundary interaction is not problematic.}
\label{fig:boundary_circle_test}
 \end{figure}

A further variant of the divergence advection test we consider replaces the periodic boundaries by a free boundary, since many applications of SPMHD involve free boundaries (e.g. the merger of two neutron stars, \citealt{pr06}, or studies of galaxy interactions, \citealt{kotarbaetal10, kotarbaetal11}).

\subsubsection{Setup}

The setup is identical to the divergence advection problem (Section~\ref{sec:divBadvection}) with $r_{0} = 1/\sqrt{8}$, except that the domain is a circular area of fluid with $\rho = 1$ for $r \le 1$ and $\rho = 0$ (no particles) for $r > 1$, set up using a total of 1976 particles placed on a square lattice.  The divergence perturbation is introduced at the centre of the circle, and the velocity field is set to zero. Rather than impose an external confining potential, we solve only Equations~\ref{eq:induction_psi}--\ref{eq:psi_evolution} without the full MHD equations, as in Section~\ref{sec:test-density-jump}.

\subsubsection{Results}
 Figure~\ref{fig:boundary_circle_test} shows the results of purely hyperbolic cleaning ($\sigma_\psi=0$) for this case. As in Figure~\ref{fig:leftright}, the top row shows the unconstrained and non-conservative difference/difference formulation, while the bottom row shows results using the conservative difference/symmetric combination. Similar results are also found in this case, with divergence errors piling up at the free boundary in the non-conservative formulation leading to numerical instability, but our constrained formulation remaining stable.

\subsection{2D Blast wave in a magnetised medium}
\label{sec:blast}
We now turn to tests that are more representative of the dynamics encountered in typical astrophysical simulations, beginning with a blast wave expanding in a magnetised medium. In this case the initial magnetic field is divergence-free, meaning that the only divergence errors are those created by numerical errors during the course of a simulation --- rather than the artificial errors we have induced in the previous tests. Based on the results from the previous tests, in this and subsequent tests we apply cleaning \emph{only} using constrained, energy-conserving formulations --- that is, with conjugate operators for $\nabla\cdot{\bf B}$ and $\nabla\psi$. We use this problem to the examine the effectiveness of the divergence cleaning in the presence of strong shocks, as well as to investigate whether cleaning should be performed using the difference or symmetric $\nabla \cdot {\bf B}$ operator.  As with the divergence advection test, a key goal is to find optimal values for the damping parameter $\sigma_\psi$.

\subsubsection{Setup}
\label{sec:blast-setup}

The implementation of the blast wave follows that of \citet{ldz00}.  The domain is a unit square with periodic boundaries, set up with $512 \times 590$ particles on a triangular lattice with $\rho = 1$.  The fluid is at rest with magnetic field $B_x = 10$.  The pressure of the fluid is set to $P = 1$, with $\gamma = 1.4$, except a region of the centre of radius $0.125$ has its pressure increased by a factor of 100 by increasing its thermal energy. An adiabatic equation of state is used.

\subsubsection{Results}

\begin{figure}
 \centering
\includegraphics[width=\textwidth]{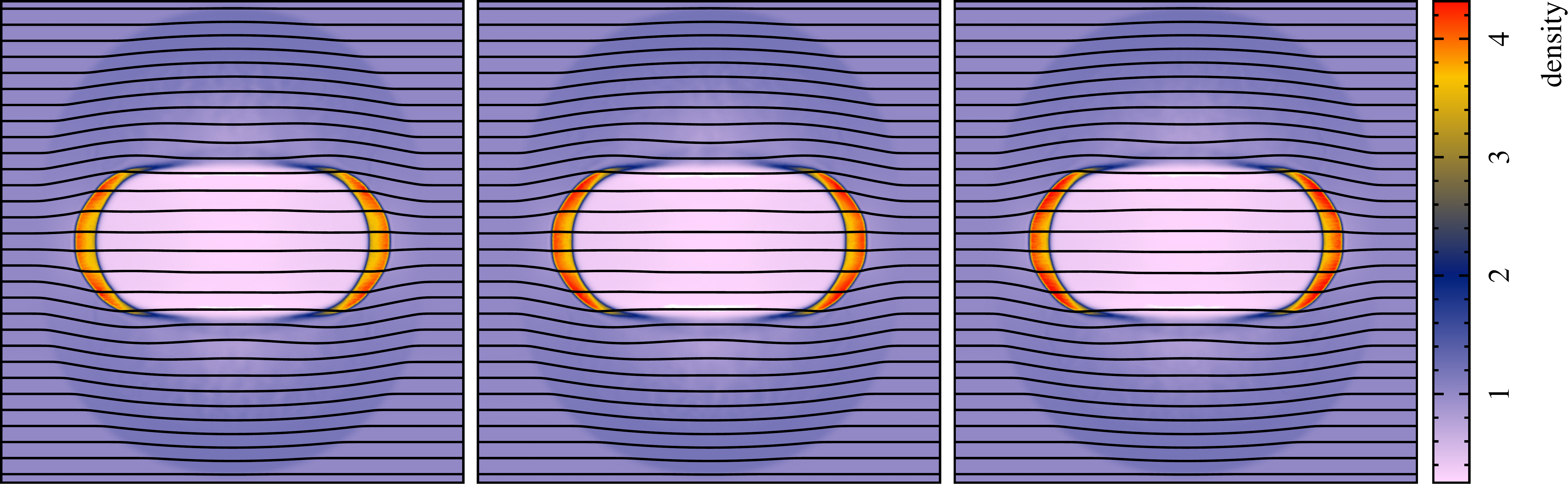}
\caption{Renderings of the density together with overlaid magnetic field lines in the MHD blast wave problem at $t=0.03$, showing the control case with no resistivity and no cleaning (left), with resistivity (centre), and with divergence cleaning (right). Only minor differences in the density evolution are evident.}
\label{fig:blast-compilation-density}
\end{figure}

\begin{figure}
\centering
\includegraphics[width=0.45\textwidth]{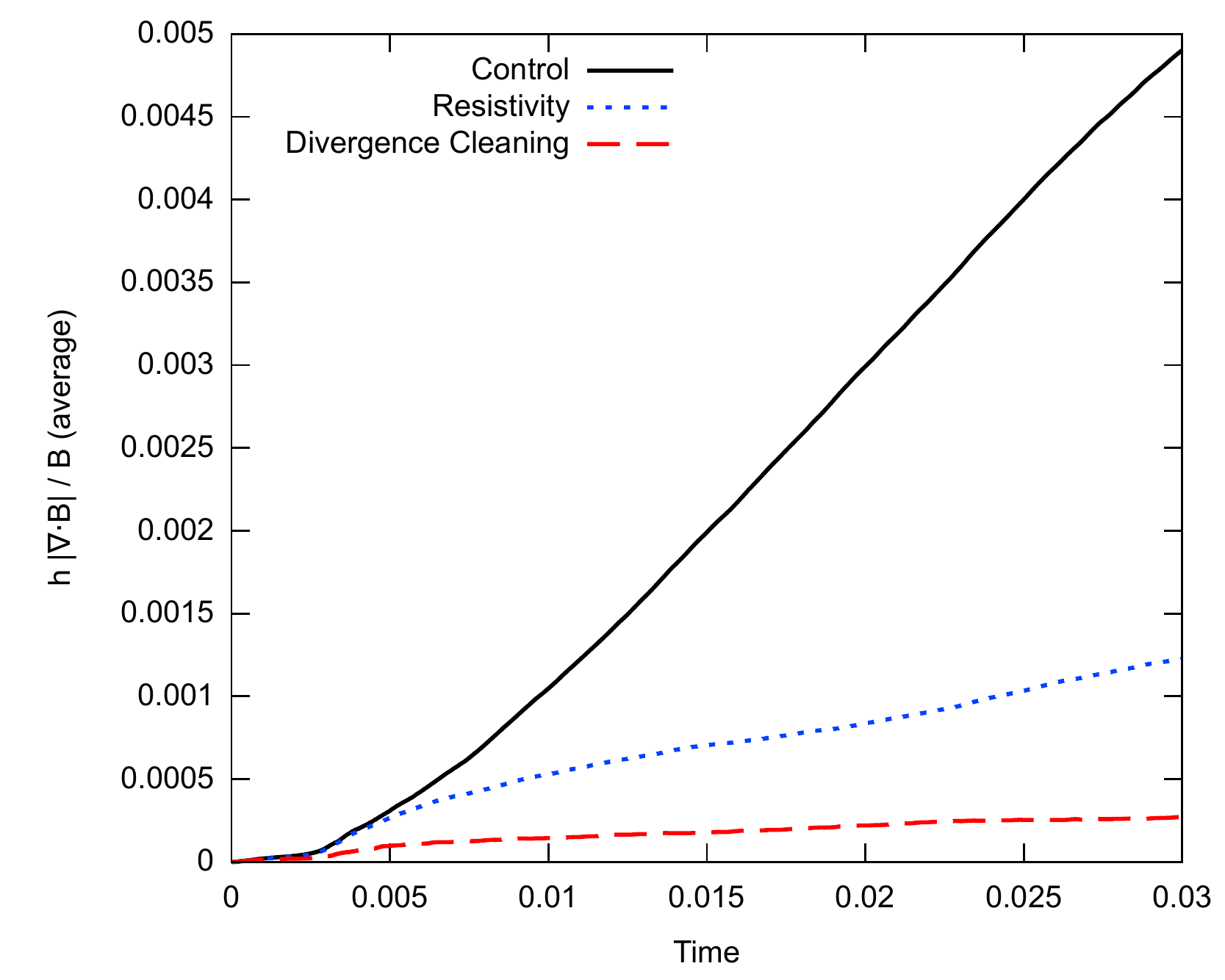}
\includegraphics[width=0.45\textwidth]{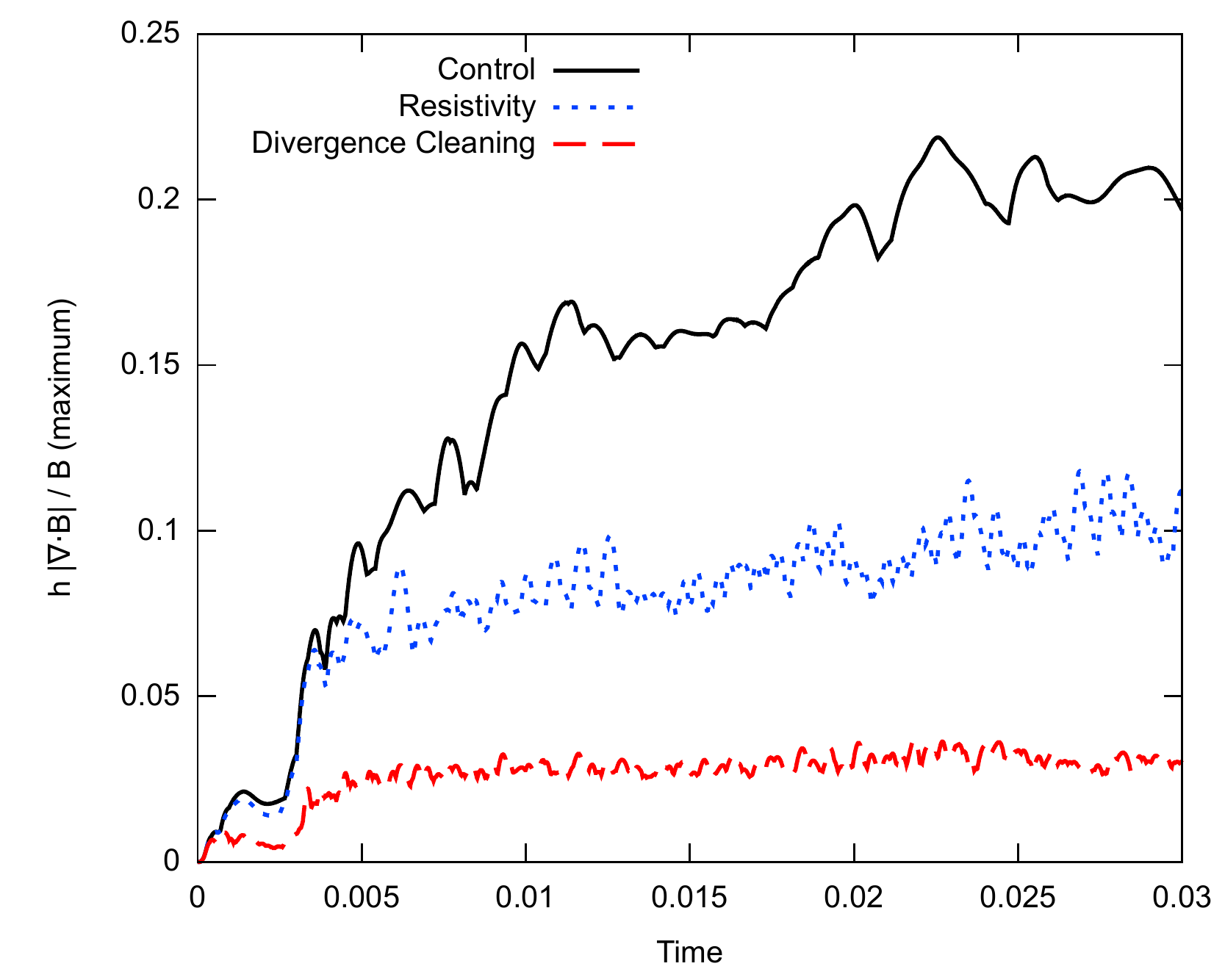}
\caption{Average and maximum of $h \vert \nabla \cdot {\bf B}\vert / \vert{\bf B}\vert$ as a function of time for the blast wave test. At $t=0.03$, resistivity has reduced the average error by a factor of 4 compared to the control case, while divergence cleaning has reduced the average divergence error by a factor of 20. The maximum error has been reduced by a factor of 2 and 8, respectively.}
\label{fig:blast-divb}
\end{figure}

Figure~\ref{fig:blast-compilation-density} shows the density and magnetic field lines at $t=0.03$ for i) the control case without cleaning and no artificial resistivity (left), ii) including artificial resistivity (centre) and iii) no resistivity, but cleaned using the difference operator (right).  At this time, the MHD fast shock has expanded to fill the domain, yet has not crossed the periodic boundaries to begin interacting with itself, and the three cases show only minimal differences in density structure. The average and maximum divergence error as a function of time are shown in Figure~\ref{fig:blast-divb}.  Although the density renderings at $t=0.03$ are quite similar, we can see that adding divergence cleaning has reduced the average and maximum divergence error by a factor of 20 and 8, respectively at $t=0.03$, compared to the control case, with factors of 5 and 4 improvement compared to the case with artificial resistivity alone. Thus, divergence cleaning is even more effective than resistivity at enforcing the divergence constraint.

\subsubsection{Operator choice for $\nabla \cdot {\bf B}$}
\label{sec:blast-symmdivb}

\begin{figure}

 %\centering
\setlength{\tabcolsep}{0.005\textwidth}
\begin{tabular}{cccl}
{\scriptsize Control} & {\scriptsize Difference Cleaned} & {\scriptsize Symmetric Cleaned} & \\
\includegraphics[height=0.305\textwidth]{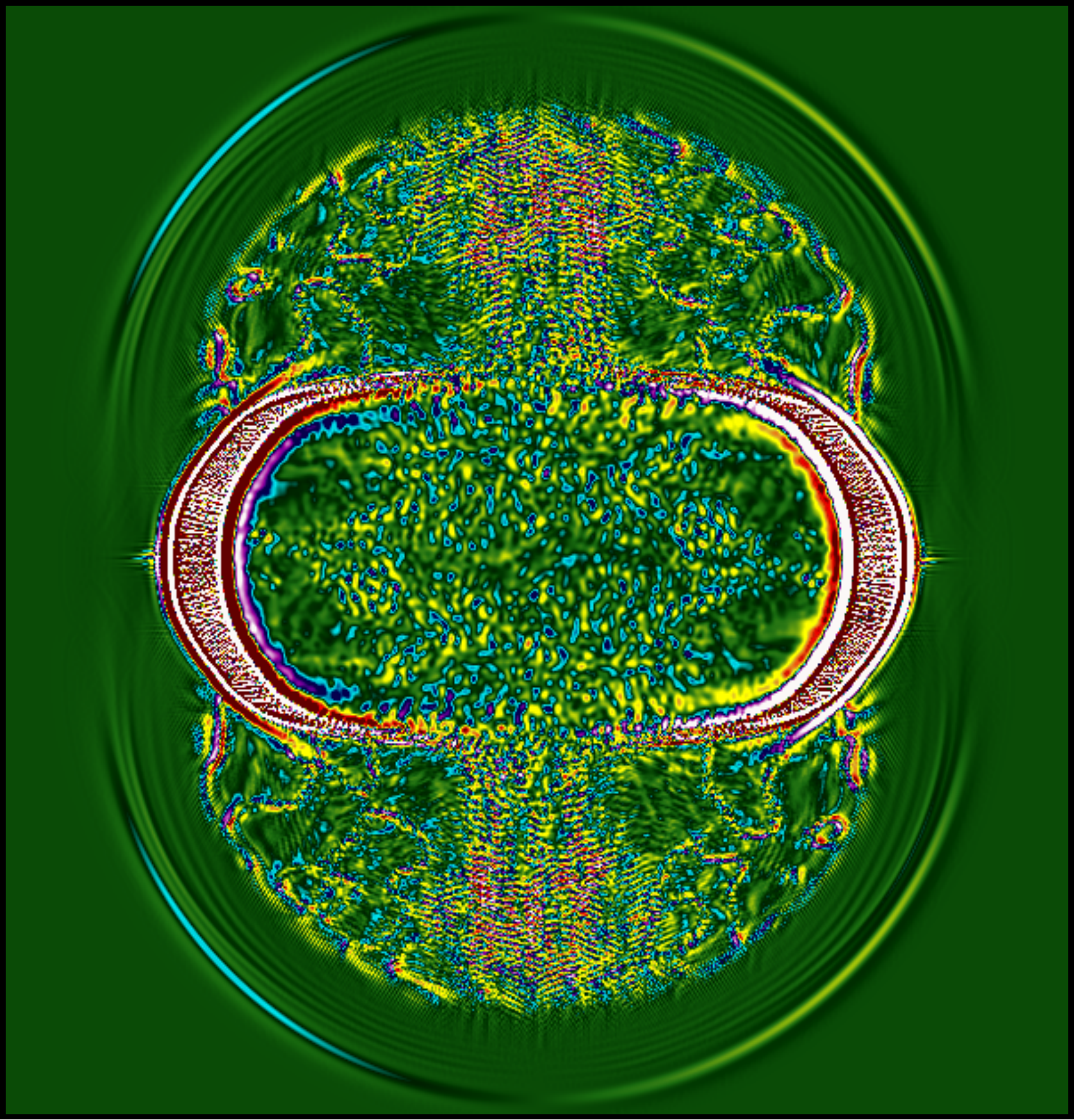} 
 & \includegraphics[height=0.305\textwidth]{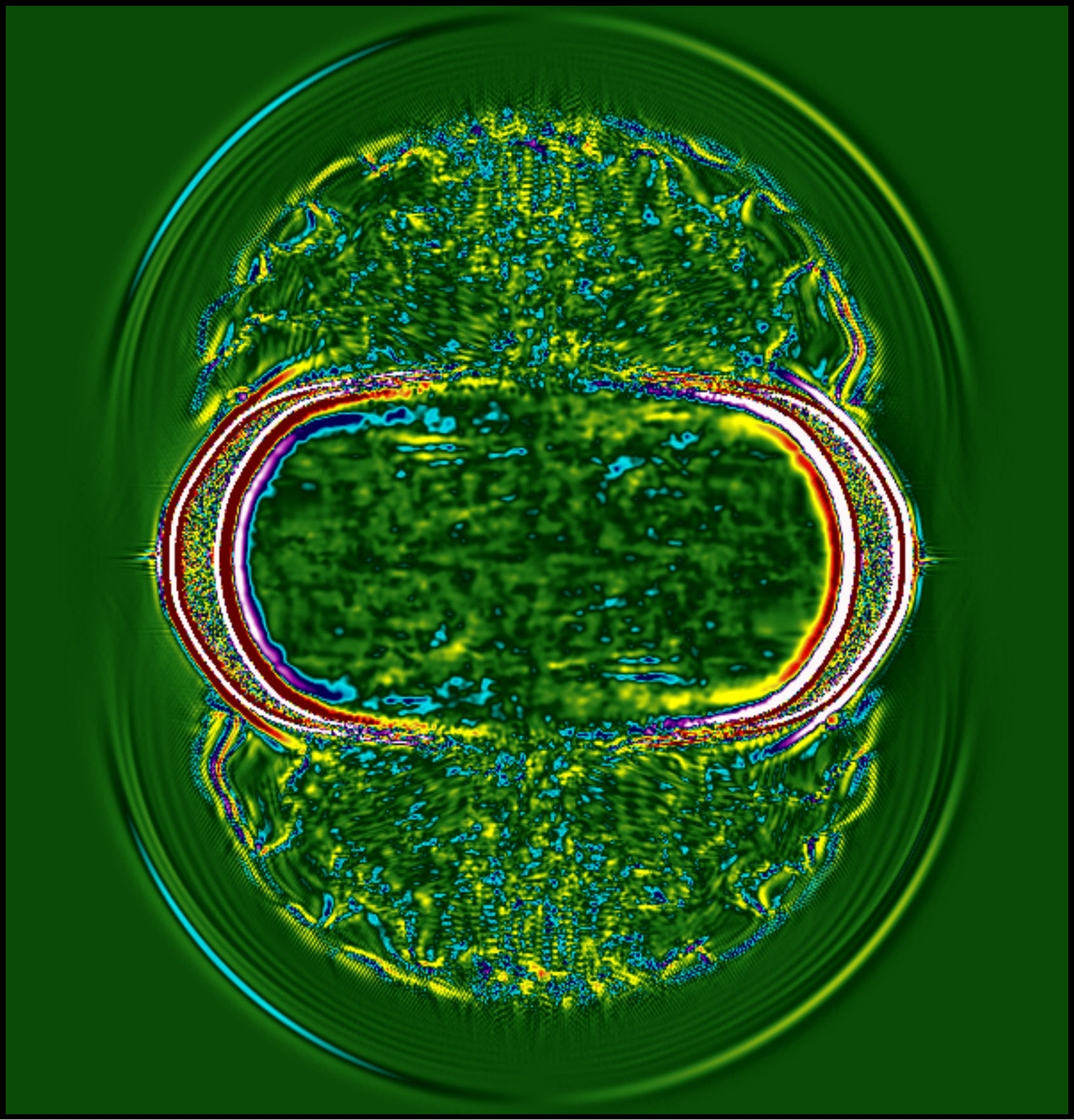}
 & \includegraphics[height=0.305\textwidth]{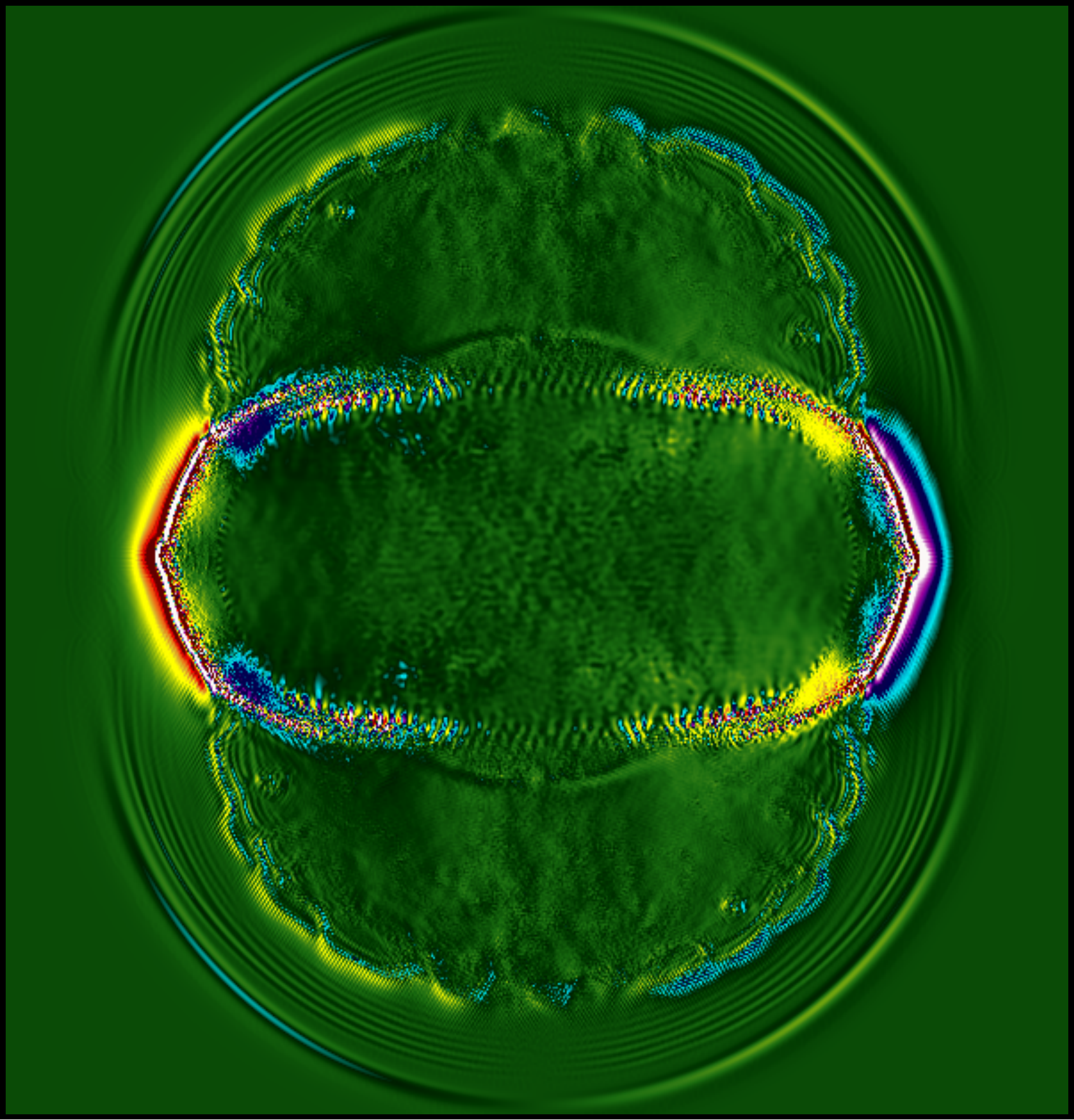}
 & \includegraphics[height=0.305\textwidth]{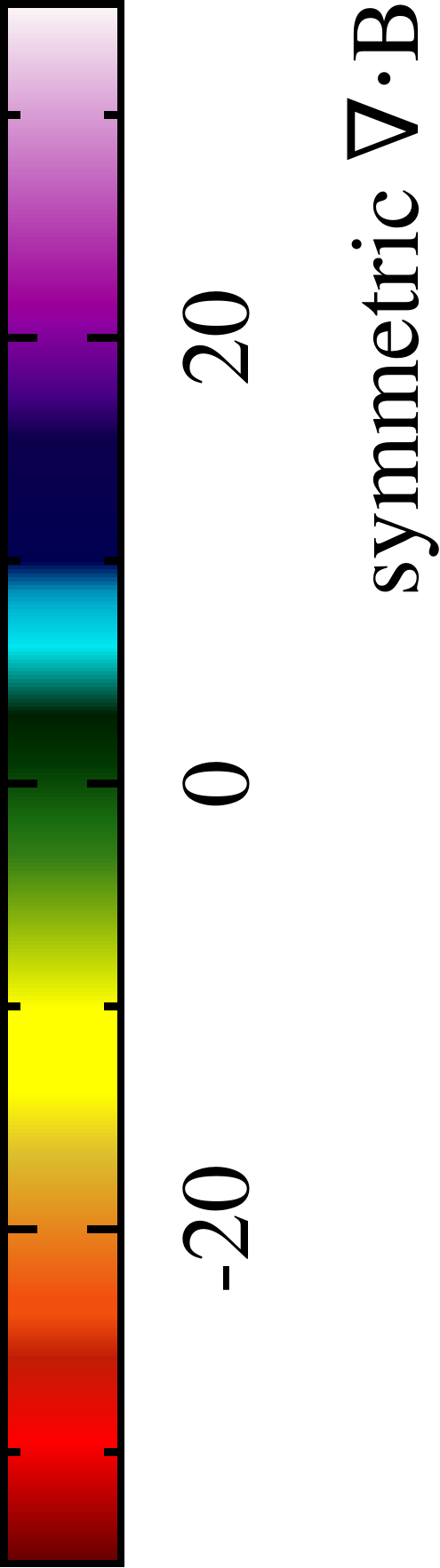}
\end{tabular}
\caption{$\nabla\cdot{\bf B}$ in the blast wave problem at $t=0.03$ measured in code units using the symmetric $\nabla \cdot {\bf B}$ operator, showing the control case (left), $\nabla\cdot{\bf B}$ measured with the opposing operator to that used in the cleaning (centre) and $\nabla\cdot{\bf B}$ measured with the same operator used in the cleaning (right).  Note in particular that the symmetric operator measures a divergence error around the leading edge of the fast MHD wave, even though the field is quite regular.}
\label{fig:blast-compilation-divb-symm}
%\end{figure}
\vspace{5mm}

%\begin{figure}
 %\centering
\setlength{\tabcolsep}{0.005\textwidth}
\begin{tabular}{cccl}
{\scriptsize Control} & {\scriptsize Symmetric Cleaned} & {\scriptsize Difference Cleaned} & \\
\includegraphics[height=0.305\textwidth]{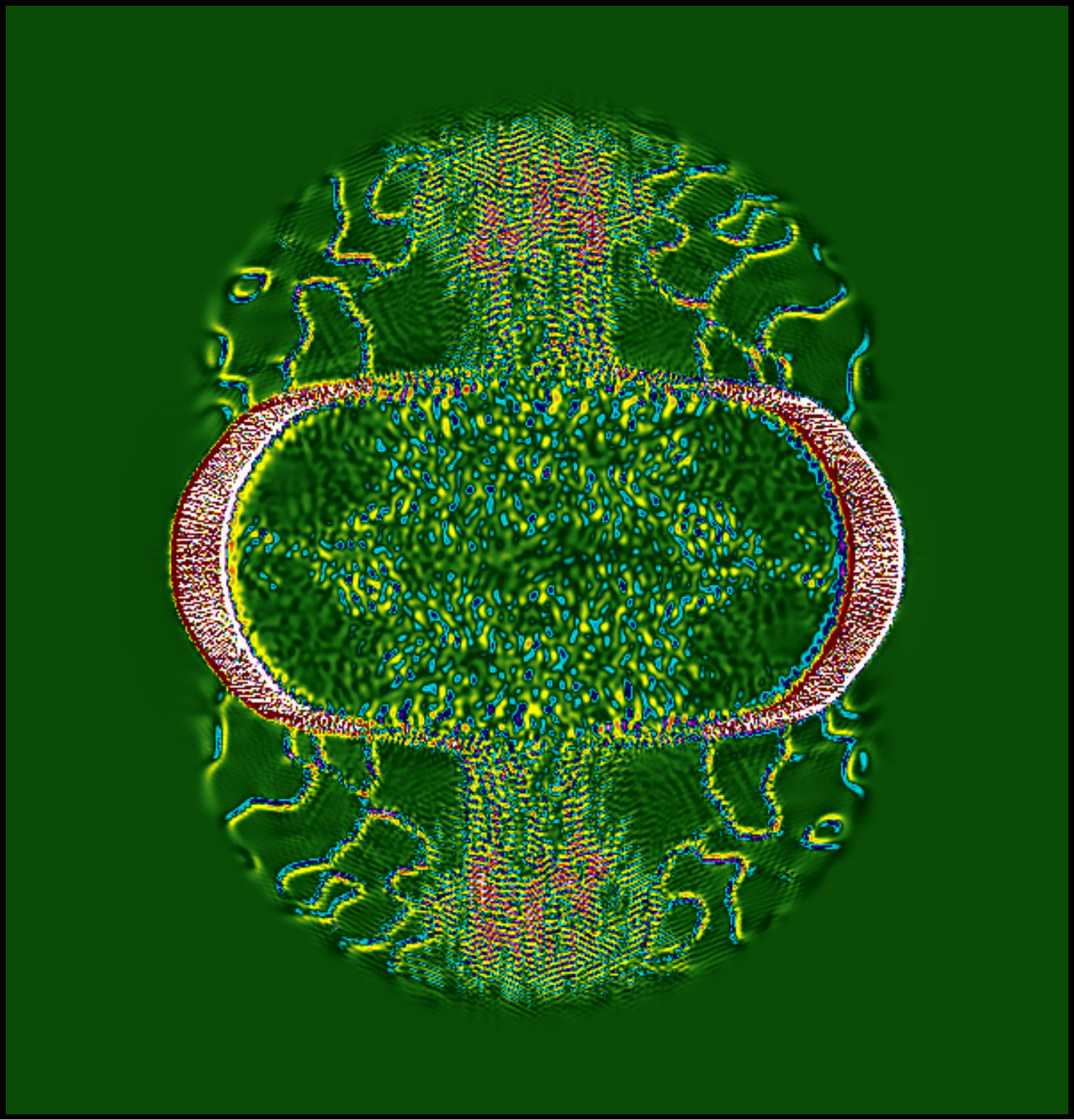} 
 & \includegraphics[height=0.305\textwidth]{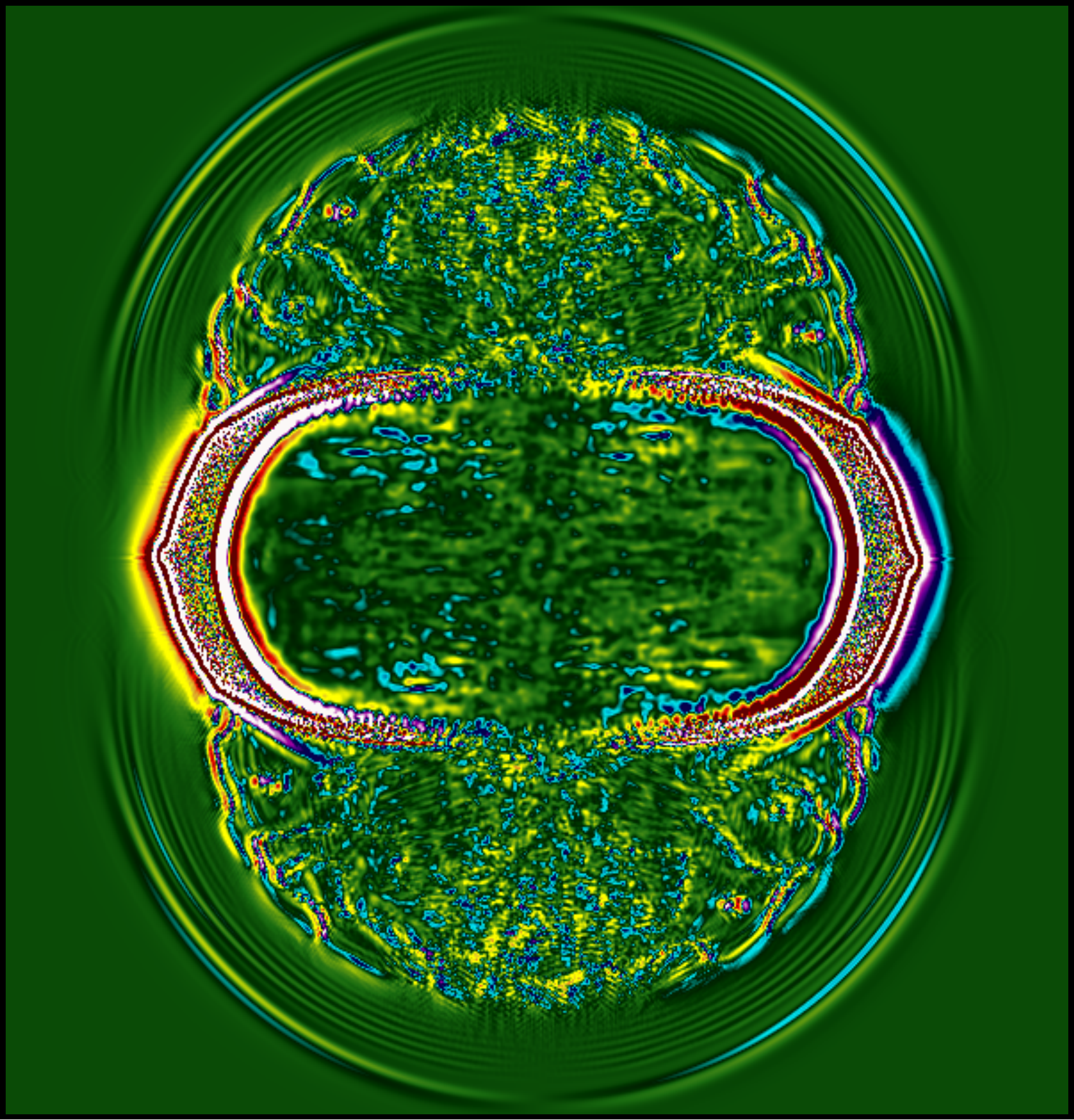}
 & \includegraphics[height=0.305\textwidth]{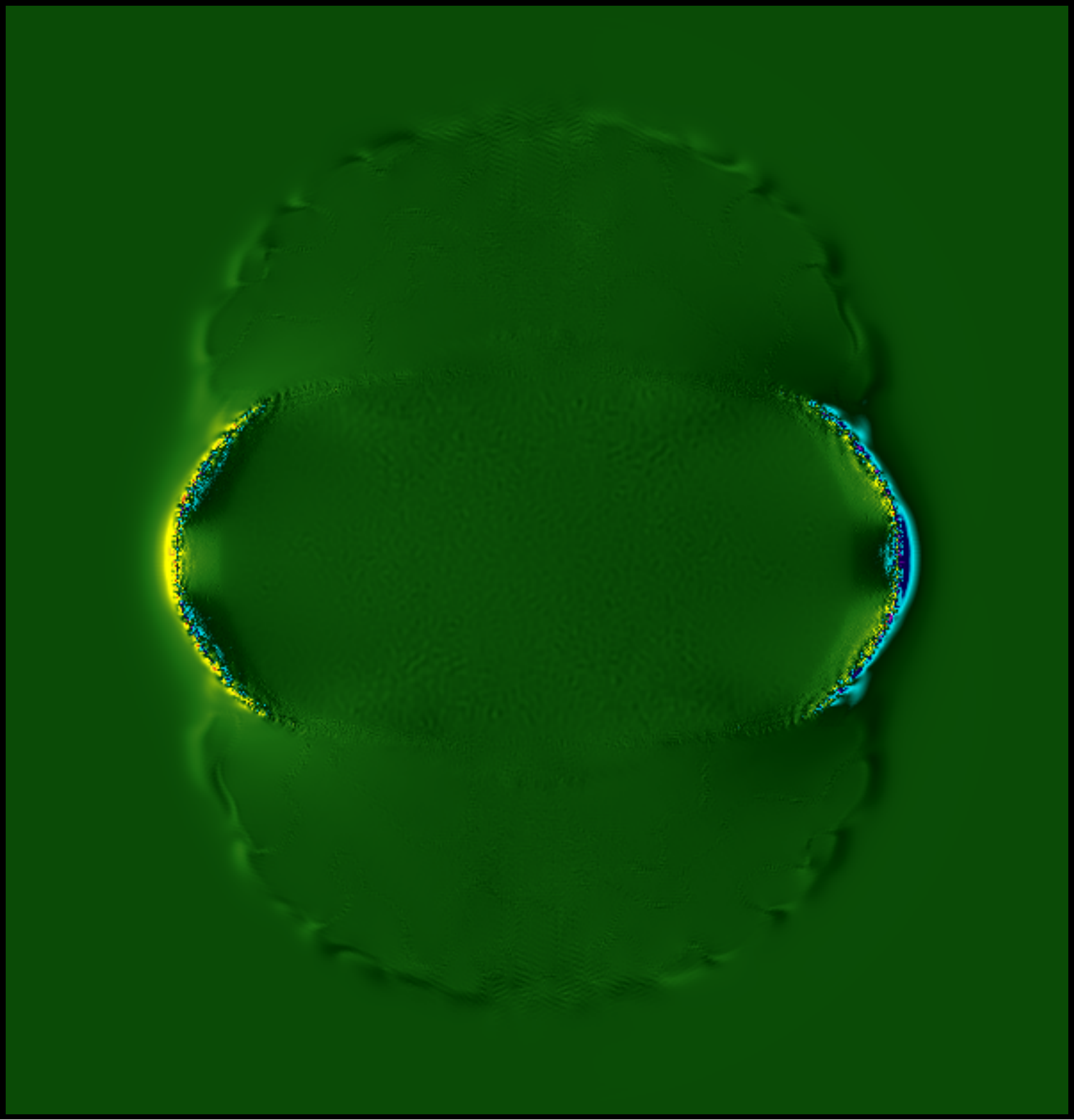}
 & \includegraphics[height=0.305\textwidth]{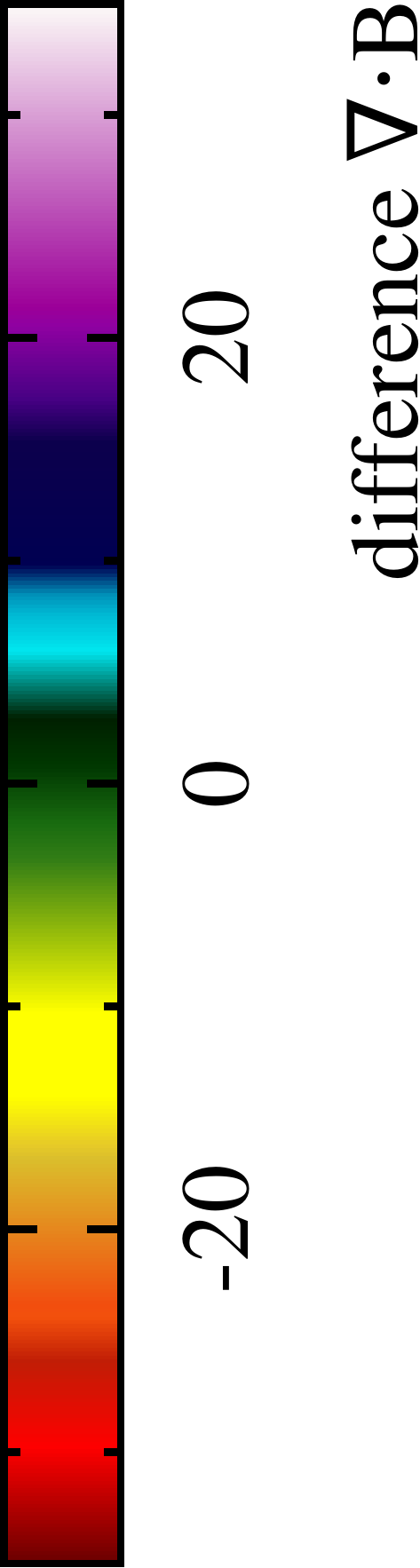}
\end{tabular}
\caption{As in Figure~\ref{fig:blast-compilation-divb-symm}, but showing $\nabla \cdot {\bf B}$ measured using the difference operator. With this operator, no $\nabla\cdot{\bf B}$ is measured along the leading edge of the magnetic edge for the control and difference-cleaned cases. However, symmetric cleaning produces spurious divergence in this region when measured with the difference operator, because changes have been induced in the magnetic field to compensate for particle disorder.}
\label{fig:blast-compilation-divb-diff}
\end{figure}
%\end{minipage}

 To answer the question of whether there is any advantage to cleaning with the symmetric $\nabla \cdot {\bf B}$ operator, the blast wave problem was simulated for three cases: no cleaning; cleaning using the difference operator for $\nabla \cdot {\bf B}$; and cleaning using the symmetric operator. The question is further complicated by fact that the operator used for cleaning may differ from the operator used to measure the error. We therefore show $\nabla \cdot {\bf B}$ for these three cases measured with both the symmetric (Figure~\ref{fig:blast-compilation-divb-symm}) and difference (Figure~\ref{fig:blast-compilation-divb-diff}) operators, so that the effect of cleaning using one operator can be seen in both.

The symmetric operator for $\nabla \cdot {\bf B}$ can be seen to pick up a non-zero divergence error on the leading edge of the magnetic wave from the blast (Figure~\ref{fig:blast-compilation-divb-symm}) despite the fact that the magnetic field shows no error in this region when measured with the difference operator (Figure~\ref{fig:blast-compilation-divb-diff}). This suggests that the symmetric operator is mainly reflecting the disordered particle arrangement. In turn, it can be seen that in this region, cleaning using the symmetric operator \emph{introduces} divergence error when measured with the difference operator as it attempts to adjust the magnetic field based on the particle arrangement (centre panel of Figure~\ref{fig:blast-compilation-divb-diff}). Nevertheless, it is true that cleaning with the symmetric operator does produce the greatest reduction in the divergence when measured in the symmetric operator, and may still have potential advantages in terms of momentum conservation (this is examined further in Section~\ref{sec:ot}). However, we conclude that cleaning is best performed with the difference operator, since it shows not only the best results when measured with the difference operator (right panel of Figure~\ref{fig:blast-compilation-divb-diff}), but also an improvement even when measured with the symmetric operator (centre panel of Figure~\ref{fig:blast-compilation-divb-symm}).

\subsubsection{Optimal damping values}

\begin{figure}
\centering
 \includegraphics[width=0.45\textwidth]{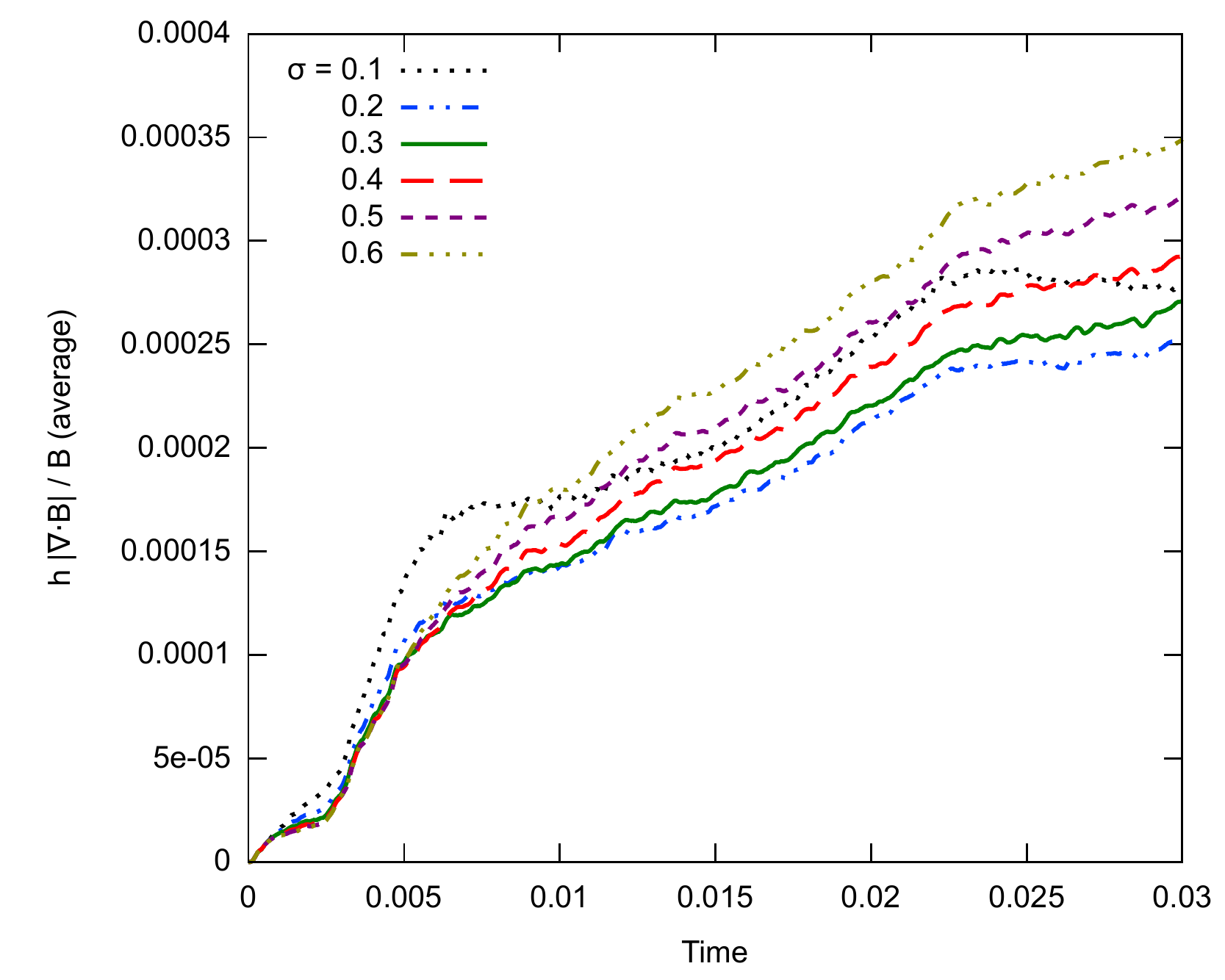}
 \includegraphics[width=0.45\textwidth]{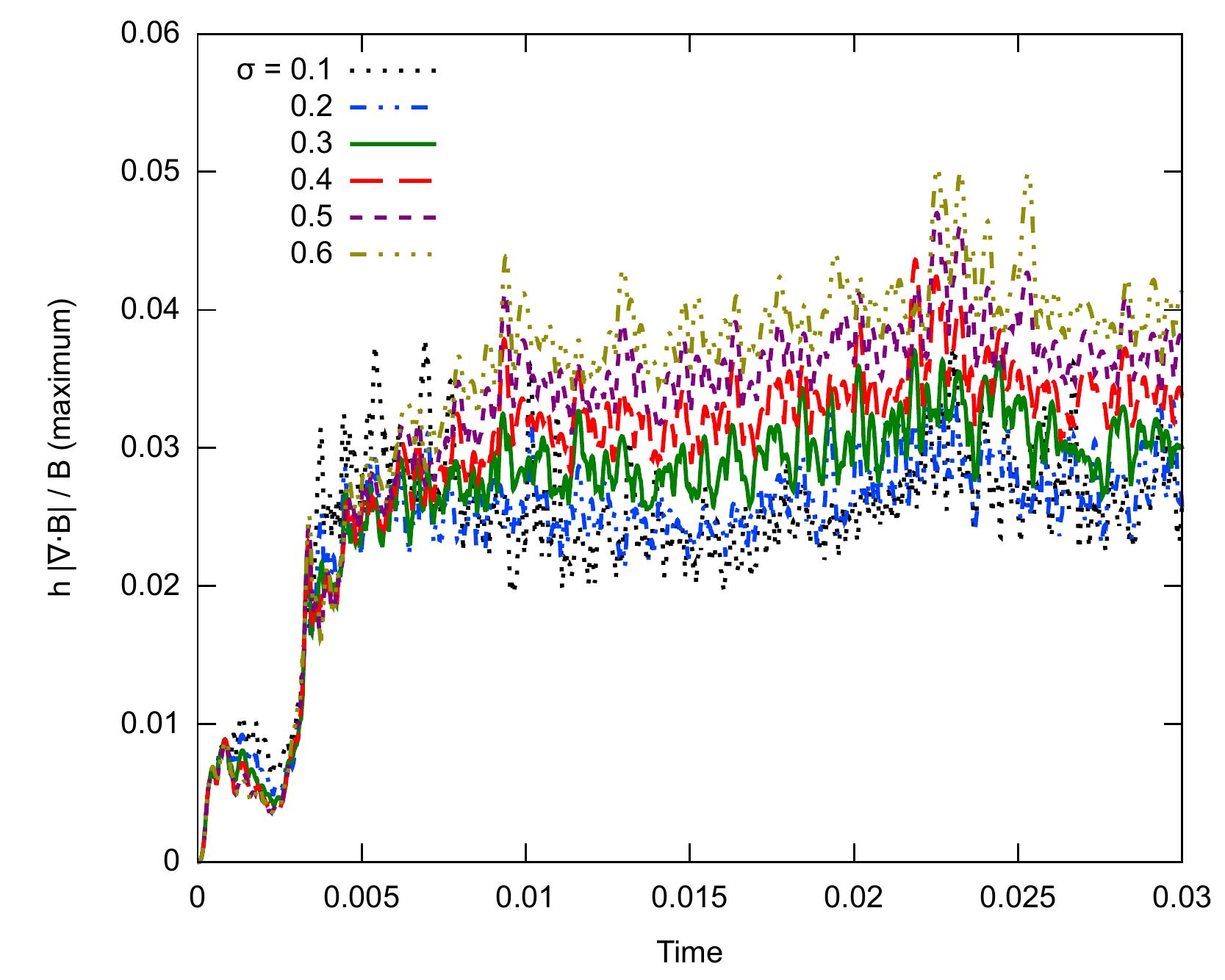}
\caption{Average and maximum $h \vert \nabla \cdot {\bf B}\vert / \vert{\bf B}\vert$ for the blast wave test with varying damping strengths.  The best results are obtained for values of $\sigma_\psi$ between 0.2--0.3.}
\label{fig:blast-sigma}
\end{figure}

Figure~\ref{fig:blast-sigma} shows the average and maximum divergence error as a function of time for differing strengths of the damping parameter $\sigma_\psi$ in the range $[0.1, 0.6]$.  The best results are obtained with $0.2 < \sigma_\psi < 0.3$, in agreement with the other two dimensional tests.

\subsubsection{Tensile instability correction}
\label{sec:halfdivb}

\begin{figure}
 \centering
\includegraphics[height=0.296\textwidth]{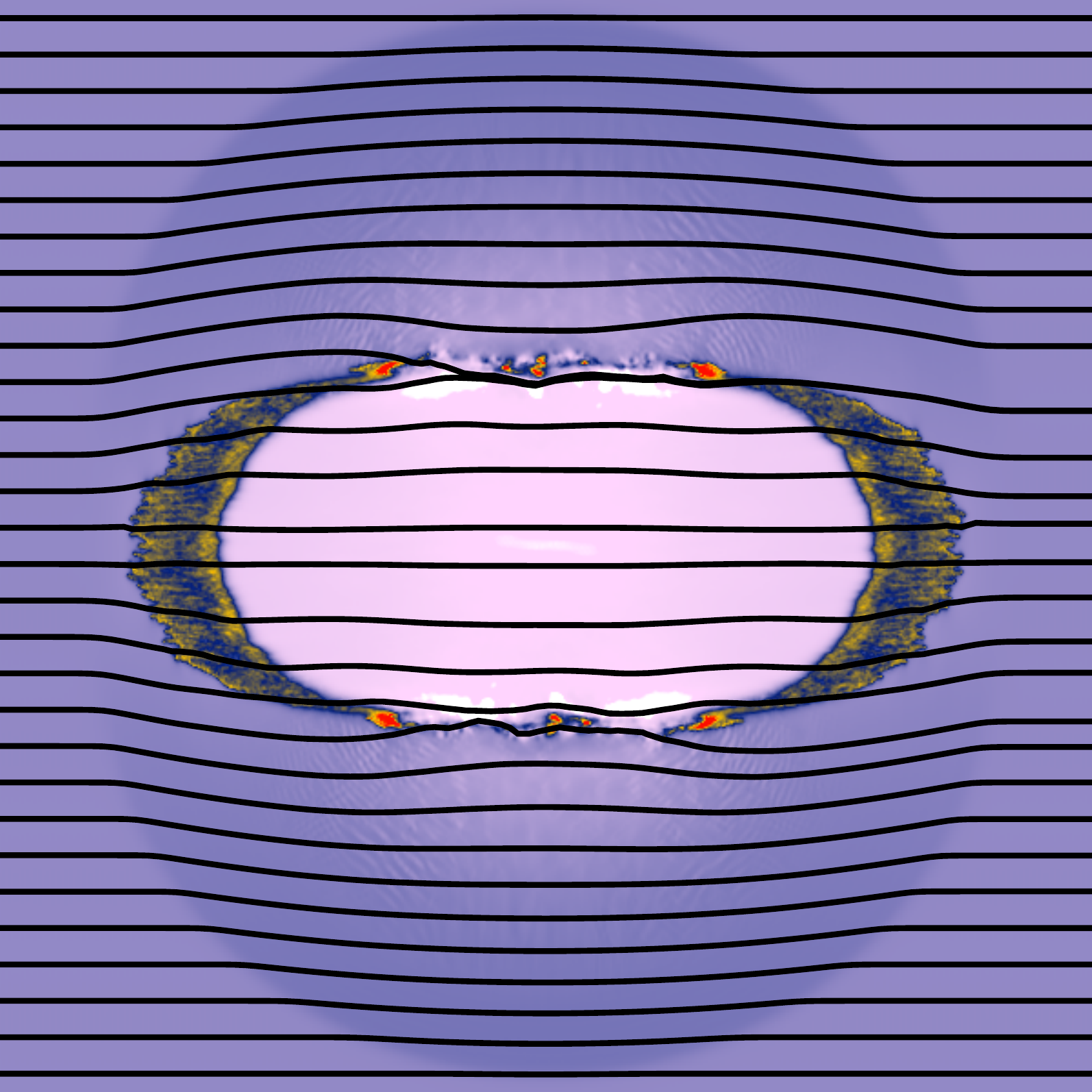}
\includegraphics[height=0.296\textwidth]{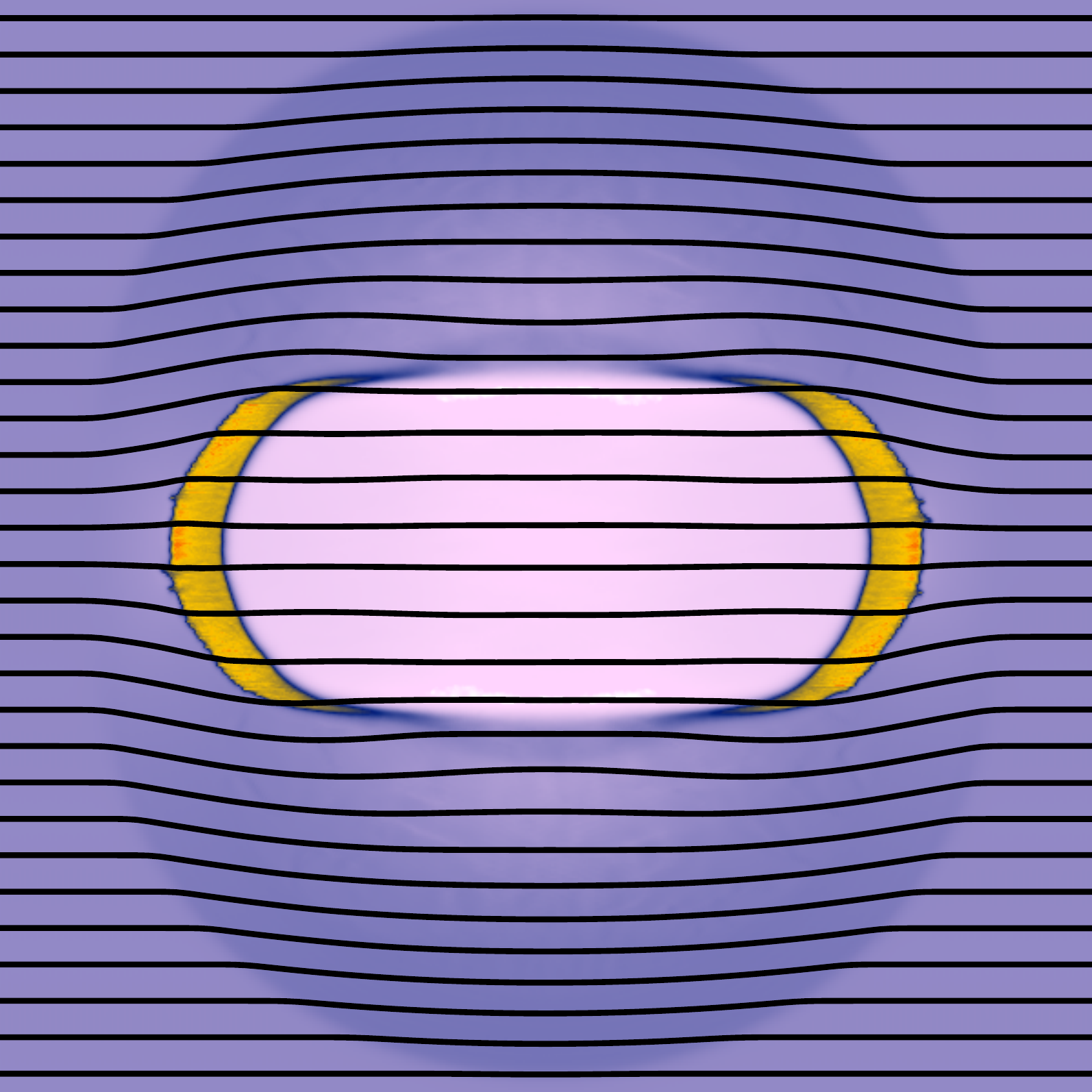}
\includegraphics[height=0.296\textwidth]{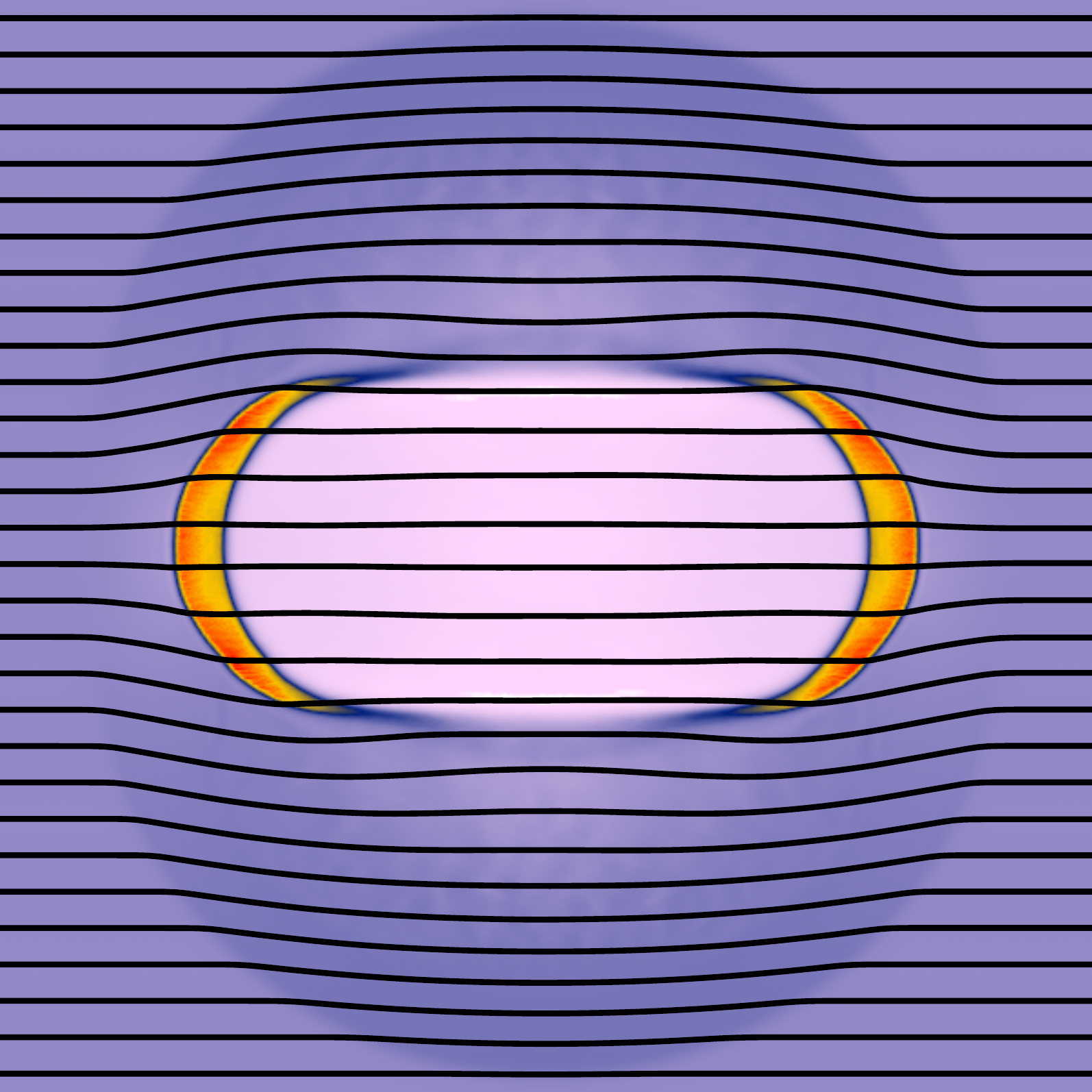}
\includegraphics[height=0.296\textwidth]{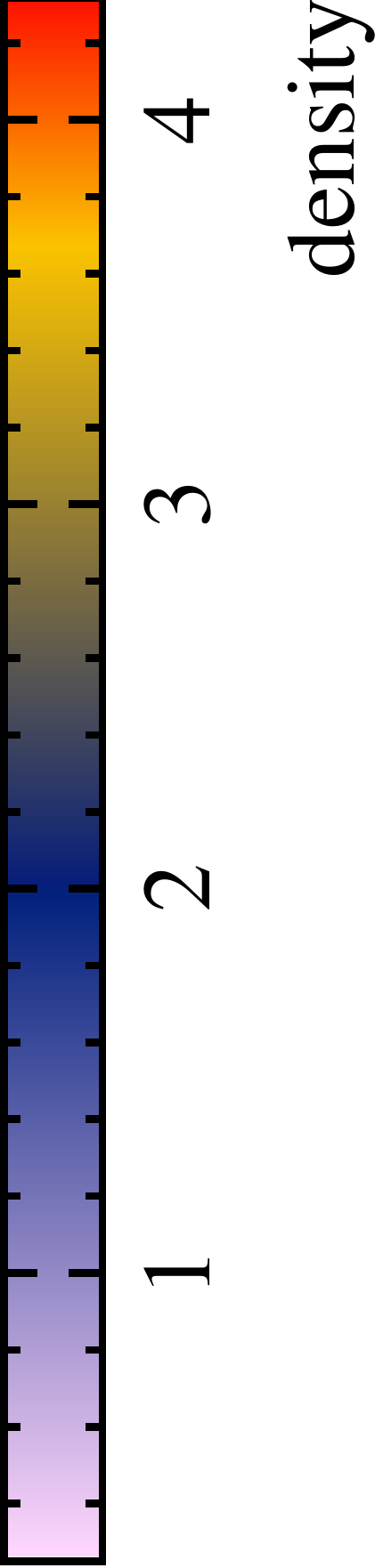}
\caption{Density of the blast wave problem with overlaid magnetic field lines, without any divergence cleaning, but examining the impact of the $-\hat{\beta} {\bf B} (\nabla \cdot {\bf B})$ term used to correct the tensile instability. Results shown use $\hat{\beta} = 0.5$ (left),  $\hat{\beta} = 0.75$ (centre) and $\hat{\beta} = 1.0$ (right).  Using only $\hat{\beta} = 0.5$ is found to result in irregularities along the shock fronts, which are not present using $\hat{\beta} = 1$. Thus, using $\hat{\beta} = 0.5$ is not recommended.}
\label{fig:halfdivb}
\end{figure}
 Finally, we noticed important consequences in this test concerning the $\hat{\beta}{\bf B} (\nabla \cdot {\bf B})$ correction for the tensile instability (Section~\ref{sec:spmhd-tensileinstability}).  Since using $\hat{\beta} = 0.5$ is in principle sufficient to prevent the instability, its use has been suggested by \citet{bot04}, \citet{bkw12}, and \citet{price12}.  However, we found this to be problematic in our simulations of the blast wave problem: Figure~\ref{fig:halfdivb} shows the density with overlaid magnetic field lines at $t=0.03$ using ${\hat{\beta}}$ = 0.5, 0.75 and 1.0 (left to right). With only $\hat{\beta} = 0.5$ (left panel), irregularities can be seen to form in the densest parts of the shockwave. These are not present when performing the full $\hat{\beta} = 1$ subtraction (right panel).

\subsection{Orszag-Tang Vortex}
\label{sec:ot}
The final two dimensional test is the Orszag-Tang vortex \citep{ot79}, which has been widely used as a test of MHD codes \citep[e.g.][]{ramses, athena, mhdgadget}.  It consists of a magnetic vortex superimposed onto a velocity vortex generating several classes of interacting shock waves. The complex dynamics provides an excellent test of the constrained hyperbolic divergence cleaning method.  To measure the effectiveness of the method in this case, the results are compared against that of simulations using artificial resistivity (with particle independent strengths as described in Section~\ref{sec:artresis-switches}) and Euler Potentials as measures of divergence control.  This test is also used to examine whether or not cleaning using the symmetric operator for $\nabla \cdot {\bf B}$ provides any advantage in terms of momentum conservation.  As previously, the damping parameter $\sigma_\psi$ is varied to find optimal values.

\subsubsection{Setup}

The problem is set up in a box with dimensions $x,y \in [0,1]$ with periodic boundary conditions.  The initial gas state is set to $\rho = 25 / (36\pi)$, $P = 5/(12\pi)$, $\gamma = 5/3$, with velocity field ${\bf v} = [-\sin(2\pi y), \sin(2\pi x)]$.  The initial magnetic field is ${\bf B} = [-\sin(2\pi y), \sin(4\pi x)]$.  All examples presented use $512 \times 590$ particles initially arranged on a triangular lattice.

\subsubsection{Results}

% density: 0.073,  0.350
% magpres: 1.0e-6,  0.280
% divb: -8,  +8

\begin{figure}
 %\centering
\setlength{\tabcolsep}{0.005\textwidth}
\begin{tabular}{ccccl}
{\scriptsize Control} & {\scriptsize Resistivity} & {\scriptsize Euler Potentials} & {\scriptsize Divergence Cleaning} & \\
   \includegraphics[height=0.22\textwidth]{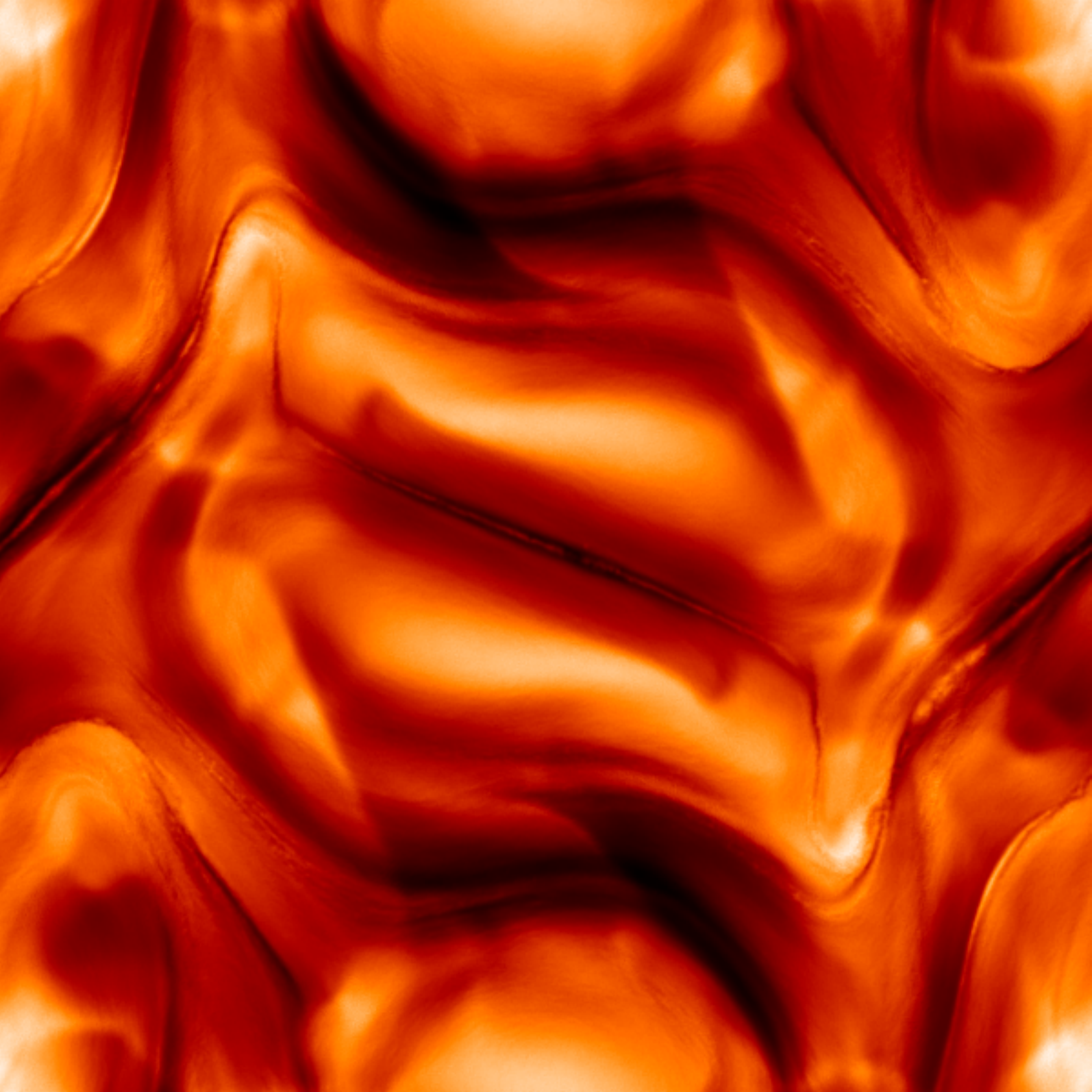} 
 & \includegraphics[height=0.22\textwidth]{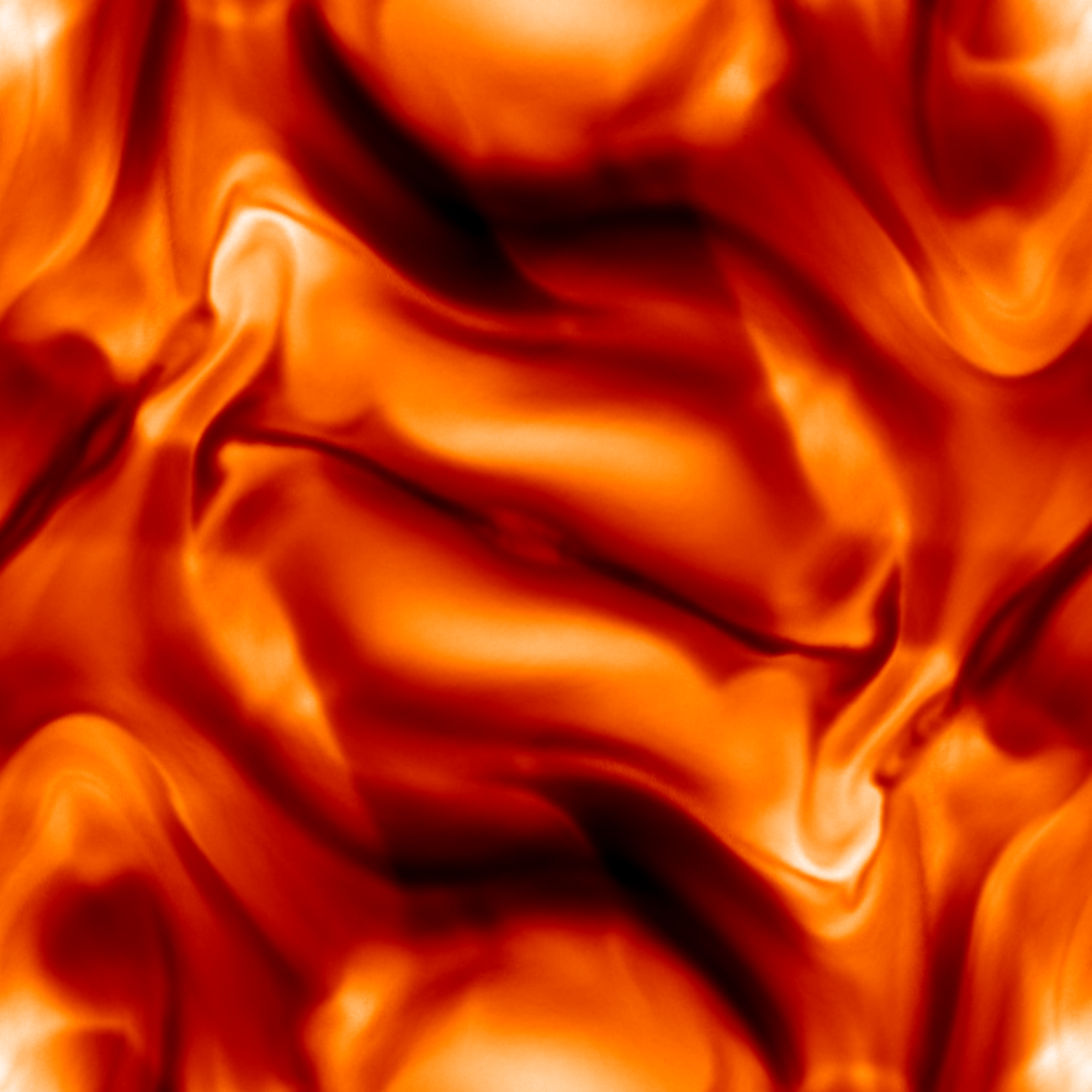}
 & \includegraphics[height=0.22\textwidth]{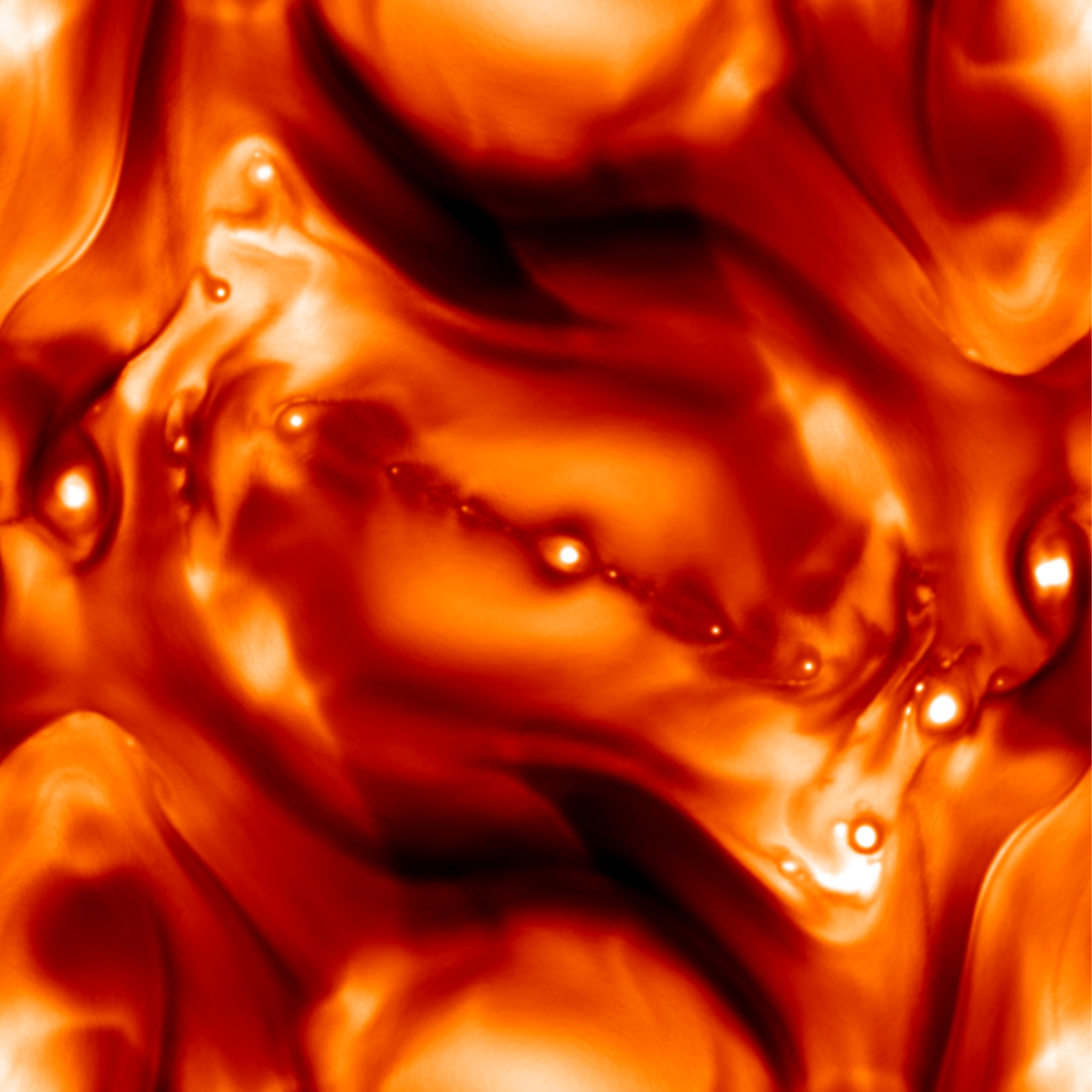}
 & \includegraphics[height=0.22\textwidth]{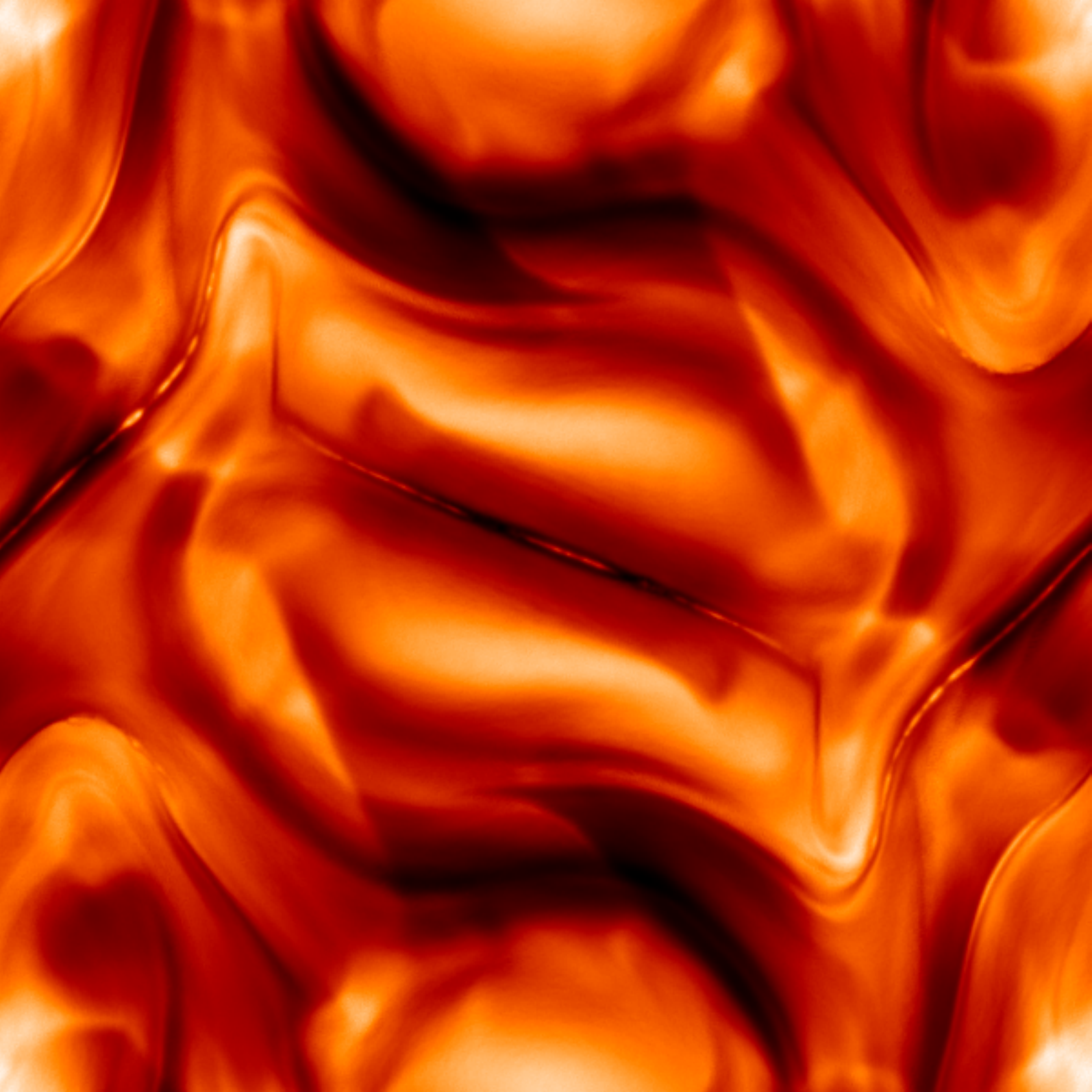}
 & \includegraphics[height=0.22\textwidth]{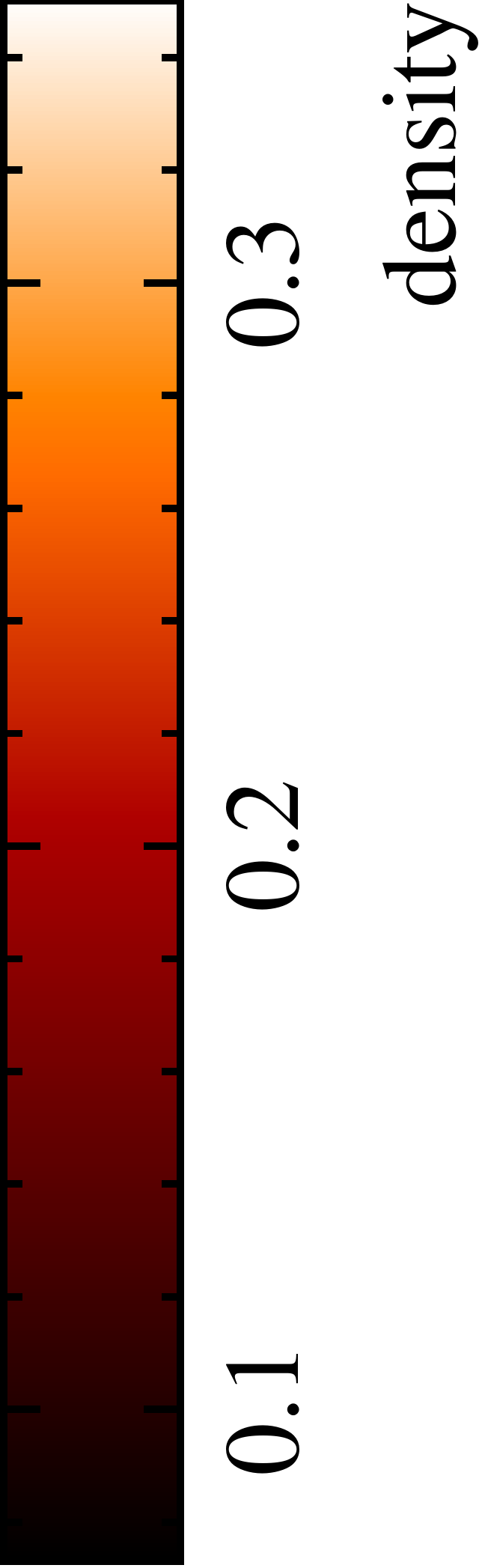}
\\
   \includegraphics[height=0.22\textwidth]{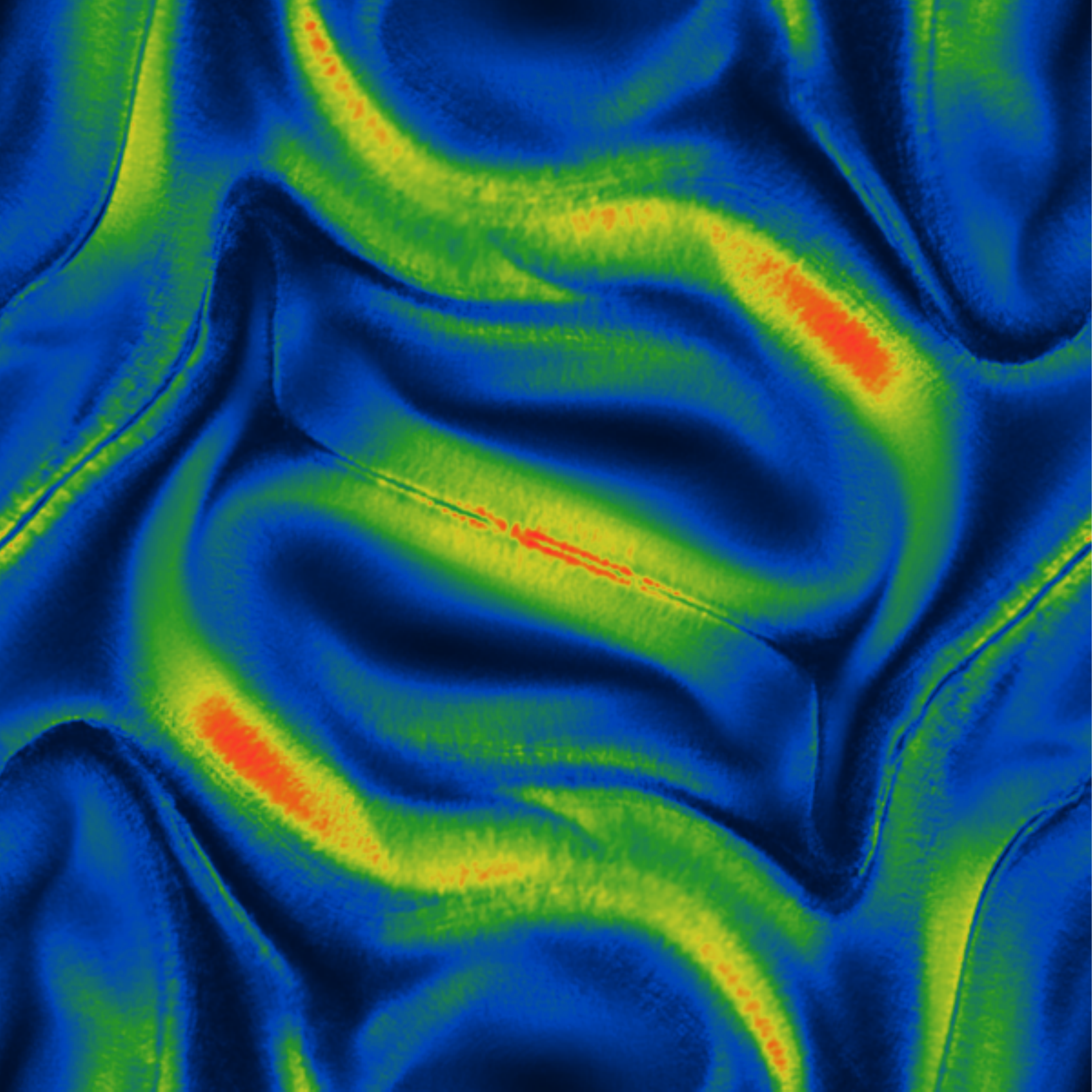}
 & \includegraphics[height=0.22\textwidth]{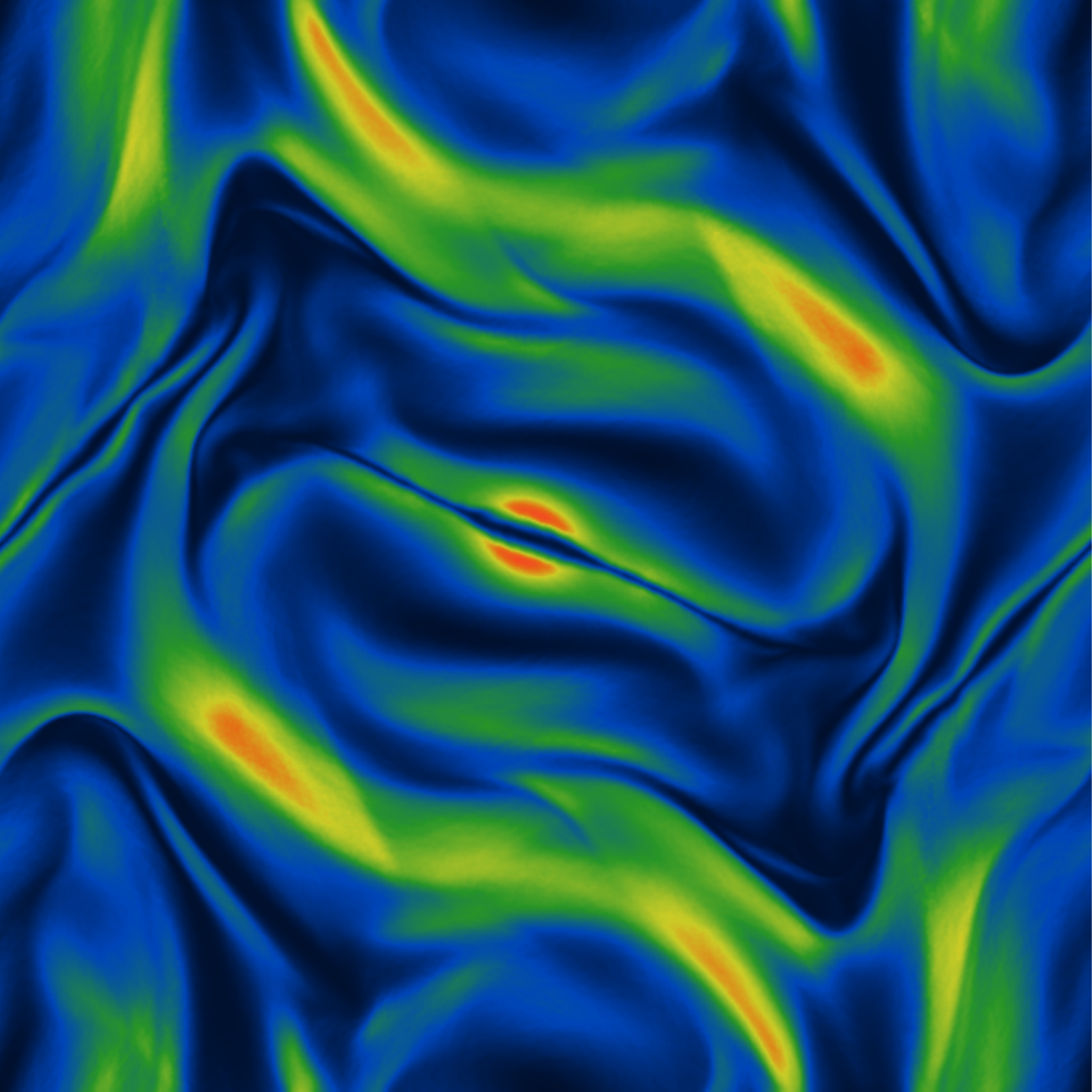}
 & \includegraphics[height=0.22\textwidth]{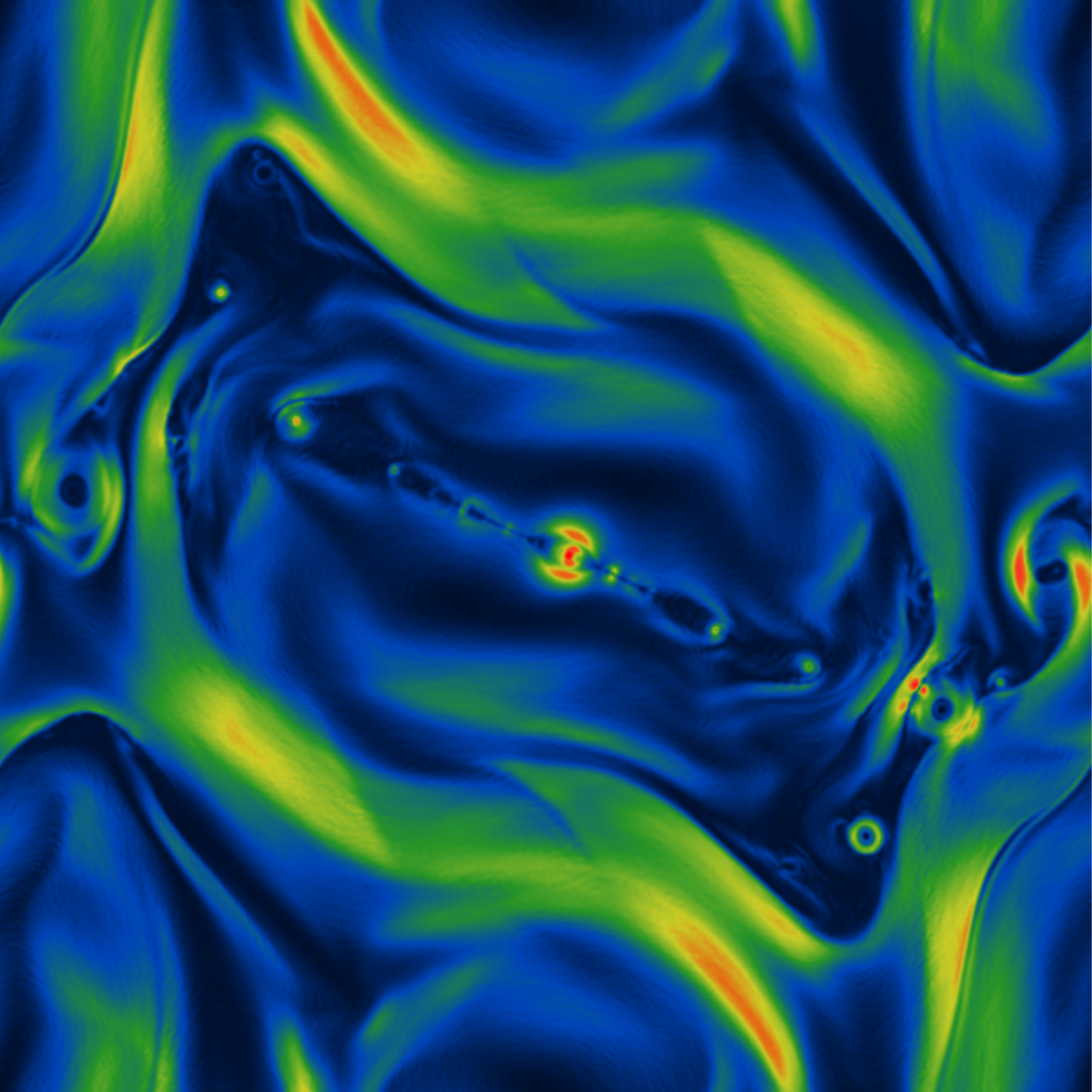}
 & \includegraphics[height=0.22\textwidth]{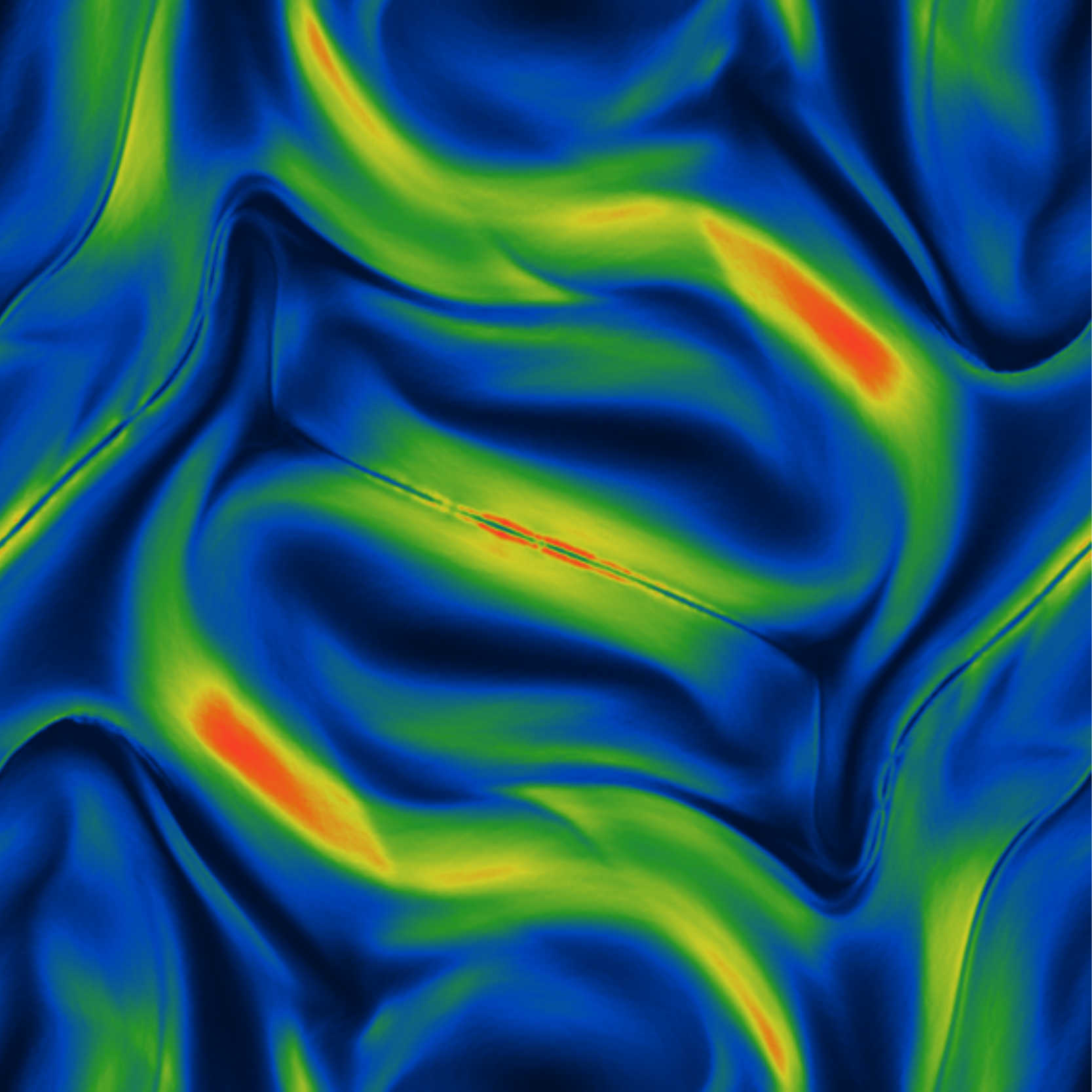}
 & \includegraphics[height=0.22\textwidth]{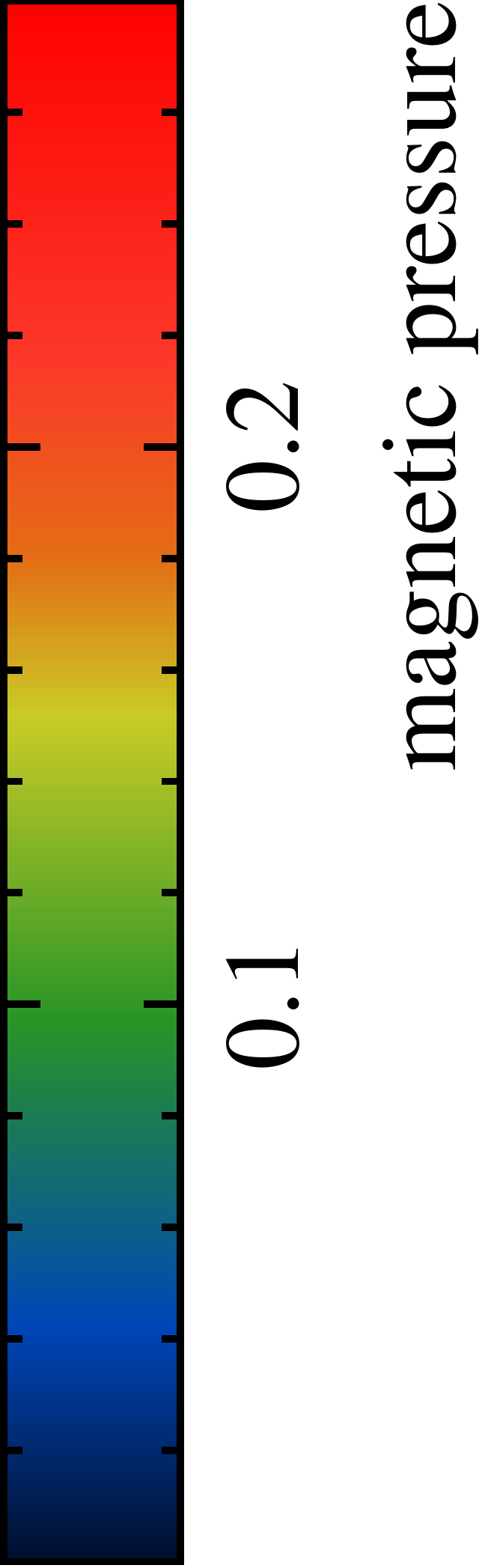}
\\
   \includegraphics[height=0.22\textwidth]{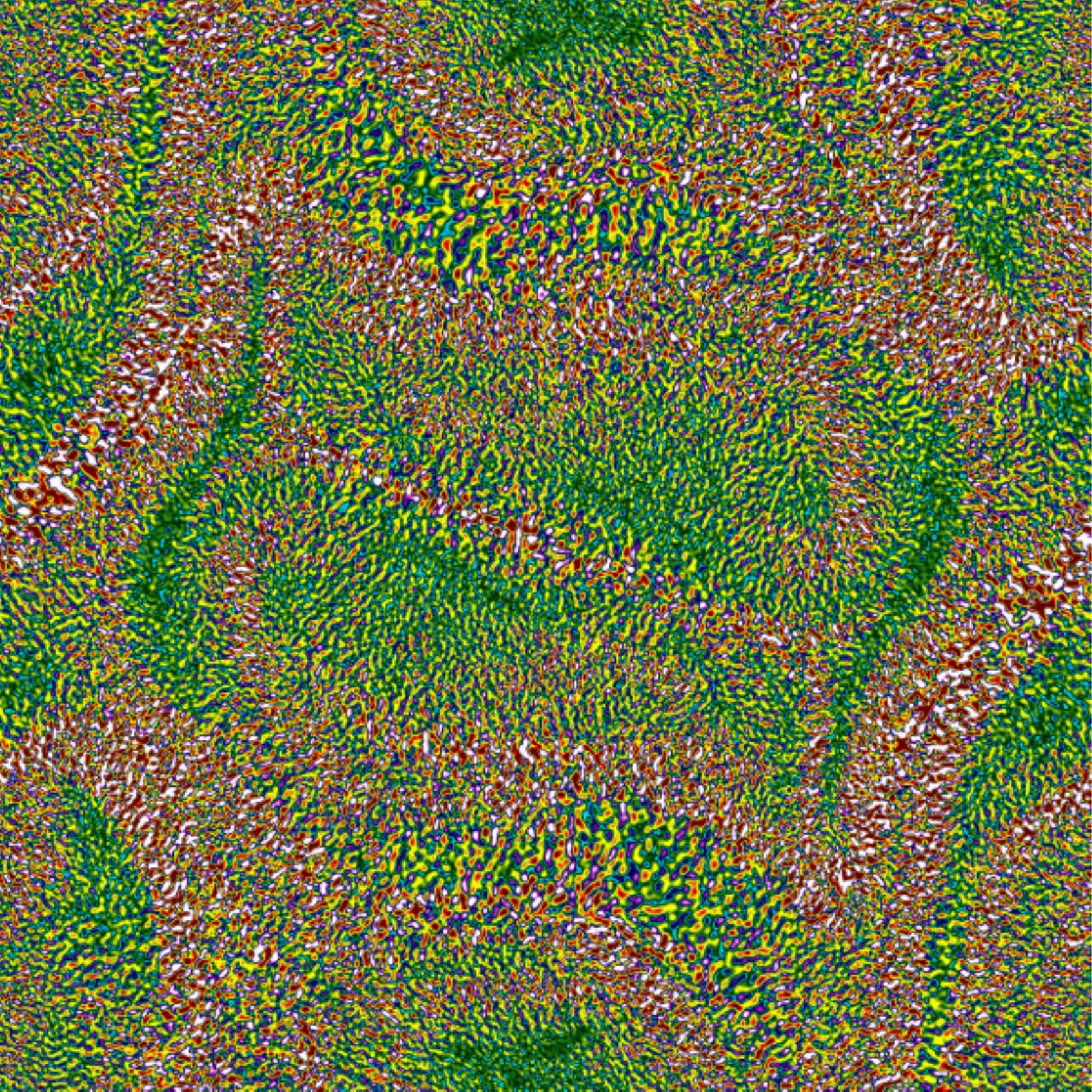}
 & \includegraphics[height=0.22\textwidth]{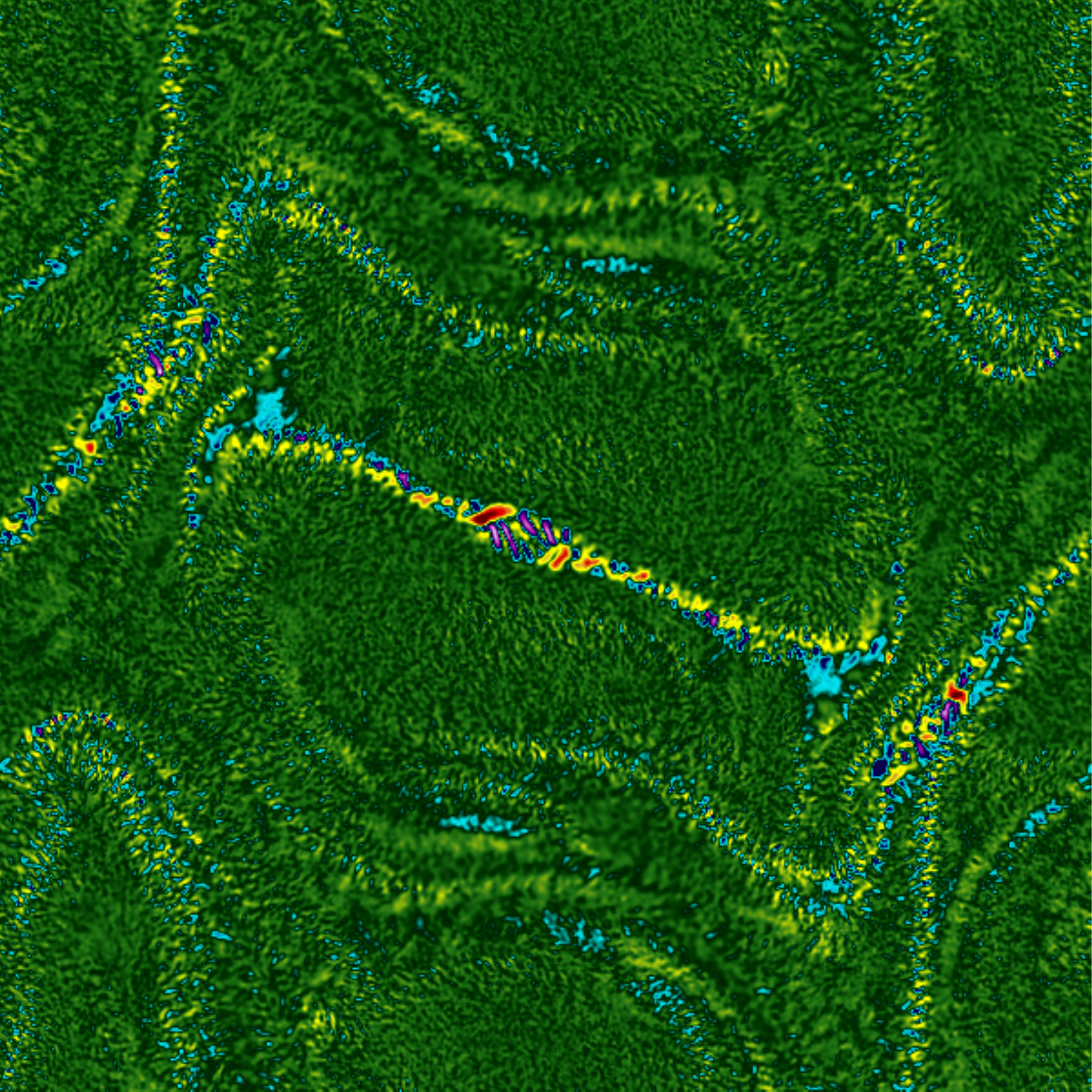}
 & \includegraphics[height=0.22\textwidth]{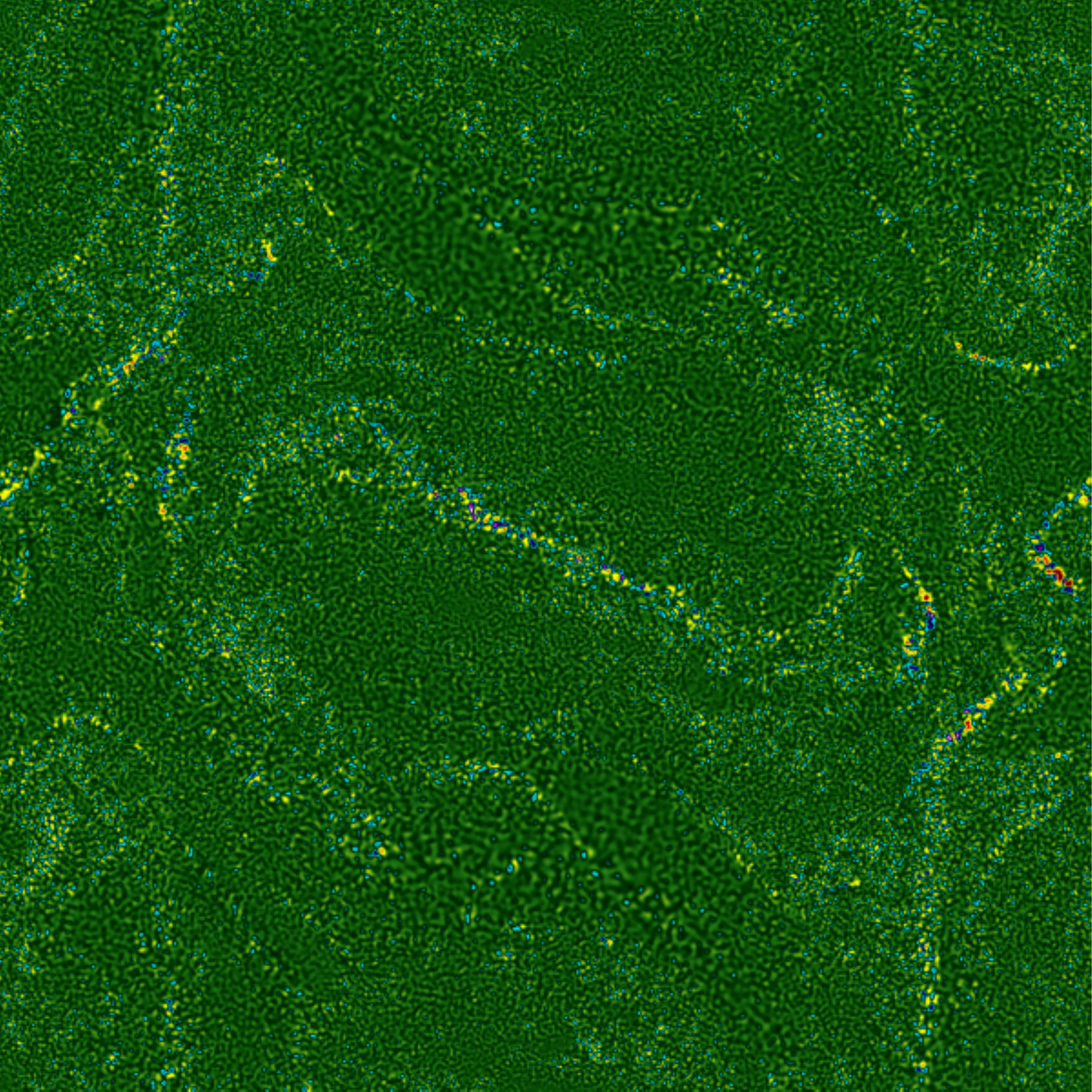}
 & \includegraphics[height=0.22\textwidth]{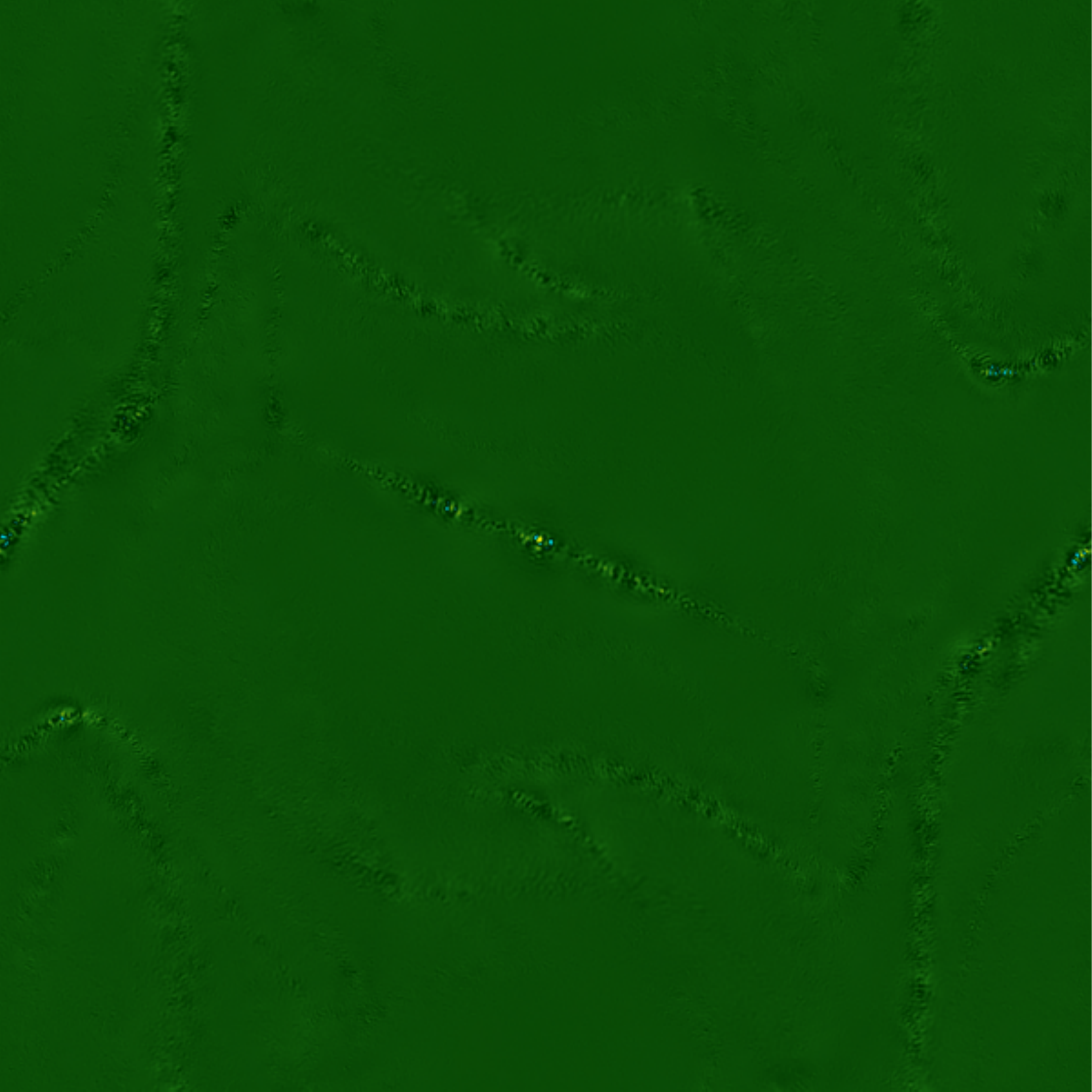}
 & \includegraphics[height=0.22\textwidth]{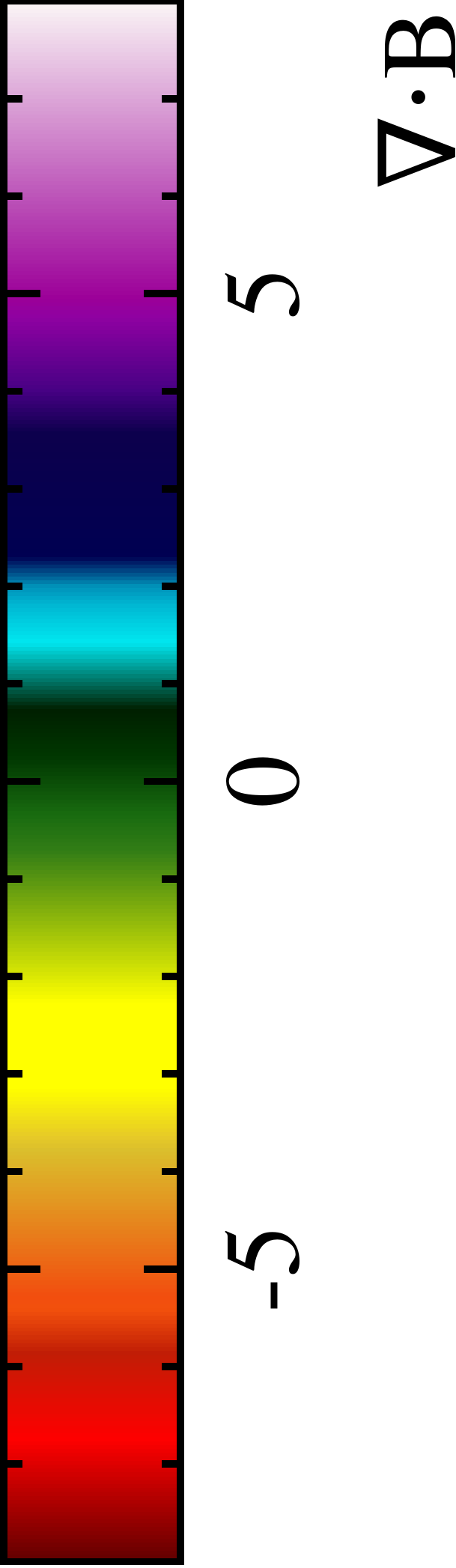}
\end{tabular}
\caption{The density (top row), magnetic pressure (middle row), and the difference measurement of $\nabla \cdot {\bf B}$ (bottom row) in the Orszag-Tang vortex at $t=1.0$ comparing the control case (far left), including artificial resistivity (centre left), evolving the magnetic field using Euler Potentials (centre right), and applying the constrained divergence cleaning method (far right).}
\label{fig:orszag-compilation}
\end{figure}

\begin{figure}
 \centering
\includegraphics[width=0.45\textwidth]{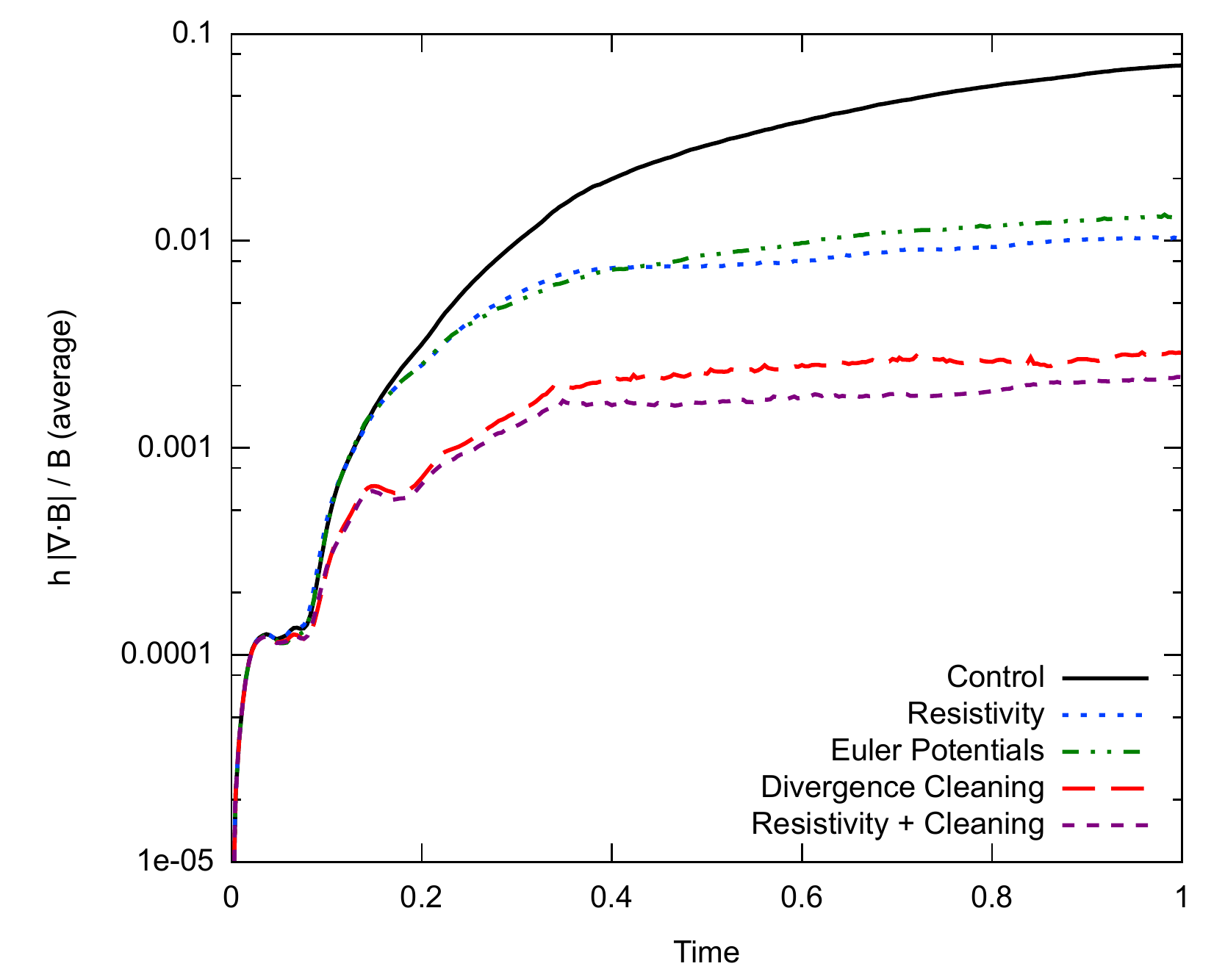}
\includegraphics[width=0.45\textwidth]{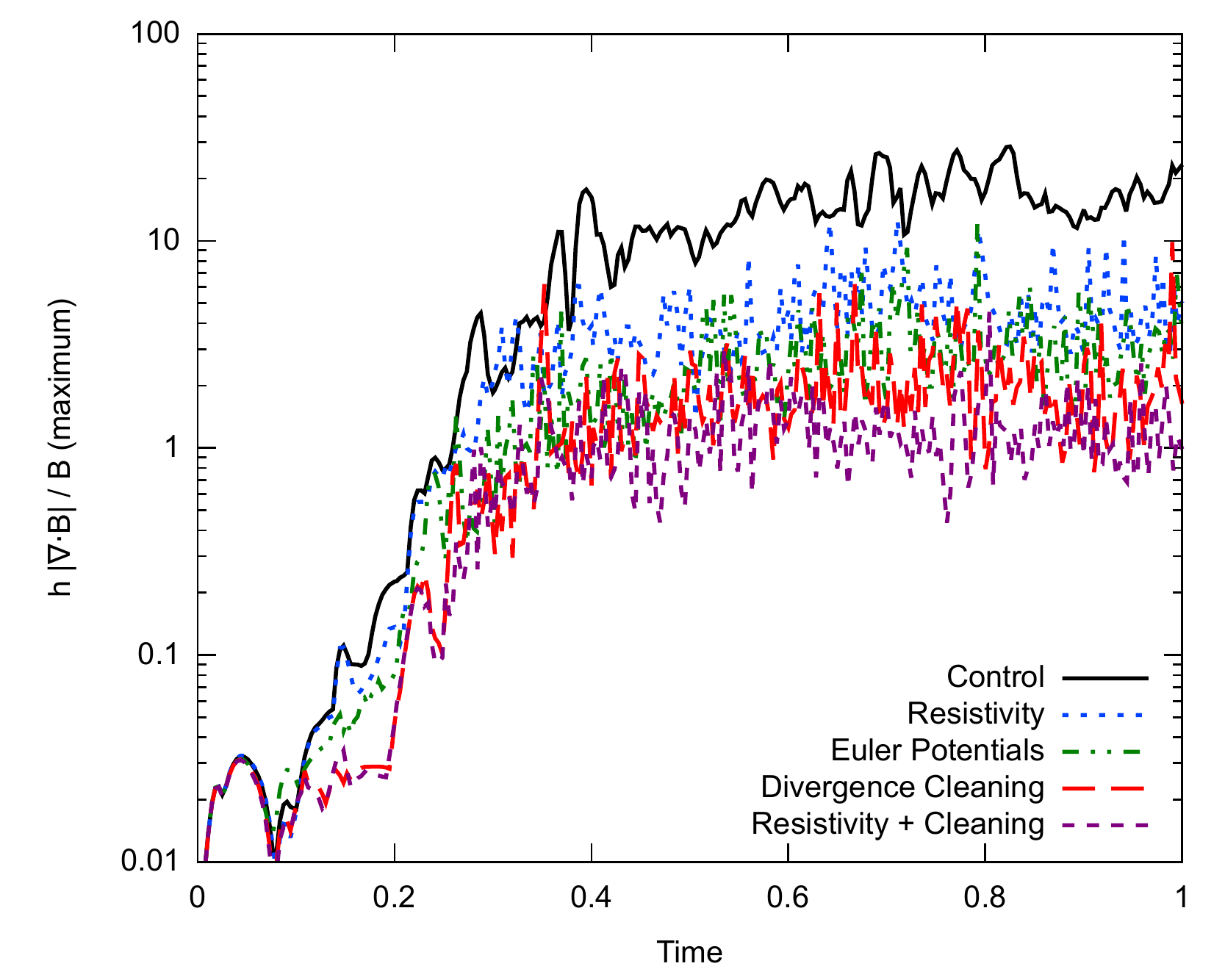}
\caption{Average (left) and maximum (right) $h \vert \nabla \cdot {\bf B}\vert / \vert{\bf B}\vert$ in the Orszag-Tang vortex problem with (top to bottom in left panel) no divergence control, using Euler Potentials, adding an artificial resistivity, using divergence cleaning, and cleaning while including resistivity.  Divergence cleaning has lower divergence error than when using Euler Potentials or artificial resistivity, and continues to reduce divergence error even when used in combination with artificial resistivity.}
\label{fig:orszag-divb}
\end{figure}

Figure~\ref{fig:orszag-compilation} shows the density (top), magnetic pressure (middle row), and $\nabla \cdot {\bf B}$ (bottom row) at $t=1.0$ for four cases: i) control, ii) using artificial resistivity, iii) employing Euler Potentials, and iv) applying divergence cleaning.  This time is chosen because the divergence errors in the control case are large enough to produce small scale disturbances in the density and magnetic pressure fields.  By adding resistivity or using Euler Potentials, the average $h |\nabla \cdot {\bf B}| / |{\bf B}|$ is decreased by an order of magnitude (c.f. second and third panels in bottom row of Figure~\ref{fig:orszag-compilation} and the left panel of Figure~\ref{fig:orszag-divb}).  When divergence cleaning is used, the average divergence error is reduced by almost two orders of magnitude (red/dashed line in left panel of Figure~\ref{fig:orszag-divb}).  In addition to the average and maximum divergence error for the above four cases, Figure~\ref{fig:orszag-divb} also presents the results from a case where artificial resistivity has been applied in tandem with divergence cleaning.  In this case, the average $h |\nabla \cdot {\bf B}| / |{\bf B}|$ is reduced by nearly an order of magnitude compared to resistivity alone, and when compared to the control case, this results in two orders of magnitude reduction in the average together with an order of magnitude reduction in the maximum.

\subsubsection{Cleaning using symmetric $\nabla \cdot {\bf B}$}

\begin{figure}
\centering
\begin{minipage}[t]{0.45\textwidth}
\includegraphics[width=\textwidth]{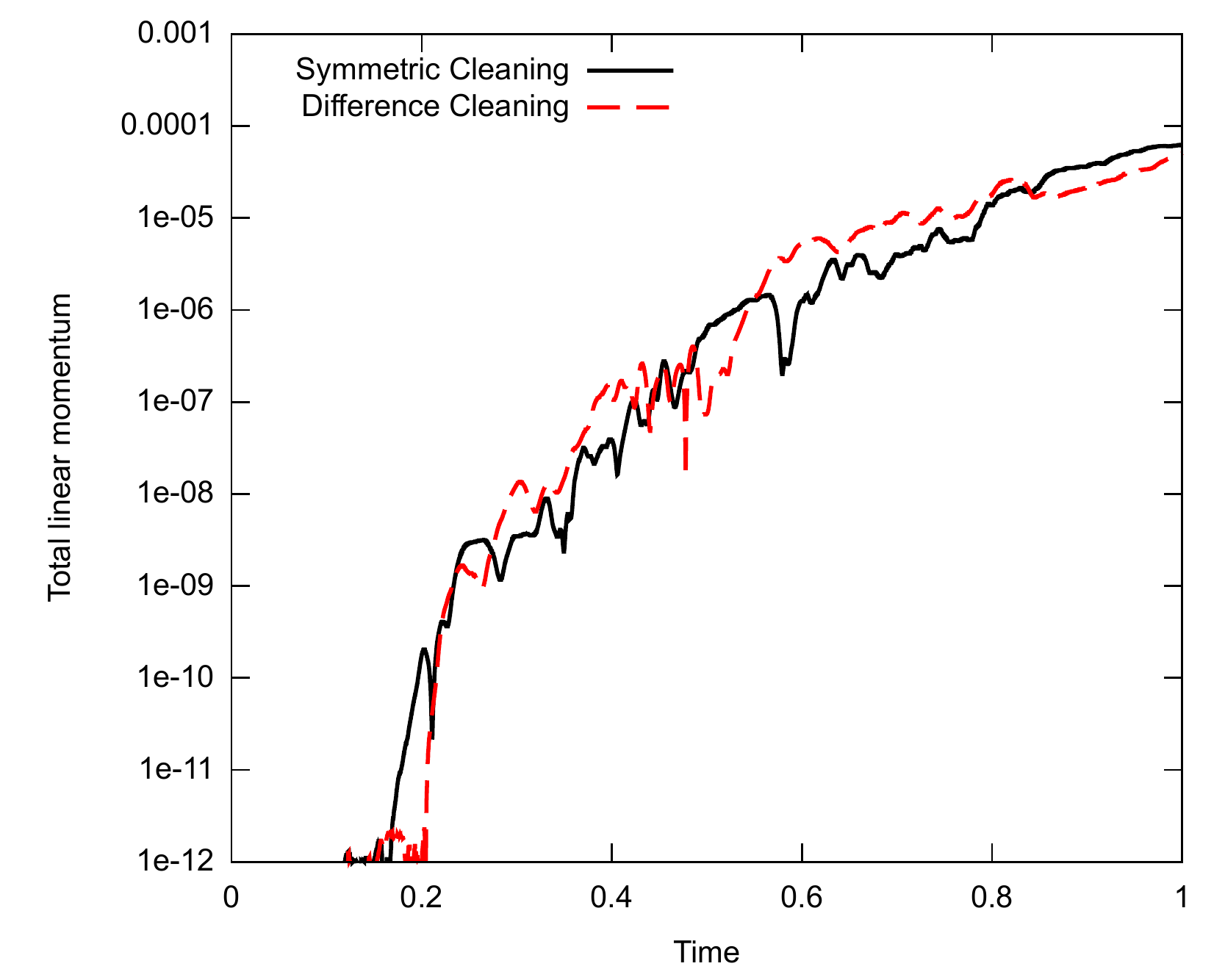}
\caption{Total linear momentum for the Orszag-Tang vortex for divergence cleaning using the difference and symmetric operators of $\nabla \cdot {\bf B}$.  There is no significant distinction between the two.}
\label{fig:orszag-mom}
\end{minipage}
\hspace{0.05\textwidth}
\begin{minipage}[t]{0.45\textwidth}
\includegraphics[width=\textwidth]{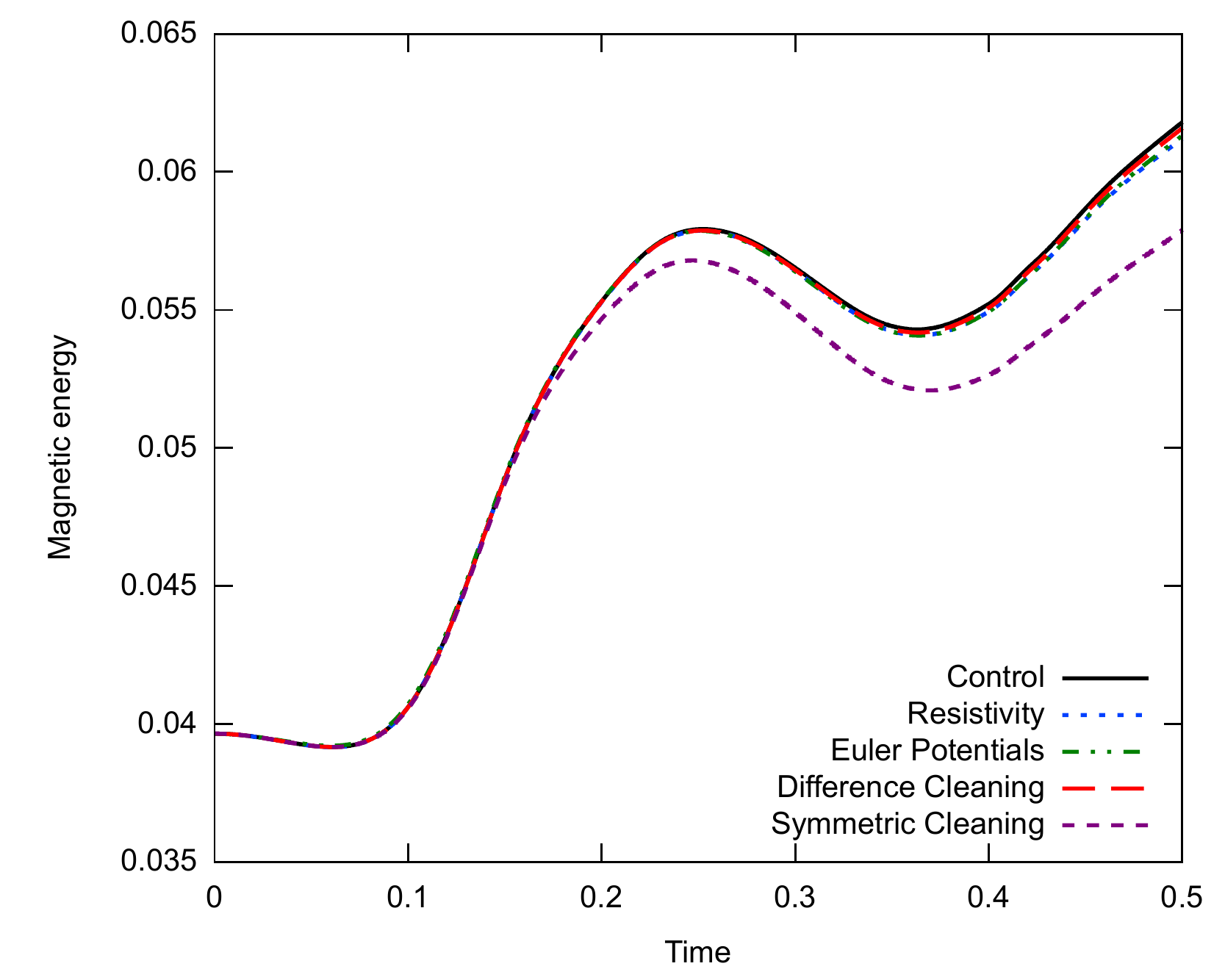}
\caption{Magnetic energy as a function of time in the Orszag-Tang vortex test.  Using the symmetric form of $\nabla \cdot {\bf B}$ for divergence cleaning leads to a $10\%$ reduction in magnetic energy by $t=0.5$ compared to the other schemes.}
\label{fig:orszag-be}
\end{minipage}
\end{figure}

Since the symmetric operator for $\nabla \cdot {\bf B}$ is used in the momentum equation and tensile instability correction term, it was hoped that its use for cleaning would confer some advantage over the difference measure by way of improved momentum conservation.  However, as shown in Figure~\ref{fig:orszag-mom}, no significant difference in the momentum is found between cleaning with the symmetric operator compared to the difference operator.   Figure~\ref{fig:orszag-be} shows the magnetic energy profile of the system for $t \le 0.5$, where all test cases (control, resistivity, Euler Potentials, difference cleaning) yield the same profile, except for symmetric cleaning which shows a $\sim 10\%$ reduction in magnetic energy compared to the other solutions. This occurs due to the symmetric operator removing magnetic energy to compensate for irregularities in particle position (which begin to occur at $t \sim 0.15$). Furthermore, although we have already shown in Section~\ref{sec:blast-symmdivb} that use of ${\hat{\beta}} = \tfrac{1}{2}$ in the tensile instability correction could result in numerical artefacts in the blast wave test, we also found large errors in the density and magnetic field profiles when ${\hat{\beta}} = \tfrac{1}{2}$ is used in combination with symmetric cleaning on the Orszag-Tang problem.  For these reasons, we recommend using $\hat{\beta} =1$ and applying cleaning only with the difference $\nabla \cdot {\bf B}$ operator.

\subsubsection{Optimal damping values}

\begin{figure}
 \centering
\includegraphics[width=0.45\textwidth]{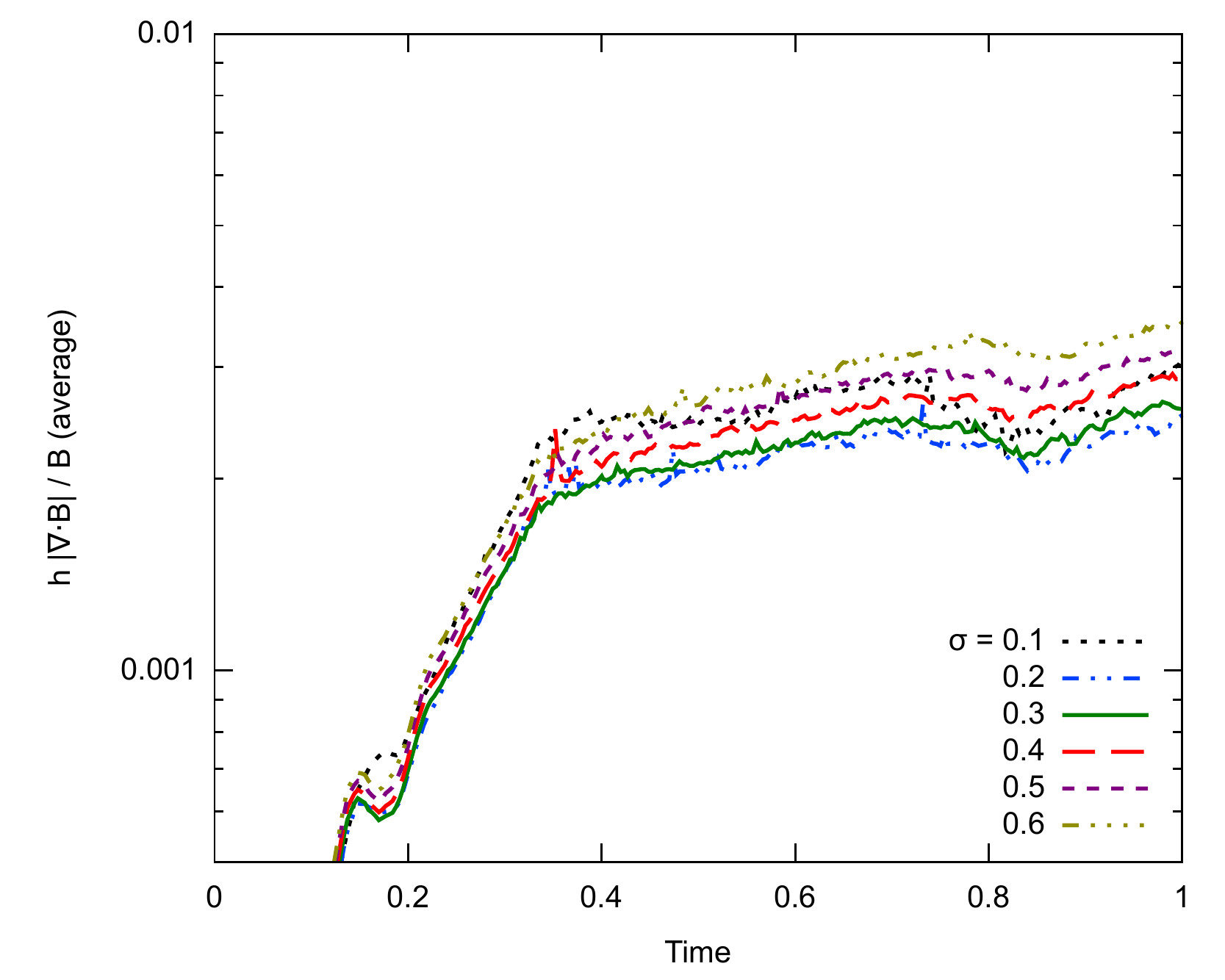}
\includegraphics[width=0.45\textwidth]{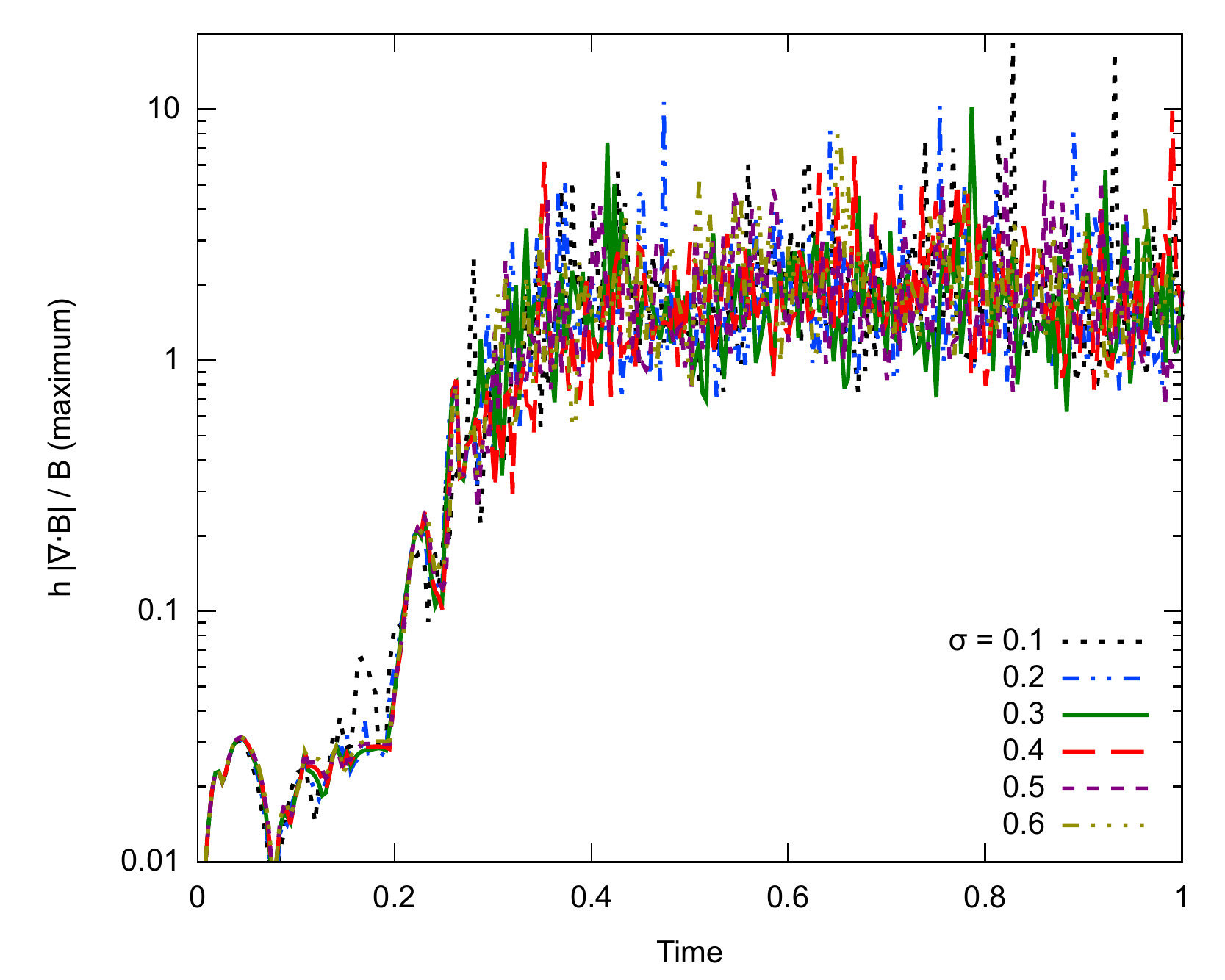}
\caption{Average (left) and maximum (right) divergence error in the Orszag-Tang vortex problem, varying the damping parameter $\sigma_\psi$.  The best results are obtained with values $\sim 0.2 - 0.3$.}
\label{fig:orszag-sigma}
\end{figure}

As with the previous tests, the damping parameter $\sigma_\psi$ was varied to find the best results (Figure~\ref{fig:orszag-sigma}), which, as previously, were obtained for $0.2 < \sigma_\psi < 0.3$ for this 2D test.

\subsubsection{Resolution study}

% Density range is 
\begin{figure}
 \centering
\setlength{\tabcolsep}{0.001\textwidth}
\begin{tabular}{ccccl}
 \includegraphics[width=0.1989\textwidth]{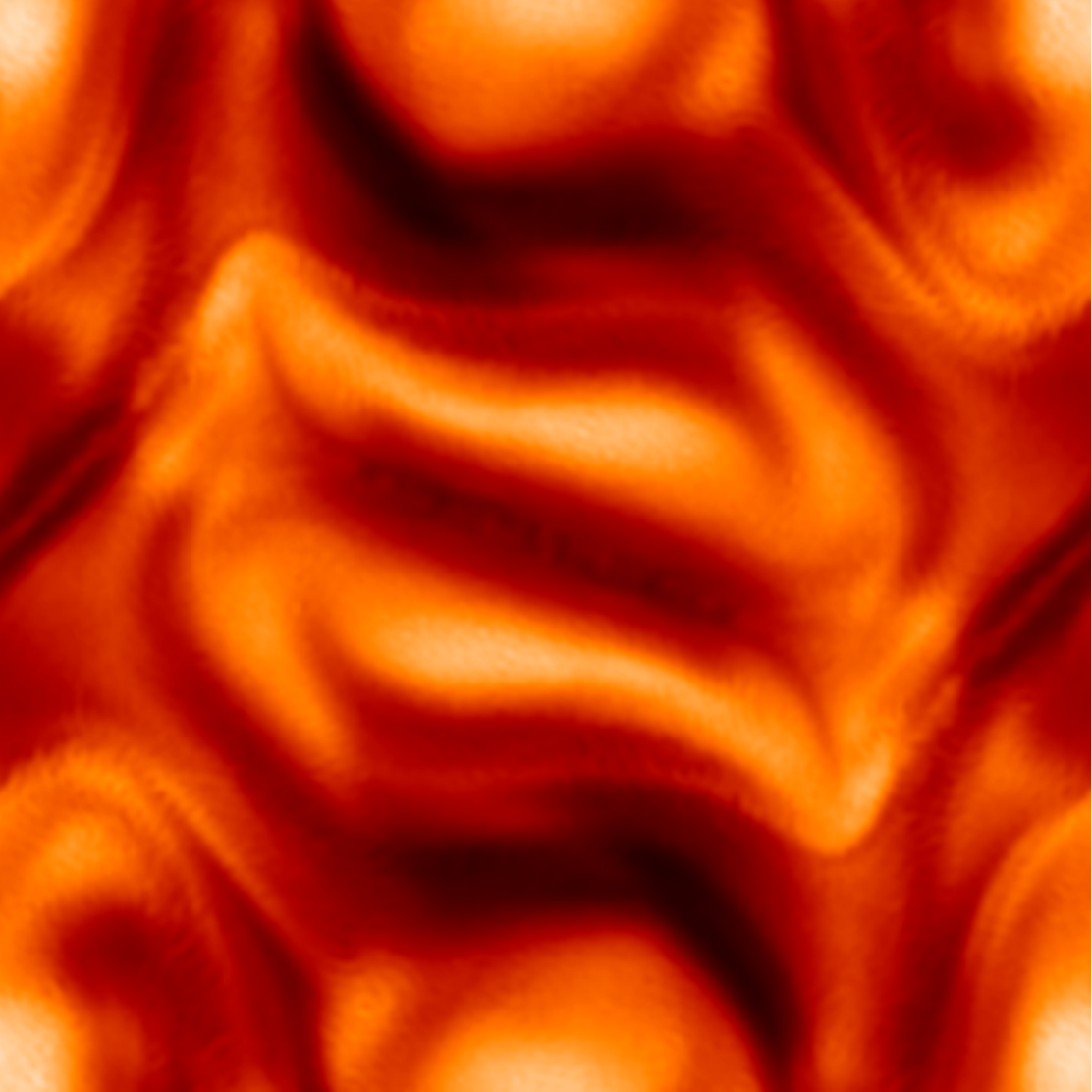}
& \includegraphics[width=0.1989\textwidth]{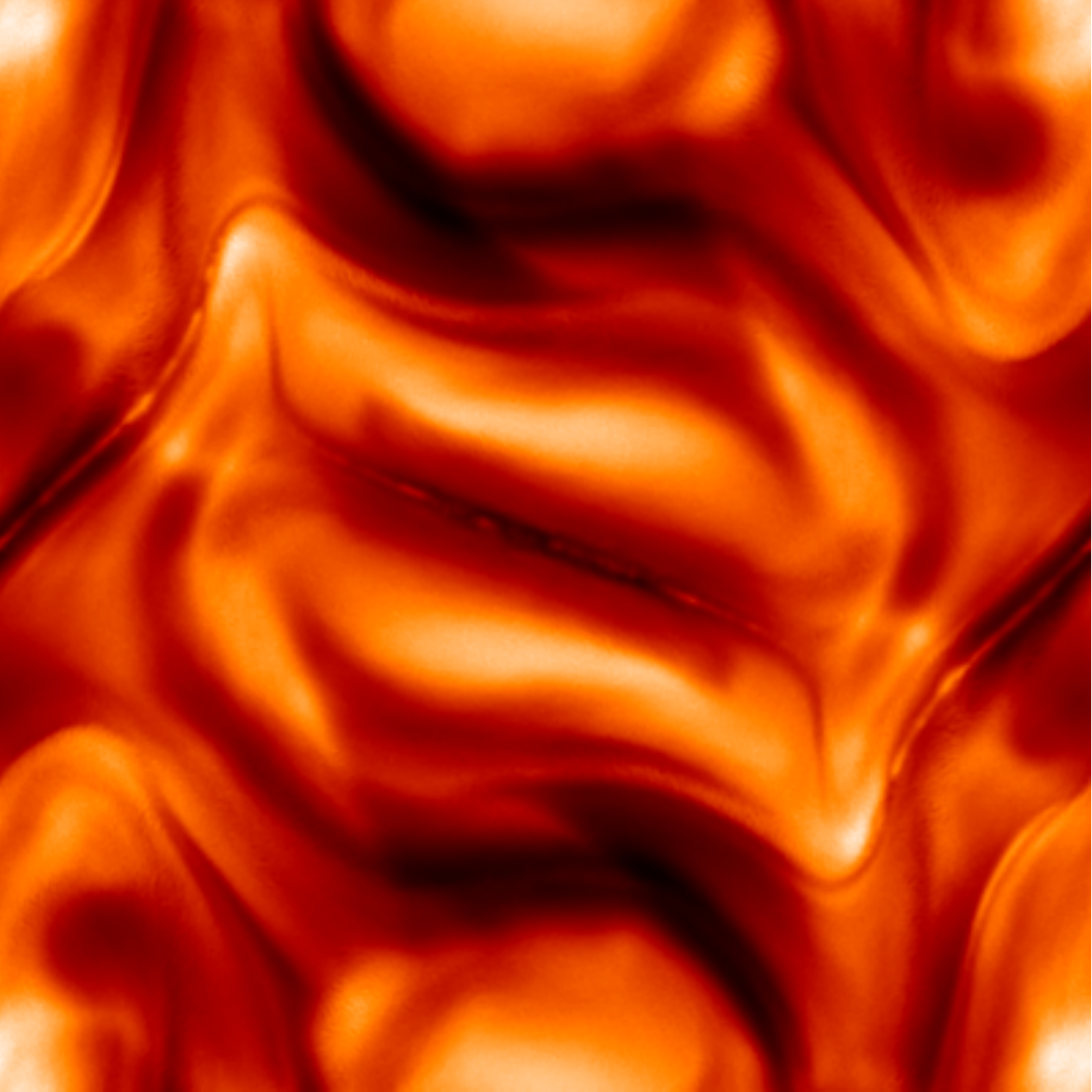}
& \includegraphics[width=0.1989\textwidth]{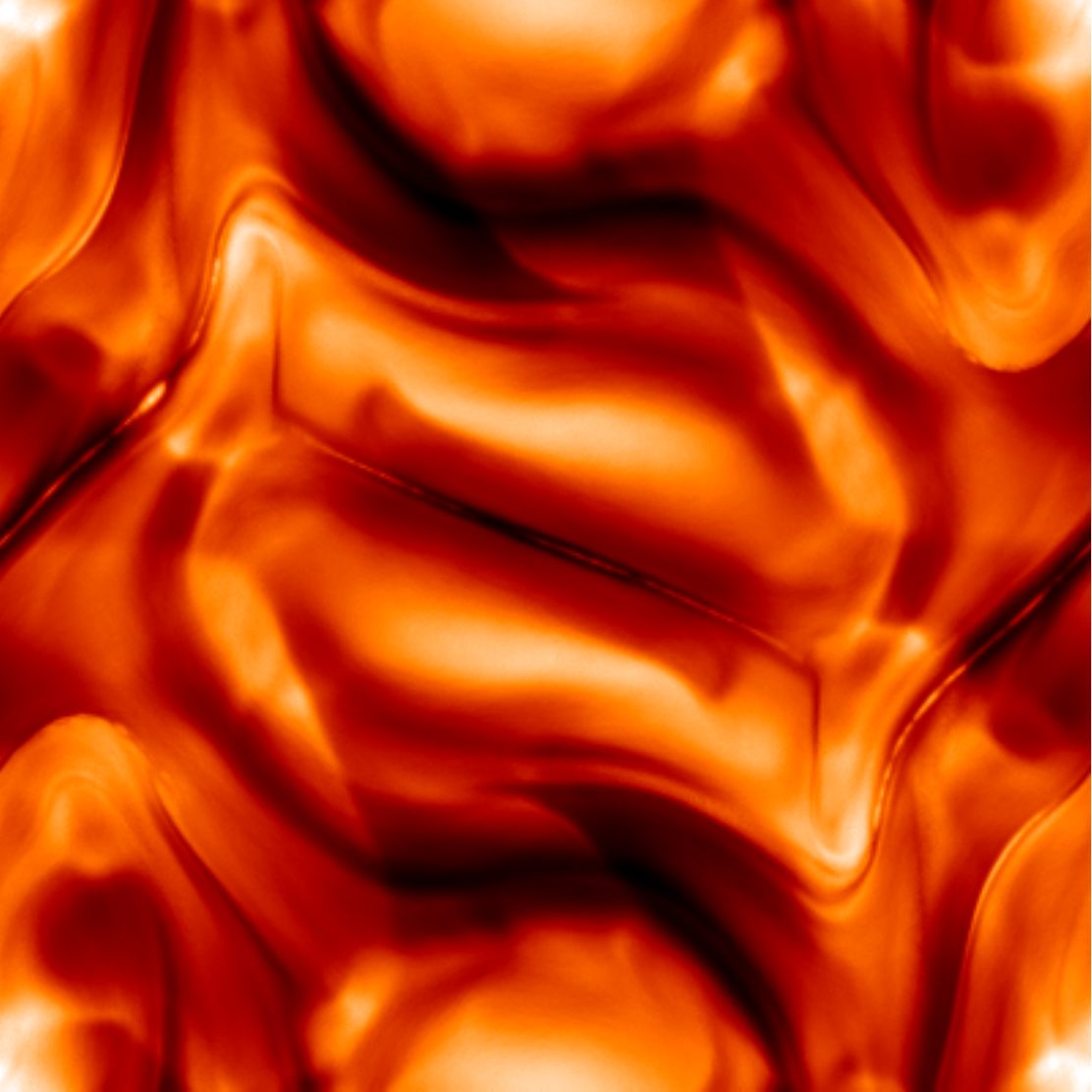}
& \includegraphics[width=0.1989\textwidth]{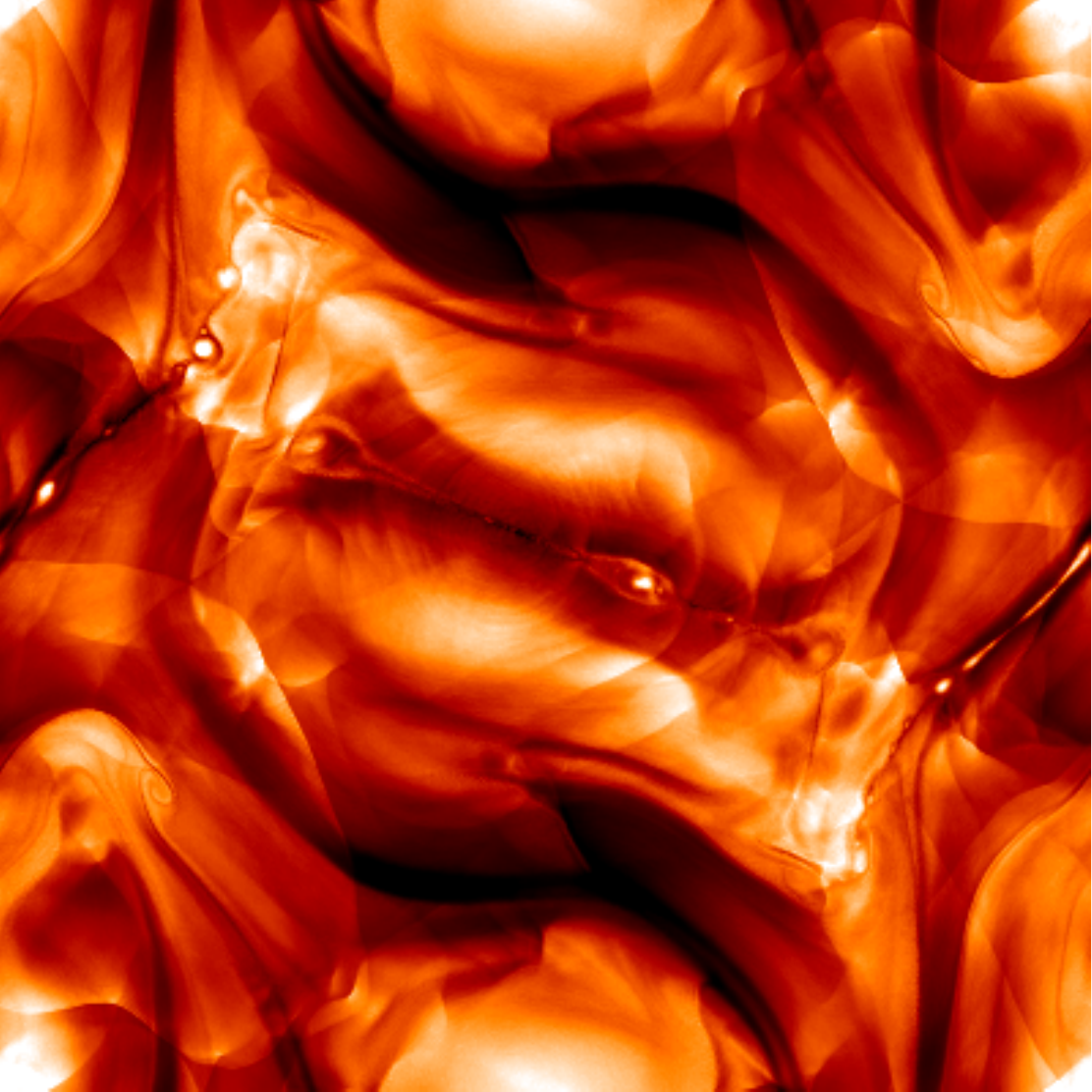}
& \includegraphics[width=0.1989\textwidth]{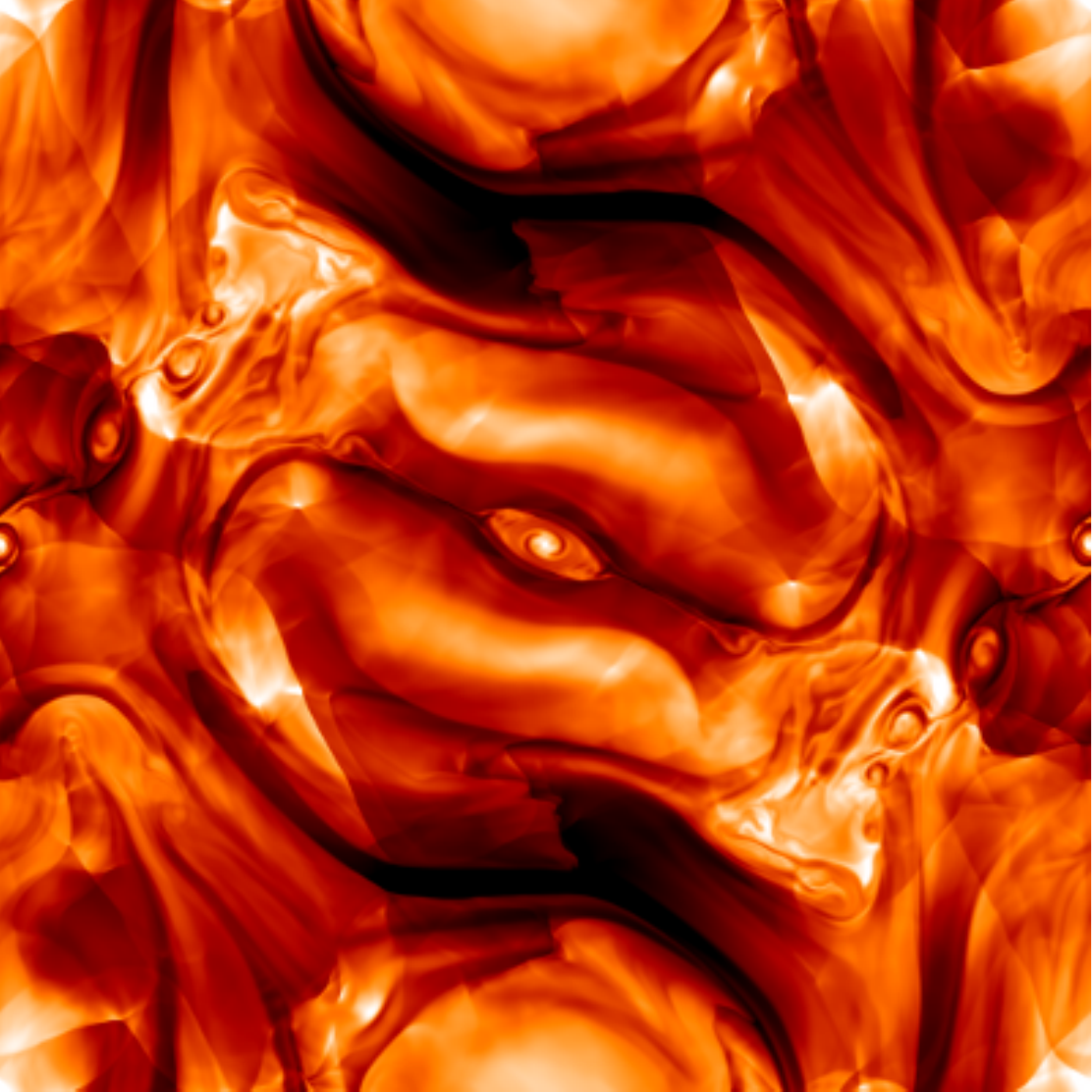}
%& \includegraphics[height=0.1855\textwidth]{cleaning/orszag/colorbar-orszag-density.pdf} 
\end{tabular}
\includegraphics[width=0.999\textwidth]{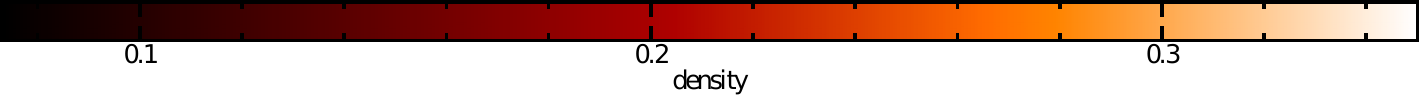}
\caption{Density of the Orszag-Tang vortex at resolutions of $128\times148$, $256\times296$, $512\times590$, and $1024\times1182$ particles (left to right), with comparison to results obtained using the {\sc Athena} code for $1024^2$ grid cells (far right).  As the resolution is increased, high density islands begin to form which is also observed in results from the {\sc Athena} code.}
\label{fig:orszag-resolution-density}
\end{figure}

\begin{figure}
 \centering
\includegraphics[width=0.45\textwidth]{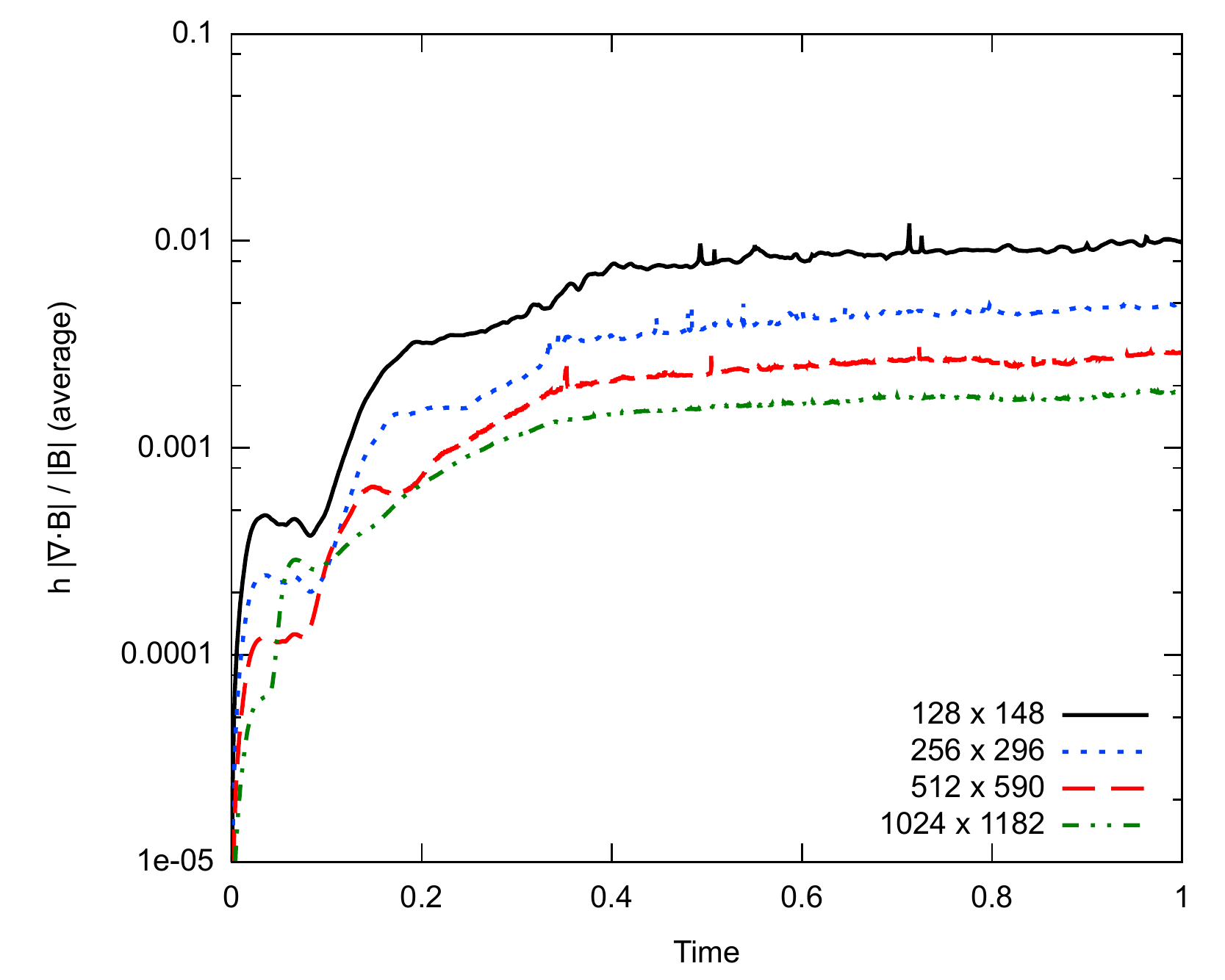}
\includegraphics[width=0.45\textwidth]{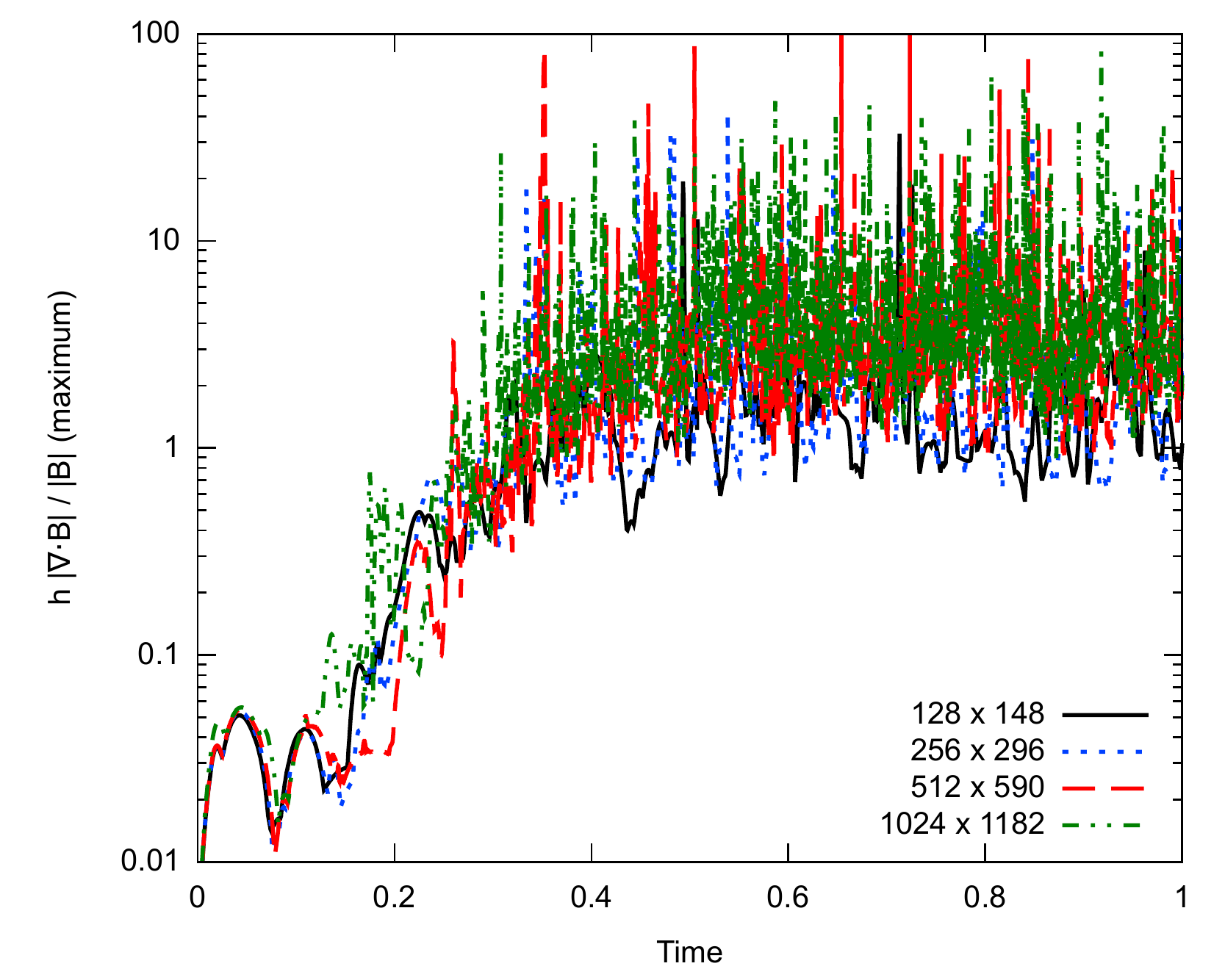}
\caption{Average (left) and maximum (right) divergence error in the Orszag-Tang vortex at resolutions of $128\times148$, $256\times296$, $512\times590$, and $1024\times1182$ particles.  The maximum divergence error remains similar for the different resolutions, but the average divergence error decreases with increasing resolution.}
\label{fig:orszag-resolution}                                            
\end{figure}

Finally, the Orszag-Tang vortex test was performed for a series of increasing resolution: $128\times148$, $256\times296$, $512\times590$, and $1024\times1182$ particles.  Divergence cleaning, without resistivity, was used for all cases.  The densities of these runs at $t=1.0$ are shown in Figure~\ref{fig:orszag-resolution-density}, along with results obtained using the {\sc Athena} code \citep{athena} with $1024^2$ grid cells.  In the largest resolution case, high density islands begin to form in the solution. These features arise from the tearing mode instability of non-ideal MHD \citep{fkr63}, caused by magnetic reconnection along a current sheet. They are also exhibited in the results from the {\sc Athena} code, and can be seen at lower resolutions in SPMHD when the Euler Potentials are used (see Figure~\ref{fig:orszag-compilation} for an example).  Despite the symmetrical initial state, the islands show artificial asymmetries resulting from momentum not being exactly conserved in SPMHD. Figure~\ref{fig:orszag-resolution} shows the average and maximum divergence error (left and right panels, respectively).  Though the maximum error remains similar for all cases, the average is seen to decrease with increasing resolution.

\subsection{Three dimensional divergence advection}
\label{sec:adv3d}
 We now turn to 3D tests, beginning with a three dimensional generalisation of the divergence advection problem.  In particular, we wish to determine the optimal values for $\sigma_\psi$ when the divergence waves propagate in three dimensions rather than two.

\subsubsection{Setup}

The principle of the test remains similar to 2D versions, except a cubic volume of fluid is used in the region $x,y,z \in [-0.5, 1.5]$.  The initial velocity field is extended to ${\bf v} = [1,1,1]$ to add drift in the $z$-direction.  The magnetic field remains as previously, $B_z = 1 / \sqrt{4 \pi}$, with a spherical perturbation introduced to the $x$-component of the field as given by Equation~\ref{eq:adv-divergence-perturbation}, except now using $r = \sqrt{x^2 + y^2 + z^2}$.  The radial extent $r_0 = h$ is chosen to mimic a divergence error at the resolution scale.  The density and pressure remain unchanged, with $\rho = 1$, $P = 6$, and $\gamma = 5/3$.  The problem is set up on a cubic lattice with $50^3$ particles.

\subsubsection{Optimal values of the damping parameter}
\label{sec:adv3d-sigma}

\begin{figure}
 \centering
\includegraphics[width=0.45\textwidth]{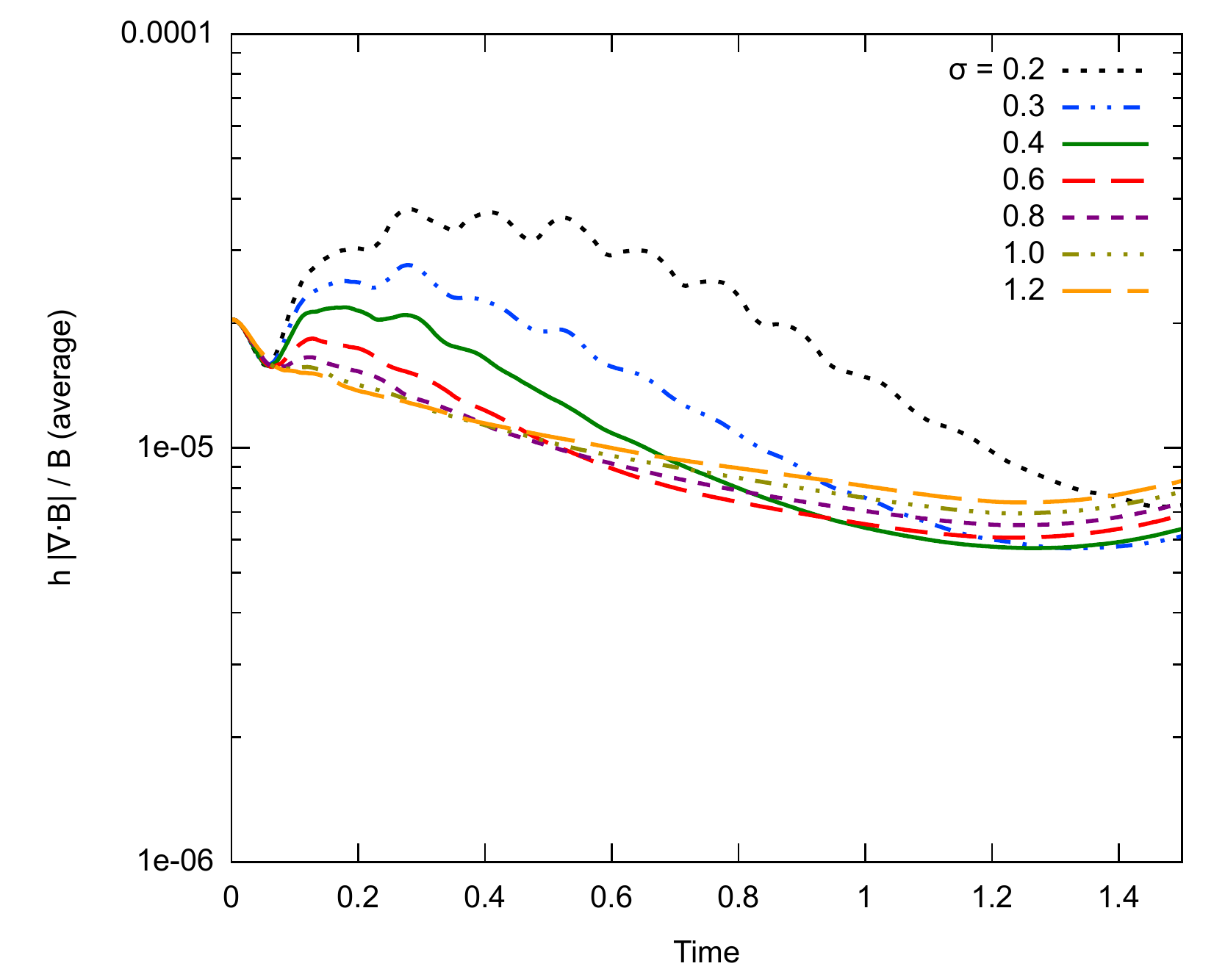}
\includegraphics[width=0.45\textwidth]{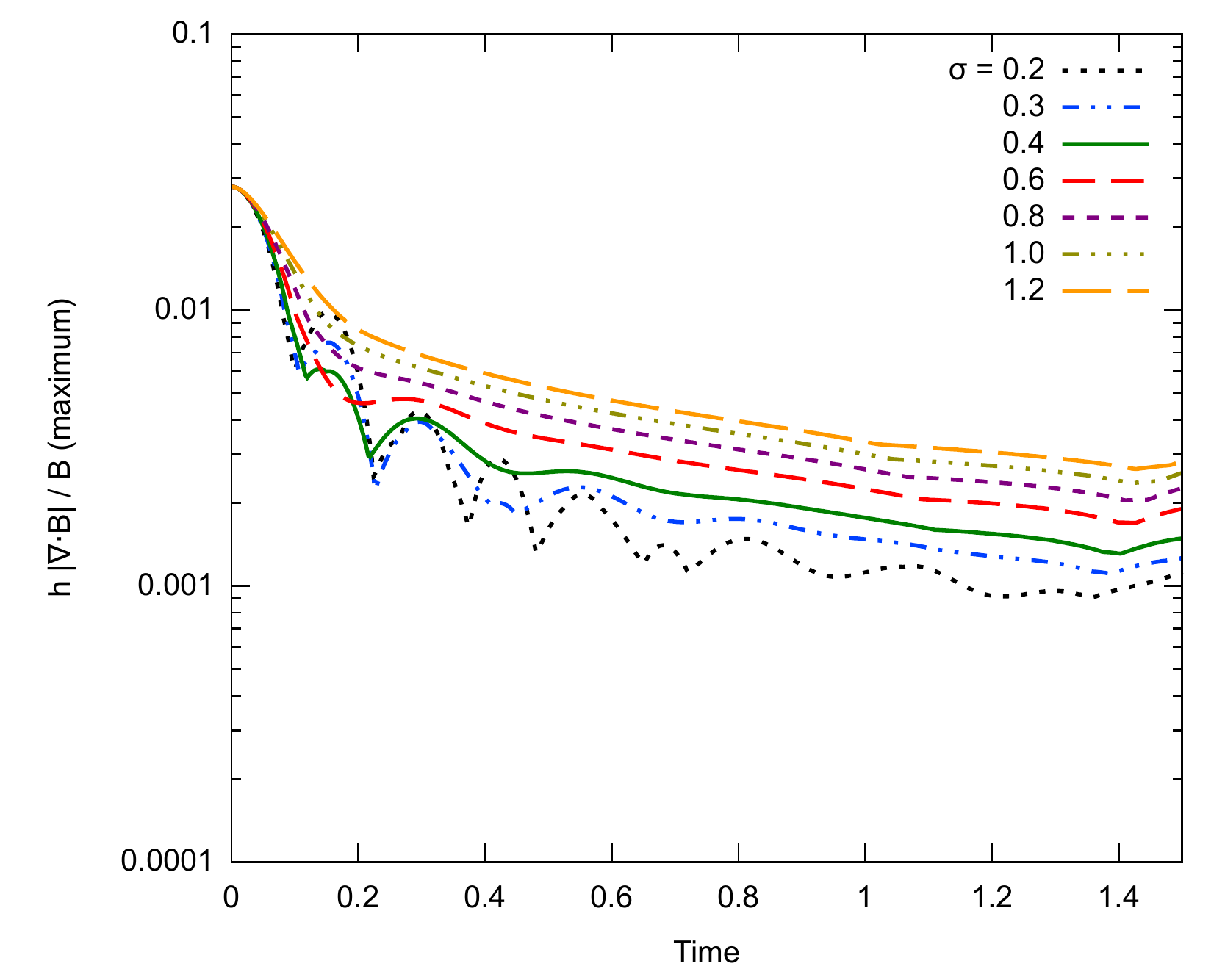}
\caption{Average and maximum divergence error in the 3D advection test for varying strengths of the damping parameter, $\sigma_\psi$.  The best results are obtained for $\sigma_\psi \sim$ 0.8--1.2.}
\label{fig:adv3d-sigma}                                            
\end{figure}

This test was performed for $\sigma_\psi \in [0.2, 1.2]$ with results of the average and maximum divergence given by Figure~\ref{fig:adv3d-sigma}.  The optimal cleaning is obtained for $\sigma_\psi \sim$ 0.8--1.2, which differs from the optimal values obtained for the 2D tests of $\sigma_\psi \sim$ 0.2--0.3.  This is attributed to the hyperbolic wave spreading over a volume instead of an area, thus being more effective than in our 2D tests, and therefore requiring a higher value of $\sigma_\psi$ to achieve critical damping.

\subsection{Gravitational collapse of a magnetised molecular cloud core}
\label{sec:jet}

Our most complex test is drawn from our intended application: simulations of star formation that involve magnetic fields \citep{ptb12}.  These simulations follow \citet{pb07}, where an initial one solar mass sphere of gas with uniform magnetic field in the $z$-direction and in solid body rotation contracts under self-gravity to form a protostar with surrounding disc.  However, at times near peak density, the magnetic field in the dense central region becomes strong and can produce high divergence errors.  This has limited the range of initial magnetic field strengths which could be simulated, as if the divergence grows too large, the tensile instability correction term injects enough momentum into the system to erroneously eject the protostar out of its disc \citep{pf10b}.  Thus, this simulation proves an excellent demonstration of the capabilities of the constrained hyperbolic divergence cleaning method to reduce divergence errors in realistic, 3D simulations.

\subsubsection{Setup}

This simulation uses the code, {\sc sphNG}. The sphere of gas has radius $R = 4\times10^{16} \text{cm}$ with uniform density $\rho = 7.43\times10^{-18}$ g $\text{cm}^{-3}$ and is set in solid body rotation with $\Omega~=~1.77~\times~10^{-13}$ rad~$\text{s}^{-1}$.  A barotropic equation of state is used, as described in \citet{pb07}.  The magnetic field strength is set to give a mass-to-magnetic flux ratio of 5 times the critical value for magnetic fields to provide support against gravitational collapse.  To avoid edge effects with the magnetic field, the sphere is embedded in a periodic box of length $4R$ containing material surrounding the sphere set in pressure equilibrium with density ratio 1:30.  This test uses only a minimal amount of resistivity, with $\alpha_B \in [0, 0.1]$. Self-gravity is simulated using a hierarchical binary tree where each node contains mutual nearest neighbours \citep{benzetal90}, with gravitational force softening using the SPH kernel as described by \citet{pm07}.  The free fall time is $\sim 24000$ years.  A sink particle \citep{bbp95} is inserted once the gas density surpasses $\rho_{\text{sink}} = 10^{-10}$ g $\text{cm}^{-3}$, and accretes particles within a radius of $6.7$ AU. The mass and momentum of accreted particles are added to the sink particle, but no information is retained about the magnetic field.

\subsubsection{Results}
\label{sec:star-results}
\begin{figure}
\centering
\includegraphics[width=0.9\textwidth]{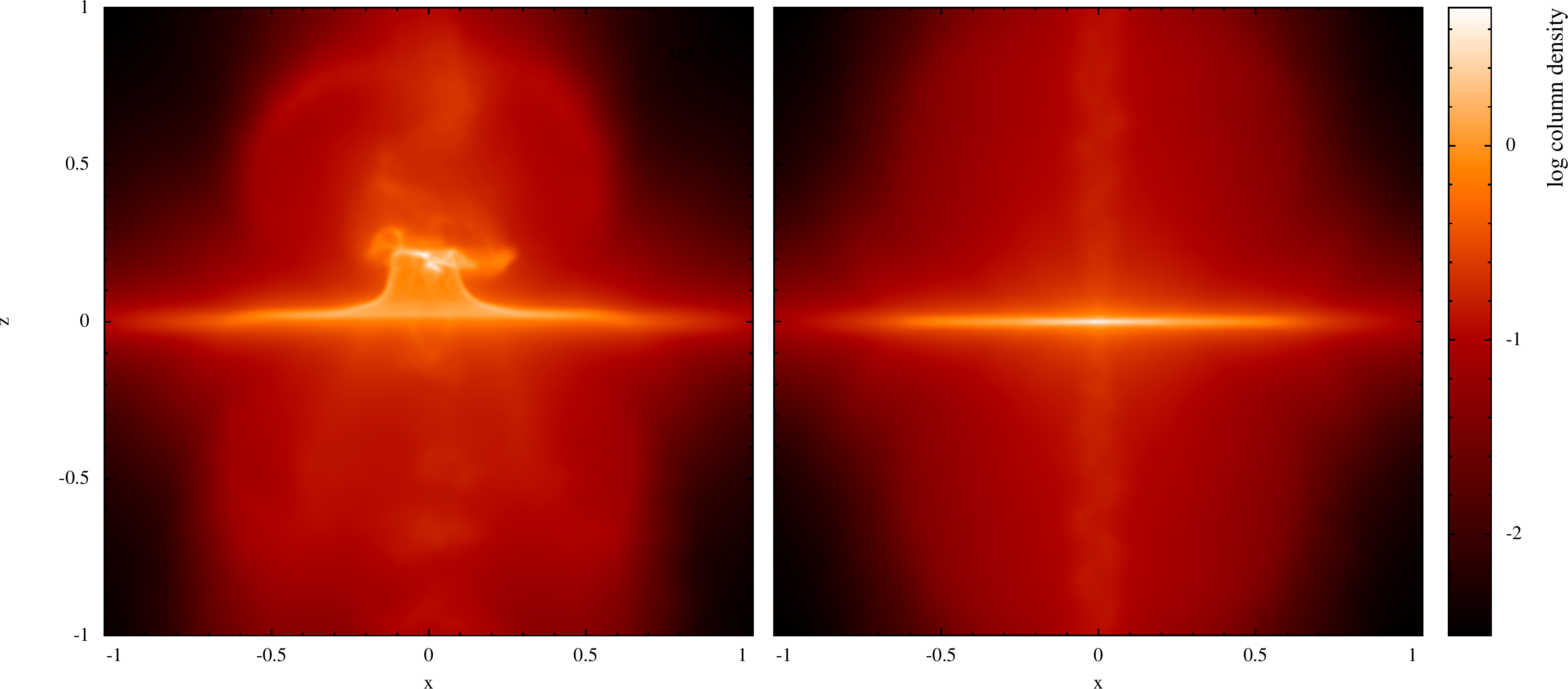}
\caption{Renderings of the column density of the star formation simulation at $t=1.1$ free fall times. The simulation without cleaning (left) suffers a dramatic loss of momentum conservation (c.f. Figure~\ref{fig:star-mom}) induced by high divergence errors (c.f. Figure~\ref{fig:star-divb}). By contrast, the simulation with our new divergence cleaning scheme applied (right) remains stable and launches a steady, collimated outflow \citep{ptb12}.}
\label{fig:star-column-density}
\end{figure}

\begin{figure}
\centering
\begin{minipage}[t]{0.45\textwidth}
\includegraphics[width=\textwidth]{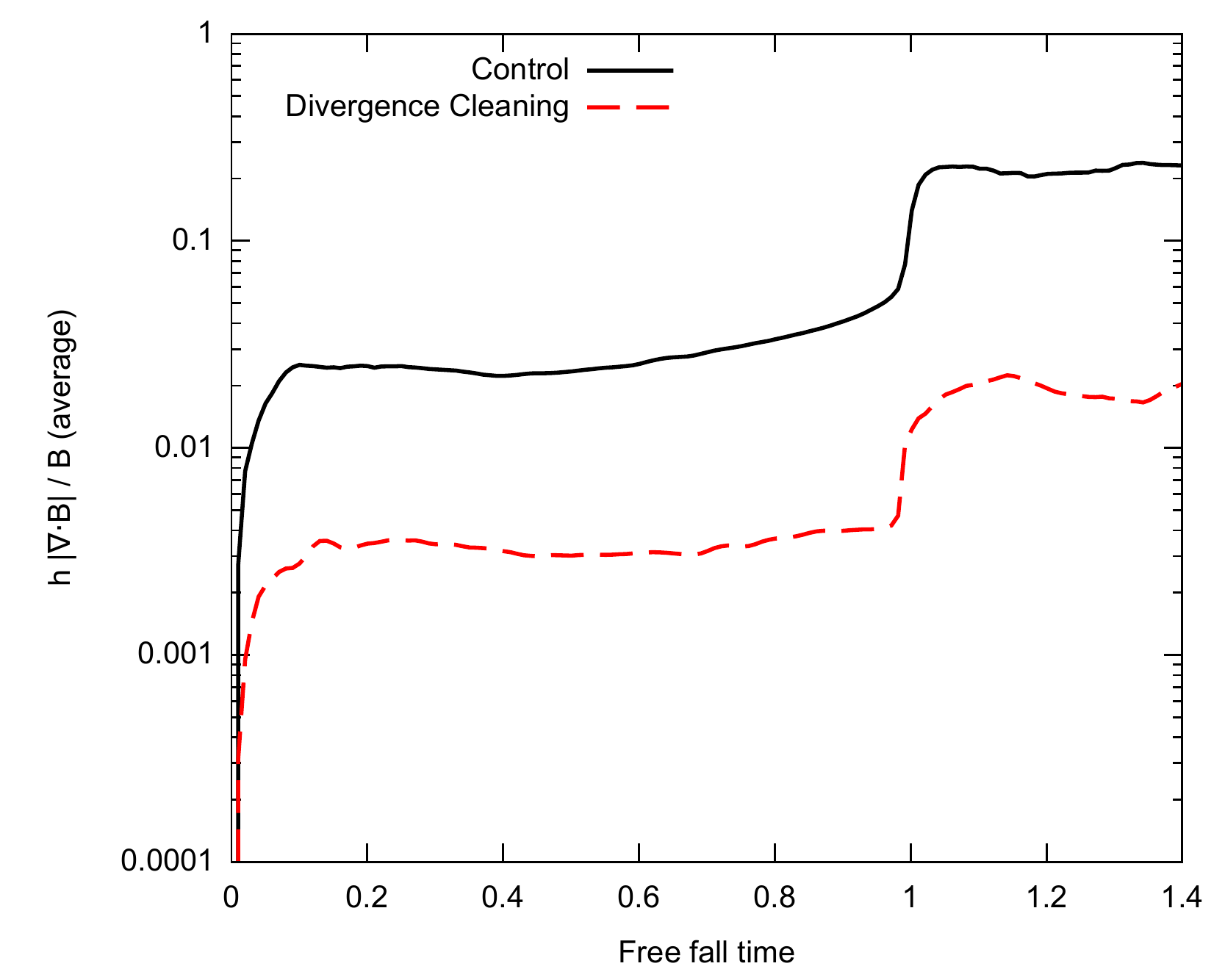}
\caption{Average divergence error as a function of time for the star formation simulation, which shows that adding divergence cleaning reduces the divergence error by an order of magnitude.}
\label{fig:star-divb}
\end{minipage}
\hspace{0.05\textwidth}
\begin{minipage}[t]{0.45\textwidth}
\includegraphics[width=\textwidth]{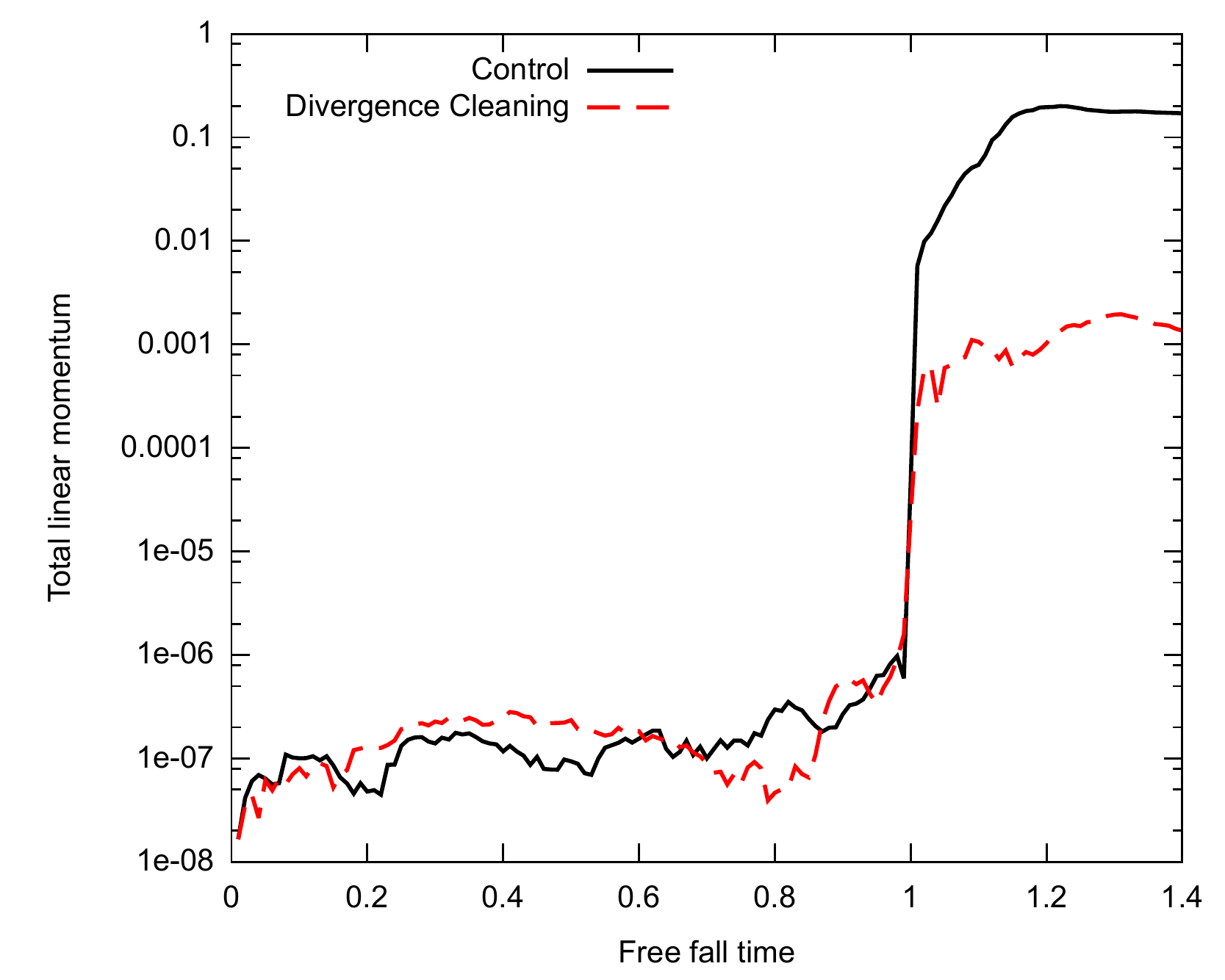}
\caption{Magnitude of the total linear momentum in the star formation simulation. The system initially has zero net momentum, which increases due to the magnetic tensile instability correction and tree-based gravitational forces.  After the protostar forms ($t=1$), the momentum conservation in the divergence cleaning case is improved by two orders of magnitude over the control case.}
\label{fig:star-mom}
\end{minipage}
\end{figure}

Figure~\ref{fig:star-column-density} shows column density comparisons of simulations with (right) and without (left) divergence cleaning at $t=1.1$ free fall time, showing that drastic improvements to the results are obtained by incorporating divergence cleaning.  Most importantly, the protostar remains stable in its disc.  The average and maximum divergence error are both reduced by roughly an order of magnitude (Figure~\ref{fig:star-divb}), and this leads to a corresponding improvement in the momentum conservation of around two orders of magnitude (Figure~\ref{fig:star-mom}).

\subsubsection{Optimal sigma values}

\begin{figure}
 \centering
\includegraphics[width=0.45\textwidth]{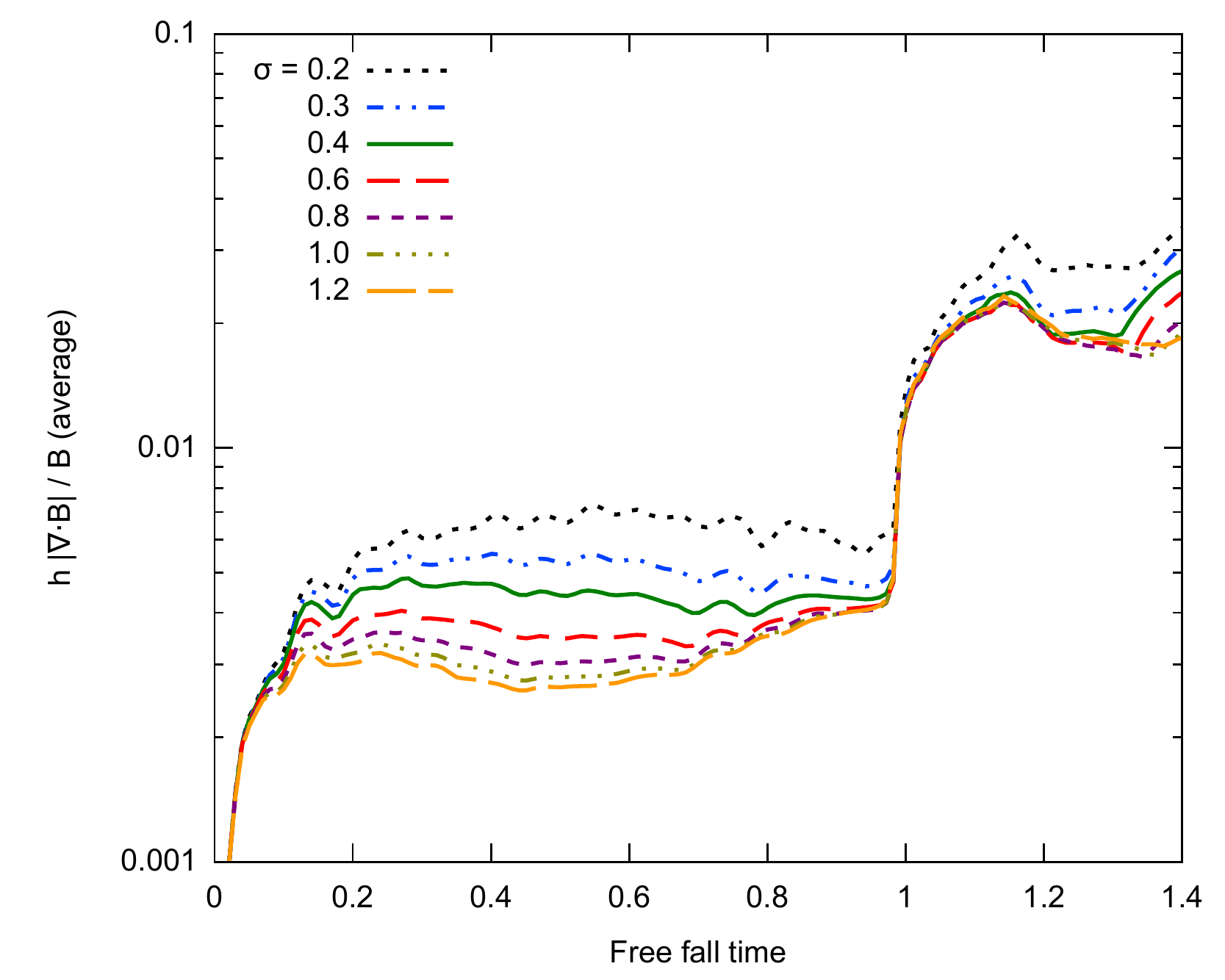}
\includegraphics[width=0.45\textwidth]{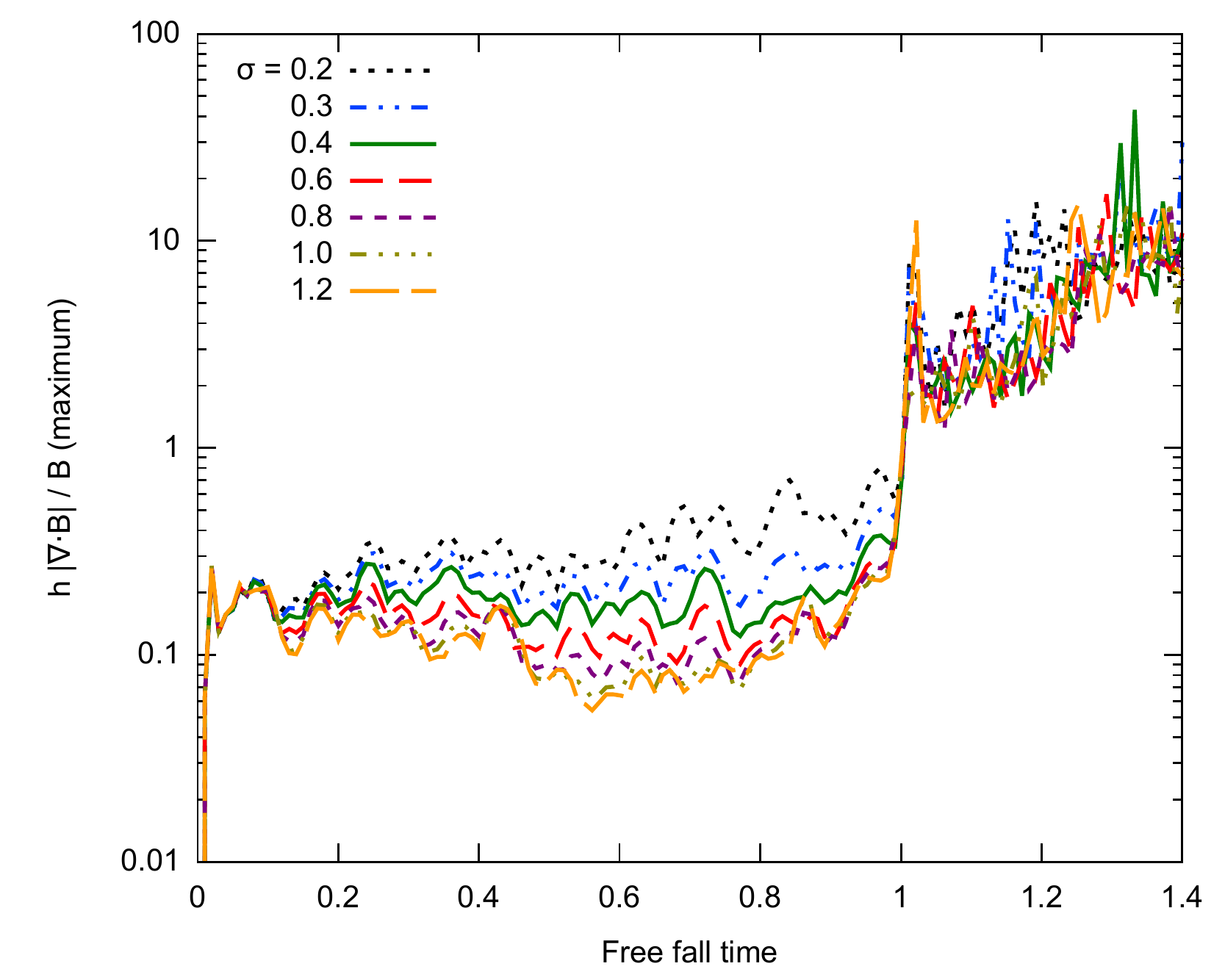}
\caption{Average and maximum divergence error in the star formation simulation, varying the damping parameter in the range $\sigma_\psi \in [0.2,1.2]$.  The best results are obtained with $\sigma_\psi \sim$ 0.8--1.2.}
\label{fig:star-cleaning-sigma}
\end{figure}

This simulation was repeated for several values of the damping parameter in the range $\sigma_\psi \in [0.2-1.2]$, with the results on average and maximum divergence error presented in Figure~\ref{fig:star-cleaning-sigma}.  Optimal results were obtained for $\sigma_\psi \sim$ 0.8--1.2, which agrees with values found for the 3D advection test (Section~\ref{sec:adv3d-sigma}).

\subsubsection{Inclusion of the $\tfrac{1}{2} \psi (\nabla \cdot {\bf v})$ term}

Adding $\tfrac{1}{2} \psi (\nabla \cdot {\bf v})$ to the evolution equation for $\psi$ was motivated by energy conservation considerations, but the resulting question is what effect this has on divergence cleaning. The star formation simulation represents the ideal test case with which to examine this, with a large $\nabla\cdot{\bf v}$ present due to the gravitational collapse of the gas.  We have performed this simulation both with and without this term, using $\sigma_\psi=0.8$, and found no distinguishable difference in the linear momentum, and average and maximum $h \vert \nabla\cdot{\bf B} \vert / \vert {\bf B} \vert$ profiles. Similar results were obtained also found in the other tests. We conclude that, although this term is necessary for strict energy conservation, it has almost zero effect on the effectiveness of the cleaning scheme.

\subsection{Magnetised Mach 10 turbulence}
\label{sec:mhdturb-cleaning}

To fully investigate the importance of the $\tfrac{1}{2} \psi (\nabla \cdot {\bf v})$ term, we turn to `turbulence in a box' simulations of driven, magnetised, Mach 10 turbulence. Such a calculation will have many interacting shocks, and for this calculation, produces density variations up to $1000\times$ the initial density. There is substantial and continual compression and rarefaction of the gas, and as such, is the perfect testbed to test the effect of a term which includes $\nabla \cdot {\bf v}$. This test utilises the same simulation setup as in Section~\ref{sec:mhdturb}, with the full details and a comparison of results between SPMHD and grid-based methods presented in Chapter~\ref{sec:chapter-mhdturb}.

\subsubsection{Setup}

The system is set in a periodic box of unit length, $L=1$. The initial density is uniform $\rho=1$, and the initial velocity is ${\bf v}=0$. The initial magnetic field is set $B_z = \sqrt{2} \times 10^{-5}$. The equation of state is isothermal, $P = c_{\rm s} \rho$, with $c_{\rm s}=1$. The resolution is $128^3$ particles. The \citet{mm97} switch for artificial viscosity has been used, along with the artificial resistivity switch developed in Chapter~\ref{sec:chapter-switch}.

The turbulence is stochastically driven at large scales ($1 < k < 3$, peaked at $k=2$) by a force obtained from the Ornsetin-Uhlenbeck process \citep{ep88, federrathetal10}. This keeps the turbulence in a statistical steady state at rms velocity Mach 10 ($\mathcal{M}=10$). The autocorrelation time of the driving motion is one turbulent turnover time, $t_{\rm c} = L/(2 \mathcal{M} c_{\rm s})$. The turbulence is driven using using only the solenoidal component of the driving force, obtained through construction in Fourier space.

The calculations have been run both with and without the $\tfrac{1}{2} \psi (\nabla \cdot {\bf v})$ term in the divergence cleaning. Both calculations use $\sigma_\psi=1.0$.

\subsubsection{Results}

\begin{figure}
 \centering
\includegraphics[width=0.45\textwidth]{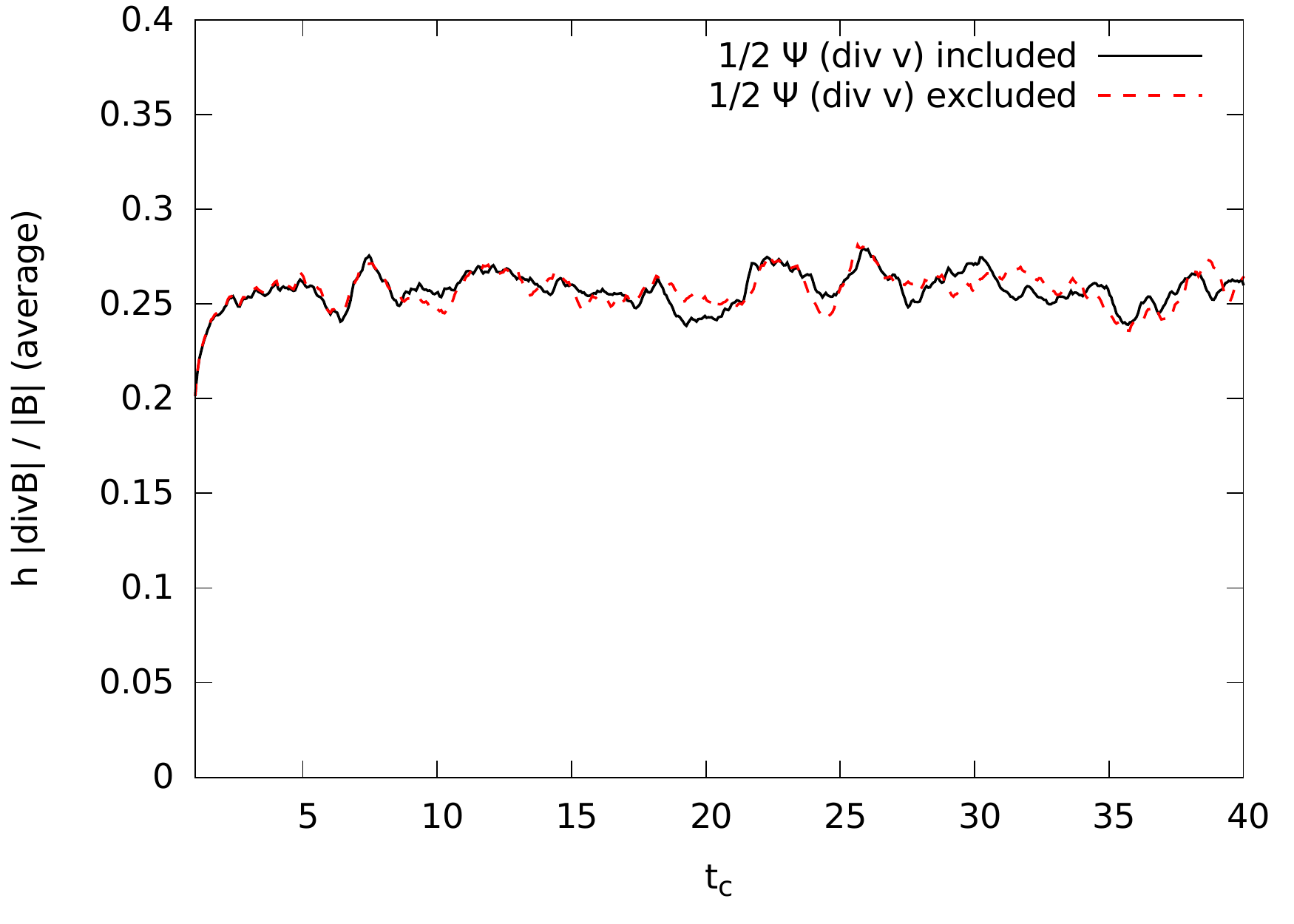}
\includegraphics[width=0.45\textwidth]{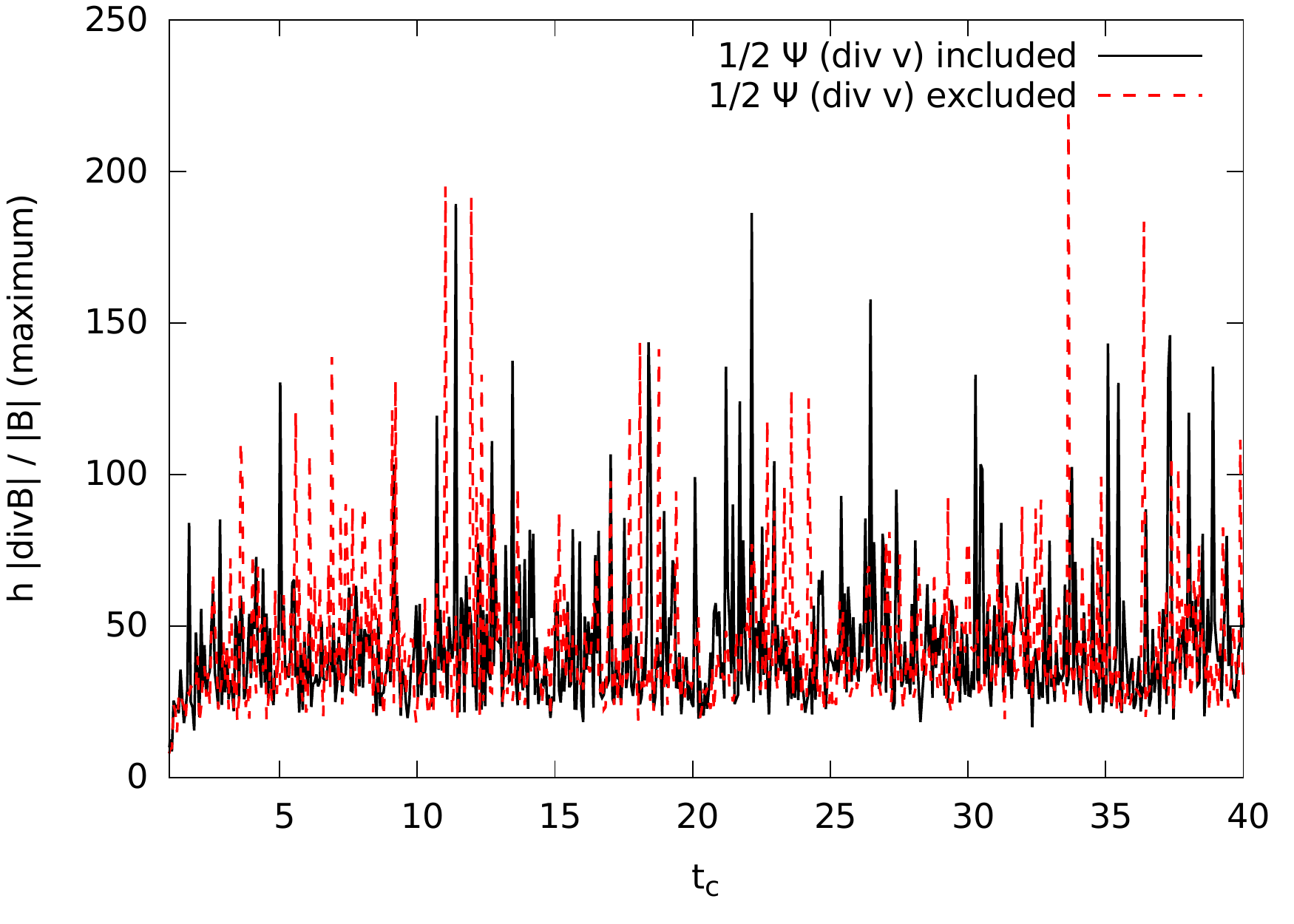}
\caption{Average and maximum divergence error in the magnetised, Mach 10 turbulence calculation with and without the $\tfrac{1}{2} \psi (\nabla \cdot {\bf v})$ term in the $\psi$ evolution equation. No distinguishable difference is present between the calculations, and we conclude that this term is not important for the effectiveness of the cleaning method.}
\label{fig:cleaning-mhdturb}
\end{figure}

Figure~\ref{fig:cleaning-mhdturb} shows the average and maximum $h \vert \nabla \cdot {\bf B} \vert / \vert {\bf B} \vert$ over a period of 40$t_{\rm c}$. No distinguishable difference is found in the average and maximum divergence error. As with the star formation calculation, we conclude that the inclusion of this term does not improve the effectiveness of the cleaning and is not needed for stability. Even in the presence of strong compression and rarefaction of the gas, this term has no practical effect on the cleaning method.

\section{Enhancing the cleaning method}
\label{sec:cleaning-enhanced}

The constrained implementation of mixed hyperbolic/parabolic divergence cleaning only approximately upholds the divergence-free constraint on the magnetic field. The maximum effectiveness of the cleaning is limited by the explicit timestep constraint, as it must obey the Courant condition in order to ensure stability of the hyperbolic waves. In the following section, we investigate two approaches to enhance the effectiveness of the cleaning method, in particular demonstrating how the method can be used to achieve $\nabla \cdot {\bf B}=0$.

\subsection{Over-cleaning}

The simplest approach to improve the effectiveness of the cleaning is to increase $c_{\rm h}$. Introducing a factor, $f_{\rm ovc} \ge 1$, the cleaning wave speed is adjusted to $c_{\rm h} \to f_{\rm ovc} c_{\rm h}$. This requires a corresponding reduction in timestep according to $\Delta t \to \Delta t / f_{\rm ovc}$. We call this method `over-cleaning'.

This incurs a significant computational cost as the {\it full} simulation is slowed down by a commensurate amount to the over-cleaning factor. If 10 times over-cleaning is used, then the simulation will run 10 times slower. This sacrifices the cheap computational cost of hyperbolic divergence cleaning, however the simplicity of this approach has found practical use by \citet{btp14} who used 30$\times$ over-cleaning to reduce divergence errors in their simulations of protostar formation.

\subsection{Sub-cycling}

% set number of iterations or set a tolerance

The second approach investigated is to cycle the cleaning equations in-between timesteps. After each timestep, the errors introduced into the magnetic field are removed by running the cleaning equations in isolation. The dynamics of the simulation halted during the sub-cycles. This is similar to the `static cleaning' tests in Section~\ref{sec:test-density-jump} and \ref{sec:test-free-boundaries}. 

This approach is significantly cheaper in computational expense compared to over-cleaning.  It does not slow the evolution of the system down (timestep size remains as normal). Since the particle positions remain static during the sub-cycles, only one neighbour search is required to be performed at the start of the iterations.

A second advantage is that sub-cycling allows for the continual cleaning of the magnetic field until the error falls below a desired tolerance. This guarantees the divergence error to be below that tolerance at all times.

Implementing sub-cycling with individual timesteps requires special consideration. Consider the situation of a subset of particles which are `active', surrounded by particles which are `inactive'. If sub-cycling is run on just the active set of particles, then the magnetic field near the boundary between the two sets of particles may be become distorted as the cleaning is unable to remove the errors on the adjacent inactive particles. Running sub-cycling on the full system is expensive (especially if the active regions constitute only a tiny fraction of the full system). One approach would be to perform iterations in a hierarchy according to the timestep bins. The particles on the smallest timestep bin could be cleaned every iteration, the particles on the second smallest timestep bin are cleaned every second iteration, and so on.

\subsection{Numerical tests}

Our tests compare the effectiveness of over-cleaning and sub-cycling. To allow direct comparison, sub-cycling is run for a fixed number of iterations so that it compares more directly to over-cleaning. However, tests are also performed where sub-cycling is run with a tolerance on the average $h \vert \nabla \cdot {\bf B} \vert / \vert {\bf B} \vert$. Following these tests, we consider a `static' test of an evolved state of the Orszag-Tang vortex where the optimal value of $\sigma_\psi$ in the damping term for sub-cycling is investigated. We will show that sub-cycling is able to clean the magnetic field all the way to $\nabla \cdot {\bf B}=0$.

\subsubsection{Orszag-Tang vortex}
\label{sec:itvovc-test}

The effectiveness of over-cleaning and sub-cycling is investigated using the Orszag-Tang vortex. The initial state of the problem is described in Section~\ref{sec:ot}. For these tests, we used $512^2$ particles arranged on a square lattice. In all cases, artificial resistivity is applied using the new resistivity switch described in Chapter~\ref{sec:chapter-switch}. 

% show orszag-tang normal, 10x ovc, 100x ovc, 10x iteration, 100x iteration

The vortex is simulated using 10, 20, and 100$\times$ over-cleaning ($f_{\rm ovc}=10$, $20$, and $100$) and sub-cycling with 10, 20, and 100 iterations ($9$, $19$, and $99$ between timesteps plus the cleaning performed on the system timestep). These are intended to provide an overall level of cleaning that is comparable between the two approaches. A calculation using `normal' hyperbolic divergence cleaning (with $\sigma_\psi=0.3$) is included to act as a reference.

\begin{figure}
\centering
\includegraphics[width=0.75\linewidth]{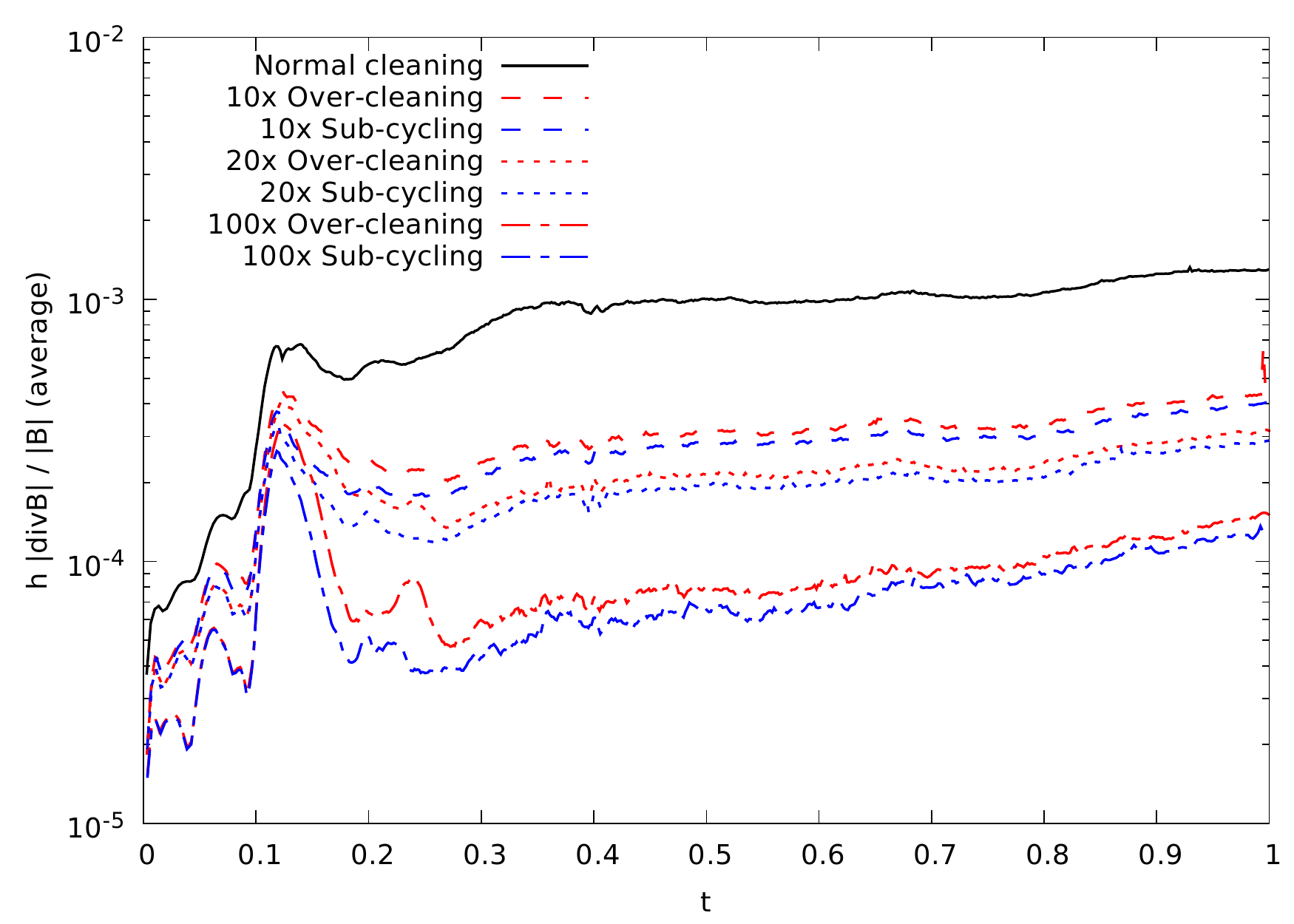}
\caption{The Orszag-Tang vortex with $10$, $20$, and $100\times$ over-cleaning and $10$, $20$, and $100$ iterations of sub-cycling. Over-cleaning and sub-cycling have similar results when the over-clean factor and iteration count are the same. Each factor of 10 increase in the over-cleaning factor and sub-cycling iteration count yield half an order of magnitude reduction in average divergence error.}
\label{fig:itvovc}
\end{figure}

Figure~\ref{fig:itvovc} shows the average divergence error in the system, showing that over-cleaning and sub-cycling yield similar average divergence error when the over-cleaning factor and number of sub-cycling iterations are equal. Each factor of 10 increase in the over-cleaning factor and iteration count yield approximately half an order of magnitude decrease in average divergence error.

\begin{table}
\caption{Computational Cost of Over-cleaning and Sub-cycling}
\label{tbl:itvovc}
\centering
\begin{tabular}{ccc}
\hline
\hline
Calculation & cpu hours & Relative expense  \\ \hline
Normal cleaning & 17.6 & 1.0$\times$ \\
10$\times$ over-cleaning & 212.5 & $12.1\times$ \\
20$\times$ over-cleaning & 460.3 & $26.2\times$ \\
100$\times$ over-cleaning & 1956.0 & $111.1\times$ \\
10$\times$ sub-cycling & 60.0 & $3.4\times$ \\
20$\times$ sub-cycling & 96.5 & $5.5\times$ \\
100$\times$ sub-cycling & 372.3 & $21.2\times$ \\
0.1\% tolerance sub-cycling & 25.3 & $1.4\times$ \\
0.05\% tolerance sub-cycling & 34.7 & $2.0\times$ \\
0.02\% tolerance sub-cycling & 125.1 & $7.1\times$ \\
\hline
\end{tabular}
\end{table}

The computational expense of the over-cleaning and sub-cycling calculations is given in Table~\ref{tbl:itvovc}. Over-cleaning is almost directly proportional to the over-cleaning factor used, with $10\times$ over-cleaning corresponding to $12\times$ increase in computation time. On the other hand, 10 iterations of sub-cycling only increases computation time by $3.4\times$. Overall, sub-cycling is ${\sim}~4$--$5$ times more computationally efficient than over-cleaning.

% show orszag-tang with iteration tolerances, along with iteration count

\begin{figure}
\centering
\includegraphics[width=0.49\linewidth]{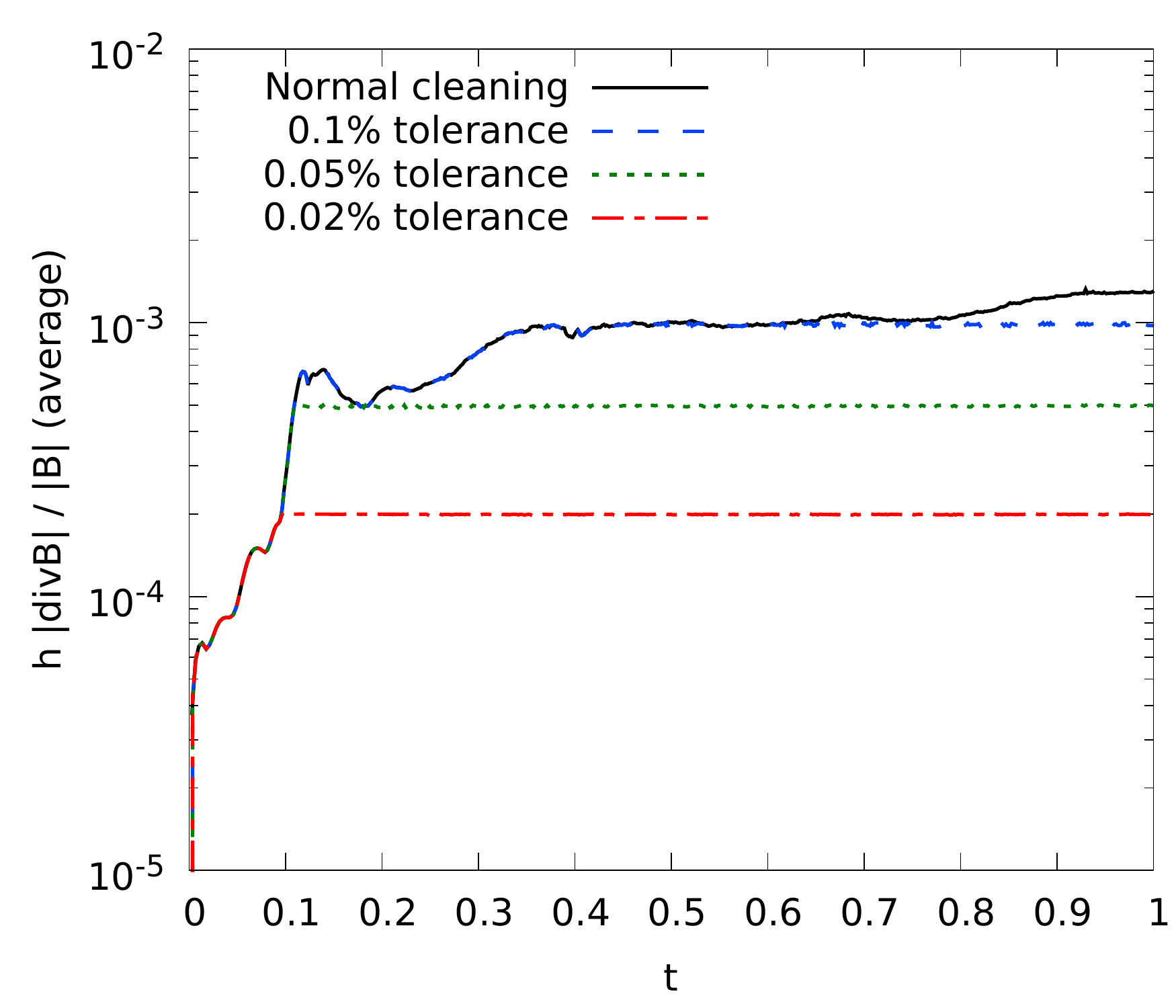} 
\includegraphics[width=0.49\linewidth]{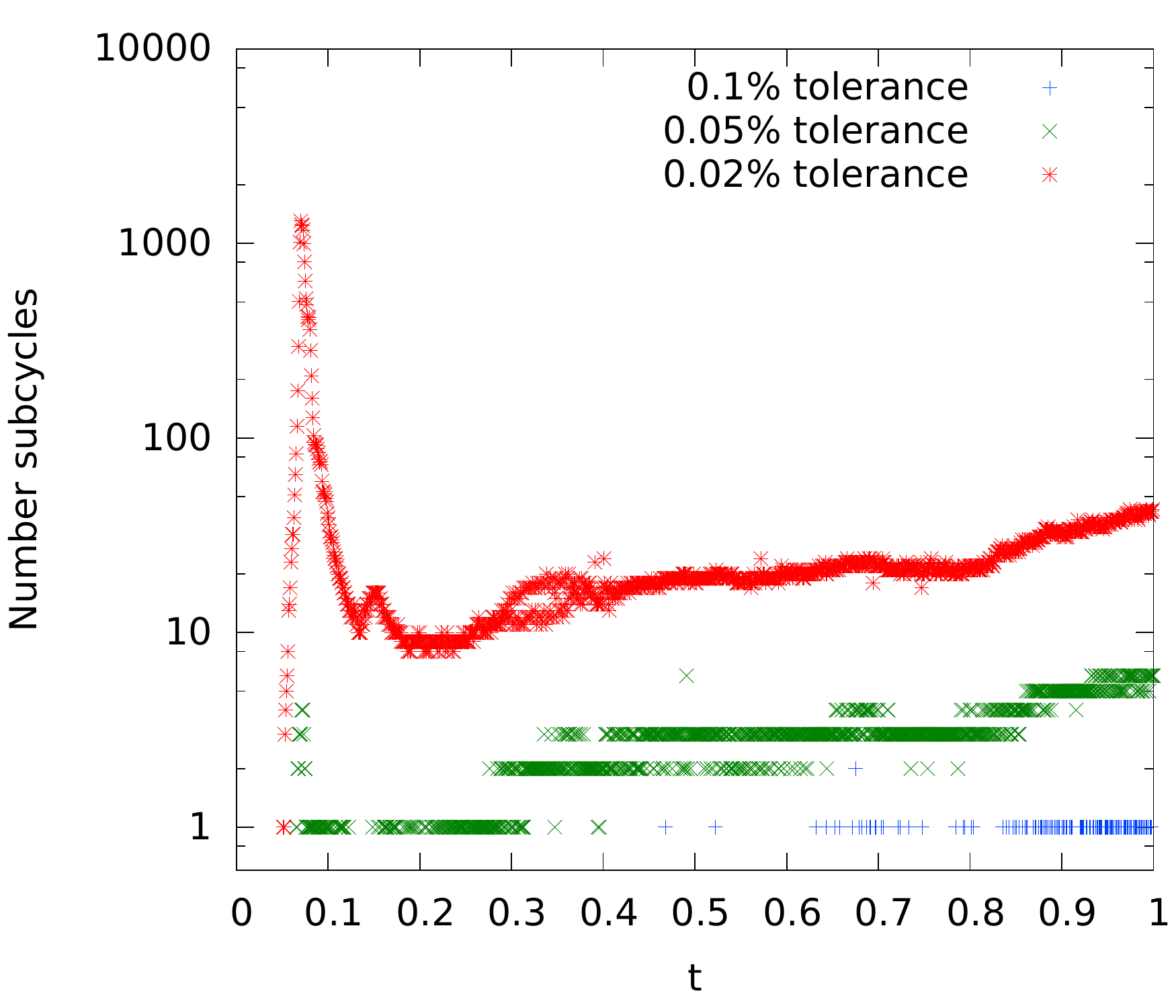}
\caption{The average divergence error in the Orszag-Tang vortex when sub-cycling is used to keep the average error below $0.1\%$, $0.05\%$, and $0.02\%$. The right panel shows the number of iterations required per timestep. The large peak in the $0.02\%$ tolerance case at $t~{\sim}~0.15$ is due to the particles breaking off their initial lattice arrangement, requiring ${\sim}~1000$ iterations for this case to treat.}
\label{fig:itvovc-tolerance}
\end{figure}

Additional simulations were performed where sub-cycling was run until the average divergence error was below $0.1\%$, $0.05\%$, and $0.02\%$. There was no limit placed on the number of iterations. Figure~\ref{fig:itvovc-tolerance} shows the average divergence error is maintained at the specified tolerance.

The number of iterations per sub-cycle step is given in the second panel of Figure~\ref{fig:itvovc-tolerance}. The $0.1\%$ average error case requires no iterations until late in the simulation, and then only an extra 1 or 2 per timestep. The $0.05\%$ average error case is under 10 iterations at all time. However, the $0.02\%$ average error case requires ${\sim}1000$ iterations early when the particles first break off their initial lattice arrangement ($t~{\sim}~0.15$), after which $20$--$40$ iterations are performed on average. A simulation was performed where the tolerance on average divergence error was set to $0.01\%$, but this caused a prohibitive number of iterations at $t~{\sim}~0.15$ ($>10^5$ iterations). It would be advisable to impose a maximum number of iterations to prevent such situations from stalling the calculation. %Note that Section~\ref{sec:itvovc-static-test} investigates the optimal value for $\sigma$ during sub-cycles, and these calculations were performed for $\sigma=0.3$.

We note that no high density islands form in any of the over-cleaning or sub-cycling calculations. These features appeared in the $512\times590$ particle calculation using Euler Potentials (Figure~\ref{fig:orszag-compilation}), as well as the high resolution $1024\times1182$ particle divergence cleaning calculation from Figure~\ref{fig:orszag-resolution-density} and the {\sc Athena} calculation in the same figure. They do not appear in the $2048^2$ particle calculations performed in Section~\ref{sec:orszag}, despite being even higher resolution. The difference in the $2048^2$ particle calculations is they included an artificial resistivity, whereas the $1024\times1182$ particle calculation did not. The formation of these features is related to the tearing mode instability of non-deal MHD \citep{fkr63}, whereby magnetic reconnection causes magnetic islands to form along current sheets. Thus, these features are related to the dissipation of the magnetic field, not from divergence errors in the magnetic field, though are sensitive to the degree of resistivity present. For example, they form in the $1024\times1182$ particle calculation when no artificial resistivity is applied, yet do not form in the higher resolution $2048^2$ calculation when a minimal amount of artificial resistivity is present. They form in the $512\times590$ particle Euler Potentials calculation, yet do not appear in the over-cleaning and sub-cycling calculations of similar resolution (but which contain artificial resistivity).

\subsubsection{Static test: cleaning to $\nabla \cdot {\bf B}=0$}
\label{sec:itvovc-static-test}

% show a static cleaning test to 10^-18

\begin{figure}
\centering
\includegraphics[width=0.49\linewidth]{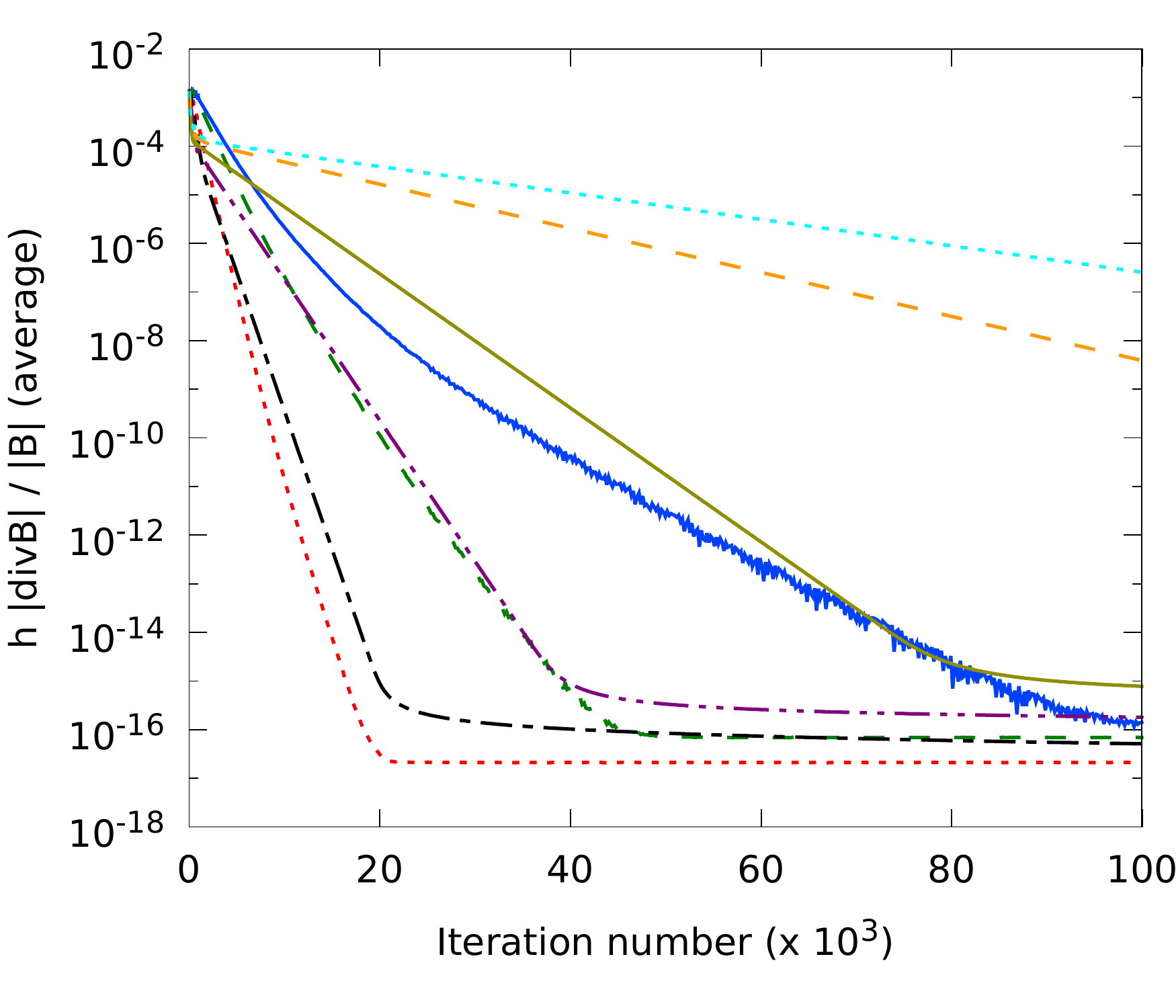} 
\includegraphics[width=0.49\linewidth]{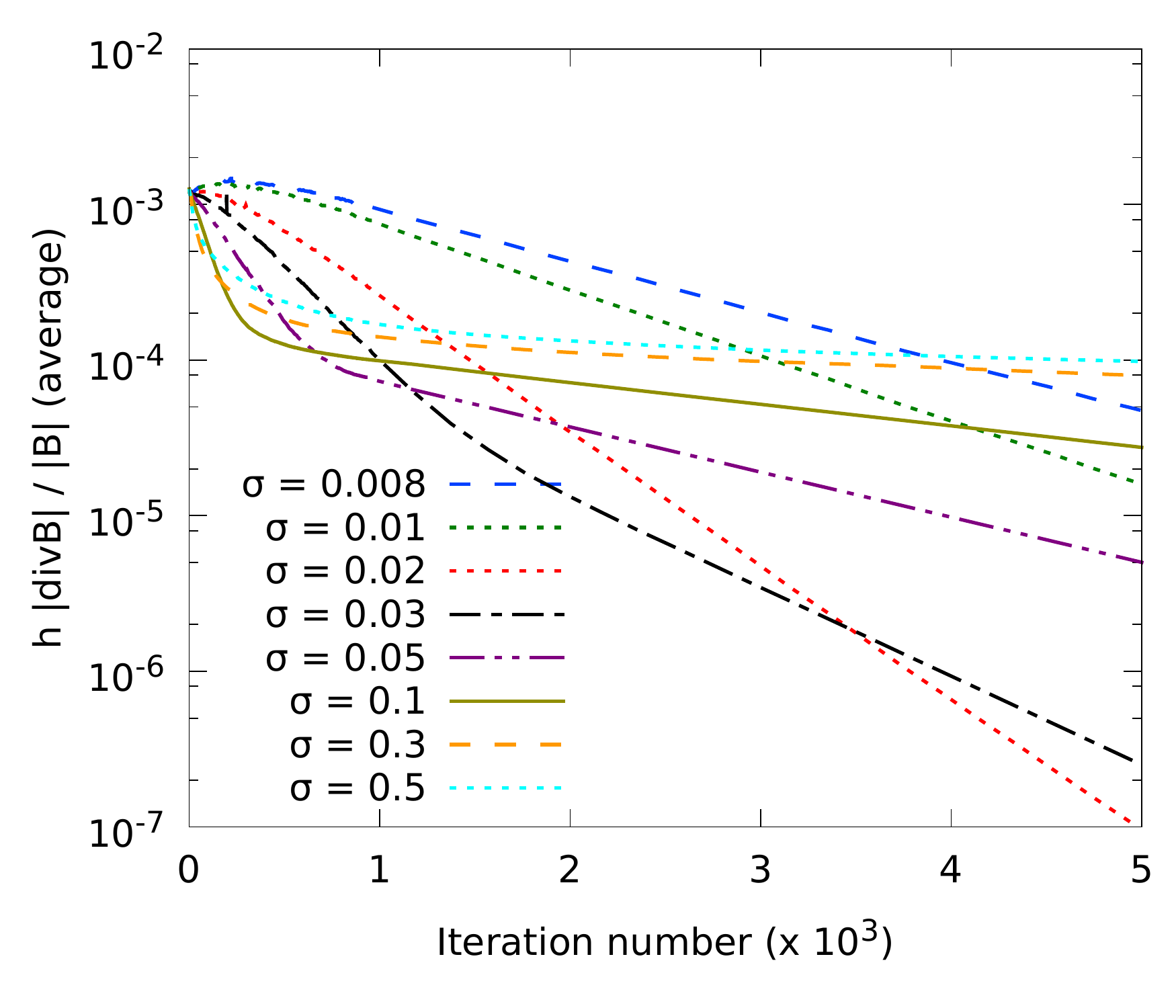}
\caption{Comparing values of $\sigma_\psi$ in the damping parameter to obtain an optimal value for sub-cycling, with the left panel for $10^5$ iterations and right panel the first $5000$. Short wavelength errors are quickly removed using $\sigma_\psi=0.3$ (right panel), though performs poorly at removing slowly long wavelength modes (left panel). Using $\sigma_\psi=0.03$, though initially worse at reducing divergence error, is found to remove long wavelength errors in the shortest number of iterations.}
\label{fig:itvovc-cleantozero}
\end{figure}

The tolerance on average divergence error may be set arbitrarily low when performing sub-cycling, and it is possible to achieve $\nabla \cdot {\bf B}=0$ as we will show. 

For this test, we take the $t~=~1$ evolved state of the Orszag-Tang vortex from the reference calculation in Section~\ref{sec:itvovc-test}, and perform sub-cycling for $10^5$ iterations. Figure~\ref{fig:itvovc-cleantozero} shows the average divergence error versus iteration count for a set of calculations varying the $\sigma_\psi$ parameter in the damping term. Our results show that $\nabla \cdot {\bf B}=0$ is achievable to arbitrarily small precision in SPMHD, provided a sufficient number of iterations are performed. 

The calculations requiring the minimum number of iterations to reach average divergence error below $10^{-15}$ are the $\sigma_\psi=0.02$--$0.03$ cases (${\sim}~10^4$ iterations). These $\sigma_\psi$ values are $10\times$ smaller than the optimal value of $\sigma_\psi$ found during the testing of the divergence cleaning method ($0.03$ to $0.3$). The $\sigma_\psi~=~0.3$ case still has not reached $<10^{-15}$ average divergence error after $10^5$ iterations. This difference may be understood through the differing rates on the removal of small and long wavelength errors. The right panel of Figure~\ref{fig:itvovc-cleantozero} shows the first 5000 iterations of these set of calculations, which shows that $\sigma_\psi~\sim~0.3$ provides the most rapid reduction of divergence error for the first ${\sim}~300$ iterations. Divergence error is introduced into the system at small wavelengths, which this level of damping is most effective at removing (and hence is optimal when the system is evolving and continually injecting divergence error). However, the reduction of average divergence error slows significantly after ${\sim}~300$ iterations because at that stage, only large wavelength errors remain which slowly decay. While $\sigma_\psi~{\sim}~0.03$ is initially worse, the hyperbolic waves are allowed to propagate more effectively, averaging the errors throughout the system, which in turn allows the diffusion term to become more effective at reducing the long wavelength modes.

To investigate this further, we performed a calculation using a static snapshot of the Orszag-Tang vortex where cleaning was performed each iteration using a range of damping values between $\sigma_\psi \in [0.01, 0.4]$, with the value of $\sigma_\psi$ that yielded the largest reduction in divergence error being accepted and the magnetic field evolved one step with that value. This process was repeated each timestep, with the aim of mapping out the optimal values of $\sigma_\psi$ for high numbers of iterations. It was found that the $\sigma_\psi$ oscillated each iteration below the maximum value (0.4) and the minimum value (0.01) for ${\sim}~500$ iterations, then remained at $\sigma_\psi=0.01$ for ${\sim}~500$ iterations, with this pattern repeating. It initially yielded reduction of divergence error on par with the fixed $\sigma_\psi=0.3$ case, with long-term reduction similar, but not as rapid, as the fixed $\sigma_\psi=0.03$ case. Considering these results, it would be best to run sub-cycling with fixed $\sigma_\psi=0.3$ for several hundred iterations to remove short wavelength errors, then switch to $\sigma_\psi=0.03$ for the reduction of long wavelength errors. 

We also make note that the choice of time integration scheme is important when taking $>10^3$--$10^4$ iterations. The errors introduced from Euler integration can become significant over this many iterations, corrupting the magnetic field. Leapfrog integration may be implemented with a predictor step for $\psi$ (as it depends upon both itself and ${\bf B}$) in a manner similar to Section~\ref{sec:leapfrog}. It does require derivative evaluations of $\nabla \psi$ (for ${\rm d}{\bf B}/{\rm d}t$) and $\nabla \cdot {\bf B}$ (for ${\rm d}\psi/{\rm d}t$) to occur out of phase with each other. The tests performed here use a second order Runge-Kutta scheme, which we find to be adequate.

\section{Velocity divergence cleaning for weakly compressible SPH}
\label{sec:velclean}

Incompressible fluids are divergence-free in the velocity field, thus approaches utilised to satisfy the magnetic divergence constraint can be adapted to work with the velocity divergence constraint. A popular method is incompressible SPH (ISPH), which uses a projection method to construct a divergence-free velocity field via the solution of a Poisson equation \citep{cr99}. Another is weakly compressible SPH (WCSPH), where, rather than directly ensure the velocity field is divergence-free, a stiff equation of state is used to limit density variations. Both have advantages and disadvantages, as highlighted in the comparison of \citet{leeetal08}. In short, WCSPH is simple to program, but ISPH tends to produce smoother velocity and pressure fields. ISPH has the computational expense of solving an elliptic equation, but WCSPH is also expensive because its high sound speed produces small timesteps due to the Courant condition. ISPH is known to produce numerical instability if particle distributions become highly disordered, though there have been attempts to modify ISPH to fix this issue \citep{sl03, ha07, lindetal12}. \citet{xsl09} tested and compared a number of ISPH approaches.

In this section, the constrained hyperbolic divergence cleaning method is adapted for use on the velocity field, and test its effectiveness at reducing density variations in the weakly compressible approximation \citep[see also][]{tp12b}. 

\subsection{Weakly compressible SPH}
\label{sec:wcsph}

A common method for modelling incompressible fluid behaviour with SPH is to use a stiff equation of state with the standard Lagrangian SPH formulation.  This sacrifices true incompressibility for simplicity of implementation.  However, this does not imply computational efficiency as the high speed of sound ($\sim 10\times$ maximum fluid velocity as a minimum) necessitates small sized time steps for stability.  Using the equation of state
\begin{equation}
 P = \frac{c_{\rm s}^2 \rho_0}{7} \left( \left(\frac{\rho}{\rho_0}\right)^7 - 1\right) ,
\label{eq:wcsph-eos}
\end{equation}
where $\rho_0$ is the reference density of the fluid and $c_{\rm s}$ is the sound speed, this typically results in density variations of $\sim 1\%$ \citep{monaghan94}.

The equations of motion which are solved are
\begin{equation}
\frac{{\rm d}{\bf v}_a}{{\rm d}t} = - \sum_b m_b \left( \frac{P_a}{\rho_a^2} + \frac{P_b}{\rho_b^2} \right) \nabla_a W_{ab} .
\label{eq:wcsph-momentum}
\end{equation}
In this case, we evolve the density using the SPH equivalent of the continuity equation (Equation~\ref{eq:discretisedcty}) rather than by summation, as this can lead to sharper profiles near free surfaces. The smoothing length of the particles is held constant, initially calculated according to Equation~\ref{eq:h}.

\subsection{Hyperbolic divergence cleaning for the velocity field}
\label{sec:velclean-continuum}

Since the continuity equation relies on $\nabla \cdot {\bf v}$ to evolve density, minimising this quantity should lead to improvements in the representation of incompressibility.  We now construct a formulation of divergence cleaning suitable for the velocity field.  The cleaning equations to be solved are modified to become
\begin{align}
\frac{{\rm d}{\bf v}}{{\rm d}t} =& - \frac{\nabla \psi}{\rho} , \label{eq:velclean-gradpsi}\\
\frac{{\rm d}\psi}{{\rm d}t} =& - c_{\rm h}^2 \rho \nabla \cdot {\bf v} - \frac{\psi}{\tau} . \label{eq:velclean-psievolve}
\end{align}
As the intended application is for incompressible fluids, we assume throughout this section that the density is uniform and constant.  Equations~\ref{eq:velclean-gradpsi} and \ref{eq:velclean-psievolve} still combine to produce the equivalent of the damped wave equation of Equation~\ref{eq:cleaning-waveeqn}.

We follow a procedure in step with that of Section~\ref{sec:hyperbolic}.  Consider the closed system of equations given by Equations~\ref{eq:velclean-gradpsi}~\&~\ref{eq:velclean-psievolve}. The total energy of this velocity-cleaning subsystem is
\begin{equation}
 E = \int \left[ \frac{v^2}{2} + \tilde{e}_\psi \right] \rho {\rm d}V ,
\label{eq:velclean-energyeq}
\end{equation}
where $\tilde{e}_\psi$ is the energy of the $\psi$ field. Constraining the energy of this subsystem to be conserved implies
\begin{equation}
 \frac{{\rm d}E}{{\rm d}t} = \int \left[ {\bf v} \cdot \frac{{\rm d}{\bf v}}{{\rm d}t} + \chi \frac{{\rm d}\psi}{{\rm d}t} \right] \rho {\rm d}V = 0,
\end{equation}
where $\chi$ is an unspecified variable to be determined. Inserting Equation~\ref{eq:velclean-gradpsi} and \ref{eq:velclean-psievolve} yields
\begin{equation}
 \int \left[ - {\bf v} \cdot \frac{\nabla \psi}{\rho} - \chi c_{\rm h}^2 \rho \nabla \cdot {\bf v} \right] \rho {\rm d} V = 0 .
\end{equation}
Integrating the first term by parts, we obtain
\begin{equation}
 \int \left[ \frac{\psi}{\rho} - \chi c_{\rm h}^2 \rho \right] (\nabla \cdot {\bf v}) \rho {\rm d}V + \int_s \psi {\bf v} \cdot {\rm d}\hat{s} = 0 ,
\end{equation}
which leads to $\chi = \psi / c_{\rm h}^2 \rho^2$ and hence
\begin{equation}
 \tilde{e}_\psi = \frac{\psi^2}{2 c_{\rm h}^2 \rho^2} .
\end{equation}
Using this energy term in Equation~\ref{eq:velclean-energyeq} will yield ${\rm d}\rho/{\rm d}t$ terms, but we neglect the addition of these terms under the assumption of incompressibility.

\subsection{Discretised hyperbolic velocity divergence cleaning}
\label{sec:velclean-wcsph}

With the appropriate energy term for this cleaning system, the constrained SPH implementation may be constructed.  We clean using the same $\nabla \cdot {\bf v}$ operator as in the continuity equation, that is,
\begin{equation}
 \nabla \cdot {\bf v}_a = - \frac{1}{\rho_a} \sum_b m_b {\bf v}_{ab} \cdot \nabla_a W_{ab} .
\label{eq:divv}
\end{equation}
The SPH discretised version of Equation~\ref{eq:velclean-energyeq} is 
\begin{equation}
 E = \sum_a m_a \left[ \frac{v_a^2}{2} + \frac{\psi_a^2}{2 c_{\rm h}^2 \rho_a^2} \right] .
\end{equation}
Differentiating with respect to time and using Equations~\ref{eq:velclean-psievolve} and \ref{eq:divv}, we obtain
\begin{equation}
 \sum_a m_a {\bf v}_a \frac{{\rm d}{\bf v}_a}{{\rm d}t} = \sum_a \frac{m_a \psi_a}{\rho_a^2} \sum_b m_b {\bf v}_{ab} \cdot \nabla_a W_{ab} .
\end{equation}
By splitting the RHS into two halves, swapping summations on one half, then combining, it is concluded that
\begin{equation}
 \frac{{\rm d}{\bf v}_a}{{\rm d}t} = - \sum_b m_b \left( \frac{\psi_a}{\rho_a^2} + \frac{\psi_b}{\rho_b^2} \right) \nabla_a W_{ab} .
\end{equation}
As before, conjugate operators for $\nabla \cdot {\bf v}$ and $\nabla \psi$ become imposed.  In addition to exactly conserving energy, this form for $\nabla \psi$ also conserves momentum. The evolution equation for $\psi$ is
\begin{equation}
\frac{{\rm d}\psi_a}{{\rm d}t} = c_{\rm h}^2 \sum_b m_b {\bf v}_{ab} \cdot \nabla_a W_{ab} - \frac{\psi}{\tau}.
\end{equation}

\subsection{Oscillating water drop test}
\label{sec:drop}

\begin{figure}
 \centering
\setlength\fboxsep{0pt}
\setlength\fboxrule{0.5pt}
\fbox{\includegraphics[height=0.22\textwidth]{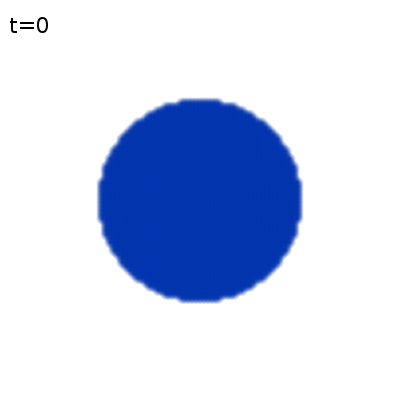}}
\fbox{\includegraphics[height=0.22\textwidth]{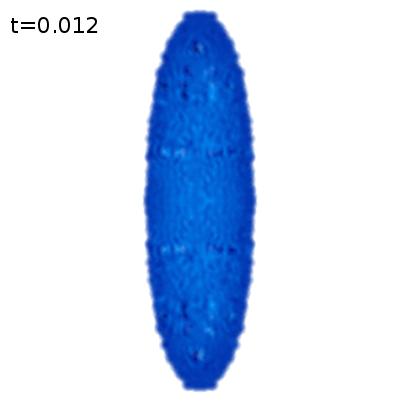}}
\fbox{\includegraphics[height=0.22\textwidth]{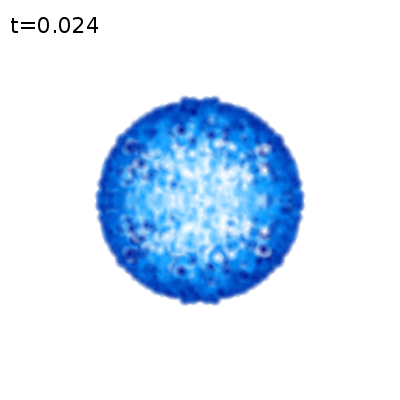}}
\fbox{\includegraphics[height=0.22\textwidth]{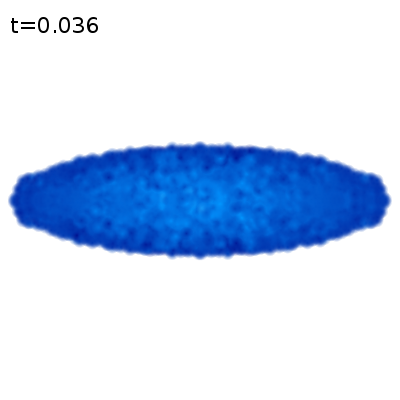}}
\includegraphics[height=0.22\textwidth]{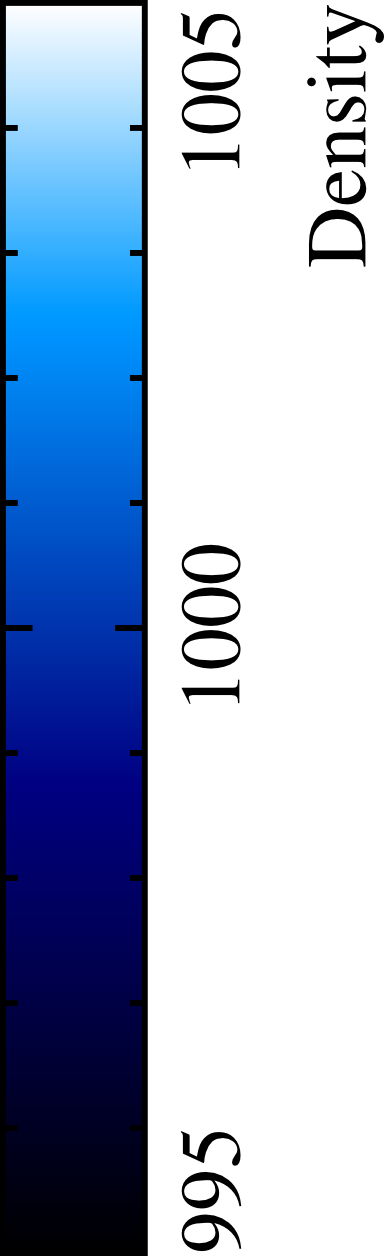}
\caption{Snapshots of the oscillating water drop test.  The circular drop has an initial velocity which squeezes it into an elliptical shape along the $y$-axis.  A radial force is present which halts the expansion of the drop, then contracts it to its original shape before expanding along the opposite axis.  This behaviour repeats causing the drop to oscillate alternately along the two axes.}
\label{fig:drop-compilation}
\end{figure}

To investigate the effectiveness of our velocity cleaning algorithm, it is applied to an oscillating elliptic water drop.  The water drop is initially circular and is free standing.  A radial force is exerted upon it, and with an initial velocity which is compressional along one axis, the drop oscillates, squeezing alternately along the two axes.  This behaviour is demonstrated in Figure~\ref{fig:drop-compilation}.

\begin{figure}
\centering
\begin{minipage}[t]{0.45\textwidth}
\includegraphics[width=\linewidth]{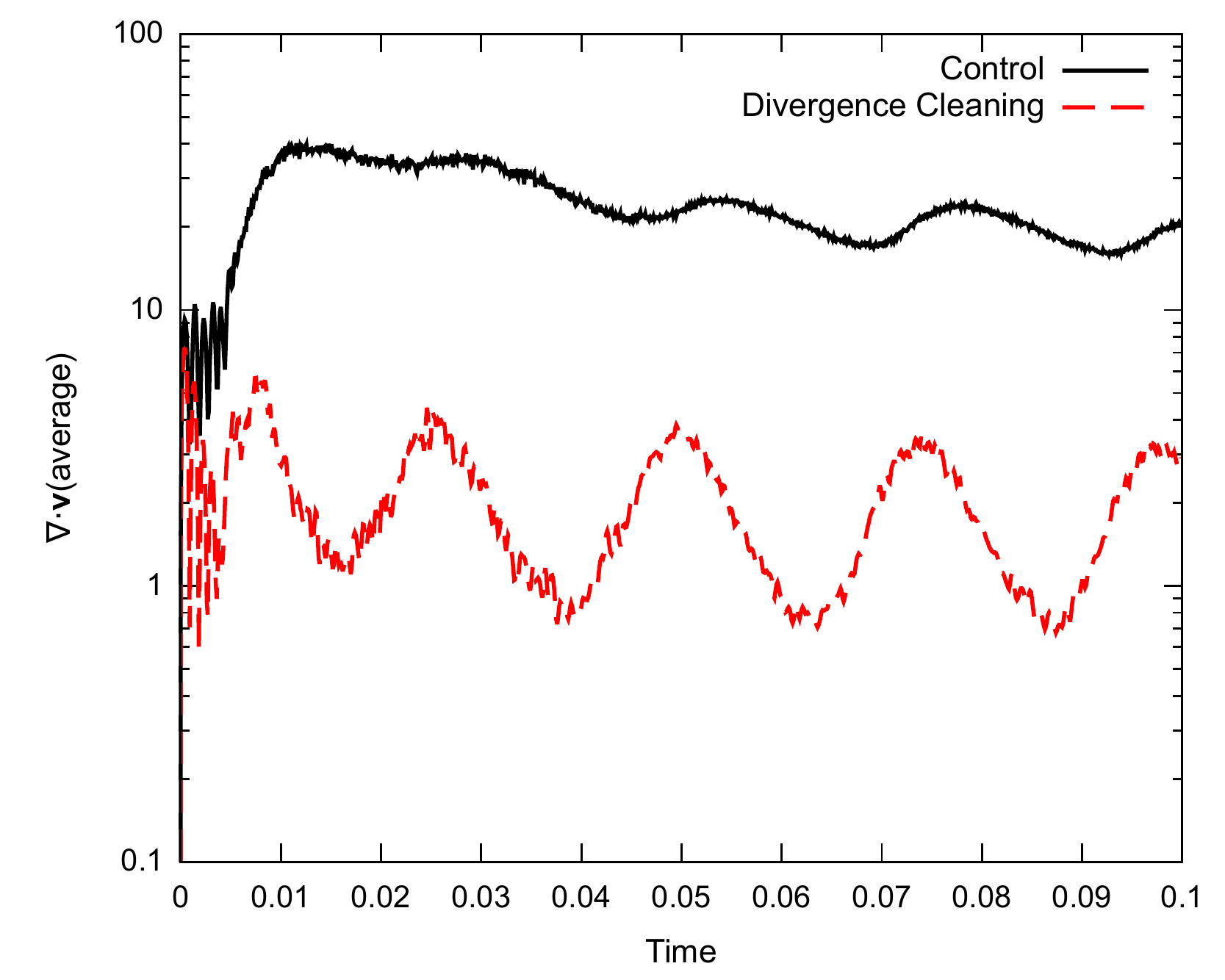} 
\caption{Average $\nabla \cdot {\bf v}$ of the elliptic water drop test.  Average velocity divergence is reduced by approximately an order of magnitude when divergence cleaning is applied.}
\label{fig:drop-avgdiv}
\end{minipage}
\hspace{0.05\textwidth}
\begin{minipage}[t]{0.45\textwidth}
\includegraphics[width=\linewidth]{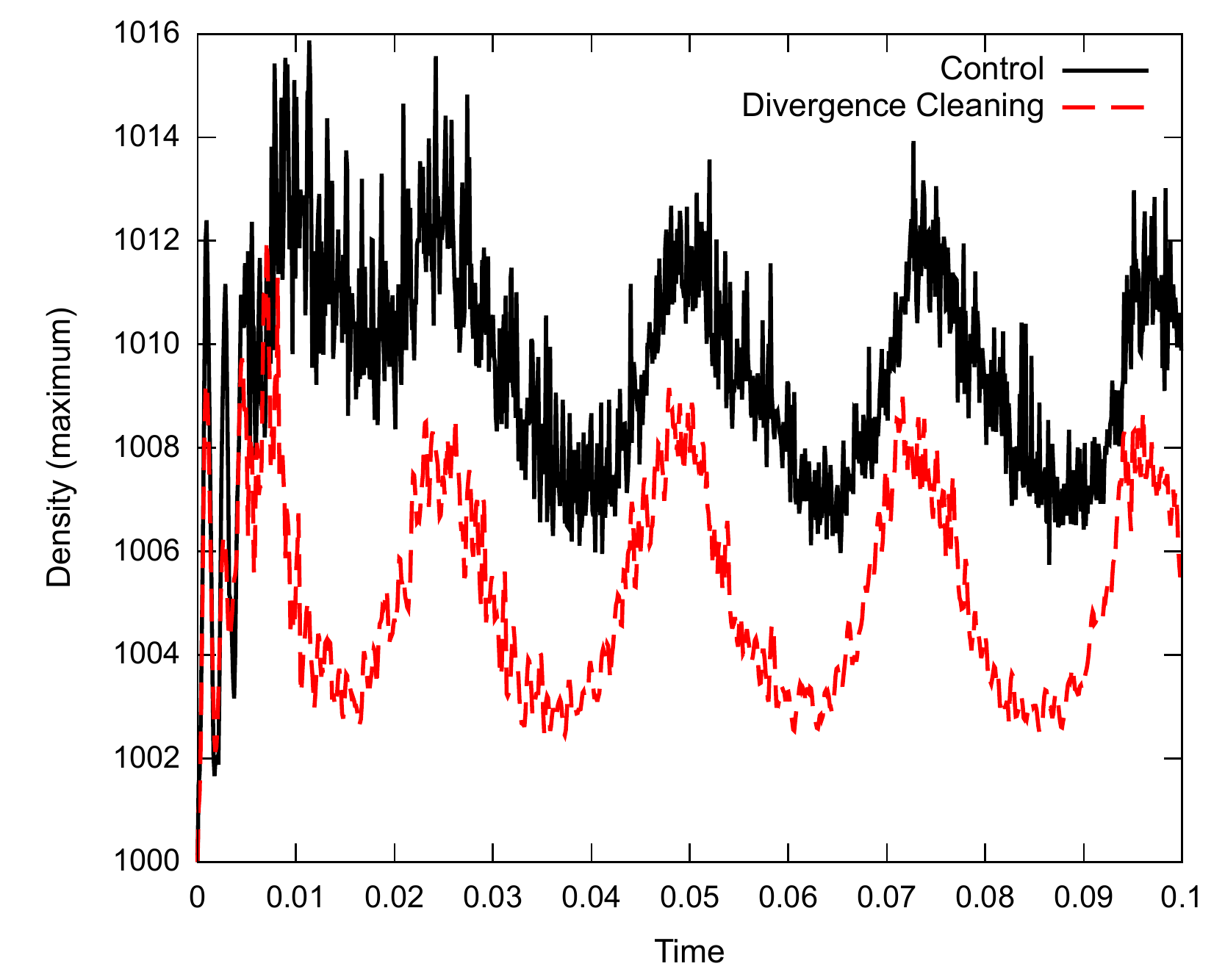}
\caption{Maximum density variation during the elliptic water drop test.  Applying divergence cleaning to the velocity field reduces density changes from the reference density by $\sim 0.5$.}
\label{fig:drop-maxrho}
\end{minipage}
\end{figure}

The drop is modelled using the weakly compressible approximation (Equations~\ref{eq:wcsph-momentum} and \ref{eq:discretisedcty} with Equation~\ref{eq:wcsph-eos} as the equation of state).  The reference density is $\rho_0 = 1000$ kg $\text{m}^{-2}$, and the initial velocity field is ${\bf v} = [-100x, 100y]$.  The radial force is $-100^2 {\bf r}$.  The drop has radius $R=1$, and a total of $1976$ particles are used arranged on a square lattice.

\begin{figure}
\captionsetup{width=0.6\textwidth}
\centering
 \includegraphics[width=0.45\textwidth]{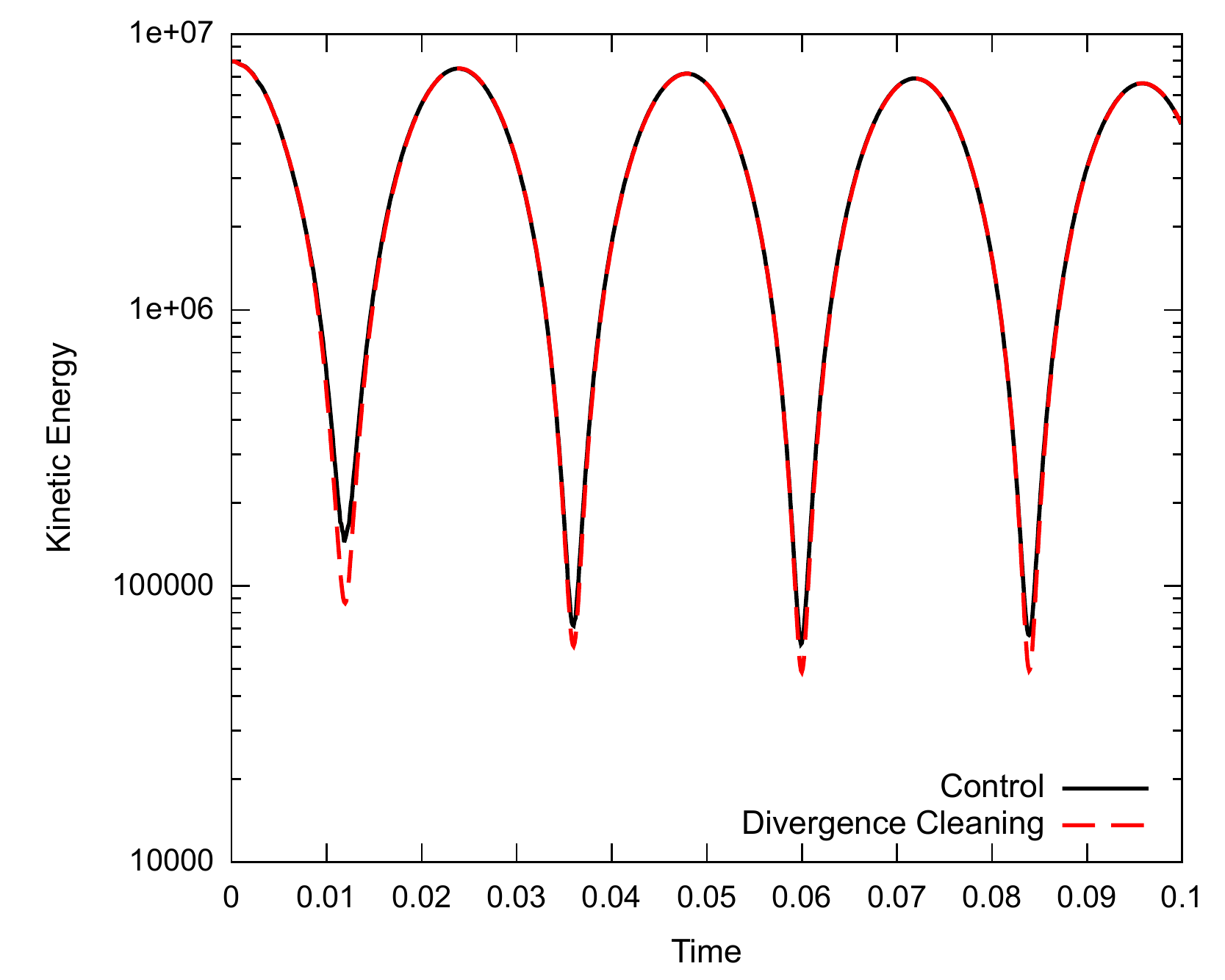}
\caption{Total kinetic energy of the elliptic water drop test.  No significant discrepancies exist between the control and divergence cleaned tests.}
\label{fig:drop-ke}
\end{figure}

The evolution of the drop is tracked until $t=0.1$, approximately two oscillation periods.  Figure~\ref{fig:drop-avgdiv} shows the average velocity divergence of the system as a function of time for both the cleaned and uncleaned systems.  Applying cleaning reduces the average divergence by nearly an order of magnitude, similar to results obtained for magnetic field cleaning.  This leads to a reduction in maximum density error by a factor of two (Figure~\ref{fig:drop-maxrho}).  The dissipation of kinetic energy by the cleaning algorithm is insignificant, as shown in Figure~\ref{fig:drop-ke}.

\section{Summary}
\label{sec:cleaningsummary}
In this chapter we have developed an implementation of \citet{dedneretal02}'s hyperbolic divergence cleaning for SPMHD that is constrained to be numerically stable and to always decrease the magnetic energy \citep[see also][]{tp12}.  To achieve this, we first defined the energy associated with the scalar $\psi$ field (Section~\ref{sec:continuum-energy-conservation}). This term was used to show that when the density varies over time, the evolution equation of $\psi$ should be modified to include a $-\tfrac{1}{2} \psi (\nabla\cdot {\bf v})$ term in order to conserve energy.

 In Section~\ref{sec:sph_energy_conserv} we derived an energy conserving formulation of divergence cleaning for SPMHD.  By using the $\psi$ energy term, we showed that if a difference operator is chosen to discretise $\nabla\cdot{\bf B}$ in the ${\rm d}\psi/{\rm d}t$ equation, then the conjugate, symmetric operator for $\nabla\psi$ should be used (Section~\ref{sec:spmhd-clean-diff}).  Similarly, with symmetric $\nabla\cdot{\bf B}$, difference $\nabla\psi$ should be used in the induction equation (Section~\ref{sec:spmhd-clean-symm}).  Use of conjugate operators was found to be the key to a numerically stable formulation.  In Section~\ref{sec:spmhdenergy}, we presented the correct SPMHD form of the $-\tfrac{1}{2}\psi(\nabla\cdot{\bf v})$ term, and in Section~\ref{sec:negdef}, demonstrated that parabolic damping will always lead to negative definite changes of energy.

Tests of our constrained hyperbolic divergence cleaning were presented in Section~\ref{sec:cleaningtests}.  The selection of tests were for both 2 and 3D, and were designed to evaluate our method in isolation using simple, idealised systems and also in more realistic applications. Our idealised 2D tests consisted of a divergence advection test (Section~\ref{sec:divBadvection}), and variants involving a density jump (Section~\ref{sec:test-density-jump}) and free boundaries (Section~\ref{sec:test-free-boundaries}).  The more complex 2D tests were an MHD blast wave (Section~\ref{sec:blast}) and the Orszag-Tang vortex (Section~\ref{sec:ot}).  A version of the divergence advection test extended to 3D was used in Section~\ref{sec:adv3d}.  Results from the gravitational collapse of a molecular cloud core, representing our most challenging test case and a gauge of divergence cleaning applied to ``real'' applications, were presented in Section~\ref{sec:jet}.  Calculations of magnetised, supersonic turbulence were examined in Section~\ref{sec:mhdturb-cleaning}. From the results of these tests, we draw the following conclusions:
\begin{enumerate}
\item[i)] Constrained hyperbolic/parabolic divergence cleaning provides an effective method of maintaining the divergence constraint in SPMHD, typically maintaining the average $h \vert \nabla\cdot{\bf B} \vert / \vert {\bf B} \vert$ to between 0.1--1\%.
\item[ii)] The constrained formulation using conjugate operators for $\nabla\cdot{\bf B}$ and $\nabla\psi$ is stable at density jumps and free boundaries, in contrast to previous implementations.
\item[iii)] We strongly recommend cleaning using the difference operator for $\nabla\cdot{\bf B}$. Cleaning using the symmetric operator was not found to provide any advantage over the difference operator in terms of momentum conservation and was found to dissipate physical components of the magnetic field as well as the divergence error.
\item[iv)] Constrained divergence cleaning is more effective than artificial resistivity at reducing the divergence error, and still reduces the divergence error further when used in combination with resistivity.
\item[v)] Divergence cleaning can provide an improvement of up to two orders of magnitude in momentum conservation when applied to realistic, 3D simulations.
\item[vi)] Optimal values for the damping parameter $\sigma_\psi$ were found to be $\sigma_\psi =$ 0.2--0.3 in 2D and $\sigma_\psi =$ 0.8--1.2 in 3D for all of the test problems considered in this thesis.
\item[vii)] Addition of the $-\tfrac{1}{2}\psi \nabla\cdot{\bf v}$ term to the ${\rm d}\psi/{\rm d}t$ equation, while necessary for strict energy conservation of the hyperbolic cleaning equations, was found to have no noticeable effect even when the gas is strongly compacted, such as in simulations of star formation and magnetised, Mach 10 turbulence.
\item[viii)] We found numerical artefacts in several problems when subtracting only $- \tfrac{1}{2}{\bf B} (\nabla \cdot{\bf B})$ in the momentum equation to counteract the tensile instability. Instead, we strongly recommend using the full $-{\bf B} (\nabla \cdot{\bf B})$ correction.
\end{enumerate}

Two approaches were investigated to enhance the cleaning method in Section~\ref{sec:cleaning-enhanced}: Over-cleaning, where the cleaning wave speed is explicitly increased with a corresponding decrease in the size of the timestep, and sub-cycling, where the cleaning equations are solved in isolation between timesteps. Both reduce the average divergence error, yielding half an order of magnitude reduction per factor of ten increase in wave speed or number of sub-cycle iterations. Sub-cycling is able to keep average divergence error below a specified tolerance at all times, and it was shown that given enough iterations, the divergence error may be reduced to arbitrarily low limits. It was found that high number of iterations ($>300$), using $\sigma_\psi=0.03$ is optimal to reduce long wavelength errors.

In Section~\ref{sec:velclean}, we constructed cleaning equations for the velocity field for use in weakly compressible SPH simulations \citep[see also][]{tp12b}. The velocity divergence cleaning equations conserve energy and momentum. When the velocity field was cleaned in weakly compressible SPH simulations of an oscillating water drop, density variations were reduced by half, with negligible kinetic energy dissipation. Though these results are encouraging, additional work is required, in particular, for cases involving boundary particles.

In summary, our constrained hyperbolic divergence cleaning scheme is a robust and effective method for enforcing the divergence constraint in SPMHD simulations, providing a pathway to accurate simulation of a wide range of magnetic phenomena in astrophysics and beyond.

%% file: declaration-chapter4.tex
\newpage
{
\chapter*{}

\vspace{-40mm}

\section*{Declaration for Chapter 4}

\subsection*{Declaration by Candidate}

\noindent In the case of Chapter 4, the nature and extent of my contribution to the work was the following:

\vspace{4mm}

\noindent \begin{tabular}{| >{\raggedright}p{11.55cm} | >{\raggedright}p{3.45cm} |}
\hline
{\bf Nature of Contribution} & {\bf Extent of Contribution (\%)} \tabularnewline
\hline
First author of 2013, ``A switch to reduce resistivity in smoothed particle magnetohydrodynamics'', {\it MNRAS} {\bf 436}, 2810--2817.
 & 90 \tabularnewline
\hline
\end{tabular}

\vspace{6mm}
\noindent The following co-authors contributed to the work:

\vspace{4mm}

{
\noindent \begin{tabular}{| >{\raggedright}p{3cm} | >{\raggedright}p{8.1cm} | >{\raggedright}p{3.45cm} |}
\hline 
{\bf Name} & {\bf Nature of Contribution} & {\bf Extent of Contribution (\%) for student co-authors} \tabularnewline
\hline
Daniel Price & PhD supervisor & \tabularnewline
\hline
\end{tabular}
}

\vspace{6mm}

\noindent The undersigned hereby certify that the above declaration reflects the nature and extent of the candidate's and co-author's contributions to this work.

% left bottom right top
\noindent \includegraphics[trim=1.2cm 14cm 0.6cm 12cm, clip, width=\textwidth, angle=-1, scale=1.03]{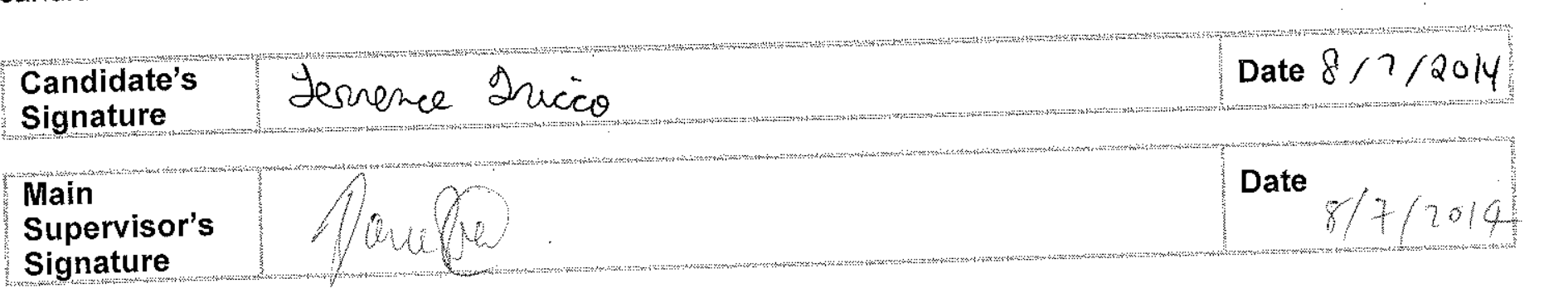}

}

%% file: switch.tex
\chapter{A switch to reduce resistivity}
\label{sec:chapter-switch}

%\section{Introduction}
Magnetised shocks and discontinuities pervade the interstellar medium \citep{es04,gaensleretal11}. Capturing these properly in numerical simulations is critical to accurately predicting the formation of stars from turbulent, magnetised, molecular clouds \citep{fk12}, and this necessitates adding numerical dissipation to simulations. On the other hand, estimates of the microscopic viscosity and resistivity in the interstellar medium suggest very high values of the kinematic and magnetic Reynolds numbers, respectively, typically orders of magnitude higher than can be achieved in numerical codes \citep[c.f.][]{es04}. Thus, it is important to minimise numerical dissipation in simulation codes. This is typically done by improving the shock capturing scheme so that dissipation away from shocks is reduced, though the Reynolds number near the shock may still be decreased.

The usual approach to shock-capturing in SPH is to treat discontinuities in fluid variables by adding dissipation terms which smooth the variable across sharp jumps in order to resolve the discontinuity (Section~\ref{sec:spmhd-shock-capturing}). Artificial viscosity for treatment of hydrodynamic shocks was developed by \citet{mg83}.  In this work, we use the form of artificial viscosity by \citet{monaghan97}, developed by analogy with Riemann solvers (Section~\ref{sec:artvisc}). In SPMHD, an artificial resistivity for the magnetic field is included to capture magnetic shocks and discontinuities (i.e., current sheets) (Section~\ref{sec:artresis}). However, the choice of signal velocity in the artificial resistivity is less clear than for the artificial viscosity. Ideal MHD has three wave solutions but without reconstructing the full Riemann state it is not possible to determine the type of shock.  Thus, this is typically chosen to be the speed of the fast MHD wave. Since this is rather dissipative, \citet{price12} instead suggested using the averaged Alfv\'en velocity as the choice of signal velocity (though see Section~\ref{sec:mhdturb})

A switch may be employed for $\alpha_\text{B}$ to reduce dissipation away from shocks.  \citet{pm05} (PM05) suggested a switch based on analogy with \citet{mm97} (see Section~\ref{sec:artresis-switches}). This switch works satisfactorily for many problems, leading to sharper jump profiles and a decrease in the overall dissipation of the magnetic field.  However, \citet*{ptb12} noted in their star formation simulations that, even with this switch, excess dissipation could suppress the formation of protostellar jets.

Our need for a new resistivity switch is motivated by the failure of the \citetalias{pm05} switch in the limit where the Alfv\'en speed is much smaller than the sound speed, as will be shown in section \ref{sec:mhdturb}.  Since $\alpha_\text{B} \propto \vert \nabla \times {\bf B}\vert $ (assuming $\nabla \cdot {\bf B}$ is negligible), this means that $\alpha_\text{B}$ is related to the magnitude of the magnetic field.  Thus, for weak fields $\alpha_\text{B}$ may remain quite small even in the presence of strong shocks.

In this chapter, we present a new switch for $\alpha_\text{B}$ that captures shocks in the magnetic field in both weak and strong fields.  This addresses the deficiencies of the previous switch and results in less overall dissipation of magnetic energy. The chapter is organised as follows: In Section~\ref{sec:formulation}, the new resistivity formulation and implementation is described. Section~\ref{sec:switchtests} contains a suite of tests designed to test the efficacy of the new switch and to compare results against the previous switch. Section~\ref{sec:switch-generalisation} extends the concept of the switch to develop switches for artificial viscosity and thermal conductivity. Results are summarised in Section~\ref{sec:switchsummary}.

\section{Formulation}
\label{sec:formulation}
Our approach is to utilise $\nabla {\bf B}$, the $3\times3$ gradient matrix of ${\bf B}$, as the shock indicator.  For each particle, $\alpha_\text{B}$ is directly set to the dimensionless quantity
\begin{equation}
\alpha_{\text{B},a} = \frac{h_a \vert \nabla {\bf B}_a \vert}{\vert {\bf B}_a \vert} .
\label{eq:alpha_B}
\end{equation}
This is artificially restricted to the range $\alpha_{\rm B} \in [0,1]$ so that zero dissipation is applied in regions away from discontinuities.

By using the norm of the gradient of the magnetic field normalised by the magnitude of the magnetic field, the dependence on magnetic field strength is removed and this gives a relative measure of the strength of the discontinuity.  This allows shocks and discontinuities to be robustly detected in both the weak and strong field regimes.  It naturally produces values of $\alpha_\text{B}$ in the desired range and of the appropriate size for the discontinuity encountered, with regions away from shocks having negligible $\alpha_\text{B}$ values. 

The numerical dissipation of the magnetic field in regions away from shocks should scale quadratically with resolution when using this switch. Consider a magnetic field which has a linear gradient, that is $\nabla {\bf B}$ is constant and resolution independent. It is clear then that the switch is proportional to the resolution length, that is, for example, $\alpha_{\rm B}$ will decrease by half when the resolution is doubled ($h$ is halved). Since artificial resistivity itself scales linearly with resolution, as is evident from Equation~\ref{eq:artificialresistivity}, it is clear that using Equation~\ref{eq:alpha_B} will yield artificial resistivity that scales quadratically.

The switch produces the same $\alpha_\text{B}$ values for multiplicative increases in magnetic field strength, important for dynamo-type problems where the magnetic field grows in strength. This represents a significant advantage over the \citetalias{pm05} switch.  Additive increases to the magnetic field, however, will yield different values of $\alpha_\text{B}$, and using this switch in relativistic contexts would require further consideration.

An obvious issue is what happens when $\vert {\bf B} \vert \to 0$.  This situation occurs in current sheets or null points where the magnetic field undergoes a reversal in direction.  In these cases, $\alpha_\text{B} \to 1$, which is the correct behaviour for current sheets since they represent a discontinuity in the magnetic field, but is not so for null points.  Dividing by zero can be avoided by adding a small parameter $\epsilon$ to $\vert {\bf B} \vert$.

\subsection{Implementation}

Each component of the gradient matrix is estimated using a standard SPH first derivative operator (e.g., Equation~\ref{eq:interp-grad-firstorder}),
\begin{equation}
\nabla{\bf B}_{a} \equiv \frac{\partial B^i_{a}}{\partial x^j_{a}} = -\frac{1}{\Omega_a \rho_a} \sum_b m_b ({B}^{i}_a - {B}^{i}_b) \nabla_a^j W_{ab}(h_{a}) .
\label{eq:sphgradb}
\end{equation}
This operator yields an estimate which is exact for constant functions.  We also investigated using an operator that is exact for linear functions, which may be obtained by performing a Taylor series expansion about ${\bf r}_a$ and solving a matrix inversion of the second error term (see Section~\ref{sec:interpolation}).  However, no difference was found for any of the tests shown in this thesis, suggesting that this is unnecessary.

The norm of $\nabla {\bf B}$ is calculated using the 2-norm,
\begin{equation}
\vert \nabla {\bf B} \vert \equiv \sqrt{ \sum_i \sum_j \left\vert  \frac{\partial B^i_{a}}{\partial x^j_{a}} \right\vert^2 } .
\end{equation}
Several choices for computing this norm were investigated, such as the 1-norm, but no significant differences were found.

We investigated using the curl of the magnetic field as the shock indicator. While tests found it to be just as effective at detecting isolated shocks, we found that it did not measure discontinuities as well as the full gradient in complicated shock interactions.  The full gradient has further advantage in that the trace of the matrix produces the divergence of the field, meaning that dissipation will be applied if large divergence errors are present.

Finally, a \citet{cd10}-like approach was also investigated, whereby a time-dependent decay term for $\alpha_\text{B}$ was added, similar to that in other artificial viscosity switches (Section~\ref{sec:artvisc-switches}) and the \citetalias{pm05} switch. In this case, $\alpha_\text{B}$ was set using Equation~\ref{eq:alpha_B} whenever this exceeded the current value, otherwise the existing value was retained and subsequently reduced on the next integration timestep using the decay term.  The aim was to let $\alpha_\text{B}$ smoothly decay after a shock had passed to improve representation of the post-shock field. We found that using Equation~\ref{eq:alpha_B} alone already gives a smooth distribution in $\alpha_\text{B}$ about the centre of the shock, indicating that a decay term is not necessary for resistivity.

\subsection{Choice of signal velocity}
\label{sec:artresis-signalvelocity}

Similar to \citet{price12} we take the signal velocity to be an average of the wave speeds between the two particles
\begin{equation}
 v^{B}_\text{sig} = 0.5 (v_{{\rm mhd},a} + v_{{\rm mhd},b}),
\end{equation}
where $v_{\rm mhd}$ is an appropriate MHD wave speed. The $-\beta {\bf v}_{ab} \cdot \hat{{\bf r}}_{ab}$ term used in the viscosity signal velocity, which corrects for the relative velocity of the particles and prevents particle interpenetration, is not included.  We find that for resistivity it is unnecessary and causes excessive dissipation. It may be noted that use of the averaged Alfv\'en speed for a signal velocity by \citet{price12} also excluded this term.

Unlike \citet{price12}, we find that the best choice is to use the fast MHD wave speed, as in the original \citet{pm04a} formulation, such that
\begin{equation}
 v_{\rm mhd}^{2} = \frac{1}{2} \left(c^2_{{\rm s},a} + v_{{\rm A},a}^2\right) + \frac12 \sqrt{ (c_{{\rm s},a}^2 + v_{{\rm A},a}^2)^2 - 4 c_{{\rm s},a}^2 v_{{\rm A},a}^2 (\hat{{\bf B}_a} \cdot \hat{{\bf r}}_{ab}) },
\label{eq:vsigfastmhd}
\end{equation}
which is a composition of the sound speed, $c_{\rm s}$, and the Alfv\'en speed, $v_{\rm A} = B/ \sqrt{\mu_0 \rho}$.  See Section~\ref{sec:mhdwaves} for more information on MHD waves. If $c_{\rm s} \gg v_{\rm A}$, we find that \citet{price12}'s suggestion to use the Alfv\'en speed in the applied resistivity is insufficient to capture fast wave shocks (see Section~\ref{sec:mhdturb}).  When $v_{\rm A} \gtrsim c_{\rm s}$, the Alfv\'en speed and the fast wave speed will differ by less than a factor of 2.

\subsection{Switches using a second derivative}

In principle, a switch constructed using a higher derivative should provide a more reliable measure of the presence of a discontinuity in the magnetic field.  One suggestion by Walter Dehnen (priv. commun.) is to use $\alpha_\text{B} = h \vert \nabla^2 {\bf B} \vert / \vert \nabla {\bf B} \vert$. Another option could be $\alpha_\text{B} = h^2 \vert \nabla^2 {\bf B} \vert / \vert {\bf B} \vert$, which would scale quadratically with resolution. 

The main difficulty in implementing higher derivative switches is calculating the second derivative in a way which is sufficiently free of noise from particle disorder.  We investigated calculating $\nabla^2 {\bf B}$ using the \citet{brookshaw85} form, that is, 
\begin{equation}
\nabla^{2} {\bf B}_{a} = \frac{2}{\Omega_a \rho_a} \sum_b m_b \left( {\bf B}_a - {\bf B}_b \right) \frac{F_{ab} (h_a)}{\vert r_{ab} \vert},
\end{equation}
where $\nabla W_{ab} \equiv \hat{\bf r}_{ab} F_{ab}$, and also by taking two first derivatives as in Equation~\ref{eq:sphgradb}, which, by taking two successive first derivatives, should lead to a more smooth estimate of the second derivative.  However, both of these simple estimates are significantly noisy when the particles are disordered, leading to high $\alpha_\text{B}$ and excessive dissipation.  The M6 quintic spline kernel was used in an attempt to reduce this noise, both by yielding a more regular particle distribution and a more accurate derivative estimate, but did not change the results.  

Therefore, a switch utilising the second derivative must use a higher order estimate in order to reduce noise from particle disorder, a conclusion similarly reached by \citet{cd10} and \citet{rh12}.  The most straightforward approach is to use two exact linear first derivatives which removes the $O(h)$ error term by taking a Taylor series expansion about ${\bf r}_a$ and performing a matrix inversion of the second error term.  Specifically, after first calculating $\nabla {\bf B}$ in such a manner, we compute
\begin{equation}
 \chi^{\sigma \gamma} \frac{\partial {\nabla {\bf B}}^{\alpha\beta}_a}{\partial {\bf x}^\sigma} = \sum_b m_b \left[ (\nabla {\bf B})_b^{\alpha \beta} - (\nabla {\bf B})_a^{\alpha \beta}\right] \nabla^{\gamma} W_{ab} 
\end{equation}
to obtain $\nabla^2 {\bf B}$, where $\chi^{\sigma\gamma} = \sum_b m_b ({\bf r}_b - {\bf r}_a)^\sigma \nabla^\gamma W_{ab}$ is the \mbox{$3\times 3$} matrix that must be inverted (see Equations~\ref{eq:interp-grad-secondorder} and \ref{eq:interp-grad-secondorder-matrix}).  This significantly improves the quality of the second derivative estimate, but requires two loops over the particles prior to the main loop where the resistivity term is calculated, meaning that it makes the overall SPMHD scheme $\sim1.5\times$ more expensive. This is a hefty price to pay for a switch that only marginally improves over Equation~\ref{eq:alpha_B}. The second derivative evaluation proposed by \citet{rh12} is even more expensive, requiring a $10\times 10$ matrix inversion, and a minimum of 400 neighbours under the kernel. 

Our overall conclusion is to prefer the simple switch of Equation~\ref{eq:alpha_B} for general use.  It performs robustly and effectively (see Section~\ref{sec:switchtests}), yet is simple to implement and cost-effective.

\section{Numerical Tests}
\label{sec:switchtests}

Our choice of tests are designed to study the ability of the switch to i) properly capture and model shock phenomena, and ii) suppress dissipation in areas away from shocks.  We have used three shocktube tests to study the former, using tests introduced by \citet{dw94} and \citet{briowu88} (corresponding to tests 1B, 2A, and 5A in \cite{rj95} (hereafter RJ95) whose naming convention we adopt).  These tests contain fast and slow shocks, fast and slow rarefactions, rotational discontinuities, and compound shock structures and are chosen to test the switch's ability to model all these separate shock types.  We then compare the new switch to the \citetalias{pm05} switch for three separate test problems: Propagation of a circularly polarised Alfv\'en wave, the Orszag-Tang vortex, and Mach 10 shocks in a fluid with an extremely weak field.  The last test is of particular interest because, as will be shown, the \citetalias{pm05} switch fails to recognise shocks in this weak field regime causing unphysical behaviour.

All our tests employ the constrained divergence cleaning algorithm of Chapter~\ref{sec:chapter-cleaning}.  The tests presented here serve to further validate this scheme.

\subsection{Shocktube 1B}
\label{sec:shock1b}

\begin{figure}
 \includegraphics[width=1.0\linewidth]{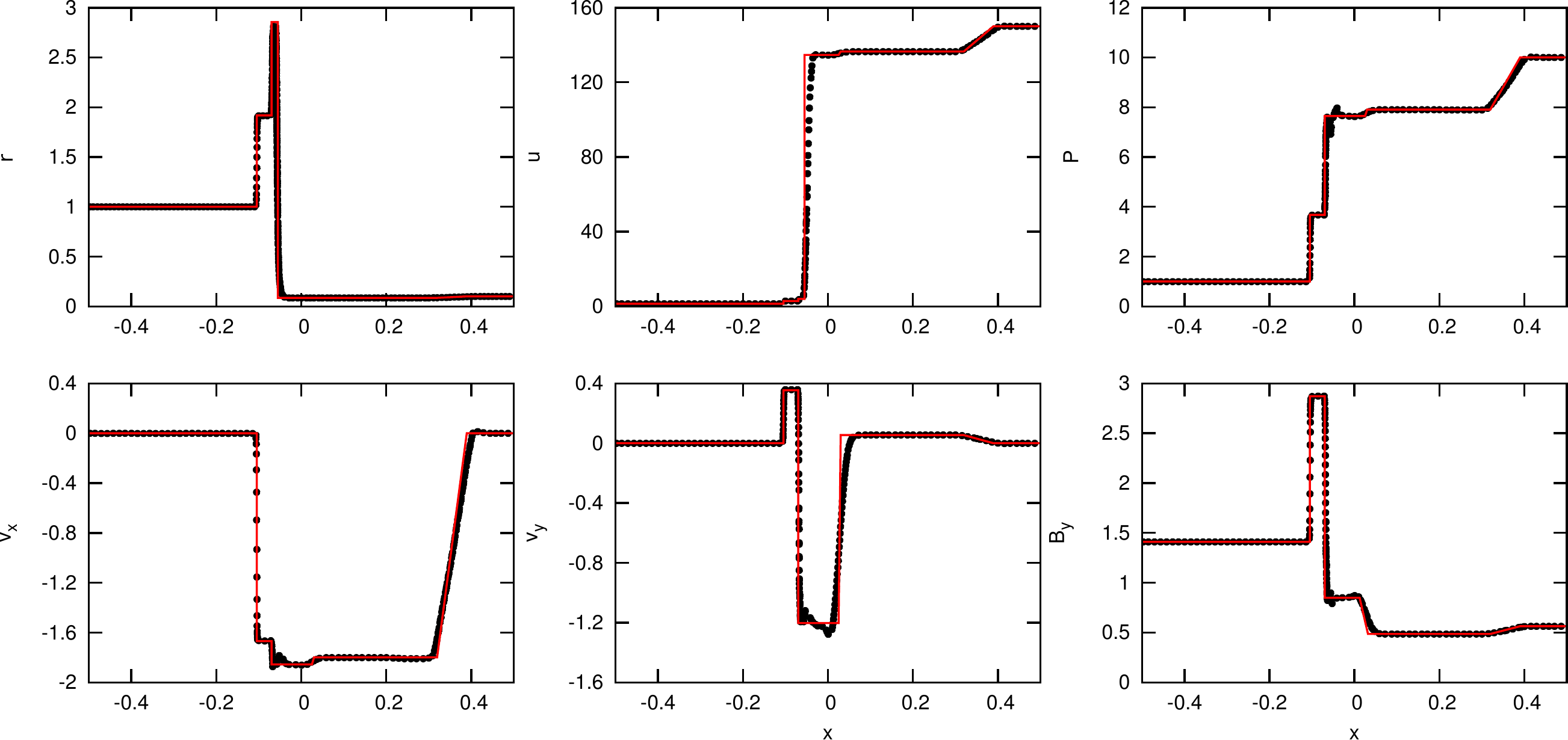}
\caption{Shocktube test 1B from \citetalias{rj95} performed in 2D with left state ($\rho$, $P$, $v_x$, $v_y$, $B_y$) $=$ (1, 1, 0, 0, $5/\sqrt{4 \pi}$) and right state ($\rho$, P, $v_x$, $v_y$, $B_y$) $=$ (0.1, 10, 0, 0, $2/\sqrt{4 \pi}$) with $B_x=3/\sqrt{4 \pi}$ at $t=0.03$.  Black circles are the particles and the red line is the solution from \citetalias{rj95}.}
\label{fig:shock1b}
\end{figure}

The first shocktube is a 2D test from \citet{dw94} which creates fast and slow shocks travelling in the -x direction, fast and slow rarefactions travelling in the +x direction, with a contact discontinuity in the centre.  The initial state for $x < 0$ (the `left state') is ($\rho$, $P$, $v_x$, $v_y$, $B_y$) $=$ (1, 1, 0, 0, $5/\sqrt{4 \pi}$), while for $x > 0$ (the `right state') is ($\rho$, P, $v_x$, $v_y$, $B_y$) $=$ (0.1, 10, 0, 0, $2/\sqrt{4 \pi}$) with $B_x=3/\sqrt{4 \pi}$ and $\gamma=5/3$.  

For this particular test, the initial density profile was used to calculate the initial thermal energy so that it forms a smooth transition between the two states.  This mitigates the presence of artificial pressure spikes in the initial conditions due to the high density contrast ($10$:$1$), seen also by \citet*{hfg13} in their studies of Kelvin-Helmholtz instabilities.

The shocktube has been simulated with 800$\times$26 particles for the left state and 260$\times$8 particles for the right state arranged on a triangular lattice.  Artificial viscosity has been applied with a constant $\alpha=1$, as has been for subsequent shocktube tests. Results at $t=0.03$ are shown in Figure~\ref{fig:shock1b} and may be compared with the \citetalias{rj95} solution for the fast and slow shock and rarefactions (red line).   The L1 error in the $B_y$ profile is $8.911\times 10^{-3}$.  This compares to $9.547\times 10^{-3}$ if the shocktube is run using the \citetalias{pm05} switch.

For this shocktube test, it is worth noting that no difficulties were found with our divergence cleaning algorithm.  Recently, \citet*{sdb13} published a different implementation and found that for this test it resulted in significant errors unless the cleaning was artificially limited. Their method is equivalent to using the \citetalias{pm05} implementation (that is, difference operators for both $\nabla \cdot {\bf B}$ and $\nabla \psi$), but with an artificial limiter that restricts corrections to the magnetic field to remain less than local changes due to the induction equation. As we have demonstrated in Sections~\ref{sec:test-density-jump}--\ref{sec:test-free-boundaries}, the \citetalias{pm05} implementation is numerically unstable and causes amplification of divergence error at sharp density contrasts. For the sharp $10:1$ density ratio in this shocktube, the artificial limiter used by \citet{sdb13} inhibits the instability from corrupting the magnetic field at the interface. Our method instead inherently fixes the numerical stability, without the need for artificial limiters, by formulating the cleaning equations so that they manifestly conserve energy (see Chapter~\ref{sec:chapter-cleaning}).

\subsection{Shocktube 2A}
\label{sec:shock2a}

\begin{figure}
\captionsetup{width=0.8\textwidth}
\centering
 \includegraphics[width=0.7\linewidth]{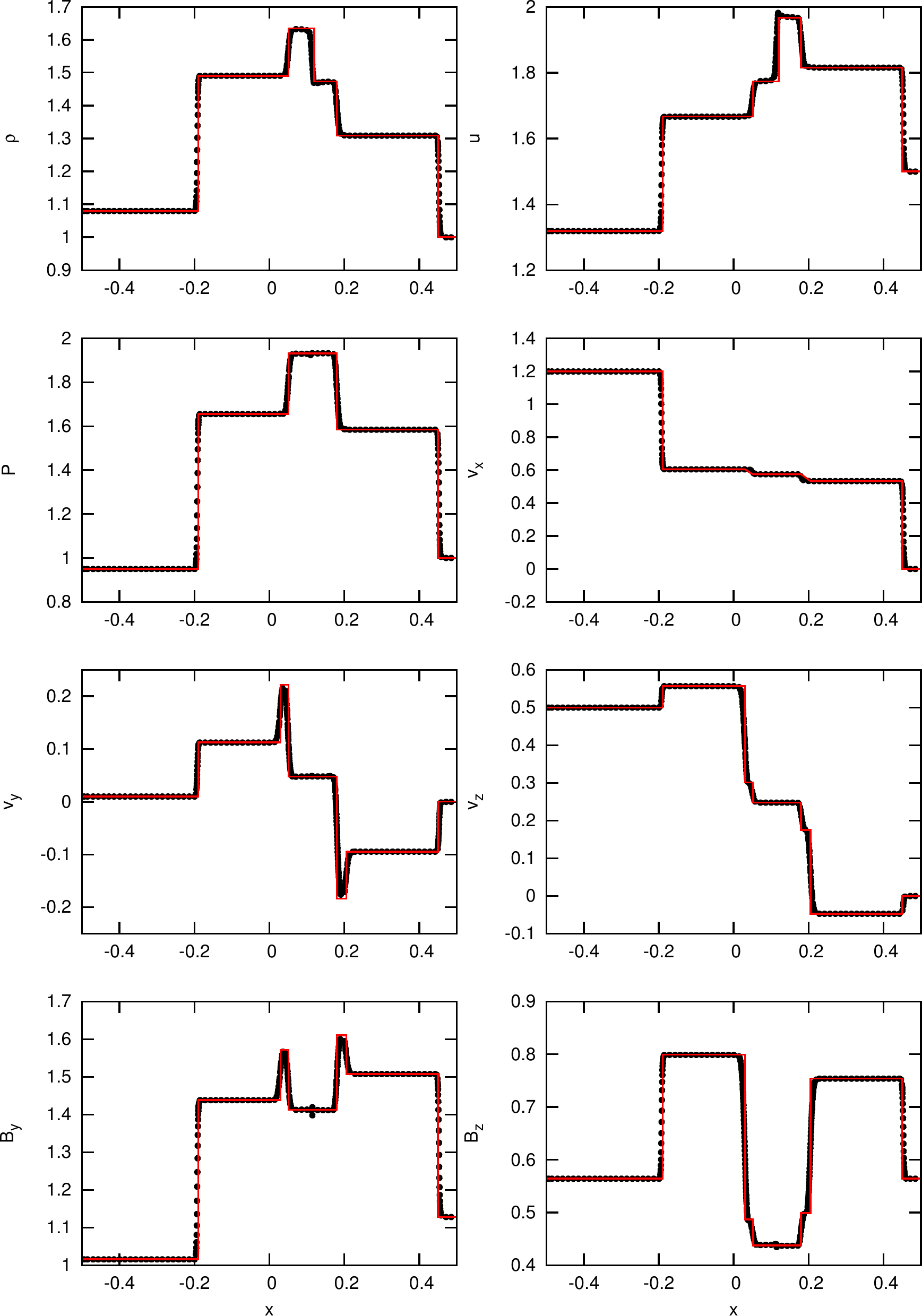}
\caption{Shocktube test 2A from \citetalias{rj95} performed in 3D with left state ($\rho$, $P$, $v_x$, $v_y$, $v_z$, $B_y$) $=$ (1.08, 0.95, 1.2, 0.01, 0.5, $3.6/\sqrt{4 \pi}$) and right state ($\rho$, P, $v_x$, $v_y$, $v_z$, $B_y$) $=$ (1, 1, 0, 0, 0, $4/\sqrt{4 \pi}$) with $B_x=B_z=2/\sqrt{4 \pi}$ at $t=0.2$. Black circles are the particles and the red line is the solution from \citetalias{rj95}.}
\label{fig:shock2a}
\end{figure}

This 3D problem originally introduced by \citet{dw94} has three dimensional velocity and magnetic fields generating two fast and slow shocks travelling in both directions, two rotational discontinuities, and a contact discontinuity in the centre.  It has left state ($\rho$, $P$, $v_x$, $v_y$, $v_z$, $B_y$) $=$ (1.08, 0.95, 1.2, 0.01, 0.5, $3.6/\sqrt{4 \pi}$) and right state ($\rho$, P, $v_x$, $v_y$, $v_z$, $B_y$) $=$ (1, 1, 0, 0, 0, $4/\sqrt{4 \pi}$) with $B_x=B_z=2/\sqrt{4 \pi}$ and $\gamma=5/3$. 

To fully capture the 3D velocity and magnetic fields, the test has been simulated in 3D with 800$\times$12$\times$12 particles on the left state and 500$\times$12$\times$12 particles on the right state arranged on close-packed triangular lattices.  Results at $t=0.2$ are presented in Figure~\ref{fig:shock2a} with all shock features, with the red line giving the solution from \citetalias{rj95}.  No post-shock noise in the magnetic field is evident, indicating that the applied artificial resistivity is sufficient.  The L1 error in the $B_y$ profile is $3.086\times 10^{-3}$, compared to $3.358\times 10^{-3}$ if the \citetalias{pm05} switch is used instead,  and for the $B_z$ profile is $5.33\times 10^{-3}$ for our new switch and $6.203\times 10^{-3}$ for the \citetalias{pm05} switch. 

\subsection{Shocktube 5A}
\label{sec:shock5a}

\begin{figure}
 \includegraphics[width=1.0\linewidth]{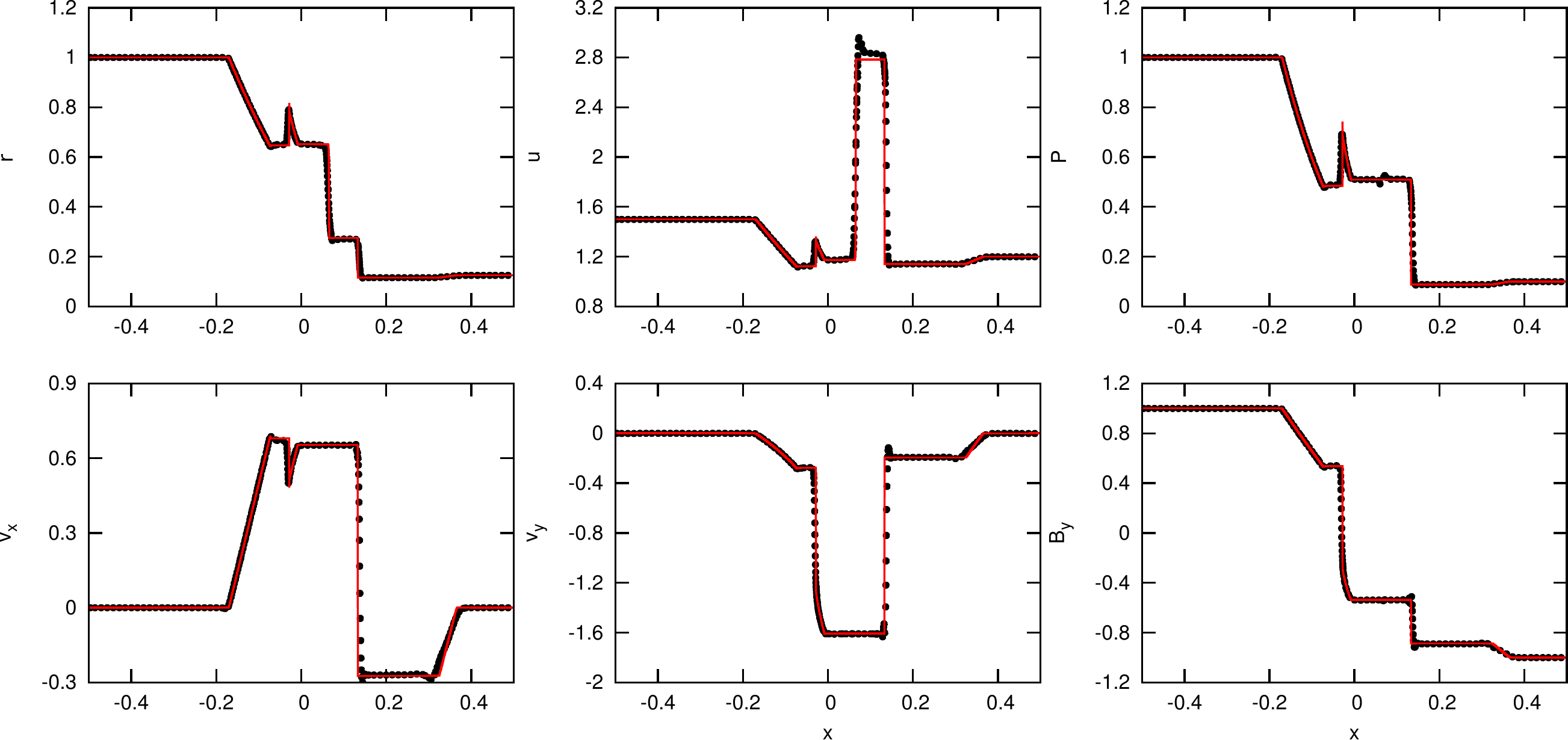}
\caption{Shocktube test 5A from \citetalias{rj95} performed in 2D with left state ($\rho$, $P$, $v_x$, $v_y$, $B_y$) $=$ (1, 1, 0, 0, 1) and right state ($\rho$, P, $v_x$, $v_y$, $B_y$) $=$ (0.125, 0.1, 0, 0, -1) with $B_x=0.75$ at $t=0.1$. Black circles are the particles and the red line is the solution obtained with the {\sc Athena} code using $10^4$ grid cells.}
\label{fig:shock5a}
\end{figure}

The final shocktube originates from \citet{briowu88}.  It is another 2D shocktube, however it is of particular interest as it contains a compound shock/rarefaction structure.  It has the same initial density and pressure profile as the standard Sod shocktube \citep{sod78}, but with the addition of a magnetic field.  The left state is ($\rho$, $P$, $v_x$, $v_y$, $B_y$) $=$ (1, 1, 0, 0, 1) and right state ($\rho$, P, $v_x$, $v_y$, $B_y$) $=$ (0.125, 0.1, 0, 0, -1) with $B_x=0.75$.  Here we use $\gamma=5/3$ instead of $2$ to follow the results of \citetalias{rj95}.  

The shock has been simulated with 800$\times$30 particles for the right state and 300$\times$10 particles for the right state.  Results at $t=0.1$ are presented in Figure~\ref{fig:shock5a}.  For this test, the Riemann solution of \citetalias{rj95} does not contain the slow compound structure, so instead we compare our results against those from the {\sc Athena} code \citep{athena} using $10^4$ grid cells.  As previously, no post-shock noise in the magnetic field is found.  The L1 error profile for $B_y$ is $4.231\times 10^{-3}$ when using our new switch, compared to $6.259 \times 10^{-3}$ if the \citetalias{pm05} switch is used.

\subsection{Polarised Alfv\'en Wave}
\label{sec:polarizedalfven}

We now examine the ability of the switch to reduce dissipation when no shocks are present.  The test problem used is a circularly polarised Alfv\'en wave travelling in a 2D periodic box, following \citet{toth00}.  This is an exact solution to the ideal MHD equations, so the wave should return to its original state after each crossing.  There are no discontinuities in the magnetic field in this test, but gradients in the magnetic field may cause the $\alpha_\text{B}$ switch to activate.

The simulation is set up using $1682$ particles arranged on a triangular lattice in a periodic domain of lengths $[x,y] = [1/\cos(\omega), 1/\sin(\omega)]$ using $\omega=\pi/6$ which sets the direction of wave motion.  The initial density and pressure are $\rho=1$ and $P=0.1$ with $\gamma=5/3$.  The velocity and magnetic fields parallel and perpendicular to the wave are $[v_\parallel, v_\perp] = [0, 0.1 \sin(2 \pi x_\xi)]$, and $[B_\parallel, B_\perp] = [1, 0.1 \sin(2 \pi x_\xi)]$ where $x_\xi = x \cos(\omega) + y \sin(\omega)$.  Velocity and magnetic field components oriented out of the plane are $v_z = B_z = 0.1 \cos(2 \pi x_\xi)$. Artificial viscosity is not applied in this test in order to minimise dissipation of energy in the system.

\begin{figure}
\centering
 \includegraphics[width=0.7\linewidth]{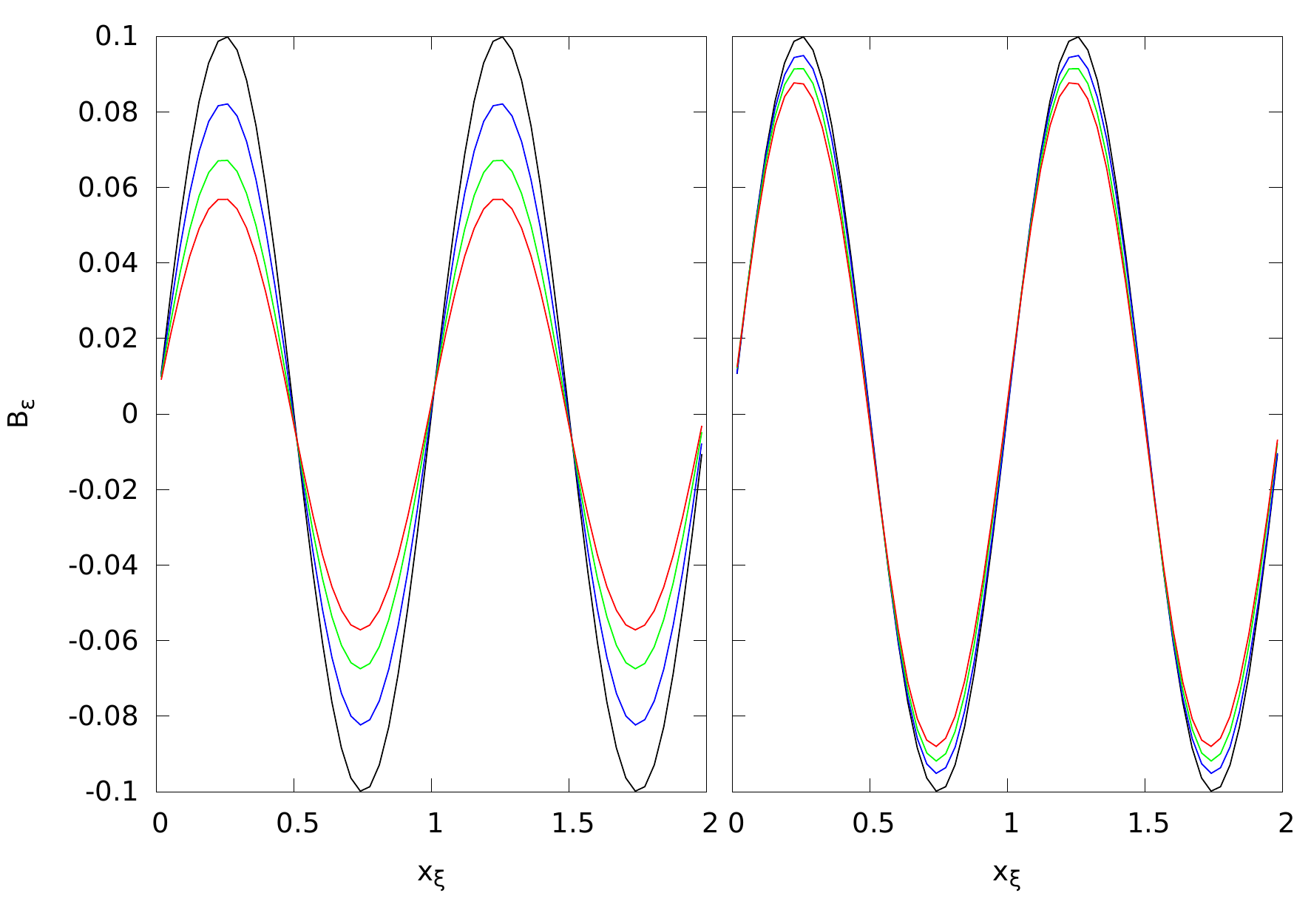}
\caption{Results of the polarised Alfv\'en wave propagation test in 2D, with the exact solution in black, and at $t=2,4,6$ corresponding to 2, 4, and 6 periods.  On the left, the \citetalias{pm05} switch has been used whereas on the right the new resistivity switch has been used.  The maximum $\alpha_\text{B}$ values are $10\times$ higher for the \citetalias{pm05} switch than the new switch, and after 6 periods the amplitude of the wave has decayed over $40\%$ for the \citetalias{pm05} switch compared to only $10\%$ for the new switch.}
\label{fig:polarizedalfven}
\end{figure}

The value of $\alpha_\text{B}$ produced using the new switch can be calculated from the initial conditions, which give $\vert \nabla {\bf B} \vert = 0.2 \pi$ and $\vert {\bf B} \vert = 1$. Thus, for a smoothing length $h=1.2 \Delta x$ where $\Delta x$ is the particle spacing, the new switch gives $\alpha_\text{B}\sim 0.02$ at this resolution. By contrast, the simulations using the \citetalias{pm05} switch produce maximum $\alpha_\text{B}$ values approximately $10 \times$ higher ($0.22$ vs $0.02$), meaning that in this case the \citetalias{pm05} switch is an order of magnitude more dissipative at $t=0$.

After 6 periods, the amplitude of the wave has decayed by over $40\%$ using the \citetalias{pm05} compared to only $\sim 10\%$ for the new switch.  Although the maximum $\alpha_\text{B}$ is $10 \times$ higher with the \citetalias{pm05} switch than the new switch, this is not reflected in the wave amplitude after 6 periods because $\vert \nabla {\bf B} \vert$ and the source term in the \citetalias{pm05} switch (Equation~\ref{eq:pm05switch}) are reduced as the wave is damped.  The rate of this reduction differs between the two switches since the \citetalias{pm05} switch damps the wave more heavily.

% rho = [0.08, 0.35]
% Bpres = [0, 0.18]
% alpha = [0, 1]

\begin{figure}
\centering
 \setlength{\tabcolsep}{0.002\textwidth}
\begin{tabular}{ccc}
% \scriptsize{PM05 switch} & \scriptsize{New switch} 
   \includegraphics[width=0.3\linewidth]{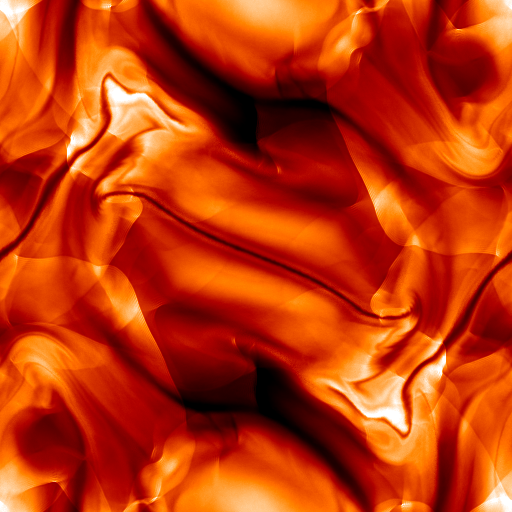} 
 & \includegraphics[width=0.3\linewidth]{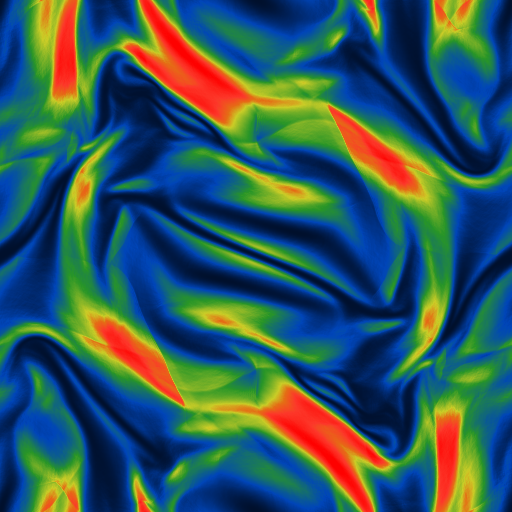}
 & \includegraphics[width=0.3\linewidth]{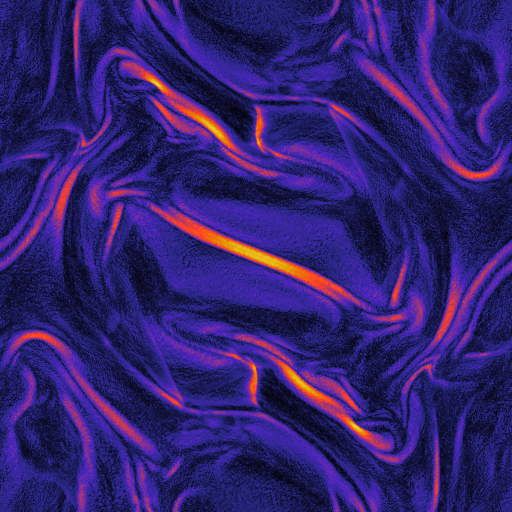}
\\
  \includegraphics[width=0.3\linewidth]{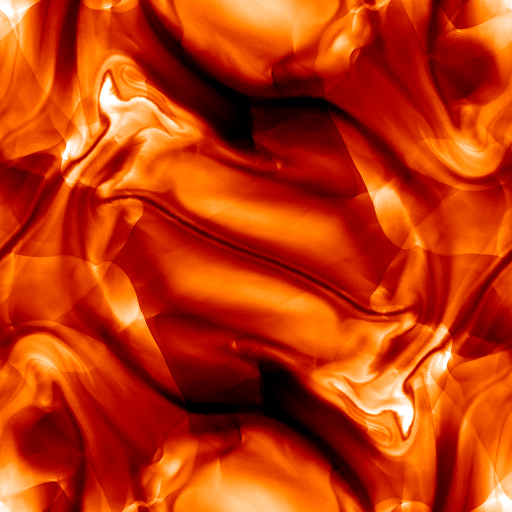} 
 & \includegraphics[width=0.3\linewidth]{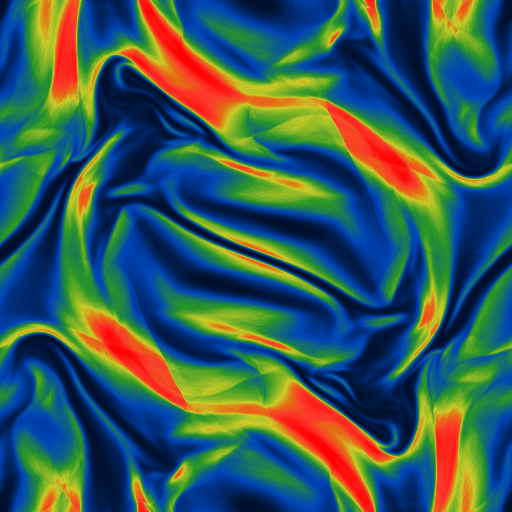}
 & \includegraphics[width=0.3\linewidth]{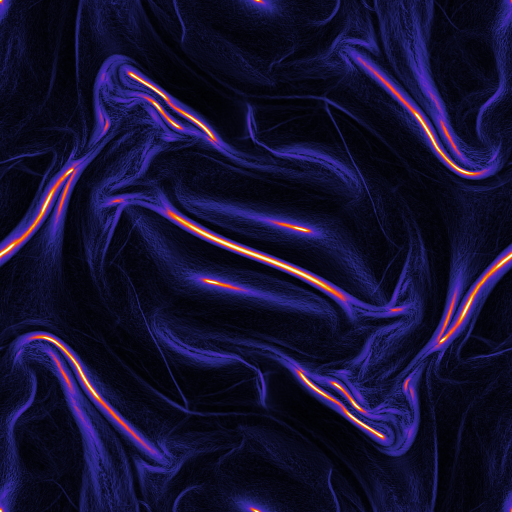}
\\
  \includegraphics[width=0.3\linewidth]{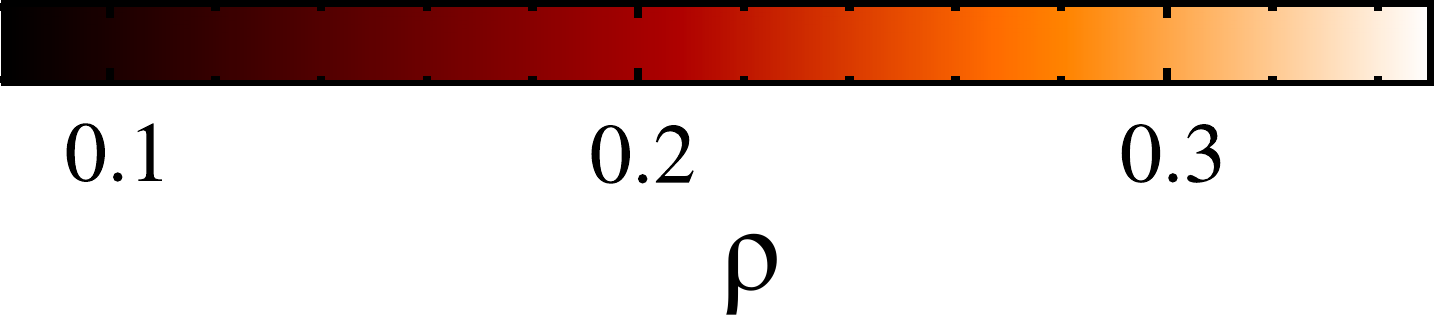}
 & \includegraphics[width=0.3\linewidth]{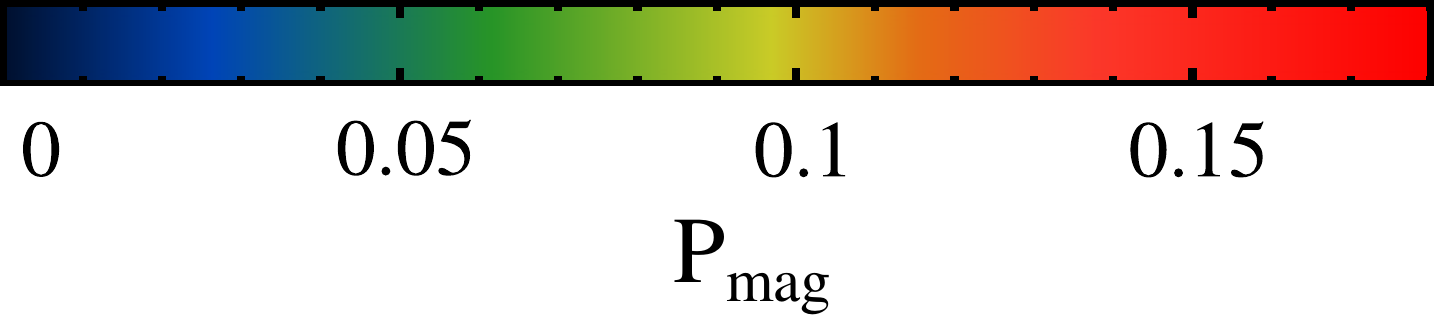}
 & \includegraphics[width=0.3\linewidth]{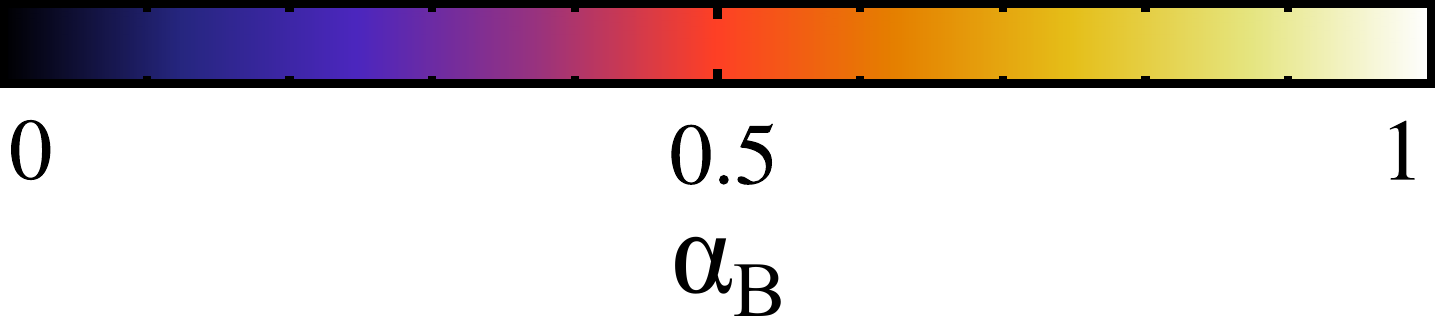}
\end{tabular}
\caption{The density, magnetic pressure, and $\alpha_\text{B}$ (left to right panels) of the Orszag-Tang vortex at $t=1$ for the old (top row) and new (bottom row) resistivity switches.  The new switch effectively traces the shock lines, with little or no dissipation between shocks.  The low density regions are more sharply defined using the new switch due to the decreased dissipation of the magnetic field structure.}
\label{fig:orszag-alphab}
\end{figure}

\subsection{Orszag-Tang vortex}
\label{sec:orszag}

The Orszag-Tang vortex \citep{ot79} is a widely used test for many astrophysical MHD codes \citep[e.g.,][]{ramses, athena, mhdgadget}.  The problem has an initial vortex structure creating several classes of interacting shock waves which evolve into turbulence, with the initial conditions as given in Section~\ref{sec:ot}.

% The initial structure has $\rho=25/(36\pi)$, $P=5/(12\pi)$, ${\bf v}=[-\sin(2\pi y), \sin(2\pi x)]$, and ${\bf B}=[-\sin(2\pi y), \sin(4\pi x)]$ with $\gamma=5/3$. 

The test has been simulated using $512^2$, $1024^2$, and $2048^2$ particles initially arranged on a square lattice.  The initial conditions are set up by first creating the particles in one quadrant of the domain, then mirroring the configuration to the other quadrants with appropriate sign changes in the velocity and magnetic fields as needed.  This removes the slight discrepancies from floating point arithmetic, retaining exact symmetry in the initial conditions. The \citet{mm97} switch for artificial viscosity has been used.

% @512: alphaB mean = 0.205 old
% @512: alphaB mean = 0.110 new

% @1024: alphaB mean = 0.145 old
% @1024: alphaB mean = 0.070 new

% @2048: alphaB mean = 0.097 old
% @2048: alphaB mean = 0.046 new

Results are presented at $t=1$ in Figure~\ref{fig:orszag-alphab} which shows renderings of the density, magnetic pressure, and $\alpha_\text{B}$ in the domain for $1024^2$ particles.  The new switch is effective at activating resistivity along the shock lines, yet keeps $\alpha_\text{B}$ minimal between shocks.  By contrast, the \citetalias{pm05} switch results in broad regions with $\alpha_\text{B}\approx 1$ near shocks and a mean $\alpha_\text{B}$ twice as high ($\sim0.2$ to $\sim0.1$).  This leads to a smoothing away of subtle magnetic features, particularly noticeable around the central magnetic feature, and in some of the low density regions which are less sharply defined.

\begin{figure}
\centering
 \includegraphics[width=0.7\linewidth]{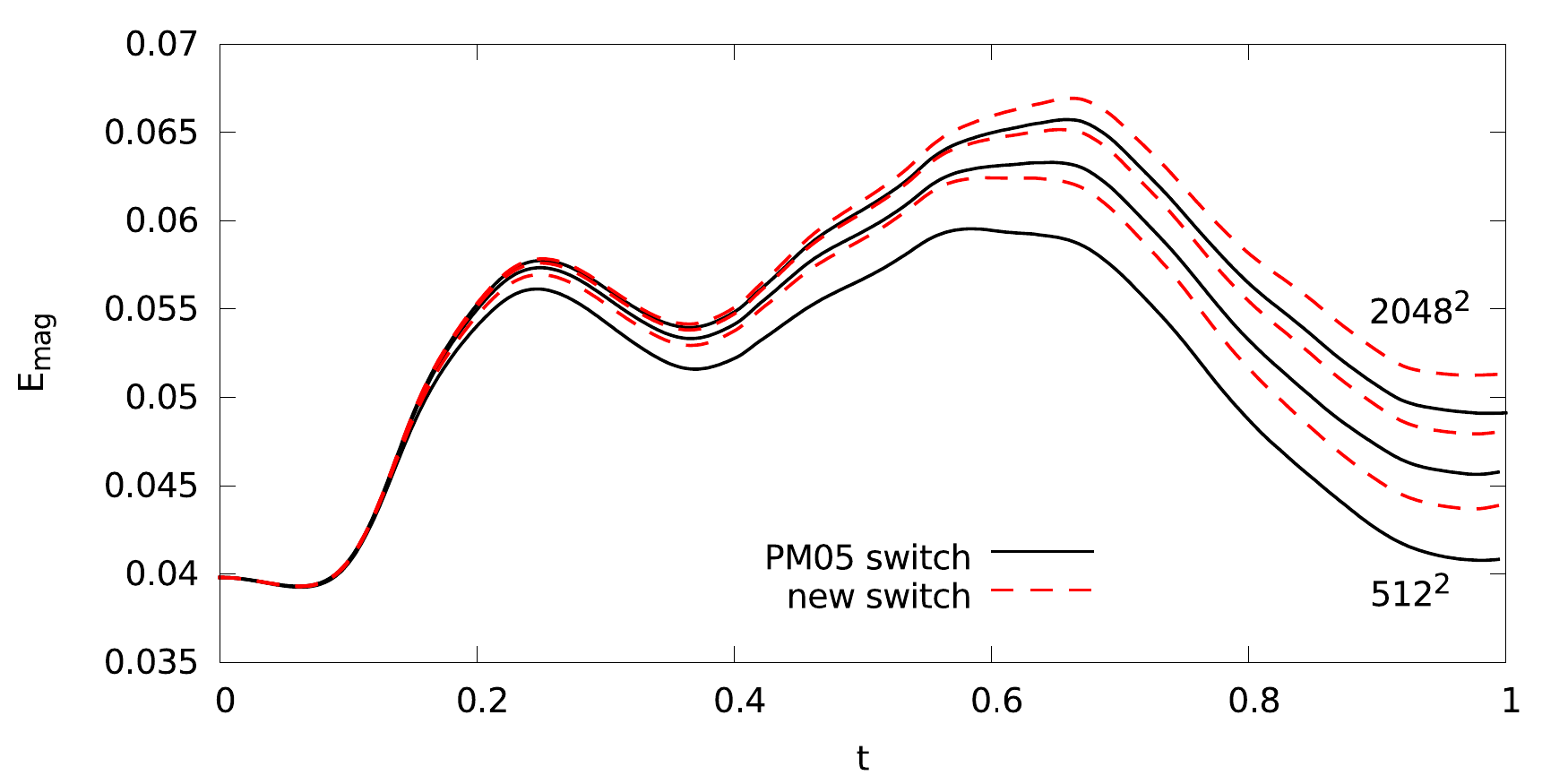}
\caption{Evolution of the magnetic energy for the Orszag-Tang vortex using the PM05 resistivity switch (black, solid lines) and the new resistivity switch (red, dashed lines) at resolutions of $512^2$, $1024^2$, and $2048^2$ particles.  The new switch is much less dissipative than the PM05 switch, producing an effect similar to increasing the resolution.}
\label{fig:orszag-be-switch}
\end{figure}

Figure~\ref{fig:orszag-be-switch} shows the evolution of the magnetic energy as a function of time for $512^2$, $1024^2$, and $2048^2$ particles.  This shows that the magnetic energy is dissipated less at higher resolution.  Using the new artificial resistivity switch also leads to a lower dissipation rate compared to the \citetalias{pm05} switch, producing an effect equivalent to running the test at higher resolution.

\subsection{Mach 10 MHD turbulence}
\label{sec:mhdturb}

Our final test is of supersonic magnetised turbulence which is representative of conditions in molecular clouds \citep[see reviews by][]{evans99,es04,mo07}.  A stochastic, solenoidal driving force is applied, generating turbulence with a root-mean-square Mach number of 10.  It has an initially weak magnetic field, with the kinetic energy approximately 10 orders of magnitude larger than magnetic energy, which grows through dynamo amplification by the conversion of kinetic to magnetic energy \citep[see review by][]{bs05}.  Our simulations follow the SPH Mach 10 turbulence study of \citet{pf10}, but in the MHD case of turbulent dynamo amplification studied by \citet{federrathetal11}. See Chapter~\ref{sec:chapter-mhdturb} for further simulation details and for a comparison of SPMHD with grid-based methods on turbulent small-scale dynamo amplification.

The simulation is set up at a resolution of $128^3$ particles.  The initial density is $\rho=1$ with an isothermal equation of state using a speed of sound of $c_{\rm s}=1$.  The gas is initially at rest, and has a uniform magnetic field $B_z = \sqrt{2}\times 10^{-5}$ such that the initial plasma $\beta = 10^{10}$.  The \citet{mm97} switch for artificial viscosity has been used.

To drive the turbulence, an acceleration based on an Ornstein-Uhlenbeck process is used \citep{ep88, federrathetal10}, which is a stochastic process with a finite autocorrelation timescale that drives motion at low wave numbers.  The driving force is constructed in Fourier space, allowing it to be decomposed into solenoidal and compressive components and for this case we only use the solenoidal component.

\begin{figure}
 \centering
\setlength{\tabcolsep}{0.002\textwidth}
\begin{tabular}{cccr}
\scriptsize{Fixed $\alpha_\text{B}=1$} & \scriptsize{PM05 switch} & \scriptsize{New switch} 
\\
   \includegraphics[height=0.28\linewidth]{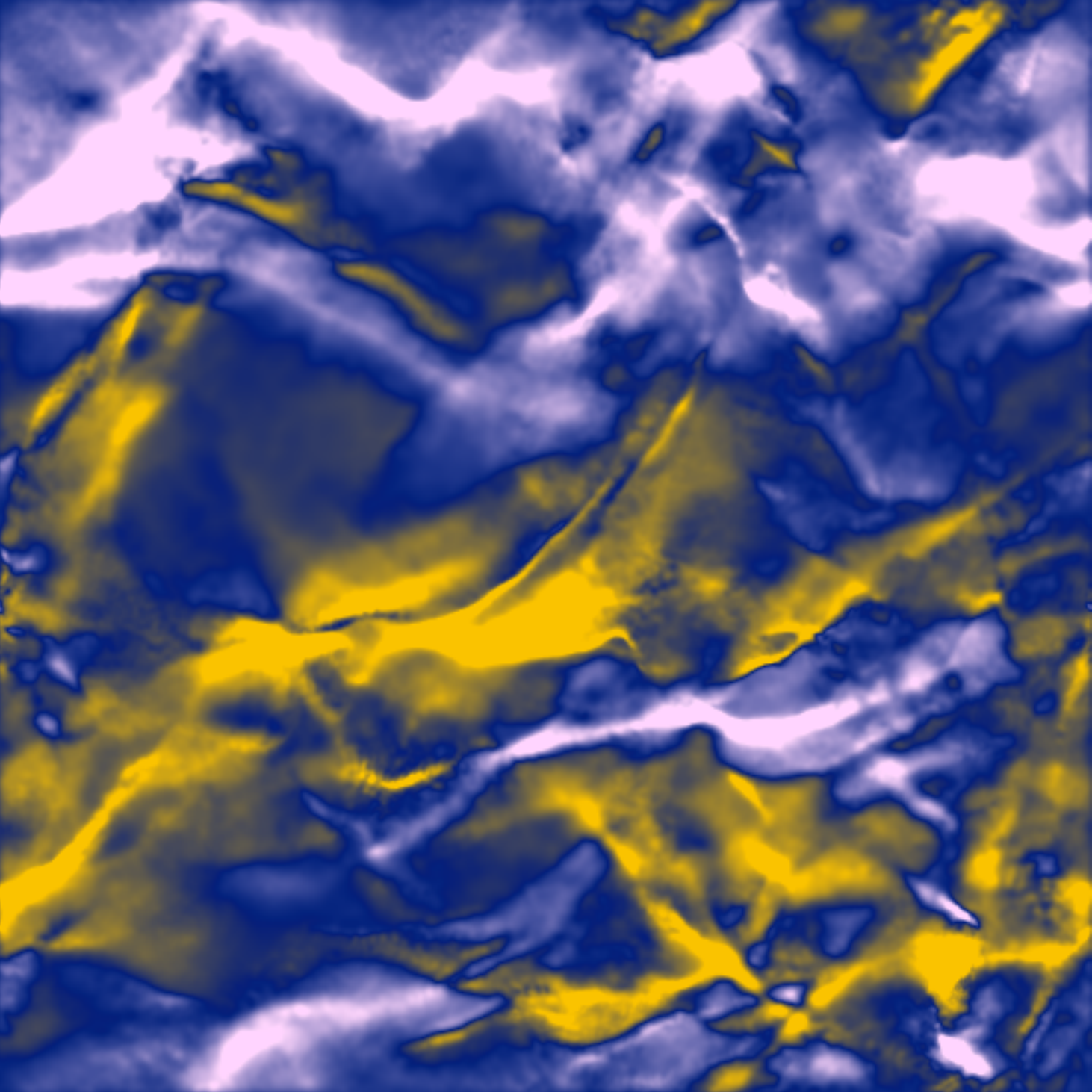}
 & \includegraphics[height=0.28\linewidth]{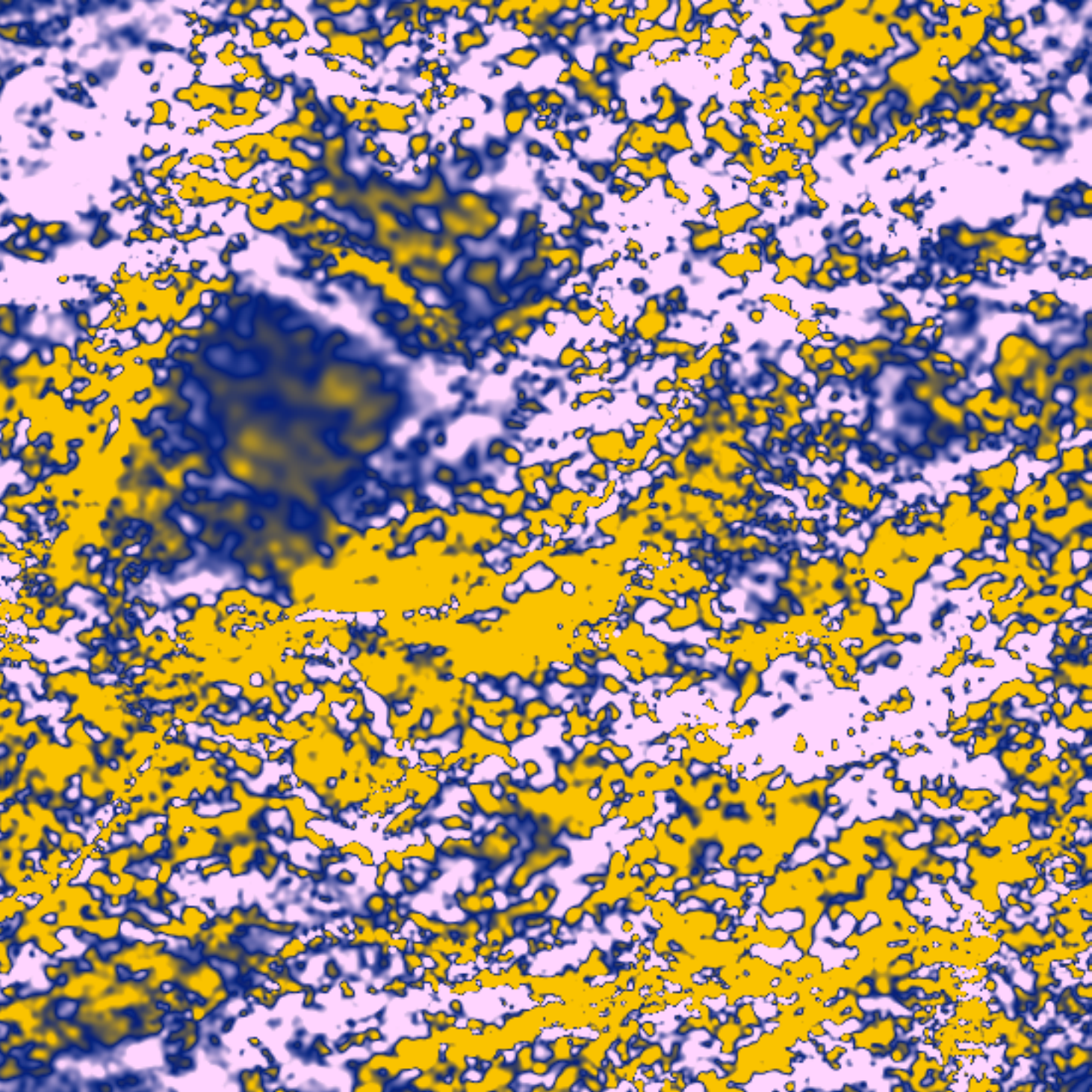}
 & \includegraphics[height=0.28\linewidth]{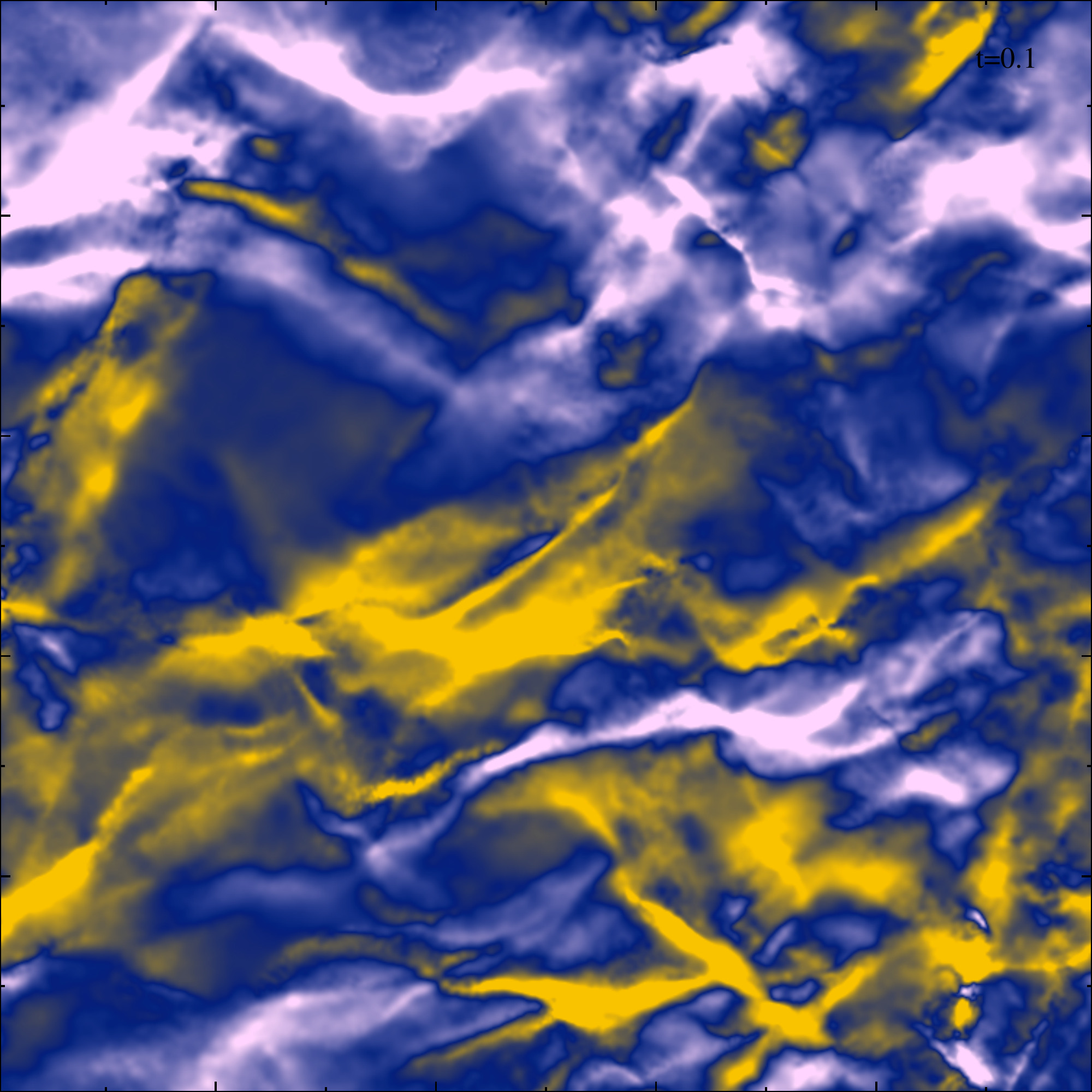}
 & \includegraphics[height=0.28\linewidth]{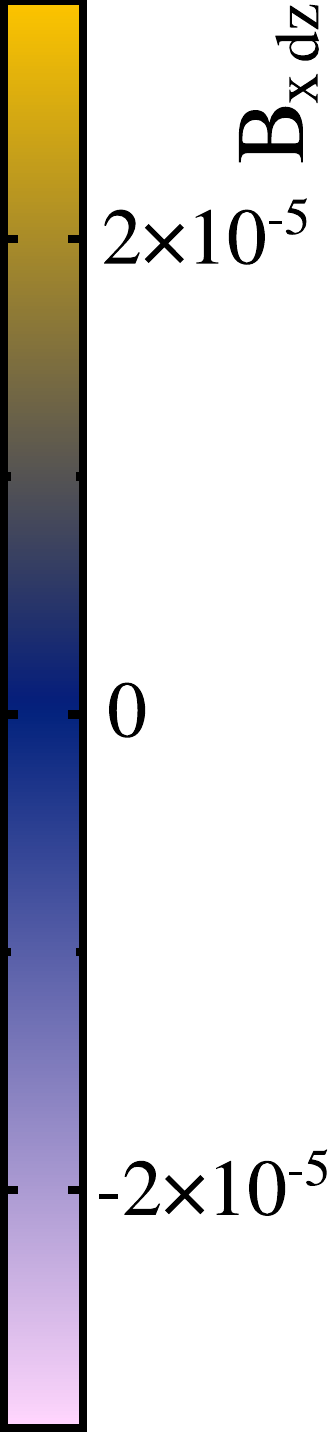}
\\
   \includegraphics[height=0.28\linewidth]{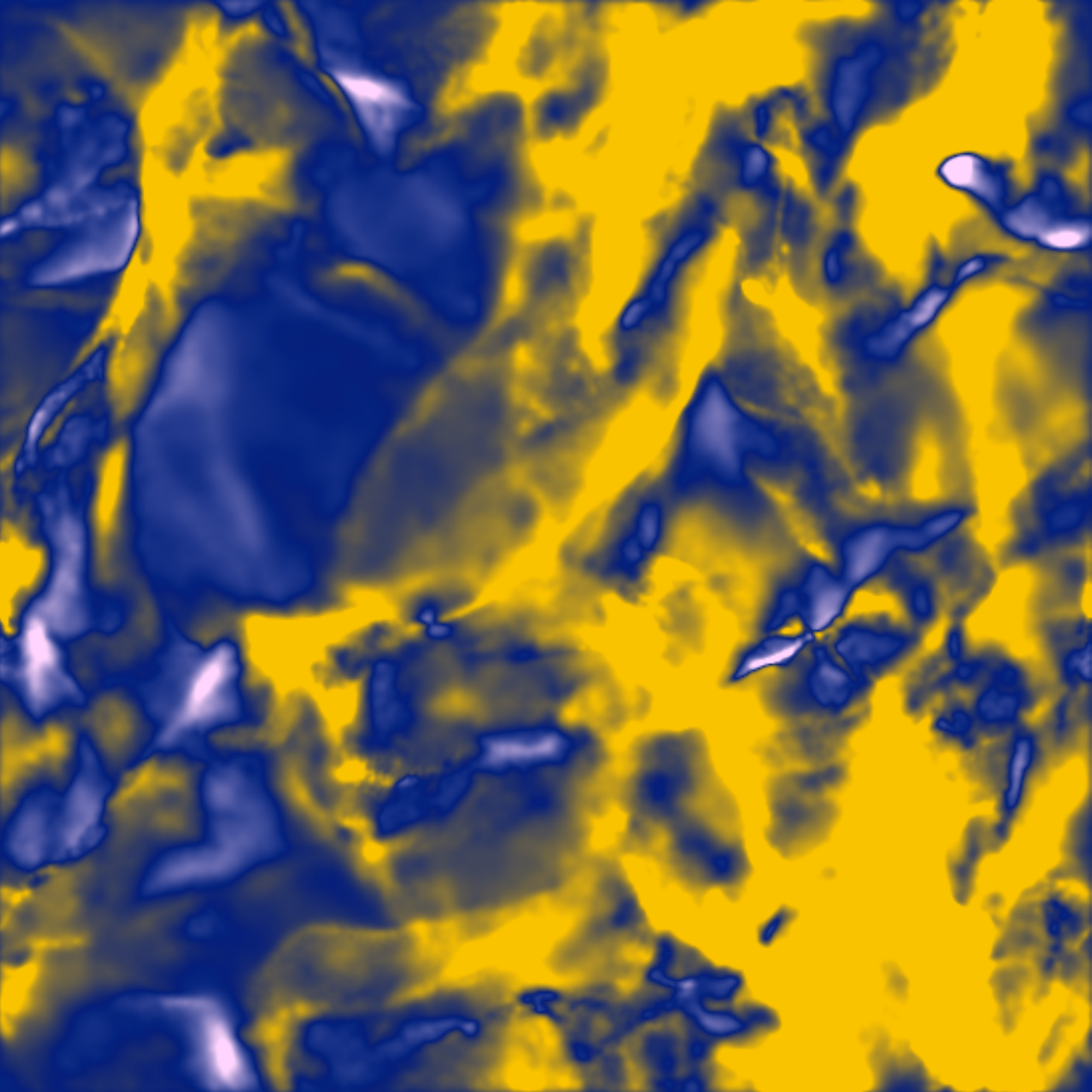}
 & \includegraphics[height=0.28\linewidth]{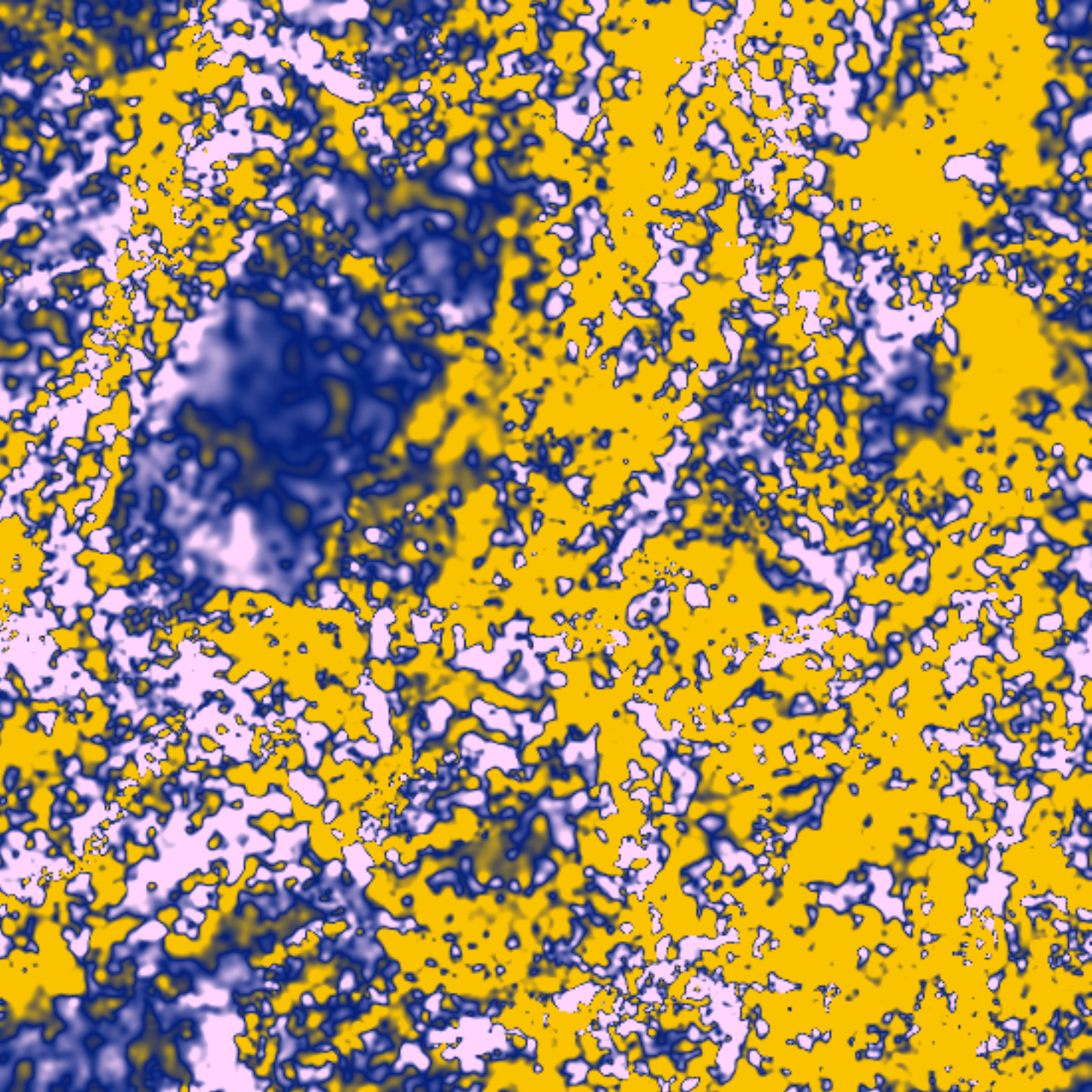}
 & \includegraphics[height=0.28\linewidth]{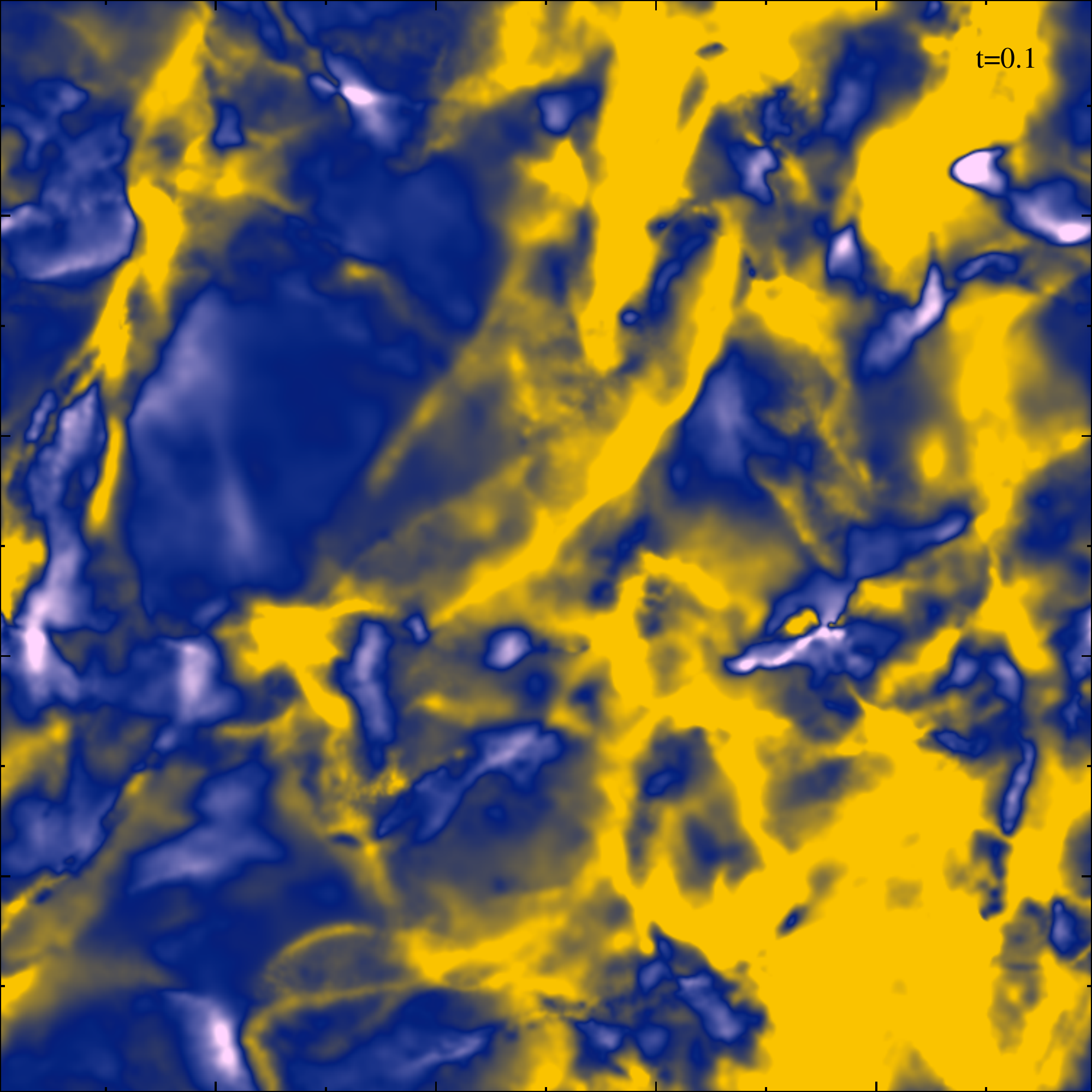}
 & \includegraphics[height=0.28\linewidth]{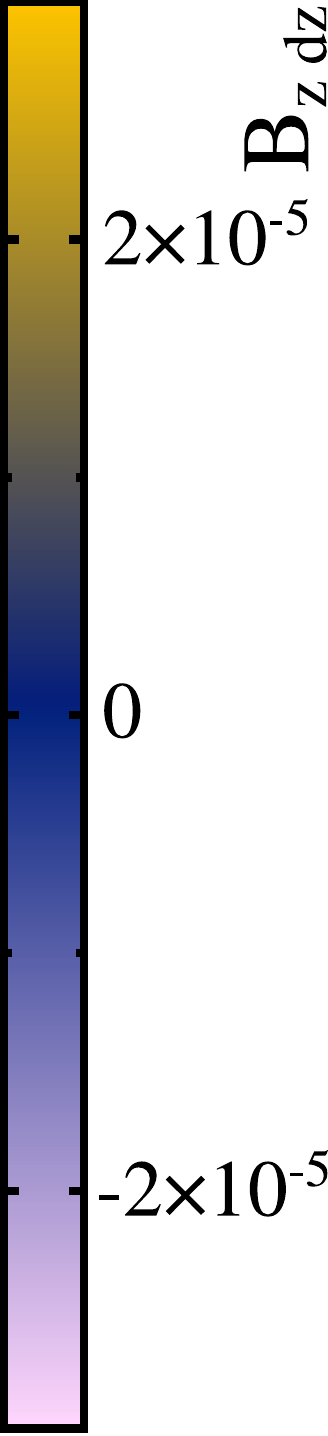}
\end{tabular}
\caption{The column integrated $x$ \& $z$ (top, bottom) magnetic field components using fixed $\alpha_\text{B}=1$ (left), the \citetalias{pm05} switch (centre), and the new switch (right) after two turbulent turnover times (i.e., the regime of fully developed turbulence).  The magnetic field structure using the previous switch is dominated by unphysical noise due to the shocks failing to be captured (centre), whereas the new switch is able to capture the shocks and the magnetic field retains its physical structure (right).}
\label{fig:mhdturb}
\end{figure}

The column integrated $x$ \& $z$ components of the magnetic field are shown in Figure~\ref{fig:mhdturb} at $t=2$ turbulent turnover times.  The \citetalias{pm05} switch fails to raise $\alpha_\text{B}$ to appreciable levels ($\alpha_\text{B} \sim 10^{-5}$), and as demonstrated in Figure~\ref{fig:mhdturb}, the shocks in the magnetic field fail to be captured.  This leads to break-up of the shocks, causing unphysical magnetic field growth until such a time as the field is strong enough to activate the switch.  By contrast, the new switch is invariant to field strength meaning that it turns on resistivity in the shocked regions and the shocks are captured.

We also found for this simulation that using the averaged Alfv\'en speed as the signal velocity for resistivity produces the same behaviour (shocks breaking apart).  In this instance, it is due to the large disparity between the Alfv\'en and sound speed meaning that the applied resistivity is too weak to capture the strong shocks properly.  With the fast MHD wave speed in the signal velocity (Equation~\ref{eq:vsigfastmhd}), the shocks are captured correctly.

\section{Generalisation to other dissipation terms}
\label{sec:switch-generalisation}

In this section, we extend the design concept of the new artificial resistivity switch to the artificial viscosity and thermal conductivity dissipation terms, constructing new switches based on the general idea of a normalised shock indicator \citep[see also][]{tp13b}. %No fine tuning of these switches has been attempted, rather the intent is to explore their possibility and create a basis upon which they could be developed.

\subsection{New artificial viscosity switch}
\label{sec:newvisc}

For the new artificial viscosity switch, we continue to use $- \nabla \cdot {\bf v}$ as the shock indicator as in the \citet{mm97} approach.  It would be unwise to use $\vert {\bf v} \vert$ for the normalisation as this would break Galilean invariance.  Instead, the fast MHD wave speed is used (or sound speed in pure hydrodynamics), relating the quantity to the Mach number.

It is also important that artificial viscosity is applied to the wake of the shock to reduce post-shock oscillations of the particles (as discussed in Section~\ref{sec:artvisc}).  Therefore, $\alpha$ is reduced over time using an integrated decay term like in the \citet{mm97}, \citet{cd10}, and \citet{rh12} approaches.

The resulting switch is therefore to set 
\begin{equation}
 \alpha_a = - \frac{h_a \nabla \cdot {\bf v}_a}{v_{\rm mhd}} 
\label{eq:newalphavisc}
\end{equation}
when greater than the current value of $\alpha_a$, otherwise $\alpha_a$ is reduced on the next time step according to
\begin{equation}
 \frac{{\rm d}\alpha_a}{{\rm d}t} = -\frac{\alpha_a}{\tau} ,
\end{equation}
where $\tau=h/\sigma_{v} v_{\rm mhd}$ has the same meaning as in the \citet{mm97} switch (Equation~\ref{eq:mm97switch}).  By following the considerations outlined above, this viscosity switch is quite similar in principle to the switch of \citet{cd10}, albeit with a simpler version for Equation~\ref{eq:newalphavisc}.

\subsection{New thermal conductivity switch}
\label{sec:newcond}

A switch for thermal conductivity can be constructed by analogy to Equation~\ref{eq:alpha_B}.  The gradient of thermal energy is chosen to detect discontinuities, setting
\begin{equation}
 \alpha_{u,a} = \frac{h_a \vert \nabla u_a \vert}{\vert u_a \vert} .
\end{equation}
As with artificial resistivity, it is expected that thermal conductivity only needs to be applied at the location of the discontinuity since there is no worry of oscillations in particle motion. Therefore, no time-integrated decay term is used.
%It may be worth considering using the second derivative to improve discontinuity detection, particularly for astrophysical applications.  During star formation, for example, situations may emerge where the system has a continuous gradient of thermal energy, yet is in hydrodynamic equilibrium due to gravitational forces.  

\subsection{Tests of artificial viscosity and thermal conductivity switches}

The efficacy of the new artificial viscosity and thermal conductivity switches are examined using a standard Sod shocktube test, and a setup producing Kelvin-Helmholtz instabilities.  %These are intended to prove the concept of the switches, not exhaustively test the capabilities.  

\subsubsection{Viscosity: Sod shocktube}
\label{sec:visctest}

The Sod shocktube \citep{sod78} has become a canonical test for hydrodynamic shocks.  It consists of a fluid with a discontinuity in the density and pressure that sends a shock wave into the low density medium and a rarefaction into the high density medium, with a contact discontinuity in the centre.  Artificial viscosity is required in this test in order to treat the shock wave.  

The simulation has left state ($x<0$) $\rho=1$ and $P=1$ in contact with a fluid of $\rho=0.125$ and $P=0.1$ ($x>0$) with $\gamma=5/3$.  Both states have zero initial motion.  The shocktube is simulated in 1D using $1000$ and $125$ particles for the two states, respectively.  Thermal conductivity is used with fixed $\alpha_u=1$, using the \citet{price08} switch (see Section~\ref{sec:thermcondswitches}). 

\begin{figure}
\centering
 \includegraphics[width=1.0\linewidth]{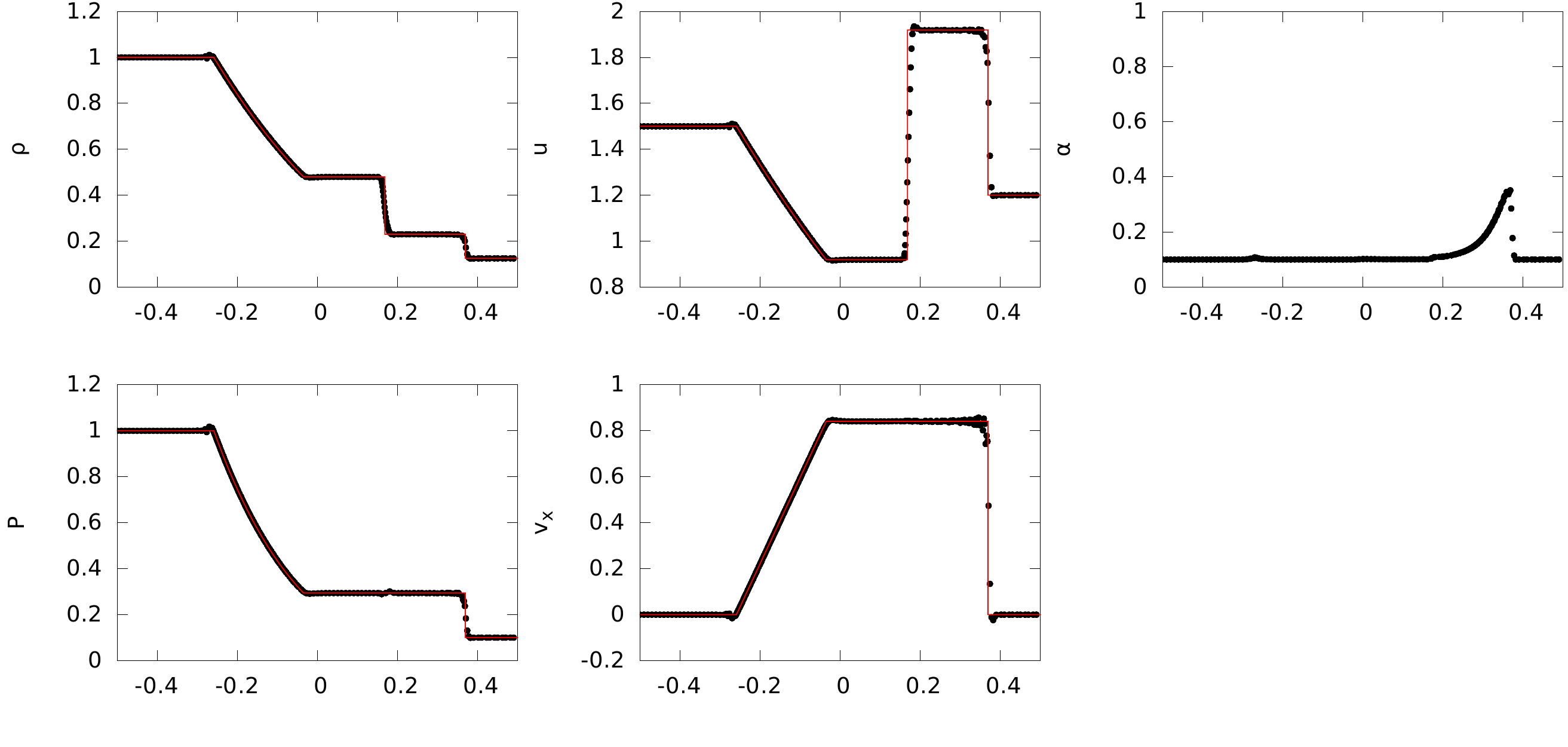}
\caption{Sod shocktube results at $t=0.2$ using the new viscosity switch described in Section~\ref{sec:newvisc}.  The black circle are values from the particles with the red line the Riemann solution.}
\label{fig:sodshock-results}
\end{figure}

Results at $t=0.2$ are presented in Figure~\ref{fig:sodshock-results} along with the solution calculated from a Riemann solver. The SPH solution agrees well with the Riemann solution, though there are small post-shock oscillations in the velocity field, which may also be noticed when using the \citet{mm97} switch (see also \citealt*{cd10}). These can be reduced by adjusting the decay timescale for $\alpha_{\rm AV}$. The contact discontinuity is spread over $\sim12$ particles, but this is not related to the artificial viscosity scheme. 

\subsubsection{Thermal conductivity: Kelvin-Helmholtz instability}
\label{sec:condtest}

Kelvin-Helmholtz instabilities have been studied many times with SPH \citep[e.g.,][]{agertzetal07, price08, valckeetal10, mlp12, hfg13}.  The instability occurs when there is a velocity shear in a fluid, causing turbulence to form along the interface.  An important aspect to simulating this correctly in SPH is application of thermal conductivity to treat thermal energy discontinuities across the interface.   If ignored, spurious pressure is generated preventing the fluid from mixing properly.  We use this test to investigate the ability of the new thermal conductivity switch to allow mixing of the fluids across the interface, and produce the ``curls'' which are emblematic of Kelvin-Helmholtz instabilities.

The test performed here follows the initial set up of \citet{price08}.  The fluid contains two regions in a 2:1 density contrast.  The domain is $x,y=[-0.5,0.5]$ and periodic boundary conditions are used creating two interfaces along which Kelvin-Helmholtz instabilities form.  The initial density profile is
\begin{equation}
 \rho =
\begin{cases}
 2 & \text{  } \vert y \vert < 0.25, \\
 1 & \text{  } \vert y \vert > 0.25.
\end{cases}
\end{equation}
The two regions are in pressure equilibrium with uniform $P=2.5$ with $\gamma=5/3$.  The $x$-velocity is $-0.5$ for the $\rho=2$ region, and $0.5$ for the $\rho=1$ region.  The $y$-velocity is zero, however the $n=4$ instability is seeded with a perturbation across the interfaces by
\begin{equation}
 v_y =
\begin{cases}
 A \sin[-2 \pi (x+ 0.5) / \lambda] & +0.225 < y < +0.275, \\
 A \sin[2 \pi (x + 0.5) / \lambda] & -0.225 < y < -0.275,
\end{cases}
\end{equation}
where $A=0.025$ and $\lambda = 1/6$.

The characteristic growth timescale of the instability depends on the densities of both fluids and the relative velocity shear. For an incompressible fluid, the timescale is \citep{chandrasekhar61}
\begin{equation}
\tau_{\rm KH} = \frac{\lambda (\rho_1 + \rho_2)}{(\rho_1 \rho_2)^{1/2} \vert v_1 - v_2 \vert} ,
\end{equation}
where $\rho_1$ and $v_1$ are the density and velocity of one fluid, with $\rho_2$ and $v_2$ the density and velocity of the other. For this calculation, the Kelvin-Helmholtz growth timescale is $\tau_{\rm KH}\approx0.35$.

The particles are initially arranged on triangular lattices.  A total of $454 184$ particles are used, with a particle spacing of $\Delta = 1/512$ in the low density region and $\Delta = 1/724$ in the high density region.  The \citet{mm97} switch (Equation~\ref{eq:mm97switch}) is used for artificial viscosity. The initial thermal energy is calculated using the density of the particles so that the pressure is constant across the interfaces. The quintic spline has been used.

The calculations were run using four different thermal conductivity switches: two based on pressure discontinuities and two on the relative velocity between particles (as motivated by methods such as \citealt{wvc08,sws10,valdarnini12, rh12}). The four thermal conductivity switches tested are: the normalised thermal gradient switch developed in Section~\ref{sec:newcond}; the divergence of the velocity, $v_{\rm sig}^u = \vert {\bf v}_{ab} \cdot \hat{\bf r}_{ab} \vert$, as used by \citet{valdarnini12}; the curl of the velocity, $v_{\rm sig}^u = \vert {\bf v}_{ab} \times \hat{\bf r}_{ab} \vert$, which should perform better at detecting shear motions; and the approach of \citet{price08} to set $v_{\rm sig}^u = \sqrt{ \vert P_a - P_b \vert / \overline{\rho}_{ab} }$. A calculation was performed with no thermal conductivity to act as a reference.
\begin{figure*}
 \centering
%\begin{minipage}{0.9\textwidth}
\setlength{\tabcolsep}{0.001\textwidth}
\begin{tabular}{cccc}
   \includegraphics[width=0.2\textwidth]{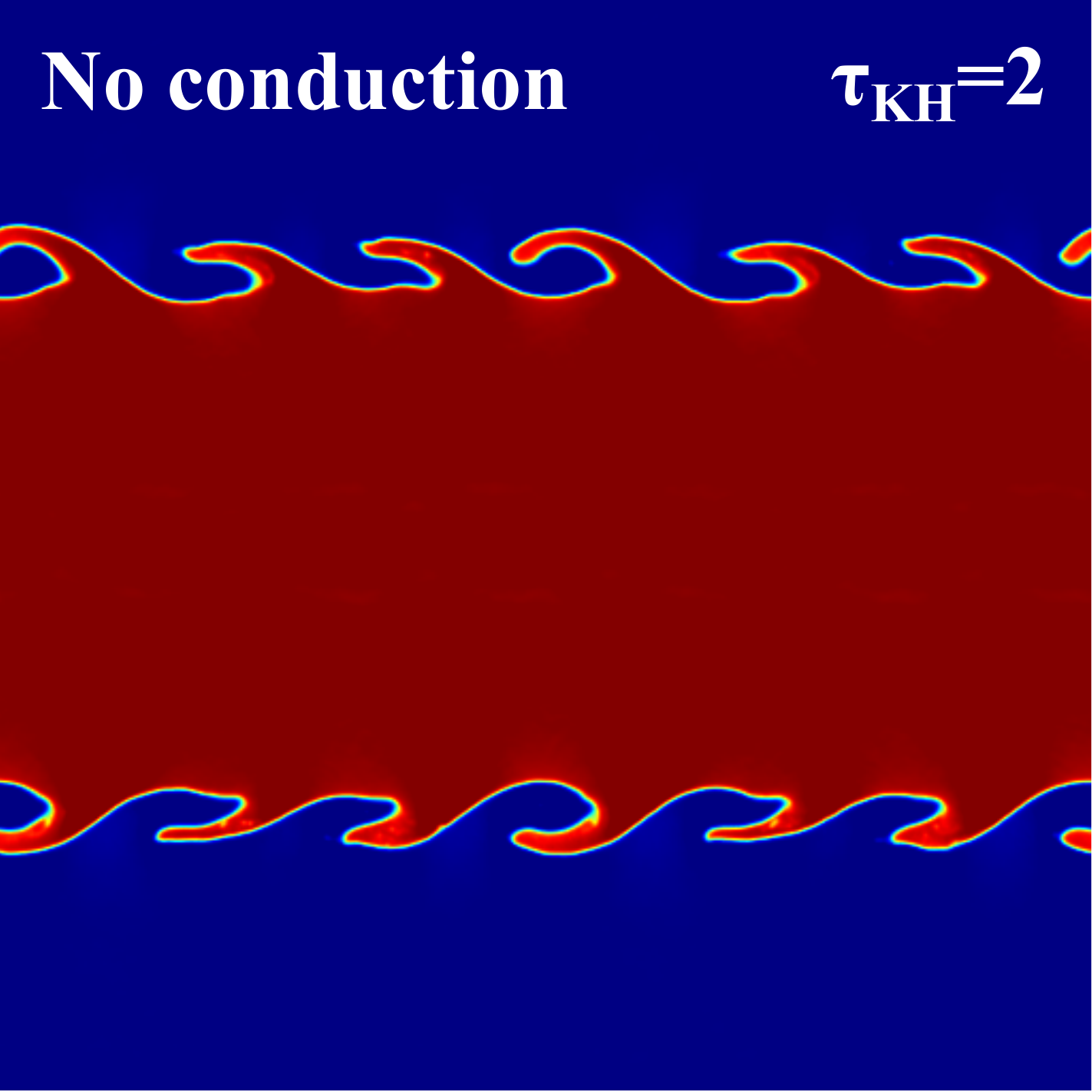}
 & \includegraphics[width=0.2\textwidth]{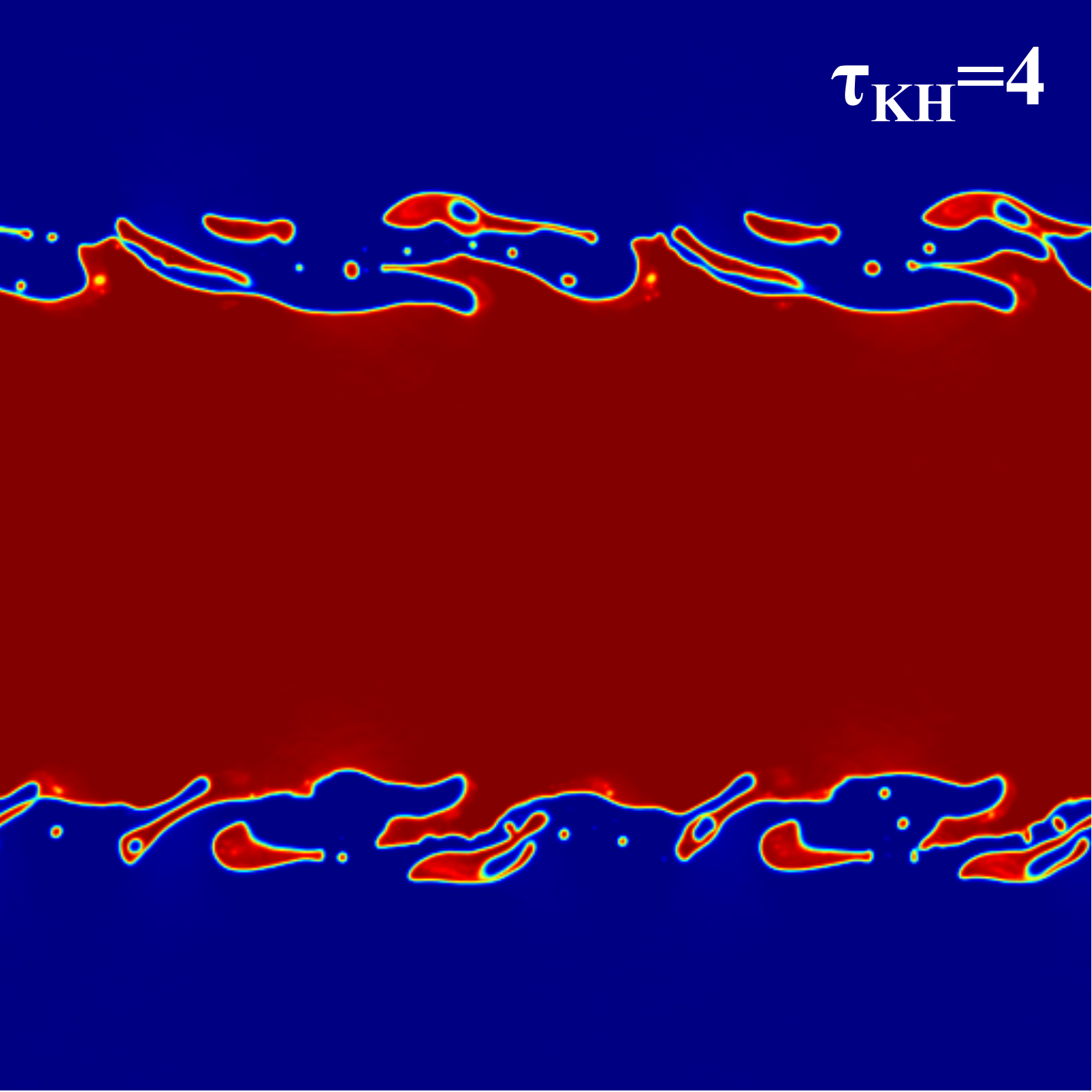}
 & \includegraphics[width=0.2\textwidth]{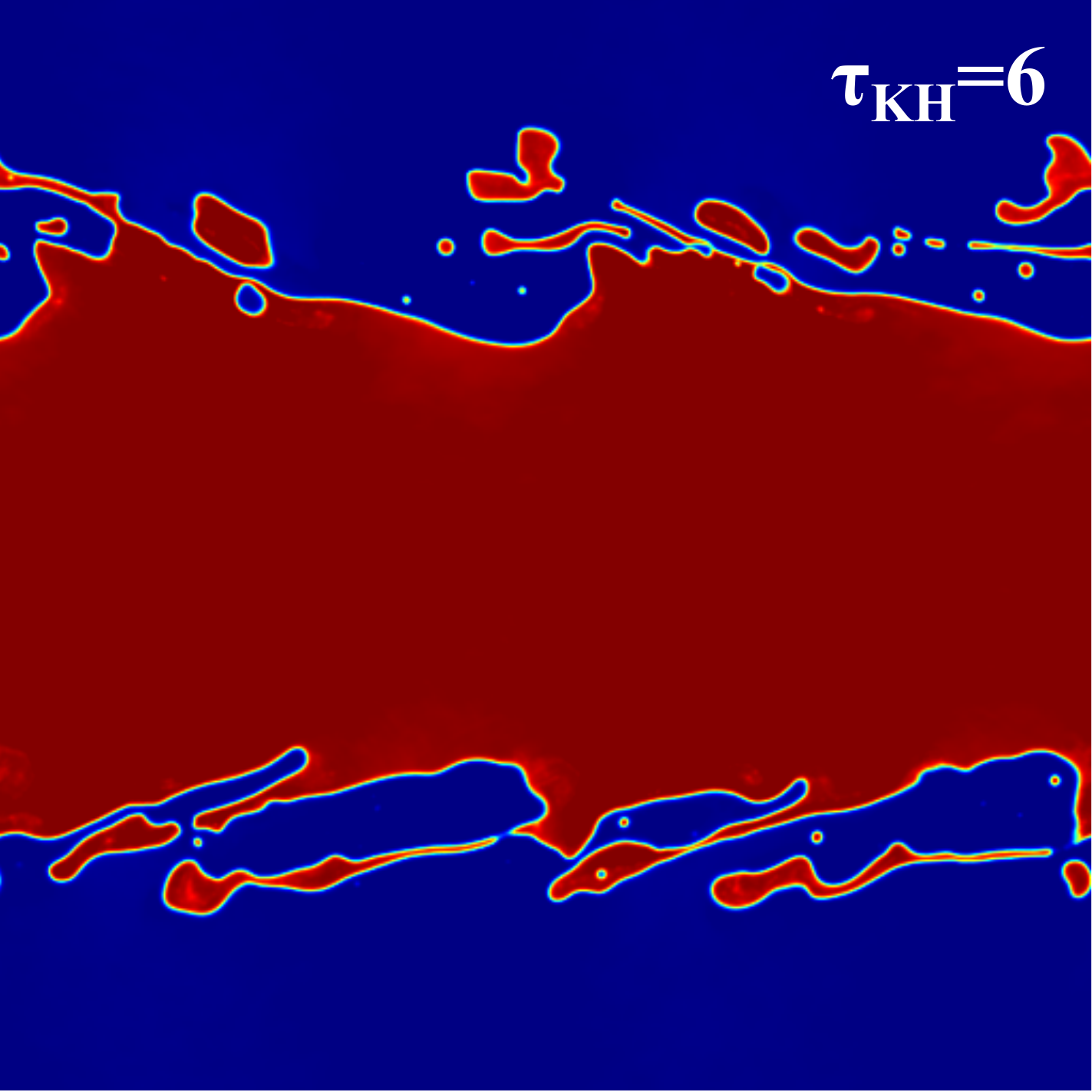}
 & \includegraphics[width=0.2\textwidth]{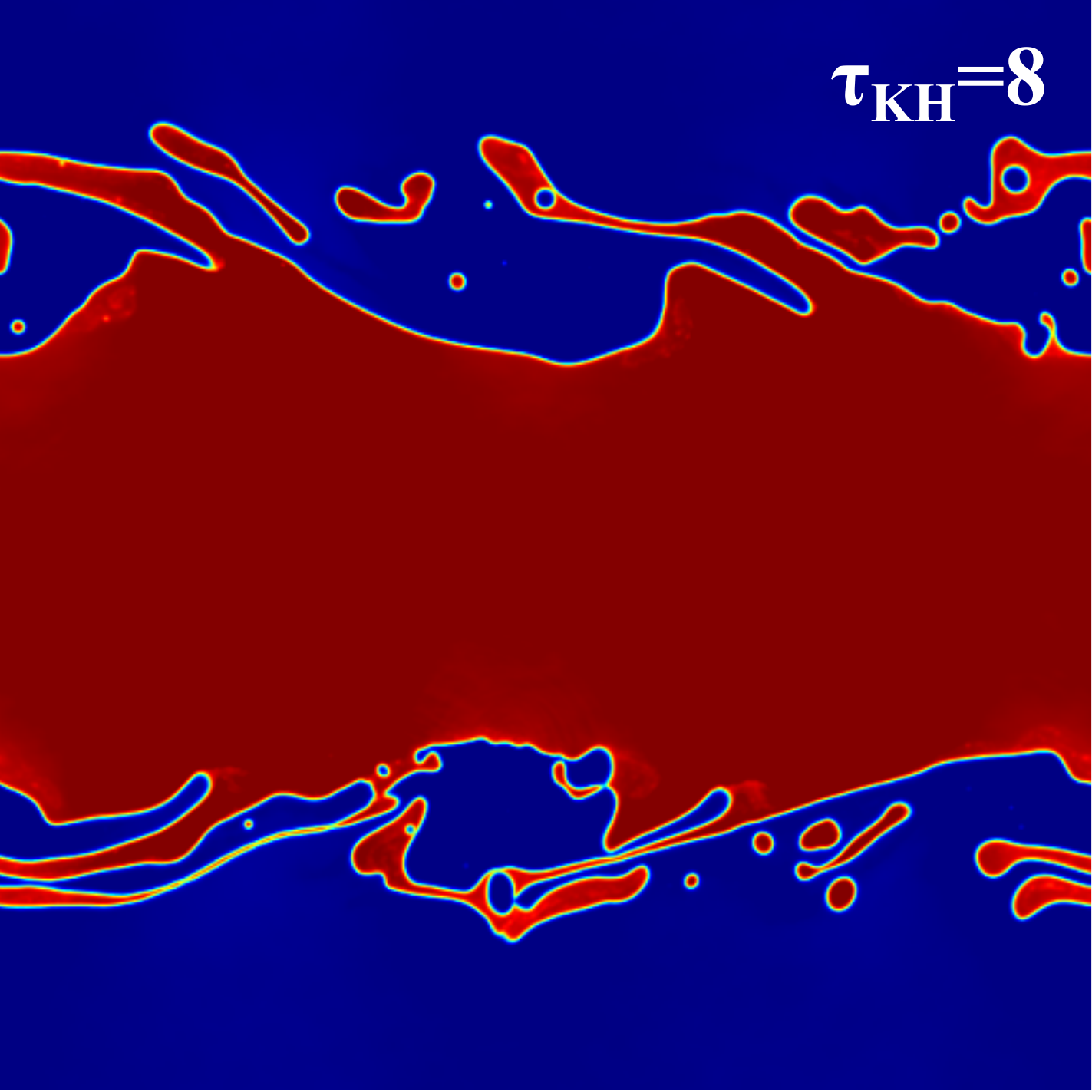}
 \\
   \includegraphics[width=0.2\textwidth]{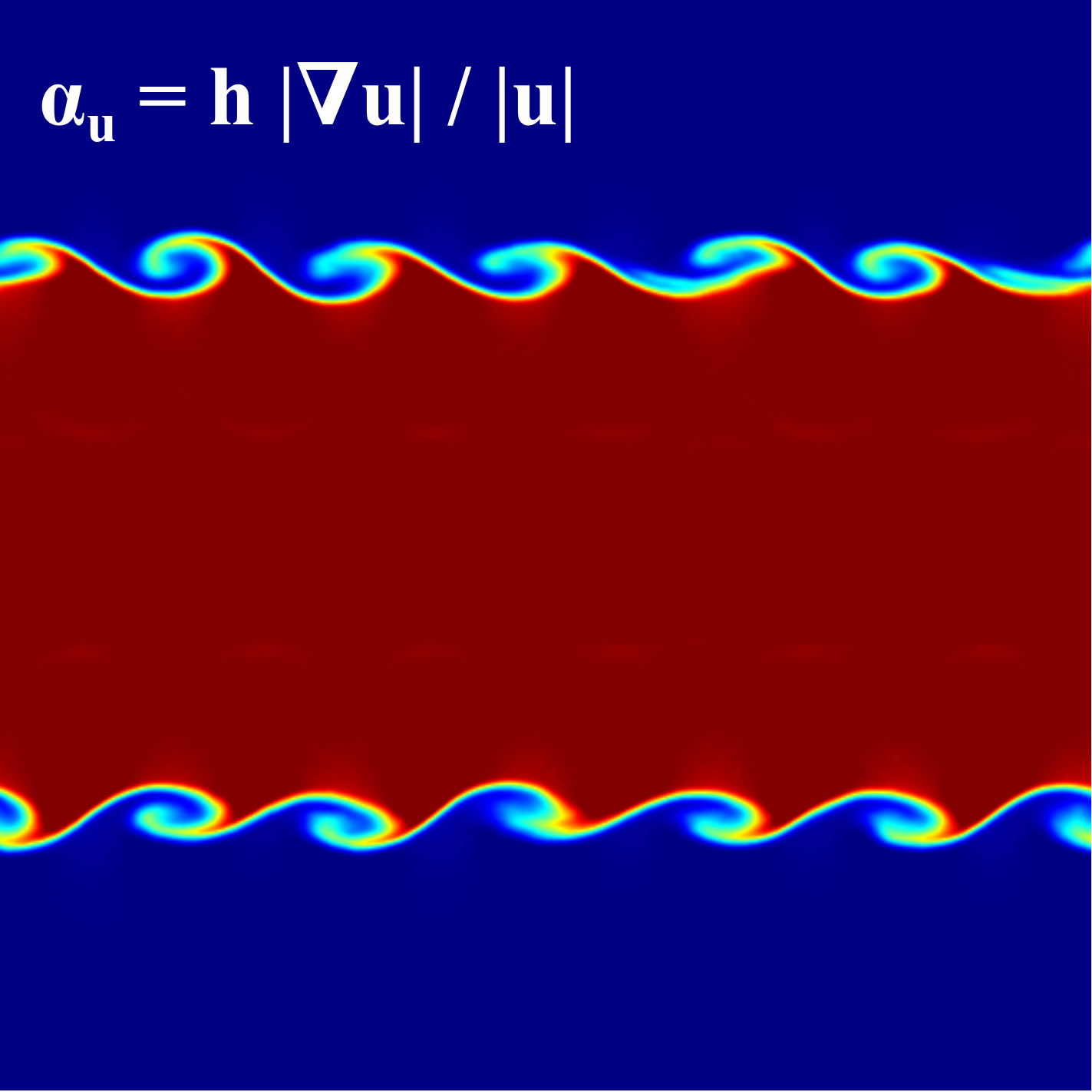}
 & \includegraphics[width=0.2\textwidth]{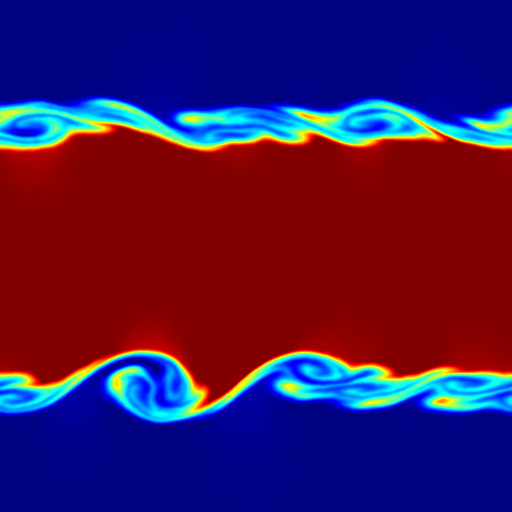}
 & \includegraphics[width=0.2\textwidth]{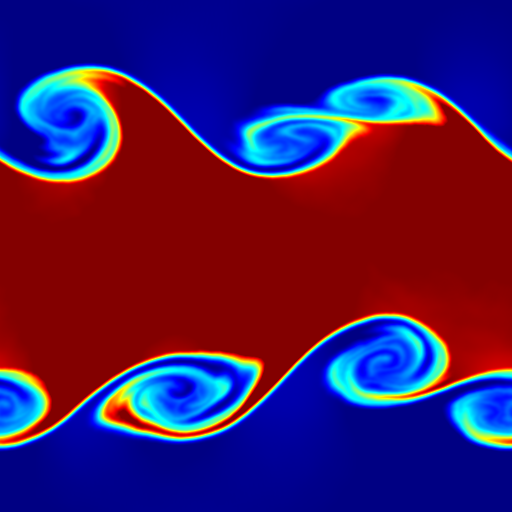}
 & \includegraphics[width=0.2\textwidth]{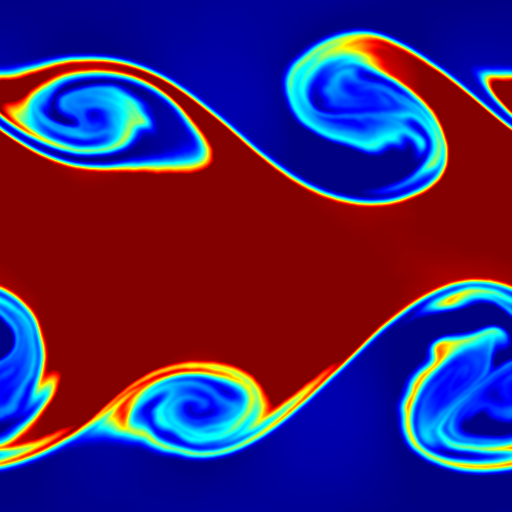}
 \\
   \includegraphics[width=0.2\textwidth]{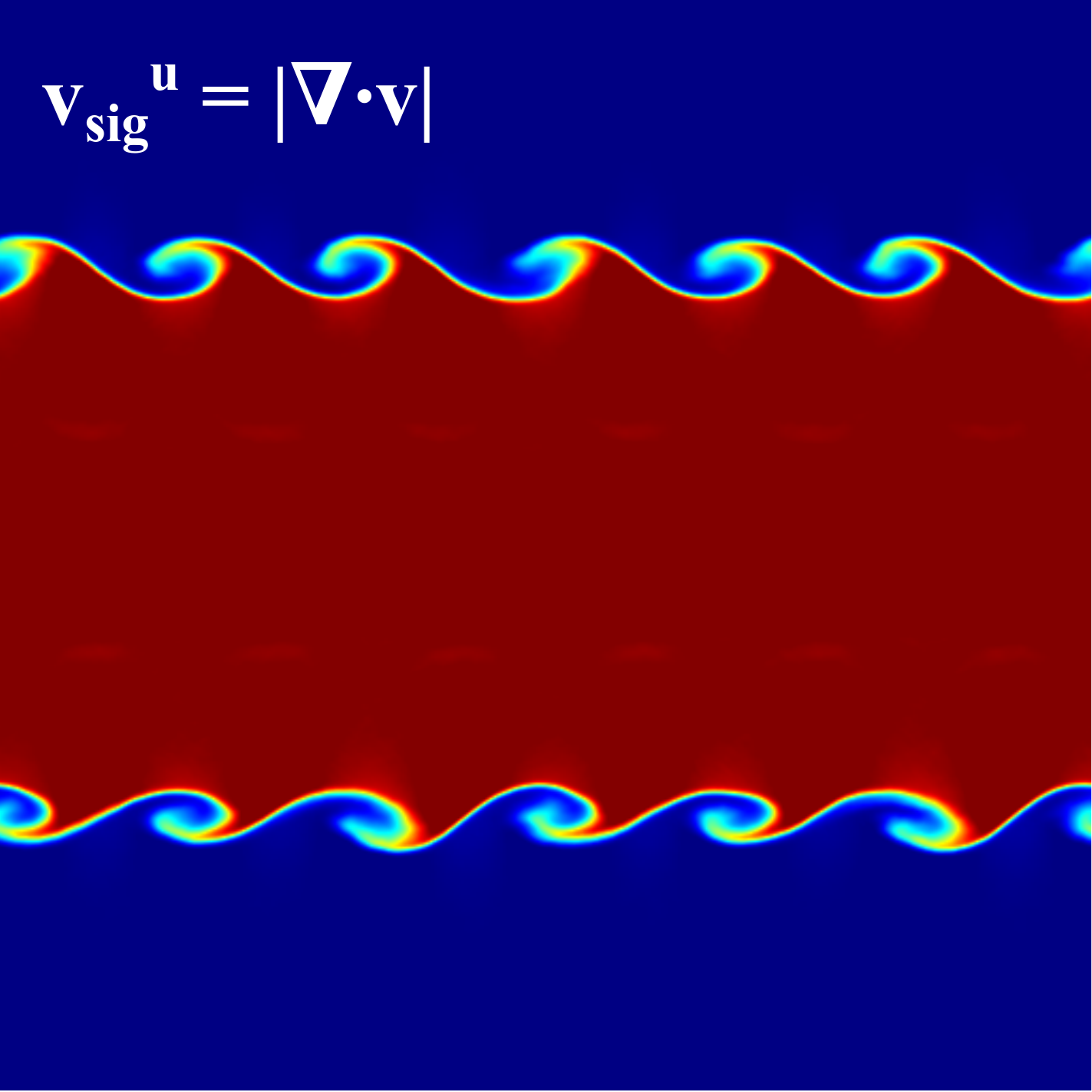}
 & \includegraphics[width=0.2\textwidth]{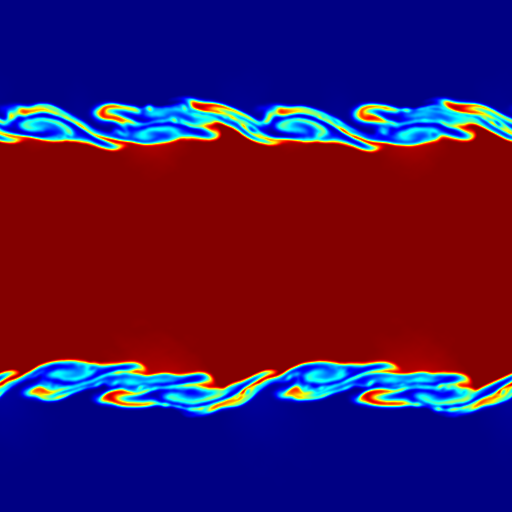}
 & \includegraphics[width=0.2\textwidth]{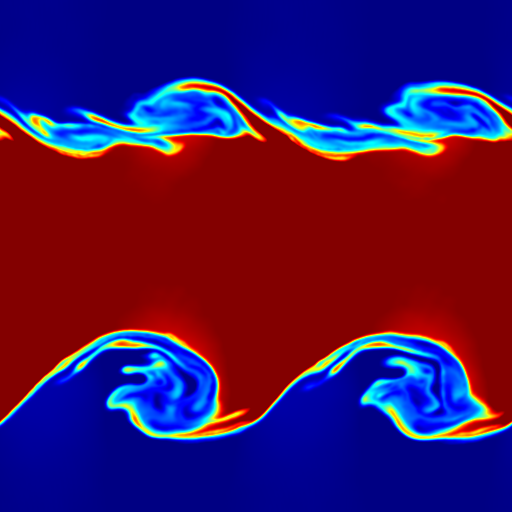}
 & \includegraphics[width=0.2\textwidth]{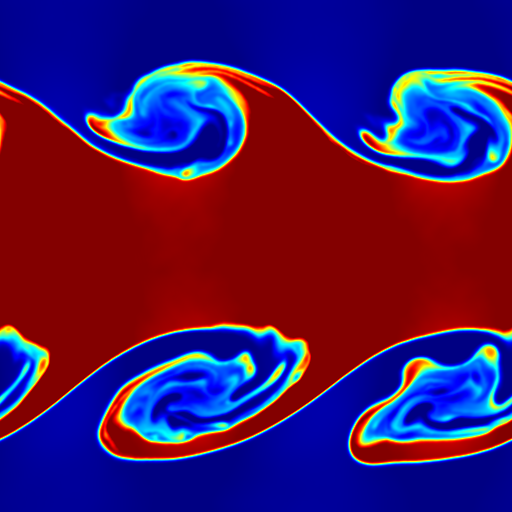}
 \\
   \includegraphics[width=0.2\textwidth]{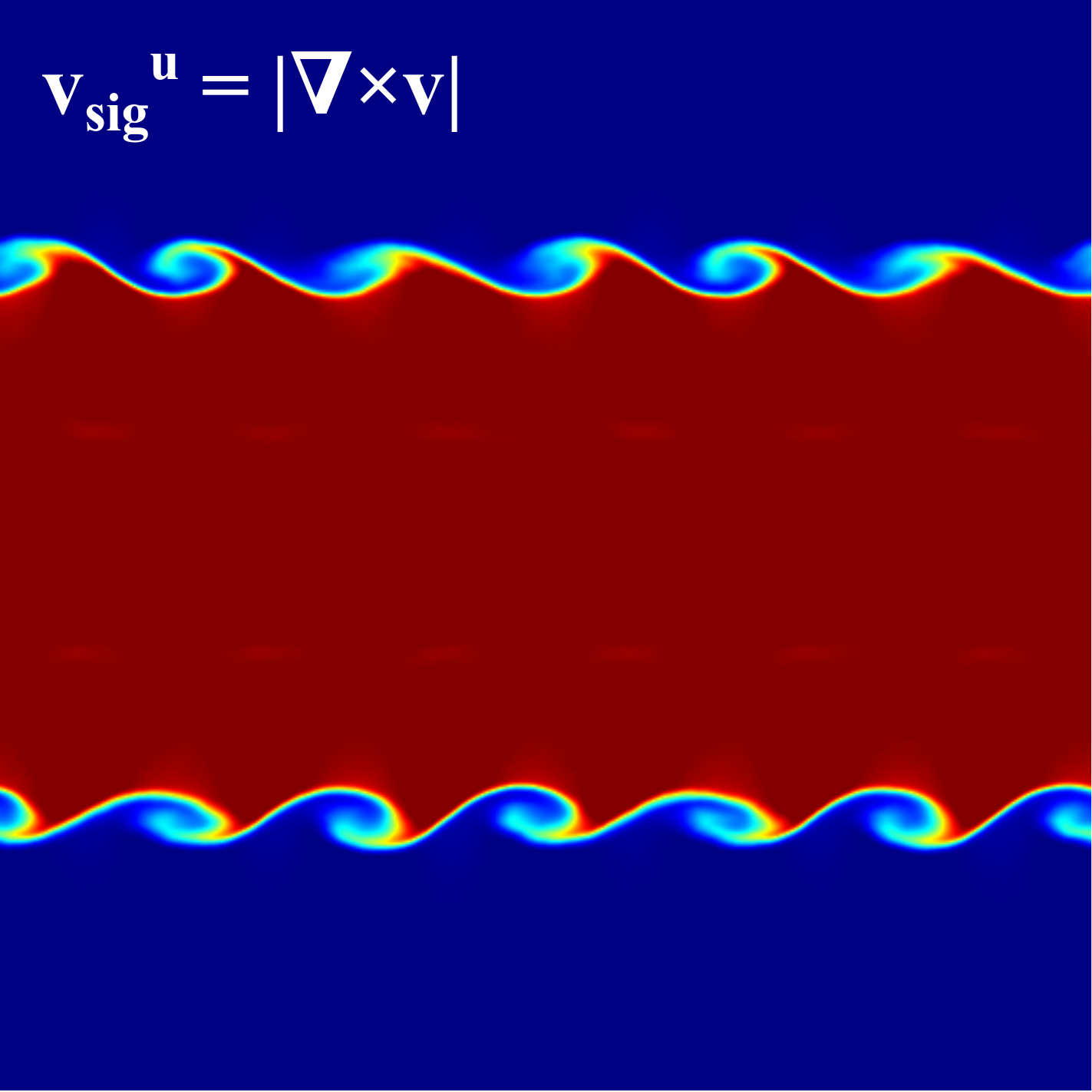}
 & \includegraphics[width=0.2\textwidth]{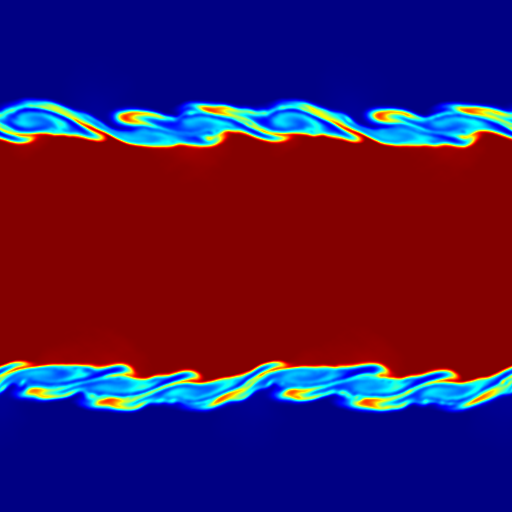}
 & \includegraphics[width=0.2\textwidth]{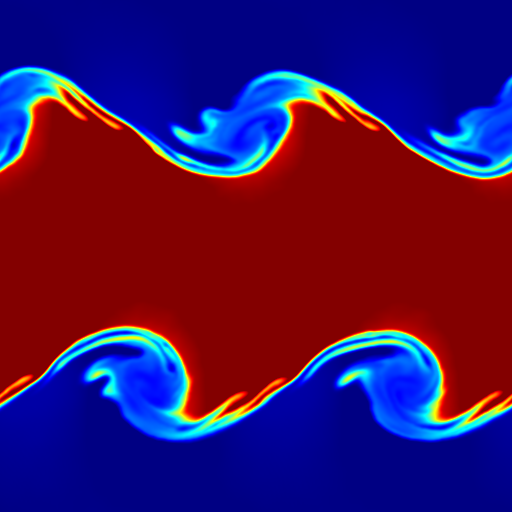}
 & \includegraphics[width=0.2\textwidth]{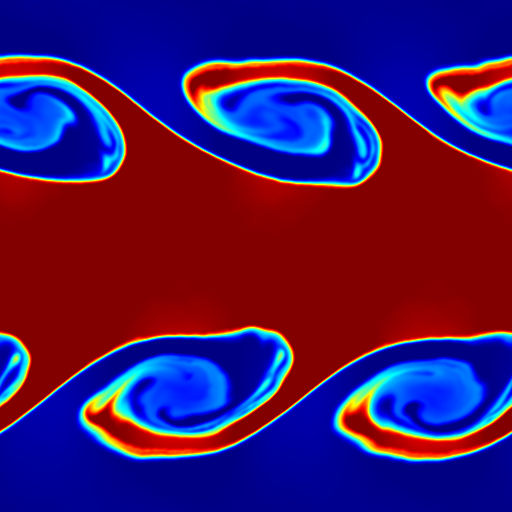}
\\
   \includegraphics[width=0.2\textwidth]{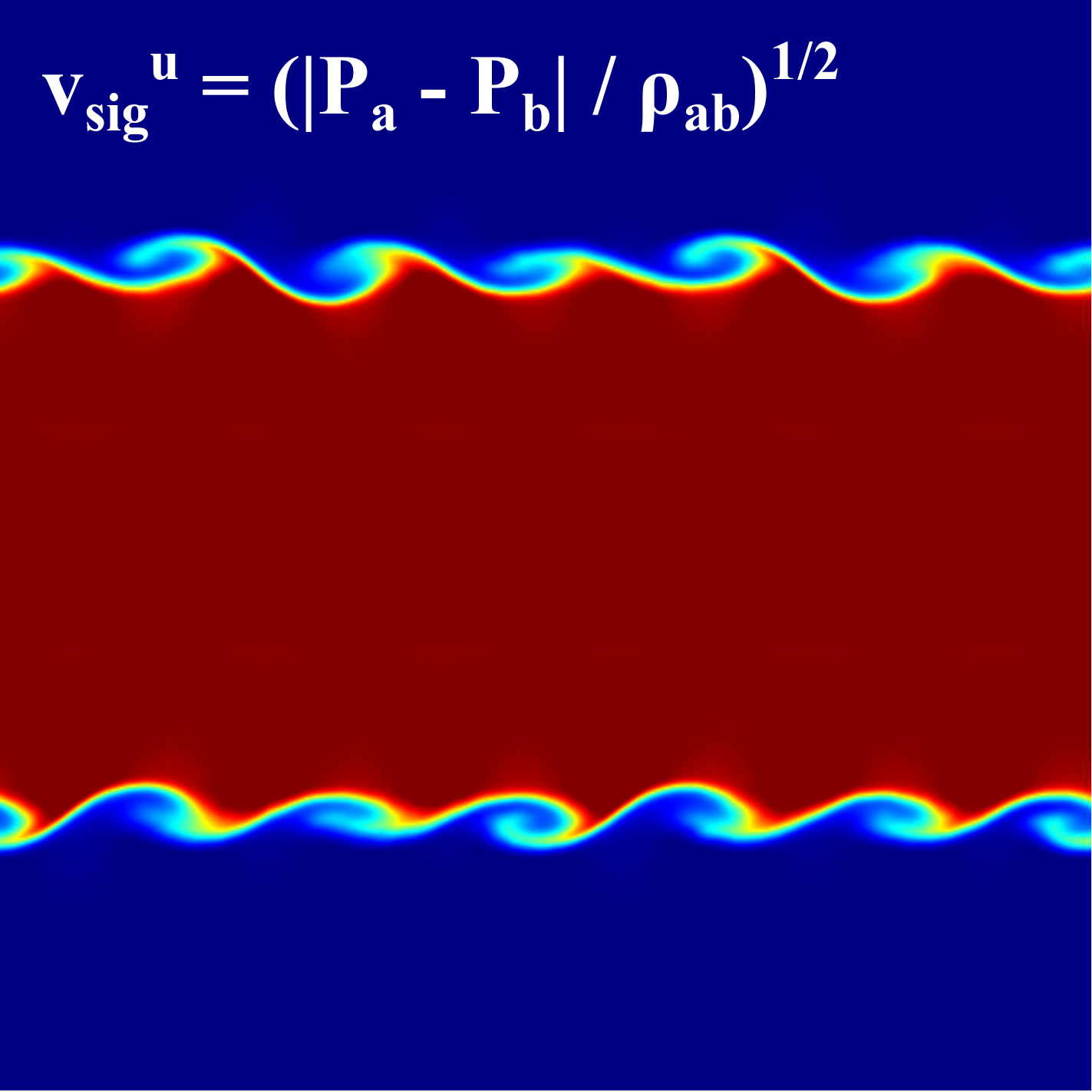}
 & \includegraphics[width=0.2\textwidth]{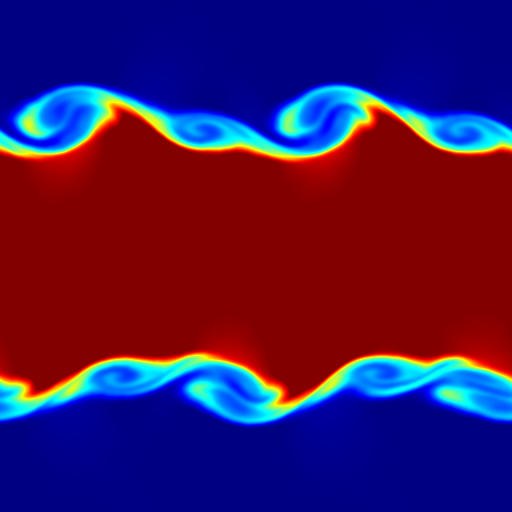}
 & \includegraphics[width=0.2\textwidth]{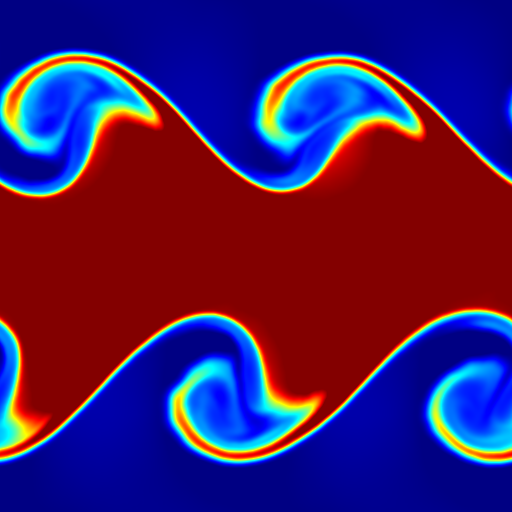}
 & \includegraphics[width=0.2\textwidth]{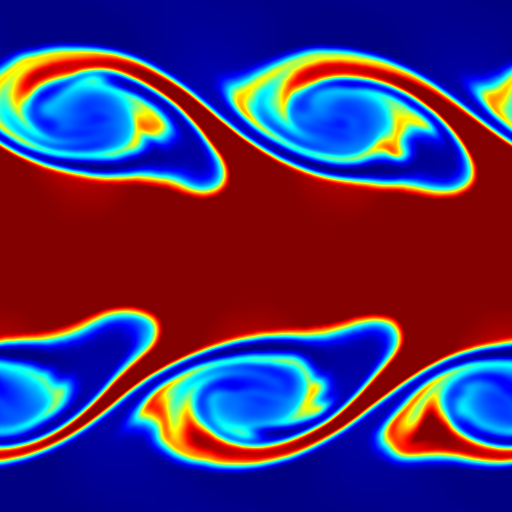}
\\
 \multicolumn{4}{c}{\includegraphics[width=0.803\textwidth]{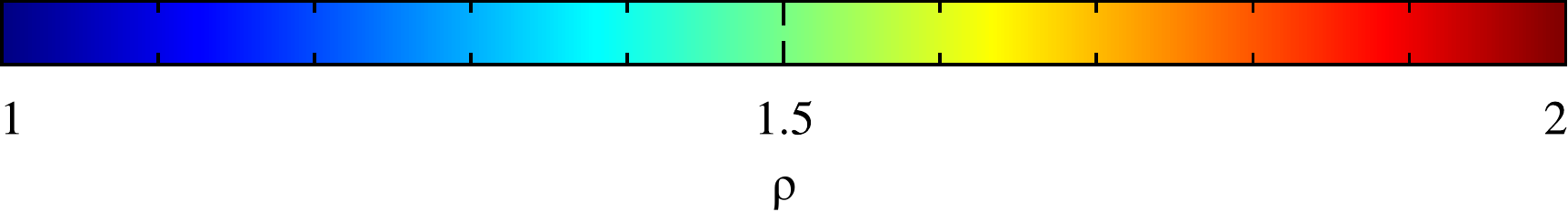}} \\
\end{tabular}

\caption{Results of the Kelvin-Helmholtz instability test using no thermal conductivity (first row), then thermal conductivity with the following switches: $\alpha_u = h \vert \nabla u \vert / \vert u \vert$ (second row), $v_{\rm sig}^u = \vert {\bf v}_{ab} \cdot \hat{\bf r}_{ab} \vert$ (third row), $v_{\rm sig}^u = \vert {\bf v}_{ab} \times \hat{\bf r}_{ab} \vert$ (fourth row), and $v_{\rm sig}^u = \sqrt{ \vert P_a - P_b \vert / \overline{\rho_{ab}} }$ (fifth row). Times are shown for $\tau_\text{KH}=2,4,6,8$ (left to right). Instabilities form along the 2:1 density contrast interfaces due to a velocity shear.  Without thermal conductivity, the discontinuity in thermal energy is untreated, leading to a spurious pressure that prevents mixing of the two regions. All conductivity switches allow the instability to form.}
\label{fig:kh}
\end{figure*}

Figure~\ref{fig:kh} shows the results of the calculations at $\tau_\text{KH}=2, 4, 6, 8$.  The results without thermal conductivity demonstrate its necessity. The spurious pressure across the interface cause globs of the high density fluid to stretch and break off without mixing and the instability fails to correctly develop. All results using thermal conductivity are qualitatively similar --- six small curls are formed along either interface, which coalesce to form two large curls. The interior structure of the large curls differ between the cases, which is to be expected due to the non-linearity of the problem. This initial configuration is known to be unstable at all wavelengths \citep{chandrasekhar61}, thus while the $\lambda = 1/6$ wavelength is seeded and most prominent, secondary instabilities at other wavelengths will occur. 

We find that the \citet{price08} switch leads to the most symmetrical result. The new thermal gradient switch is effective at promoting mixing along the interface, though the curls at $\tau_{\rm KH}=8$ are all of different sizes. The $\vert {\bf v}_{ab} \cdot \hat{\bf r}_{ab} \vert$ is similar in that the curls are dissimilar in shape. In contrast, the $\vert {\bf v}_{ab} \times \hat{\bf r}_{ab} \vert$ has curls which are more uniform and like that of the \citet{price08} switch case. We conclude that using a switch based on the thermal gradient is effective at promoting mixing, but performing further tests, particularly ones that have quantitative convergence, would be desirable. Our results do suggest, though, that a switch based on the curl of the velocity field performs better than one based on the divergence of the velocity.

\section{Summary}
\label{sec:switchsummary}
We have developed a switch to dynamically regulate the amount of artificial resistivity applied to the magnetic field in smoothed particle magnetohydrodynamics simulations \citep[see also][]{tp13}. Since the purpose of artificial resistivity is to model magnetic shocks and discontinuities, the key is to minimise spurious dissipation in smooth parts of the field.  Our switch accomplishes this by setting the artificial resistivity parameter $\alpha_\text{B}$ equal to the dimensionless quantity $h \vert \nabla {\bf B} \vert / \vert {\bf B} \vert$. This yields a simple, powerful and robust method for reducing magnetic dissipation away from shocks with no loss in shock-capturing ability. Importantly, it responds appropriately at all magnetic field strengths, a particular improvement over the PM05 switch which was found to inadequately capture shocks in weak fields.

Alternative switches using the second derivative of the magnetic field were also investigated, in particular $h \vert \nabla^2 {\bf B} \vert / \vert \nabla {\bf B} \vert$ and $h^2 \vert \nabla^2 {\bf B} \vert / \vert {\bf B} \vert$.  The key requirement to their success is a high order estimate of the second derivative, otherwise noise from particle disorder overwhelms the derivative estimate and causes excessive dissipation.  Obtaining this higher order estimate, however, adds significant computational expense.  In general, we recommend our first derivative switch for normal use since it is simple, yet performs robustly and effectively.

Three shocktube tests (Sections~\ref{sec:shock1b}, \ref{sec:shock2a}, and \ref{sec:shock5a}) were used to establish that the switch correctly models a range of shock phenomena, including fast and slow shocks, fast and slow rarefactions, rotational discontinuities, and compound shock structures. The L1 error of the magnetic field profiles for all tests was lower when using this new switch compared to using the \citetalias{pm05} switch. These tests also demonstrated that the our new divergence cleaning algorithm is stable and robust in shocktube problems, in contrast to the version proposed by \citet{sdb13}.

In Section~\ref{sec:polarizedalfven}, the propagation of a travelling Alfv\'en wave was used to gauge the switch's ability to reduce unwanted dissipation in situations not involving discontinuities, and was found to result in maximum $\alpha_\text{B}$ values 10$\times$ smaller than the \citetalias{pm05} switch ($\sim0.02$ compared to $\sim0.22$). After 6 periods, the amplitude of the wave using the \citetalias{pm05} switch was four times lower than using the new switch.
 
The Orszag-Tang vortex was used in Section~\ref{sec:orszag} to examine the performance of the new switch when there are multiple interacting shocks, producing regions of $\alpha_\text{B}\sim 1$ that closely traced the shock lines. The new switch was found to decrease the spurious dissipation in smooth regions compared to the \citetalias{pm05} switch, leading to the subtle magnetic features being more sharply defined, equivalent to running the test at higher resolution.

Finally, in Section~\ref{sec:mhdturb}, a simulation of Mach 10 MHD turbulence was used to demonstrate the switch's ability to capture magnetic shocks when a weak magnetic field is combined with strong hydrodynamic shocks.  The \citetalias{pm05} switch was found to fail for the low field strengths present in this problem, causing the magnetic field to be dominated by unphysical noise. With the new switch the magnetic shocks remain coherent.

We found that it is very important to use the fast MHD wave speed as the characteristic signal velocity for artificial resistivity. Using the Alfv\'en speed as the characteristic signal velocity, as proposed by \citet{price12}, was found to inadequately capture fast MHD shocks in the highly super-Alfv\'enic regime, leading to unphysical effects (Figure~\ref{fig:mhdturb}).

In Section~\ref{sec:switch-generalisation}, the design concept of the switch was generalised for use with artificial viscosity and thermal conductivity \citep[see also][]{tp13b}.  The new artificial viscosity switch needed an integrated decay term to treat post-shock oscillations of particle motion, yielding a switch similar to that of \citet{cd10}.  Performance on a Sod shocktube was in agreement with the Riemann solution.  The thermal conductivity switch was tested with a simulation of the Kelvin-Helmholtz instability.  It was able to mitigate the spurious pressure force across the interface, forming the characteristic billowing curls, however had noticeable asymmetries. Results were compared against other thermal conductivity switches: the \citet{price08} pressure difference formulation, and the divergence and curl of the velocity field. We found that the \citet{price08} and the velocity curl switches performed best for this problem, in that they had the most mixing and retained the symmetry of the problem.

Our new switch is widely applicable to astrophysical SPMHD simulations, in particular for simulations involving weak fields such as in galaxy and cosmological simulations, and also for dynamo processes. In every case we tested, it produced lower magnetic dissipation than the \citetalias{pm05} switch, making it possible to achieve higher magnetic Reynolds numbers in simulations of the interstellar and intergalactic medium. The new switch thus supercedes the \citetalias{pm05} in every respect.

%% file: declaration-chapter5.tex
\newpage
{
\chapter*{}

\vspace{-40mm}

\section*{Declaration for Chapter 5}

\subsection*{Declaration by Candidate}

\noindent In the case of Chapter 5, the nature and extent of my contribution to the work was the following:

\vspace{4mm}

\noindent \begin{tabular}{| >{\raggedright}p{11.55cm} | >{\raggedright}p{3.45cm} |}
\hline
{\bf Nature of Contribution} & {\bf Extent of Contribution (\%)} \tabularnewline
\hline
First author of 2014, ``A comparison between grid and particle methods on small-scale dynamo amplification of magnetic fields in supersonic turbulence'', {\it Submitted to ApJ}.
 & 90 \tabularnewline
\hline
\end{tabular}

\vspace{6mm}
\noindent The following co-authors contributed to the work:

\vspace{4mm}

{
\noindent \begin{tabular}{| >{\raggedright}p{3cm} | >{\raggedright}p{8.1cm} | >{\raggedright}p{3.45cm} |}
\hline 
{\bf Name} & {\bf Nature of Contribution} & {\bf Extent of Contribution (\%) for student co-authors} \tabularnewline
\hline
Daniel Price & Second author, PhD supervisor & \tabularnewline
\hline
Christoph Federrath & Third author, performed grid based calculations & \tabularnewline
\hline
\end{tabular}
}

\vspace{6mm}

\noindent The undersigned hereby certify that the above declaration reflects the nature and extent of the candidate's and co-author's contributions to this work.

% left bottom right top
\noindent \includegraphics[trim=1.2cm 13.1cm 0.6cm 12.9cm, clip, width=\textwidth, angle=-1.2, scale=1.03]{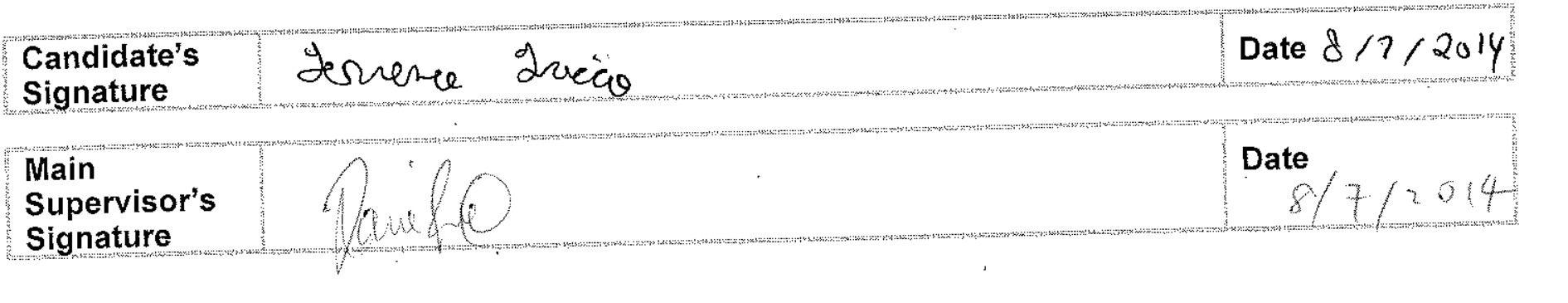}

}

%% file: turbcomp.tex
\chapter{Turbulent dynamo amplification of magnetic fields}
\label{sec:chapter-mhdturb}

%% B differs between the two codes between \mu_0 in SPH=1, but in Flash=4pi
%\section{Introduction}

Supersonic turbulence regulates star formation \citep{mk04,mo07}, producing the dense filaments that permeate molecular clouds along which dense cores and protostars form \citep[e.g.,][]{larson81, hartmann02, es04, hatchelletal05, andreetal10, perettoetal12}. That the turbulence is magnetised cannot be ignored. Magnetic fields are no longer thought to prevent gravitational collapse altogether, but may still determine the rate and efficiency of star formation, even with weak magnetic fields i.e., super-Alfv\'enic turbulence \citep{nl08, lunttilaetal09, pb08, pb09, pn11, fk12}.

Magnetic fields grow in a turbulent environment by the conversion of kinetic energy into magnetic energy. This type of dynamo is small-scale, operating near the dissipation scale. It is there that the smallest motions can efficiently grow the magnetic field through rapid winding and twisting of the magnetic field lines, with the magnetic field as a whole growing exponentially through a reverse cascade from small to large scales \citep[see review by][]{bs05}. The magnetic field will saturate first on small scales due to the back-reaction of the Lorentz force on the turbulent flow, after which it enters a linear or quadratic growth phase, until the field finally reaches saturation on all scales \citep{choetal09, schleicheretal13}. The growth rate is determined by the physical viscosity and magnetic resistivity of the plasma, which can be expressed as dimensionless numbers: the kinematic Reynolds number ($\mathrm{Re}$), the magnetic Reynolds number ($\mathrm{Rm}$), and the ratio of the two, the `magnetic Prandtl number', ${\rm Pm}={\rm Rm}/{\rm Re}$ \citep{schekochihinetal04a, bs05, schoberetal12b, schoberetal12a, bss13}. Using a large set of numerical simulations, \citet{federrathetal11} found that the dynamo growth rate also depends sensitively on the compressibility of the plasma, parameterised by the turbulent Mach number, and is more efficient for turbulence driven by solenoidal (rotational) flows rather than compression.

These processes involve highly non-linear dynamics and complex behaviours, making analytic study difficult (with notable exceptions; \citealt{sg94,gs95}). Furthermore, observations of magnetic fields in molecular clouds are time consuming and only yield field directions in the plane of the sky and magnitudes along the line of sight \citep[e.g.,][]{crutcher99,bourkeetal01,ht05,tc08}. Numerical simulations complement analytics and observations. It is therefore important to compare results from different codes and to establish the conditions under which those results are representative of the physical processes involved. 

There have been several major code comparison projects related to supersonic turbulence. \citet{taskeretal08} compared two grid codes ({\sc Enzo}, {\sc Flash}) and two particle-based codes ({\sc Gadget}2, {\sc Hydra}) on simple test problems involving strong hydrodynamic shocks, finding comparable results when the number of particles were roughly equal to the number of grid cells. \citet{kitsionasetal09} studied decaying, supersonic, hydrodynamic (non-magnetised) turbulence, comparing four grid codes ({\sc Enzo}, {\sc Flash}, {\sc TVD}, {\sc Zeus}) and three particle codes ({\sc Gadget}, {\sc Phantom}, {\sc Vine}). They found similar velocity power spectra and density probability distribution functions (PDF) when the number of resolution elements was comparable, though the particle codes were found to be more dissipative. \citet{kritsuketal11} compared decaying, supersonic turbulence with magnetohydrodynamics (MHD) using nine different grid codes: {\sc Enzo}, {\sc Flash}, {\sc KT-MHD}, {\sc LL-MHD}, {\sc Pluto}, {\sc PPML}, {\sc Ramses}, {\sc Stagger}, and {\sc Zeus}. They found that all methods produced physically consistent results, with the quality of results improved with higher-order numerical solvers, and by exactly rather than approximately maintaining the divergence-free constraint on the magnetic field.

A key shortcoming of both the \citet{kitsionasetal09} and \citet{kritsuketal11} comparisons was that they studied decaying turbulence. Interpolating the initial conditions obtained by driving the turbulence in one code introduced discrepancies between codes before the numerical experiments even started. Those discrepancies in the initial conditions were most severe between grid and particle methods, but also for different grid discretisations (e.g. staggered vs. unstaggered meshes), and is problematic in the MHD case since one must enforce $\nabla \cdot {\bf B}=0$. Furthermore, it is difficult to obtain a statistically significant sample of simulation snapshots in the absence of a statistical steady-state, given that supersonic turbulence decays within a few crossing times. This limitation means that intermittent, intrinsic fluctuations of the turbulence largely exceeded systematic differences in the numerical schemes, which we want to quantify.

\citet{pf10} (hereafter PF10) addressed these issues in a hydrodynamic comparison by using driven instead of decaying turbulence, enabling the calculations to start from a well-defined initial state and allowing time-averaged statistical comparisons. They compared two codes: the grid code, {\sc Flash}, and the smoothed particle hydrodynamics (SPH) code, {\sc Phantom}, taken as broadly representative of the two classes of hydrodynamical methods used in astrophysics: grid-based versus particle-based. Both codes used exactly the same turbulence driving routine and sequence to prevent any bias from different implementation of driving or from different initial conditions: both codes start with a gas of uniform density at rest. They found similar resolution requirements to previous studies, but that grid-based methods were better at resolving volumetric statistics at a given resolution, while SPH better sampled density-weighted quantities. However, this comparison was limited to hydrodynamic turbulence.

{\sc Phantom} was initially entered for the \citet{kritsuketal11} comparison, but was withdrawn\footnote{\textsc{Phantom} was entered for the hydrodynamic comparison, the results of which were not analysed or used in the published paper.} because, at the time, the best approach to maintaining $\nabla\cdot{\bf B} = 0$ in smoothed particle magnetohydrodynamics (SPMHD) used the Euler potentials, ${\bf B} = \nabla \alpha \times \nabla \beta$ \citep{pb07,rp07}. This formulation was incompatible with the initial conditions used in the comparison (see discussion in \citealt{rp07}) and excludes dynamo processes by construction because the Euler potentials method cannot represent and follow wound-up magnetic field structures \citep{pb08, brandenburg10, price12}. Results evolving the magnetic field directly were poor (no $\nabla \cdot {\bf B}$ control). These difficulties have now been resolved. In Chapter~\ref{sec:chapter-cleaning}, we developed a new constrained hyperbolic divergence cleaning method for SPMHD that maintains $\nabla\cdot{\bf B} = 0$ to sufficient accuracy for a wide range of problems, without the topological restrictions associated with the Euler potentials. In particular, this has enabled successful simulations of jets and outflows during protostar formation \citep{ptb12, btp14} which involve winding up of magnetic fields. In Chapter~\ref{sec:chapter-switch}, we have further improved the magnetic shock-capturing algorithm, particularly when dealing with weak magnetic fields. Hence, it is now possible to simulate magnetic dynamos with SPMHD.
 
This chapter presents a comparison between {\sc Flash} and {\sc Phantom} on the direct numerical simulation of small-scale dynamo amplification of a weak magnetic field from driven, supersonic turbulence. We investigate the dependence of the growth rate on the numerical resolution, a measure of the numerical dissipation properties of each code. We also study the statistical properties of super-Alfv\'enic turbulence after the magnetic field has saturated. The comparison is otherwise identical to \citetalias{pf10}, using the same driving routine and Mach number, so that any differences are from the MHD implementations only. In order to capture the growth of the magnetic field and obtain time-averaged statistics after it has saturated, the calculations are evolved for one hundred crossing times, in contrast to only ten in \citetalias{pf10}. This is another motivation for performing driven turbulence calculations, rather than revisit the decaying turbulence simulations of the \citet{kritsuketal11} comparison. Such simulations can only be evolved for a few turbulent crossing times before the turbulence decays.

 In Section~\ref{sec:turbdetails}, we describe the equations to be solved, the initial state, and the driving routine. Section~\ref{sec:codes} introduces the numerical details of {\sc Flash} and {\sc Phantom}. Results of the simulations are analysed in Section~\ref{sec:turbresults}, focusing on the transient phase during which the turbulence is formed (Section~\ref{sec:transientphase}), the growth of magnetic energy from the turbulent dynamo (Section~\ref{sec:growth}), and after the magnetic energy reaches its saturation value (Section~\ref{sec:sat}).  A summary of our results is presented in Section~\ref{sec:turbsummary}.

\section{Comparison details}
\label{sec:turbdetails}

The aim is to compare grid and particle methods on the growth and saturation behaviour of a magnetic field that is amplified by a turbulent dynamo. We solve the ideal MHD equations (Section~\ref{sec:ideal-mhd-summary}). The continuum equations have zero viscous and resistive dissipation (hence ideal). Since the growth rate of the small-scale dynamo is set by the dissipation of the system, it may seem strange to perform a comparison using the ideal MHD equations. However, it is common to use the ideal MHD equations for astrophysical simulations, and some amount of dissipation is inevitably introduced when solving these equations numerically. In grid-based methods, numerical dissipation is introduced by the discretisation of the advection term in the material derivative. By contrast, in Lagrangian particle-based methods the material derivative is computed exactly. The shock-capturing scheme is the other source of numerical dissipation. Modern grid-based methods use Riemann solvers which introduce a numerical dissipation related to the accuracy of the shock reconstruction. The approach in particle methods is to explicitly add viscous and resistive terms in order to capture shocks, using switches to tune the dissipation to the relevant discontinuity. Since the small-scale turbulent dynamo is sensitive to both viscous and resistive dissipation, and the ratio of these two (${\rm Pm}=\nu/\eta$), it can be used to quantify and compare the numerical dissipation between codes.

The initial state is simple so that both codes start from the same initial conditions. The system is initialised with a uniform density field, $\rho_0=1$, and zero velocity field.  The system is contained in a periodic box of length $L = 1$. An isothermal equation of state, $P~=~c_{\rm s}^2 \rho$, is used to calculate the pressure with sound speed $c_{\rm s} = 1$. The magnetic field is set to $\sqrt{2}\times 10^{-5}$ in the $z$~-~direction. With $\mu_0=1$, this yields an initial plasma beta, the ratio of thermal to magnetic pressure, of $\beta=P/P_{\rm mag}=10^{10}$.  

As in \citetalias{pf10}, supersonic turbulence is initiated and sustained at a root mean square (rms) Mach number of $\mathcal{M} = 10$ by an imposed driving force generated from an Ornstein-Uhlenbeck process \citep{ep88, schmidtetal09, federrathetal10}. This is a stochastic process with a finite autocorrelation timescale. By using this approach, the driving force can be decomposed in Fourier space into longitudinal and solenoidal modes. For this comparison, only the solenoidal component of the force is used, and therefore the turbulence is driven primarily by vorticity rather than compression (see \citealt{federrathetal11, federrath13} for a study on the effect of different driving). However, 1/3 of the kinetic energy will still be contained in compressive modes due to the high Mach number of the turbulence \citep{ps10, federrathetal10}. 

Consistency of the driving pattern between codes is achieved by pre-generating the time-sequence of the Ornstein-Uhlenbeck modes, with both codes reading the pattern from file. The driving is at large scales, with a parabolic weighting of modes between $k_{\rm min} = 1$ and $k_{\rm max} = 3$, with smaller structures forming through turbulent cascade. The autocorrelation timescale is $1t_{\rm c}$, with $t_{\rm c}$ as defined in Equation~\ref{eq:turbtimescale}. The input parameters used to generate the pattern file are specified in Table~\ref{tbl:stir}. The stirring energy is used to obtain the variance of the Ornstein-Uhlenbeck process, corresponding to the autocorrelation time and energy input rate.

\begin{table}
\caption{Stirring Routine Input Parameters.}
\label{tbl:stir}
\centering
\begin{tabular}{ll}
\hline
\hline
Parameter & Value \\
\hline
spectral form & 1 (Parabola) \\
solenoidal weight & 1 \\
%st\_solweightnorm & $\sqrt{3/2}$ \\
stirring energy & 8.0 \\
autocorrelation time & 0.05 \\
minimum wavenumber & 6.28 \\
maximum wavenumber & 18.90 \\
original random seed & 1 \\
\hline
\end{tabular}
\end{table}

The relevant physical timescale is the turbulent crossing time, which we define according to
\begin{equation}
t_{\rm c} \equiv \frac{L}{2 \mathcal{M} c_{\rm s}},
\label{eq:turbtimescale}
\end{equation}
corresponding to $t_{\rm c} = 0.05$ in code units. The turbulence is simulated for 100 crossing times, covering the full growth phase of the dynamo up until the magnetic energy reaches its saturation level, with at least half of the total time spent in the saturation phase.  The set of simulations use $128^3$, $256^3$, and $512^3$ resolution elements (grid points and particles, respectively).

\section{Numerical codes and methods}
\label{sec:codes}

We compare the codes {\sc Flash} and {\sc Phantom}. Both solve the ideal MHD equations but with fundamentally different numerical approaches: {\sc Flash} discretises all fluid variables into fixed grid points, whereas {\sc Phantom} discretises the mass of the fluid into a set of Lagrangian particles that move with the fluid velocity. We take these two codes to be representative of the general class of Eulerian, grid-based methods ({\sc Flash}) and Lagrangian, particle-based methods ({\sc Phantom}).  

\subsection{{\sc Flash}}
\label{sec:flash}

{\sc Flash} is a grid-based code using a finite volume scheme for solving the MHD equations \citep{fryxelletal00, dubeyetal08}. Although {\sc Flash} can be used with adaptive mesh refinement \citep[AMR,][]{bc89}, our simulations employ a fixed and uniform cartesian grid for simplicity. We here use {\sc Flash} with the HLL3R approximate Riemann solver for ideal MHD, based on a MUSCL-Hancock scheme \citep*{wfk11}. This is a predictor-corrector scheme and is second-order accurate in both space and time. \citet{wfk11} further show that this MHD scheme maintains $\nabla\cdot{\bf B}\sim0$ to within negligible errors, by using divergence cleaning in the form of the parabolic cleaning method of \citet{marder87} \citep[see also][]{dedneretal02}. The MHD solver is particularly efficient and robust, because it uses a relaxation technique that guarantees positive density and gas pressure and thus avoids unphysical states, by construction.

\begin{figure}
\centering
 \includegraphics[width=0.9\textwidth]{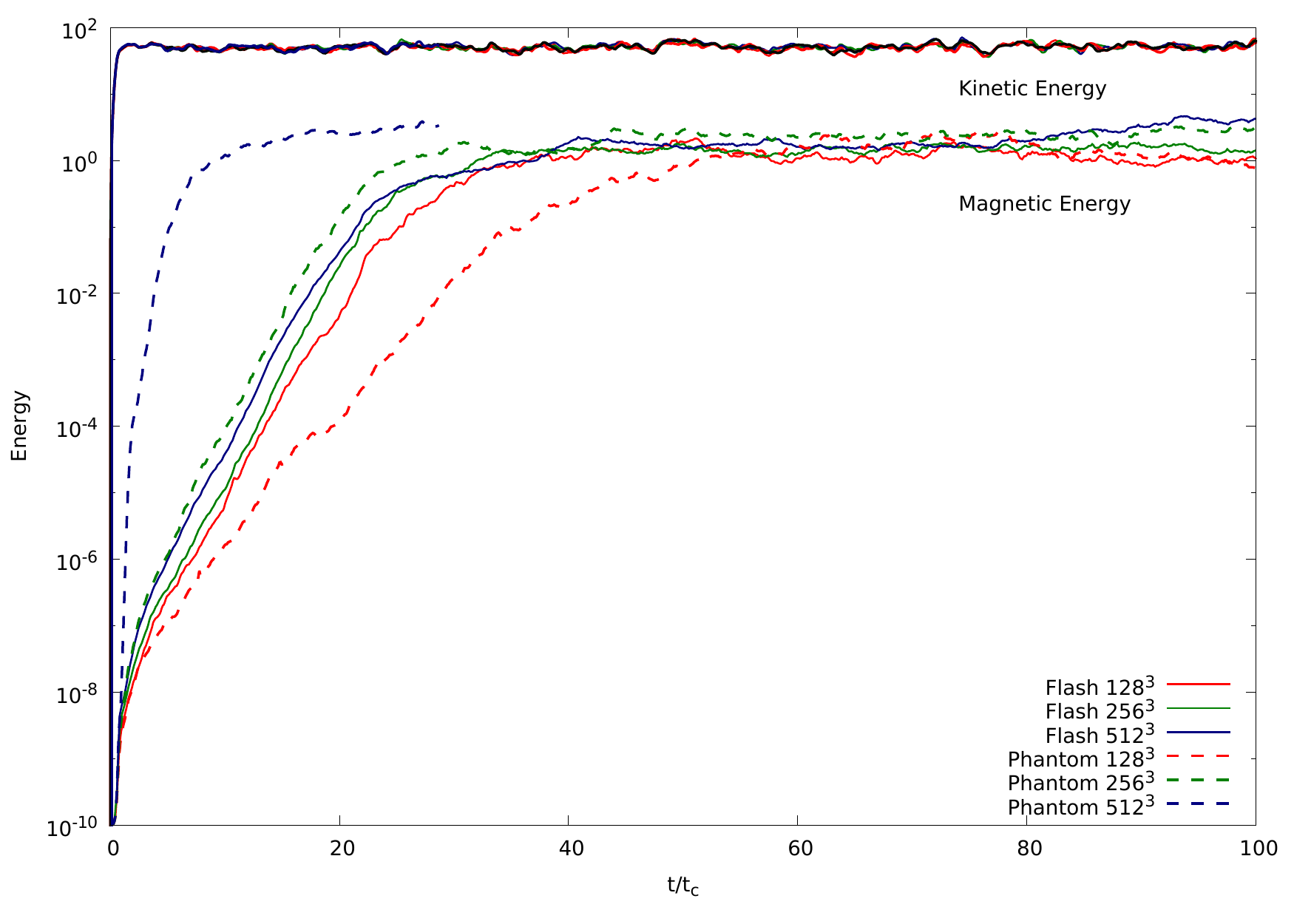}
\caption{Growth and saturation of the magnetic energy for {\sc Flash} and {\sc Phantom} at resolutions of $128^3$, $256^3$, and $512^3$. The top lines are the kinetic energy for the six calculations. {\sc Flash} has similar growth rates across the resolutions simulated, while {\sc Phantom} exhibits faster growth rates with increasing resolution. This resolution dependence is a consequence of the artificial dissipation terms. Both codes saturate the magnetic energy at similar levels.}\label{fig:en_mag}
\end{figure}

\subsection{{\sc Phantom}}
\label{sec:phantom}

{\sc Phantom} is a smoothed particle magnetohydrodynamics (SPMHD) code. The MHD equations (Equations \ref{eq:mhdcty}--\ref{eq:induction}) are implemented as described in \citet{pm04a, pm04b, pm05} and \citet{price12}, using \citet*{bot01}'s method of subtracting ${\bf B} (\nabla \cdot {\bf B})$ from the momentum equation to keep the magnetic tensional force stable. This implementation of momentum and induction equations resolves issues related to non-zero $\nabla \cdot {\bf B}$ in a manner that is equivalent to the Powell 8-wave approach \citep{powell94, powelletal99}. 

The divergence of the magnetic field is kept close to zero by using the constrained hyperbolic divergence cleaning method developed in Chapter~\ref{sec:chapter-cleaning}, which is an SPMHD adaptation and improvement of the cleaning algorithm by \citet{dedneretal02}. The cleaning wave speed is set to the local fast MHD wave speed. During the course of this work, it was found that if the wave speed included the term involving the relative velocity of particles (as in the artificial viscosity), then the individual timestepping scheme could introduce significant errors to the magnetic field. This occurred when particles were interacting on timestep bins that were spaced too far apart. Using a timestep limiter \citep[i.e.,][]{sm09} can prevent these errors, but for these calculations we instead chose the simpler solution of just reducing the cleaning speed by excluding the relative velocity.

Shocks are captured by adding an artificial viscosity, as described by \citet{pm04a, pm05} and based on the \citet{monaghan97} formulation. It is important that the signal velocity, defining the characteristic speed of information propagation, include a term involving the relative motion of particles to prevent particle interpenetration, and it was found by \citet{pf10} that for Mach 10 shocks, it was necessary to increase this by setting the dimensionless constant $\beta_{\rm AV}=4$ (as opposed to the common $\beta_{\rm AV}=2$). We use the \citet{mm97} switch to reduce dissipation away from shocks.

Discontinuities in the magnetic field are treated with an artificial resistivity. {\sc Phantom} uses the new switch developed in Chapter~\ref{sec:chapter-switch} to reduce dissipation of the magnetic field away from discontinuities. This switch solves problems with the switch proposed by \citet{pm05}, namely that it is able to capture shocks when the sound speed is significantly higher than the Alfv\'en speed (i.e., in the super-Alfv\'enic regime when the magnetic field is very weak).  This is done by using the dimensionless quantity $h\vert \nabla {\bf B} \vert / \vert {\bf B} \vert$, which measures the relative strength of the discontinuity in the magnetic field. 

The smoothing length (resolution length), $h$, of each particle is calculated in the usual manner by iteration of the density summation with $h = 1.2 (m / \rho)^{1/3}$ using a Newton-Raphson solver (Section~\ref{sec:spmhd-h}). This means that the numerical resolution scales with the density. For these set of simulations, the resolution increases by $4$--$8\times$ in the highest density regions, with a decrease in resolution of about $2\times$ in the lowest density regions.  Timesteps are set individual to each particle in a scheme that is block hierarchical in powers of two, with each particle setting its timestep based on its local Courant condition.  Second order leapfrog time integration is used (Section~\ref{sec:leapfrog}).

\subsection{Computational cost}

% Flash timings:
%
% simulation terry/mhdcodecomp/MHD128M10sIdeal/ (which is the 128^3, I believe):
% run on 8 cores and took 11.3 hours, i.e., 9.06e1 cpu-hours; n_timesteps=10842
% (run until t=5.0)
% 
% simulation terry/mhdcodecomp/MHD256M10sIdeal/ (which is the 256^3, I believe):
% run on 64 cores and took 25.2 hours, i.e., 1.61e3 cpu-hours, n_timesteps=23696
% (run until t=5.0)
% 
% simulation terry/mhdcodecomp/MHD512M10sIdeal/ (which is the 512^3, I believe):
% run on 512 cores and took 79.6 hours, i.e., 4.07e4 cpu-hours, n_timesteps=73887
% (run until t=5.0)

% Phantom timings:
%
% 128r: 2.445e5 s (67.9 hours)
%       9.66e6 s (2683 cpu-hours)
%
% 256r: 3.8253e6 s (1062.6 hours)
%       1,579e8 s (43 861 cpu-hours)
%
% 512r: guesstimating that each run took 4.85e6 s cpu time, ~93 runs to t=20

The {\sc Flash} calculations used $90$, $1600$, and $40~000$ cpu-hours for the $128^3$, $256^3$, and $512^3$ simulations. The {\sc Phantom} calculations used $2700$ and $44~000$ cpu-hours for the $128^3$ and $256^3$ simulations, with the $512^3$ calculation using $125~000$ cpu-hours for $t=0\to20$. It is expected that each factor of 2 increase in resolution should increase the computational expense by $16\times$, since there are $8\times$ more resolution elements and the Courant condition should reduce the timestep by half, and both codes exhibit a scaling behaviour that is close to this. For {\sc Phantom}, the particles are spread over $\sim6$, $7$, and $8$ individual timestep bins for the $128^3$, $256^3$, and $512^3$ resolution calculations, respectively.  Approximately $35\%$ of the computational expense in the {\sc Phantom} calculations is spent on neighbour finding. The driving routine adds negligible computational expense ($\sim2\%$ of overall cpu-hours). As in \citetalias{pf10}, we find that the $256^3$ {\sc Phantom} calculation takes approximately an equivalent amount of computational time as the $512^3$ {\sc Flash} calculation.

\subsection{Analysis methods}

\subsubsection{Power spectra}

Power spectra are calculated using the same analysis tool for both codes to ensure that results are comparable. The {\sc Flash} data is directly analysed with this tool, while the power spectrum of {\sc Phantom} data is obtained by interpolating the particles to a grid of double the particle resolution (i.e., $256^3$ particles are interpolated to $512^3$ grid points). A higher resolution grid is chosen in order to represent the energy contained in the highest density structures, which are up to $4$--$8\times$ higher than the initial resolution. Appendix~\ref{sec:gridinterp} investigates the effect of the resolution of the interpolated grid, in addition to the difference between mass and volume weighted interpolation. We found that the magnetic field was satisfactorily represented by a grid which has twice the resolution of the particle calculation.

\subsubsection{Probability distribution functions}
Computing a volume-weighted PDF from grid methods simply involves binning the cells according to the value of the quantity and normalising such that the integral under the PDF is unity. For SPH this is more complicated since the resolution is tied to the mass rather than the volume. \citetalias{pf10} computed the PDF directly from the SPH particles by weighting the contribution of each particle, $i$, by the volume element $m_{i}/\rho_{i}$. \citet*{pfb11} later found that this was inaccurate at high Mach number because $\sum_{i} m_{i}/\rho_{i}$ has no requirement that it equals the total volume. Instead, one should interpolate to a fixed volume using the SPH kernel $W$, since $m_{i}/\rho_{i}$ is only meaningful when multiplied by the kernel (since SPH is derived assuming $\sum_{i} m_{i}/\rho_{i} W({\bf r} - {\bf r}_{i}, h) = 1$). However, interpolation to a fixed grid \citep[e.g.][]{kitsionasetal09} is also problematic since the resolution in our simulations is 4--8$\times$ higher in the densest regions compared to a fixed grid with the same number of resolution elements. Hence, sampling the high density tail of the SPH calculation would require a commensurably high resolution grid. We follow \citet{pfb11} in using an adaptive mesh to compute the PDF from the SPH particles, where the mesh is refined until the cell size is smaller than the smoothing length. The SPH PDF is then computed and normalised directly from this adaptive mesh.

%%
%% medium square page size
%% 3x1 tiling on
%% character height of 1.1
%% 0.94 length colour bar
%% 1.6 character height width colorbar

\begin{figure}
\centering
\setlength{\tabcolsep}{0.0025\textwidth}
\renewcommand{\arraystretch}{1.0}
 \begin{tabular}{ccc}
 \scriptsize{\sc $128^3$} & \scriptsize{\sc $256^3$} & \scriptsize{\sc $512^3$} \\
  \includegraphics[width=0.16\textwidth]{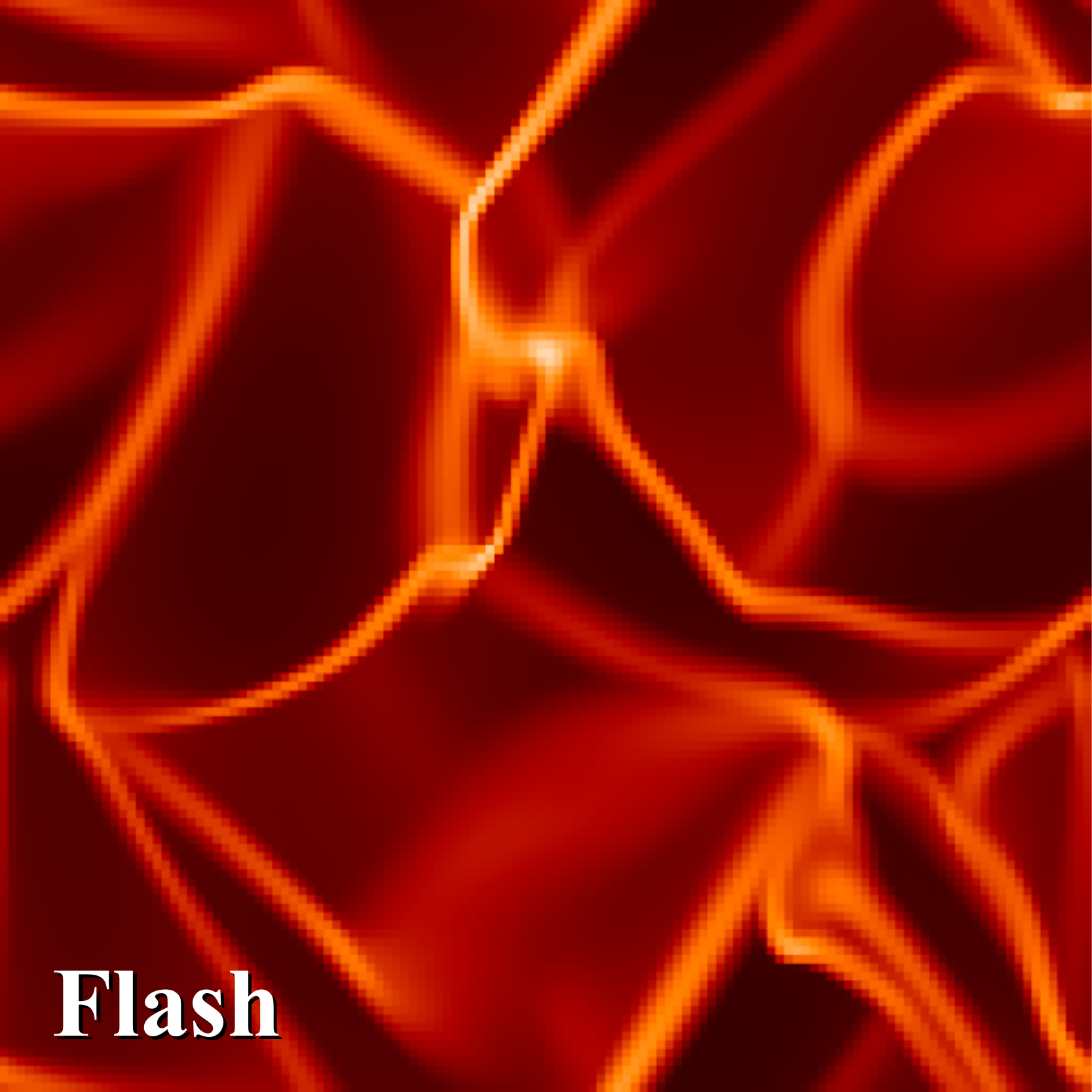} 
& \includegraphics[width=0.16\textwidth]{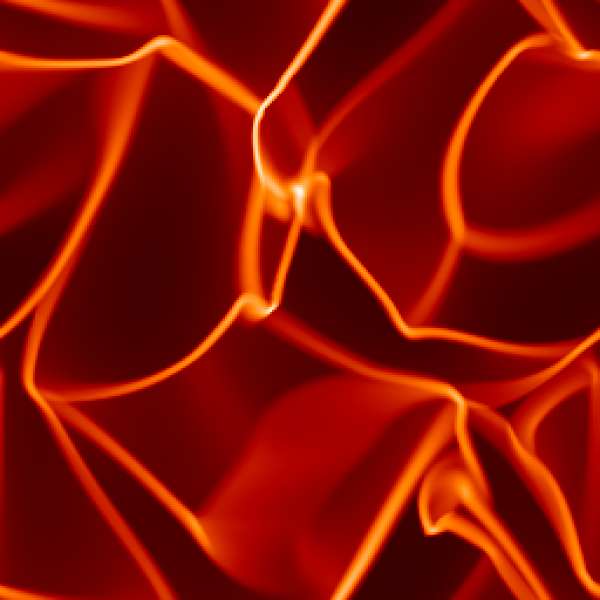} 
& \includegraphics[width=0.16\textwidth]{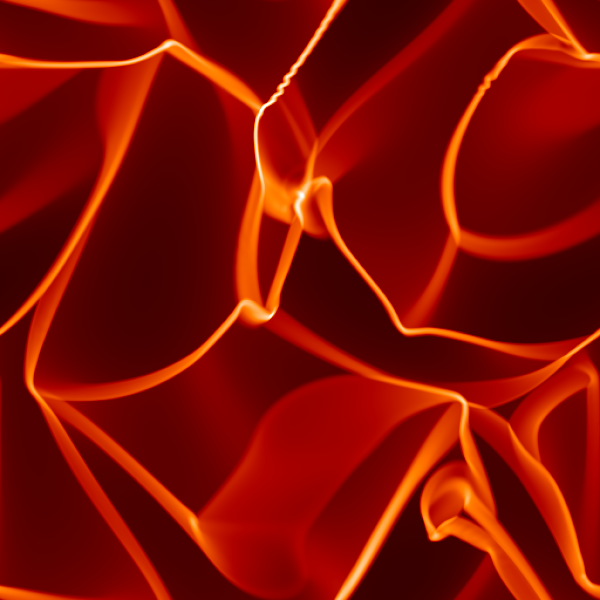} \\ 
  \includegraphics[width=0.16\textwidth]{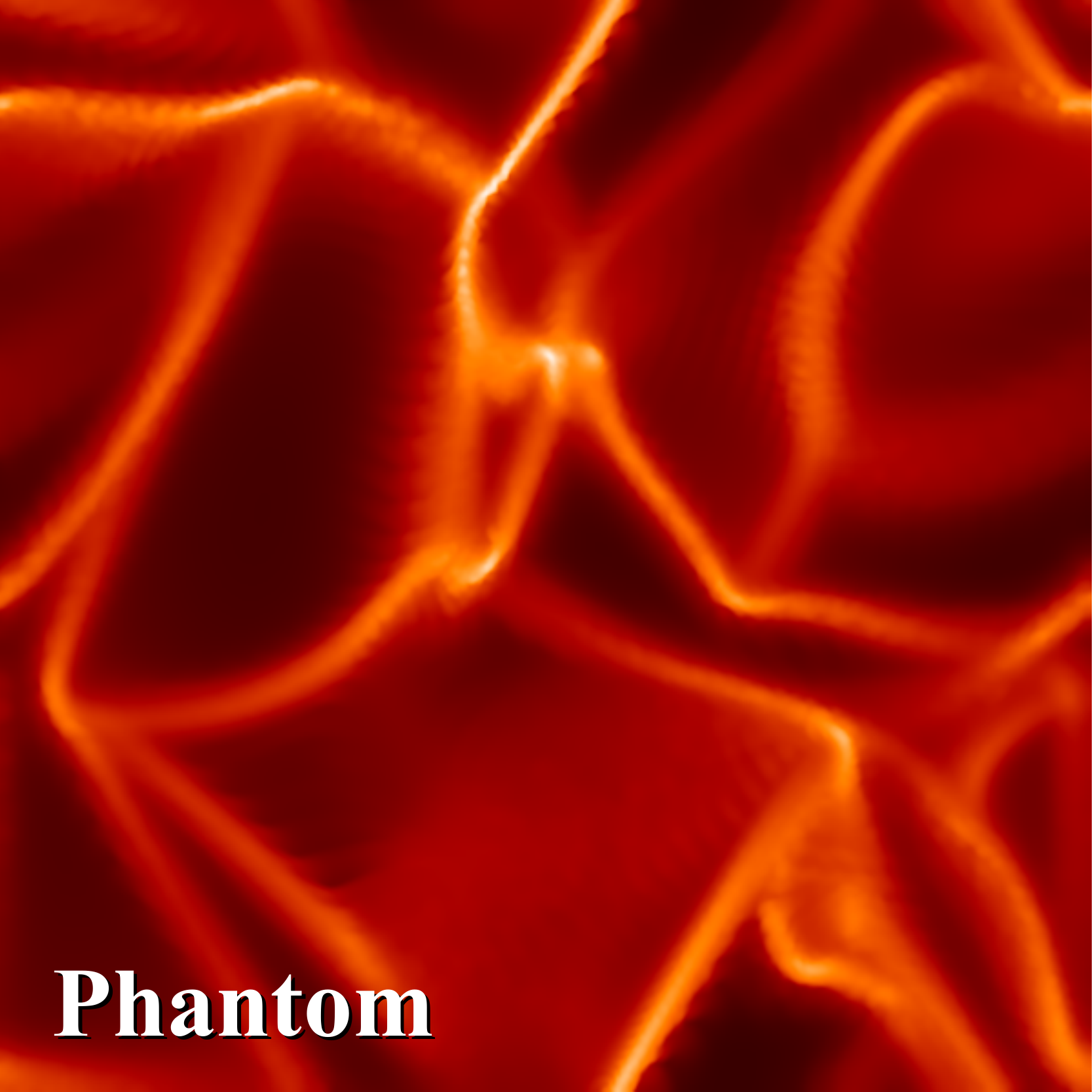} 
& \includegraphics[width=0.16\textwidth]{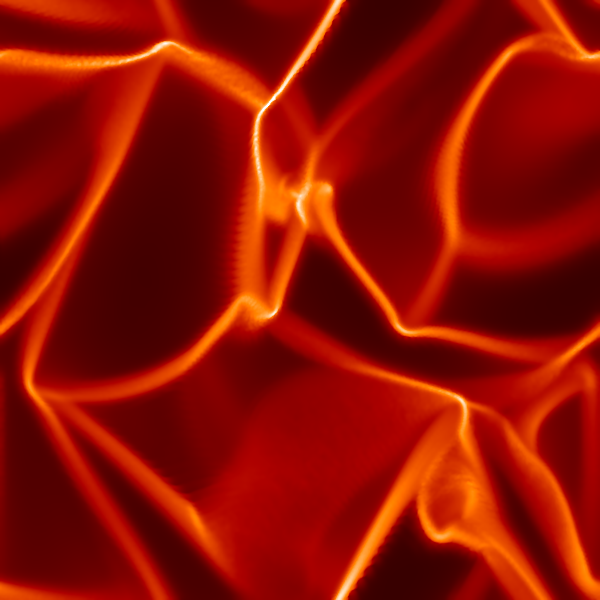} 
& \includegraphics[width=0.16\textwidth]{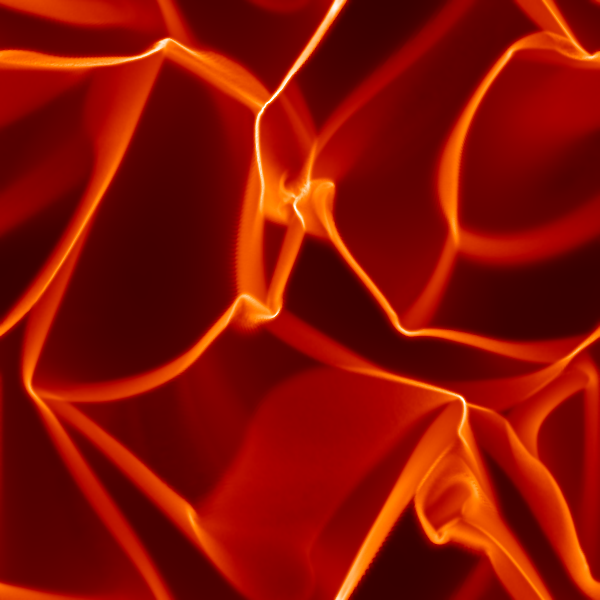} 
\end{tabular}
 \begin{tabular}{ccc}
 \scriptsize{\sc $128^3$} & \scriptsize{\sc $256^3$} & \scriptsize{\sc $512^3$} \\
  \includegraphics[height=0.16\textwidth]{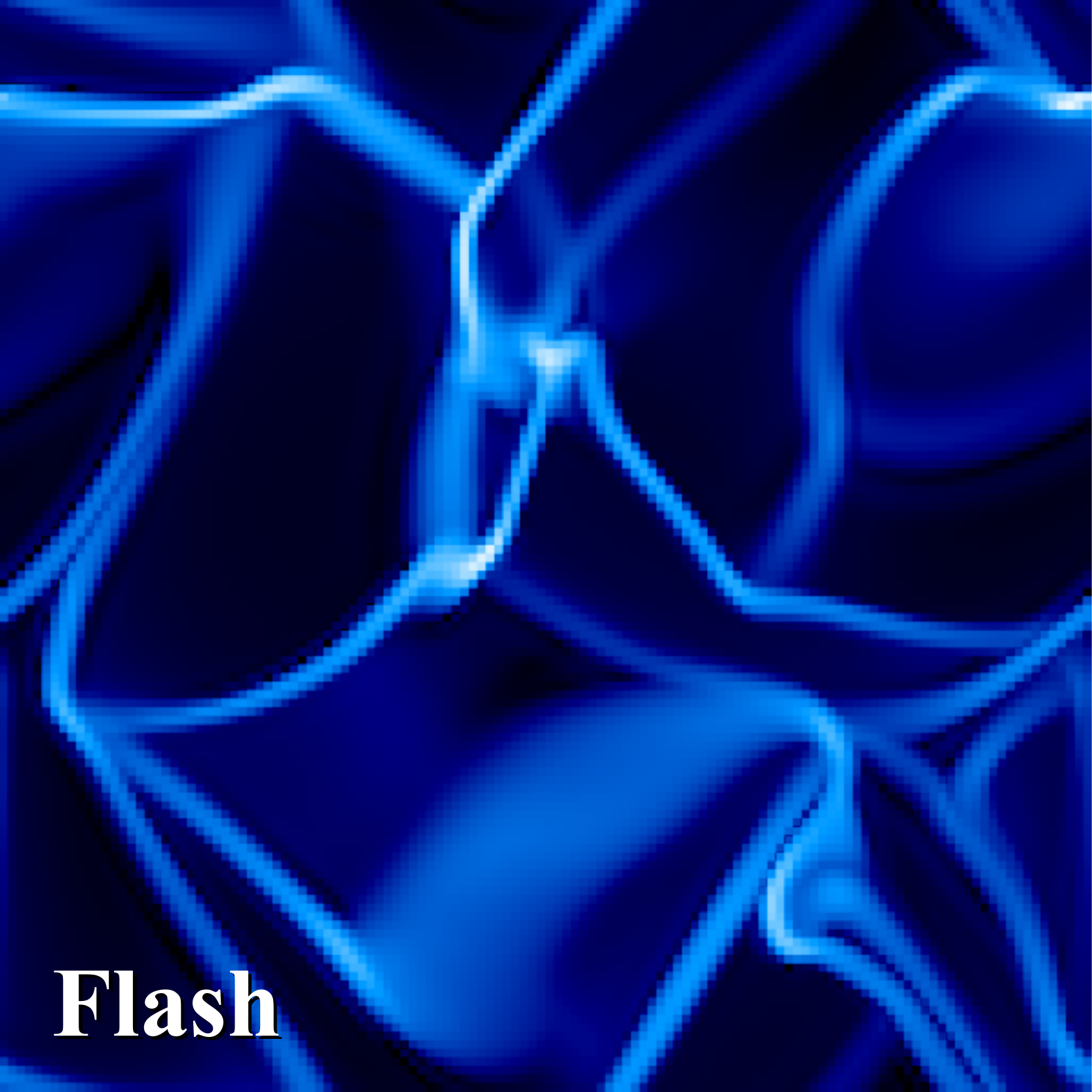} 
& \includegraphics[height=0.16\textwidth]{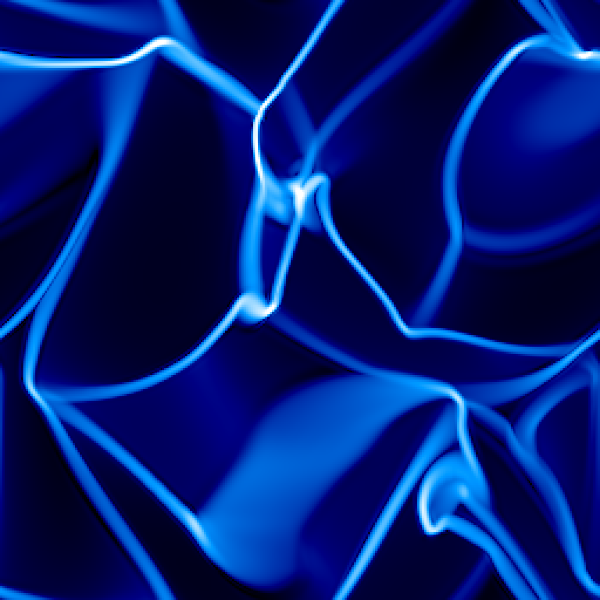} 
& \includegraphics[height=0.16\textwidth]{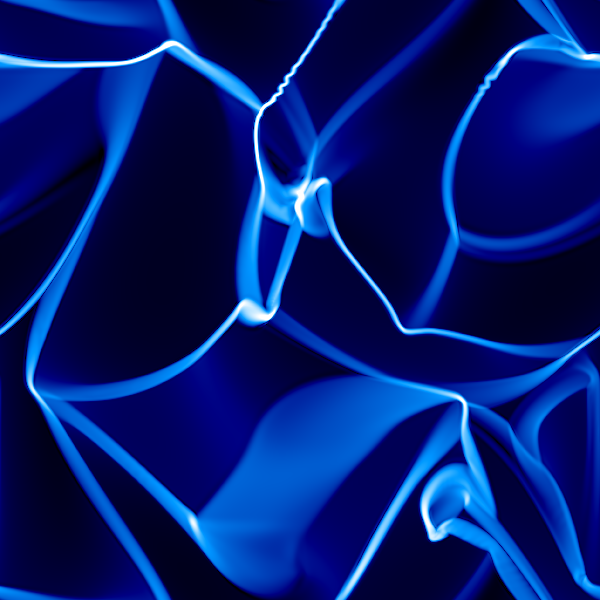} \\
  \includegraphics[height=0.16\textwidth]{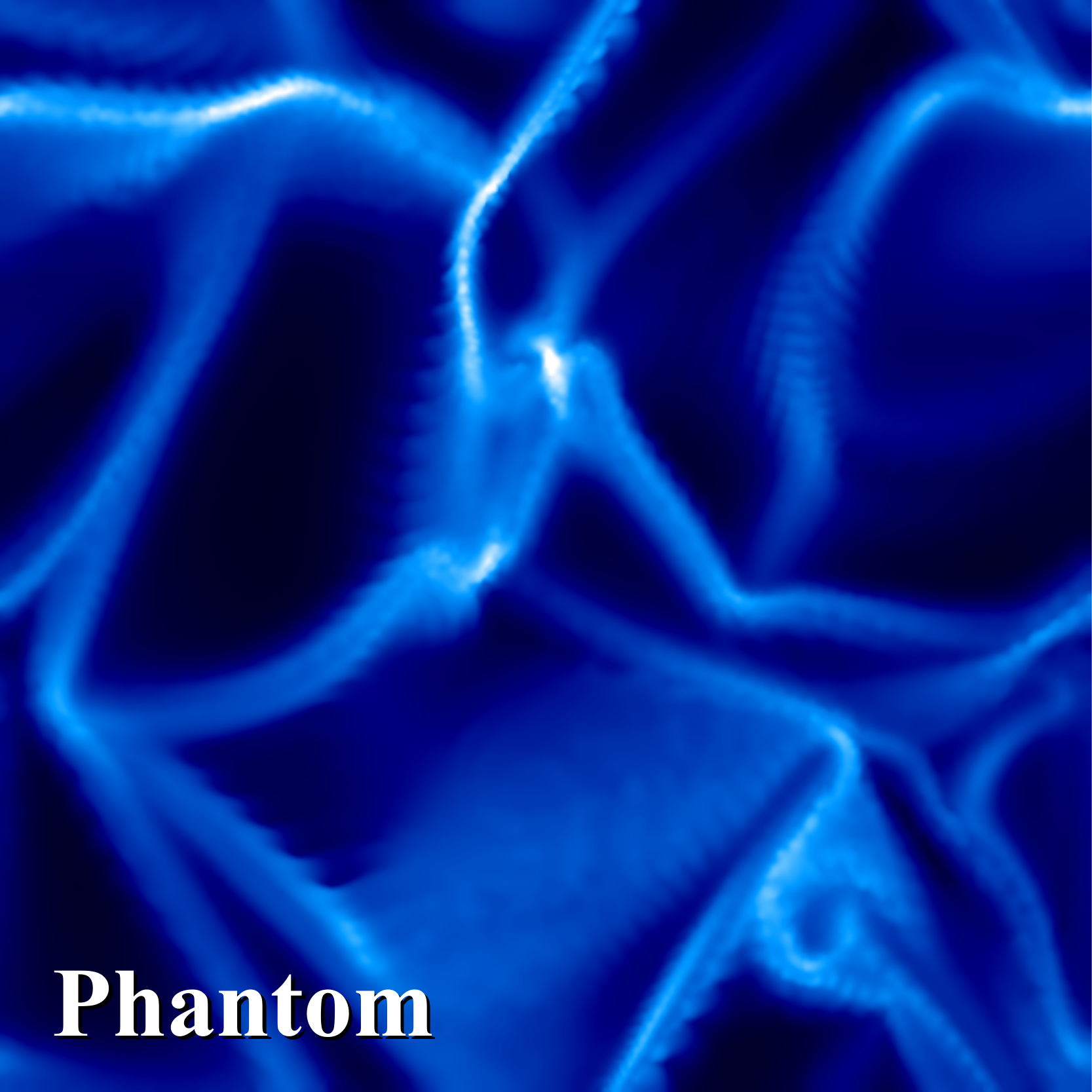} 
& \includegraphics[height=0.16\textwidth]{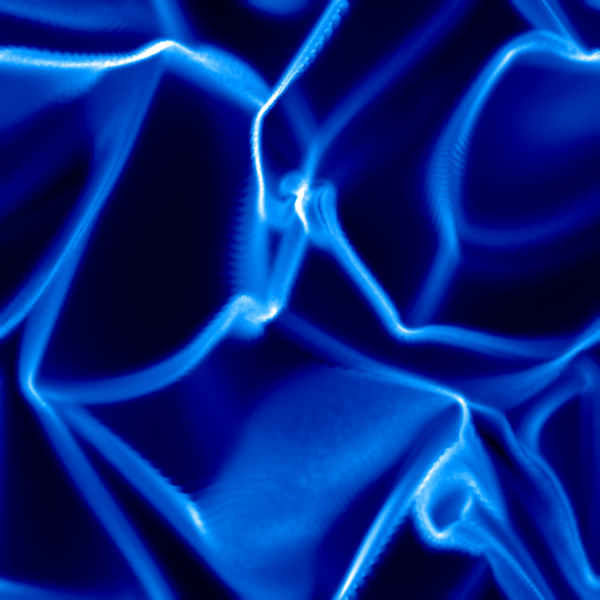} 
& \includegraphics[height=0.16\textwidth]{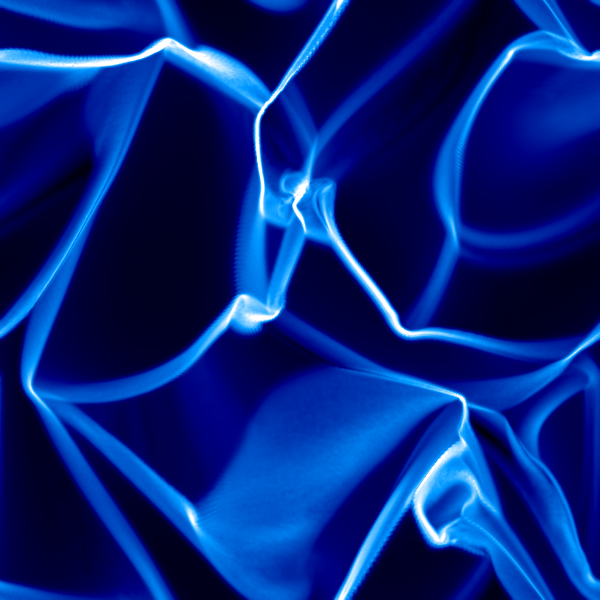} 
\end{tabular}
\begin{tabular}{cc}
\includegraphics[width=0.491\textwidth]{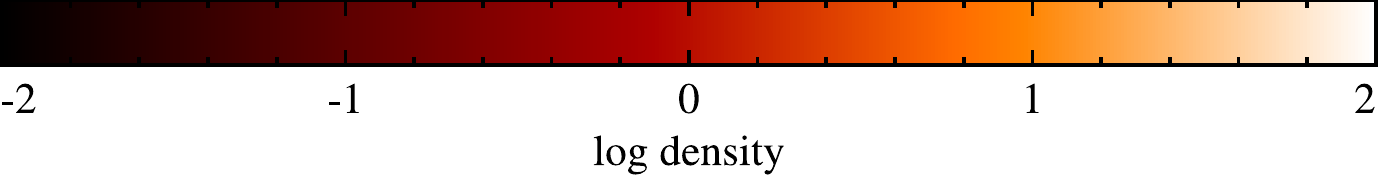}
&
\hspace{1.4mm}\includegraphics[width=0.491\textwidth]{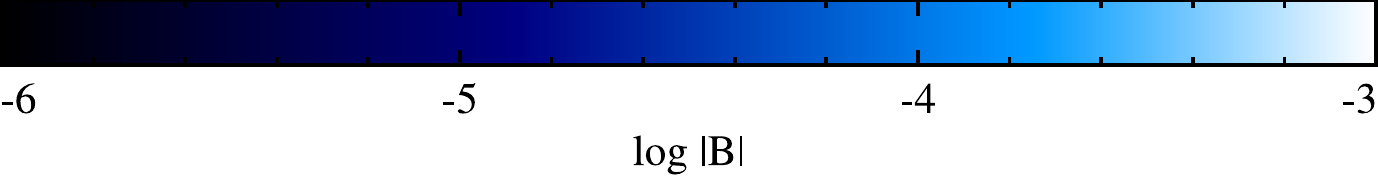}
\end{tabular}
\caption{Slices of $\rho$ and $\vert B \vert$ in the $z=0.5$ mid-plane at $t/t_{\rm c}=1$ during the transient phase. The results from {\sc Flash} (top row) and {\sc Phantom} (bottom row) are shown for resolutions of $128^3$, $256^3$, and $512^3$ (left to right). As the resolution is increased, the shock lines become more well defined. The regions with highest magnetic field strength are in the dense shocks.}
\label{fig:slices}
\end{figure}

\section{Results}
\label{sec:turbresults}

Since this comparison uses the same codes, initial conditions, and turbulent driving routine as the hydrodynamic turbulence comparison of \citetalias{pf10}, our analysis focuses on the magnetic properties of the turbulence. Hence, analyses performed by \citetalias{pf10} have only been repeated where the addition of magnetic fields would be expected to alter the result (i.e., for the density PDF).

A focus of our analysis is the effect of numerical resolution on the dynamo. Since we assume ideal MHD, the kinetic and magnetic Reynolds numbers vary with resolution. This affects the growth rate and saturation level of magnetic energy, enabling us to contrast the scaling behaviour of the two methods.

The evolution of the magnetic field in the simulations may be divided into three different phases --- transient growth, exponential growth, and saturation. These phases can be seen in Figure~\ref{fig:en_mag} which shows the magnetic and kinetic energy as a function of time for the calculations from both \textsc{Phantom} and \textsc{Flash} using $128^{3}$, $256^{3}$ and $512^{3}$ resolution elements (see legend). We analyse each of these phases in detail below.

\subsection{Initial transient growth; $t/t_{\rm c} \lesssim 2$}
\label{sec:transientphase}

The simulations begin with a brief transient phase while turbulence is generated by the driving routine. Slices of $\rho$ and $\vert B \vert$ at $z=0.5$ for $t/t_{\rm c}=1$ are shown in Figure~\ref{fig:slices}, shortly after the large shocks created by the driving routine have begun to interact. Magnetic fields are strongest where the density is highest, due to compression of the magnetic field in the shocks. Conversely, the low density regions exhibit relatively weaker magnetic fields.

Approximately half a crossing time is required for the kinetic energy to saturate (see Figure~\ref{fig:en_mag}), though it takes another turbulent crossing time before the turbulence is fully developed. The magnetic energy is amplified by two orders of magnitude during this phase (Figure~\ref{fig:en_mag}). This occurs in two steps: The first ($t/t_{\rm c} \lesssim 1$) is a sharp rise in magnetic energy caused by the formation of large-scale shocks (Figure~\ref{fig:slices}). The second occurs during the generation of small-scale structure in the density and magnetic fields caused by the interaction of the shocks. During the second step, from $t/t_{\rm c} \approx 1$--2, the magnetic energy increases exponentially similar to the growth phase (Section~\ref{sec:growth}), but at a rate higher by a factor of 2--3.

The initial transient growth of the magnetic field is resolution dependent, with higher resolutions yielding higher magnetic energy. For example, the magnetic energy in the $512^3$ {\sc Phantom} calculation increases by an additional $3$--$4$ orders of magnitude compared to the other calculations. We have investigated whether this is a numerical artefact of the timestepping by re-doing the initial phase with a reduction in the Courant factor, and also by using global timesteps instead of individual timesteps. These did not alter our results. Additionally, we have checked if this growth is driven by spurious generation of divergence of the magnetic field by both turning off the hyperbolic divergence cleaning, and conversely by increasing the hyperbolic cleaning wave speed by a factor of $10$ ($10\times$ over-cleaning, see Section~\ref{sec:cleaning-enhanced}). These showed the same fast transient magnetic field growth, so this is not caused by a high $\nabla\cdot{\bf B}$. Hence, the growth of magnetic energy in the {\sc Phantom} simulations appears to be physical, originating from the explicitly added dissipation terms rather than occurring due to numerical error or instability.

% Flash lines of best fit:
% 128 = 10**(0.296*x - 8.07)
% 256 = 10**(0.326*x - 8.11)
% 512 = 10**(0.319*x - 7.56)
%
% SPH lines of best fit:
% 128 = 10**(0.204*x - 7.85)
% 256 = 10**(0.344*x - 7.53)
% 512 = 

%% Saturation average energies:
%
%% Flash: 
% 128^3: 51.11 +/- 5.51    1.20 +/- 0.31
% 256^3: 51.19 +/- 4.81    1.46 +/- 0.20
% 512^3: 52.17 +/- 5.15    2.36 +/- 1.01
%
%% Phantom:
% 128^3: 50.30 +/- 4.80    1.52 +/- 0.48
% 256^3: 51.17 +/- 5.34    2.31 +/- 0.47 
% 512^3: 49.86 +/- 3.78

\begin{table}
\caption{Slope of Magnetic Energy Growth and Saturated Energy Values}
\label{tbl:energies}
\centering
\begin{tabular}{cccccc}
\hline
\hline
Calculation & $E_m$ growth rate & $\langle E_{\rm k} \rangle_\text{sat}$ & $\langle E_{\rm m} \rangle_\text{sat}$ \\ \hline
{\sc Flash} $128^3$ & 0.30 & 51.11 $\pm$ 5.51 & 1.20 $\pm$ 0.31  \\ %% 1.21 \pm 0.27
{\sc Flash} $256^3$ & 0.32 & 51.19 $\pm$ 4.81 & 1.46 $\pm$ 0.20  \\ %% 1,50 \pm 0.16
{\sc Flash} $512^3$ & 0.32 & 52.17 $\pm$ 5.15 & 2.36 $\pm$ 1.02  \\ %% 2.60 \pm 1.00
{\sc Phantom} $ 128^3$ & 0.20 & 50.30 $\pm$ 4.80 & 1.52 $\pm$ 0.48  \\ %% 1.52 \pm 0.48
{\sc Phantom} $ 256^3$ & 0.34 & 51.17 $\pm$ 5.34 & 2.31 $\pm$ 0.47  \\ %% 2.47 \pm 0.30
{\sc Phantom} $ 512^3$ & 0.71 & 50.65 $\pm$ 4.20 & 2.62 $\pm$ 0.19  \\
\hline
\end{tabular}
\end{table}

%% medium square page size
%% 1.7 character height
%% 1.9 character width color bar
\begin{figure}
\centering
\setlength{\tabcolsep}{0.0025\columnwidth}
\renewcommand{\arraystretch}{1.0}
\begin{tabular}{ccccl}
 \multicolumn{4}{c}{\sc Flash} & \\
  \includegraphics[height=0.225\textwidth]{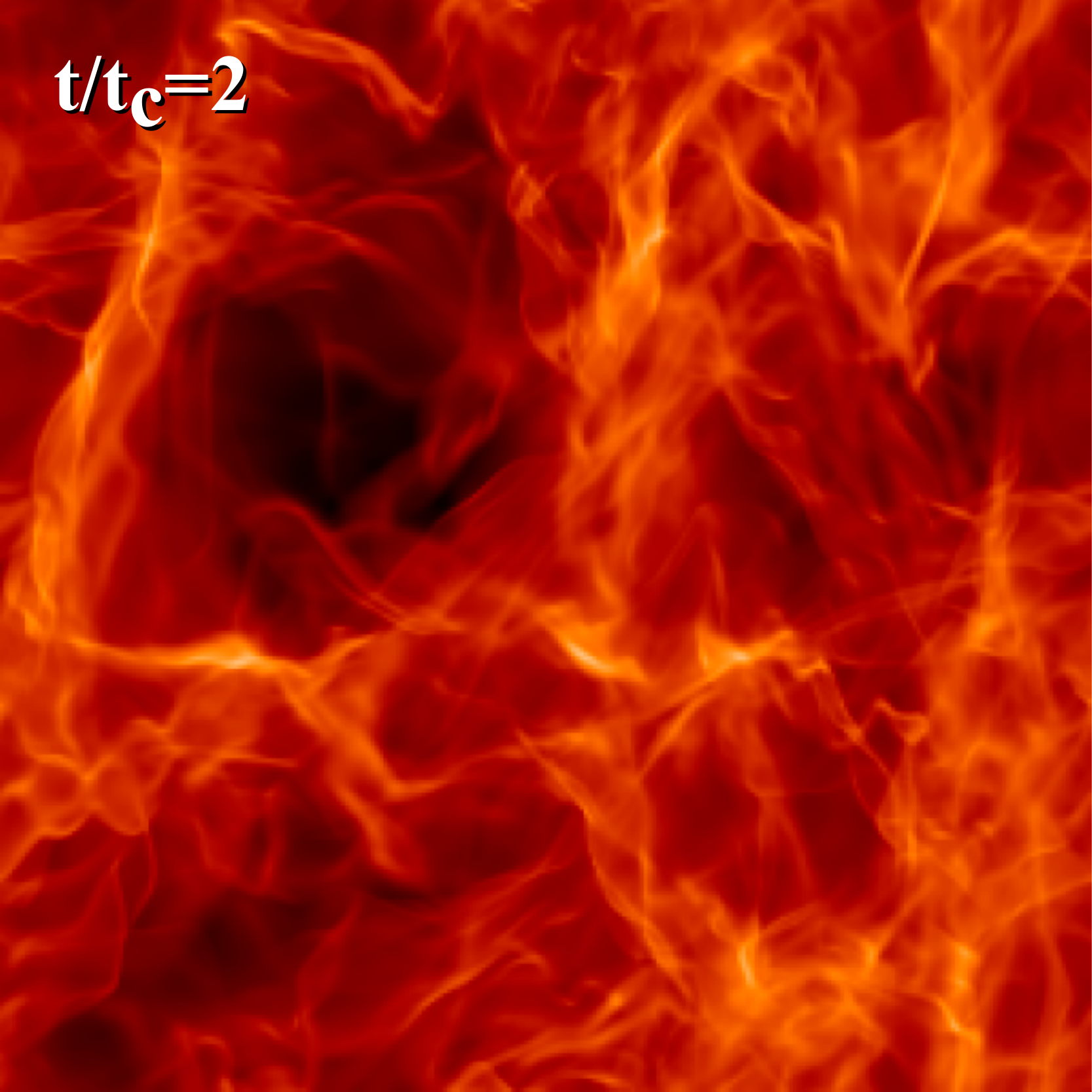}
& \includegraphics[height=0.225\textwidth]{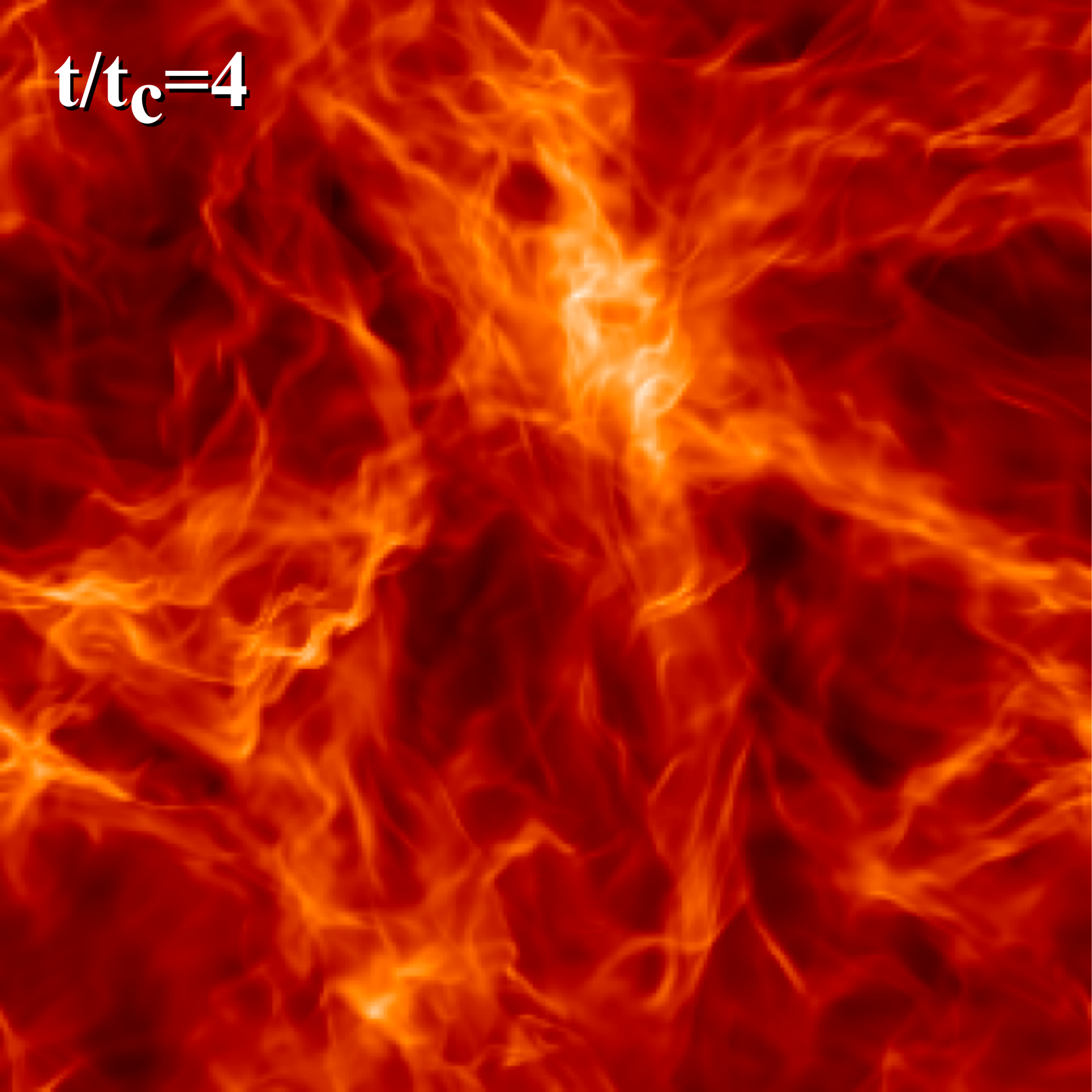}
& \includegraphics[height=0.225\textwidth]{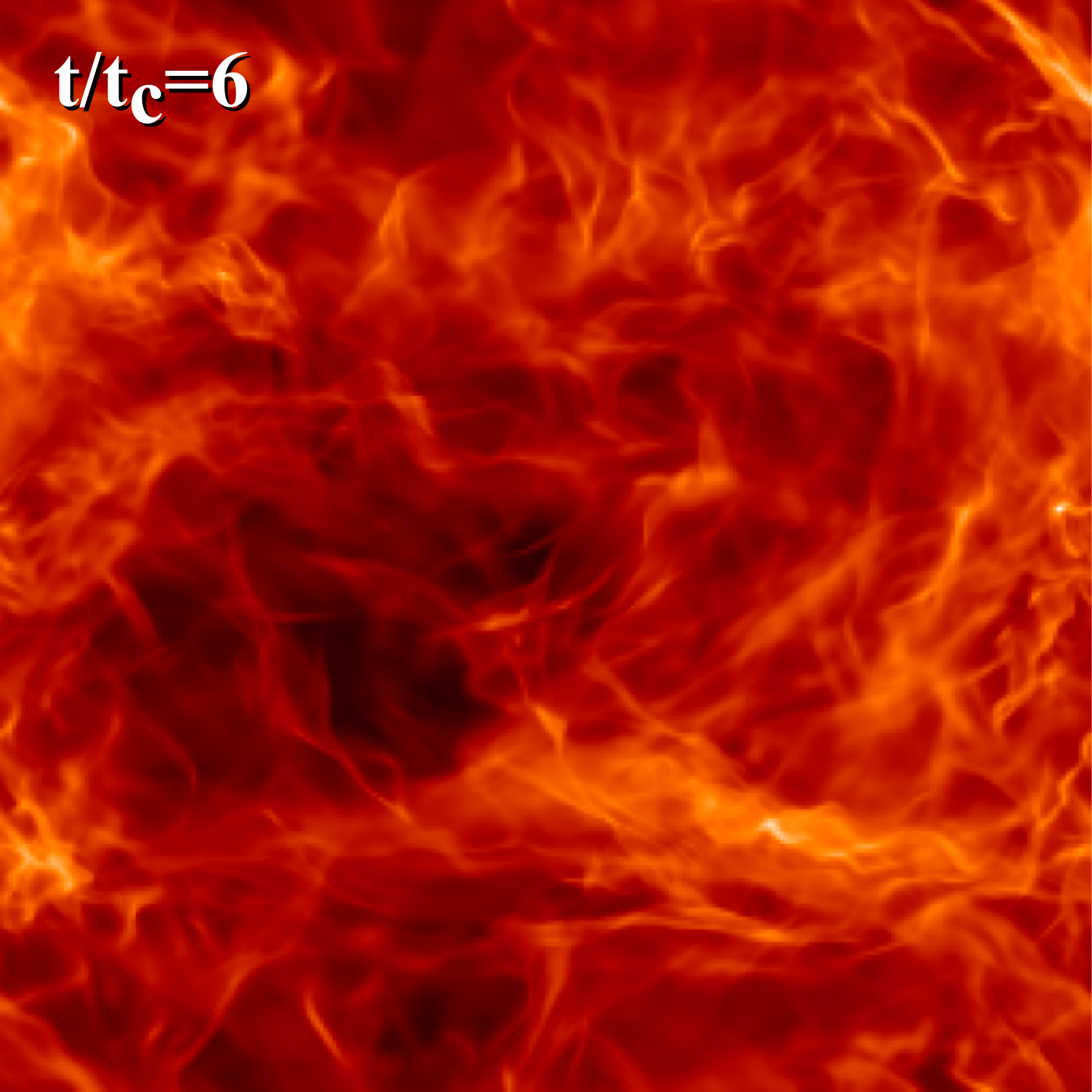}
& \includegraphics[height=0.225\textwidth]{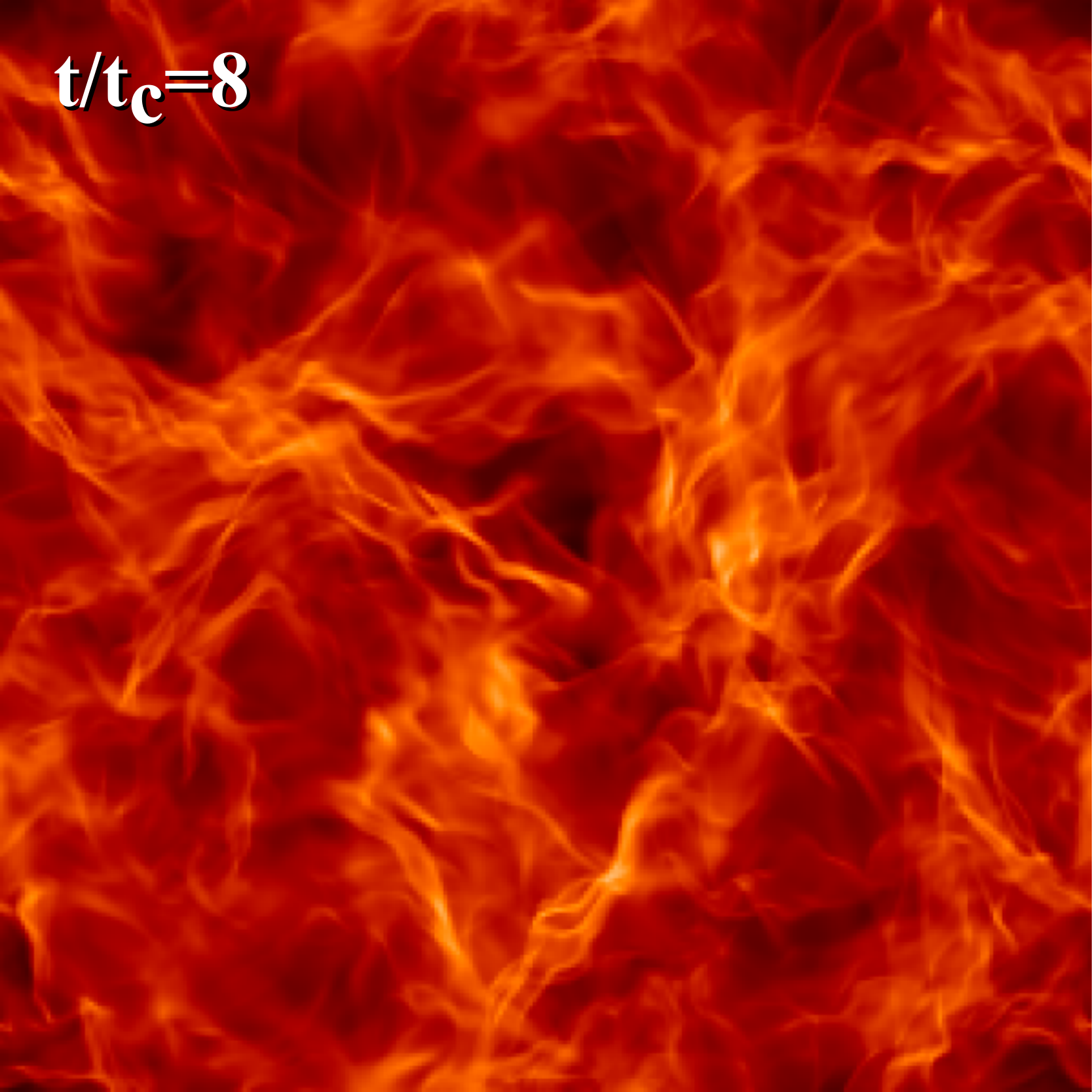} 
& \includegraphics[height=0.225\textwidth]{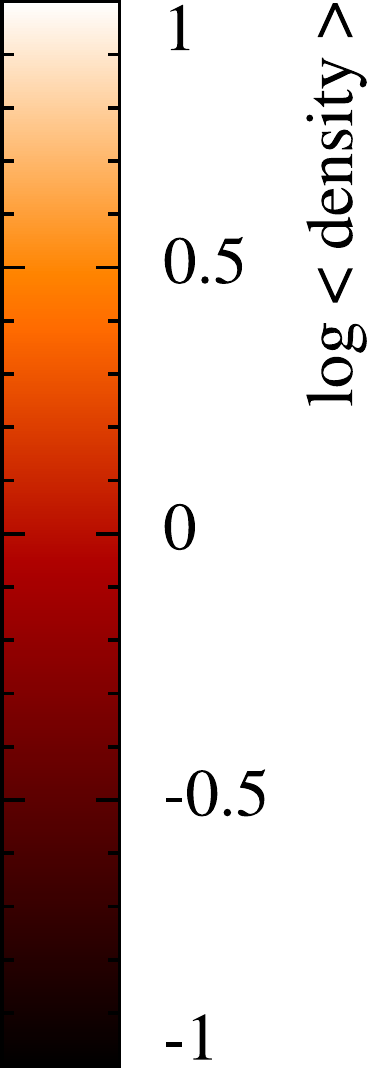} \\
  \includegraphics[height=0.225\textwidth]{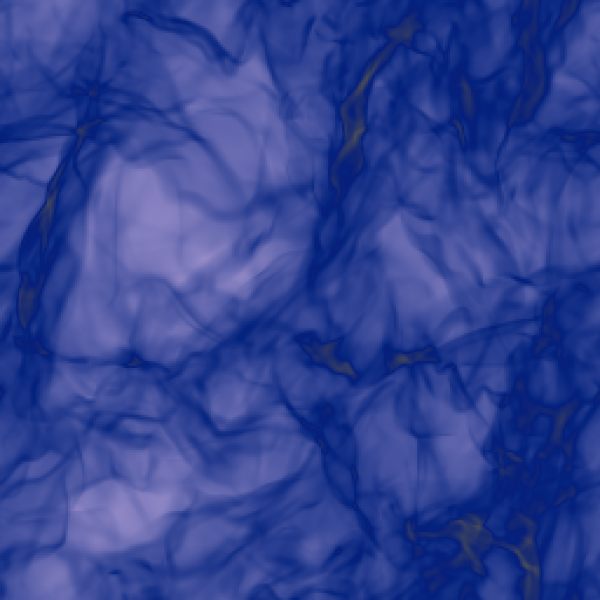}
& \includegraphics[height=0.225\textwidth]{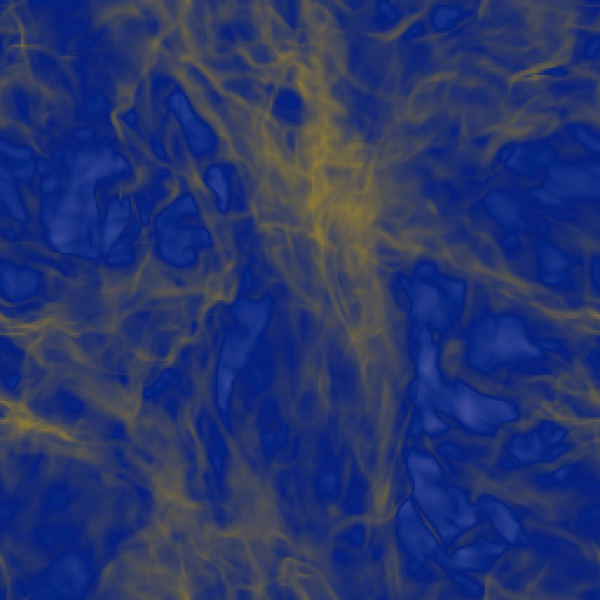}
& \includegraphics[height=0.225\textwidth]{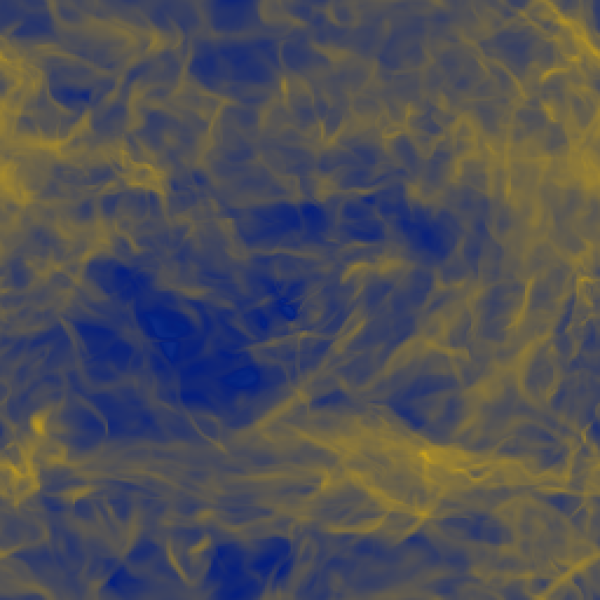}
& \includegraphics[height=0.225\textwidth]{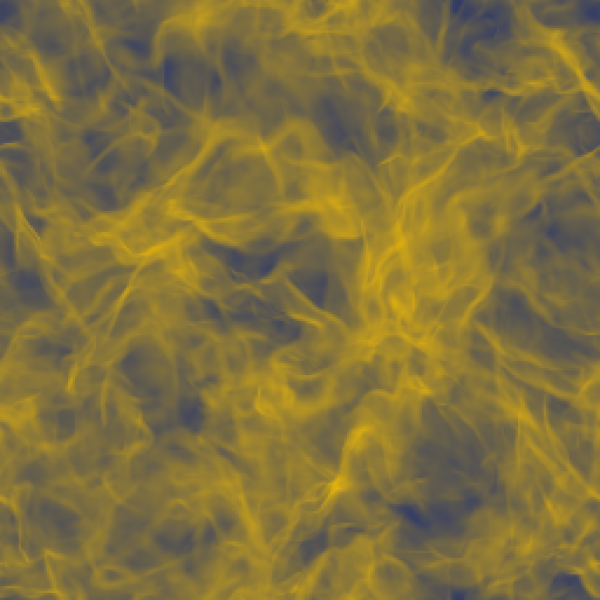}
& \includegraphics[height=0.225\textwidth]{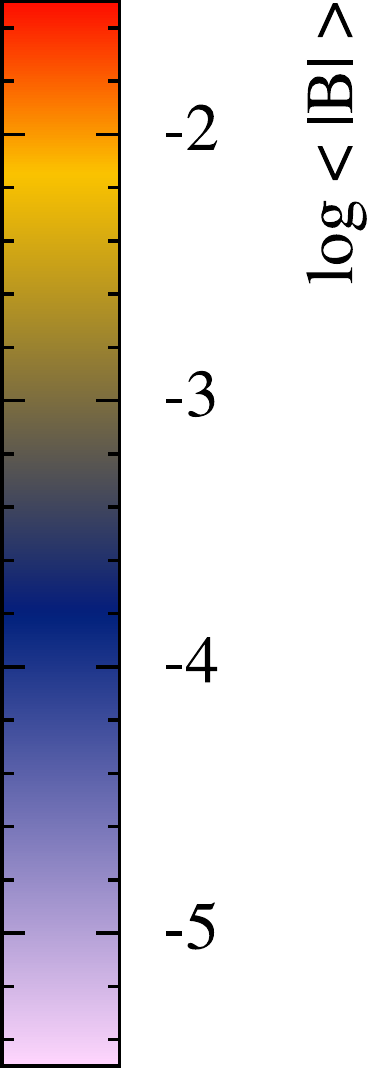}
\end{tabular} \\
\vspace{0.5cm}
\begin{tabular}{ccccl}
 \multicolumn{4}{c}{\sc Phantom} & \\
  \includegraphics[height=0.225\textwidth]{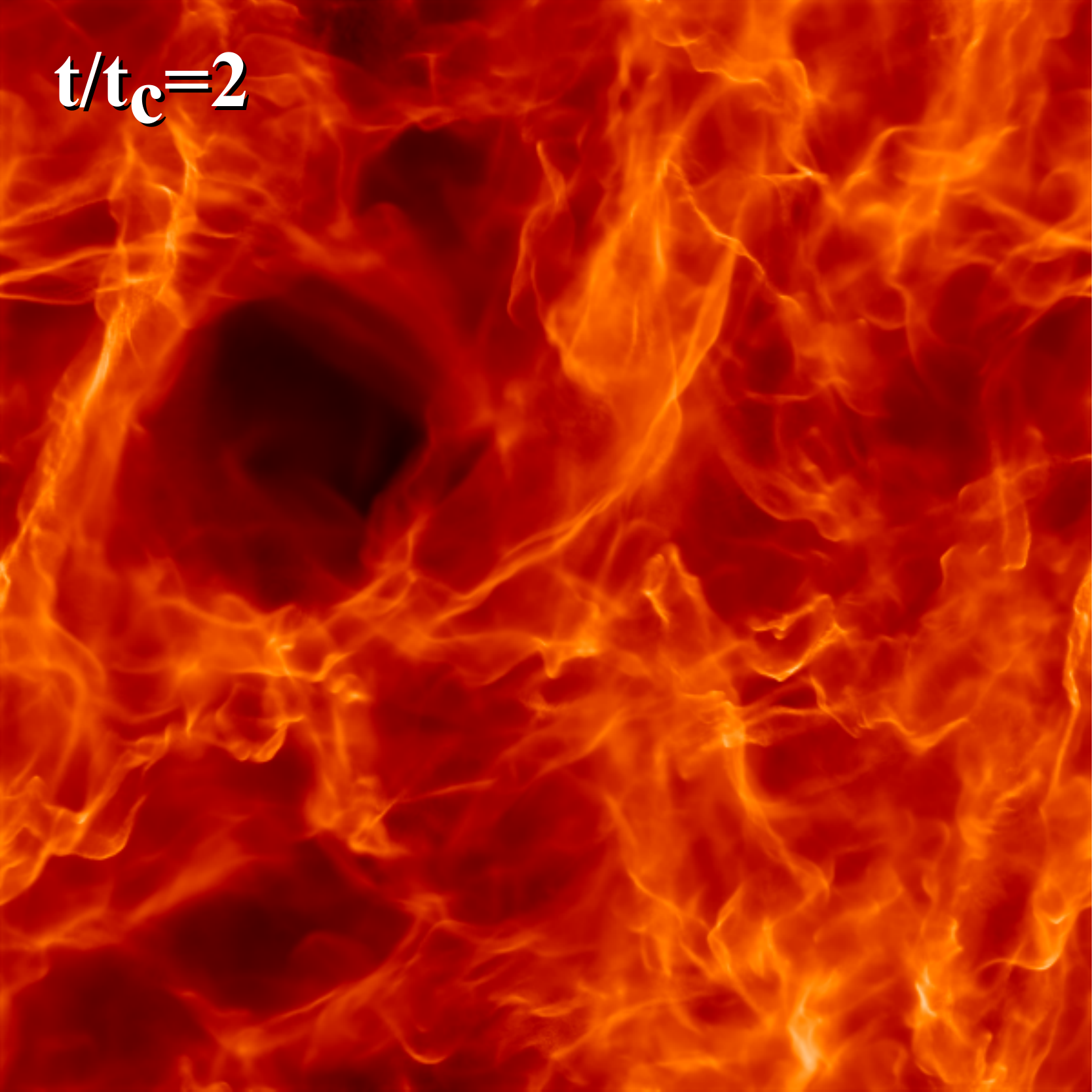}
& \includegraphics[height=0.225\textwidth]{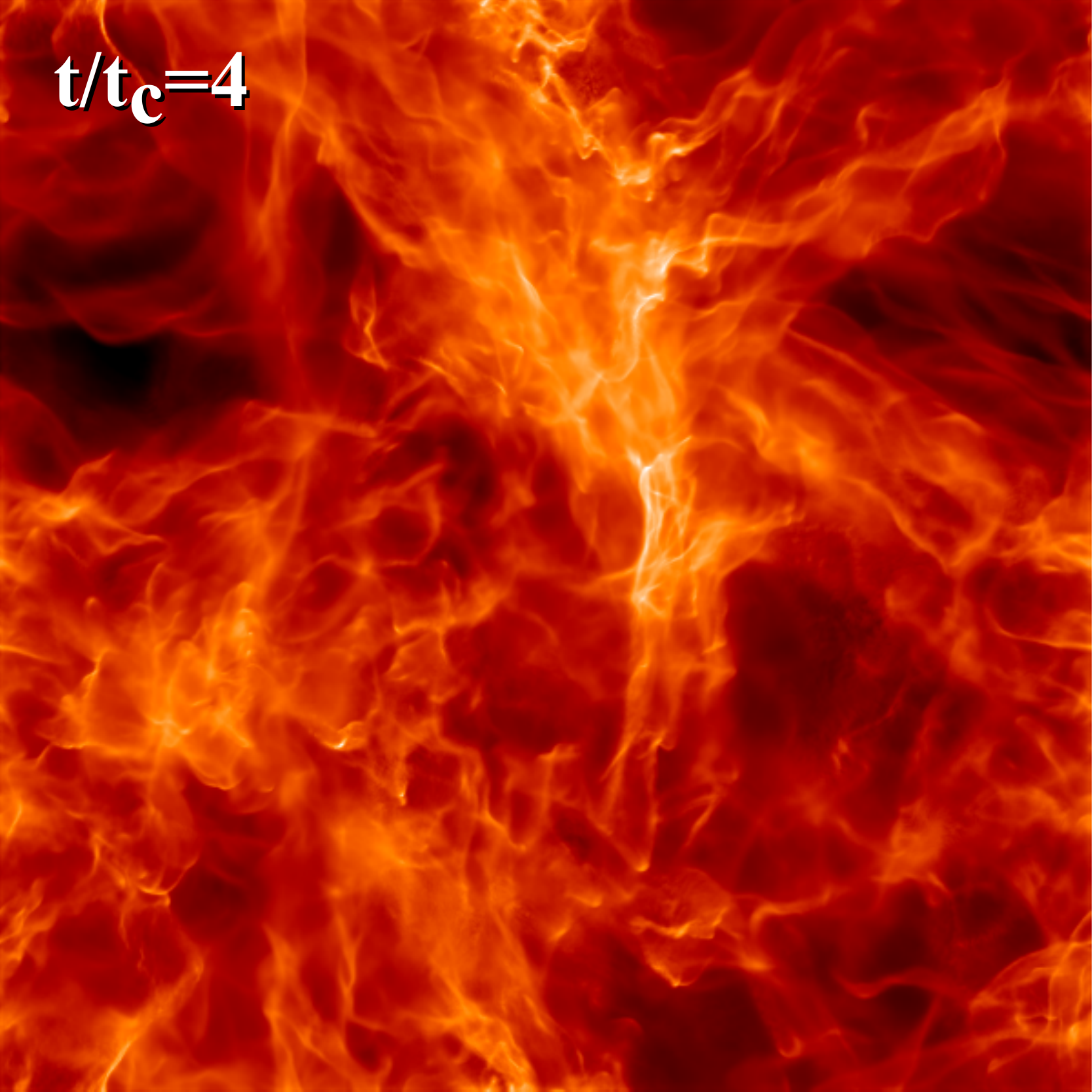}
& \includegraphics[height=0.225\textwidth]{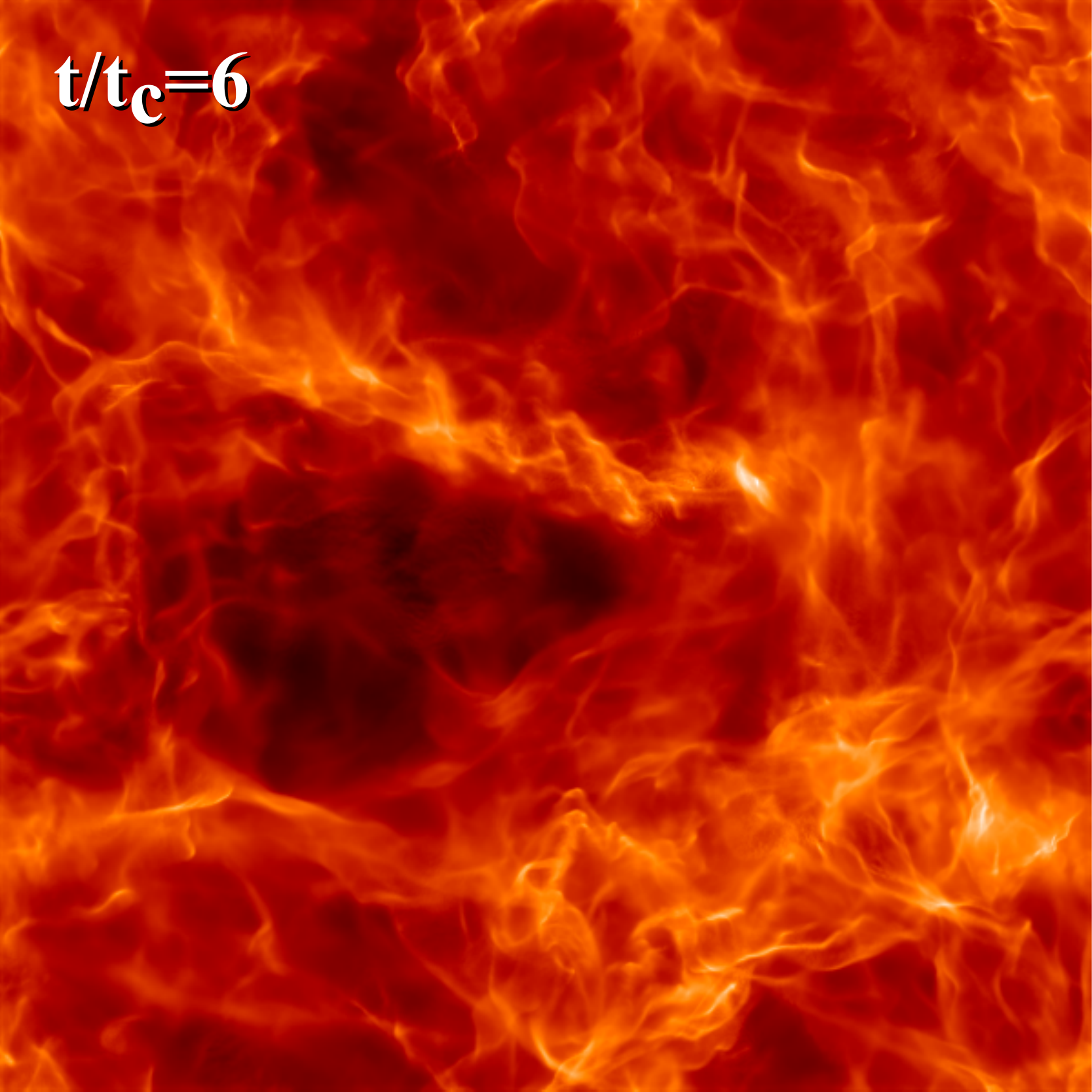}
& \includegraphics[height=0.225\textwidth]{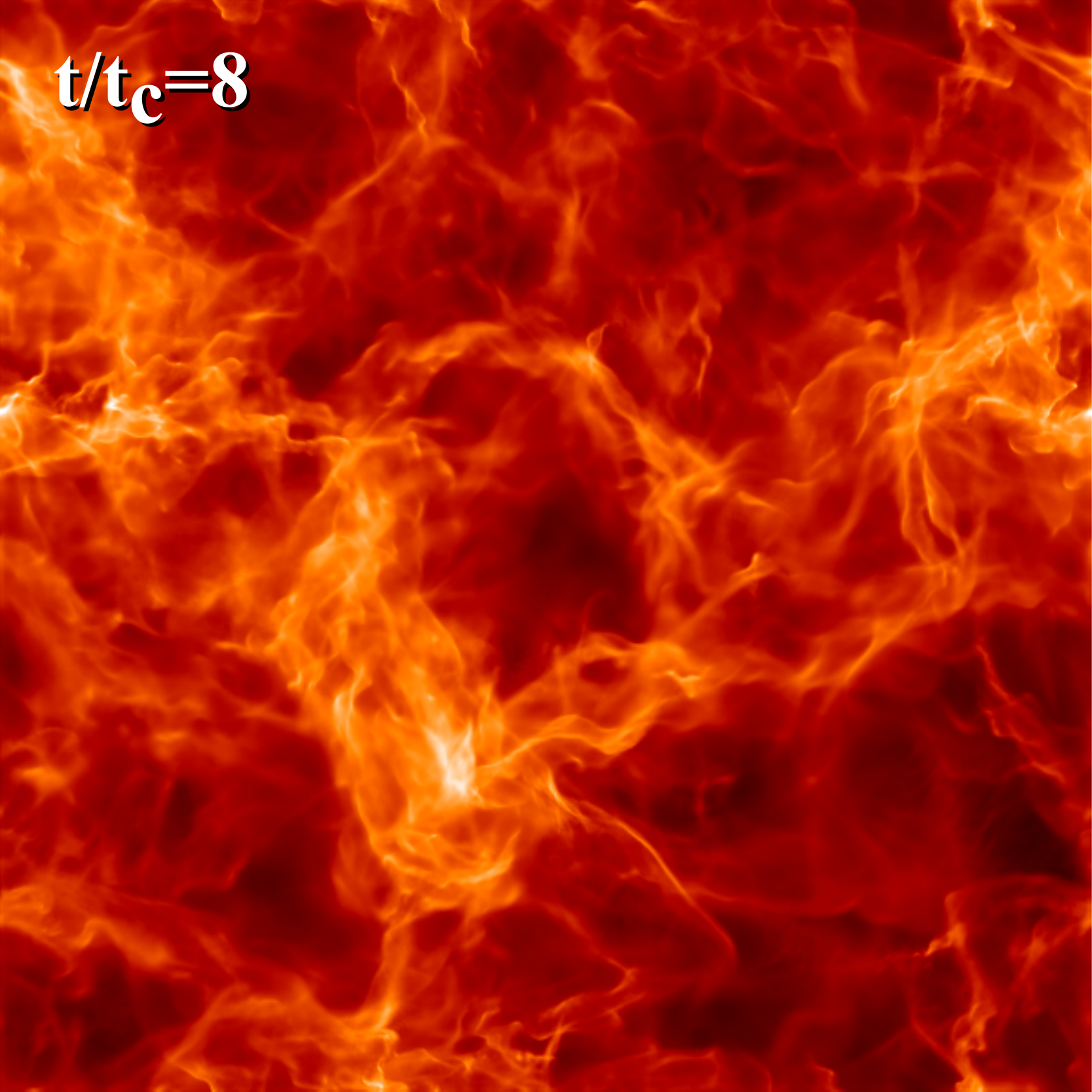}
& \includegraphics[height=0.225\textwidth]{turbcomp/cobar-column-rho-red.pdf} \\
  \includegraphics[height=0.225\textwidth]{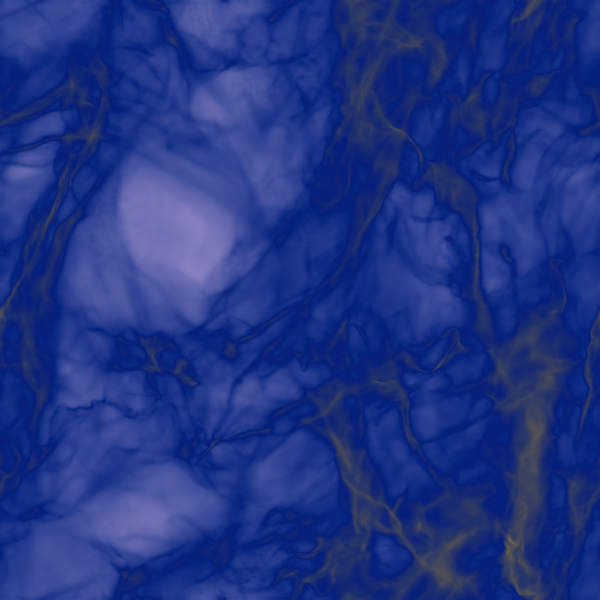}
& \includegraphics[height=0.225\textwidth]{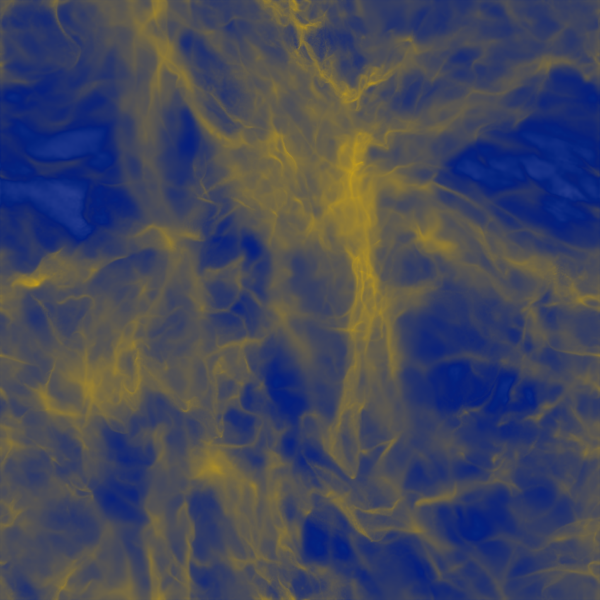}
& \includegraphics[height=0.225\textwidth]{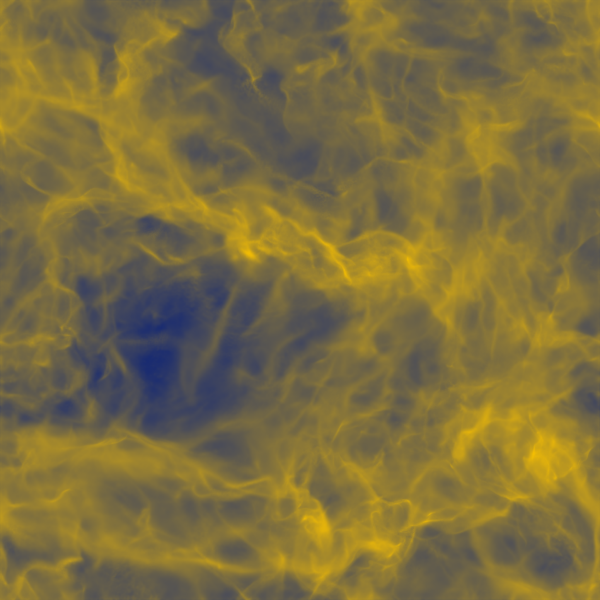} 
& \includegraphics[height=0.225\textwidth]{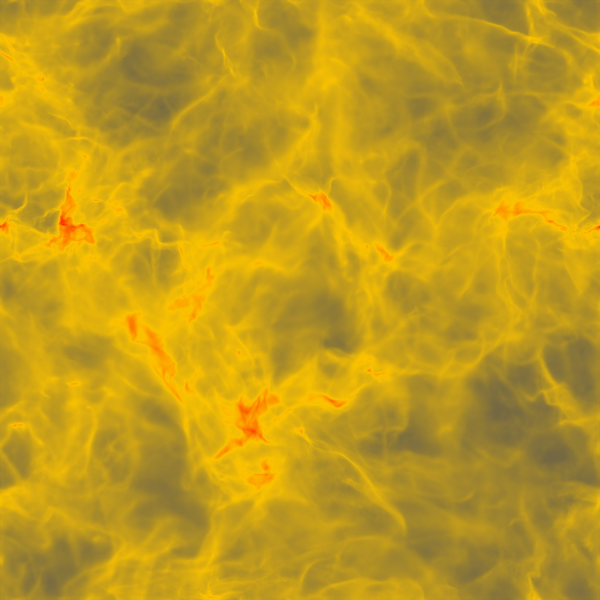} 
& \includegraphics[height=0.225\textwidth]{turbcomp/cobar-column-B-wbyr.pdf} 
\end{tabular}
\caption{$z$-column integrated $\rho$ and $\vert B \vert$, defined $<B> = \int \vert B \vert {\rm d}z / \int {\rm d}z$, for {\sc Flash} (top) and {\sc Phantom} (bottom) at resolutions of $256^3$ for $t/t_{\rm c}=2,4,6,8$. The density field has similar structure in both codes at early times, but diverge at late times due to the non-linear behaviour of the turbulence. The magnetic field is strongest in the densest regions, while the mean magnetic field strength also increases with time.}
\label{fig:column-integrated}
\end{figure}

\subsection{Growth phase;  $2 \lesssim t/t_{\rm c} \lesssim 10$--$40$}
\label{sec:growth}

Once the hydrodynamic turbulence is fully developed, the magnetic field is exponentially amplified at a steady rate via the small-scale dynamo. During this phase the magnetic energy is amplified by approximately $8$ orders of magnitude until it reaches its saturation level, occurring when the conversion of kinetic to magnetic energy is balanced by the dissipation of magnetic energy and the back-reaction by the Lorentz force resists further winding of the field.  The reservoir of kinetic energy is maintained by continual driving of large-scale motions via the driving routine. The magnetic energy saturates at $t/t_{\rm c} \sim 30$ for all three {\sc Flash} calculations, but the time of saturation in the {\sc Phantom} calculations varies from $t/t_{\rm c}=12$ to $t/t_{\rm c}=45$ depending on the resolution. In all cases, the saturation occurs when $v_{\rm A} \sim c_{\rm s}$.

\subsubsection{Correlation with the density field}

Figure~\ref{fig:column-integrated} shows a time sequence of column density and column integrated $\vert B \vert$ from $t/t_{\rm c}=2$--$8$, comparing {\sc Flash} (top figure) and {\sc Phantom} (bottom figure) calculations at $256^3$ since the growth rates are similar at this resolution (c.f. Figure~\ref{fig:en_mag} and Table~\ref{tbl:energies}). Both codes show similar patterns in column density and magnetic field for the first few crossing times (left two columns), but eventually the patterns diverge due to the chaotic nature of turbulence (right two columns; this was also found in \citetalias{pf10}). Nevertheless, there exists a definite correlation between the density and the magnetic field when compared at a fixed time for each code individually. The mean magnetic field strength can be seen to increase with time in both the low and high density regions.

\subsubsection{Magnetic energy growth rates}

% Table 2 shows growth rates
Table~\ref{tbl:energies} compares the slope of a line fitted to the magnetic energy for each of the six calculations during the growth phase (defined between $t/t_{\rm c}=3$ and the onset of the slow growth phase). 

Analytic studies of the small-scale dynamo have shown that for $\text{Pm} \ll 1$, the growth rate scales with $\text{Rm}^{1/2}$, while for $\text{Pm} \gg 1$, it scales with $\text{Re}^{1/2}$ \citep*{bss13}. Theoretical predictions of the growth rate for $\text{Pm}\sim 1$, which is the Prandtl number regime for numerical codes in the absence of explicit dissipation terms, are more uncertain. \citet{federrathetal11} measured the effective Prandtl number in {\sc Flash} through comparison with calculations with physical dissipation terms, finding that $\text{Pm}\sim2$. This is in agreement with similar experiments by \citet{lb07}. For {\sc Phantom}, the effective Prandtl number can be estimated analytically from the artificial dissipation terms, for which we find that $\text{Pm}\sim1$ for these calculations (see Appendix~\ref{sec:prandtl} for further discussion).

\begin{figure}
 \centering
\includegraphics[width=0.49\columnwidth]{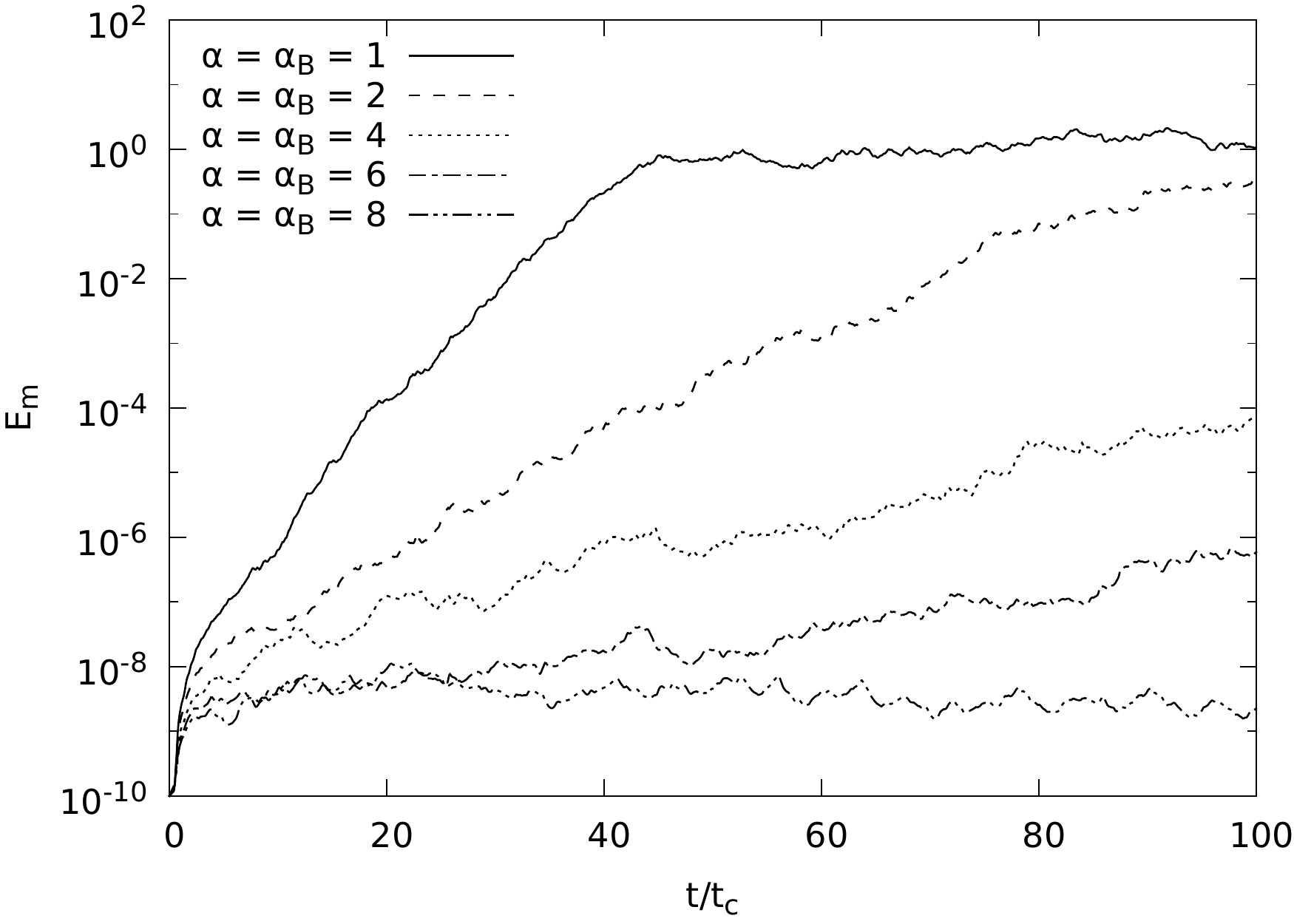}
\includegraphics[width=0.49\columnwidth]{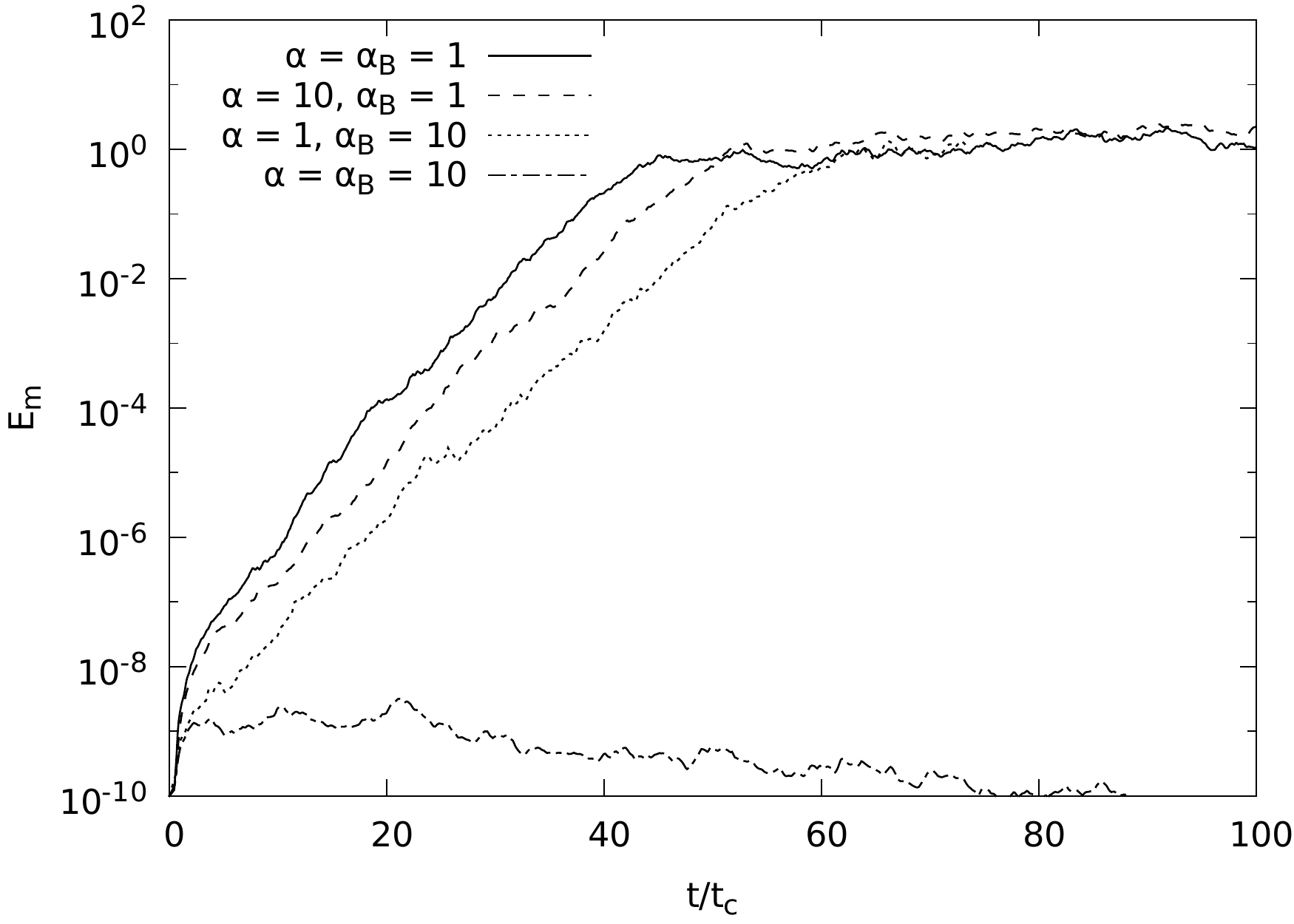}
\caption{{\it Left panel:} A $128^3$ {\sc Phantom} calculation where the artificial viscosity and resistivity parameters are systematically increased (no switches are used). With increasing dissipation, the growth rate decreases, producing the same behaviour as changing the resolution. {\it Right panel:} Comparing $128^3$ {\sc Phantom} calculations where the viscous and resistive dissipation parameters scale equally ($\alpha = \alpha_{\rm B}$ set to $1$ and $10$) to calculations where one is scaled independent of the other ($\alpha=1$, $\alpha_{\rm B}=10$ and $\alpha=10$, $\alpha_{\rm B}=1$). The growth rate appears to depend upon both Reynolds numbers --- specifically, being set by whichever is higher.}
\label{fig:en_alphas}
\end{figure}

% varying alpha runs, shows that dependence is due to dissipation terms.
To investigate the growth rate dependence on resolution for {\sc Phantom}, we performed a series of calculations where the dimensionless parameters $\alpha$ and $\alpha_{\rm B}$ in the artificial viscosity and resistivity terms were fixed to different values. We found that the growth rate depended sensitively on the amount of artificial dissipation applied, producing an effect equivalent to changing the resolution (left panel of Figure~\ref{fig:en_alphas}). Since the dissipation in {\sc Phantom} is proportional to resolution, we conclude that the growth rates obtained in our comparison are consistent with the expected resolution scaling of the artificial dissipation terms. Interestingly, the growth rate only changed when both dissipation parameters were varied. Changing only one left the growth rate largely unchanged (right panel of Figure~\ref{fig:en_alphas}), suggesting that the growth depends upon the higher of ${\rm Re}$ and ${\rm Rm}$. A worthwhile follow-up would be to compare growth rates with physical dissipation terms that are resolution independent.

% We use the 256^3 runs for a quantitative analysis since they are most comparable in their growth rates
Given that the growth rates in the $256^3$ calculations are comparable (see Table~\ref{tbl:energies} and Figure~\ref{fig:en_mag}), we perform quantitative analysis of our results during the growth phase using the $256^3$ resolution calculations. This allows for direct comparison of results.

\subsubsection{Magnetic energy power spectra}

\begin{figure}
 \centering
\includegraphics[width=0.75\columnwidth]{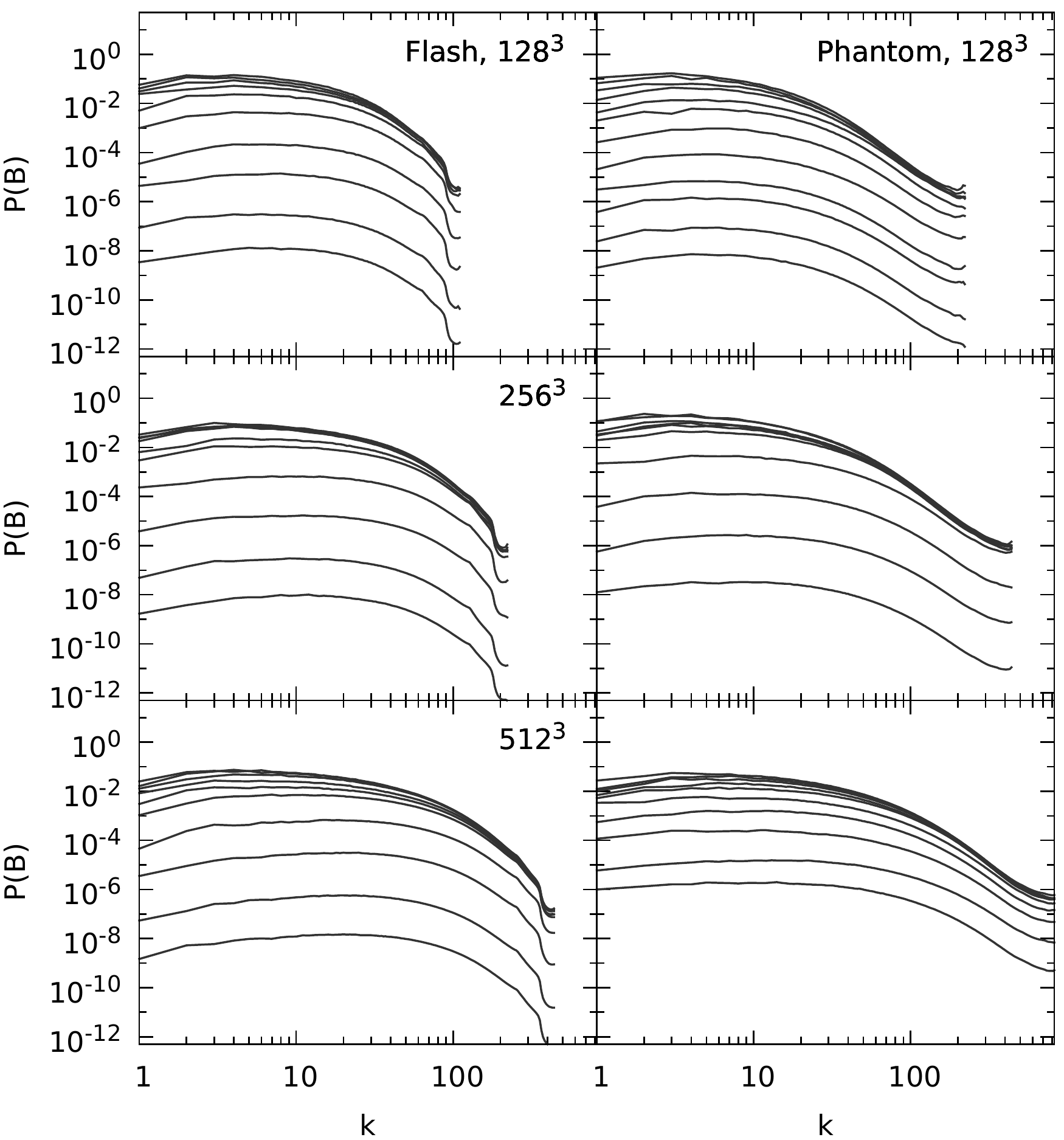}
\caption{Spectra of the magnetic energy during the growth phase for {\sc Flash} (left) and {\sc Phantom} (right) for resolutions of $128^3$, $256^3$, and $512^3$ (top to bottom). Each spectral line is sampled at intervals of $5t/t_{\rm c}$ up to $t/t_{\rm c}=50$, except for the $512^3$ {\sc Phantom} run which is sampled every $t/t_{\rm c}$ (from $t/t_{\rm c}=2$--$12$). The magnetic field grows equally at all spatial scales for all calculations.}
\label{fig:growthspectra}
\end{figure}

\begin{figure}
 \centering
\includegraphics[width=0.6\columnwidth]{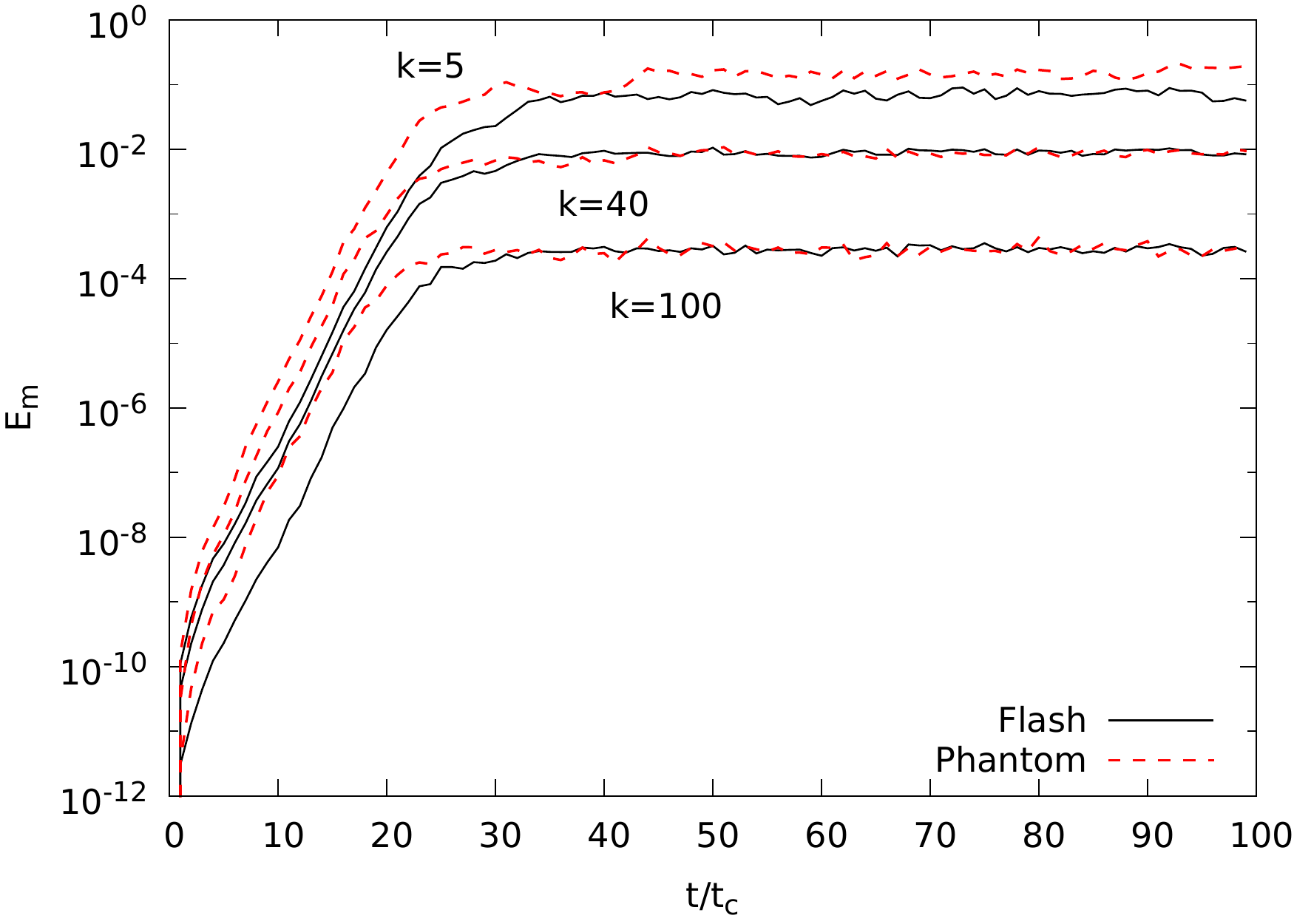}
\caption{Spectra of the magnetic energy at the $k=5$, $40$, and $100$ bands as a function of time for the $256^3$ resolution calculations of {\sc Flash} (black lines) and {\sc Phantom} (red dashed lines). The growth rate at these different wavenumbers is nearly identical. The saturation level is the same between the two codes for $k=40$ and $100$, with {\sc Phantom} containing ${\sim}2$ times as much energy in the large-scale $k=5$ band.}
\label{fig:kbands}
\end{figure}

That the total magnetic field is growing in strength --- and not just in isolated regions --- may be quantified by examining the power spectra of the magnetic energy, ${\rm P}(B)$. The magnetic energy spectra during the growth phase for the six calculations is shown in Figure~\ref{fig:growthspectra}. The magnetic energy can be seen to grow uniformly at all spatial scales in all six calculations (indicated by the translation of the power spectrum along the y-axis in the plots with minimal change in the shape), behaviour consistent with the small-scale dynamo \citep{bs05}. All of the spectra have the same general shape, with a decrease in spectral energy at and above the driving scale ($k\le3$) and a more-or-less flat spectrum (${\rm P}(B) \approx$ constant) between $3<k<10$ for the $128^3$ calculations, extending to $k\sim20$ and $k\sim40$ for the $256^3$ and $512^3$ calculations. The dissipation range in the {\sc Phantom} results extends further to smaller scales than the {\sc Flash} results for a particular resolution. The maximum in the magnetic energy spectrum in both codes occurs at high wavenumbers, as expected for small-scale dynamos \citep{cv00,bss12}, occurring around the high $k$ end of the `relatively' flat region of the spectra.

Figure~\ref{fig:kbands} shows a cross section of the power spectrum evolution at $k=5$, $40$, and $100$ for the $256^3$ calculations. These scales were chosen to represent the large, medium, and small-scale structure. This shows that the magnetic field grows in the same manner at all scales in both codes.

\subsubsection{Approach to saturation}

Figure~\ref{fig:growthspectra} shows that the magnetic energy saturates first at small scales. This is characteristic of the small-scale dynamo since this is where magnetic energy is being generated \citep{choetal09}. It is expected that the magnetic energy will grow linearly at this stage, though for Burgers turbulence, which is closer to the regime our simulations are in, it is expected that the magnetic energy growth will be closer to quadratic \citep{schleicheretal13}.  This slow growth phase lasts until the reverse cascade of magnetic energy saturates all spatial scales. This turnover in magnetic energy growth may be clearly seen in the $128^3$ and $512^{3}$ {\sc Phantom} growth curves in Figure~\ref{fig:en_mag}. It is also evident from Figure~\ref{fig:kbands} that the magnetic field enters the slow growth phase first at high wavenumbers.

\subsubsection{PDFs of $B^{2}$}

\begin{figure}
\centering
\includegraphics[width=0.99\textwidth]{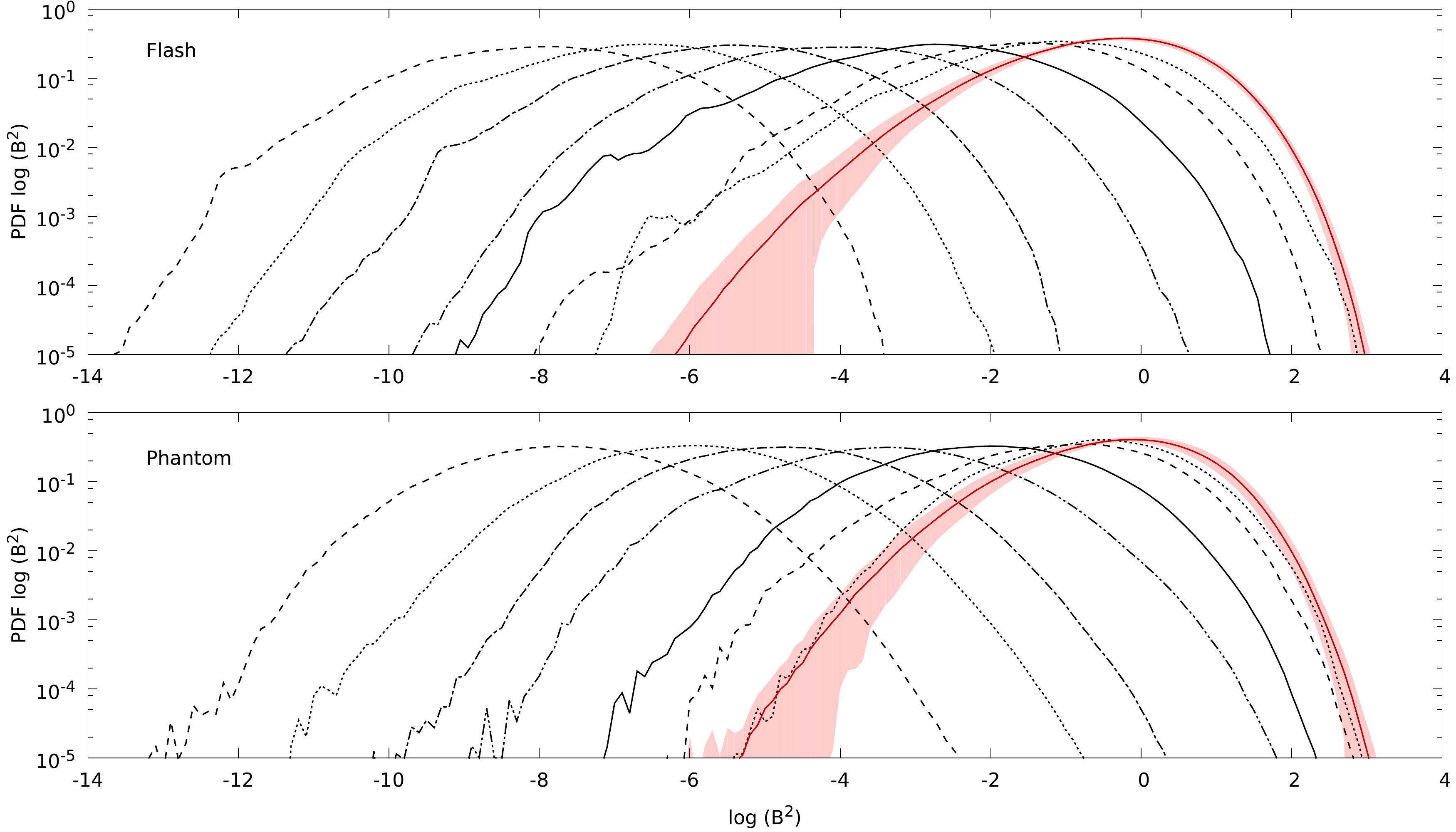}
\caption{PDF of $\log(B^2)$ during the growth phase, with the red line time averaged during the saturation phase. The top panel shows the {\sc Flash} calculation, with the bottom panel the {\sc Phantom} calculation. The PDF has a log-normal distribution during the growth phase, maintaining its width while the peak smoothly translates to higher magnetic field strengths. During the saturation phase, the PDFs of both codes have the similar peaks and high-end tails, with {\sc Flash} exhibiting a slightly extended low-end tail.}
\label{fig:bsqpdf}
\end{figure}

\begin{figure}
\centering
\includegraphics[width=0.6\columnwidth]{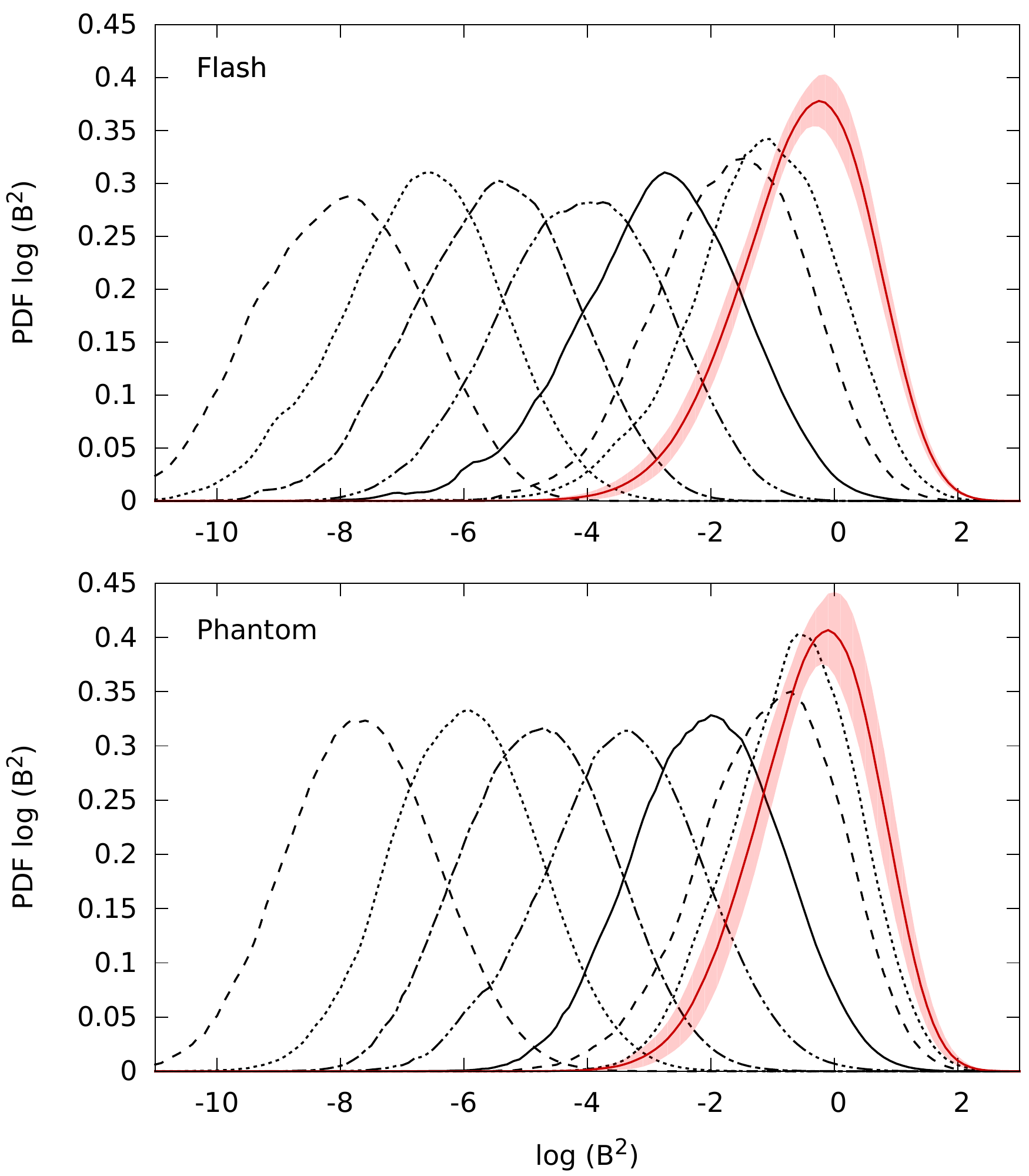}
\caption{PDF of $\log(B^2)$ during the growth phase, with the red line time averaged during the saturation phase. This is equivalent to Figure~\ref{fig:bsqpdf} but on a linearly scaled plot. In the saturation phase, the distribution is skewed with smaller deviation of magnetic field strengths.}
\label{fig:bsqpdf-linear}
\end{figure}

Figure~\ref{fig:bsqpdf} shows the time evolution of the PDF of $B^2$ for the $256^3$ calculations. The instantaneous PDFs are shown from $t/t_{\rm c}=4$--$28$ at intervals of $\Delta t = 4$, with the time-averaged PDF during the saturation phase given by the red line. The shape of the PDF remains mostly log-normal during the growth phase. As the dynamo amplifies the magnetic field, the PDF maintains its width and shape, with the peak simply translating to higher magnetic field strengths \cite[see also][]{schekochihinetal04b}. In other words, the PDF does not become distorted during the exponential growth phase but only during the slow growth phase as it approaches saturation (see below). This may also be seen in Figure~\ref{fig:bsqpdf-linear} which shows the PDF of $B^2$ on a linear scale. Figure~\ref{fig:bsqpdf} additionally shows that {\sc Flash} is able to sample lower magnetic field strengths compared to \textsc{Phantom}, which was noted by \citetalias{pf10} in the density PDFs and was attributed to the better weighting of resolution elements towards low density regions in the grid code.

The approach to saturation changes the shape of the PDF of $B^2$. It follows a log-normal distribution during the exponential growth phase, but once the dynamo enters the slow growth regime, it is no longer able to amplify the magnetic field on small-scales. Thus, the high-end tail of the distribution remains anchored, and is ``squeezed'' as the peak and low-end tail continue increasing. This produces a lop-sided distribution \citep{schekochihinetal04b}. Both codes show this behaviour as the magnetic field saturates.

\subsection{Saturation phase; $15 \lesssim t/t_{\rm c} \lesssim 100$}
\label{sec:sat}

The magnetic energy plateaus once the injection of energy balances its dissipation. While the magnetic field topology changes due to the turbulence during this phase, the magnetic energy remains in a statistical steady state.

\subsubsection{Magnetic energy saturation level}

The mean magnetic energy in the saturation phase is approximately $2$--$4\%$ of the mean kinetic energy (Table~\ref{tbl:energies}). The mean magnetic energy shows a trend of increasing with resolution, with the $512^3$ calculations twice as high as the corresponding calculations at $128^3$ (for both {\sc Flash} and {\sc Phantom}), though remains within the standard deviation. We note that the $512^3$ {\sc Phantom} calculation is averaged over a shorter time (${\sim}7t_{\rm c}$ compared to $50$--$70t_{\rm c}$), which is reflected by its smaller standard deviation. The $512^3$ {\sc Flash} calculation shows a long-term variation, with a $50\%$ increase in mean energy above $80t_{\rm c}$. This is reflected in the wider standard deviation in this calculation (${\sim}1.0$ compared to $0.2$--$0.3$ in the $128^3$ and $256^3$ calculations). Overall, while the statistical ranges of mean energy overlap between resolutions, it does appear that {\sc Phantom} yields higher mean magnetic energy during the saturation phase than {\sc Flash} at comparable resolution.

The mean magnetic energy in both codes increases with resolution. Given that the Prandtl number in {\sc Flash} does not scale with resolution (Section~\ref{sec:growth}), this suggests that the saturation level of the magnetic energy does not depend exclusively on the Prandtl number \citep{federrathetal14}. Instead, our results suggest that it depends on either the kinetic or magnetic Reynolds numbers.

 To investigate this further, we performed a set of {\sc Phantom} simulations, keeping the same artificial viscosity parameters but turning off the artificial resistivity switch developed in Chapter~\ref{sec:chapter-switch} (i.e., using a constant artificial resistivity parameter, $\alpha_{\rm B} = 1$), thereby increasing the amount of resistive dissipation.  This reduced the mean magnetic energy in the saturation phase at all three resolutions ($128^3$: 1.52 to 1.01, $256^3$: 2.31 to 1.32, $512^3$: 2.62 to 1.40). This, along with the {\sc Flash} results, suggests that the magnetic Reynolds number is primarily responsible for determining the saturation level of the magnetic field.

\subsubsection{Alfv\'enic Mach number}

\begin{figure}
 \centering
\includegraphics[width=0.6\columnwidth]{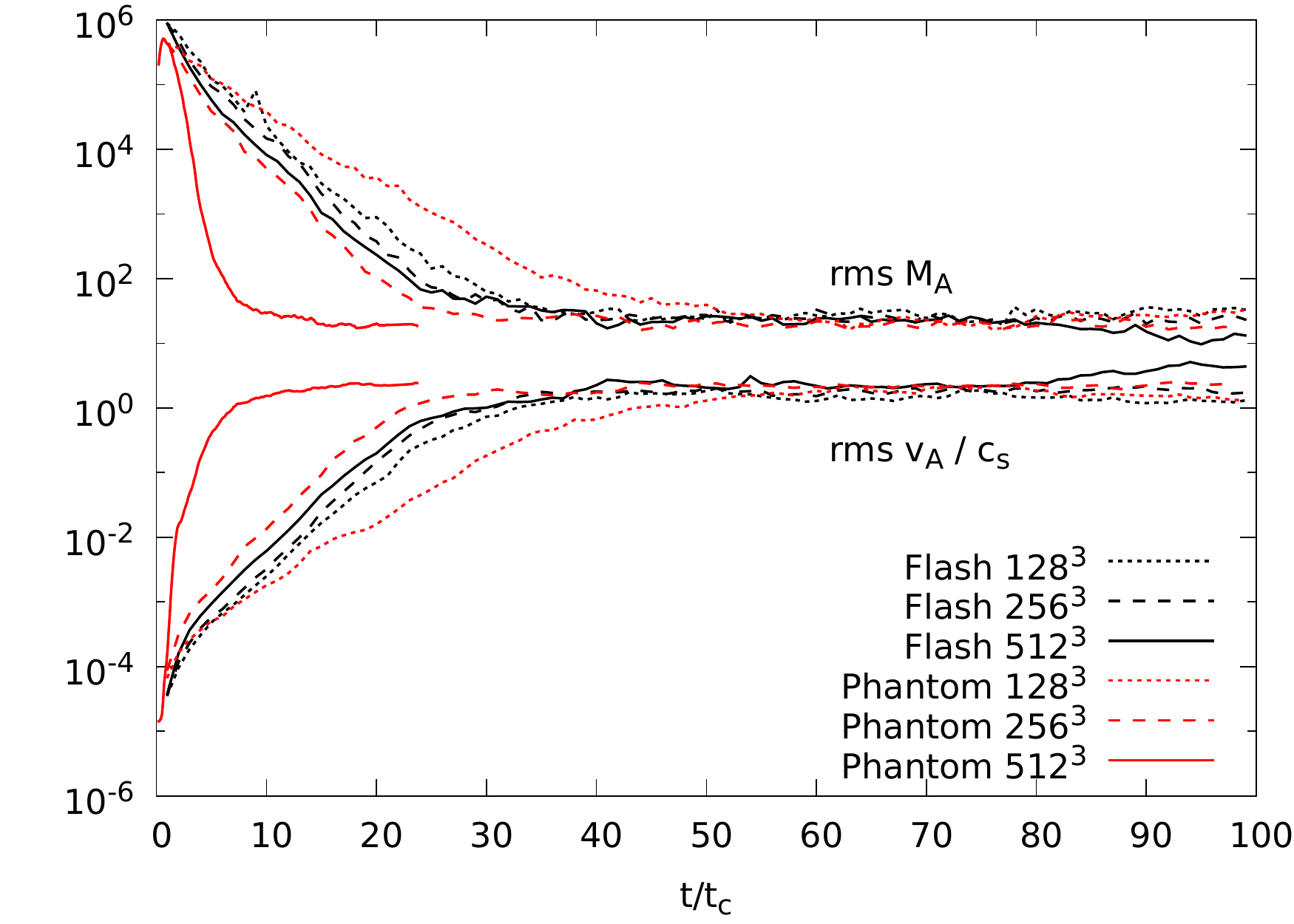}
\caption{Time evolution of the rms Alfv\'en speed and Alfv\'enic Mach number. For all calculations, the time averaged rms Alfv\'en speed in the saturation phase is $v_\text{A}\sim2 c_{\rm s}$. The rms Alfv\'enic Mach numbers is $M_\text{A}\sim20$. This differs noticeably from the rms Alfv\'en speed divided by the rms velocity (${\sim}5$).}
\label{fig:alfvenmach}
\end{figure}

Figure~\ref{fig:alfvenmach} shows the time evolution of the rms Alfv\'en speed, $v_{\rm A}$, and rms Alfv\'enic Mach number. The rms Alfv\'en speed in the saturation phase is approximately twice the sound speed. In other words, the turbulence remains super-Alfv\'enic even once the magnetic field has reached saturation. It is worth noting that the rms $\mathcal{M}_{\rm A}$ in Figure~\ref{fig:alfvenmach} is calculated by taking the rms of the local $\mathcal{M}_{\rm A}$ as calculated per grid cell or particle. In the saturation phase, calculating the rms in this manner yields $\mathcal{M}_{\rm A}\sim20$, which is different (by a factor of 4) to that calculated by dividing the rms velocity ($10$) by the rms $v_{\rm A}$ (yielding ${\sim} 5$).  

\subsubsection{Power spectra}

\begin{figure}
 \centering
\includegraphics[width=0.6\columnwidth]{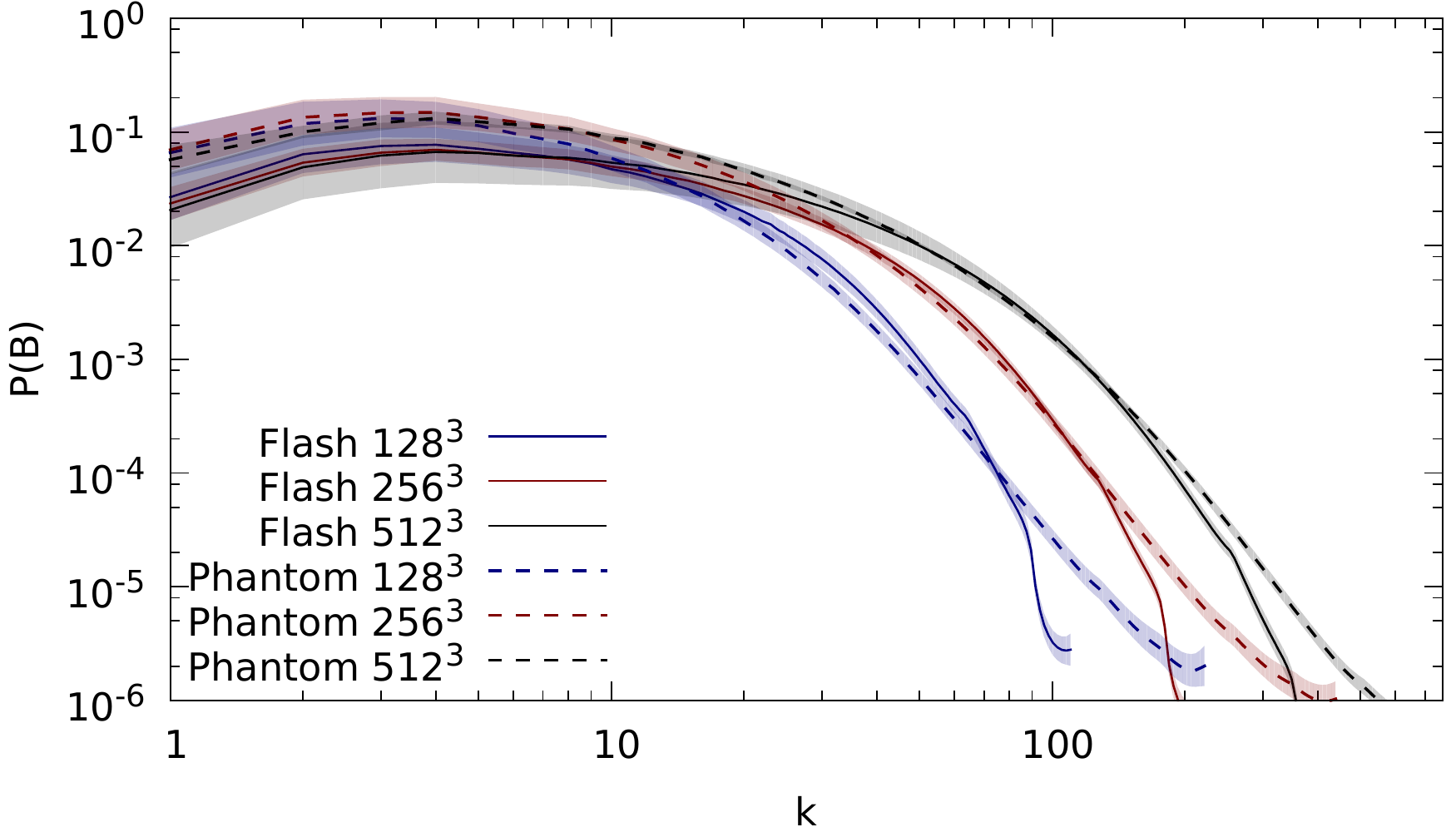}
\caption{Time averaged spectra of the magnetic energy in the saturation phase for {\sc Flash} (solid lines) and {\sc Phantom} (dashed lines) at resolutions of $128^3$ (blue), $256^3$ (red), and $512^3$ (black). Shaded regions represent the standard deviation. The {\sc Phantom} calculations systematically contain more magnetic energy (approximately $2\times$) in large-scale structure ($k<10$) compared to {\sc Flash}, and have an extended tail at high $k$ due to the adaptive resolution.}
\label{fig:satspect}
\end{figure}

Figure~\ref{fig:satspect} shows the time-averaged spectra of the magnetic energy from all six calculations in the saturation phase, with the shaded regions showing one standard deviation of the time-average. In each case 50 spectra have been averaged over a minimum of $50 t_{\rm c}$. The spectra of {\sc Flash} and {\sc Phantom} can be seen to be similar in shape, except that the {\sc Phantom} calculations contain approximately twice as much magnetic energy in large-scale structure ($k<10$). This is consistent with the higher mean magnetic energy in the {\sc Phantom} calculations in Table~\ref{tbl:energies}, indicating that this energy is stored in the largest scales of the field.

The peak of the magnetic energy spectra for both codes is at $k\sim3$--$4$, occurring just above the driving scale. As the resolution is increased, both codes extend the spectra further towards small scales. The {\sc Flash} power spectra drop sharply at the Nyquist frequency, while the {\sc Phantom} power spectra reaches higher wavenumbers than {\sc Flash} for the same number of resolution elements. While, in SPH, the smoothing kernel will distribute power to higher wavenumbers, the {\sc Phantom} calculations have adaptive resolution that reach $4$--$8\times$ that of the {\sc Flash} calculation in the densest regions. For that reason, the {\sc Phantom} power spectra has been analysed on a grid that is twice the resolution of the {\sc Flash} grid (see Appendix~\ref{sec:gridinterp} for why this resolution was chosen), and it is expected that these power spectra correspond to resolved structures.

\begin{figure}
 \centering
\includegraphics[width=0.6\columnwidth]{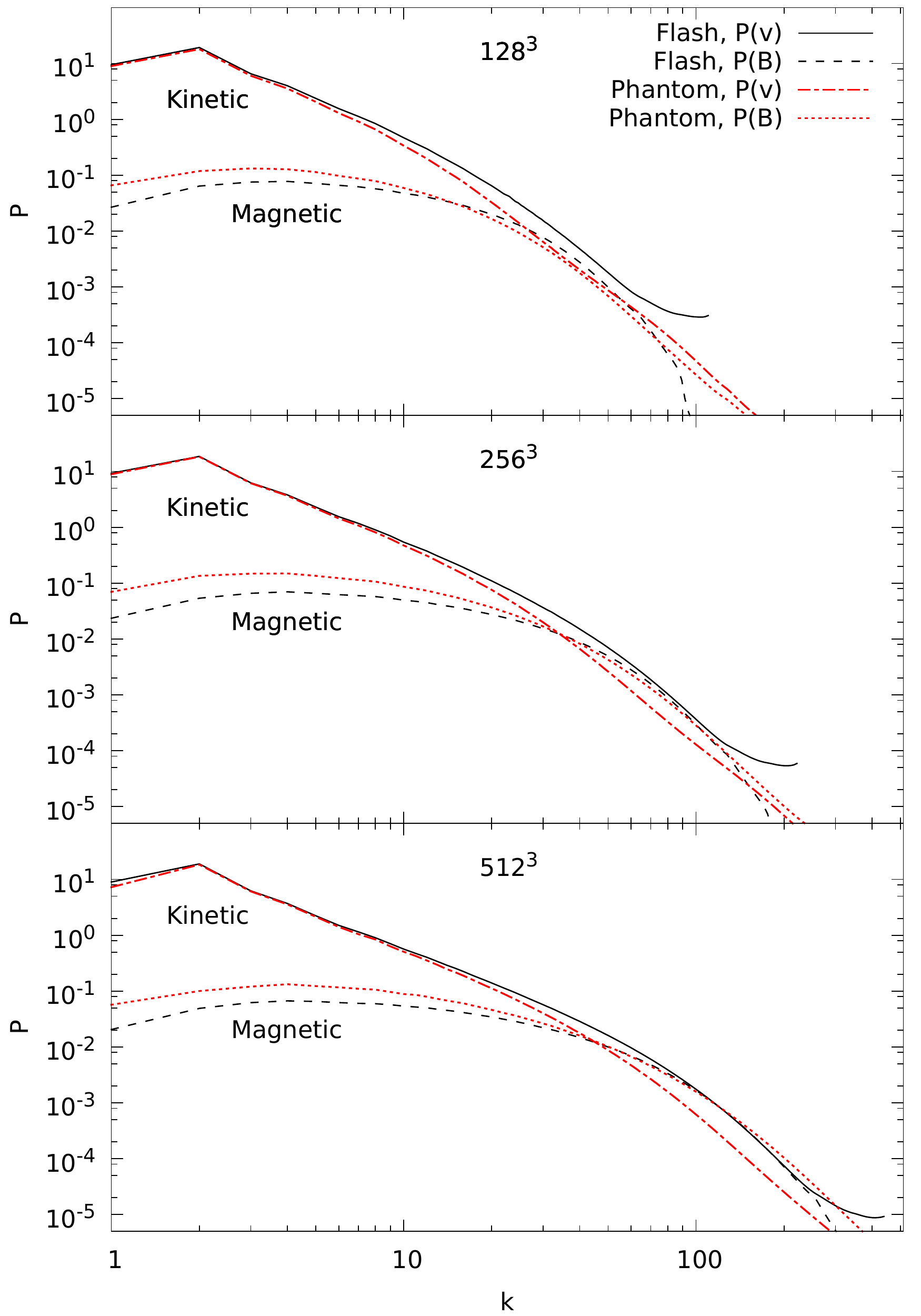}
\caption{Time averaged kinetic and magnetic spectra in the saturated phase for {\sc Flash} (black lines) and {\sc Phantom} (red lines). As the resolution is increased, the magnetic energy spectra begins to overtake the kinetic energy spectra at high $k$ as found by \citet{hbd04}.}
\label{fig:spectmagkin}
\end{figure}

Figure~\ref{fig:spectmagkin} compares the magnetic spectra to the kinetic energy spectra. It is characteristic for the small-scale dynamo for the peak in the magnetic energy spectrum to be at a wavenumber just above the peak in the kinetic energy spectrum \citep{cv00, bss12}. This is clearly seen in Figure~\ref{fig:spectmagkin}. The sharp peak at $k=2$ in the kinetic energy spectra is due to the driving force, and at all resolutions the peak of the magnetic energy spectra occurs just above this scale ($k\sim3$--$4$).
 
The relative amount of energy in the magnetic field increases with resolution in both codes. For {\sc Phantom}, the magnetic energy exceeds the kinetic energy at high $k$ at all three resolutions. For {\sc Flash}, the magnetic energy spectra approaches the kinetic energy spectra with increasing resolution, overtaking it only in the $512^{3}$ calculation. \citet*{hbd04} found that the magnetic energy spectra overtook the kinetic energy spectra at high $k$ in their ${\rm Pm}=1$ simulations, consistent with the {\sc Phantom} results, though we note that the exact nature of this behaviour will certainly depend upon the magnetic Prandtl number. 

\subsubsection{PDFs of $B^{2}$}

The red line in Figure~\ref{fig:bsqpdf} shows the time-averaged PDF of $B^2$ during the saturation phase, with the shaded region representing the standard deviation of the time-averaging. The distributions for {\sc Flash} and {\sc Phantom} peak at similar magnetic field strengths, agreeing to within 10\% on the maximum of the peak, with similar ranges on the high-end tail of the distribution. The maximum magnetic field achievable agrees to within 10\%. Similar to the growth phase, the low-end tail extends further for {\sc Flash}, and has a larger standard deviation. From the linearly scaled plots of Figure~\ref{fig:bsqpdf-linear}, it is seen that the probability of being at the mean magnetic field strength increases by ${\sim}20\%$ once the magnetic field has saturated, corresponding to a reduced variance in the distribution of magnetic field strengths. This occurs due to the saturation of the strongest magnetic fields \citep{schekochihinetal04b}. 

\subsubsection{Density PDFs}

\begin{figure}
 \centering
\includegraphics[width=0.49\columnwidth]{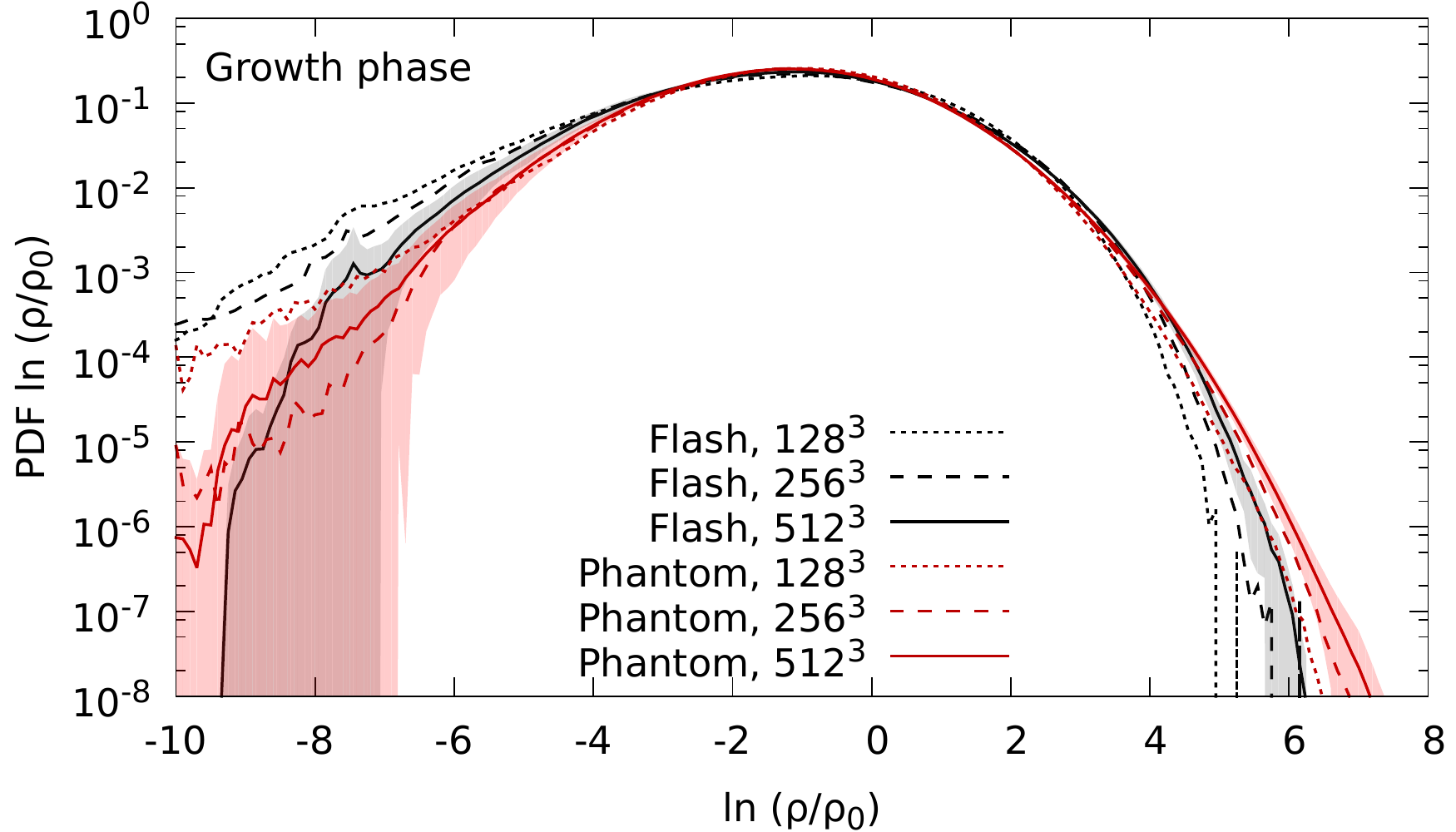} 
\includegraphics[width=0.49\columnwidth]{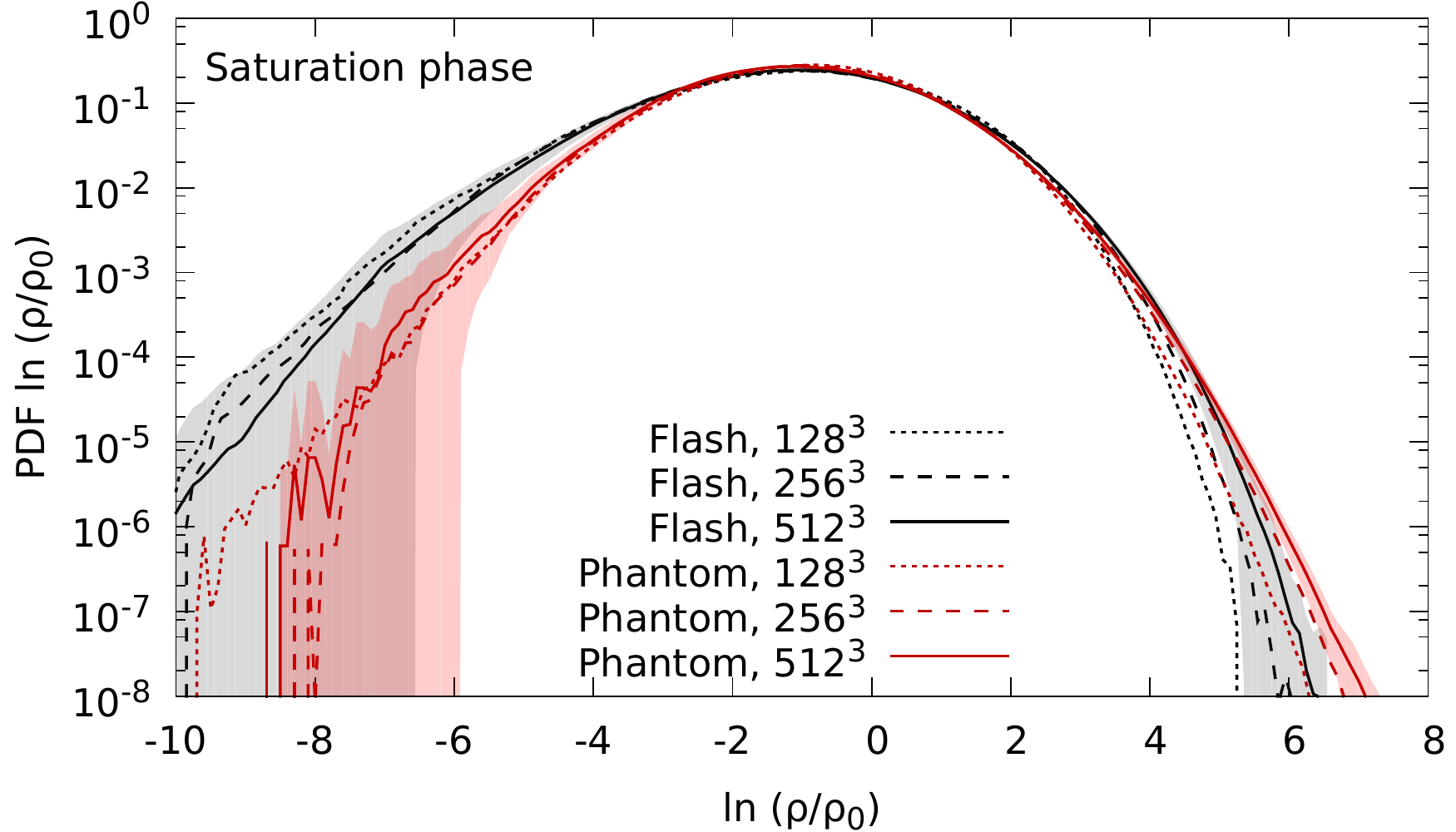}
\caption{Time averaged density PDFs during the growth phase (top panel, for $t/t_{\rm c}=2$--$10$) and during the saturation phase (bottom panel, for $t/t_{\rm c}=30$--$100$, only $t\ge50$ for the $128^3$ {\sc Phantom} calculation).  The peaks and high end tail of the PDF are similar for both cases, but the low density tail is less extended when the magnetic field has reached saturation.}
\label{fig:densitypdfs}
\end{figure}

Figure~\ref{fig:densitypdfs} compares the PDFs of the density contrast, $s\equiv \ln (\rho / \rho_0)$ (for a motivation of this choice of variable, see \citealt{vs94, fks08}), during the growth phase, while the magnetic field is dynamically weak, to the saturation phase when the magnetic field is at its strongest. The PDFs in the growth phase were time-averaged during the first half of the growth phase while $E_{\rm m} < 10^{-4}$ (excluding the initial transient growth). This allows for statistical averaging of a number of crossing times while the magnetic field is still dynamically weak. The PDFs in the saturation phase were averaged over at least $50 t_{\rm c}$. The standard deviation from the time averaging is shown for the highest resolution calculations by the shaded regions (black for {\sc Flash}, red for {\sc Phantom}).
 
It has been noted many times that, for supersonic turbulence, the PDF of $s$ follows a log-normal distribution \citep[e.g.,][]{vs94, pnj97, pvs98, np99, klessen00, ls08, fks08, federrathetal10, pf10, fk13}. This is a consequence of the density at a location being perturbed randomly and independently over time. According to the central limit theorem, the PDF will converge to a log-normal distribution \citep{papoulis84, vs94}. Other processes may affect the shape of the PDF. Self-gravity has been demonstrated to add power-law tails at high densities \citep[e.g.,][]{klessen00, lkmm03, fk12}, magnetic fields can narrow the width of the distribution effectively decreasing the compressibility of the gas \citep{collinsetal12, molinaetal12, fk13}, non-isothermal equations of state can introduce power-law tails at high and low densities \citep{pvs98, lkmm03}, and different forcing mechanisms (compressive vs solenoidal) influence the shape of the distribution \citep{fks08, federrathetal10}.

\begin{figure}
 \centering
\includegraphics[width=0.49\columnwidth]{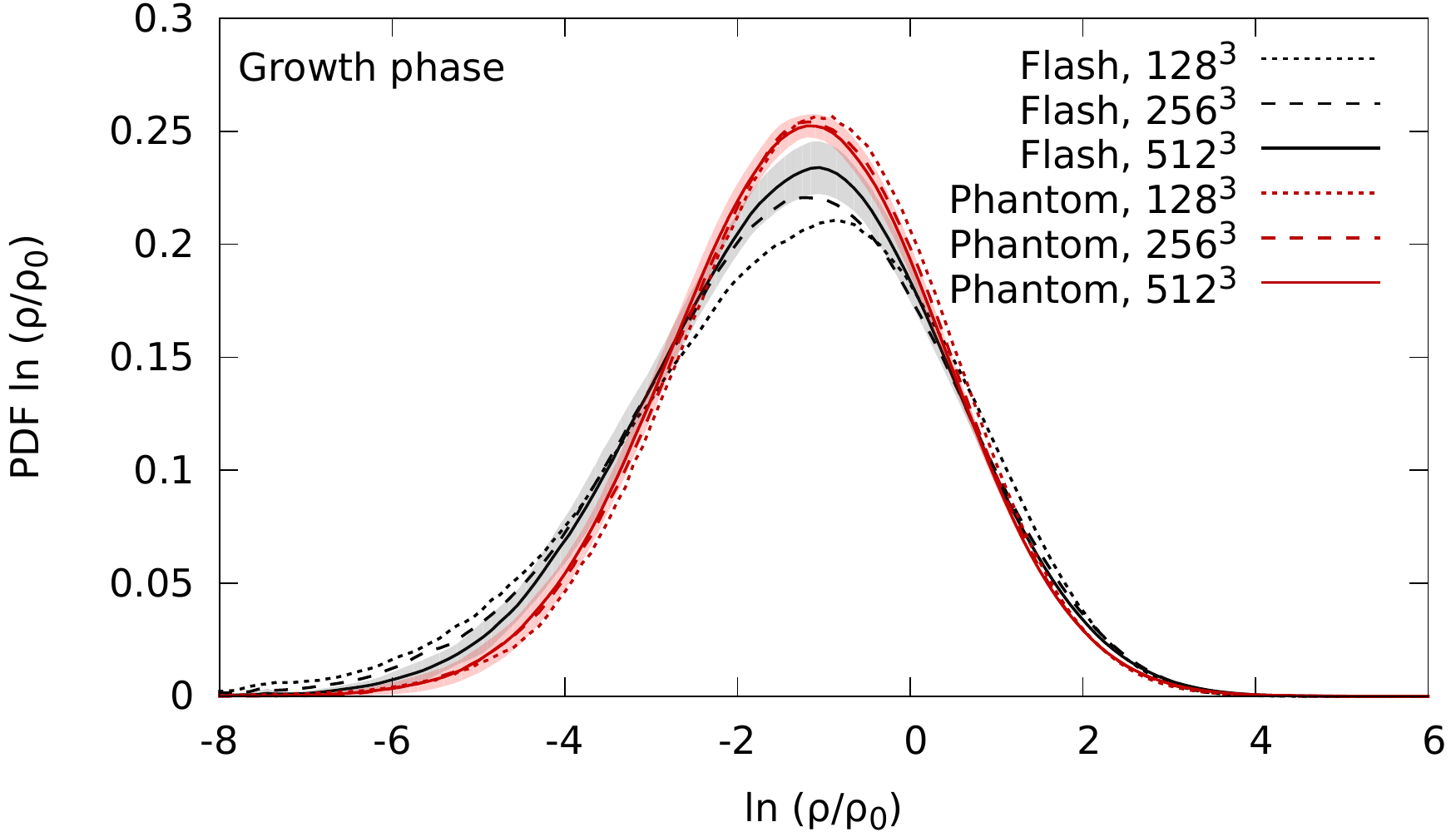} 
\includegraphics[width=0.49\columnwidth]{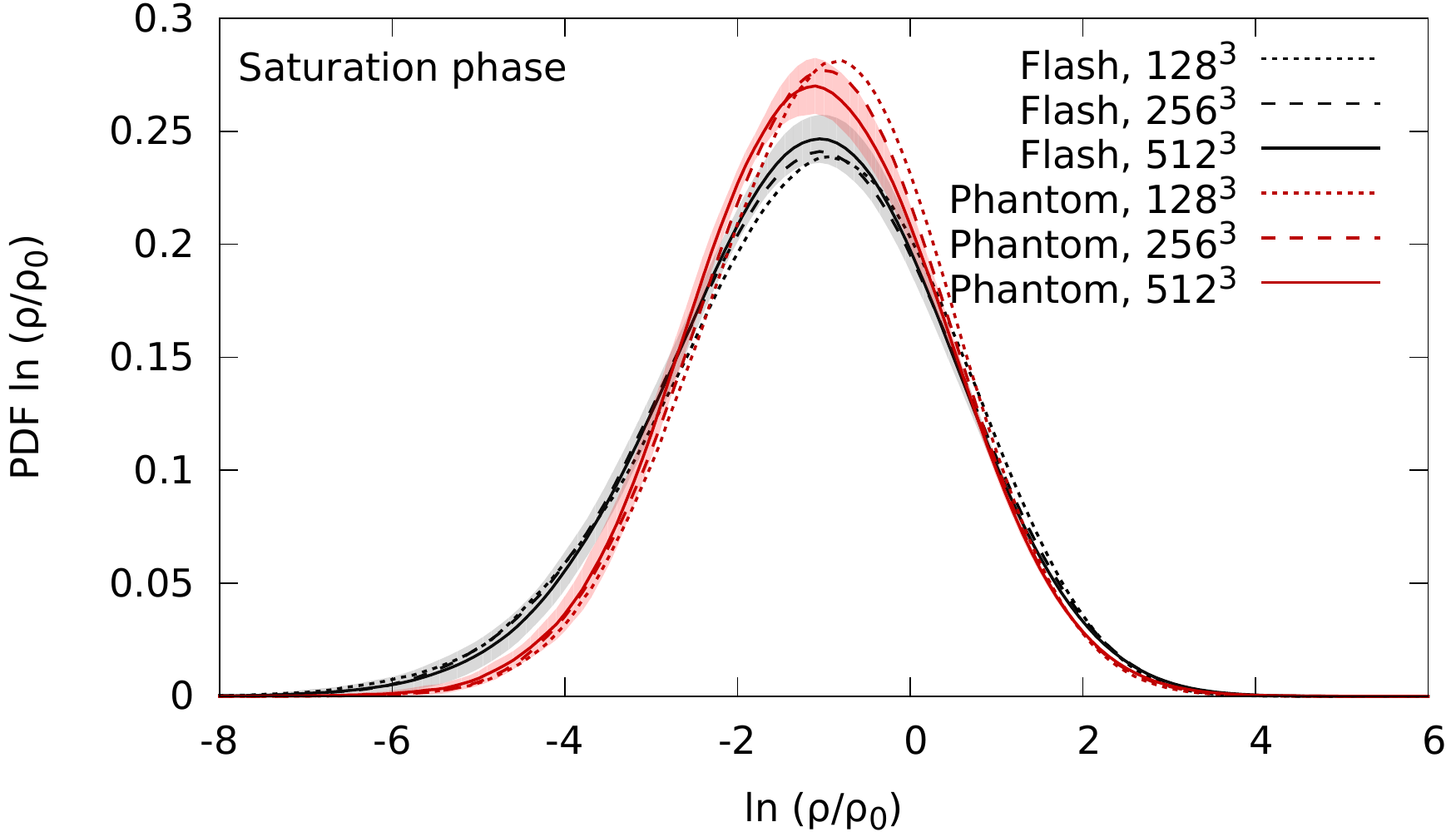}
\caption{Time averaged density PDFs during the growth phase (top panel, for $t/t_{\rm c}=2$--$20$) and during the saturation phase (bottom panel, for $t/t_{\rm c}=30$--$100$, only $t/t_{\rm c}\ge50$ for the $128^3$ {\sc Phantom} calculation).  This is equivalent to Figure~\ref{fig:densitypdfs} but on a linearly scaled plot.  The peaks and high end tail of the PDF are similar for both cases, but the low density tail is less extended when the magnetic field has reached saturation.}
\label{fig:densitypdfs-linear}
\end{figure}

Both codes show PDFs close to a log-normal distribution in both the growth and saturation phases. As expected, {\sc Flash} can be seen to sample a lower range of densities, while {\sc Phantom} samples a higher range. This behaviour is similar to that found by \citetalias{pf10}, and occurs because {\sc Phantom} uses adaptive resolution based on the density. The stronger magnetic field in the saturation phase reduces the low-end tail of the distribution, making it more log-normally distributed, consistent with previous findings \citep{klb07, ls08, pfb11, molinaetal12, fk13}. The peak and high-end tail of the distribution remain quite similar during both the growth and saturation phases. Figure~\ref{fig:densitypdfs-linear} shows the PDFs of $s$ on a linear scale. As before, the extended low-end tail for {\sc Flash} is visible, but from this it is also clear that the mean of the distribution is higher for {\sc Phantom}. This was also noted by \citetalias{pf10}.

\section{Conclusions}
\label{sec:turbsummary}

% - 
We have performed a comparison between grid-based ({\sc Flash}) and particle-based ({\sc Phantom}) MHD methods on the small-scale dynamo amplification of magnetic fields. The calculations used supersonic turbulence driven at rms Mach 10 in an isothermal fluid contained in a periodic box. The initial magnetic field was uniform and had an energy $12$ orders of magnitude smaller than the mean kinetic energy. 

Our conclusions are as follows:

i) Both codes exhibited similar qualitative behaviour. The initially weak magnetic field was exponentially amplified at a steady rate over a period of tens of turbulent crossing times, saturating when the magnetic energy was $2$--$4$\% of the kinetic energy.  

ii) The growth rate of magnetic energy in the {\sc Flash} calculations varies only slightly with resolution (5--10\%), while the {\sc Phantom} calculations are sensitive to resolution (nearly doubling with each factor of two increase in resolution). This is due to the resolution scaling of the artificial dissipation terms. 

iii) For ${\rm Pm}\sim1$, the saturation level of magnetic energy appears to primarily depend on the magnetic Reynolds number, with little variation as the magnetic Prandtl number varies. 

iv) Although {\sc Phantom} is more computationally expensive, with the $256^3$ {\sc Phantom} calculation taking as many cpu-hours as the $512^3$ {\sc Flash} calculation, our results indicate the mean magnetic energy at saturation is comparable to {\sc Flash} calculations at higher resolution.

v) Both codes show magnetic energy spectra at saturation that is relatively flat at large-scale, peaking around $k\sim3$--$5$.

vi) During the growth phase, both codes produce a log-normal PDF of $B^2$ which linearly translates to higher magnetic field strengths over time.

vii) During the saturation phase, the PDF of $B^2$ in both codes becomes skewed, sampling a smaller range of magnetic field strengths.

This comparison has shown that SPMHD is capable of simulating the small-scale dynamo amplification of magnetic fields. We have performed this comparison using the ideal MHD equations, as many astrophysical simulations are performed under the assumption of ideal MHD. The most pressing follow-up to this work would be to perform a comparison using fixed physical dissipation terms in order to obtain agreement on the physical scaling of the growth rates.

%% file: conclusion.tex
%%
%% Conclusion chapter
%%

\chapter{Conclusion}
\label{sec:chapter-conclusion}

In this thesis, methods have been developed to improve the representation of magnetic fields in smoothed particle magnetohydrodynamics (SPMHD). The ideal MHD equations can be straightforwardly added to SPH with naive simplicity, however the resulting algorithm has numerical difficulties in simulating real astrophysical applications. The difficulty stems from the divergence-free constraint of the magnetic field. Maxwell's equations cleanly state $\nabla \cdot {\bf B}=0$, but the induction equation, which describes the evolution of a magnetic field, does not manifestly enforce this.

\section{Summary}

\subsection{Chapter 2 --- Smoothed particle magnetohydrodynamics}
Chapter~\ref{sec:chapter-spmhd} provided an overview of SPMHD. Insight into the numerical challenges surrounding $\nabla \cdot {\bf B}$ was obtained by deriving the ideal MHD equations in the continuum limit from the Euler fluid equations, Maxwell's equations, and the Lorentz force law. Developing the discretisation of these equations into SPMHD started with the density estimate and variable smoothing length formulation. Using basic interpolation theory, the discretised version of the induction, energy, and momentum equations were obtained. Through understanding of the continuum equations, the instability present in the momentum equations was understood as arising from the inclusion of a term proportional to $\nabla \cdot {\bf B}$. Strategies to address this instability were presented. Methods for capturing shocks and discontinuities were also discussed. 

\subsection{Chapter 3 --- Constrained hyperbolic divergence cleaning}

In Chapter~\ref{sec:chapter-cleaning}, we developed a constrained implementation of mixed hyperbolic/parabolic divergence cleaning based on the \citet{dedneretal02} method \citep[see also][]{tp12}. The premise is to couple a scalar field, $\psi$, to the magnetic field via a set of hyperbolic equations. These are used to propagate divergence through the magnetic field as a series of waves, with a parabolic diffusion term removing divergence error from the magnetic field. The hyperbolic waves make the parabolic diffusion more effective, as it increases the volume on which it can act.

In developing the numerical implementation of the cleaning system, it was useful to consider the physical picture of how it operates. Energy is transferred from the magnetic field into the $\psi$ field, then transported to a different location in a wavelike manner. It is key that energy be conserved in this process, otherwise, the waves might contain more energy than they did initially, leading to an increase in divergence error. This behaviour was indeed found by \citet{pm05} in their implementation. One circumstance where this may occur is at sharp density contrasts, where it can lead to a catastrophic increase in energy (see shocktube test 1B in \citealt{sdb13}; also Section~\ref{sec:test-density-jump}). \citet{sdb13} used an artificial limiter to restrict the cleaning scheme in an attempt to reduce the severity of this error, but this does not address the fundamental issue of spurious energy production and reduces the overall effectiveness of the scheme.

Our method is `constrained' in that it manifestly conserves the energy in the hyperbolic system of equations, inherently preventing spurious increases in the divergence of the magnetic field \citep[consider again shocktube test 1B using our method in Section~\ref{sec:shock1b} and in][]{tp13}. This was accomplished by defining the energy content of the $\psi$ field, then including it in the system Lagrangian. This allowed the discretised version of the hyperbolic equations to be obtained in the same manner as the SPMHD equations of motion --- yielding versions which are guaranteed to conserve energy and to always decrease divergence error. This implementation was found to significantly increase the stability and robustness of the method, in particular solving issues with sharp density contrasts and free boundaries. This divergence cleaning scheme was found to yield an order of magnitude reduction of average divergence error on most test problems and astrophysical applications, and was responsible for the successful simulation of jets during the first hydrostatic core phase of star formation (\citealt{ptb12}, see also \citealt{tpfb13, bpt14}), and outflows during stellar core formation (\citealt{btp14}, see also \citealt{pbt13}).

Enhancing the effectiveness of the divergence cleaning method was tested in two ways. One, by explicitly increasing the hyperbolic wave speed, with a corresponding reduction in timestep (over-cleaning). Two, by iterating the cleaning equations in-between timesteps (sub-cycling). We found that both approaches yield similar reductions in divergence error for the same number of steps (i.e., a factor of 10 increase in wave speed yields results similar to 10 iterations of sub-cycling the cleaning equations). Using sub-cycling, the divergence of the magnetic field may be set arbitrarily low provided a sufficient number of iterations are taken. 

A formulation of the constrained hyperbolic divergence cleaning method was developed for the velocity field, for use in weakly compressible SPH simulations \citep[see also][]{tp12b}. The aim is to improve the representation of incompressibility. Tests on an oscillating water drop reduced the magnitude of density variations by half, with negligible kinetic energy dissipation.

\subsection{Chapter 4 --- A switch to reduce resistivity}

In Chapter~\ref{sec:chapter-switch}, a new switch was developed and tested that dynamically regulates the amount of artificial resistivity applied to the magnetic field \citep[see also][]{tp13}. The motivation for a new switch arose during the course of a comparison of supersonic magnetised turbulence (Chapter~\ref{sec:chapter-mhdturb}). It was discovered that the \citetalias{pm05} switch completely failed to work for highly super-Alfv\'enic shocks. Therefore, a new switch was required which could detect discontinuities in the magnetic field regardless of the magnetic field strength. The new switch accomplished this by measuring the gradient of the magnetic field at the resolution scale normalised by the magnitude of the magnetic field. This measures the relative size of the jump in magnetic field strength, allowing it to detect discontinuities regardless of the absolute magnetic field strength.

This switch not only solved problems in detecting shocks in very weak magnetic fields, but was found to out-perform the \citetalias{pm05} switch in every respect. It reduced the L1 error by $7$--$45\%$ in a number of shocktube tests (containing a variety of magnetic shock types) compared to the \citetalias{pm05} switch. It resulted in lower magnetic energy dissipation in simulations of a circularly polarised Alfv\'en wave and the Orszag-Tang vortex problem. We concluded that it should be adopted for general use in SPMHD simulations.
% fixed the deficiencies of the \citetalias{pm05} in the super-Alfv\'enic regime, lead to more accurate shock solutions, and is less dissipative.

The concept of the switch --- using a normalised shock indicator --- was used to construct switches for artificial viscosity and thermal conductivity \citep[see also][]{tp13b}. The new artificial viscosity switch required a slow decay of $\alpha$ in order to treat post-shock oscillations, and was found to reproduce correct shock profiles in the Sod shocktube test.

\subsection{Chapter 5 --- Turbulent dynamo amplification of magnetic fields}

In Chapter~\ref{sec:chapter-mhdturb}, we compared SPMHD (using the code {\sc Phantom}) with grid-based methods ({\sc Flash}) on the simulation of the turbulent small-scale dynamo amplification of a magnetic field (see also \citealt{tpf14}, and some early work which appeared in \citealt{tpf13, tpfb13}). The turbulent motions on the smallest length scale efficiently twist and wind the magnetic field, imparting kinetic energy from the motion of the fluid into magnetic energy. The magnetic field as a whole grows through a reverse cascade, with energy being transported from small into large scales. This dynamo is an example of a fast dynamo, in that it leads to exponential amplification of the magnetic field. Since the dynamo operates near the smallest scales in the system, its growth rate is set by the combination of kinetic and resistive dissipation (kinetic and magnetic Reynolds numbers, and the ratio of the two, the magnetic Prandtl number). Once the magnetic field saturates on small scales, it enters a linear or quadratic growth phase as the reverse cascade continues to amplify large scales. The magnetic field fully saturates once the conversion of kinetic energy to magnetic energy is balanced by the dissipation of energy.

The simulations used supersonic turbulence to drive the dynamo. The conditions were representative of the interior of molecular clouds --- the gas was isothermal, and had root mean square velocity of Mach 10. The turbulence was sustained through a stochastic driving force, with the pattern pre-generated so that both codes followed the same pattern as closely as possible. The initial conditions --- uniform density, zero velocity, and a uniform magnetic field --- were simple for the same reason. The simulations used a box of unit length with periodic boundary conditions, and were performed at three resolutions: $128^3$, $256^3$, and $512^3$.

Both methods had similar qualitative behaviour, with the magnetic energy being steadily amplified at an exponential rate until saturation. The saturation level of the mean magnetic energy was consistent at around $2$--$4\%$ of the mean kinetic energy, increasing with higher magnetic Reynolds numbers (i.e., higher resolution), though the saturation energy in the {\sc Phantom} calculations was approximately equal to the {\sc Flash} calculations at double the resolution. 

The spectra of the magnetic energy had similar shape between the two methods, showing a relatively flat spectrum at low wavenumbers ($k<10$).  In all calculations, the dynamo saturated the magnetic energy at smallest scales first. This marked the onset of the linear growth phase, during which the reverse cascade of magnetic energy slowly saturated the large scales. The saturated level of magnetic energy was similar at medium and small scales between the methods, though with the {\sc Phantom} calculations containing twice as much magnetic energy at large scales ($k<10$) compared to {\sc Flash}. 

%As similarly found by \citet{hbd04}, the magnetic energy spectra overtook the kinetic energy spectra at high wavenumbers, which was resolved in the $128^3$ {\sc Phantom} calculations (and higher) but only in the highest resolution ($512^3$) {\sc Flash} calculations.

The rate of magnetic energy amplification differed between the two methods. For {\sc Flash}, similar growth rates were obtained for the three resolutions simulated (within 5--10\% difference). However, {\sc Phantom} showed a stronger resolution dependence, with the growth rate nearly doubling with each factor of two increase in resolution. This effect was determined to be caused by the artificial viscosity and resistivity. Changing the dimensionless parameters in the dissipation terms ($\alpha$ and $\alpha_{\rm B}$) lead to the same effect as changing the resolution.

The probability distribution function (PDF) of magnetic field strengths was log-normal during the amplification phase. Its width remained constant, with the mean smoothly translating to higher field strengths over time. As the magnetic field approached saturation, the dynamo was unable to continue amplifying the strongest regions of the fields and the high-end tail of the distribution anchored in place. However, the low-end tail continued to increase, leading to a skewed distribution as the high-end tail became `squeezed'. Both methods exhibited this behaviour, agreeing to within 10\% on the width and peak of the distributions, and on the location of the `anchor' point of the high-end tail.

\section{Future work}

The direct extension to this thesis would be to continue developing the algorithms of SPMHD. The need for continued development will be best determined through application of the current methods to astrophysical problems. While the methods developed here have given a path forward on many problems, stress testing them through application will illuminate any weaknesses that need be addressed. 

There are several aspects of the method that could be thought about further. The tensile instability in the equations of motion is understood as arising from the term proportional to $\nabla \cdot {\bf B}$ in the continuum equations. Subtracting this term solves the instability, and is attractive for its simplicity and physical motivation, though creates equations of motions which do not conserve momentum. It is not clear how to improve upon this aside from employing better methods to uphold the divergence-free constraint on the magnetic field. 

In addition, the shock capturing methods used in SPMHD are rather crude in comparison to grid-based methods. As such, the numerical dissipation in SPMHD is typically higher than `standard' grid-based methods. One issue is that artificial resistivity uses the fast MHD wave speed for the capture of all shock types, introducing a level of dissipation in excess of what is required for pure Alfv\'en waves. \citet{ii11} developed a Godunov style shock solver for SPMHD that yields lower dissipation, and it would be interesting to investigate such approaches further.

The resolution of the magnetic field is tied to the mass. This is typically not a problem, as for the majority of astrophysical systems the regions of high mass are the most relevant. However consider the case of a neutron star. The magnetic field outside the neutron star is important, yet the density contrast at the surface is so large that the resolution of the magnetic field outside will be essentially absent if using equal mass particles. Also, consider the case of an accretion disc. SPH can simulate free boundaries easily, however, if the magnetic field is vertically aligned, then the field lines on the top and bottom boundaries of the disc are unphysically cut.

The immediate research plans for the future are for the following investigations:

\subsection{Exactly upholding $\nabla \cdot {\bf B}=0$}

The first is to find a suitable method to exactly maintain the divergence-free constraint of the magnetic field. The constrained hyperbolic divergence cleaning method only approximately upholds the divergence-free constraint, and it would be better to {\it exactly} maintain it. The Euler potentials and vector potential approaches have been already investigated as possible solutions, but neither are tenable in practice, and it is not clear how to adapt constrained transport to a particle method. Revisiting projection methods offer the best hope. \citetalias{pm05} have already developed scalar and vector projection methods, and testing these further would be of considerable interest. The difficulty is the computational cost involved, which only becomes trickier when dealing with individual timesteps.

% t1 = +/- 0.025
% t20 = 
% t25 = +/- 0.25
\begin{figure}
 \centering
\includegraphics[width=0.25\columnwidth]{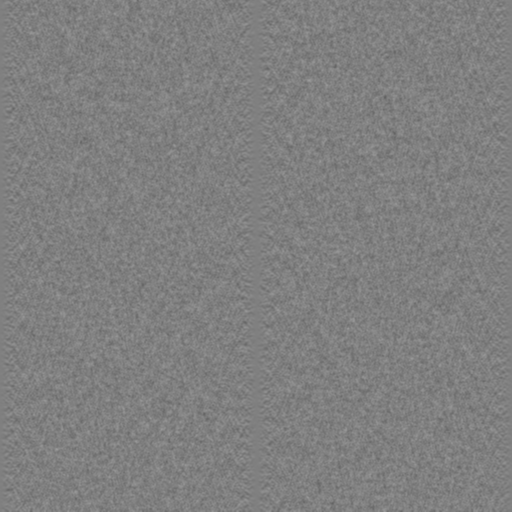}
\includegraphics[width=0.25\columnwidth]{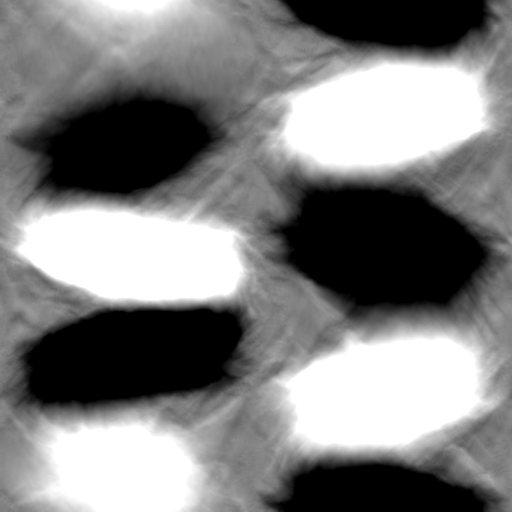}
\includegraphics[width=0.25\columnwidth]{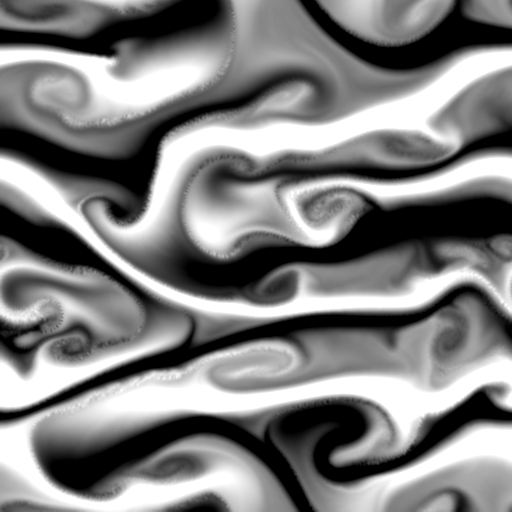} \\
\includegraphics[width=0.25\columnwidth]{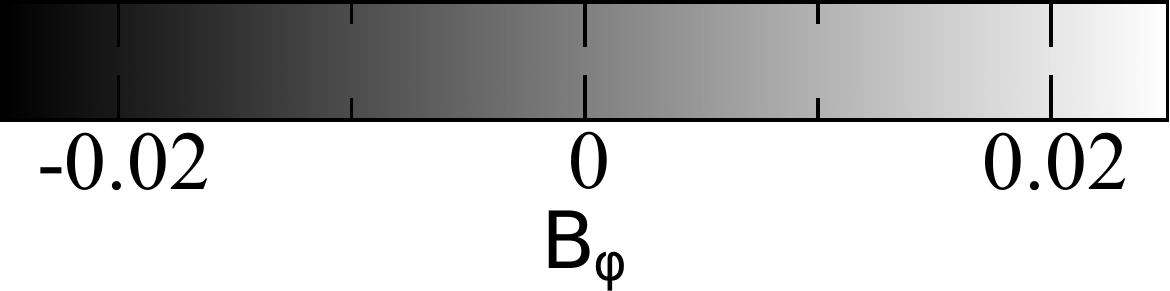}
\includegraphics[width=0.25\columnwidth]{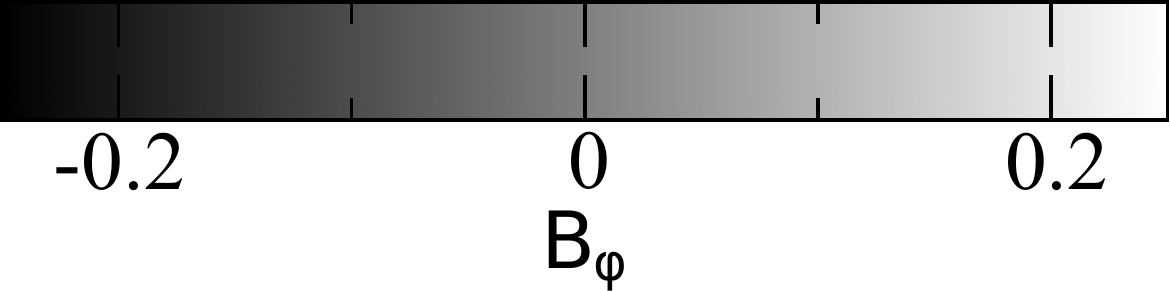}
\includegraphics[width=0.25\columnwidth]{mri/mri-cobar2.pdf}
\caption{Snapshots of $B_\phi$ at $t=1$, $20$, and $25\Omega$ for the $512^2$ 2D shearing box MRI test. Random small motions in the velocity lead to perturbations in the magnetic field ($t=1\Omega$). These coalesce to form large structures ($t=20\Omega$), which lead to the generation of turbulence ($t=25\Omega$). Renderings are not all on the same scale.}
\label{fig:mri-evolution}
\end{figure}

\subsection{Magneto-rotational instability}

The second research plan is to investigate SPMHD's ability to simulate the magneto-rotational instability (MRI). A significant amount of work has been performed with grid-based methods on local simulations of accretion discs using the shearing box approximation. However, this has been shown to be deficient in capturing the global physics of accretion discs \citep[e.g.,][]{fp07, kpl07, pb13}. Due to its excellent angular momentum conservation and ability to simulate complex geometries and free boundaries, SPH is quite natural for simulating global accretion discs. Demonstrating that SPMHD can simulate the MRI would open a wide range of physical problems for future study.

\begin{figure}
 \centering
\includegraphics[width=0.65\columnwidth]{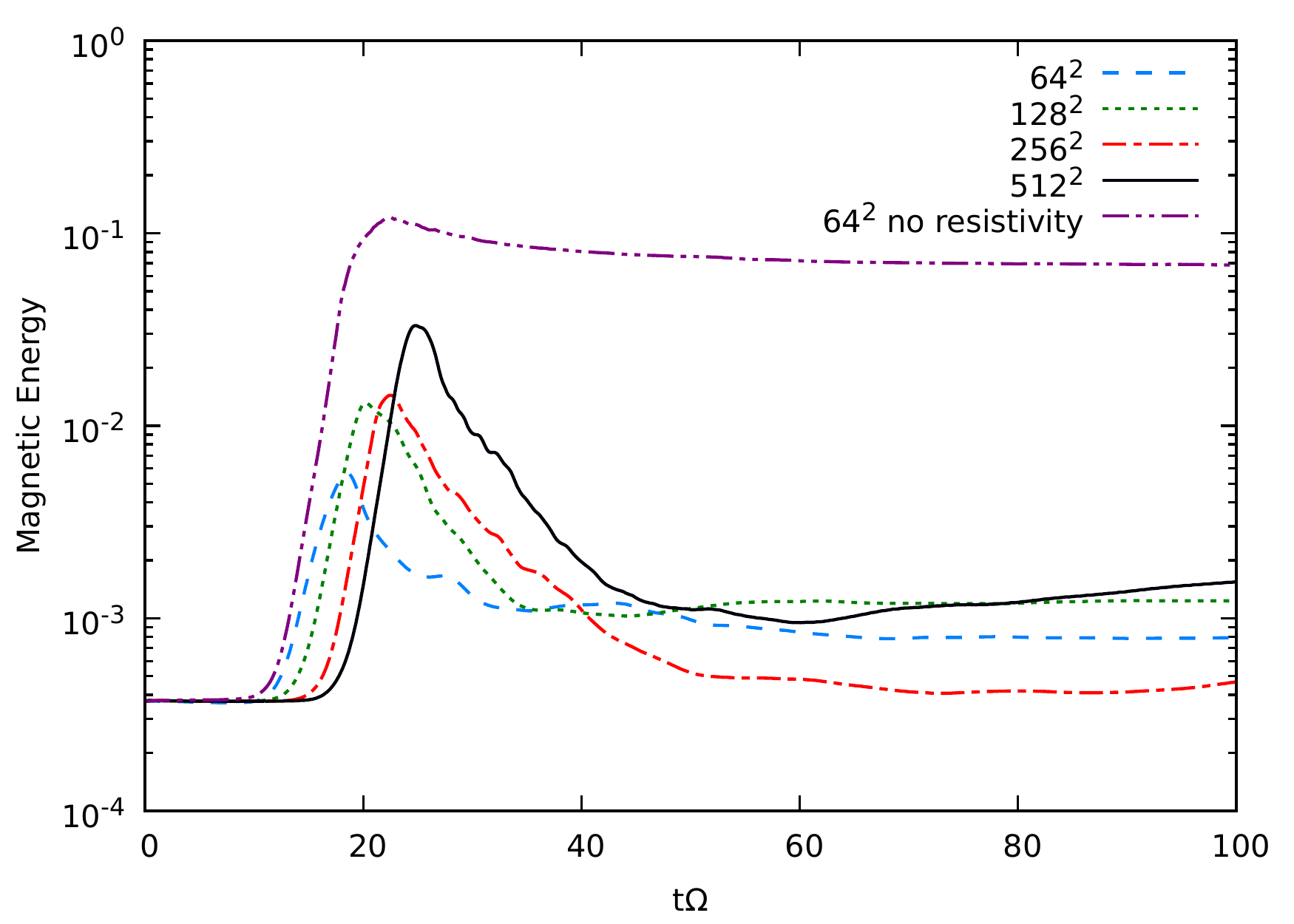}
\caption{The evolution of magnetic energy for the 2D shearing box test. After several orbits, the stretching of the magnetic field lines triggers the MRI leading to exponential growth of magnetic energy.}
\label{fig:mri-energy}
\end{figure}

Preliminary work has been performed on simulating the MRI in SPMHD. \citet*{vkp14} used simulations of the MRI to test their SPMHD code, finding their results to have qualitative agreement with the grid-based simulations of \citet{gg08}. In the following, we describe the results of simulations of the MRI in 2D shearing boxes. The initial density is uniform $\rho=1$. The equation of state is isothermal. The initial magnetic field is a sine wave in the $z$-direction, with amplitude defined to have plasma beta $\beta=1348$ ($B_z\approx0.0358$). Random perturbations are introduced to the velocity field on the scale of $0.01~c_{\rm s}$. We note that these simulations use the quintic spline kernel in order to adequately resolve these small perturbations. These conditions mimic the fiducial model of \citet{gg08} and model S1 of \citet{hb92}. We perform the simulations for $100$ orbital periods ($t=100\Omega$ where $\Omega$ is the orbital frequency) at resolutions of $64^2$, $128^2$, $256^2$, and $512^2$.  Since the flow is subsonic, we used the averaged Alfv\'en speed in the artificial resistivity signal velocity, rather than the fast MHD wave speed. We found that at low resolutions, the dissipation from using the fast MHD wave speed can prevent the instability from activating. Similarly, \citet{vkp14} comment that they use only the $\nabla \cdot {\bf B}$ source term when using the \citetalias{pm05} switch in order to avoid excessive dissipation.

Snapshots of the $\phi$ component of the magnetic field at $t=1$, $20$, and $25\Omega$ are presented in Figure~\ref{fig:mri-evolution}. The correct qualitative behaviour is obtained: random motions in the initial conditions induce small perturbations in the magnetic field, which coalesce leading to turbulent motion. The magnetic energy as a function of time is given by Figure~\ref{fig:mri-energy}. The MRI leads to an exponential increase in magnetic energy which decays once turbulence has formed. The location of the peak is in agreement with \citet{gg08}. It occurs later for higher resolutions because the energy contained in the perturbations is at smaller scales, requiring more time for the perturbations to coalesce into large structures. The maximum magnetic energy achieved increases with resolution due to the reduction in numerical dissipation.  Notably, the decay in magnetic energy is more rapid than in the grid-based results. This is due to the level of magnetic energy dissipation. The switch developed in Chapter~\ref{sec:chapter-switch} will activate when the field undergoes reversals in direction (${\bf B}\to0$), which is the correct behaviour at current sheets, but adds an unnecessary level of dissipation for this problem. We have performed a $64^2$ resolution calculation with no artificial resistivity applied, the results of which are included in Figure~\ref{fig:mri-energy}. The maximum energy level achieved is over an order of magnitude higher than the same resolution with artificial resistivity, with only a slow decay in magnetic energy after. This preliminary work shows promise that SPMHD can activate the MRI, though further work is needed.

\section{Conclusion}

The methods developed in this thesis have significantly advanced SPMHD as a numerical method capable of performing astrophysical MHD simulation. The proof is in the pudding: they have already lead to published work on the role of magnetic fields in the formation of protostars \citep{ptb12, btp14}. To quote \citet{pm05},
\begin{quote}
Our results suggest that the method is ripe for application to problems of current theoretical interest, such as that of star formation.
\end{quote}

%% file: cleaning-appendix.tex
\chapter{Artificial $\psi$-dissipation term}
\label{sec:appendix-cleaning-dissipation}

\begin{figure}
\centering
\includegraphics[width=0.45\textwidth]{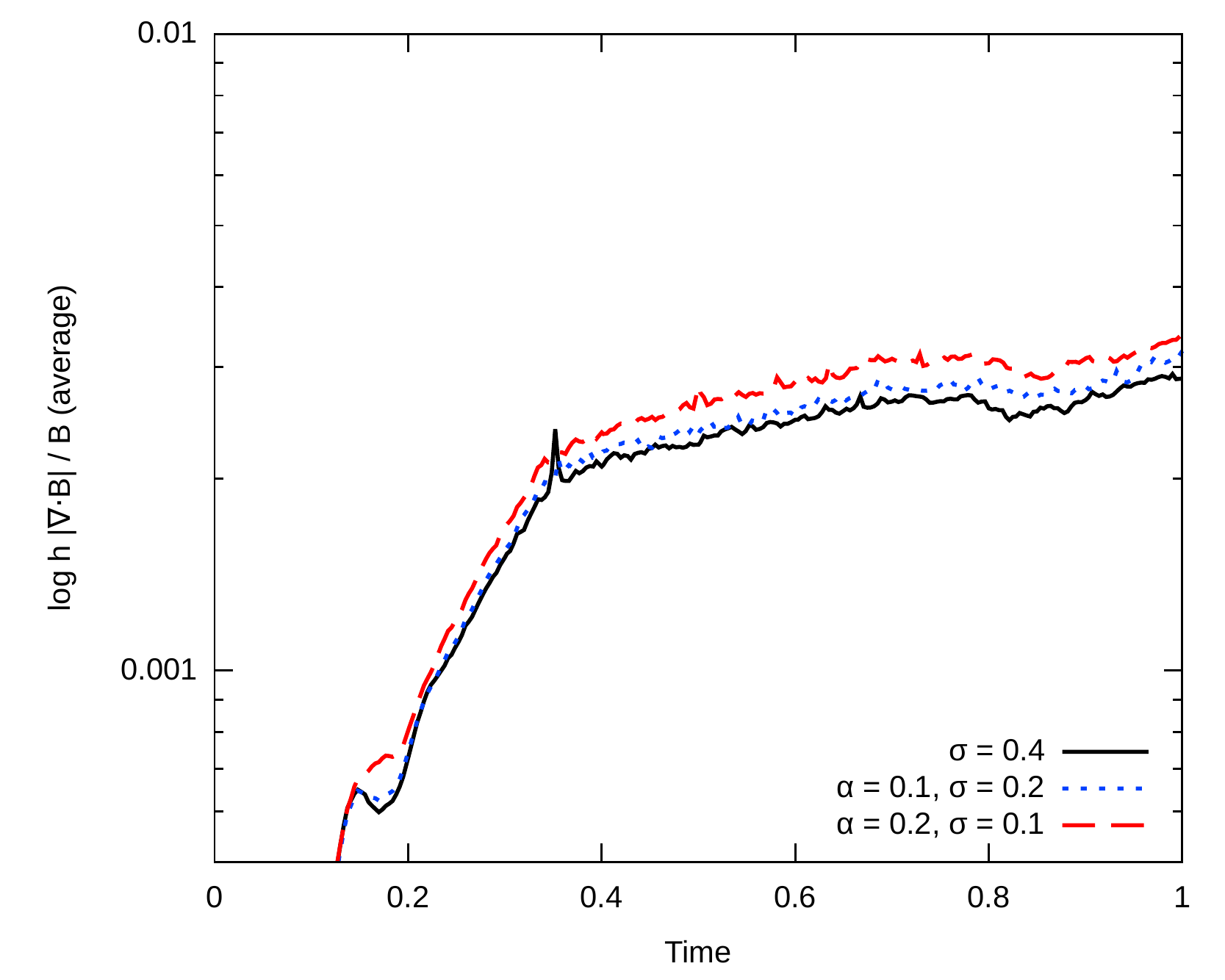}
\includegraphics[width=0.45\textwidth]{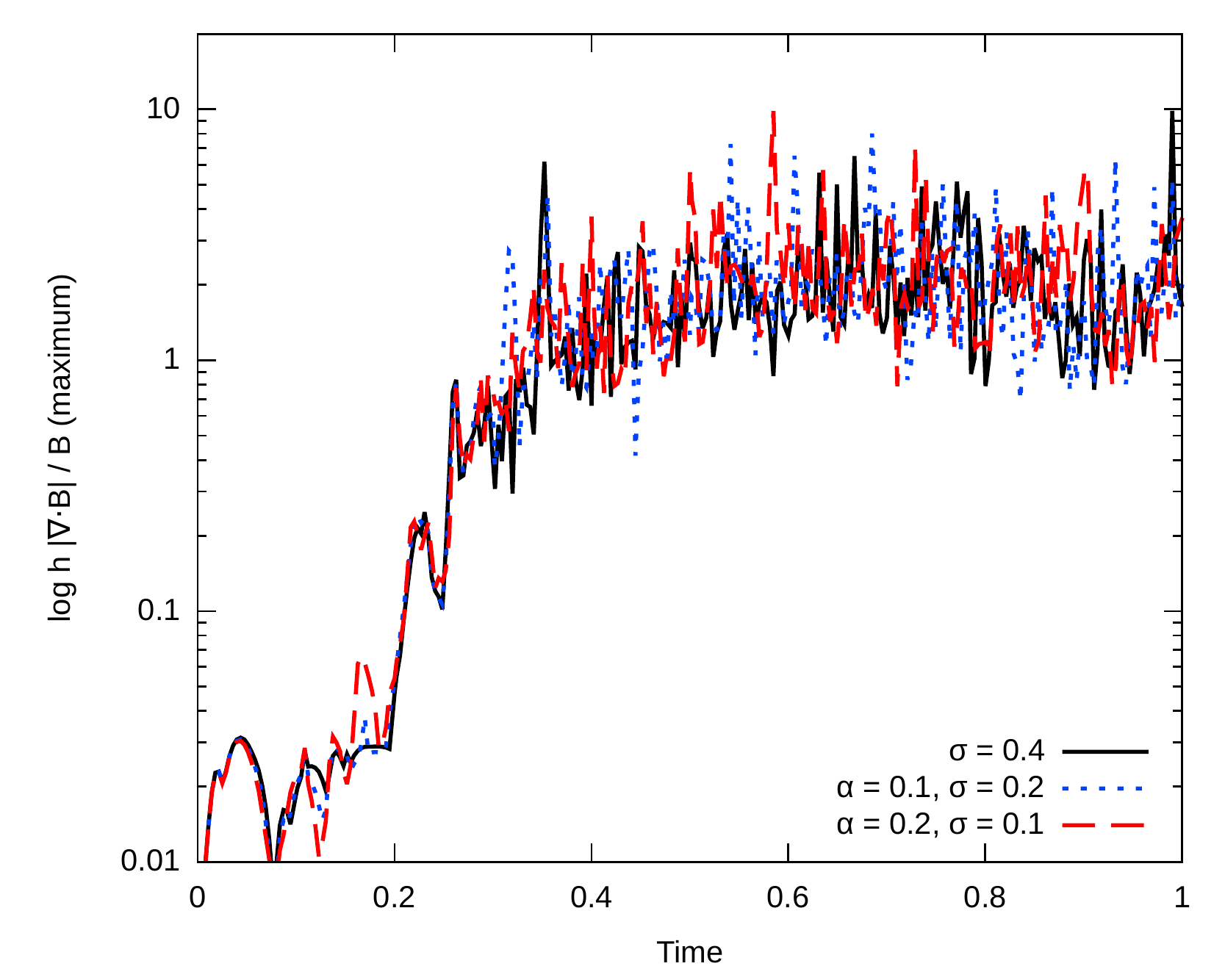}
\caption{Average and maximum divergence error when including the new, artificial $\psi$ dissipation term in the Orszag-Tang vortex test.  Values of $\alpha_\psi$ and $\sigma_\psi$ are chosen so that the combination is close to critical damping, however no benefit is noted over use of the regular damping term.}
\label{fig:dissipation}
\end{figure}

Although the hyperbolic divergence cleaning method developed in Chapter~\ref{sec:chapter-cleaning} already includes a damping term to reduce $\psi$, we have investigated the addition of a new dissipation term, analogous to artificial resistivity or viscosity, of the form
\begin{equation}
\left( \frac{{\rm d}\psi_a}{{\rm d}t} \right)_\text{diss} = \rho_a \sum_b m_b \frac{c_{\rm h} \alpha_\psi}{\overline{\rho}_{ab}^2} \left( \psi_a - \psi_b \right) F_{ab},
\label{eq:artificial-psi-dissipation}
\end{equation}
where $\nabla W_{ab} = \hat{\bf r}_{ab} F_{ab}$. This dissipation term is mainly designed to capture discontinuities in the $\psi$ field, motivated by our neglect of the surface integral term in Equation~\ref{eq:psi_energy_derivation}. The term is essentially an SPH expression for a diffusion term of the form $\eta_{\psi} \nabla^{2} \psi$, where $\eta_{\psi} \propto \alpha_{\psi} c_{\rm h} h$, which in comparison to the damping term, acts more strongly to smooth relative differences in $\psi$. This artificial $\psi$-dissipation can be used in conjunction with the damping term, however since both the damping and diffusion terms dissipate $\psi$, it is important that values of $\alpha_\psi$ and $\sigma_\psi$ be chosen carefully to avoid overdamping the system.  For example, we found that in our two dimensional tests that propagation of divergence waves were damped too severely with $\alpha_\psi=1$, and that using $\alpha_\psi, \sigma_\psi = [0.1, 0.2]$ or $[0.2, 0.1]$ yielded near critical damping (see Figure~\ref{fig:dissipation}).

For this dissipation term, the energy loss is given by
\begin{equation}
\sum_a m_a \frac{\psi_a}{\mu_0 \rho_a c_{\rm h}^2} \left( \frac{{\rm d}\psi_a}{{\rm d}t} \right)_{\text{diss}} = \sum_a m_a \frac{\psi_a}{\mu_0 c_{\rm h}} \sum_b m_b \frac{\alpha_\psi}{\overline{\rho}_{ab}^2} \left(\psi_a - \psi_b\right) F_{ab}.
\end{equation}
This can be shown to be negative definite by splitting the RHS into two halves, performing a change of summation indices on the second half, then rejoining to obtain
\begin{equation}
-\frac{1}{2} \sum_a m_a \frac{\alpha_\psi}{\mu_0 c_{\rm h}} \sum_b m_b \frac{\left(\psi_a - \psi_b\right)^2}{\overline{\rho}_{ab}^2} F_{ab},
\end{equation}
which, since $F_{ab}$ is negative for positive kernels, gives a negative definite contribution to the total energy (and conversely would give a positive definite heat contribution).

Inclusion of the dissipation term was tried with all test cases presented in Section~\ref{sec:cleaningtests}.  Similar reductions in the divergence error were obtained, however no results were improved beyond that of using the damping alone (Fig.~\ref{fig:dissipation}).

%% file: turbcomp-appendix1.tex
\chapter{Interpolating particle data to a grid}
\label{sec:gridinterp}

The power spectra of kinetic and magnetic energy of the {\sc Phantom} calculations (Chapter~\ref{sec:chapter-mhdturb}) are computed by interpolating the particle data to a grid using an SPH kernel weighted summation over neighbouring particles. We have investigated whether using mass weighted or volume weighted interpolation changes the results. Furthermore, we have tested grids of varying resolution to find the optimal grid resolution to properly represent the magnetic field.

\begin{figure}
\centering
 \includegraphics[width=0.6\columnwidth]{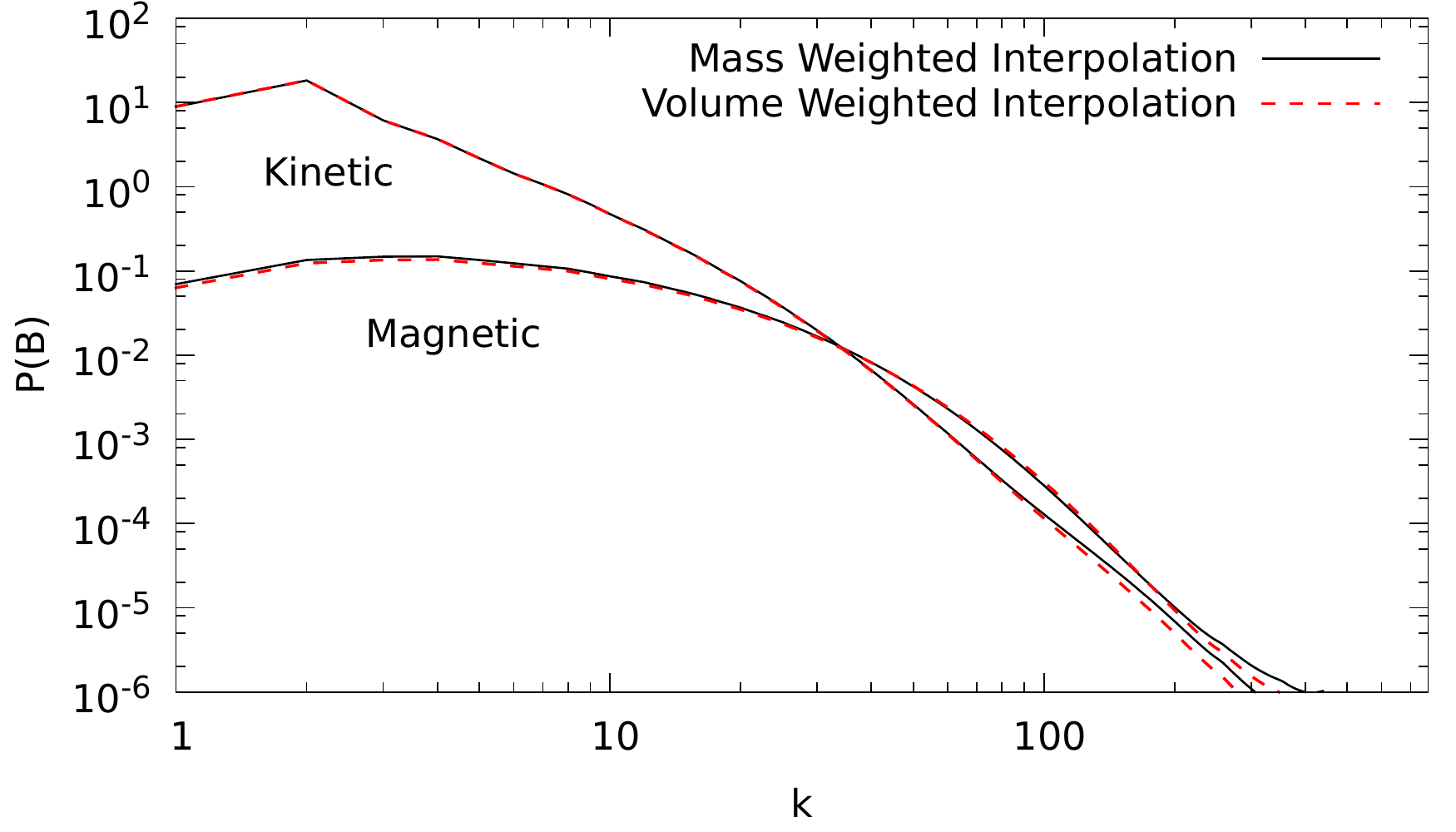}
\caption{Time averaged kinetic and magnetic spectra of the $256^3$ particle {\sc Phantom} calculation interpolated to a grid using mass weighted and volume weighted interpolation. Both approaches yield the same result.}
\label{fig:interpspect}
\end{figure}

\begin{figure}
\centering
 \includegraphics[width=0.6\columnwidth]{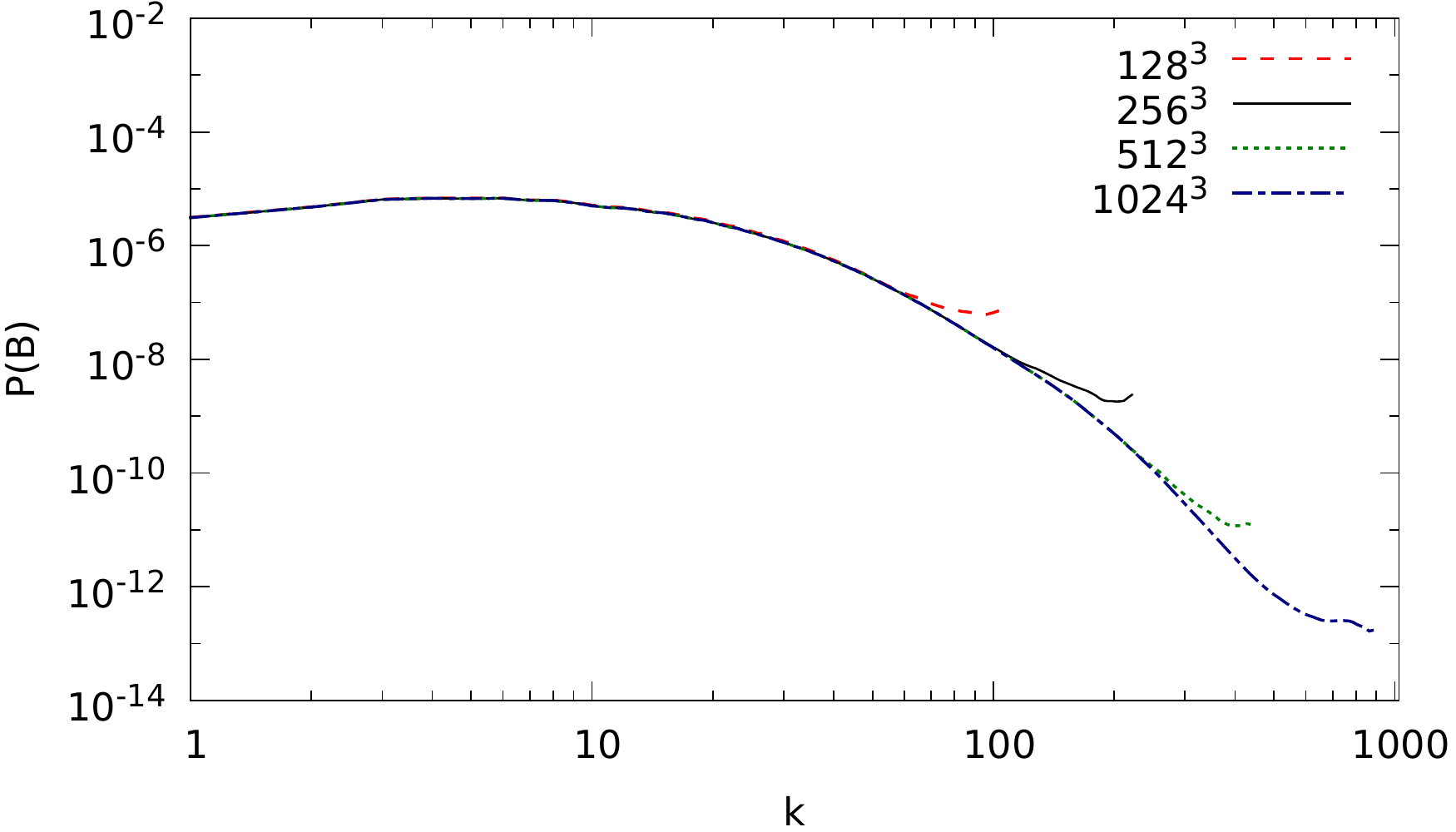}
\caption{Conversion of a $128^3$ particle {\sc Phantom} snapshot to grids of resolutions from $128^3$ to $1024^3$ grid points. Each resolution agrees well on the large-scale structure, but captures more of the small-scale structure as the resolution is increased. We find that the $256^3$ sufficiently the total magnetic energy, therefore choose grids with double the number the grid points as particles for our analysis.}
\label{fig:gridreso}
\end{figure}

The volume weighted interpolation of a quantity $A$ (in this case, the magnetic field ${\bf B}$) may be computed according to
\begin{equation}
 {\bf B}({\bf r}) = \frac{\sum_b \dfrac{m_b}{\rho_b} {\bf B}_b W(\vert {\bf r} - {\bf r}_b\vert, h_b)}{\sum_c \dfrac{m_c}{\rho_c} W(\vert {\bf r} - {\bf r}_c \vert, h_c)} .
\end{equation}
The denominator is the normalisation condition. A mass weighted interpolation may be computed as
\begin{equation}
 {\bf B}({\bf r}) = \frac{\sum_b m_b {\bf B}_b W(\vert {\bf r} - {\bf r}_b \vert, h_b)}{\sum_c m_c W(\vert {\bf r} - {\bf r}_c \vert, h_c)} .
\end{equation}
%where similarly the denominator is a normalization. For the mass weighted summation, the normalization is equivalent using the standard SPH summation to calculate density. 

Figure~\ref{fig:interpspect} shows the kinetic and magnetic energy spectra for the $128^3$ {\sc Phantom} calculation computed from a grid using volume weighted and mass weighted interpolations, respectively. The spectra between the two interpolation methods are nearly indistinguishable, differing from each other by less than 1\% at all $k$ and deviating only near the resolution scale. We conclude that either approach is acceptable, and for the spectra generated in Chapter~\ref{sec:chapter-mhdturb}, we have used the mass weighted interpolation.

The smoothing length in the calculations in Chapter~\ref{sec:chapter-mhdturb} can decrease by up to $8\times$ in the highest density regions, therefore we have tested the effect of different grid resolutions on the magnetic spectra. Figure~\ref{fig:gridreso} shows magnetic energy spectra from a $128^3$ particle {\sc Phantom} calculation interpolated to grids with resolutions of $128^3$ to $1024^3$. Our results show that the large-scale structure ($k<50$) is nearly identical at all grid resolutions, with the spectra differing on the order of $0.1\%$ at each $k$-band. The only difference is that the spectra extends to higher $k$ as the resolution is increased. We find that the magnetic energy contained on the $128^3$ grid differs by $1\%$ of the energy contained on the particles, while the $256^3$ grid resolution differs by only $0.1\%$. Higher resolutions only minimally change the energy content of the magnetic field. Our conclusion is that a grid with double the resolution of the {\sc Phantom} calculation is sufficient for computing the magnetic energy spectra accurately.

%% file: turbcomp-appendix2.tex
\chapter{Effective magnetic Prandtl numbers in grid and particle methods}
\label{sec:prandtl}

\section{Prandtl numbers in Eulerian schemes}

The primary source of numerical dissipation in Eulerian schemes is from the discretisation of advection terms. Consider a simple example of the contents of one grid cell advecting into an adjacent grid cell. If only a partial amount is transferred into the adjacent cell, then the contents must be reconstructed from the flux across the boundary. This approximation introduces diffusion due to its truncation error \citep[e.g.,][]{robertsonetal10}. The diffusion term in the first-order upwind scheme of \citet{cir52}, for example, scales according to $\propto v \Delta x (1 - \vert C \vert)$, where $C=v \Delta t / \Delta x$ is the Courant number. Higher order methods will change the scaling of the diffusion, but in all schemes it depends upon the resolution, time step size, and fluid velocity. 

Quantifying the effective numerical dissipation may be done by comparing simulations against analytic solutions. \citet{lb07} compared the analytic solution of a linear mode of the magneto-rotational instability (MRI) to shearing box simulations in order to calibrate their version of {\sc Zeus3D}. They varied the size of the time step and investigated resolutions from $32^3$ to $128^3$, determining that the total numerical dissipation (viscous and resistive) scaled linearly with time and quadratically with resolution. They found the magnetic Prandtl number to be approximately 2 (though in the context of this comparison, these simulations are for subsonic flows). In a similar manner, \citet{fromangetal07} performed simulations of the MRI with and without physical viscous and resistive dissipation terms. They found that the results of their ideal MHD simulations (dissipation is purely numerical) corresponded to ${\rm Pm}\approx2$, though cautioned that this depends upon the nature of the flow.

The effective Prandtl number for the version of {\sc Flash} used in Chapter~\ref{sec:chapter-mhdturb} was calibrated by \citet{federrathetal11}. Using simulations of the small-scale dynamo amplification of a magnetic field, they compared results from ideal MHD simulations to simulations employing a fixed dissipation (at varying resolution). They found that ${\rm Pm}\approx2$ for flows of Mach numbers $0.4$ and $2$. Thus, it is expected that the {\sc Flash} calculations in our comparison will have a similar Prandtl number.

\section{Prandtl numbers in smoothed particle magnetohydrodynamics}

In SPMHD, the equations of motion are derived from the discretised Lagrangian \citep{pm04b, price12}. Advection is computed exactly. Hence, the only sources of numerical dissipation are from the explicit sources of artificial viscosity and resistivity, which can be used to estimate the Reynolds and Prandtl numbers.

Artificial viscosity and resistivity in SPMHD are discretisations of physical dissipation terms, but with diffusion parameters that depend on resolution. \citet{al94} and \citet{murray96} analytically derived the amount of corresponding physical dissipation from the \citet{mg83} form of artificial viscosity (see also \citealt{monaghan05, lp10}). The artificial viscosity acts as both a shear and bulk viscosity. In these calculations, we use the \citet{monaghan97} form of artificial viscosity, which is similar except for the absence of a factor $h / \vert r_{ab} \vert$. \citet{mb12} calculated the amount of viscosity this adds in the continuum limit, and have shown that for the \citet{monaghan97} form of viscosity, it is approximately 18\% stronger for the $\alpha$ term. Using this approach, they also derived the coefficients for the $\beta_{\rm AV}$ term in the signal velocity. Hence, the shear viscosity in the simulations in Chapter~\ref{sec:chapter-mhdturb} corresponds to
\begin{equation}
\nu_{\rm AV} = \frac{62}{525} \alpha v_{\rm sig} h + \frac{9}{35 \pi} \beta_{\rm AV} \vert \nabla \cdot {\bf v} \vert h^2.
\label{eq:physicalnu}
\end{equation}
where
\begin{equation}
v_{\rm sig} = \sqrt{c_{\rm s}^{2} + v_{\rm A}^{2}}.
\end{equation}
The bulk viscosity will be $5/3\times$ this value \citep{lp10}. 

We note that these coefficients are twice the values quoted by \citet{mb12}. Their work is derived in the context of a Keplerian accretion disc, in which they safely assume that half the particles inside a particle volume are approaching while the other half are receding. It is standard in SPH to apply artificial viscosity only to approaching particles. In this paper, we calculate Reynolds numbers for particles where $\nabla~\cdot~{\bf v}~<~0$, and use the full value of the coefficient as it is expected that inside a shock, nearly all particles will be approaching. The bulk viscosity will be important for supersonic flows, which will affect the Reynolds number. In order to compare Reynolds numbers with {\sc Flash}, we compute the Reynolds numbers using only the shear viscosity. Including this term in the estimate of the Reynolds number would lead to larger ${\rm Pm}$ values.

The corresponding physical dissipation from the artificial resistivity can be calculated in a similar manner (Section~\ref{sec:artresis}). Artificial resistivity corresponds to a physical resistivity given by
\begin{equation}
\eta_{\rm AR} = \frac{1}{2} \alpha_{\rm B} v_{\rm sig} h.
\end{equation}
We note, as concluded in Chapter~\ref{sec:chapter-switch}, that the $\beta_{\rm AV}$ term in the artificial viscosity is not required for artificial resistivity. It is added to artificial viscosity to prevent particle interpenetration in high Mach number shocks, and otherwise leads to unnecessary dissipation if added to artificial resistivity.

\begin{figure*}
\centering
\includegraphics[width=0.85\textwidth]{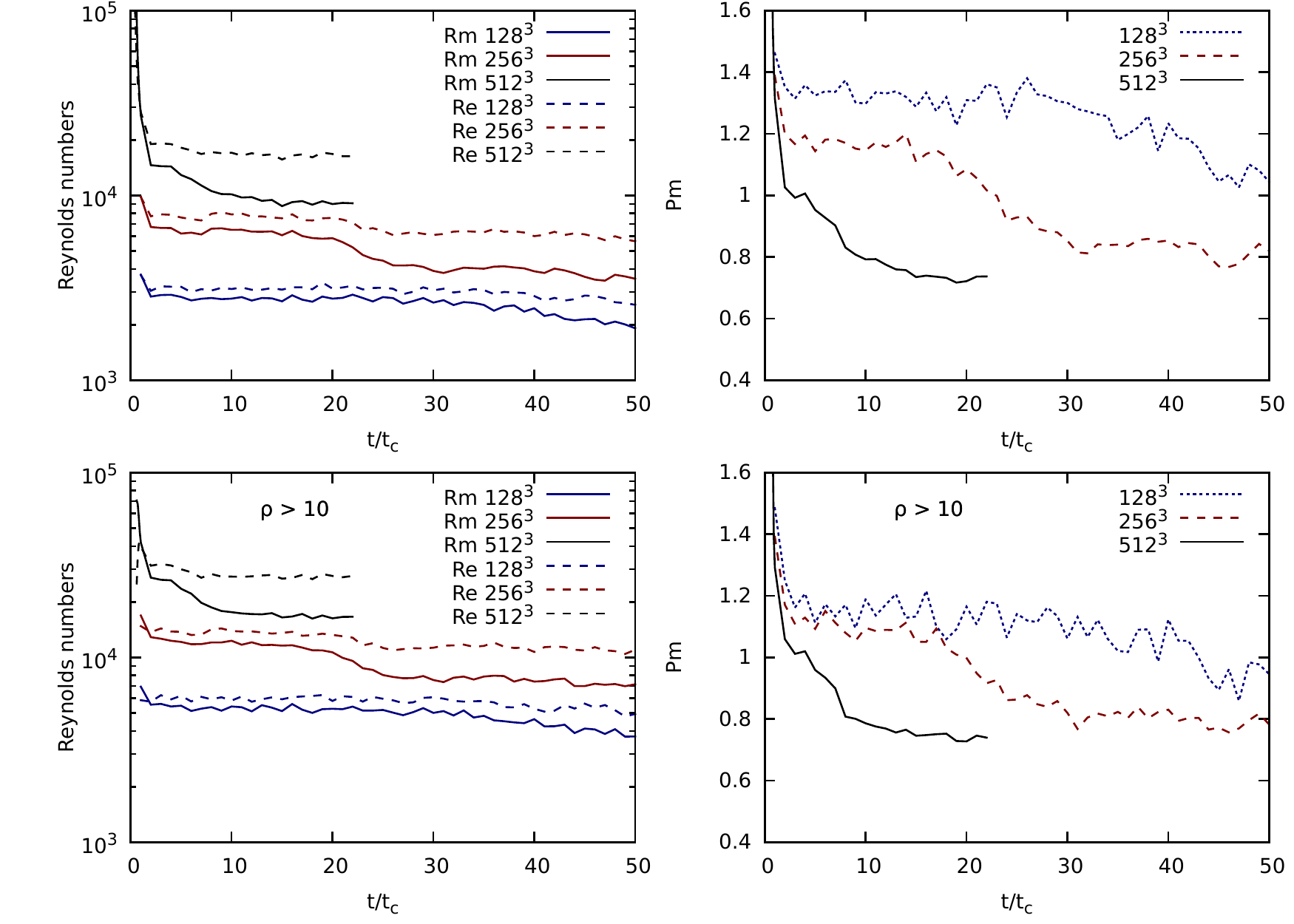}
\caption{The kinetic and magnetic Reynolds numbers (left plots) and Prandtl numbers (right plots) for {\sc Phantom}. The top row shows the averaged numbers for particles which have $\nabla \cdot {\bf v} < 0$, while the bottom row is averaged for regions where $\rho > 10 \rho_0$. The higher density regions have approximately double the kinetic and magnetic Reynolds numbers. The drop in Reynolds and Prandtl numbers over time is due to the fast MHD wave speed increasing in the signal velocity of the artificial dissipation terms. The Prandtl numbers are about unity, though decrease with resolution.}
\label{fig:ReRmPrandtl}
\end{figure*}

Since the dissipation terms use the local signal velocity, and our simulations use switches to dynamically adjust the values of $\alpha$ and $\alpha_{\rm B}$ for each particle, $\nu_{\rm AV}$ and $\eta_{\rm AR}$ are calculated per particle. In Figure~\ref{fig:ReRmPrandtl}, we show the average kinetic Reynolds, magnetic Reynolds, and magnetic Prandtl numbers on the particles for our simulations. We find that the mean Prandtl number in these set of SPMHD calculations is approximately unity. The Prandtl number decreases with resolution, a consequence of the quadratic scaling of the $\beta_{\rm AV}$ term, which is present in the artificial viscosity but not artificial resistivity. The Prandtl number also decreases with time. This results from the signal velocity scaling behaviour, as the dissipation from the $\alpha$ term increases as the magnetic field is amplified ($v_{\rm A}$ increasing). The $\beta_{\rm AV}$ term is unaffected by this. For high-density regions ($\rho > 10$), we note that the Reynolds numbers are increased by approximately a factor of 2, directly corresponding to the reduction in $h$.

%% file: phdthesis.bbl
\begin{thebibliography}{}

\bibitem[\protect\citeauthoryear{{Agertz} et~al.}{2007}]{agertzetal07}
{Agertz}, O., B. {Moore}, J. {Stadel}, D. {Potter}, F. {Miniati}, J. {Read}, L.
  {Mayer}, A. {Gawryszczak}, A. {Kravtsov}, {\AA}. {Nordlund}, F. {Pearce}, V.
  {Quilis}, D. {Rudd}, V. {Springel}, J. {Stone}, E. {Tasker}, R. {Teyssier},
  J. {Wadsley}, and R. {Walder}: 2007, `{Fundamental differences between SPH
  and grid methods}'.
\newblock {\em \mnras} {\bf 380}, 963--978.

\bibitem[\protect\citeauthoryear{{Andr{\'e}} et~al.}{2010}]{andreetal10}
{Andr{\'e}}, P., A. {Men'shchikov}, S. {Bontemps}, V. {K{\"o}nyves}, F.
  {Motte}, N. {Schneider}, P. {Didelon}, V. {Minier}, P. {Saraceno}, D.
  {Ward-Thompson}, J. {di Francesco}, G. {White}, S. {Molinari}, L. {Testi}, A.
  {Abergel}, M. {Griffin}, T. {Henning}, P. {Royer}, B. {Mer{\'{\i}}n}, R.
  {Vavrek}, M. {Attard}, D. {Arzoumanian}, C.~D. {Wilson}, P. {Ade}, H.
  {Aussel}, J.-P. {Baluteau}, M. {Benedettini}, J.-P. {Bernard}, J.~A.~D.~L.
  {Blommaert}, L. {Cambr{\'e}sy}, P. {Cox}, A. {di Giorgio}, P. {Hargrave}, M.
  {Hennemann}, M. {Huang}, J. {Kirk}, O. {Krause}, R. {Launhardt}, S. {Leeks},
  J. {Le Pennec}, J.~Z. {Li}, P.~G. {Martin}, A. {Maury}, G. {Olofsson}, A.
  {Omont}, N. {Peretto}, S. {Pezzuto}, T. {Prusti}, H. {Roussel}, D. {Russeil},
  M. {Sauvage}, B. {Sibthorpe}, A. {Sicilia-Aguilar}, L. {Spinoglio}, C.
  {Waelkens}, A. {Woodcraft}, and A. {Zavagno}: 2010, `{From filamentary clouds
  to prestellar cores to the stellar IMF: Initial highlights from the Herschel
  Gould Belt Survey}'.
\newblock {\em \aap} {\bf 518}, L102.

\bibitem[\protect\citeauthoryear{{Artymowicz} and {Lubow}}{1994}]{al94}
{Artymowicz}, P. and S.~H. {Lubow}: 1994, `{Dynamics of binary-disk
  interaction. 1: Resonances and disk gap sizes}'.
\newblock {\em \apj} {\bf 421}, 651--667.

\bibitem[\protect\citeauthoryear{{Balbus} and {Hawley}}{1991}]{bh91}
{Balbus}, S.~A. and J.~F. {Hawley}: 1991, `{A powerful local shear instability
  in weakly magnetized disks. I - Linear analysis. II - Nonlinear evolution}'.
\newblock {\em \apj} {\bf 376}, 214--233.

\bibitem[\protect\citeauthoryear{{Balsara}}{1995}]{balsara95}
{Balsara}, D.~S.: 1995, `{von Neumann stability analysis of smooth particle
  hydrodynamics--suggestions for optimal algorithms}'.
\newblock {\em J. Comput. Phys.} {\bf 121}, 357--372.

\bibitem[\protect\citeauthoryear{{Barnes} et~al.}{2012}]{bkw12}
{Barnes}, D.~J., D. {Kawata}, and K. {Wu}: 2012, `{Cosmological simulations
  using GCMHD+}'.
\newblock {\em \mnras} {\bf 420}, 3195--3212.

\bibitem[\protect\citeauthoryear{Batchelor}{2000}]{batchelor00}
Batchelor, G.: 2000, {\em An Introduction to Fluid Dynamics}, Cambridge
  Mathematical Library.
\newblock Cambridge University Press.

\bibitem[\protect\citeauthoryear{{Bate} et~al.}{1995}]{bbp95}
{Bate}, M.~R., I.~A. {Bonnell}, and N.~M. {Price}: 1995, `{Modelling accretion
  in protobinary systems}'.
\newblock {\em \mnras} {\bf 277}, 362--376.

\bibitem[\protect\citeauthoryear{{Bate} et~al.}{2014a}]{bpt14}
{Bate}, M.~R., D.~J. {Price}, and T.~S. {Tricco}: 2014a, `Modelling Magnetised
  Protostellar Jets with SPH'.
\newblock In: D. Stamatellos, S. Goodwin, and D. Ward-Thompson (eds.): {\em The
  Labyrinth of Star Formation}, Vol.~36 of {\em Astrophysics and Space Science
  Proceedings}.
\newblock Springer International Publishing, pp. 101--104.

\bibitem[\protect\citeauthoryear{{Bate} et~al.}{2014b}]{btp14}
{Bate}, M.~R., T.~S. {Tricco}, and D.~J. {Price}: 2014b, `{Collapse of a
  molecular cloud core to stellar densities: stellar-core and outflow formation
  in radiation magnetohydrodynamic simulations}'.
\newblock {\em \mnras} {\bf 437}, 77--95.

\bibitem[\protect\citeauthoryear{{Beck} et~al.}{2013}]{becketal13}
{Beck}, A.~M., K. {Dolag}, H. {Lesch}, and P.~P. {Kronberg}: 2013, `{Strong
  magnetic fields and large rotation measures in protogalaxies from supernova
  seeding}'.
\newblock {\em \mnras} {\bf 435}, 3575--3586.

\bibitem[\protect\citeauthoryear{{Beck} et~al.}{2012}]{becketal12}
{Beck}, A.~M., H. {Lesch}, K. {Dolag}, H. {Kotarba}, A. {Geng}, and F.~A.
  {Stasyszyn}: 2012, `{Origin of strong magnetic fields in Milky Way-like
  galactic haloes}'.
\newblock {\em \mnras} {\bf 422}, 2152--2163.

\bibitem[\protect\citeauthoryear{Bellan}{2006}]{bellan06}
Bellan, P.: 2006, {\em Fundamentals of Plasma Physics}.
\newblock Cambridge University Press.

\bibitem[\protect\citeauthoryear{{Benz} et~al.}{1990}]{benzetal90}
{Benz}, W., A.~G.~W. {Cameron}, W.~H. {Press}, and R.~L. {Bowers}: 1990,
  `{Dynamic mass exchange in doubly degenerate binaries. I - 0.9 and 1.2 solar
  mass stars}'.
\newblock {\em \apj} {\bf 348}, 647--667.

\bibitem[\protect\citeauthoryear{{Berger} and {Colella}}{1989}]{bc89}
{Berger}, M.~J. and P. {Colella}: 1989, `{Local adaptive mesh refinement for
  shock hydrodynamics}'.
\newblock {\em J. Comput. Phys.} {\bf 82}, 64--84.

\bibitem[\protect\citeauthoryear{{Biermann}}{1950}]{biermann50}
{Biermann}, L.: 1950, `{{\"U}ber den Ursprung der Magnetfelder auf Sternen und
  im interstellaren Raum (miteinem Anhang von A. Schl{\"u}ter)}'.
\newblock {\em Zeitschrift Naturforschung Teil A} {\bf 5}, 65.

\bibitem[\protect\citeauthoryear{{Blandford} and {Payne}}{1982}]{bp82}
{Blandford}, R.~D. and D.~G. {Payne}: 1982, `{Hydromagnetic flows from
  accretion discs and the production of radio jets}'.
\newblock {\em \mnras} {\bf 199}, 883--903.

\bibitem[\protect\citeauthoryear{{Bonafede} et~al.}{2011}]{bonafedeetal11}
{Bonafede}, A., K. {Dolag}, F. {Stasyszyn}, G. {Murante}, and S. {Borgani}:
  2011, `{A non-ideal magnetohydrodynamic GADGET: simulating massive galaxy
  clusters}'.
\newblock {\em \mnras} {\bf 418}, 2234--2250.

\bibitem[\protect\citeauthoryear{{Bonet} and {Lok}}{1999}]{bl99}
{Bonet}, J. and T.-S.~L. {Lok}: 1999, `Variational and momentum preservation
  aspects of smooth particle hydrodynamic formulations'.
\newblock {\em Comput. Methods Appl. Mech. Eng.} {\bf 180}(1-2), 97--115.

\bibitem[\protect\citeauthoryear{{B{\o}rve} et~al.}{2001}]{bot01}
{B{\o}rve}, S., M. {Omang}, and J. {Trulsen}: 2001, `{Regularized Smoothed
  Particle Hydrodynamics: A New Approach to Simulating Magnetohydrodynamic
  Shocks}'.
\newblock {\em \apj} {\bf 561}, 82--93.

\bibitem[\protect\citeauthoryear{{B{\o}rve} et~al.}{2004}]{bot04}
{B{\o}rve}, S., M. {Omang}, and J. {Trulsen}: 2004, `{Two-dimensional MHD
  Smoothed Particle Hydrodynamics Stability Analysis}'.
\newblock {\em \apjs} {\bf 153}, 447--462.

\bibitem[\protect\citeauthoryear{{Bourke} et~al.}{2001}]{bourkeetal01}
{Bourke}, T.~L., P.~C. {Myers}, G. {Robinson}, and A.~R. {Hyland}: 2001, `{New
  OH Zeeman Measurements of Magnetic Field Strengths in Molecular Clouds}'.
\newblock {\em \apj} {\bf 554}, 916--932.

\bibitem[\protect\citeauthoryear{{Bovino} et~al.}{2013}]{bss13}
{Bovino}, S., D.~R.~G. {Schleicher}, and J. {Schober}: 2013, `{Turbulent
  magnetic field amplification from the smallest to the largest magnetic
  Prandtl numbers}'.
\newblock {\em New J. Phys.} {\bf 15}(1), 013055.

\bibitem[\protect\citeauthoryear{{Brackbill} and {Barnes}}{1980}]{bb80}
{Brackbill}, J.~U. and D.~C. {Barnes}: 1980, `{The effect of nonzero product of
  magnetic gradient and B on the numerical solution of the magnetohydrodynamic
  equations}'.
\newblock {\em J. Comput. Phys.} {\bf 35}, 426--430.

\bibitem[\protect\citeauthoryear{{Brandenburg}}{2010}]{brandenburg10}
{Brandenburg}, A.: 2010, `{Magnetic field evolution in simulations with Euler
  potentials}'.
\newblock {\em \mnras} {\bf 401}, 347--354.

\bibitem[\protect\citeauthoryear{{Brandenburg} et~al.}{2012}]{bss12}
{Brandenburg}, A., D. {Sokoloff}, and K. {Subramanian}: 2012, `{Current Status
  of Turbulent Dynamo Theory. From Large-Scale to Small-Scale Dynamos}'.
\newblock {\em \ssr} {\bf 169}, 123--157.

\bibitem[\protect\citeauthoryear{{Brandenburg} and {Subramanian}}{2005}]{bs05}
{Brandenburg}, A. and K. {Subramanian}: 2005, `{Astrophysical magnetic fields
  and nonlinear dynamo theory}'.
\newblock {\em \physrep} {\bf 417}, 1--209.

\bibitem[\protect\citeauthoryear{{Brio} and {Wu}}{1988}]{briowu88}
{Brio}, M. and C.~C. {Wu}: 1988, `{An upwind differencing scheme for the
  equations of ideal magnetohydrodynamics}'.
\newblock {\em J. Comput. Phys.} {\bf 75}, 400--422.

\bibitem[\protect\citeauthoryear{{Brookshaw}}{1985}]{brookshaw85}
{Brookshaw}, L.: 1985, `{A method of calculating radiative heat diffusion in
  particle simulations}'.
\newblock {\em Proc. Astron. Soc. Aust.} {\bf 6}, 207--210.

\bibitem[\protect\citeauthoryear{{B{\"u}rzle} et~al.}{2011a}]{burzleetal11b}
{B{\"u}rzle}, F., P.~C. {Clark}, F. {Stasyszyn}, K. {Dolag}, and R.~S.
  {Klessen}: 2011a, `{Protostellar outflows with smoothed particle
  magnetohydrodynamics}'.
\newblock {\em \mnras} {\bf 417}, L61--L65.

\bibitem[\protect\citeauthoryear{{B{\"u}rzle} et~al.}{2011b}]{burzleetal11a}
{B{\"u}rzle}, F., P.~C. {Clark}, F. {Stasyszyn}, T. {Greif}, K. {Dolag}, R.~S.
  {Klessen}, and P. {Nielaba}: 2011b, `{Protostellar collapse and fragmentation
  using an MHD GADGET}'.
\newblock {\em \mnras} {\bf 412}, 171--186.

\bibitem[\protect\citeauthoryear{{Chandrasekhar}}{1961}]{chandrasekhar61}
{Chandrasekhar}, S.: 1961, {\em {Hydrodynamic and hydromagnetic stability}},
  Dover Books on Physics Series.
\newblock Dover Publications.

\bibitem[\protect\citeauthoryear{{Cho} and {Vishniac}}{2000}]{cv00}
{Cho}, J. and E.~T. {Vishniac}: 2000, `{The Generation of Magnetic Fields
  through Driven Turbulence}'.
\newblock {\em \apj} {\bf 538}, 217--225.

\bibitem[\protect\citeauthoryear{{Cho} et~al.}{2009}]{choetal09}
{Cho}, J., E.~T. {Vishniac}, A. {Beresnyak}, A. {Lazarian}, and D. {Ryu}: 2009,
  `{Growth of Magnetic Fields Induced by Turbulent Motions}'.
\newblock {\em \apj} {\bf 693}, 1449--1461.

\bibitem[\protect\citeauthoryear{Choudhuri}{1998}]{choudhuri98}
Choudhuri, A.: 1998, {\em The Physics of Fluids and Plasmas: An Introduction
  for Astrophysicists}.
\newblock Cambridge University Press.

\bibitem[\protect\citeauthoryear{{Chow} and {Monaghan}}{1997}]{cm97}
{Chow}, E. and J.~J. {Monaghan}: 1997, `{Ultrarelativistic SPH}'.
\newblock {\em J. Comput. Phys.} {\bf 134}, 296--305.

\bibitem[\protect\citeauthoryear{{Collins} et~al.}{2012}]{collinsetal12}
{Collins}, D.~C., A.~G. {Kritsuk}, P. {Padoan}, H. {Li}, H. {Xu}, S.~D.
  {Ustyugov}, and M.~L. {Norman}: 2012, `{The Two States of Star-forming
  Clouds}'.
\newblock {\em \apj} {\bf 750}, 13.

\bibitem[\protect\citeauthoryear{{Courant} et~al.}{1928}]{cfl28}
{Courant}, R., K. {Friedrichs}, and H. {Lewy}: 1928, `{{\"U}ber die partiellen
  Differenzengleichungen der mathematischen Physik}'.
\newblock {\em Mathematische Annalen} {\bf 100}, 32--74.

\bibitem[\protect\citeauthoryear{{Courant} et~al.}{1952}]{cir52}
{Courant}, R., E. {Isaacson}, and M. {Rees}: 1952, `On the solution of
  nonlinear hyperbolic differential equations by finite differences'.
\newblock {\em Comm. Pure Appl. Math.} {\bf 5}(3), 243--255.

\bibitem[\protect\citeauthoryear{{Crutcher}}{1999}]{crutcher99}
{Crutcher}, R.~M.: 1999, `{Magnetic Fields in Molecular Clouds: Observations
  Confront Theory}'.
\newblock {\em \apj} {\bf 520}, 706--713.

\bibitem[\protect\citeauthoryear{{Cullen} and {Dehnen}}{2010}]{cd10}
{Cullen}, L. and W. {Dehnen}: 2010, `{Inviscid smoothed particle
  hydrodynamics}'.
\newblock {\em \mnras} {\bf 408}, 669--683.

\bibitem[\protect\citeauthoryear{{Cummins} and {Rudman}}{1999}]{cr99}
{Cummins}, S.~J. and M. {Rudman}: 1999, `{An SPH Projection Method}'.
\newblock {\em J. Comput. Phys.} {\bf 152}, 584--607.

\bibitem[\protect\citeauthoryear{{Dai} and {Woodward}}{1994}]{dw94}
{Dai}, W. and P.~R. {Woodward}: 1994, `{Extension of the Piecewise Parabolic
  Method to Multidimensional Ideal Magnetohydrodynamics}'.
\newblock {\em J. Comput. Phys.} {\bf 115}, 485--514.

\bibitem[\protect\citeauthoryear{{Dedner} et~al.}{2002}]{dedneretal02}
{Dedner}, A., F. {Kemm}, D. {Kr{\"o}ner}, C.-D. {Munz}, T. {Schnitzer}, and M.
  {Wesenberg}: 2002, `{Hyperbolic Divergence Cleaning for the MHD Equations}'.
\newblock {\em J. Comput. Phys.} {\bf 175}, 645--673.

\bibitem[\protect\citeauthoryear{{Dobbs} and {Price}}{2008}]{dp08}
{Dobbs}, C.~L. and D.~J. {Price}: 2008, `{Magnetic fields and the dynamics of
  spiral galaxies}'.
\newblock {\em \mnras} {\bf 383}, 497--512.

\bibitem[\protect\citeauthoryear{{Dolag} and {Stasyszyn}}{2009}]{mhdgadget}
{Dolag}, K. and F. {Stasyszyn}: 2009, `{An MHD GADGET for cosmological
  simulations}'.
\newblock {\em \mnras} {\bf 398}, 1678--1697.

\bibitem[\protect\citeauthoryear{{Donnert} et~al.}{2009}]{donnertetal09}
{Donnert}, J., K. {Dolag}, H. {Lesch}, and E. {M{\"u}ller}: 2009, `{Cluster
  magnetic fields from galactic outflows}'.
\newblock {\em \mnras} {\bf 392}, 1008--1021.

\bibitem[\protect\citeauthoryear{{Dubey} et~al.}{2008}]{dubeyetal08}
{Dubey}, A., R. {Fisher}, C. {Graziani}, G.~C. {Jordan}, IV, D.~Q. {Lamb},
  L.~B. {Reid}, P. {Rich}, D. {Sheeler}, D. {Townsley}, and K. {Weide}: 2008,
  `{Challenges of Extreme Computing using the FLASH code}'.
\newblock In: N.~V. {Pogorelov}, E. {Audit}, and G.~P. {Zank} (eds.): {\em
  Numerical Modeling of Space Plasma Flows}, Vol. 385 of {\em Astronomical
  Society of the Pacific Conference Series}. p. 145.

\bibitem[\protect\citeauthoryear{{Elmegreen} and {Scalo}}{2004}]{es04}
{Elmegreen}, B.~G. and J. {Scalo}: 2004, `{Interstellar Turbulence I:
  Observations and Processes}'.
\newblock {\em \araa} {\bf 42}, 211--273.

\bibitem[\protect\citeauthoryear{{Eswaran} and {Pope}}{1988}]{ep88}
{Eswaran}, V. and S.~B. {Pope}: 1988, `{An examination of forcing in direct
  numerical simulations of turbulence}'.
\newblock {\em Comput. Fluids} {\bf 16}, 257--278.

\bibitem[\protect\citeauthoryear{{Evans} and {Hawley}}{1988}]{eh88}
{Evans}, C.~R. and J.~F. {Hawley}: 1988, `{Simulation of magnetohydrodynamic
  flows - A constrained transport method}'.
\newblock {\em \apj} {\bf 332}, 659--677.

\bibitem[\protect\citeauthoryear{{Evans}}{1999}]{evans99}
{Evans}, II, N.~J.: 1999, `{Physical Conditions in Regions of Star Formation}'.
\newblock {\em \araa} {\bf 37}, 311--362.

\bibitem[\protect\citeauthoryear{{Federrath}}{2013}]{federrath13}
{Federrath}, C.: 2013, `{On the universality of supersonic turbulence}'.
\newblock {\em \mnras} {\bf 436}, 1245--1257.

\bibitem[\protect\citeauthoryear{{Federrath} et~al.}{2011}]{federrathetal11}
{Federrath}, C., G. {Chabrier}, J. {Schober}, R. {Banerjee}, R.~S. {Klessen},
  and D.~R.~G. {Schleicher}: 2011, `{Mach Number Dependence of Turbulent
  Magnetic Field Amplification: Solenoidal versus Compressive Flows}'.
\newblock {\em \prl} {\bf 107}(11), 114504.

\bibitem[\protect\citeauthoryear{{Federrath} and {Klessen}}{2012}]{fk12}
{Federrath}, C. and R.~S. {Klessen}: 2012, `{The Star Formation Rate of
  Turbulent Magnetized Clouds: Comparing Theory, Simulations, and
  Observations}'.
\newblock {\em \apj} {\bf 761}, 156.

\bibitem[\protect\citeauthoryear{{Federrath} and {Klessen}}{2013}]{fk13}
{Federrath}, C. and R.~S. {Klessen}: 2013, `{On the Star Formation Efficiency
  of Turbulent Magnetized Clouds}'.
\newblock {\em \apj} {\bf 763}, 51.

\bibitem[\protect\citeauthoryear{{Federrath} et~al.}{2008}]{fks08}
{Federrath}, C., R.~S. {Klessen}, and W. {Schmidt}: 2008, `{The Density
  Probability Distribution in Compressible Isothermal Turbulence: Solenoidal
  versus Compressive Forcing}'.
\newblock {\em \apjl} {\bf 688}, L79--L82.

\bibitem[\protect\citeauthoryear{{Federrath} et~al.}{2010}]{federrathetal10}
{Federrath}, C., J. {Roman-Duval}, R.~S. {Klessen}, W. {Schmidt}, and M.-M.
  {Mac Low}: 2010, `{Comparing the statistics of interstellar turbulence in
  simulations and observations. Solenoidal versus compressive turbulence
  forcing}'.
\newblock {\em \aap} {\bf 512}, A81.

\bibitem[\protect\citeauthoryear{{Federrath} et~al.}{2014}]{federrathetal14}
{Federrath}, C., J. {Schober}, S. {Bovino}, and D.~R.~G. {Schleicher}: 2014,
  `{The Turbulent Dynamo in Highly Compressible Supersonic Plasmas}'.
\newblock {\em \apjl} {\bf 797}, L19.

\bibitem[\protect\citeauthoryear{{Frenk} et~al.}{1999}]{frenketal99}
{Frenk}, C.~S., S.~D.~M. {White}, P. {Bode}, J.~R. {Bond}, G.~L. {Bryan}, R.
  {Cen}, H.~M.~P. {Couchman}, A.~E. {Evrard}, N. {Gnedin}, A. {Jenkins}, A.~M.
  {Khokhlov}, A. {Klypin}, J.~F. {Navarro}, M.~L. {Norman}, J.~P. {Ostriker},
  J.~M. {Owen}, F.~R. {Pearce}, U.-L. {Pen}, M. {Steinmetz}, P.~A. {Thomas},
  J.~V. {Villumsen}, J.~W. {Wadsley}, M.~S. {Warren}, G. {Xu}, and G. {Yepes}:
  1999, `{The Santa Barbara Cluster Comparison Project: A Comparison of
  Cosmological Hydrodynamics Solutions}'.
\newblock {\em \apj} {\bf 525}, 554--582.

\bibitem[\protect\citeauthoryear{{Fromang} et~al.}{2006}]{ramses}
{Fromang}, S., P. {Hennebelle}, and R. {Teyssier}: 2006, `{A high order Godunov
  scheme with constrained transport and adaptive mesh refinement for
  astrophysical magnetohydrodynamics}'.
\newblock {\em \aap} {\bf 457}, 371--384.

\bibitem[\protect\citeauthoryear{{Fromang} and {Papaloizou}}{2007}]{fp07}
{Fromang}, S. and J. {Papaloizou}: 2007, `{MHD simulations of the
  magnetorotational instability in a shearing box with zero net flux. I. The
  issue of convergence}'.
\newblock {\em \aap} {\bf 476}, 1113--1122.

\bibitem[\protect\citeauthoryear{{Fromang} et~al.}{2007}]{fromangetal07}
{Fromang}, S., J. {Papaloizou}, G. {Lesur}, and T. {Heinemann}: 2007, `{MHD
  simulations of the magnetorotational instability in a shearing box with zero
  net flux. II. The effect of transport coefficients}'.
\newblock {\em \aap} {\bf 476}, 1123--1132.

\bibitem[\protect\citeauthoryear{{Fryxell} et~al.}{2000}]{fryxelletal00}
{Fryxell}, B., K. {Olson}, P. {Ricker}, F.~X. {Timmes}, M. {Zingale}, D.~Q.
  {Lamb}, P. {MacNeice}, R. {Rosner}, J.~W. {Truran}, and H. {Tufo}: 2000,
  `{FLASH: An Adaptive Mesh Hydrodynamics Code for Modeling Astrophysical
  Thermonuclear Flashes}'.
\newblock {\em \apjs} {\bf 131}, 273--334.

\bibitem[\protect\citeauthoryear{{Furth} et~al.}{1963}]{fkr63}
{Furth}, H.~P., J. {Killeen}, and M.~N. {Rosenbluth}: 1963,
  `{Finite-Resistivity Instabilities of a Sheet Pinch}'.
\newblock {\em Phys. Fluids} {\bf 6}, 459--484.

\bibitem[\protect\citeauthoryear{{Gaburov} and {Nitadori}}{2011}]{gn11}
{Gaburov}, E. and K. {Nitadori}: 2011, `{Astrophysical weighted particle
  magnetohydrodynamics}'.
\newblock {\em \mnras} {\bf 414}, 129--154.

\bibitem[\protect\citeauthoryear{{Gaensler} et~al.}{2011}]{gaensleretal11}
{Gaensler}, B.~M., M. {Haverkorn}, B. {Burkhart}, K.~J. {Newton-McGee}, R.~D.
  {Ekers}, A. {Lazarian}, N.~M. {McClure-Griffiths}, T. {Robishaw}, J.~M.
  {Dickey}, and A.~J. {Green}: 2011, `{Low-Mach-number turbulence in
  interstellar gas revealed by radio polarization gradients}'.
\newblock {\em \nat} {\bf 478}, 214--217.

\bibitem[\protect\citeauthoryear{{Gingold} and {Monaghan}}{1977}]{gm77}
{Gingold}, R.~A. and J.~J. {Monaghan}: 1977, `{Smoothed particle hydrodynamics
  - Theory and application to non-spherical stars}'.
\newblock {\em \mnras} {\bf 181}, 375--389.

\bibitem[\protect\citeauthoryear{{Goldreich} and {Sridhar}}{1995}]{gs95}
{Goldreich}, P. and S. {Sridhar}: 1995, `{Toward a theory of interstellar
  turbulence. 2: Strong alfvenic turbulence}'.
\newblock {\em \apj} {\bf 438}, 763--775.

\bibitem[\protect\citeauthoryear{Griffiths}{1999}]{griffiths99}
Griffiths, D.: 1999, {\em Introduction to Electrodynamics}.
\newblock Prentice Hall.

\bibitem[\protect\citeauthoryear{{Guan} and {Gammie}}{2008}]{gg08}
{Guan}, X. and C.~F. {Gammie}: 2008, `{Axisymmetric Shearing Box Models of
  Magnetized Disks}'.
\newblock {\em \apjs} {\bf 174}, 145--157.

\bibitem[\protect\citeauthoryear{{Hairer} et~al.}{2006}]{hlw06}
{Hairer}, E., C. {Lubich}, and G. {Wanner}: 2006, {\em {Geometric Numerical
  Integration: Structure-Preserving Algorithms for Ordinary Differential
  Equations}}, Springer Series in Computational Mathematics.
\newblock Springer.

\bibitem[\protect\citeauthoryear{{Hartmann}}{2002}]{hartmann02}
{Hartmann}, L.: 2002, `{Flows, Fragmentation, and Star Formation. I. Low-Mass
  Stars in Taurus}'.
\newblock {\em \apj} {\bf 578}, 914--924.

\bibitem[\protect\citeauthoryear{{Hatchell} et~al.}{2005}]{hatchelletal05}
{Hatchell}, J., J.~S. {Richer}, G.~A. {Fuller}, C.~J. {Qualtrough}, E.~F.
  {Ladd}, and C.~J. {Chandler}: 2005, `{Star formation in Perseus. Clusters,
  filaments and the conditions for star formation}'.
\newblock {\em \aap} {\bf 440}, 151--161.

\bibitem[\protect\citeauthoryear{{Haugen} et~al.}{2004}]{hbd04}
{Haugen}, N.~E., A. {Brandenburg}, and W. {Dobler}: 2004, `{Simulations of
  nonhelical hydromagnetic turbulence}'.
\newblock {\em \pre} {\bf 70}(1), 016308.

\bibitem[\protect\citeauthoryear{{Hawley} and {Balbus}}{1992}]{hb92}
{Hawley}, J.~F. and S.~A. {Balbus}: 1992, `{A powerful local shear instability
  in weakly magnetized disks. III - Long-term evolution in a shearing sheet}'.
\newblock {\em \apj} {\bf 400}, 595--609.

\bibitem[\protect\citeauthoryear{{Hayward} et~al.}{2014}]{haywardetal13}
{Hayward}, C.~C., P. {Torrey}, V. {Springel}, L. {Hernquist}, and M.
  {Vogelsberger}: 2014, `{Galaxy mergers on a moving mesh: a comparison with
  smoothed particle hydrodynamics}'.
\newblock {\em \mnras} {\bf 442}, 1992--2016.

\bibitem[\protect\citeauthoryear{{Heiles} and {Troland}}{2005}]{ht05}
{Heiles}, C. and T.~H. {Troland}: 2005, `{The Millennium Arecibo 21 Centimeter
  Absorption-Line Survey. IV. Statistics of Magnetic Field, Column Density, and
  Turbulence}'.
\newblock {\em \apj} {\bf 624}, 773--793.

\bibitem[\protect\citeauthoryear{{Hopkins}}{2013}]{hopkins13}
{Hopkins}, P.~F.: 2013, `{A general class of Lagrangian smoothed particle
  hydrodynamics methods and implications for fluid mixing problems}'.
\newblock {\em \mnras} {\bf 428}, 2840--2856.

\bibitem[\protect\citeauthoryear{{Hu} and {Adams}}{2007}]{ha07}
{Hu}, X.~Y. and N.~A. {Adams}: 2007, `{An incompressible multi-phase SPH
  method}'.
\newblock {\em J. Comput. Phys.} {\bf 227}, 264--278.

\bibitem[\protect\citeauthoryear{{Hubber} et~al.}{2013}]{hfg13}
{Hubber}, D.~A., S.~A.~E.~G. {Falle}, and S.~P. {Goodwin}: 2013, `{Convergence
  of AMR and SPH simulations - I. Hydrodynamical resolution and convergence
  tests}'.
\newblock {\em \mnras} {\bf 432}, 711--727.

\bibitem[\protect\citeauthoryear{{Hut} et~al.}{1995}]{hmm95}
{Hut}, P., J. {Makino}, and S. {McMillan}: 1995, `{Building a better
  leapfrog}'.
\newblock {\em \apjl} {\bf 443}, L93--L96.

\bibitem[\protect\citeauthoryear{{Iwasaki} and {Inutsuka}}{2011}]{ii11}
{Iwasaki}, K. and S.-I. {Inutsuka}: 2011, `{Smoothed particle
  magnetohydrodynamics with a Riemann solver and the method of
  characteristics}'.
\newblock {\em \mnras} {\bf 418}, 1668--1688.

\bibitem[\protect\citeauthoryear{{King} et~al.}{2007}]{kpl07}
{King}, A.~R., J.~E. {Pringle}, and M. {Livio}: 2007, `{Accretion disc
  viscosity: how big is alpha?}'.
\newblock {\em \mnras} {\bf 376}, 1740--1746.

\bibitem[\protect\citeauthoryear{{Kitsionas} et~al.}{2009}]{kitsionasetal09}
{Kitsionas}, S., C. {Federrath}, R.~S. {Klessen}, W. {Schmidt}, D.~J. {Price},
  L.~J. {Dursi}, M. {Gritschneder}, S. {Walch}, R. {Piontek}, J. {Kim}, A.-K.
  {Jappsen}, P. {Ciecielag}, and M.-M. {Mac Low}: 2009, `{Algorithmic
  comparisons of decaying, isothermal, supersonic turbulence}'.
\newblock {\em \aap} {\bf 508}, 541--560.

\bibitem[\protect\citeauthoryear{{Klessen}}{2000}]{klessen00}
{Klessen}, R.~S.: 2000, `{One-Point Probability Distribution Functions of
  Supersonic Turbulent Flows in Self-gravitating Media}'.
\newblock {\em \apj} {\bf 535}, 869--886.

\bibitem[\protect\citeauthoryear{{Kotarba} et~al.}{2010}]{kotarbaetal10}
{Kotarba}, H., S.~J. {Karl}, T. {Naab}, P.~H. {Johansson}, K. {Dolag}, H.
  {Lesch}, and F.~A. {Stasyszyn}: 2010, `{Simulating Magnetic Fields in the
  Antennae Galaxies}'.
\newblock {\em \apj} {\bf 716}, 1438--1452.

\bibitem[\protect\citeauthoryear{{Kotarba} et~al.}{2011}]{kotarbaetal11}
{Kotarba}, H., H. {Lesch}, K. {Dolag}, T. {Naab}, P.~H. {Johansson}, J.
  {Donnert}, and F.~A. {Stasyszyn}: 2011, `{Galactic m{\'e}nage {\`a} trois:
  simulating magnetic fields in colliding galaxies}'.
\newblock {\em \mnras} {\bf 415}, 3189--3218.

\bibitem[\protect\citeauthoryear{{Kotarba} et~al.}{2009}]{kotarbaetal09}
{Kotarba}, H., H. {Lesch}, K. {Dolag}, T. {Naab}, P.~H. {Johansson}, and F.~A.
  {Stasyszyn}: 2009, `{Magnetic field structure due to the global velocity
  field in spiral galaxies}'.
\newblock {\em \mnras} {\bf 397}, 733--747.

\bibitem[\protect\citeauthoryear{{Kowal} et~al.}{2007}]{klb07}
{Kowal}, G., A. {Lazarian}, and A. {Beresnyak}: 2007, `{Density Fluctuations in
  MHD Turbulence: Spectra, Intermittency, and Topology}'.
\newblock {\em \apj} {\bf 658}, 423--445.

\bibitem[\protect\citeauthoryear{{Kritsuk} et~al.}{2011}]{kritsuketal11}
{Kritsuk}, A.~G., {\AA}. {Nordlund}, D. {Collins}, P. {Padoan}, M.~L. {Norman},
  T. {Abel}, R. {Banerjee}, C. {Federrath}, M. {Flock}, D. {Lee}, P.~S. {Li},
  W.-C. {M{\"u}ller}, R. {Teyssier}, S.~D. {Ustyugov}, C. {Vogel}, and H. {Xu}:
  2011, `{Comparing Numerical Methods for Isothermal Magnetized Supersonic
  Turbulence}'.
\newblock {\em \apj} {\bf 737}, 13.

\bibitem[\protect\citeauthoryear{{Larson}}{1981}]{larson81}
{Larson}, R.~B.: 1981, `{Turbulence and star formation in molecular clouds}'.
\newblock {\em \mnras} {\bf 194}, 809--826.

\bibitem[\protect\citeauthoryear{{Lee} et~al.}{2008}]{leeetal08}
{Lee}, E.-S., C. {Moulinec}, R. {Xu}, D. {Violeau}, D. {Laurence}, and P.
  {Stansby}: 2008, `{Comparisons of weakly compressible and truly
  incompressible algorithms for the SPH mesh free particle method}'.
\newblock {\em J. Comput. Phys.} {\bf 227}, 8417--8436.

\bibitem[\protect\citeauthoryear{{Lemaster} and {Stone}}{2008}]{ls08}
{Lemaster}, M.~N. and J.~M. {Stone}: 2008, `{Density Probability Distribution
  Functions in Supersonic Hydrodynamic and MHD Turbulence}'.
\newblock {\em \apjl} {\bf 682}, L97--L100.

\bibitem[\protect\citeauthoryear{{Lesaffre} and {Balbus}}{2007}]{lb07}
{Lesaffre}, P. and S.~A. {Balbus}: 2007, `{Exact shearing box solutions of
  magnetohydrodynamic flows with resistivity, viscosity and cooling}'.
\newblock {\em \mnras} {\bf 381}, 319--333.

\bibitem[\protect\citeauthoryear{{Li} et~al.}{2003}]{lkmm03}
{Li}, Y., R.~S. {Klessen}, and M.-M. {Mac Low}: 2003, `{The Formation of
  Stellar Clusters in Turbulent Molecular Clouds: Effects of the Equation of
  State}'.
\newblock {\em \apj} {\bf 592}, 975--985.

\bibitem[\protect\citeauthoryear{{Lind} et~al.}{2012}]{lindetal12}
{Lind}, S.~J., R. {Xu}, P.~K. {Stansby}, and B.~D. {Rogers}: 2012,
  `{Incompressible smoothed particle hydrodynamics for free-surface flows: A
  generalised diffusion-based algorithm for stability and validations for
  impulsive flows and propagating waves}'.
\newblock {\em J. Comput. Phys.} {\bf 231}, 1499--1523.

\bibitem[\protect\citeauthoryear{{Lodato} and {Price}}{2010}]{lp10}
{Lodato}, G. and D.~J. {Price}: 2010, `{On the diffusive propagation of warps
  in thin accretion discs}'.
\newblock {\em \mnras} {\bf 405}, 1212--1226.

\bibitem[\protect\citeauthoryear{{Londrillo} and {Del Zanna}}{2000}]{ldz00}
{Londrillo}, P. and L. {Del Zanna}: 2000, `{High-Order Upwind Schemes for
  Multidimensional Magnetohydrodynamics}'.
\newblock {\em \apj} {\bf 530}, 508--524.

\bibitem[\protect\citeauthoryear{{Lucy}}{1977}]{lucy77}
{Lucy}, L.~B.: 1977, `{A numerical approach to the testing of the fission
  hypothesis}'.
\newblock {\em \aj} {\bf 82}, 1013--1024.

\bibitem[\protect\citeauthoryear{{Lunttila} et~al.}{2009}]{lunttilaetal09}
{Lunttila}, T., P. {Padoan}, M. {Juvela}, and {\AA}. {Nordlund}: 2009, `{The
  Super-Alfv{\'e}nic Model of Molecular Clouds: Predictions for Mass-to-Flux
  and Turbulent-to-Magnetic Energy Ratios}'.
\newblock {\em \apjl} {\bf 702}, L37--L41.

\bibitem[\protect\citeauthoryear{{Lynden-Bell}}{1996}]{lynden-bell96}
{Lynden-Bell}, D.: 1996, `{Magnetic collimation by accretion discs of quasars
  and stars}'.
\newblock {\em \mnras} {\bf 279}, 389--401.

\bibitem[\protect\citeauthoryear{{Lynden-Bell}}{2003}]{lynden-bell03}
{Lynden-Bell}, D.: 2003, `{On why discs generate magnetic towers and collimate
  jets}'.
\newblock {\em \mnras} {\bf 341}, 1360--1372.

\bibitem[\protect\citeauthoryear{{Mac Low} and {Klessen}}{2004}]{mk04}
{Mac Low}, M.-M. and R.~S. {Klessen}: 2004, `{Control of star formation by
  supersonic turbulence}'.
\newblock {\em Rev. Mod. Phys.} {\bf 76}, 125--194.

\bibitem[\protect\citeauthoryear{{Marder}}{1987}]{marder87}
{Marder}, B.: 1987, `{A Method for Incorporating Gauss' Law into
  Electromagnetic PIC Codes}'.
\newblock {\em J. Comput. Phys.} {\bf 68}, 48--55.

\bibitem[\protect\citeauthoryear{{McKee} and {Ostriker}}{2007}]{mo07}
{McKee}, C.~F. and E.~C. {Ostriker}: 2007, `{Theory of Star Formation}'.
\newblock {\em \araa} {\bf 45}, 565--687.

\bibitem[\protect\citeauthoryear{{McNally} et~al.}{2012}]{mlp12}
{McNally}, C.~P., W. {Lyra}, and J.-C. {Passy}: 2012, `{A Well-posed
  Kelvin-Helmholtz Instability Test and Comparison}'.
\newblock {\em \apjs} {\bf 201}, 18.

\bibitem[\protect\citeauthoryear{{Meglicki} et~al.}{1995}]{mwd95}
{Meglicki}, Z., D. {Wickramasinghe}, and R.~L. {Dewar}: 1995, `{Gravitational
  collapse of a magnetized vortex: application to the Galactic Centre}'.
\newblock {\em \mnras} {\bf 272}, 717--729.

\bibitem[\protect\citeauthoryear{{Meru} and {Bate}}{2012}]{mb12}
{Meru}, F. and M.~R. {Bate}: 2012, `{On the convergence of the critical cooling
  time-scale for the fragmentation of self-gravitating discs}'.
\newblock {\em \mnras} {\bf 427}, 2022--2046.

\bibitem[\protect\citeauthoryear{{Mignone} and {Tzeferacos}}{2010}]{mt10}
{Mignone}, A. and P. {Tzeferacos}: 2010, `{A second-order unsplit Godunov
  scheme for cell-centered MHD: The CTU-GLM scheme}'.
\newblock {\em J. Comput. Phys.} {\bf 229}, 2117--2138.

\bibitem[\protect\citeauthoryear{{Mocz} et~al.}{2014}]{mvh14}
{Mocz}, P., M. {Vogelsberger}, and L. {Hernquist}: 2014, `{A constrained
  transport scheme for MHD on unstructured static and moving meshes}'.
\newblock {\em \mnras} {\bf 442}, 43--55.

\bibitem[\protect\citeauthoryear{{Molina} et~al.}{2012}]{molinaetal12}
{Molina}, F.~Z., S.~C.~O. {Glover}, C. {Federrath}, and R.~S. {Klessen}: 2012,
  `{The density variance-Mach number relation in supersonic turbulence - I.
  Isothermal, magnetized gas}'.
\newblock {\em \mnras} {\bf 423}, 2680--2689.

\bibitem[\protect\citeauthoryear{{Monaghan}}{1985}]{monaghan85}
{Monaghan}, J.~J.: 1985, `{Extrapolating B. Splines for Interpolation}'.
\newblock {\em J. Comput. Phys.} {\bf 60}, 253--262.

\bibitem[\protect\citeauthoryear{{Monaghan}}{1989}]{monaghan89}
{Monaghan}, J.~J.: 1989, `{On the problem of penetration in particle methods}'.
\newblock {\em J. Comput. Phys.} {\bf 82}, 1--15.

\bibitem[\protect\citeauthoryear{{Monaghan}}{1994}]{monaghan94}
{Monaghan}, J.~J.: 1994, `{Simulating Free Surface Flows with SPH}'.
\newblock {\em J. Comput. Phys.} {\bf 110}, 399--406.

\bibitem[\protect\citeauthoryear{{Monaghan}}{1997}]{monaghan97}
{Monaghan}, J.~J.: 1997, `{SPH and Riemann Solvers}'.
\newblock {\em J. Comput. Phys.} {\bf 136}, 298--307.

\bibitem[\protect\citeauthoryear{{Monaghan}}{2002}]{monaghan02}
{Monaghan}, J.~J.: 2002, `{SPH compressible turbulence}'.
\newblock {\em \mnras} {\bf 335}, 843--852.

\bibitem[\protect\citeauthoryear{{Monaghan}}{2005}]{monaghan05}
{Monaghan}, J.~J.: 2005, `{Smoothed particle hydrodynamics}'.
\newblock {\em Rep. Prog. Phys.} {\bf 68}, 1703--1759.

\bibitem[\protect\citeauthoryear{{Monaghan} and {Gingold}}{1983}]{mg83}
{Monaghan}, J.~J. and R.~A. {Gingold}: 1983, `{Shock Simulation by the Particle
  Method SPH}'.
\newblock {\em J. Comput. Phys.} {\bf 52}, 374.

\bibitem[\protect\citeauthoryear{{Monaghan} and {Lattanzio}}{1985}]{ml85}
{Monaghan}, J.~J. and J.~C. {Lattanzio}: 1985, `{A refined particle method for
  astrophysical problems}'.
\newblock {\em \aap} {\bf 149}, 135--143.

\bibitem[\protect\citeauthoryear{{Morris}}{1996}]{morris96}
{Morris}, J.~P.: 1996, `{Analysis of Smoothed Particle Hydrodynamics with
  Applications}'.
\newblock Ph.D. thesis, Monash University.

\bibitem[\protect\citeauthoryear{{Morris} and {Monaghan}}{1997}]{mm97}
{Morris}, J.~P. and J.~J. {Monaghan}: 1997, `{A Switch to Reduce SPH
  Viscosity}'.
\newblock {\em J. Comput. Phys.} {\bf 136}, 41--50.

\bibitem[\protect\citeauthoryear{{Murray}}{1996}]{murray96}
{Murray}, J.~R.: 1996, `{SPH simulations of tidally unstable accretion discs in
  cataclysmic variables}'.
\newblock {\em \mnras} {\bf 279}, 402--414.

\bibitem[\protect\citeauthoryear{{Nakamura} and {Li}}{2008}]{nl08}
{Nakamura}, F. and Z.-Y. {Li}: 2008, `{Magnetically Regulated Star Formation in
  Three Dimensions: The Case of the Taurus Molecular Cloud Complex}'.
\newblock {\em \apj} {\bf 687}, 354--375.

\bibitem[\protect\citeauthoryear{{Nordlund} and {Padoan}}{1999}]{np99}
{Nordlund}, {\AA}.~K. and P. {Padoan}: 1999, `{The Density PDFs of Supersonic
  Random Flows}'.
\newblock In: J. {Franco} and A. {Carraminana} (eds.): {\em Interstellar
  Turbulence}. p. 218.

\bibitem[\protect\citeauthoryear{{Orszag} and {Tang}}{1979}]{ot79}
{Orszag}, S.~A. and C.-M. {Tang}: 1979, `{Small-scale structure of
  two-dimensional magnetohydrodynamic turbulence}'.
\newblock {\em J. Fluid Mech.} {\bf 90}, 129--143.

\bibitem[\protect\citeauthoryear{{Padoan} and {Nordlund}}{2011}]{pn11}
{Padoan}, P. and {\AA}. {Nordlund}: 2011, `{The Star Formation Rate of
  Supersonic Magnetohydrodynamic Turbulence}'.
\newblock {\em \apj} {\bf 730}, 40.

\bibitem[\protect\citeauthoryear{{Padoan} et~al.}{1997}]{pnj97}
{Padoan}, P., A. {Nordlund}, and B.~J.~T. {Jones}: 1997, `{The universality of
  the stellar initial mass function}'.
\newblock {\em \mnras} {\bf 288}, 145--152.

\bibitem[\protect\citeauthoryear{{Pakmor} et~al.}{2011}]{pbs11}
{Pakmor}, R., A. {Bauer}, and V. {Springel}: 2011, `{Magnetohydrodynamics on an
  unstructured moving grid}'.
\newblock {\em \mnras} {\bf 418}, 1392--1401.

\bibitem[\protect\citeauthoryear{{Pan} and {Scannapieco}}{2010}]{ps10}
{Pan}, L. and E. {Scannapieco}: 2010, `{Mixing in Supersonic Turbulence}'.
\newblock {\em \apj} {\bf 721}, 1765--1782.

\bibitem[\protect\citeauthoryear{{Papoulis}}{1984}]{papoulis84}
{Papoulis}, A.: 1984, {\em {Probability, random variables and stochastic
  processes}}, McGraw-Hill Series in Electrical Engineering.
\newblock McGraw-Hill.

\bibitem[\protect\citeauthoryear{{Parkin} and {Bicknell}}{2013}]{pb13}
{Parkin}, E.~R. and G.~V. {Bicknell}: 2013, `{Global simulations of
  magnetorotational turbulence - I. Convergence and the quasi-steady state}'.
\newblock {\em \mnras} {\bf 435}, 2281--2298.

\bibitem[\protect\citeauthoryear{{Passot} and
  {V{\'a}zquez-Semadeni}}{1998}]{pvs98}
{Passot}, T. and E. {V{\'a}zquez-Semadeni}: 1998, `{Density probability
  distribution in one-dimensional polytropic gas dynamics}'.
\newblock {\em \pre} {\bf 58}, 4501--4510.

\bibitem[\protect\citeauthoryear{{Peretto} et~al.}{2012}]{perettoetal12}
{Peretto}, N., P. {Andr{\'e}}, V. {K{\"o}nyves}, N. {Schneider}, D.
  {Arzoumanian}, P. {Palmeirim}, P. {Didelon}, M. {Attard}, J.~P. {Bernard}, J.
  {Di Francesco}, D. {Elia}, M. {Hennemann}, T. {Hill}, J. {Kirk}, A.
  {Men'shchikov}, F. {Motte}, Q. {Nguyen Luong}, H. {Roussel}, T. {Sousbie}, L.
  {Testi}, D. {Ward-Thompson}, G.~J. {White}, and A. {Zavagno}: 2012, `{The
  Pipe Nebula as seen with Herschel: formation of filamentary structures by
  large-scale compression?}'.
\newblock {\em \aap} {\bf 541}, A63.

\bibitem[\protect\citeauthoryear{{Phillips}}{1986a}]{phillips86a}
{Phillips}, G.~J.: 1986a, `{Three-dimensional numerical simulations of
  collapsing, isothermal magnetic gas clouds}'.
\newblock {\em \mnras} {\bf 221}, 571--587.

\bibitem[\protect\citeauthoryear{{Phillips}}{1986b}]{phillips86b}
{Phillips}, G.~J.: 1986b, `{Three-dimensional numerical simulations of
  collapsing isothermal magnetic gas clouds - Non-uniform initial fields}'.
\newblock {\em \mnras} {\bf 222}, 111--119.

\bibitem[\protect\citeauthoryear{{Phillips} and {Monaghan}}{1985}]{pm85}
{Phillips}, G.~J. and J.~J. {Monaghan}: 1985, `{A numerical method for
  three-dimensional simulations of collapsing, isothermal, magnetic gas
  clouds}'.
\newblock {\em \mnras} {\bf 216}, 883--895.

\bibitem[\protect\citeauthoryear{{Powell}}{1994}]{powell94}
{Powell}, K.~G.: 1994, `{Approximate Riemann solver for magnetohydrodynamics
  (that works in more than one dimension)}'.
\newblock Technical Report ICASE 94-24, NASA Langley Research center (available
  from http://www.sti.nasa.gov).

\bibitem[\protect\citeauthoryear{{Powell} et~al.}{1999}]{powelletal99}
{Powell}, K.~G., P.~L. {Roe}, T.~J. {Linde}, T.~I. {Gombosi}, and D.~L. {De
  Zeeuw}: 1999, `{A Solution-Adaptive Upwind Scheme for Ideal
  Magnetohydrodynamics}'.
\newblock {\em J. Comput. Phys.} {\bf 154}, 284--309.

\bibitem[\protect\citeauthoryear{{Preto} and {Tremaine}}{1999}]{pt99}
{Preto}, M. and S. {Tremaine}: 1999, `{A Class of Symplectic Integrators with
  Adaptive Time Step for Separable Hamiltonian Systems}'.
\newblock {\em \aj} {\bf 118}, 2532--2541.

\bibitem[\protect\citeauthoryear{{Price} et~al.}{2013}]{pbt13}
{Price}, D., M. {Bate}, and T. {Tricco}: 2013, `{Collapse to stellar densities
  with radiation and magnetic fields: Outflows from the first and second
  hydrostatic cores}'.
\newblock In: {\em Protostars and Planets VI Posters}. p.~96.

\bibitem[\protect\citeauthoryear{{Price}}{2004}]{price04}
{Price}, D.~J.: 2004, `{Magnetic fields in Astrophysics}'.
\newblock Ph.D. thesis, Institute of Astronomy, Madingley Rd, Cambridge, CB2
  0HA, UK.

\bibitem[\protect\citeauthoryear{{Price}}{2008}]{price08}
{Price}, D.~J.: 2008, `{Modelling discontinuities and Kelvin-Helmholtz
  instabilities in SPH}'.
\newblock {\em J. Comput. Phys.} {\bf 227}, 10040--10057.

\bibitem[\protect\citeauthoryear{{Price}}{2010}]{price10}
{Price}, D.~J.: 2010, `{Smoothed Particle Magnetohydrodynamics - IV. Using the
  vector potential}'.
\newblock {\em \mnras} {\bf 401}, 1475--1499.

\bibitem[\protect\citeauthoryear{{Price}}{2012}]{price12}
{Price}, D.~J.: 2012, `{Smoothed Particle Hydrodynamics and
  Magnetohydrodynamics}'.
\newblock {\em J. Comput. Phys.} {\bf 231}, 759--794.

\bibitem[\protect\citeauthoryear{{Price} and {Bate}}{2007}]{pb07}
{Price}, D.~J. and M.~R. {Bate}: 2007, `{The impact of magnetic fields on
  single and binary star formation}'.
\newblock {\em \mnras} {\bf 377}, 77--90.

\bibitem[\protect\citeauthoryear{{Price} and {Bate}}{2008}]{pb08}
{Price}, D.~J. and M.~R. {Bate}: 2008, `{The effect of magnetic fields on star
  cluster formation}'.
\newblock {\em \mnras} {\bf 385}, 1820--1834.

\bibitem[\protect\citeauthoryear{{Price} and {Bate}}{2009}]{pb09}
{Price}, D.~J. and M.~R. {Bate}: 2009, `{Inefficient star formation: the
  combined effects of magnetic fields and radiative feedback}'.
\newblock {\em \mnras} {\bf 398}, 33--46.

\bibitem[\protect\citeauthoryear{{Price} and {Federrath}}{2010a}]{pf10}
{Price}, D.~J. and C. {Federrath}: 2010a, `{A comparison between grid and
  particle methods on the statistics of driven, supersonic, isothermal
  turbulence}'.
\newblock {\em \mnras} {\bf 406}, 1659--1674.

\bibitem[\protect\citeauthoryear{{Price} and {Federrath}}{2010b}]{pf10b}
{Price}, D.~J. and C. {Federrath}: 2010b, `{Smoothed Particle Hydrodynamics:
  Turbulence and MHD}'.
\newblock In: {N.~V.~Pogorelov, E.~Audit, \& G.~P.~Zank} (ed.): {\em Numerical
  Modeling of Space Plasma Flows, Astronum-2009}, Vol. 429 of {\em ASP Conf.
  Ser.} p. 274.

\bibitem[\protect\citeauthoryear{{Price} et~al.}{2011}]{pfb11}
{Price}, D.~J., C. {Federrath}, and C.~M. {Brunt}: 2011, `{The Density
  Variance-Mach Number Relation in Supersonic, Isothermal Turbulence}'.
\newblock {\em \apjl} {\bf 727}, L21.

\bibitem[\protect\citeauthoryear{{Price} and {Monaghan}}{2004a}]{pm04a}
{Price}, D.~J. and J.~J. {Monaghan}: 2004a, `{Smoothed Particle
  Magnetohydrodynamics - I. Algorithm and tests in one dimension}'.
\newblock {\em \mnras} {\bf 348}, 123--138.

\bibitem[\protect\citeauthoryear{{Price} and {Monaghan}}{2004b}]{pm04b}
{Price}, D.~J. and J.~J. {Monaghan}: 2004b, `{Smoothed Particle
  Magnetohydrodynamics - II. Variational principles and variable
  smoothing-length terms}'.
\newblock {\em \mnras} {\bf 348}, 139--152.

\bibitem[\protect\citeauthoryear{{Price} and {Monaghan}}{2005}]{pm05}
{Price}, D.~J. and J.~J. {Monaghan}: 2005, `{Smoothed Particle
  Magnetohydrodynamics - III. Multidimensional tests and the $\nabla\cdot{\bf
  B}= 0$ constraint}'.
\newblock {\em \mnras} {\bf 364}, 384--406.

\bibitem[\protect\citeauthoryear{{Price} and {Monaghan}}{2007}]{pm07}
{Price}, D.~J. and J.~J. {Monaghan}: 2007, `{An energy-conserving formalism for
  adaptive gravitational force softening in smoothed particle hydrodynamics and
  N-body codes}'.
\newblock {\em \mnras} {\bf 374}, 1347--1358.

\bibitem[\protect\citeauthoryear{{Price} and {Rosswog}}{2006}]{pr06}
{Price}, D.~J. and S. {Rosswog}: 2006, `{Producing Ultrastrong Magnetic Fields
  in Neutron Star Mergers}'.
\newblock {\em Science} {\bf 312}, 719--722.

\bibitem[\protect\citeauthoryear{{Price} et~al.}{2012}]{ptb12}
{Price}, D.~J., T.~S. {Tricco}, and M.~R. {Bate}: 2012, `{Collimated jets from
  the first core}'.
\newblock {\em \mnras} {\bf 423}, L45--L49.

\bibitem[\protect\citeauthoryear{{Quinn} et~al.}{1997}]{quinnetal97}
{Quinn}, T., N. {Katz}, J. {Stadel}, and G. {Lake}: 1997, `{Time stepping
  N-body simulations}'.
\newblock {\em ArXiv Astrophysics e-prints}.

\bibitem[\protect\citeauthoryear{{Read} and {Hayfield}}{2012}]{rh12}
{Read}, J.~I. and T. {Hayfield}: 2012, `{SPHS: smoothed particle hydrodynamics
  with a higher order dissipation switch}'.
\newblock {\em \mnras} {\bf 422}, 3037--3055.

\bibitem[\protect\citeauthoryear{{Read} et~al.}{2010}]{rha10}
{Read}, J.~I., T. {Hayfield}, and O. {Agertz}: 2010, `{Resolving mixing in
  smoothed particle hydrodynamics}'.
\newblock {\em \mnras} {\bf 405}, 1513--1530.

\bibitem[\protect\citeauthoryear{{Ritchie} and {Thomas}}{2002}]{rt02}
{Ritchie}, B.~W. and P.~A. {Thomas}: 2002, `{Hydrodynamic simulations of
  merging clusters of galaxies}'.
\newblock {\em \mnras} {\bf 329}, 675--688.

\bibitem[\protect\citeauthoryear{{Robertson} et~al.}{2010}]{robertsonetal10}
{Robertson}, B.~E., A.~V. {Kravtsov}, N.~Y. {Gnedin}, T. {Abel}, and D.~H.
  {Rudd}: 2010, `{Computational Eulerian hydrodynamics and Galilean
  invariance}'.
\newblock {\em \mnras} {\bf 401}, 2463--2476.

\bibitem[\protect\citeauthoryear{{Rosswog} and {Price}}{2007}]{rp07}
{Rosswog}, S. and D. {Price}: 2007, `{MAGMA: a three-dimensional, Lagrangian
  magnetohydrodynamics code for merger applications}'.
\newblock {\em \mnras} {\bf 379}, 915--931.

\bibitem[\protect\citeauthoryear{{Ryu} and {Jones}}{1995}]{rj95}
{Ryu}, D. and T.~W. {Jones}: 1995, `{Numerical magetohydrodynamics in
  astrophysics: Algorithm and tests for one-dimensional flow`}'.
\newblock {\em \apj} {\bf 442}, 228--258.

\bibitem[\protect\citeauthoryear{{Saitoh} and {Makino}}{2009}]{sm09}
{Saitoh}, T.~R. and J. {Makino}: 2009, `{A Necessary Condition for Individual
  Time Steps in SPH Simulations}'.
\newblock {\em \apjl} {\bf 697}, L99--L102.

\bibitem[\protect\citeauthoryear{{Saitoh} and {Makino}}{2013}]{sm13}
{Saitoh}, T.~R. and J. {Makino}: 2013, `{A Density-independent Formulation of
  Smoothed Particle Hydrodynamics}'.
\newblock {\em \apj} {\bf 768}, 44.

\bibitem[\protect\citeauthoryear{{Schekochihin}
  et~al.}{2004a}]{schekochihinetal04a}
{Schekochihin}, A.~A., S.~C. {Cowley}, J.~L. {Maron}, and J.~C. {McWilliams}:
  2004a, `{Critical Magnetic Prandtl Number for Small-Scale Dynamo}'.
\newblock {\em \prl} {\bf 92}(5), 054502.

\bibitem[\protect\citeauthoryear{{Schekochihin}
  et~al.}{2004b}]{schekochihinetal04b}
{Schekochihin}, A.~A., S.~C. {Cowley}, J.~L. {Maron}, and J.~C. {McWilliams}:
  2004b, `{Self-Similar Turbulent Dynamo}'.
\newblock {\em \prl} {\bf 92}(6), 064501.

\bibitem[\protect\citeauthoryear{{Schleicher} et~al.}{2013}]{schleicheretal13}
{Schleicher}, D.~R.~G., J. {Schober}, C. {Federrath}, S. {Bovino}, and W.
  {Schmidt}: 2013, `{The small-scale dynamo: breaking universality at high Mach
  numbers}'.
\newblock {\em New J. Phys.} {\bf 15}(2), 023017.

\bibitem[\protect\citeauthoryear{{Schmidt} et~al.}{2009}]{schmidtetal09}
{Schmidt}, W., C. {Federrath}, M. {Hupp}, S. {Kern}, and J.~C. {Niemeyer}:
  2009, `{Numerical simulations of compressively driven interstellar
  turbulence. I. Isothermal gas}'.
\newblock {\em \aap} {\bf 494}, 127--145.

\bibitem[\protect\citeauthoryear{{Schober} et~al.}{2012a}]{schoberetal12b}
{Schober}, J., D. {Schleicher}, S. {Bovino}, and R.~S. {Klessen}: 2012a,
  `{Small-scale dynamo at low magnetic Prandtl numbers}'.
\newblock {\em \pre} {\bf 86}(6), 066412.

\bibitem[\protect\citeauthoryear{{Schober} et~al.}{2012b}]{schoberetal12a}
{Schober}, J., D. {Schleicher}, C. {Federrath}, R. {Klessen}, and R.
  {Banerjee}: 2012b, `{Magnetic field amplification by small-scale dynamo
  action: Dependence on turbulence models and Reynolds and Prandtl numbers}'.
\newblock {\em \pre} {\bf 85}(2), 026303.

\bibitem[\protect\citeauthoryear{Sch{\"o}nberg}{1946}]{schoenberg46}
Sch{\"o}nberg, I.~J.: 1946, `Contributions to the problem of approximation of
  equidistant data by analytic functions'.
\newblock {\em Quart. Appl. Math} {\bf 4}(2), 45--99.

\bibitem[\protect\citeauthoryear{{Shao} and {Lo}}{2003}]{sl03}
{Shao}, S. and E.~Y.~M. {Lo}: 2003, `{Incompressible SPH method for simulating
  Newtonian and non-Newtonian flows with a free surface}'.
\newblock {\em Adv. in Water Resour.} {\bf 26}, 787--800.

\bibitem[\protect\citeauthoryear{{Shen} et~al.}{2010}]{sws10}
{Shen}, S., J. {Wadsley}, and G. {Stinson}: 2010, `{The enrichment of the
  intergalactic medium with adiabatic feedback - I. Metal cooling and metal
  diffusion}'.
\newblock {\em \mnras} {\bf 407}, 1581--1596.

\bibitem[\protect\citeauthoryear{{Sijacki} et~al.}{2012}]{sijackietal12}
{Sijacki}, D., M. {Vogelsberger}, D. {Kere{\v s}}, V. {Springel}, and L.
  {Hernquist}: 2012, `{Moving mesh cosmology: the hydrodynamics of galaxy
  formation}'.
\newblock {\em \mnras} {\bf 424}, 2999--3027.

\bibitem[\protect\citeauthoryear{{Smagorinsky}}{1963}]{smagorinsky63}
{Smagorinsky}, J.: 1963, `{General Circulation Experiments with the Primitive
  Equations}'.
\newblock {\em Mon. Weather Rev.} {\bf 91}, 99.

\bibitem[\protect\citeauthoryear{{Sod}}{1978}]{sod78}
{Sod}, G.~A.: 1978, `{A survey of several finite difference methods for systems
  of nonlinear hyperbolic conservation laws}'.
\newblock {\em J. Comput. Phys.} {\bf 27}, 1--31.

\bibitem[\protect\citeauthoryear{{Springel}}{2005}]{gadget2}
{Springel}, V.: 2005, `{The cosmological simulation code GADGET-2}'.
\newblock {\em \mnras} {\bf 364}, 1105--1134.

\bibitem[\protect\citeauthoryear{{Springel} and {Hernquist}}{2002}]{sh02}
{Springel}, V. and L. {Hernquist}: 2002, `{Cosmological smoothed particle
  hydrodynamics simulations: the entropy equation}'.
\newblock {\em \mnras} {\bf 333}, 649--664.

\bibitem[\protect\citeauthoryear{{Sridhar} and {Goldreich}}{1994}]{sg94}
{Sridhar}, S. and P. {Goldreich}: 1994, `{Toward a theory of interstellar
  turbulence. 1: Weak Alfvenic turbulence}'.
\newblock {\em \apj} {\bf 432}, 612--621.

\bibitem[\protect\citeauthoryear{{Stasyszyn} et~al.}{2013}]{sdb13}
{Stasyszyn}, F.~A., K. {Dolag}, and A.~M. {Beck}: 2013, `{A divergence-cleaning
  scheme for cosmological SPMHD simulations}'.
\newblock {\em \mnras} {\bf 428}, 13--27.

\bibitem[\protect\citeauthoryear{{Stone} et~al.}{2008}]{athena}
{Stone}, J.~M., T.~A. {Gardiner}, P. {Teuben}, J.~F. {Hawley}, and J.~B.
  {Simon}: 2008, `{Athena: A New Code for Astrophysical MHD}'.
\newblock {\em \apjs} {\bf 178}, 137--177.

\bibitem[\protect\citeauthoryear{{Tasker} et~al.}{2008}]{taskeretal08}
{Tasker}, E.~J., R. {Brunino}, N.~L. {Mitchell}, D. {Michielsen}, S. {Hopton},
  F.~R. {Pearce}, G.~L. {Bryan}, and T. {Theuns}: 2008, `{A test suite for
  quantitative comparison of hydrodynamic codes in astrophysics}'.
\newblock {\em \mnras} {\bf 390}, 1267--1281.

\bibitem[\protect\citeauthoryear{{T{\'o}th}}{2000}]{toth00}
{T{\'o}th}, G.: 2000, `{The $\nabla$.B=0 Constraint in Shock-Capturing
  Magnetohydrodynamics Codes}'.
\newblock {\em J. Comput. Phys.} {\bf 161}, 605--652.

\bibitem[\protect\citeauthoryear{{Tricco} et~al.}{2013a}]{tpf13}
{Tricco}, T., D. {Price}, and C. {Federrath}: 2013a, `{Rapid Growth of ISM
  Magnetic Fields via a Fast Turbulent Dynamo}'.
\newblock In: {\em Protostars and Planets VI Posters}. p.~58.

\bibitem[\protect\citeauthoryear{{Tricco} and {Price}}{2012a}]{tp12}
{Tricco}, T.~S. and D.~J. {Price}: 2012a, `{Constrained hyperbolic divergence
  cleaning for smoothed particle magnetohydrodynamics}'.
\newblock {\em J. Comput. Phys.} {\bf 231}, 7214--7236.

\bibitem[\protect\citeauthoryear{{Tricco} and {Price}}{2012b}]{tp12b}
{Tricco}, T.~S. and D.~J. {Price}: 2012b, `{Hyperbolic Divergence Cleaning for
  SPH}'.
\newblock {\em Proceedings of the 7th International SPHERIC Workshop}.
\newblock {arXiv:1207.5291}.

\bibitem[\protect\citeauthoryear{{Tricco} and {Price}}{2013a}]{tp13b}
{Tricco}, T.~S. and D.~J. {Price}: 2013a, `{A Switch for Artificial Resistivity
  and Other Dissipation Terms}'.
\newblock {\em Proceedings of the 8th International SPHERIC Workshop}.
\newblock {arXiv:1310.4260}.

\bibitem[\protect\citeauthoryear{{Tricco} and {Price}}{2013b}]{tp13}
{Tricco}, T.~S. and D.~J. {Price}: 2013b, `{A switch to reduce resistivity in
  smoothed particle magnetohydrodynamics}'.
\newblock {\em \mnras} {\bf 436}, 2810--2817.

\bibitem[\protect\citeauthoryear{{Tricco} et~al.}{2014}]{tpf14}
{Tricco}, T.~S., D.~J. {Price}, and C. {Federrath}: 2014, `{A comparison
  between grid and particle methods on small-scale dynamo amplification of
  magnetic fields in supersonic turbulence}'.
\newblock {\em Submitted to ApJ}.

\bibitem[\protect\citeauthoryear{{Tricco} et~al.}{2013b}]{tpfb13}
{Tricco}, T.~S., D.~J. {Price}, C. {Federrath}, and M.~R. {Bate}: 2013b,
  `{Smoothed Particle Magnetohydrodynamics Simulations of Protostellar Jets and
  Turbulent Dynamos}'.
\newblock {\em Proceedings of Magnetic Fields in the Universe IV}.
\newblock {arXiv:1310.4567}.

\bibitem[\protect\citeauthoryear{{Troland} and {Crutcher}}{2008}]{tc08}
{Troland}, T.~H. and R.~M. {Crutcher}: 2008, `{Magnetic Fields in Dark Cloud
  Cores: Arecibo OH Zeeman Observations}'.
\newblock {\em \apj} {\bf 680}, 457--465.

\bibitem[\protect\citeauthoryear{{Valcke} et~al.}{2010}]{valckeetal10}
{Valcke}, S., S. {de Rijcke}, E. {R{\"o}diger}, and H. {Dejonghe}: 2010,
  `{Kelvin-Helmholtz instabilities in smoothed particle hydrodynamics}'.
\newblock {\em \mnras} {\bf 408}, 71--86.

\bibitem[\protect\citeauthoryear{{Valdarnini}}{2012}]{valdarnini12}
{Valdarnini}, R.: 2012, `{Hydrodynamic capabilities of an SPH code
  incorporating an artificial conductivity term with a gravity-based signal
  velocity}'.
\newblock {\em \aap} {\bf 546}, A45.

\bibitem[\protect\citeauthoryear{{Vanaverbeke} et~al.}{2014}]{vkp14}
{Vanaverbeke}, S., R. {Keppens}, and S. {Poedts}: 2014, `{GRADSPMHD: A parallel
  MHD code based on the SPH formalism}'.
\newblock {\em Comput. Phys. Commun.} {\bf 185}, 1053--1073.

\bibitem[\protect\citeauthoryear{{Vazquez-Semadeni}}{1994}]{vs94}
{Vazquez-Semadeni}, E.: 1994, `{Hierarchical Structure in Nearly Pressureless
  Flows as a Consequence of Self-similar Statistics}'.
\newblock {\em \apj} {\bf 423}, 681.

\bibitem[\protect\citeauthoryear{{Waagan} et~al.}{2011}]{wfk11}
{Waagan}, K., C. {Federrath}, and C. {Klingenberg}: 2011, `{A robust numerical
  scheme for highly compressible magnetohydrodynamics: Nonlinear stability,
  implementation and tests}'.
\newblock {\em J. Comput. Phys.} {\bf 230}, 3331--3351.

\bibitem[\protect\citeauthoryear{{Wadsley} et~al.}{2008}]{wvc08}
{Wadsley}, J.~W., G. {Veeravalli}, and H.~M.~P. {Couchman}: 2008, `{On the
  treatment of entropy mixing in numerical cosmology}'.
\newblock {\em \mnras} {\bf 387}, 427--438.

\bibitem[\protect\citeauthoryear{{Wang} and {Abel}}{2009}]{wa09}
{Wang}, P. and T. {Abel}: 2009, `{Magnetohydrodynamic Simulations of Disk
  Galaxy Formation: The Magnetization of the Cold and Warm Medium}'.
\newblock {\em \apj} {\bf 696}, 96--109.

\bibitem[\protect\citeauthoryear{{Widrow} et~al.}{2012}]{widrowetal12}
{Widrow}, L.~M., D. {Ryu}, D.~R.~G. {Schleicher}, K. {Subramanian}, C.~G.
  {Tsagas}, and R.~A. {Treumann}: 2012, `{The First Magnetic Fields}'.
\newblock {\em \ssr} {\bf 166}, 37--70.

\bibitem[\protect\citeauthoryear{{Wurster} et~al.}{2014}]{wpa14}
{Wurster}, J., D. {Price}, and B. {Ayliffe}: 2014, `{Ambipolar diffusion in
  smoothed particle magnetohydrodynamics}'.
\newblock {\em \mnras} {\bf 444}, 1104--1112.

\bibitem[\protect\citeauthoryear{{Xu} et~al.}{2009}]{xsl09}
{Xu}, R., P. {Stansby}, and D. {Laurence}: 2009, `{Accuracy and stability in
  incompressible SPH (ISPH) based on the projection method and a new
  approach}'.
\newblock {\em J. Comput. Phys.} {\bf 228}, 6703--6725.

\end{thebibliography}
